\documentclass[useAMS]{mn2e}
\usepackage{psfig}
 \usepackage{url}
\usepackage{times}
\def\spose#1{\hbox to 0pt{#1\hss}}
\newcommand\lsim{\mathrel{\spose{\lower 3pt\hbox{$\mathchar"218$}}
     \raise 2.0pt\hbox{$\mathchar"13C$}}}
\newcommand\gsim{\mathrel{\spose{\lower 3pt\hbox{$\mathchar"218$}}
     \raise 2.0pt\hbox{$\mathchar"13E$}}}
\def\ltsima{$\; \buildrel < \over \sim \;$}
\def\lsim{\lower.5ex\hbox{\ltsima}}
\def\gtsima{$\; \buildrel > \over \sim \;$}
\def\gsim{\lower.5ex\hbox{\gtsima}}

\def\sch{Schwarzschild}

\title[Fermi/LAT broad emission line blazars]
{Fermi/LAT broad emission line blazars}
\author[G. Ghisellini \& Tavecchio]
{G. Ghisellini \thanks{E--mail: gabriele.ghisellini@brera.inaf.it}, F. Tavecchio\\
 INAF -- Osservatorio Astronomico di Brera, via E. Bianchi 46, I--23807 Merate, Italy \\
}

\begin{document}

\pagerange{\pageref{firstpage}--\pageref{lastpage}} \pubyear{2012}

\maketitle
\label{firstpage}

\begin{abstract}
We study the broad emission line blazars detected in the 
$\gamma$--ray band by the Large Area Telescope onboard the {\it Fermi}
satellite and with the optical spectrum studied by Shaw et al. (2012, 2013).
The observed broad line strength provides a measure of the ionizing luminosity of 
the accretion disk, while the $\gamma$--luminosity is a proxy for the 
bolometric non--thermal beamed jet emission.
The resulting sample, composed by 217 blazars, is the best suited to study 
the connection between  accretion and  jet properties.
We compare the broad emission line properties of these blazars
with those of radio--quiet and radio--loud quasars present in the
Sloan Digital Sky Survey, to asses differences and similarities
of the disc luminosity and the virial black hole mass.
For most sources, we could derive the black hole mass by reproducing the 
IR--optical--UV data with a standard accretion disc spectrum, and
we compared the black hole masses derived with the two methods.
The distributions of the masses estimated in the two ways agree satisfactorily.
We then apply a simple, one--zone, leptonic model to all the 217 objects 
of our sample. 
The knowledge of the black hole mass and disc luminosity helps to constrain the jet parameters.
On average they are similar to what found by previous studies
of smaller samples of sources.
\end{abstract}


\section{Introduction}

Blazars are extragalactic radio--loud sources whose jet is pointing toward us
(for recent reviews, see e.g.
B\"ottcher 2007; Ghisellini 2011; Dermer 2014).
To be more precise, we may define blazars as the sources whose jet axis is at an angle
$\sin \theta_{\rm v} \le 1/\Gamma$ with respect to the line of sight, where $\Gamma$ is the bulk Lorentz factor.
This implies that for each observed blazars, there are other $2\Gamma^2$ intrinsically
identical sources, but pointing in other directions.
Blazars are classically divided in two subclasses:
Flat Spectrum Radio Quasars (FSRQs) and BL Lacs.
The usual divide between the two subclasses is based on the Equivalent width (EW) of the optical
broad emission lines: BL Lacs have EW$<$5\AA\ (rest frame, Urry \& Padovani 1995;
see Ghisellini et al. 2011 and Sbarrato et al. 2012 for an alternative definition).

Blazars are strong $\gamma$--ray emitters as a class.
After the detection of 3C 273 by the {\it COS B} satellite (Swanenburg et al. 1978), 
the {\it Compton Gamma Ray Observatory}
discovered nearly 100 blazars (Nandikotkur et al. 2007).
Now, in the era of the {\it Fermi} satellite, nearly 1,000 blazars (2LAC; Ackermann et al. 2011)
have been detected in the 0.1--100 GeV energy range.
This number is sufficiently large to start population studies 
(e.g. Ajello et al. 2012; 2014), to derive
the average properties of these sources, and to compare them to other
radio--loud -- but not necessarily so extreme -- sources and to the
general population of radio--quiet quasars.
To this aim, we can benefit from the study of Shaw et al. (2012)
concerning FSRQs detected by {\it Fermi} and systematically observed spectroscopically.
To these sources, we added the few BL Lacs in Shaw et al. (2013) that do 
show broad emission lines.
Together, they form an homogeneous sample of blazars detected in $\gamma$--rays.
Due to the uniform all sky exposure of the {\it Fermi}/LAT instrument,
these blazars form a $\gamma$--ray flux limited sample of blazars with broad 
emission lines. 
We first compare their optical properties with the radio--loud and the radio--quiet
quasars of the compilation made by Shen et al. (2011),
comprising about 105,000 quasars of the Sloan Digital Sky Survey (SDSS).

Then we collect multiwavelength data for all blazars, and apply a simple one zone
leptonic model to find the physical parameters of the sources.
With respect to our earlier studies of blazars, there are four important additions: 
i) the knowledge, for all sources, of the luminosity of the broad emission lines;
ii) the knowledge of the black hole mass derived through the virial method
(these two information are given in the study of Shaw et al. 2012); 
iii) the high frequency radio data given by the {\it WMAP} and {\it Planck} satellite observations;
iv) the  far--IR data given by the {\it WISE} satellite, that detects almost all sources.
These data at low frequency allow to determine the synchrotron peak of powerful blazars much
better than previously.

We make use of the known broad emission line
luminosity to derive the luminosity emitted by the accretion disc.
By assuming a standard Shakura \& Sunyaev (1973) disc, whose emitted spectrum depends
only on the black hole mass and accretion rate (or equivalently, on the emitted
luminosity), we are able to infer the black hole mass for 
the majority of FSRQs, and
compare this value with the mass derived with the virial method
(Peterson \& Wandel 2000; McLure \& Dunlop 2004, Peterson 2014).
Two important parameters of the model (black hole mass and disc luminosity)
are therefore determined by the observations of the emission line luminosity
and width and by disc fitting.
When the spectral energy distribution (SED) are sufficiently sampled, 
(i.e. such that the two spectral peaks of the non--thermal emission are robustly determined)
the parameters of the model 
well constrained, although some extra assumptions are needed
to univocally specify them.

Since blazars are rapidly variable, and often with large amplitude,
one should use simultaneous multiwavelength data, which in our case
is impossible, given the very large number of sources.
Therefore the results of each individual source can be affected by the
non--simultaneity of data. 
On the other hand we argue that, in a statistical sense, the derived 
distributions of the parameters are reliable.
Furthermore, consider that the accretion disc should vary with timescale longer
than the non--thermal jet continuum, and with smaller amplitudes.
Results on the black hole mass and the accretion luminosity should therefore 
be valid also for individual sources.

The paper is organized as follows: in\S 2 we present the samples 
of {\it Fermi}/LAT detected blazars and the comparison sample
of radio--loud and radio--quiet quasars.
In \S 3 we discuss the general observational properties of the blazars
of our sample in comparison with radio--loud and radio--quiet 
quasars (namely the redshift and broad emission line strength
distributions).
We also describe how we derive the black hole masses using
the disc fitting method, and compare our results with the
ones derived through the virial method.
In \S 4 we describe the blazar model we use to derive the physical
parameters of the jet emitting region, and discuss
how these parameters can be univocally determined by
knowing the disc luminosity and the black hole mass,
and in \S 5 we present the corresponding results.
In \S 6 we draw our conclusions.
The results concerning the jet power and
its relation to the accretion luminosity have been already
discussed in Ghisellini et al. (2014).

In this work, we adopt a flat cosmology with $H_0=70$ km s$^{-1}$ Mpc$^{-1}$ and
$\Omega_{\rm M}=0.3$. 

\section{The samples}

We selected our sources from the FSRQs of Shaw et al. (2012) and
the BL Lacs of Shaw et al. (2013).
We compare them with the radio--quiet and radio--loud quasars of
Shen et al. (2011).
Tab. \ref{numbers} shows a break down of the samples.

\subsection{FSRQs in Shaw et al. (2012)}

All our {\it Fermi}/LAT FSRQs come from the sample of Shaw et al. (2012,
hereafter S12).
The sample includes 229 objects, present in the 1LAC sample
of Abdo  et al. (2010), that have been 
spectroscopically observed by Shen et al. (2012; 165 sources) or 
by the SDSS Data Release 7 (DR7, Shen et al. 2011, 64 objects).
Of these, we have studied the 191 objects with enough multiwavelength information 
necessary to apply our model.
The S12 sample of FSRQs does not include several bright and famous blazars
with historical spectroscopic classifications in the literature.

\subsection{BL Lacs in Shaw et al. (2013)}

Shaw et al. (2013, hereafter S13) studied a very large sample of
BL Lac objects present on the second {\it Fermi} catalog of AGN
(2LAC, Ackermann et al. 2011).
In the original 2LAC sample, there are 410 BL Lacs, 357 FSRQs and 28 AGN of
other known types, and 326 AGN of unknown type.  
S13 themselves were able to classify some of the several ``unknown
type" AGNs present in the 2LAC catalog, increasing the number of BL Lacs
to 475, and decreasing the number of the sources of unknown type to 215.
By spectroscopically observing a large number of BL Lacs, and by
adding BL Lacs of already known redshift, they assembled a sample of 209
BL Lac with the redshift spectroscopically measured.
In addition, they could constrain the redshift of other 241 BL Lacs
(finding a lower limit on the redshift, see S13 for details and
their Tab. 1 for a break down on the number of known redshift for source type).
By visually inspecting all SED of the BL Lacs with redshifts, we
selected the 26 objects with a clear presence of broad emission lines.
Although they are classified as BL Lacs according to the classical
definition (rest frame equivalent width of the emission line less than 5 \AA),
they should be rather considered to belong to the tail, at low accretion luminosities, of FSRQs,
and for this reason we include them in our sample.

\subsection{Quasars in Shen et al. (2011)}

Shen et al. (2011, heafter S11) studied a large number
of quasars selected from the SDSS (Schneider et al. 2010) 
according to the following criteria:
i) the (rest frame) FWHM of the broad lines greater than 1,000 km s$^{-1}$, and
ii) the absolute magnitude is brighter than $M_i$=--22.0
The quasars selected in this way are 105,783.
We then selected the region of the sky covered by the FIRST survey.
In this region there are 89,783 quasars that have been observed, 
but not detected by the FIRST (with 1 mJy flux limit at 1.4 GHz).
For brevity, we call these sources ``radio--quiet".
Moreover, there are 9,393 radio--detected quasars.
Of these, there are 8,257 radio--loud sources, i.e.
objects with a radio--loudness $R_{\rm L}>10$.
The radio--loudness is defined as in S11, i.e.
the ratio of the rest  frame 5 GHz flux and the 2500 \AA\ flux,
where the 5 GHz flux is extrapolated from the observed 1.4 GHz flux
(assuming a power law slope $F_\nu \propto \nu^{-0.5}$).

\begin{table} 
\centering
\begin{tabular}{lr}
\hline
\hline
Sample                  &Number     \\
\hline 
SDSS RQ+RL (S11)                   &105,783 \\
SDSS+FIRST RQ (S11)                &89,783    \\
SDSS+FIRST Radio det. (S11)        &9,393      \\
SDSS+FIRST RL ($R_{\rm L}>$10)     &8,257       \\
FSRQ (S12) total                   &229        \\
FSRQ (S12) studied in this paper   &191        \\
BL Lacs (S13) total                 &475 \\
BL Lacs (S13) with $z$              &209       \\
BL Lacs (S13) studied in this paper &26         \\
Total blazars studied in this paper &217 \\
\hline
\hline 
\end{tabular}
\vskip 0.1 true cm
\caption{
Number of sources in the different samples. 
Note that the objects studied by S12 and S13
are all detected in $\gamma$--rays by {\it Fermi}/LAT.
}
\label{numbers}
\end{table}

\begin{figure}
\vskip -0.6 cm
\hskip -0.4cm
\psfig{file=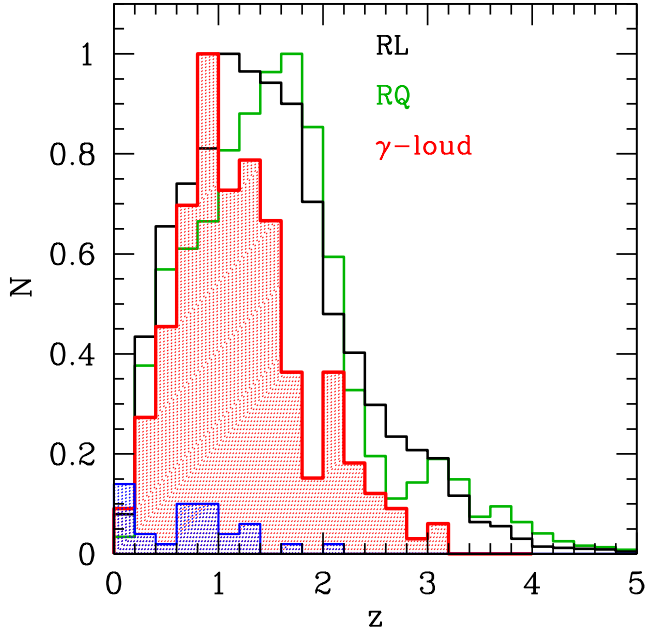,height=10.cm,width=9.5cm}
\vskip -0.6 cm
\caption{Redshift distribution of the FSRQs in our sample (red hatched), compared with 
radio--loud sources ($R_{\rm L}>10$, black) and radio--quiet (green)
quasars of S11. 
The distributions are ``normalized" (i.e. the number of the sources
at the peak of the distribution has been set equal to one).
The blue hatched distribution is for the 26 BL Lacs in our sample.
$\gamma$--loud FSRQs have a slightly smaller redshift of both radio--loud and radio--quiet
quasars, that share a similar redshift distribution.
} 
\label{iz}
\end{figure}

\section{General observed properties}

\subsection{Redshift distribution}

In Fig. \ref{iz} the redshift distribution of the FSRQs 
in our sample is compared to those of radio--loud sources 
(radio--loudness $R>10$) and radio--quiet quasars.
These distribution are normalized (namely, the number
of the sources at the peak of the distribution has been set to one).
We also show the redshift distribution for the 26 BL Lacs in our sample.
It can be seen that the {\it Fermi}/LAT blazars have, on average, smaller redshifts.
This is likely due to the still high sensitivity threshold of
the $\gamma$--ray flux: to enter the $\gamma$--ray catalog, the typical blazar must 
be closer than a critical redshift. 

\begin{figure}
\vskip -0.6 cm
\hskip -0.4cm
\psfig{file=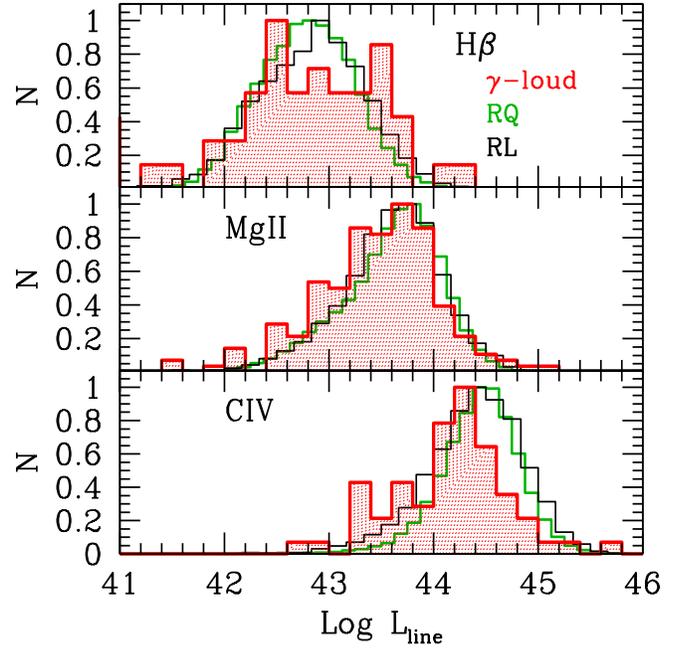,height=10.cm,width=9.5cm}
\vskip -0.4 cm
\caption{Distribution of the luminosities of the three most prominent
broad emission lines of the blazars in our sample (red hatched), compared with radio--loud
blazars (radio--loudness $R_{\rm L}>10$, black) and radio--quiet (green)
quasars of the S11 sample.
The distributions are almost the same.
Note that the trend of increasing luminosity (from the H$\beta$ to the CIV luminosity 
distribution) is an effect of the increasing average redshift.
 } 
\label{ilines}
\end{figure}

\subsection{Broad line luminosity distributions}

The three more prominent broad emission lines considered
by S12 in FSRQs are H$\beta$, MgII and CIV.
S12 provide the Full Width
Half Maximum (FWHM) and the luminosity of these lines.
According to the redshift range, one can find in the optical spectrum 
only one or two of these lines, but never all three together.
The method of line fitting is very similar to the one used by S11
for analyzing his sample of quasars (both radio--loud and radio--quiet).
The comparison between the line luminosity distributions
of the S12 $\gamma$--ray blazars and the entire S11 population
of quasars is shown in S12 (their Fig. 2).
We show in Fig. \ref{ilines} how the line luminosity distributions
of $\gamma$--loud blazars compare with radio--loud and radio--quiet
quasars.
One can see that the H$\beta$ luminosity distributions are very similar,
while the MgII and especially the CIV luminosities for $\gamma$--loud blazars
tend to be slightly under--luminous. 
The line luminosity distribution of radio--loud and radio--quiet are
instead always similar. 
This reflects the different redshift distribution of the blazars in our sample, 
that extends to lower values than the quasars in the S11 sample (see Fig. \ref{iz}).

\begin{figure}
\vskip -0.6 cm
\hskip -0.4cm
\psfig{file=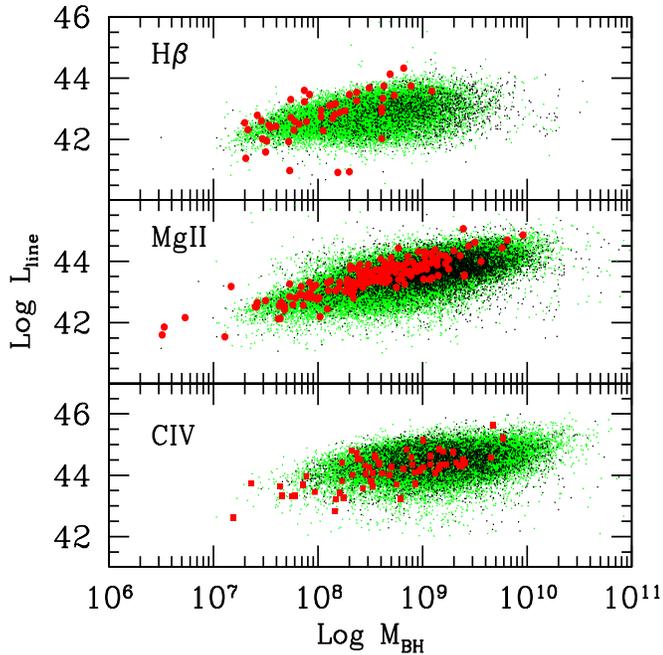,height=10.cm,width=9.5cm}
\vskip -0.4 cm
\caption{
Luminosities of the three main broad emission lines as a function
of the black hole mass estimated by S11 for the radio--quiet (green)
and radio--loud (black) AGN, and by S12 for the $\gamma$--ray blazars
(red circles).
Both studies base the black hole mass estimate on virial arguments,
using the FWHM of the lines and the radius of the BLR estimated through 
the radius--ionizing luminosity correlation. 
} 
\label{mlines}
\end{figure}

An estimate of the black hole mass based on the virial method exists for all FSRQs in
our sample  (see \S \ref{bhm}).
All values are reported in Tab. \ref{sample}, together with the black hole mass
we have derived by fitting the optical--UV continuum with a disc spectrum (see \S\ref{bhm}).
Both S11 and S12 use the same virial method. 
In Fig. \ref{mlines} we compare
the  line luminosities as a function of the virial mass.
When more than one broad emission line is
used to derive more than one value of the black hole mass,
we took the logarithmic average of the different values.
Green (black) dots are the radio--quiet (radio--loud) quasars,
red circles are the $\gamma$--loud blazars.
The $\gamma$--loud blazars tend to have smaller (virial) black hole masses
than the rest of the sources.
This effect was noted also by S12, that suggested a possible
selection effect: blazars are highly aligned sources, and if the broad line
region is not spherically symmetric, but flattened toward the disk (see e.g. Shen \& Ho 2014), we
should observe lines systematically narrower than observed in misaligned
objects (Decarli et al. 2011), leading to a smaller estimate of the black hole mass
(see \S \ref{bhm}).

\subsection{Black hole mass}
\label{bhm}

{\bf Virial mass ---} The virial method to estimate the black hole mass is 
now the most widely used, allowing to measure the black hole mass
of hundred thousand objects in a automatic or semi--automatic way.
However, the measurement, besides being based on the virial assumption
(the motion of the broad line clouds is governed by gravity)
is not direct, but is necessarily based on correlations
with their own dispersions that are not simply due to measurement errors.
The uncertainty associated to these estimates is large, of the order of 0.5--0.6 dex.
Vestergard et al. (2006) and Park et al. (2012) 
estimated that the black hole mass derived in this way 
has an uncertainty of a factor $\sim$3--4.
Besides this, there are two additional concerns:
i) the geometry of the broad line region can 
influence the observed FWHM of the lines.
Decarli et al. (2008) pointed out that a flattened BLR
(i.e. not spherical, but with clouds distributed closer
to the accretion disc) observed close to the normal of
the accretion disc shows lines of narrower FWHM.
ii) Marconi et al. (2008) noted
that if the accretion disc is emitting close to the
Eddington rates, one should account for the 
radiation pressure force exerted on the broad line clouds:
to be balanced, one needs more gravity, hence a bigger black hole mass.
Tab. \ref{sample} (in the Appendix) reports the black hole masses derived by Shaw et al. (2012)
through the virial method.

\begin{figure}
\vskip -0.6 cm
\psfig{file=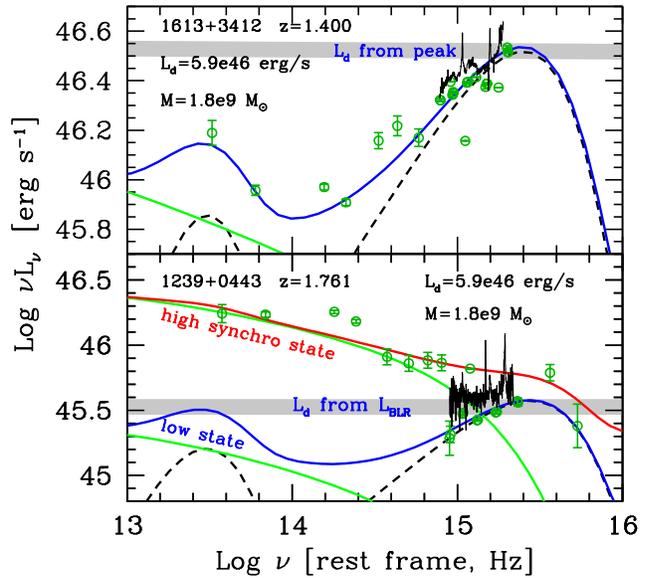,height=9.cm,width=9cm}
\vskip -0.5 cm
\caption{
Two examples for deriving the black hole mass.
In the top panel the disc emission is well defined, and $\dot M$ can be derived directly,
even without the knowledge of the broad line luminosities (that should in any 
case give consistent results).
Changing the black hole mass implies different peak frequencies, but same peak luminosity (grey stripe). 
The black hole mass is the one best accounting for the optical--UV data. 
The bottom panel shows that the jet emission can sometimes dominate the optical--UV
band, hiding the disc emission (and the observed optical spectrum, taken at a different time).
In this case the disc luminosity is found through the luminosity of the broad emission lines.
The black hole mass is found requiring that the jet+disc luminosity matches the
optical--UV data, especially in the low state.
} 
\label{zoom}
\end{figure}

\vskip 0.3 cm
\noindent
{\bf Disc--fitting mass ---} 
The spectrum of the emission produced by a standard, Shakura \& Sunyaev (1973) disc is
a superposition of black--body spectra with temperature
distribution $T(R)$, where $R$ is the distance form the black hole, 
depending only on the black hole mass $M$ and the accretion rate $\dot M$.
Assuming an efficiency $\eta$ defined by $L_{\rm d}=\eta \dot M c^2$,
where $L_{\rm d}$ is the bolometric disc luminosity, the observations
allow us to directly fix $\dot M$, in two possible ways (see also Calderone et al. 2013;
Castignani et al. 2013).
We could in fact directly observe the optical--UV hump produced
by the disc, corresponding to its maximum.
The peak of the $\nu L_\nu$ disc spectrum is $\sim L_{\rm d} /2$.
We can then infer $\dot M$ directly, once a value
for the efficiency $\eta$ is assumed (we here assume $\eta$= 0.08).

If the peak of the disc emission is not well sampled, because it lies 
outside the observable range or because it is ``contaminated" by 
the jet emission or by the host galaxy, we can infer $L_{\rm d}$
through the luminosity of the broad emission lines.
According to the template of Francis et al. (1991), setting
the relative weight of the Ly$\alpha$ luminosity equal to 100, we have that the 
weight of the luminosity of all broad lines  is 556, with
the broad hydrogen H$\alpha$, H$\beta$, MgII and CIV contributing 77, 22, 34 and 63,
respectively (see also Celotti, Padovani \& Ghisellini 1997; vanden Berk et al. 2001).
If more of one line is present, we take the logarithmic average of the
broad line region luminosity $L_{\rm BLR}$ derived by the single lines.
We then assume that $L_{\rm BLR}$ is a fixed fraction -- 10\% -- of $L_{\rm d}$.
This fixes $\dot M$. 
For a given $\dot M$ (i.e. for a given $L_{\rm d}$),
the black hole mass regulates the peak frequency of the 
disk emission (heavier black holes have larger \sch\ radii, and
thus colder discs).
In other words, the knowledge of $L_{\rm d}$ fixes the peak 
value of the $\nu L_\nu$ emission of the disc, but a change 
of the black hole mass corresponds to an horizontal shift
in a $\nu L_\nu$ plot.
Therefore, in principle, even one data point is enough to fix the 
black hole mass, if we are sure that it belong to the disc emission.
This of course is often questionable.
In some (16) cases, there is only one optical point to constrain
the black hole mass, with no indications of an upturn of the SED,
characteristic of the presence of the accretion disc emission.
These blazars are marked with an asterisk in the last column of 
Tab. \ref{sample}.
These are the values used for the jet model (see below).

The uncertainty on the resulting black hole mass therefore depends on
the quality of the data.
If the maximum of the disc spectrum is visible, the uncertainties are
less than a factor 2, better than the virial method.
Fig. \ref{zoom} illustrates two examples of how we derive the black hole mass
through the disc fitting method.
In the first case (top panel) the disc contribution is well defined,
and its luminosity  can be inferred directly. 
Sometimes it can disagree with $L_{\rm d}$ derived
from the broad line luminosity. 
In this case we prefer the value directly observed.
On the contrary, when the disc emission is diluted by
the jet flux (as in the bottom panel of Fig. \ref{zoom}), we set $L_{\rm d}=10L_{\rm BLR}$.
We then find $M$ by fitting the disc+jet emission to the data.
As long as there is some sign of emission disc flux 
(typically, an upturn at high optical--UV frequencies)
the estimate is reliable.
Less so when there is no sign of disc emission.
A broad limiting range is set by requiring that 
$10^{-2} L_{\rm Edd} \lsim L_{\rm d}\lsim L_{\rm Edd}$:
the lower limit is given by requiring that
the accretion disc is radiatively efficient, 
and so it can photo--ionize the BLR,  
while the upper limit requires the source to be sub--Eddington.
Within the corresponding range of masses, the accretion disk cannot overproduce
any existing data, and often this requirement narrows down the
black hole mass range.

\begin{figure}
\vskip -0.6 cm
\psfig{file=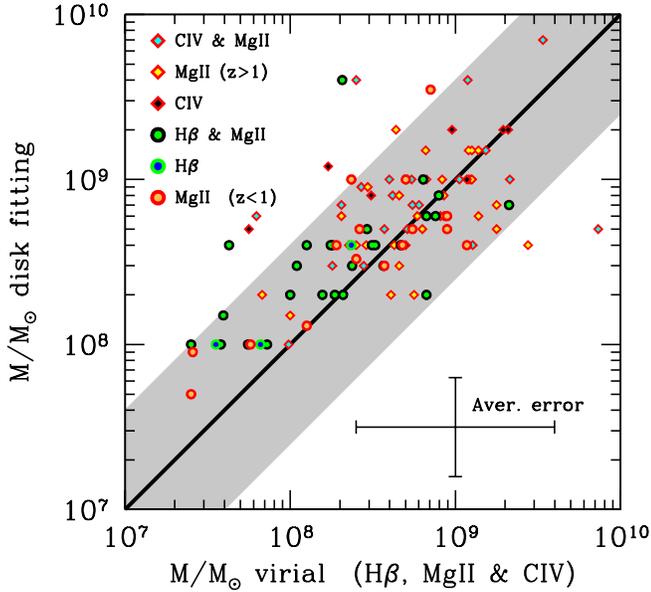,height=9.cm,width=9cm}
\vskip -0.5 cm
\caption{
Black hole mass estimated through the disk fitting method 
(in this paper) as a function of the black hole mass estimated with the virial method by S12, 
for all the FSRQs studied in this paper.
Only blazars with independent disc--fitting values
are included (i.e. we have excluded all blazars with values of the disc--fitting mass
indicated between parentheses in Tab. \ref{sample}, and all BL Lacs).
Different symbols correspond to the different lines used for the virial mass.
The diagonal line is the equality line.
The grey stripe indicates a factor 4 uncertainty on the virial mass. 
} 
\label{mm}
\end{figure}

\begin{figure}
\vskip -0.6 cm
\psfig{file=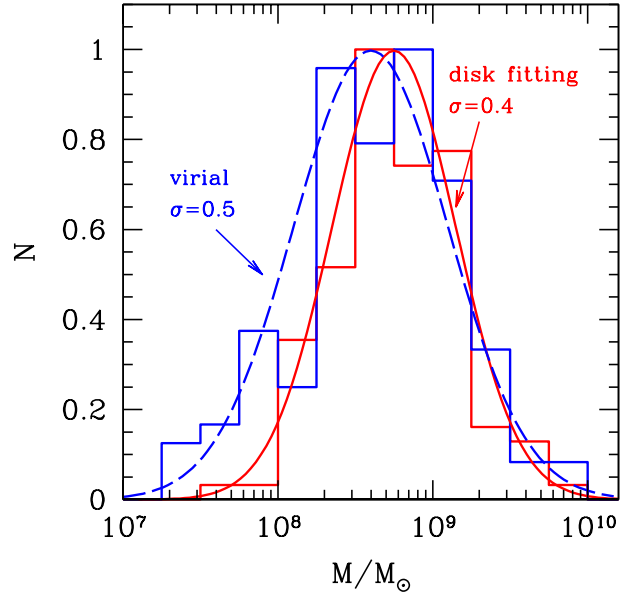,height=9.cm,width=9cm}
\vskip -0.5 cm
\caption{
Distribution of the black hole masses of $\gamma$--ray loud blazars
estimated through disk fitting (red hatched) and through the virial
method (blue hatched).  
We have excluded all blazars whose disc--fitting black hole mass is
given between parentheses in Tab. \ref{sample}, and all BL Lacs.
The masses estimated by the two methods, on average, 
differ only slightly (less than $\sim 0.2$ dex).
Both distributions are fitted with a log--normal function, whose width is indicated.
} 
\label{im}
\end{figure}

We admittedly use a rough simplification for the disc model, 
by using a Shakura \& Sunyaev (1973) disc.
In reality, it is likely that the black hole is spinning, perhaps
rapidly, and this implies a greater overall efficiency, since the 
last stable orbit moves inwards by increasing the black hole spin.
On the other hand the spectrum produced by a disc surrounding a Kerr hole
is different from the one of a Shakura--Sunyaev disk only at the highest frequencies (innermost
orbits), where it emits a greater luminosity.
Consider a standard Shakura--Sunyaev disc and a disc around a Kerr hole emitting the same $L_{\rm d}$.
The spectrum of the disc around the Kerr hole would peak at higher UV frequencies, and therefore 
it will be dimmer at IR frequencies.
To match those with the data, we will be obliged to increase the black hole
mass (i.e. to assume a large surface, hence a smaller temperature).
We thus conclude that it is possible that the mass derived here are slightly underestimated,
but by an amount which is smaller than the overall uncertainties (which are
around a factor 2--3).

Bearing in mind this caveat, we could apply the disc--fitting method 
to 116 FSRQs, while for the remaining 75 FSRQs the data were too poor,
or the synchrotron jet component too dominant.
Tab. \ref{sample} reports the value of the derived black hole mass.
The values in parenthesis could not be evaluated through the disc--fitting method. 
In most cases these values have been set equal to the virial masses (within a factor $\sim$2--3, 
denoted by ``--" in the last column). 
In a minority of cases they differ from the 
virial values for the following reasons, as flagged in the last column: 
1) a value larger than the virial one has been adopted to avoid super Eddington or nearly Eddington disc luminosities;
2) a value smaller than the virial one has been adopted to avoid to overproduce the NIR--optical flux; 
3) a value larger than the virial one has been adopted to avoid to overproduce the optical--UV flux.

In the following we will consider, as black hole masses derived through the disc--fitting method, only
the 116 values for which we could independently derived the mass with this method.
Also, we do not consider the values derived for BL Lacs.

In Fig. \ref{mm} we compare the black hole mass derived through the disc fitting method
with the virial masses of the same objects.
There is a general agreement, with a large dispersion.
The horizontal width of the grey stripe indicates a factor 4 uncertainty for the
virial masses.
The virial masses are shown with different symbols according to the line
used, but there seems to be no systematic trend.

Fig. \ref{im} compares the distributions of the 116 values of the black hole masses 
derived by the two methods.
Fitting the (logarithmic) distributions with a log--normal, we derived 
$\sigma=0.4$ for the blazar masses derived with disc fitting, and 
$\sigma=0.5$ for the virial masses.
Also the average values differ, but slightly (less than $\sim$0.2 dex).

\begin{figure}
\vskip -0.6 cm
\hskip -0.4cm
\psfig{file=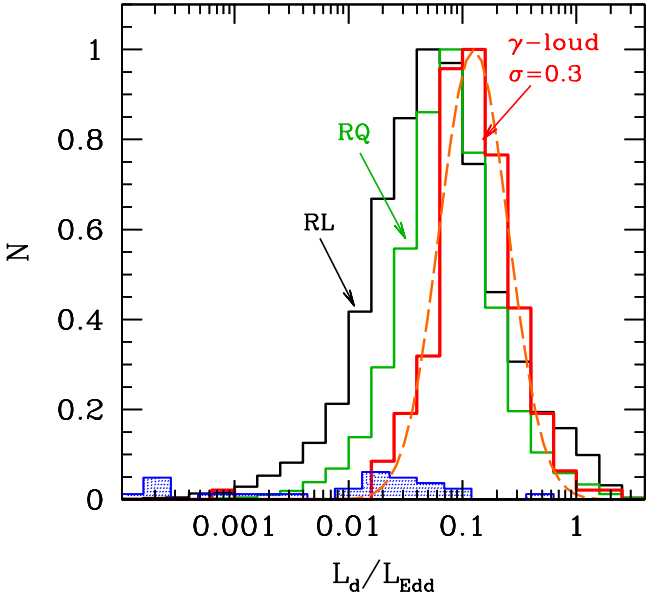,height=9.cm,width=9.5cm}
\vskip -0.4 cm
\caption{Normalized distribution of the disk luminosity on Eddington units for our blazars
(red), radio--loud (black line) and radio--quiet (green) AGN in S11.
For all sources the disc luminosity and the black hole mass are 
estimated through the line luminosity and the mass through the virial method
(see Fig. \ref{mm} for a comparison of the 
two estimates for the $\gamma$--ray loud blazars).
The blue hatched distribution is for the BL Lac objects studied in this paper.
} 
\label{ildedd}
\end{figure}


\subsection{Disc luminosity}

Fig. \ref{ildedd} shows the distribution of disc luminosities
measured in Eddington units for our blazars (normalized at
the peak, set equal to one), compared with radio--loud and radio--quiet
quasars of S11. 
The accretion disc luminosity $L_{\rm d}$ and the black hole mass $M$
of all objects, including the $\gamma$--ray loud blazars, 
have been found through the broad line luminosities
and the virial method, respectively.
The blazar distribution peaks at $L_{\rm d}/L_{\rm Edd}\sim 0.1$,
with a small dispersion.
Fitting this distribution with a log--normal function we derive $\sigma=0.3$,
with a tail towards small values made by BL Lac objects.
The distributions of the S11 radio--quiet and radio--loud objects
peak at a value slightly less than one tenth Eddington
and their  dispersions are both broader than the 
one of blazars.
This is likely due to the fact that $\langle L_{\rm d}/L_{\rm Edd}\rangle$
for all distributions is similar, but that radio--loud sources with discs
accreting below some value ($L_{\rm d}\lsim $0.1--0.03 $L_{\rm Edd}$) 
go undetected in the $\gamma$--ray band, even with the sensitivity of {\it Fermi}/LAT.

Usually, we set $L_{\rm d}$ equal to the value derived through the broad
emission lines.
There are a few cases when this value conflicts with existing data,
as made evident by the accretion disc spectrum we construct. 
In these cases we have changed $L_{\rm d}$, taking the value
derived through disc fitting.
Roughly 10\% of the sources have values of $L_{\rm d}$ 
calculated through the broad lines and through disc--fitting that disagree by
a factor between 2 and 5. 
This can be explained, since when we calculate $L_{\rm d}$ through the broad lines
we use a template that has a factor 2 (1$\sigma$) uncertainty (Calderone et al. 2013).
Therefore, statistically, $\sim$10\% of outliers are expected.


\section{The jet model}

To interpret the overall SED of our blazars, we 
use the one--zone leptonic model of Ghisellini \& Tavecchio (2009).
While we refer the reader to that paper for a full discussion
of the features of the model, we here emphasize that the
knowledge of the black hole mass and of the luminosity of the accretion disc 
considerably helps to constrain the free parameters of the model itself.
In the following we summarize the model parameters and the
observables that we can use to fix them.

\subsection{Parameters of the model}

\begin{itemize}

\item $R_{\rm diss}$: distance  of the emitting region from the black hole.
Since we assume a conical jet with fixed semi--aperture angle $\psi=0.1$,
this fixes the size of the emitting (assumed spherical) region $R=\psi R_{\rm diss}$.
 
\item $M$: black hole mass. 

\item $B$: magnetic field of the emitting region.

\item $\Gamma$, $\theta_{\rm v}$: bulk Lorentz factor and viewing angle.
With rare exceptions, we fix $\theta_{\rm v}\sim 3^\circ$ (i.e. of the
order of $1/\Gamma$).
The model assumes that $\Gamma\sim (R/3R_{\rm S})^{1/2}$ up to a final value, and
constant thereafter.
$R_{\rm S}$ is the \sch\ radius.

\item $P^\prime_{\rm e}$: injected power in relativistic electrons as measured in the jet frame.  It regulates
the jet bolometric luminosity, but not in a linear way, since the observed luminosity
depends upon $\Gamma$, $\theta_{\rm v}$, and the cooling regime.

\item $s_1$, $s_2$, $\gamma_{\rm min}$, $\gamma_{\rm b}$, $\gamma_{\rm max}$: 
slopes of the injected distribution of electrons (smoothly joining broken power law), 
and minimum, break and maximum random Lorentz factors of the SED.
The injected distribution is assumed to be:
\begin{equation}
Q(\gamma)  \, = \, Q_0\, { (\gamma/\gamma_{\rm b})^{-s_1} \over 1+
(\gamma/\gamma_{\rm b})^{-s_1+s_2} } \;\; \gamma_{\rm min} < \gamma < \gamma{\rm max}
\label{qgamma}
\end{equation}
The normalization $Q_0$ is set through $P^\prime_{\rm e}=(4\pi /3) R^3\int Q(\gamma)\gamma m_{\rm e}c^2 d\gamma$.
$\gamma_{\rm min}$ is set equal to one, while $\gamma_{\rm max}$, for $s_2>2$, is unimportant.
The slopes $s_1$ and $s_2$ are associated (by means of the continuity equation)
to the observed slopes before and after the two broad peaks of the SED.
Note that $\gamma_{\rm b}$ not necessarily coincides with $\gamma_{\rm peak}$ 
(the Lorentz factor of the electrons emitting at the peak of the SED),
that depends also on the cooling energy $\gamma_{\rm cool}$ and the slopes $s_1$and $s_2$.
For FSRQs, we almost always are in the fast cooling regime 
($\gamma_{\rm cool}<\gamma_{\rm b}$) and so $\gamma_{\rm peak}\sim \gamma_{\rm b}$.
The energy distribution of the density of particles is found through the continuity equation 
\begin{equation}
{\partial N(\gamma) \over \partial \gamma }  \, = {\partial\over \partial\gamma} 
\left[ \dot\gamma N(\gamma) \right]  +Q(\gamma) +P(\gamma)
\label{continuity}
\end{equation}
where $P(\gamma)$ corresponds to the electron--positron pair production rate and
$\dot\gamma$ is the synchrotron+inverse Compton cooling rate.
The injection rate is assumed constant and homogeneous throughout the source.
The particle distribution is calculated at the dynamical time $R/c$.
This allows to neglect adiabatic losses, and allows us to use a constant (in time)
magnetic field and volume.

\item $L_{\rm d}$: bolometric luminosity of the disc. 
A first estimate is given by the broad emission lines.
Whenever possible, we find the final value through disc--fitting.

\item $R_{\rm BLR}$: size of the BLR. We assume it scales as:
\begin{equation}
R_{\rm BLR} \, =\,  10^{17} L_{\rm d, 45}^{1/2} \,\,\,{\rm cm}
\end{equation}

\item $L_{\rm torus}$: luminosity reprocessed (and re--emitted in the infrared) 
by the molecular torus. 
$L_{\rm torus}/L_{\rm d}\sim 0.3$, with a rather narrow distribution (see e.g. Calderone et al. 2012). 

\item $R_{\rm torus}$: size of the torus. It is assumed to scale as:
\begin{equation}
R_{\rm torus}\, =\, 2\times 10^{18} L_{\rm d, 45}^{1/2} \,\,\,{\rm cm}
\end{equation}

\item $\nu_{\rm ext}$: typical frequency of the seed photons for the external Compton scattering.
This is different according if $R_{\rm diss}$ is inside the BLR or outside it, but inside the torus distance:
\begin{eqnarray}
\nu_{\rm ext}\, &=& \, \nu_{\rm Ly\alpha}=2.46\times 10^{15}\,\, {\rm Hz}, \quad {\rm BLR}
\nonumber \\
\nu_{\rm ext}\, &=&\, 7.7\times 10^{13}\,\, {\rm Hz},\qquad \quad \qquad{\rm torus}
\end{eqnarray}
Both contributions are taken into account.

\item $L_{\rm xc}$,  $\alpha_X$, $h \nu_{\rm c}$: 
luminosity, spectral index and cut--off energy of the spectrum of the accretion disk X--ray corona.
We always use $L_{\rm xc}(\nu)  \propto \nu^{-1}  \exp (- h\nu / 150\, {\rm keV} )$.
With rare exception, the total corona luminosity is assumed to be 30\% of $L_{\rm d}$.

\end{itemize}
 
Of these parameters, $\psi$, $\theta_{\rm v}$, $L_{\rm xc}$, $\alpha_X$, $h \nu_{\rm c}$, $\gamma_{\rm min}$ 
are kept fixed (with rare exceptions).
The exact value of $\gamma_{\rm max}$ is unimportant (for $s_2>2$).
$L_{\rm d}$ is found through direct fitting or through the broad line luminosities,
and fixes $R_{\rm BLR}$, $L_{\rm torus}$, $R_{\rm torus}$, $\nu_{\rm ext}$.
The black hole mass $M$ is found through the disc fitting method
(when possible), or from the virial method.

We are left with 7 relevant free parameters:
$R_{\rm diss}$ (or equivalently, $R$), 
$B$, $\Gamma$, $P^\prime_{\rm e}$, $s_1$, $s_2$, $\gamma_{\rm b}$.
The observables used to constrain these parameters are:

\begin{itemize}

\item $L_{\rm S}$, $L_{\rm C}$:
the synchrotron and the inverse Compton luminosity.

\item $\nu_{\rm S}$, $\nu_{\rm C}$: the synchrotron and the inverse Compton frequency peaks.

\item $\alpha_0$, $\alpha_1$: the spectral indices pre and post peak
(they can be different for the synchrotron and IC peak, according to the
relative importance of the SSC process and/or the importance of Klein--Nishina effects).

\item $t_{\rm var}$: the minimum variability timescales.

\end{itemize}
The radio is never fitted by these compact one--zone models, since
at the assumed scales the radio synchrotron emission is self--absorbed.
For very powerful FSRQ, the synchrotron absorption peak occurs in the mm band.
The radio data can be used as a consistency check when the spectral coverage
is poor in the mm--far IR, by assuming a flat (i.e. $F_\nu \propto \nu^0$)
spectrum joining the GHz region of the radio data to the mm band.

\subsection{Is the interpolating model unique?}

Since the model needs several input parameters, we can wonder whether
the fitting set of parameters is unique, or else there are multiple solutions.
The answer depends on the richness and quality of data.
Tavecchio et al. (1998) discussed the case of the synchrotron self Compton 
(SSC) model in one--zone leptonic models, concluding that 
in this case we have enough observables to determine all the free parameters
of the model.
In the case of a source for which the scattering with seed photons
produced externally to the jet (External Compton, EC for short) is
important, there are more parameters to deal with.
Anyway, for the sources studied in this paper,
we know the luminosity of the broad lines.
Therefore we can determine the radius of the BLR and
the molecular torus, thus the corresponding radiative energy densities
within $R_{\rm BLR}$ and $R_{\rm torus}$.
If the SED is sufficiently sampled, 
the variability timescales limits the source size
and the data are of good quality,
then we can univocally determine the entire parameter set.
In the following we discuss how the observables
are linked to the input parameters of the model.

\vskip 0.2 cm
\noindent
{\it Simultaneity of the data ---} 
Blazars are variable, so, in principle, the simultaneity of the data  should be a must.
But dealing with large samples of sources it is not possible
to have simultaneous data for all sources.
We {\it assume} that the existing data are a good representation
of a typical source state.
As mentioned, this is likely not true for any specific source, but
it is a reasonable assumption when treating several sources and if the aim
is to characterize the source population as a whole, and not the single source.

\vskip 0.2 cm
\noindent
{\it Location of the emission region ---}
In the framework of the model we use, at each location of
the emitting region ($R_{\rm diss}$), the energy densities of radiation and magnetic
field are well defined (see Figg. 2;  9; 13 and 14 
in Ghisellini \& Tavecchio 2009; see also Sikora et al. 2009).
The ratio of the two energy densities varies along the jet, and
so does the inverse Compton to synchrotron luminosity
ratio (called ``Compton dominance").
If the data show a well defined Compton dominance (larger then unity), the
possible values of $R_{\rm diss}$ are limited, usually, to
within $R_{\rm BLR}$ or outside it but within $R_{\rm torus}$.
The first case corresponds to a more compact region, that can
vary more rapidly, so it is preferred when there are indications
of fast variability.
The second case is preferred when $\nu_{\rm C}$ is particularly small:
even if $\nu_{\rm C}$ is rarely observed, it can be estimated by 
extrapolating the X--ray and the $\gamma$--ray spectrum.
Consider also that in our scheme the value of $\Gamma$ fixes a minimum distance
$R_{\rm diss}$, because of the dependence $\Gamma= (R/3 R_{\rm S})^{/1/2}$
(see the discussion in \S 4.3).
There are cases in which a small $\nu_{\rm C}$ implies a
small $\nu_{\rm ext}$, hence the prevalence of photons from the torus.
This is also accompanied by a small $\nu_{\rm S}$ (if it is not due 
to synchrotron self--absorption), due to the smaller magnetic field at larger distances. 
Marscher et al. (2008; 2010) and Sikora, Moderski, \& Madejski (2008)
proposed that the dissipation region is much farther
than what assumed here, at 10--20 pc. 
As discussed in Ghisellini et al. (2014), this possibility requires 
a much larger jet power, due to the inefficiency of the radiative
process, in turn due to the paucity of seed photons for the
inverse Compton scattering.

\vskip 0.2 cm
\noindent
{\it Importance of the SSC process ---} 
A very hard soft X--ray spectrum (i.e. $\alpha_X \lsim 0.5$)
indicates a prominent EC process, and that the SSC process
is unimportant.
The SSC luminosity depends on the square of the particle density and the magnetic energy density, 
while the EC process depends on the particle density and the external radiation field (in the jet frame).
For a distribution of relativistic electrons of density $n$ and mean square
energy $\langle \gamma^2 \rangle$ we define the Comptonization $y$ parameter as
\begin{equation}
y \, \equiv \, {4\over 3} \sigma_{\rm T} n \langle \gamma^2 \rangle 
\end{equation}
where $\sigma_{\rm T}$ is the Thomson cross section. 
In the comoving frame we have:
\begin{equation}
{L^\prime_{\rm SCC} \over L^\prime_{\rm EC}} \, = \, { y L_{\rm syn } \over L_{\rm EC} }\, =\, 
{ y  \int N(\gamma) \dot\gamma_{\rm syn} d\gamma  \over    \int N(\gamma) \dot\gamma_{\rm ext} d\gamma }
\, \sim \, y\,  { U^\prime_{\rm B} \over \Gamma^2 U_{\rm ext}}
\end{equation}
Therefore a small SSC luminosity implies either a small magnetic field or else  
a small $y$, in turn implying a  
small electron density (i.e. a small injected power enhanced by a large
Doppler boost).
The opposite of course is required when there are indications that the
SSC component contributes to the X--ray emission.
A rather clear indication is a relatively soft X--ray spectrum
(due to SSC) hardening at high energies (where the EC prevails).
This helps to robustly constrain the injected power and the magnetic field,
together with the bulk Lorentz factor.
This complex spectrum could be produced also by more than one emitting region:
the more compact one, perhaps inside the BLR, could be responsible
for the hard X--ray spectrum, while a larger region
possibly outside the BLR and/or the torus, could produce
X--rays mainly through the SSC process.
To discriminate, we have to observe if hard and soft X--rays
vary together or not, but keeping in mind that some
different variability can be introduced by the different
cooling timescales of electrons of different energies, 
even if they radiate by the same process.

\vskip 0.2 cm
\noindent
{\it Peak frequencies ---}
The synchrotron ($\nu_{\rm S}$), SSC ($\nu_{\rm SSC}$), and EC ($\nu_{\rm EC}$) observed peak frequencies are:
\begin{eqnarray}
\nu_{\rm S}  \,  =& {4\over 3}\, \gamma^2_{\rm peak} \, \nu_{\rm B} \, {\delta \over 1+z}  ~~~  \nonumber \\
\nu_{\rm SSC} \, =& {4\over 3}\, \gamma^2_{\rm peak} \, \nu_{\rm S}   ~~~~~~   ~~~~~    \nonumber  \\ 
\nu_{\rm EC}  \, =& {4\over 3}\, \gamma^2_{\rm peak} \, \nu_{\rm ext} \,{\Gamma \delta \over 1+z}
\label{nu}
\end{eqnarray}
where $\nu_{\rm B}\equiv eB/(2\pi m_{\rm e}c)=2.8$ MHz.
In the comoving frame, any monochromatic line is seen as coming from a narrow distribution of angles,
all within $1/\Gamma$.
Photons coming exactly from the forward direction are blueshifted by $2\Gamma$, photons
coming from an angle $1/\Gamma$ are blueshifted by $\Gamma$.
In this narrow range of frequencies, the observed spectrum follows $F^\prime_{\nu^\prime}\propto (\nu^\prime)^2$,
that well approximates a blackbody spectrum (this is what we use, see Tavecchio \& Ghisellini 2008).
For EC dominated sources the ratio $\nu_{\rm S}/\nu_{\rm EC}$ gives:
\begin{equation}
{B \over \Gamma} \, =\,  {\nu_{\rm S} \nu_{\rm ext} \over 2.8\times 10^6 \nu_{\rm EC}}
\end{equation}
The peak EC frequency gives (setting $\Gamma=\delta$):
\begin{equation}
\Gamma \gamma_{\rm peak} \, = \, \left[ {3\over 4} \, {\nu_{\rm EC}\over \nu_{\rm ext} } (1+z) \right]^{1/2}
\end{equation}
%

\vskip 0.2 cm
\noindent
{\it Spectral slopes ---}
The particle distribution resulting from the continuity equation and the assumed
$Q(\gamma)$ injection is a smoothly broken power law, with possible deviations
caused by the energy dependence of the Klein--Nishina cross section (when important)
and the production of electron--positron pairs (that are accounted for;
see also Dermer et al. 2009 for a detailed treatment of these processes).
For our blazars, these effects are marginal.
There are in any case several effects that can shape the observed spectrum
in addition to the slopes of the particle distribution:
\begin{itemize}

\item {\it Synchrotron self absorption.}
The synchrotron spectrum, in high power blazars, self--absorbs in the 
observed range 100--1000 GHz. 
Often the self--absorption frequency $\nu_{\rm a}$ can be greater than $\nu_{\rm S}$,
making the ``real" peak synchrotron frequency unobservable, as well
as the ``real" peak synchrotron flux.
In this cases the low frequency spectral index cannot be determined by the
synchrotron spectrum, nor the ``real" $L_{\rm S}$.

\item {\it Incomplete cooling.}
For illustration, consider $R_{\rm diss}<R_{\rm BLR}$.
The particle distribution is harder below $\gamma_{\rm cool}$
(which is a few in many cases), and  this implies a hard to soft
break at 
\begin{equation}
\nu_{\rm break}\, \sim \, \gamma^2_{\rm cool} { \Gamma\delta \nu_{\rm Ly\alpha}  \over 1+z }\, 
\sim \, 13 \,  {\Gamma \delta \gamma_{\rm cool}^2 \over 50 } \, {2\over 1+z} \,\,\,{\rm keV}
\end{equation}
This break can be used to estimate $\gamma_{\rm cool}$. 
This is particularly important if we want to calculate the total number
of emitting electrons, when calculating the jet power. 
The electron distribution, in fact, is $N(\gamma) \propto \gamma^{-2}$ between
$\gamma_{\rm cool}$ and $\gamma_{\rm b}$, and $N(\gamma)\propto \gamma^{-s_1}$ below
$\gamma_{\rm cool}$.
When $s_1 \le 1$, most of the electrons are at $\gamma_{\rm cool}$, 
that we can estimate.

\item{\it External photon starving.}
The number of seed photons available for up--scattering, in the rest frame of the source,
depends on frequency.
Assume again that $R_{\rm diss}<R_{\rm BLR}$.
The peak of the ``black--body" emission corresponding to the lines observed in the 
comoving frame is at $\nu^\prime \sim 2 \Gamma \nu_{\rm Ly\alpha} \sim 5\times 10^{16}(\Gamma/10)$ Hz.
Cold electrons (i.e. with $\gamma\sim 1$) scatter these photons leaving their frequency
unchanged.
These scattered photons are seen Doppler boosted by $\delta$ by the external observer:
\begin{equation}
\nu_{\rm obs} \, = \, {2 \delta \, \Gamma\,  \nu_{\rm Ly\alpha} \over 1+z}\, 
\sim \, 2 \, { (\delta\, \Gamma/100) \over 1+z }\,\, {\rm keV}
\end{equation}
Below this energy, the seed photons contributing to the X--ray spectrum have
smaller frequencies, and are less in number.
Therefore the scattered spectrum cannot use the full distribution of seed photons:
as a consequence the scattered spectrum becomes very hard (i.e. more and more
``photons starved" at lower and lower frequencies).

This effect is particularly important when using a single power law
as a model for spectral fitting the X--ray data.
Inevitably, since the real spectrum is intrinsically curved,
this leads to grossly overestimate the column densities $N_{\rm H}$, 
to account for the curvature as due to absorption (and particularly
so for high redshift sources) 
(see e.g. Tavecchio et al. 2007).

\item {\it EC+SSC.} 
As mentioned, the contribution of the SSC process can be
relevant in the soft X--ray band.
In general, the SSC spectrum is softer than the EC one at these energies.
Therefore the presence of an SSC component has an effect just opposite to 
what discussed above, making the spectrum softer.

\item {\it X--ray corona.} 
The effect on the total observed spectrum is to soften it
at low energies, since the assumed slope is softer than
both the EC and the SSC slopes.

\end{itemize}

The high energy slope of the particle distribution can be derived from the high energy tails 
(i.e. post--peak) both of the synchrotron and the Compton spectrum.
In general, being produced by the same electrons, the
two slopes must be equal.
This is indeed a consistency check that the one--zone model
can be applicable.
One must bear in mind, however, that in the IR--UV part
of the spectrum there can be the contribution of the torus
and disc emission.

\begin{figure}
\vskip -0.7 cm
\hskip -0.5 cm
\psfig{file=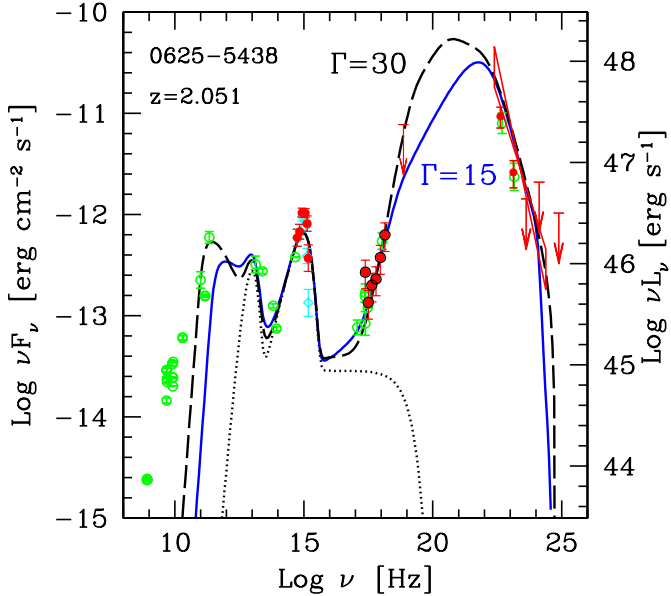,height=9.2cm,width=9.2cm}\\
\vskip -1 cm
\caption{The SED of 0625+5438 together with 2 different
fitting models, whose parameters are reported in Tab. \ref{0625para}.
The dotted line indicates the contribution of the accretion disc, the torus, 
and the X--ray corona.
Both the $\Gamma=15$ (solid line) and the $\Gamma=30$ (dashed) 
models describe the overall SED 
(except the radio fluxes, produced by a more extended region) satisfactorily well.
The main difference between them is the bulk Lorentz factor $\Gamma$,
the viewing angle $\theta_{\rm v}$ and the location of the emitting region $R_{\rm diss}$.
The zoom in Fig. \ref{0625zoom} shows that the $\Gamma=30$ model 
gives a worse representation of the X--ray data, but that it is still acceptable.
To discriminate between them we have to use other information besides the
pure SED, such as the minimum variability timescales and considerations
about the number density of the parent population, that must be more numerous 
for the $\Gamma=30$ model.
}
\label{0625test}
\end{figure} 
\begin{figure}
\vskip -0.7 cm
\hskip -0.4 cm
\psfig{file=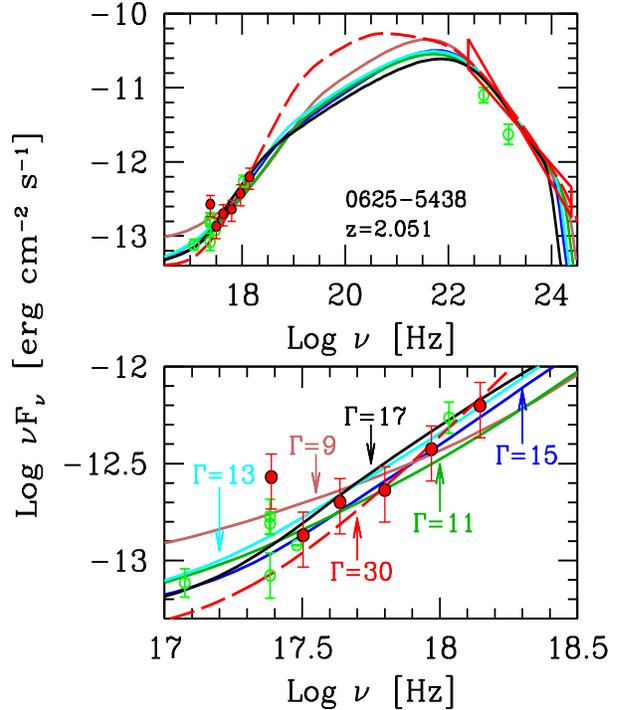,height=11cm,width=10cm} 
\vskip -0.7 cm
\caption{
Zoom on the X and $\gamma$--ray band (top panel) and on only the
X--ray data and models (bottom panel).
The different fitting models, whose parameters are reported in Tab. \ref{0625para}, 
correspond to different values of $\Gamma$, as labeled.
Values above $\Gamma=17$ requires that the dissipation region is beyond the BLR.
We can see that models with $\Gamma$ between 11 and 17 are acceptable,
but the models with $\Gamma=9$ and $\Gamma=30$ give a poor representation 
of the X--ray data.
}
\label{0625zoom}
\end{figure} 

\subsection{An illustrative example}

Fig. \ref{0625test} shows the SED of 0625--5438 ($z$=2.051), together with 
2 different models, whose parameters are listed in Tab. \ref{0625para}.
The $\Gamma=15$ model (solid blue line) is our preferred model, using $\Gamma=15$
and a dissipation region located inside the BLR.

We then investigate several model, with different values of $\Gamma$.
First we consider the case of $\Gamma=30$.
This corresponds to one of the largest values of the superluminal speeds
of the blazar components studied by Lister et al. (2013).
Since $\beta_{\rm app}$ is maximized for $\theta_{\rm v}=1/\Gamma$, 
we adopt this viewing angle, i.e. $\theta_{\rm v}=1.9^\circ$.
This implies $\delta=\Gamma$.
Since $\Gamma\propto R^{1/2}$, we increased $R_{\rm diss}$ by a factor 4 
(with respect to the $\Gamma=15$ model).
We kept the Poynting flux $P_{\rm B} \propto R^2\Gamma^2 B^2$ constant, and consequently decrease 
the magnetic field (by a factor 8).
A good fit is obtained decreasing the injected power by a factor 3,
to compensate for the increased Doppler boost.
Note that we do not change the spectral parameters ($\gamma_{\rm b}$,
$\gamma_{\rm max}$, $s_1$ and $s_2$ are kept constant).
The larger $R_{\rm diss}$ of the $\Gamma=30$ model implies that the seed external photons
are produced by the torus.
The external energy density in the comoving frame is reduced with respect to the $\Gamma=15$ model,
and this makes $\gamma_{\rm cool}$ larger.
Furthermore the Compton peak frequency $\nu_{\rm C}$ is now smaller
even if it is more blue--shifted, because $\nu_{\rm ext}$ is smaller.
Overall, 
this model fits the SED as well as the $\Gamma =15$ model.
The observed data therefore are not enough to discriminate between them.
Consider also that the jet powers are 
similar.
Therefore we cannot choose one of the two models on the basis of
the total power budget.

On the other hand the two models differ for the following properties:
\begin{itemize}

\item As shown on Fig. \ref{0625zoom}, the X--ray spectrum of the 
$\Gamma=30$ model is slightly harder than indicated by the data.

\item 
The variability timescale for the $\Gamma=30$ model is 2.5 times larger than for 
the $\Gamma=15$ model.

\item 
The number of blazars with large values of $\Gamma$ should be rare,
according to the distribution of superluminal speeds (see below and Fig. \ref{gamma}).

\item
Having large values of $\Gamma$ and very small values of $\theta_{\rm v}$ implies 
a large number of parent sources (sources pointing away from us).

\end{itemize}
These arguments are strong indications, but admittedly not solid proofs.
The knowledge gained so far on the variability timescales is based on the
few sources that are bright enough to allow a meaningful monitoring
of their flux variations.
Indications are that, indeed, timescales are short, of the order of hours
(see Ulrich \& Maraschi \& Urry 1997 for a review, and e.g. Bonnoli et al. 2011
for 3C 454.3 as a well studied specific source, and Nalewajko 2013 for a systematic study,
indicating a typical variability timescale in the {\it Fermi}/LAT band of $\approx$1 day).
Studies of the luminosity functions of blazars and their parent populations
allow for a distributions of $\Gamma$--factors consistent with the
distribution of superluminal speeds, but extreme values must be rare
(e.g. Padovani \& Urry 1992; 
see also Ajello et al. 2012 finding an average $\langle \Gamma\rangle \approx 11$).
We conclude that although our sources, lacking variability timescale
information, can be fitted with a large value of $\Gamma$ and 
a relatively large $R_{\rm diss}$, this cannot be the rule, but can
be the case in a limited number of sources, not affecting the average values
of the distributions of the parameters.

We now consider the predicted SED when changing $\Gamma$ from $\Gamma=9$ to $\Gamma=17$.
The corresponding SEDs are shown in Fig \ref{0625zoom}.
The used parameters are listed in Tab. \ref{0625para}.
All these models share the same jet dissipation location $R_{\rm diss}$ and viewing
angle $\theta_{\rm v}$, but the other parameters have been adjusted to obtain
the best representation of the data.
The top panel of Fig. \ref{0625zoom} shows the X to $\gamma$--ray SED,
while the bottom panel focusses on the X--ray band only.
Models with $\Gamma$ between 11 and 17 are all acceptable.  
The models with $\Gamma=9$ and $\Gamma=30$ do not represent well 
the slope of the X--ray data.
No model is able to reproduce the upturn given by the first (low energy)
X--ray spectrum.
Although all models with $11\le \Gamma\le 17$ are acceptable,
the one with $\Gamma=15$ seems to give the best representation of the X--ray data.
This is what led us to chose the $\Gamma=15$ model as the best one. 
We are aware that the non--simultaneity of the data can lead to a slightly wrong
choice of the parameters. 
Accounting for this would lead to a larger degree of uncertainty.
On the other hand, the large number of analysed sources should
give an unbiased distribution of parameter values.
In other words: the adopted parameters of each specific source 
could be slightly wrong, but the distribution of the parameters
for all sources should be much more reliable.

\begin{table*} 
\centering
\begin{tabular}{lllllllllllllllllll}
\hline
\hline
$\Gamma$ &$\theta_{\rm v}$ &$\delta$ &$R_{\rm diss}$ &$P^\prime_{\rm e, 45}$ &$B$ &$\gamma_{\rm peak}$                                                     
     &$\gamma_{\rm cool}$ &$U^\prime$  &$\dot M_{\rm out}$   &$\log P_{\rm r}$ &$\log P_{\rm e}$ &$\log P_{\rm B}$ 
     &$\log P_{p}$  &$t^{\rm obs}_{\rm var}$\\
~[1]    &[2]    &[3]              &[4]      &[5]                &[6]                    &[7] &[8]                       
     &[9]  &[10]          &[11]        &[12]                 &[13]        &[14]        &[15]       \\
\hline 
9   &3   &14.7 &900  &0.060  &1.60  &144  &11   &2.5  &0.15   &46.1 &44.9 &44.8 &46.9  &52 \\
11  &3   &16.5 &900  &0.030  &1.60  &156  &7.4  &3.6  &0.11   &46.0 &44.7 &44.9 &46.9  &46  \\
13  &3   &17.8 &900  &0.029  &1.60  &151  &5.4  &5.0  &0.13   &46.2 &44.7 &45.1 &47.0  &43 \\
15  &3   &18.6 &900  &0.028  &1.75  &144  &4    &6.6  &0.12   &46.3 &44.7 &45.3 &47.0  &41 \\  
17  &3   &19.0 &900  &0.032  &1.81  &161  &3.2  &8.5  &0.17   &46.0 &44.8 &45.4 &47.2  &40 \\
30  &1.9 &30   &3600 &0.015  &0.19  &171  &40   &0.2  &0.08   &46.5 &45.6 &45.2 &47.2  &102  \\
\hline
\hline 
\end{tabular}
\caption{
Parameters used to model the SED of 0625--5438, at $z=2.051$.
For both models we have used a black hole mass $M=10^9 M_\odot$ and an accretion 
luminosity of $L_{\rm d}=3.6\times 10^{46}$ erg s$^{-1}$ (corresponding to an Eddington ratio of 0.24 and to 
$\dot M_{\rm in}=7.9$  solar masses per year). 
With this disc luminosity, $R_{\rm BLR}=6\times 10^{17}$ cm and $R_{\rm torus}=1.5\times 10^{19}$ cm.
For both models, the spectral parameters are unchanged: $s_1=0$, $s_2=4$, $\gamma_{\rm max}=5\times 10^3$,
$\gamma_{\rm b}=240$.
Col. [1]: bulk Lorentz factor $\Gamma$ ar $R_{\rm diss}$;
Col. [2]: jet semi--aperture angle $\theta_{\rm v}$ in degreees;
Col. [3]: Doppler factor $\delta$;
Col. [4]: dissipation radius in units of $R_{\rm S}$;
Col. [5]: power injected in the blob calculated in the comoving frame, in units of $10^{45}$ erg s$^{-1}$; 
Col. [6]: magnetic field in Gauss;
Col. [7]: random Lorentz factors of the electrons emitting at the peak of the SED;
Col. [8]: random Lorentz factors of the electrons cooling in one dynamical time $R/c$;
Col. [9]: total (magnetic+radiative) energy density in the comoving frame in erg cm$^{-3}$;
Col. [10]: mass outflowing rate, in solar masses per year;
Col. [11]--[14]: Logarithm of the jet power in the for of produced radiation ($P_{\rm r}$),
emitting electrons ($P_{\rm e}$), magnetic field ($P_{\rm B}$), and cold protons ($P_{\rm p}$), 
assuming one proton per emitting electron;
Col. [15] the minimum observable variability timescale, defined as $t^{\rm obs}_{\rm var}\equiv (R/c)(1+z)/\delta$,
in hours.
}
\label{0625para}
\end{table*}

\section{Physical properties}

The list of parameters resulting from the application of the model
to all sources is given in Tab. \ref{para} and the distributions of several of them
are shown in Figg. \ref{isto1}--\ref{isto3}.
All the SEDs of FSRQs, together with the models, are shown in Fig. \ref{sed1}.
The SEDs of the 26 BL Lacs studied here have been already shown in Sbarrato et al. (2014),
with the purpose of discussing their classification.

\begin{figure}
\vskip -0.7 cm
\hskip -1.7 cm
\psfig{file=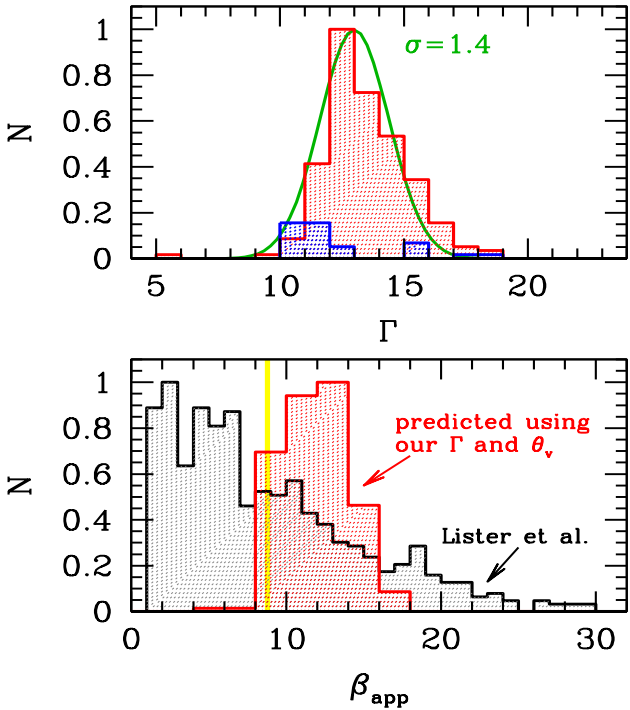,height=11cm,width=12cm}\\
\vskip -1 cm
\caption{ 
Top panel: distribution of the bulk Lorentz factors  for the blazar in our sample. 
The average value is $\langle \Gamma \rangle \sim 13$, and the width is $\sigma=1.4$,
if fitted with a Gaussian. 
Consider that the detection in the $\gamma$--ray band favours larger beaming Doppler factors
$\delta$ (hence larger $\Gamma$ and smaller $\theta_{\rm v}$; see e.g. Savolainen et al. 2010).
Blue hatched areas correspond to BL Lacs.
Bottom panel:
distribution of the apparent superluminal speed $\beta_{\rm app}$ of the sample of blazars 
of Lister et al. (2013), excluding the values $\beta_{\rm app}<1$ (grey hatched histogram).
The vertical yellow line indicates the value of $\langle \beta_{\rm app}\rangle$=8.79,
calculated excluding all values of $\beta_{\rm app}<1$.
The $\beta_{\rm app}$ distribution of Lister et al. (2013)
can be be compared with the solid (red) line, corresponding to the 
$\beta_{\rm app}$ distribution obtained by our values of $\Gamma$ and $\theta_{\rm v}$.
}
\label{gamma}
\end{figure} 

\vskip 0.2 cm
\noindent
{\it  Bulk Lorentz factors ---}
The top panel of
Fig. \ref{gamma} shows the distribution of the bulk Lorentz factor peaks at $\Gamma\sim 13$
and is rather narrow. 
A Gaussian fit returns a dispersion of $\sigma=1.4$.
The studied BL Lac objects do not show any difference with FSRQs, but they
are too few to draw any conclusion.
The bottom panel reports the distribution (black line and grey hatched histogram)
of the value of the superluminal speeds given by Lister et al. (2013) 
(these values refer to individual components, but excluding subluminal values).
The vertical yellow line indicates the average $\langle \beta_{\rm app}\rangle=8.8$
of the Lister et al. (2013) values.
Of the 645 components with  $\beta_{\rm app}>1$, only 38 (5.9\%)
have $\beta_{\rm app}>20$, 11 (1.7\%) have $\beta_{\rm app}>25$, and 
1 (0.3\%) have $\beta_{\rm app}>30$.
Fig. \ref{gamma} shows also (red line) the distribution of $\beta_{\rm app}$ corresponding
to the values of $\Gamma$ and $\theta_{\rm v}$ adopted to fit the blazars in our sample.
While most of the superluminal components have small values of $\beta_{\rm app}$,
the distribution $\Gamma$ found for the blazars in our sample lacks a low and high--$\Gamma$ tail.
This can be readily explained by a selection effect: the sources
that are not substantially beamed towards the Earth (either because they are
misaligned or have a small $\Gamma$) cannot be detected by {\it Fermi}/LAT
(see e.g. Savoilanen et al. 2010), 
while the number of sources with very large $\Gamma$ is limited.
Furthermore, consider that our study concerns blazars whose {\it one--year averaged}
$\gamma$--ray flux was detectable by {\it Fermi}/LAT, and that, by construction, 
Shaw et al. (2012; 2013) excluded the most known blazars (hence, the brightest and more
active, presumably the ones that sometimes have extreme $\Gamma$ values)
from their samples.

The distribution of $\Gamma$ and $\beta_{\rm app}$ for the blazars in our sample can be compared
with the distributions found by Savoilanen et al. (2010).

They first measured the observed brightness temperature 
using the shortest radio variability time--scale (see also Hovatta et al. 2009) and compared it 
with the theoretically expected brightness temperature assuming equipartition 
(namely an intrinsic brightness temperature $T_{\rm B}=5\times 10^{10}$ K; Readhead et al. 1994).
This gives the Doppler factor $\delta$.
Together with the knowledge of the superluminal speed $\beta_{\rm app}$ (Lister et al. 2009), 
of the fastest component in the jet, they could derive both $\Gamma$ and $\theta_{\rm v}$.
Given that 
i) the sample of Savoilanen et al. (2010) was composed of the brightest  {\it Fermi}/LAT 
blazars (while ours is composed of blazars having a 1--year {\it averaged} $\gamma$--ray flux detectable
by {\it Fermi}/LAT); 
ii) that the shortest variability event coupled with the fastest knot in the jet
is bound to give the largest $\Gamma$; 
iii) that it is very likely that the jet has components
moving with different $\Gamma$, we can conclude that the range of $\Gamma$--values 
derived by us and by Savoilanen et al. (2010) are consistent.

We again stress that the narrowness of the distribution of $\Gamma$--values
is partly due to a selection effect: to be detected by {\it Fermi}/LAT, the bulk
Lorentz factor  $\Gamma$ 
should be relatively large, although very large values are not present i) because they are rare;
ii) because in our sample we study {\it one--year averaged} $\gamma$--ray fluxes,
and iii) because sources like 3C 454.3 (the brightest and more active in the {\it Fermi}/LAT
band) are excluded by Shaw et al. (2013; 2013).

\begin{figure}
\vskip -0.7 cm
\hskip -0.8 cm
\psfig{file=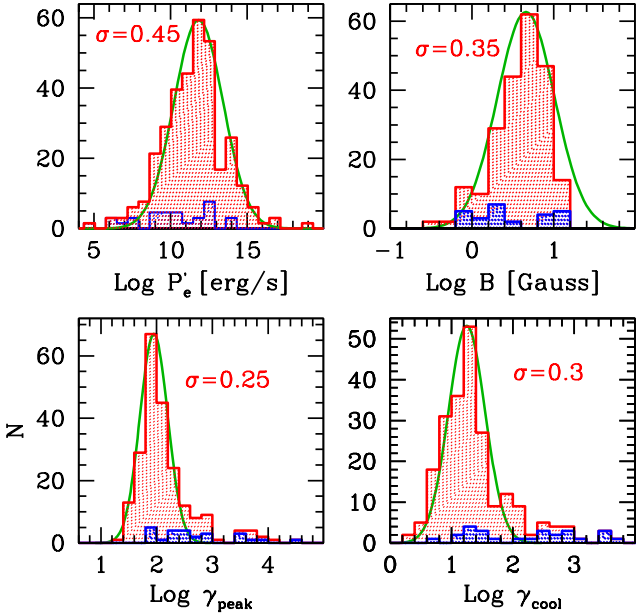,height=9.8cm,width=9.7cm}\\
\vskip -1 cm
\caption{
Distribution of the total power injected in relativistic electrons, as measured
in the comoving frame $P^\prime_e$ (top left);
of the magnetic field (top right);
of the random Lorentz factor $\gamma_{\rm peak}$ of the electrons emitting at the peaks of the SEDs
(bottom right) and of the values of the random Lorentz factor of electrons cooling in one
dynamical time $R/c$.
Blue hatched areas correspond to BL Lacs.
}
\label{isto1}
\end{figure} 
\begin{figure}
\vskip -0.7 cm
\hskip -0.8 cm
\psfig{file=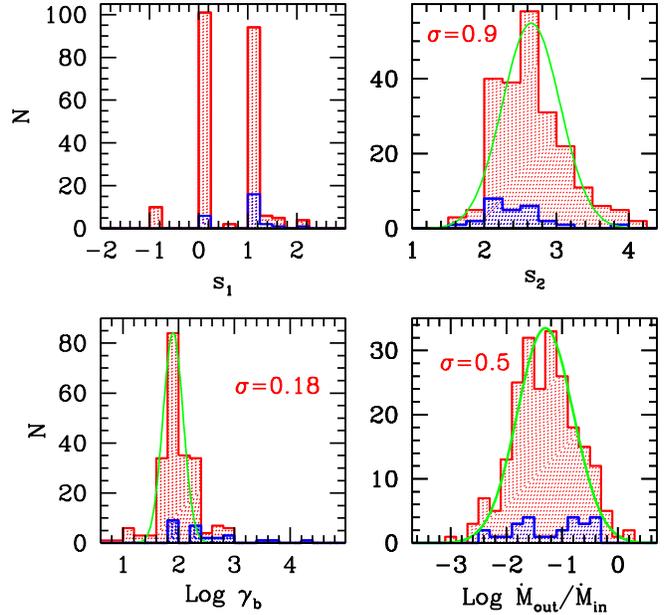,height=9.8cm,width=9.7cm}
\vskip -0.8 cm
\caption{
Distribution of the injection spectral index at low ($s_1$, top left)
and at high energies ($s_2$, top right), break Lorentz factor $\gamma_{\rm b}$
(bottom left), and of the ratio between the outflowing and inflowing mass rate
 (bottom right).
}
\label{isto2}
\end{figure} 

\begin{figure}
\vskip -0.7 cm
\hskip -0.8 cm
\psfig{file=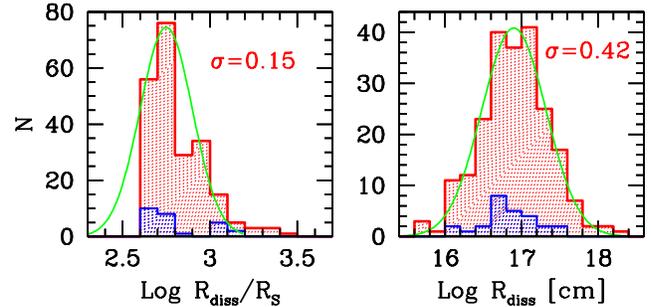,height=9.8cm,width=9.7cm}\\
\vskip -5.5 cm
\caption{
Distribution of the location of the emitting region (i.e. its distance from the black hole)
in units of the \sch\ radius (top left) and in cm (top right).
}
\label{isto3}
\end{figure} 

\vskip 0.2 cm
\noindent
{\it  Magnetic field ---}
The (logarithmic) distribution is centered on $\langle B \rangle=4.6$ Gauss
(Fig. \ref{isto1}).
If fitted with a log--normal, the dispersion is $\sigma=0.35$ dex.
There is no difference between FSRQs and BL Lacs.
                                                                                                                                                                                 
\vskip 0.2 cm
\noindent
{\it Spectral parameters ---}
The distributions of the low energy spectral slope $s_1$, shown in 
Fig. \ref{isto2} is almost bimodal, with most of the sources
with $s_1=0$ or $s_1=1$.
As long as $s_1\le 1$, the exact values is not very important for the
spectral fitting, since in these cases the electron distribution is always
$\propto \gamma^{-2}$ down to $\gamma_{\rm cool}$.
On the other hand, the slope impacts on the total amount of
emitting particles that are present in the source, and therefore
on the total amount of protons, that regulate the kinetic power of the jet.

The distribution of $s_2$ peaks at $s_2=2.65$ and its dispersion (if fitted with a Gaussian)
is $\sigma=0.9$. 
BL Lacs tend to have smaller $s_2$.
The mean value is rather steep, and the distribution rather broad, 
in agreement with recent numerical simulations of shock acceleration
(Sironi \& Spitkovsky 2011).
The break energy logarithmic distribution peaks at $\gamma_{\rm b} \sim 80$,
with a width $\sigma \sim 0.18$.
BL Lacs tend to have larger $\gamma_{\rm b}$.
The logarithmic distribution of power injected in the form of relativistic electrons
(in the comoving frame) peaks at $P^\prime_{\rm e}\sim 5\times 10^{42}$ erg s$^{-1}$,
with a width $\sigma=0.45$ dex.
BL Lacs tends to have lower $P^\prime_{\rm e}$, but they are too few to firmly assess it.
Note that the values of $P^\prime_{\rm e}$ refer to one blob. 
On average, since there are {\it two} jets, we should require twice as much.

\vskip 0.2 cm
\noindent
{\it Mass outflowing rate ---}
The bottom right panel of Fig. \ref{isto2} shows the ratio
$\dot M_{\rm out}/\dot M_{\rm in}$ between the outflowing mass rate and
the accretion rate.
In this case we calculate $\dot M_{\rm out}$ assuming that an equal 
amount of matter is flowing out in each of the two jet, and 
we consider the total value.
The distribution peaks around a value of 6\%, and is rather narrow,
with a dispersion of 0.5 dex.
This suggests that the matter content of the jet is related to the disc,
not to entrainment, that presumably would be dependent on
the properties of the ambient medium, that can be different in systems of equal disc
and jet power.
This should be related to the fact that also the distribution of $\Gamma$ is very narrow.
We can easily understand the average value through (see Ghisellini \& Celotti 2002):
\begin{equation}
{\dot M_{\rm out} \over \dot M_{\rm in} } \, =\, {P_{\rm jet} \over L_{\rm d} } \, {\eta \over  \Gamma }
\, =\, 10^{-2}\, { P_{\rm jet}\over L_{\rm d}} \, {\eta \over 0.1}\,  {10\over \Gamma}
\end{equation}
Since $\langle \Gamma\rangle\sim 13$, the above equation, together with 
the average value of $\dot M_{\rm out}/\dot M_{\rm in}=0.06$ indicates that,
on average, $P_{\rm jet}\sim 10 L_{\rm d}$, as found in Ghisellini et al. (2014).

\vskip 0.2 cm
\noindent
{\it Location of the emitting region ---}
The location of the emitting region (Fig. \ref{isto3}) has a rather sharp cut--off
at $R_{\rm diss}=400 R_{\rm s}$, and a tail extending up to 3000 $R_{\rm s}$.
The low end cut--off is partly due to the requirement of having a sufficiently large $\Gamma$:
since $\Gamma\propto R_{\rm diss}^{1/2}$, $R_{\rm diss}$ must be sufficiently large.

\begin{figure}
\vskip -0.6 cm
\psfig{file=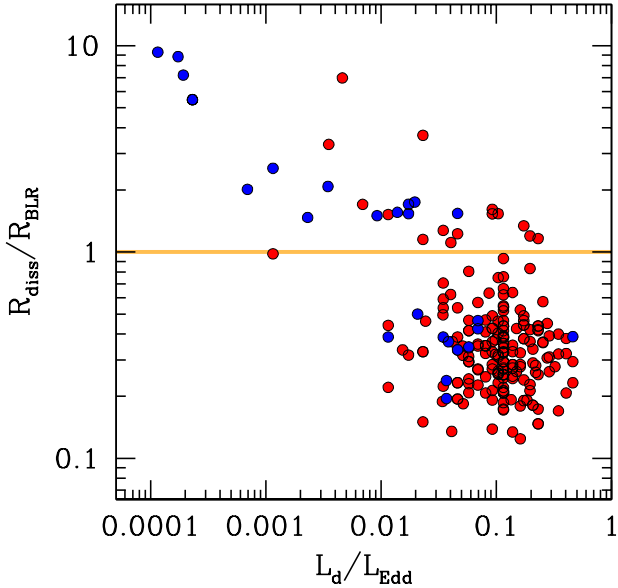,height=9.cm,width=9cm}
\vskip -0.5 cm
\caption{The ratio $R_{\rm diss}/R_{\rm BLR}$ of the location of the emitting region and
the radius of the BLR as a function of the disc luminosity measured in Eddington units.
Red symbols are for FSRQs, blue circles for BL Lacs.
} 
\label{rdissrblr}
\end{figure}

In absolute units  (right panel) the logarithmic distribution is more log normal--like, 
and peaks at $R_{\rm diss}\sim 10^{17}$ cm.
Fig. \ref{rdissrblr} shows $R_{\rm diss}$ in units of the BLR radius
as a function of the disc luminosity in Eddington units.
Above the horizontal line the emitting region is beyond the BLR region,
so that there is an important lack of seed photons, causing a less severe cooling.
This region is populated mainly by BL Lacs and a few FSRQs.

\begin{figure}
\vskip -0.6 cm
\psfig{file=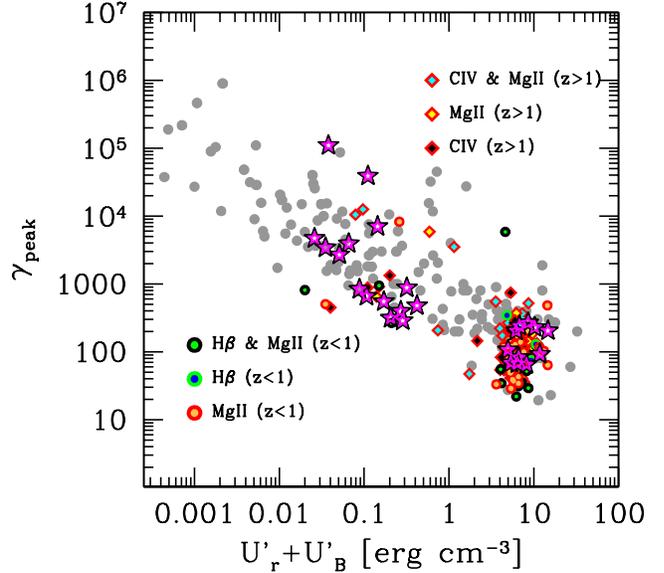,height=9.cm,width=9cm}
\vskip -0.5 cm
\caption{The energy $\gamma_{\rm peak}$ of the electrons emitting at the peaks of the SED
as a function of the energy density (magnetic+radiative) as seen in the comoving frame.
Gray symbols refer to blazars studied previously (Ghisellini et al. 2010; Celotti \& Ghisellini 2008).
} 
\label{gpeak}
\end{figure}

\vskip 0.2 cm
\noindent
{\it Blazar type and cooling ---}
Fig. \ref{gpeak} shows $\gamma_{\rm peak}$ as a function 
of the energy density (radiative+magnetic) as seen in the comoving frame.
The grey symbols corresponds to previous works, that studied many more BL Lacs,
and are shown for comparison.
We confirm the same trend found before: $\gamma_{\rm peak}$
is a function of the cooling rate suffered by the injected electrons.
This explains the so called ``blazar sequence" (Fossati et al. 1998; Ghisellini et al. 1998;
Donato et al. 2001): electrons in low power objects can attain high energies
and give rise to a SED with two peaks in the UV--soft X--rays (synchrotron) and GeV--TeV
(SSC), while high power sources peak in the sub--mm and MeV bands.
The clustering of the points in Fig. \ref{gpeak} is due to the fact that
the vast majority of the sources analyzed here belongs to the same type
(relatively high power FSRQs).

\section{Discussion}

We have studied  a large sample of $\gamma$--ray loud
quasars observed spectroscopically, hence with measured broad emission lines
and optical continuum. 
To the sample of FSRQs of S12, we added a small number of BL Lacs presented
in S13, having broad emission lines, but of small equivalent width.
These few  ``BL Lacs" (26 out of 475) should not be considered as
classical lineless BL Lac objects, but transition objects, characterized
by small accretion rates just above the threshold for radiatively efficient 
disk, or even as true FSRQs with strong lines, high accretion rates, but
particularly enhanced synchrotron flux.
In other words, we alert the reader that we cannot make a robust 
distinction between the FSRQs and the BL Lacs studied here.
The BL Lacs in our sample should be considered as the
low luminosity tail of FSRQs.
For a discussion on the classification issue of blazars we remand the reader
to Sbarrato et al. (2014) and reference therein.

We stress the specific characteristics of the studied sample, highlighting the 
obtained results:

\begin{itemize}
\item Due to the rather uniform sky coverage of {\it Fermi}/LAT, it is a 
flux limited complete sample. 
Comparing $\gamma$--ray detected and radio--loud sources, we know that
only a minority of flat spectrum radio--loud sources have been detected
in the $\gamma$--ray band (Ghirlanda et al. 2011).
This is due to the still limited sensitivity of {\it Fermi}/LAT
if compared with radio telescopes: only the most luminous sources
have a chance to be detected in $\gamma$--rays.
On the other hand, the 20--fold improvement with respect to {\it CGRO}/EGRET
allows to do population studies.

\item It is the largest sample so far of $\gamma$--ray detected blazars
observed spectroscopically in a uniform and complete way in the sky area accessible to
the telescopes used by Shaw et al. (2012; 2013).
This allows to compare the emitting line properties of blazars with
radio--quiet and other radio--loud sources.
The H$\beta$ and MgII line luminosities of the blazars in our sample
have similar distributions and average values of the ones of the
radio--quiet and radio--loud sample selected from the S11 SDSS quasar sample.
The CIV line luminosities of the blazar on our sample are slightly smaller (Fig. \ref{ilines}),
reflecting the different redshift distributions (Fig. \ref{iz}).
 
\item It is the largest blazar sample for which we can have {\it two independent} estimates of the black hole 
mass: one through the virial method, and the second through disc fitting.
The two estimates agree within the uncertainties, and this gives confidence
that both methods are reliable (Fig. \ref{mm} and Fig. \ref{im}).
However, the uncertainties remain large.
They are of different nature, though.
The virial methods has uncertainties that are independent of the quality of the data,
being related to the dispersion of the correlations required to estimate the mass.
The disc fitting method, instead, does depend on the quality of the data,
and if the peak of the disc emission is observed, errors becomes small,
of the order of a factor 1.5--2.
On the other hand, the assumption of a disc perfectly described by a 
standard, Shakura \& Sunyaev (1973) disc may be questionable (see Calderone et al. 2013
for discussion), especially concerning the effect of a Kerr hole on the disc
properties.
To improve, we should directly observe the peak of the disc emission  in many sources. 
This implies either to have nearby sources with large black hole masses observed in UV
and accreting in a radiatively efficient mode,
or high--$z$ and high mass sources observed in the optical. 
But above $z=2$ absorption by intervening matter can be important, limiting
the redshift range.
We think that these constraints have so far limited our knowledge of 
the disc emission in quasars. 
Potentially, the disc--fitting method can result on a good technique to estimate the black hole 
spin, especially if one has a good independent estimate of the black hole mass.

\item It is the largest blazar sample for which a theoretical model has been
applied to derive the physical properties of jets. 
With respect to earlier studies (e.g. Ghisellini et al. 2010)
we also benefit from a better knowledge of the high frequency radio, mm, and far IR
continuum, given by the {\it WMAP}, {\it Planck} and {\it WISE} satellites.
The new data better show the peak of the synchrotron component, that is
slightly more luminous than previously thought (in, e.g. Ghisellini et al. 2010).
Average parameters, and their distributions, are similar to what found
previously, for the blazars detected (at more than 10$\sigma$
in the first 3 months of {\it Fermi}; Ghisellini et al. 2010).

\end{itemize}
The characteristics of the samples here studied make them ideal to study the 
relation between the disc and the jet, since the knowledge  
of the broad emission line luminosities and velocity widths tell us about
the disc luminosity and the black hole mass, while the knowledge of the 
non--thermal continuum (peaking on the $\gamma$--ray band) tell us about
the jet properties, including its power.
The issue of the connection between the jet power and the accretion rate
of the same objects studied here has been discussed in Ghisellini et al. (2014).

\section*{Acknowledgements}
We thank the referee for his/her constructive criticism.
This research has made use of the NASA/IPAC Extragalactic
Database (NED) which is operated by the Jet Propulsion Laboratory, 
California Institute of Technology, under contract with the
National Aeronautics and Space Administration. 
Part of this work is based on archival data software or online services 
provided by the ASDC. 
This publication makes use of data products from the
Wide-field Infrared Survey Explorer, which is a joint project of
the University of California, Los Angeles, and the Jet Propulsion
Laboratory/California Institute of Technology, funded by the 
National Aeronautics and Space Administration.

\vskip 2 cm

\noindent
{\bf APPENDIX}

\begin{figure*}
\begin{tabular}{cccc}
\psfig{file=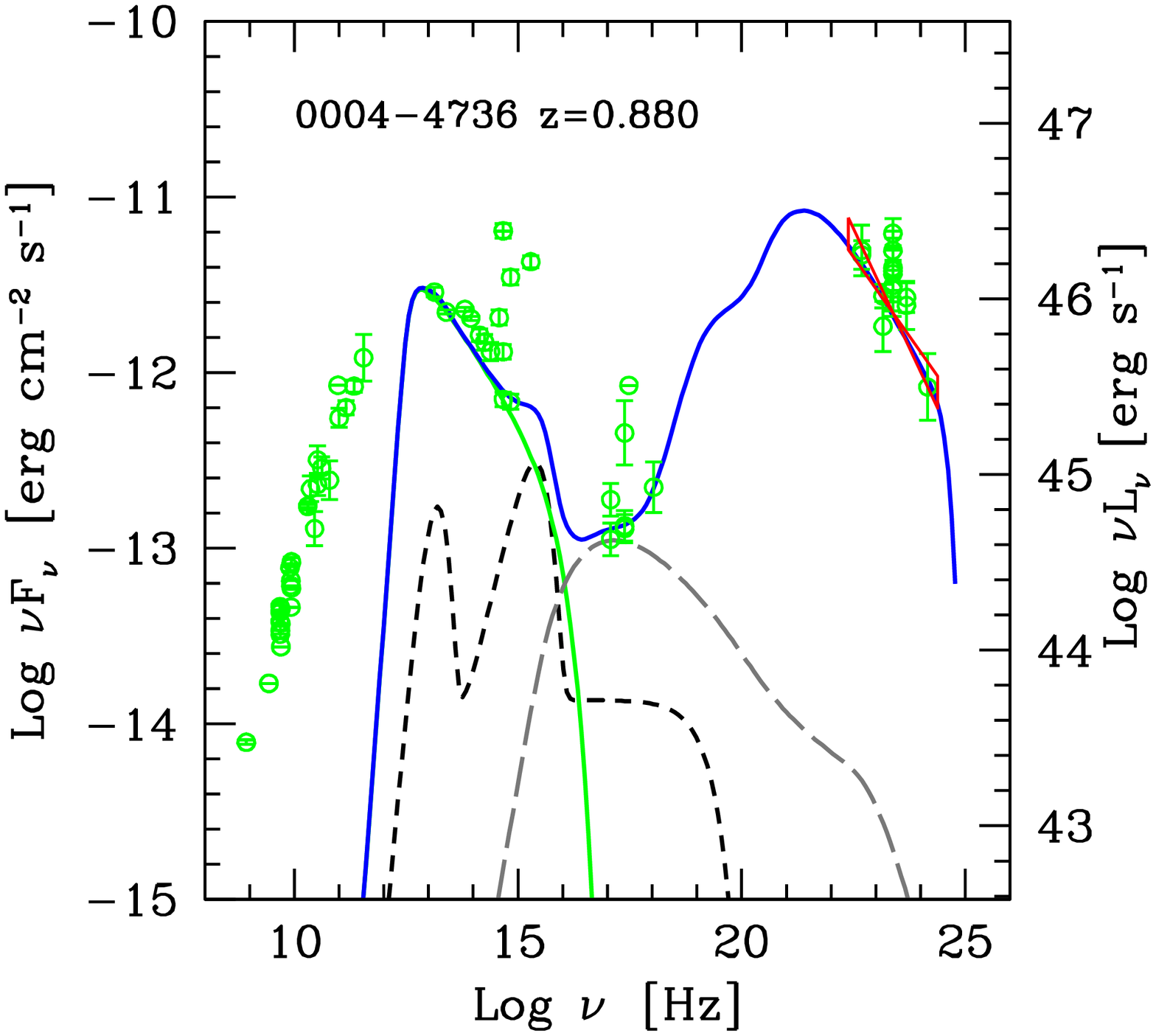,width=4.3cm,height=3.7cm}  
&\psfig{file=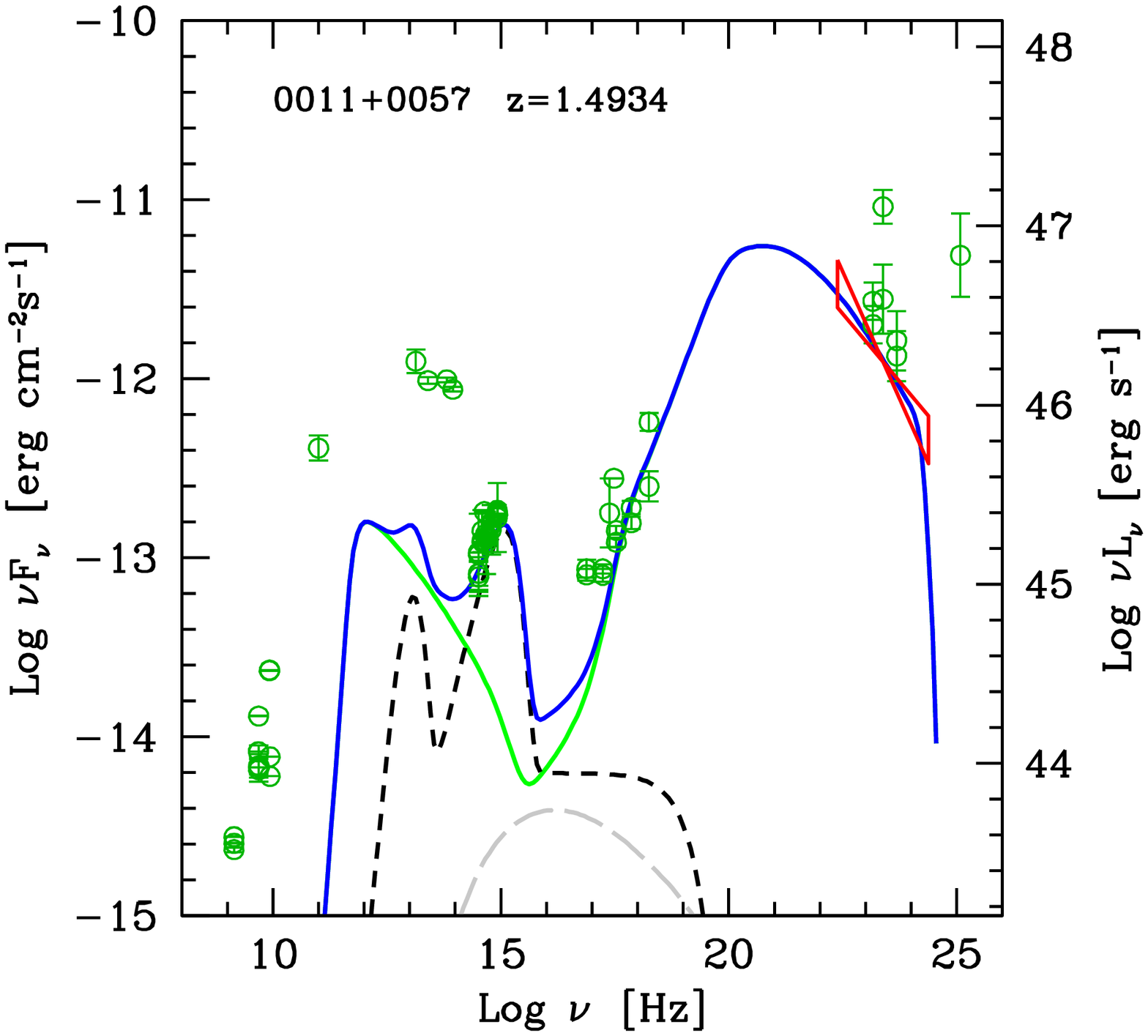,width=4.3cm,height=3.7cm}  
&\psfig{file=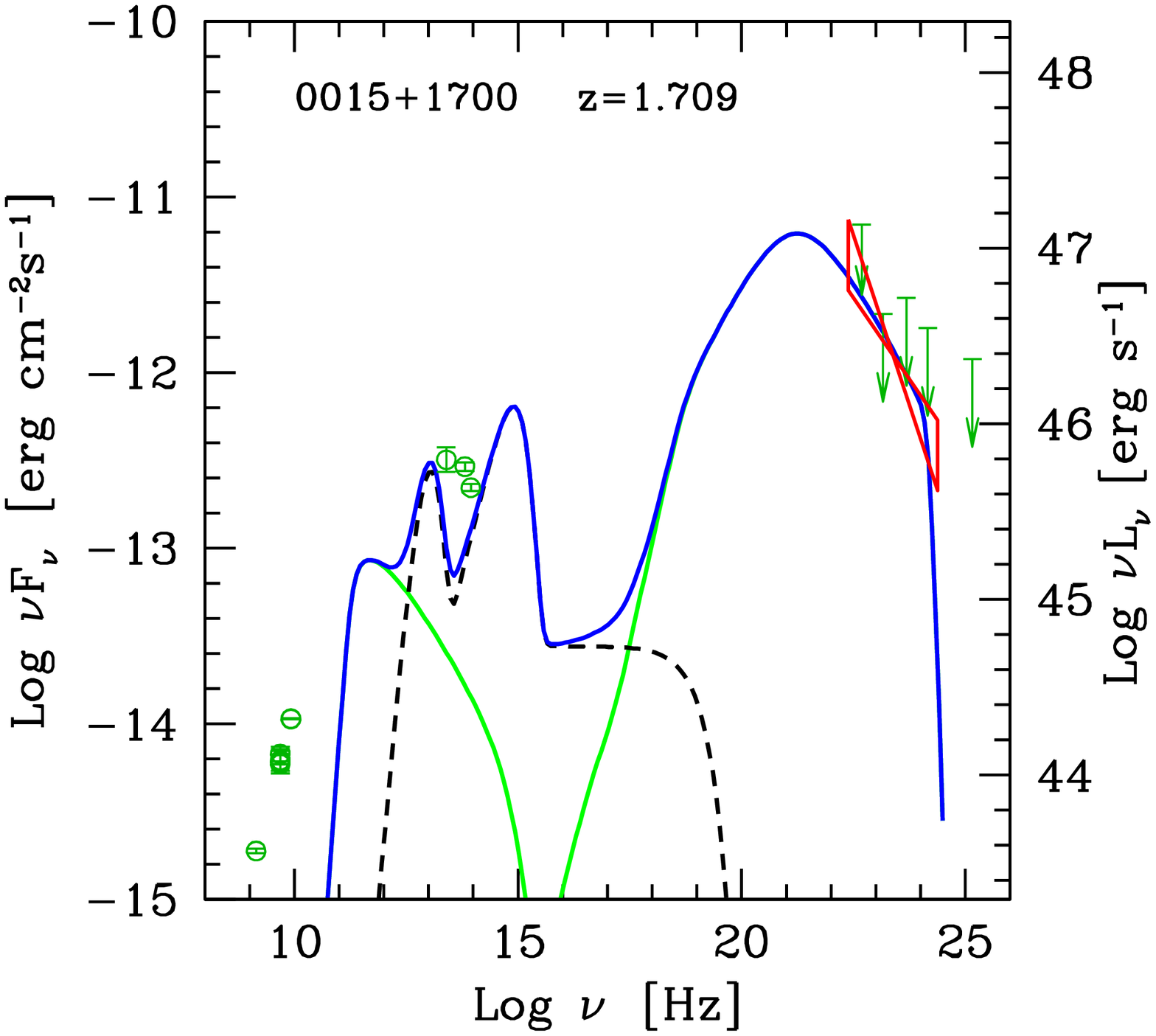,width=4.3cm,height=3.7cm} 
&\psfig{file=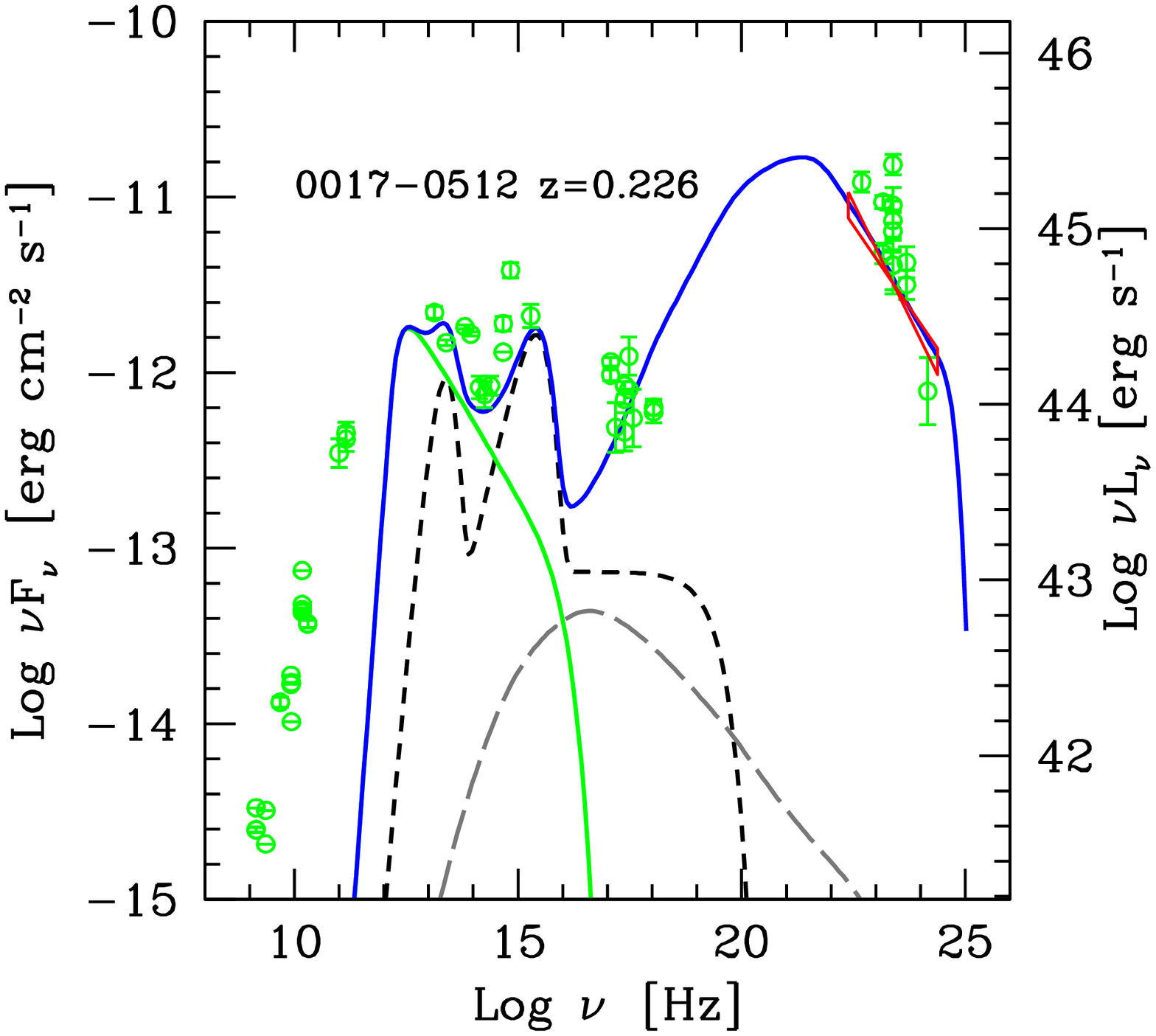,width=4.3cm,height=3.7cm} \\
\psfig{file=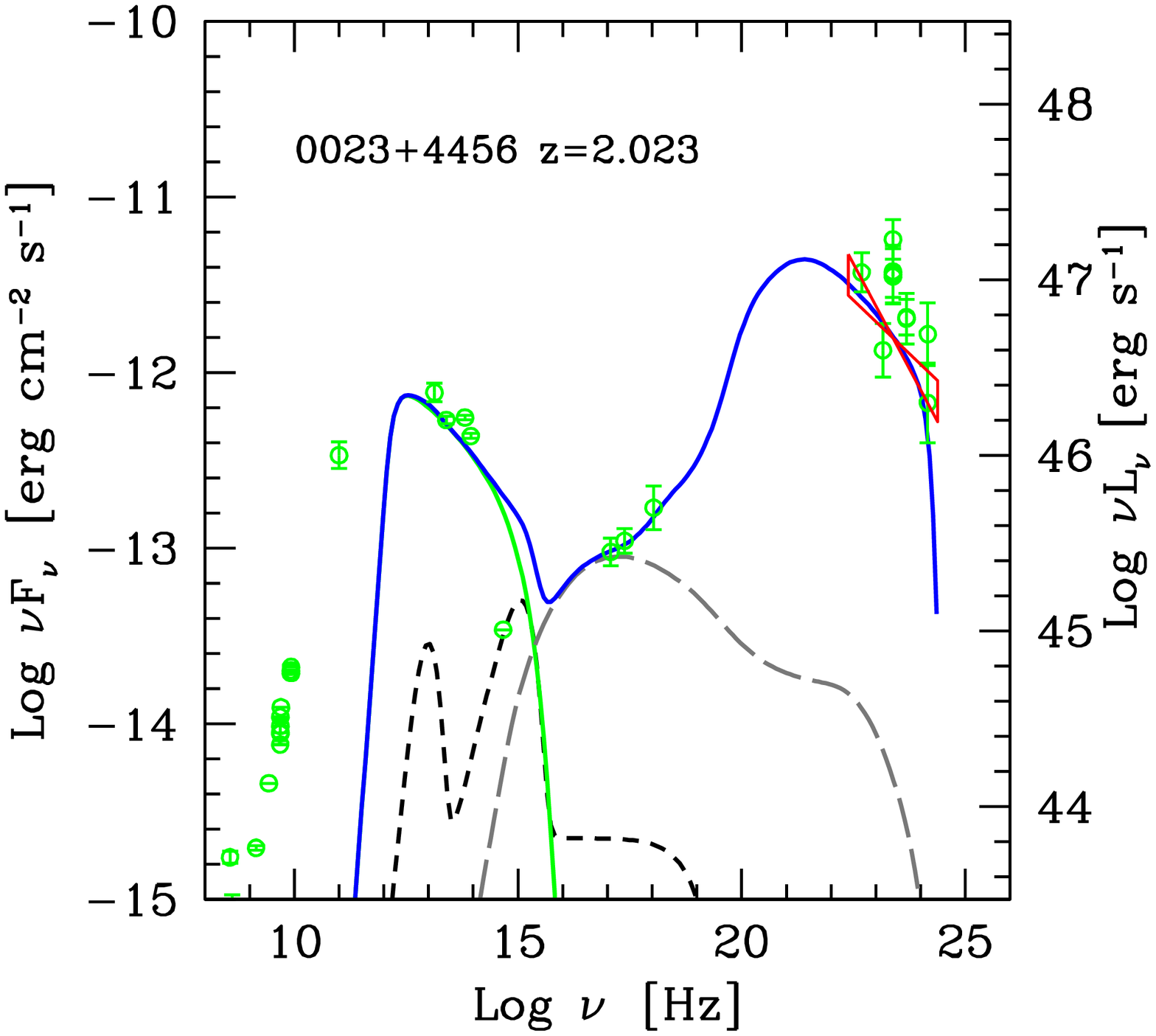,width=4.3cm,height=3.7cm} 
&\psfig{file=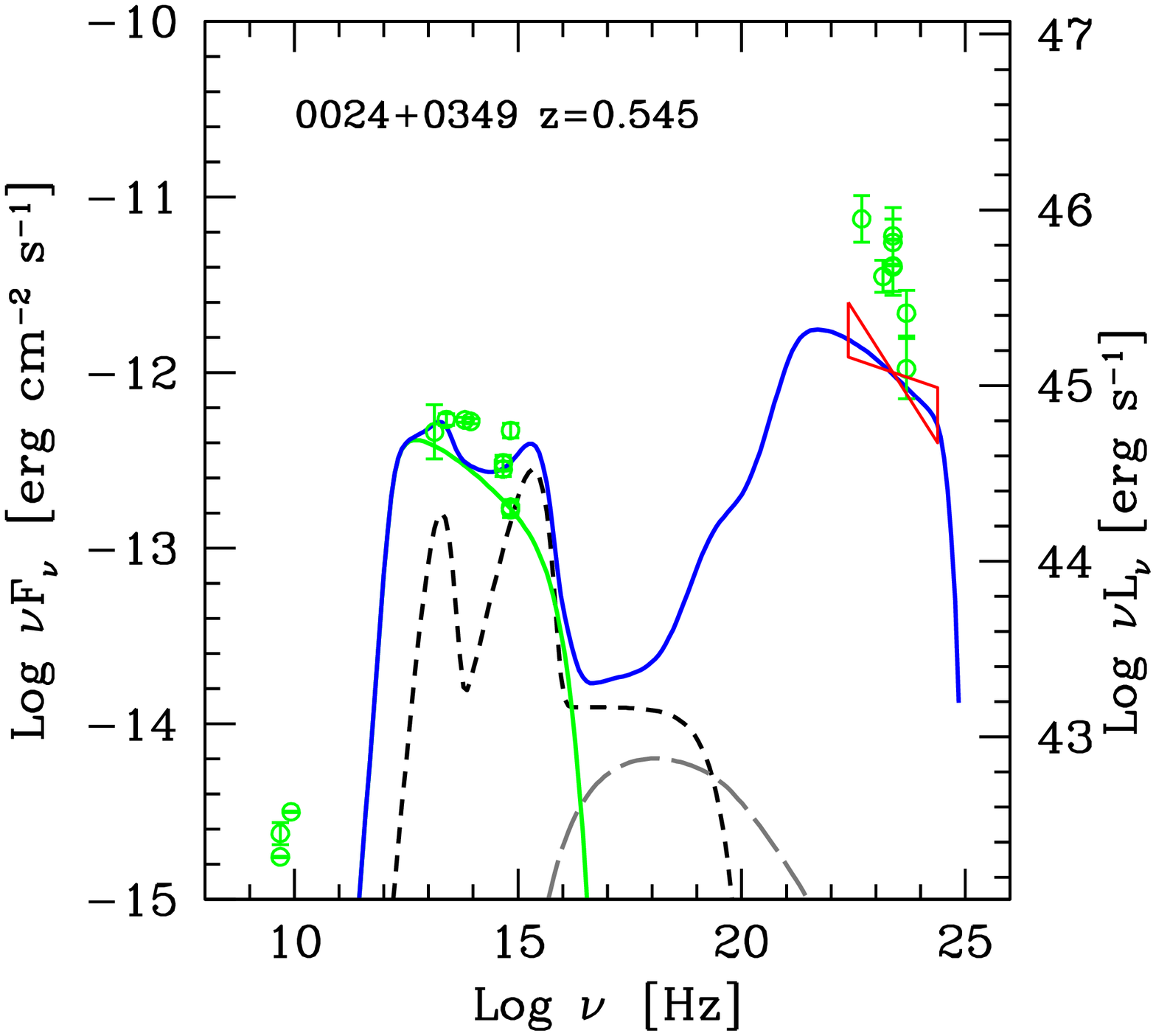,width=4.3cm,height=3.7cm} 
&\psfig{file=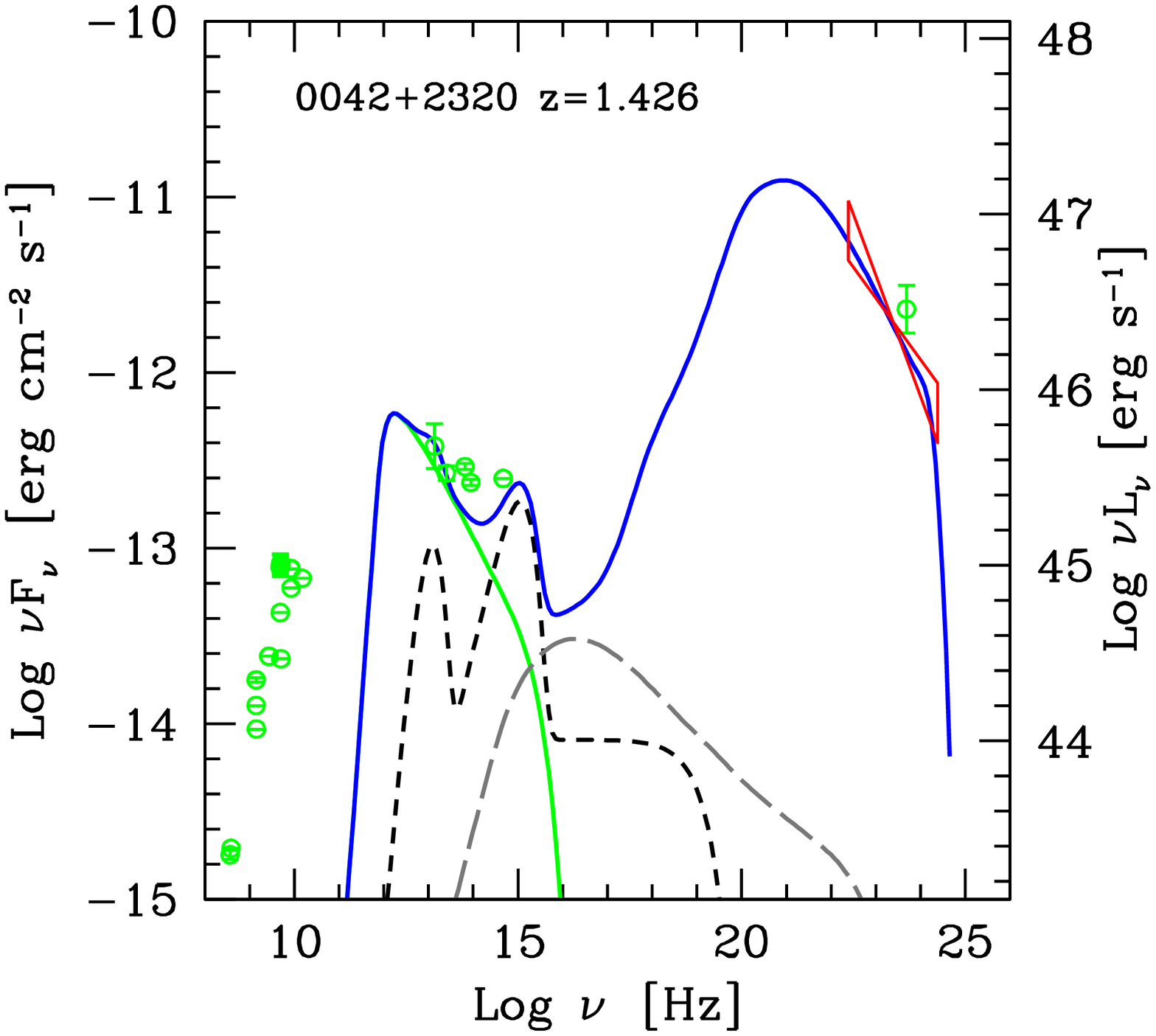,width=4.3cm,height=3.7cm} 
&\psfig{file=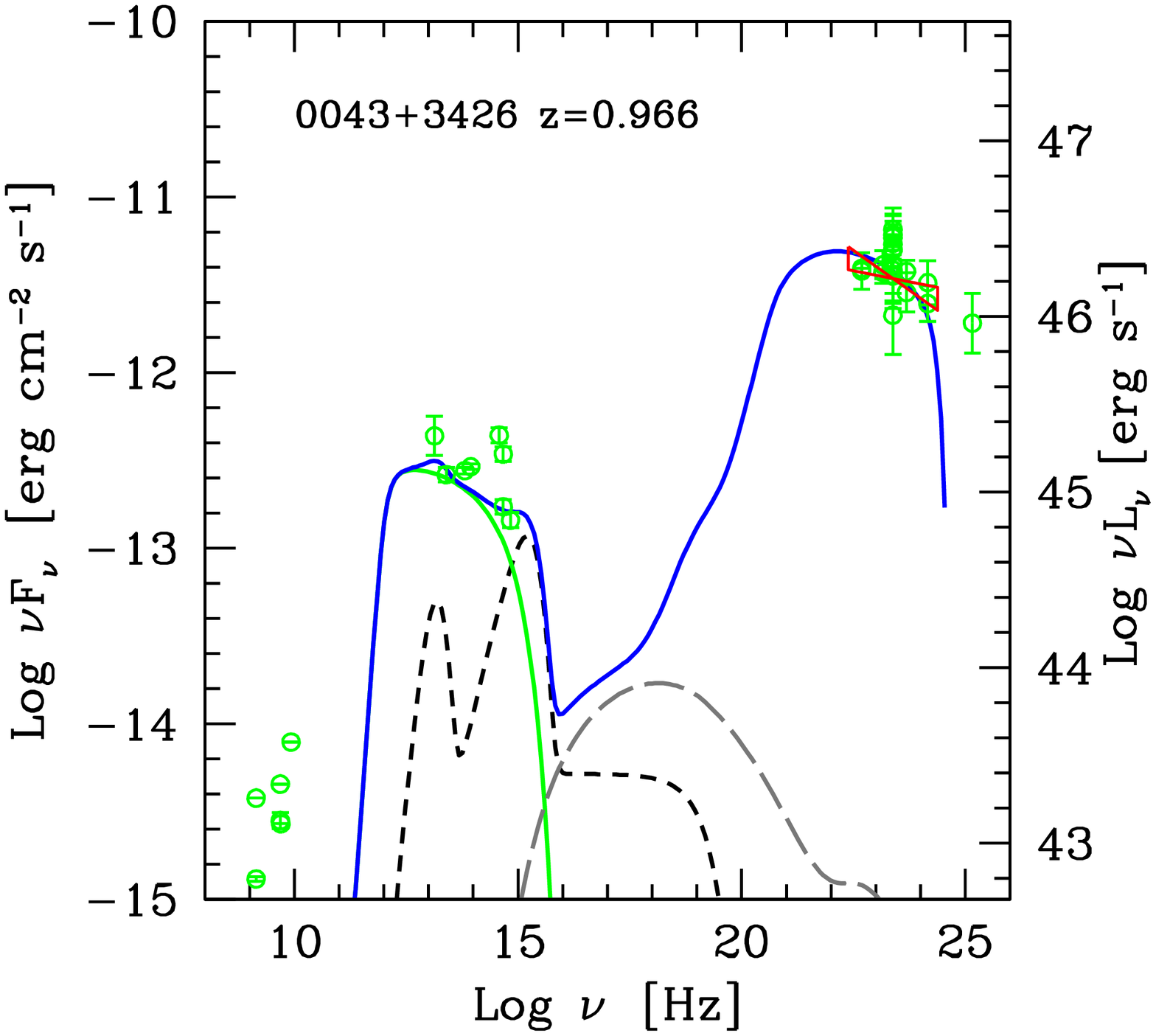,width=4.3cm,height=3.7cm} \\
\psfig{file=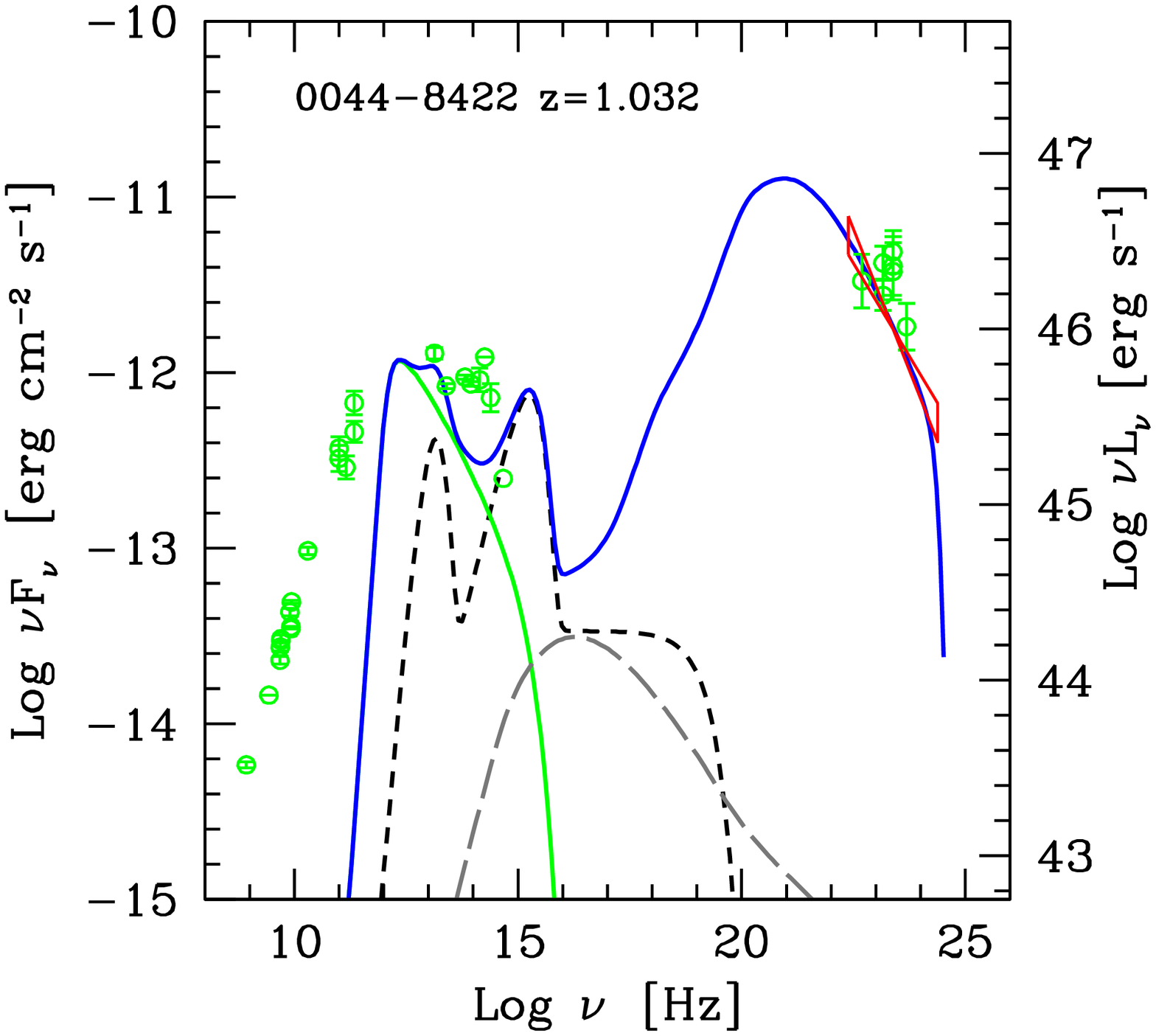,width=4.3cm,height=3.7cm} 
&\psfig{file=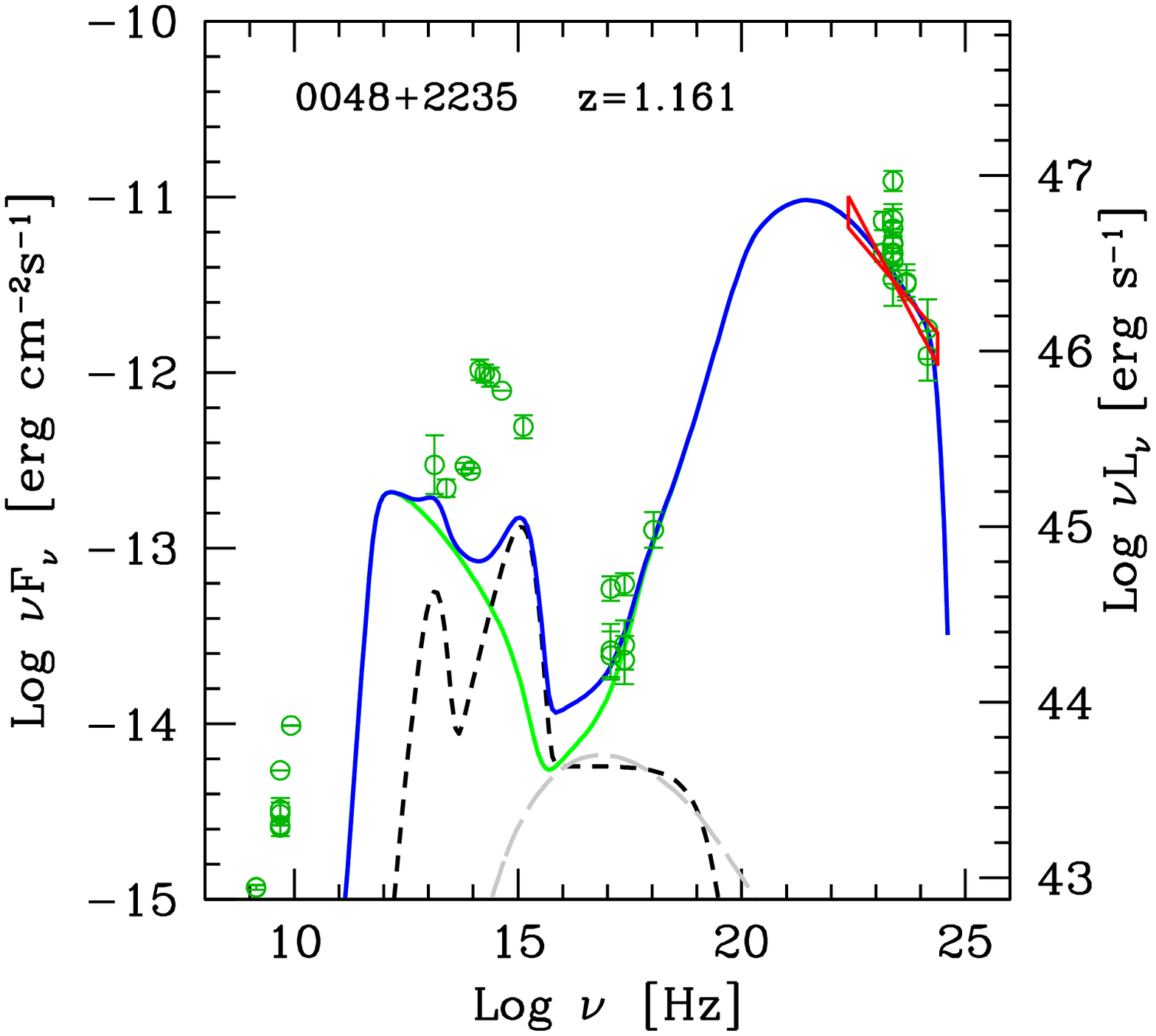,width=4.3cm,height=3.7cm}  
&\psfig{file=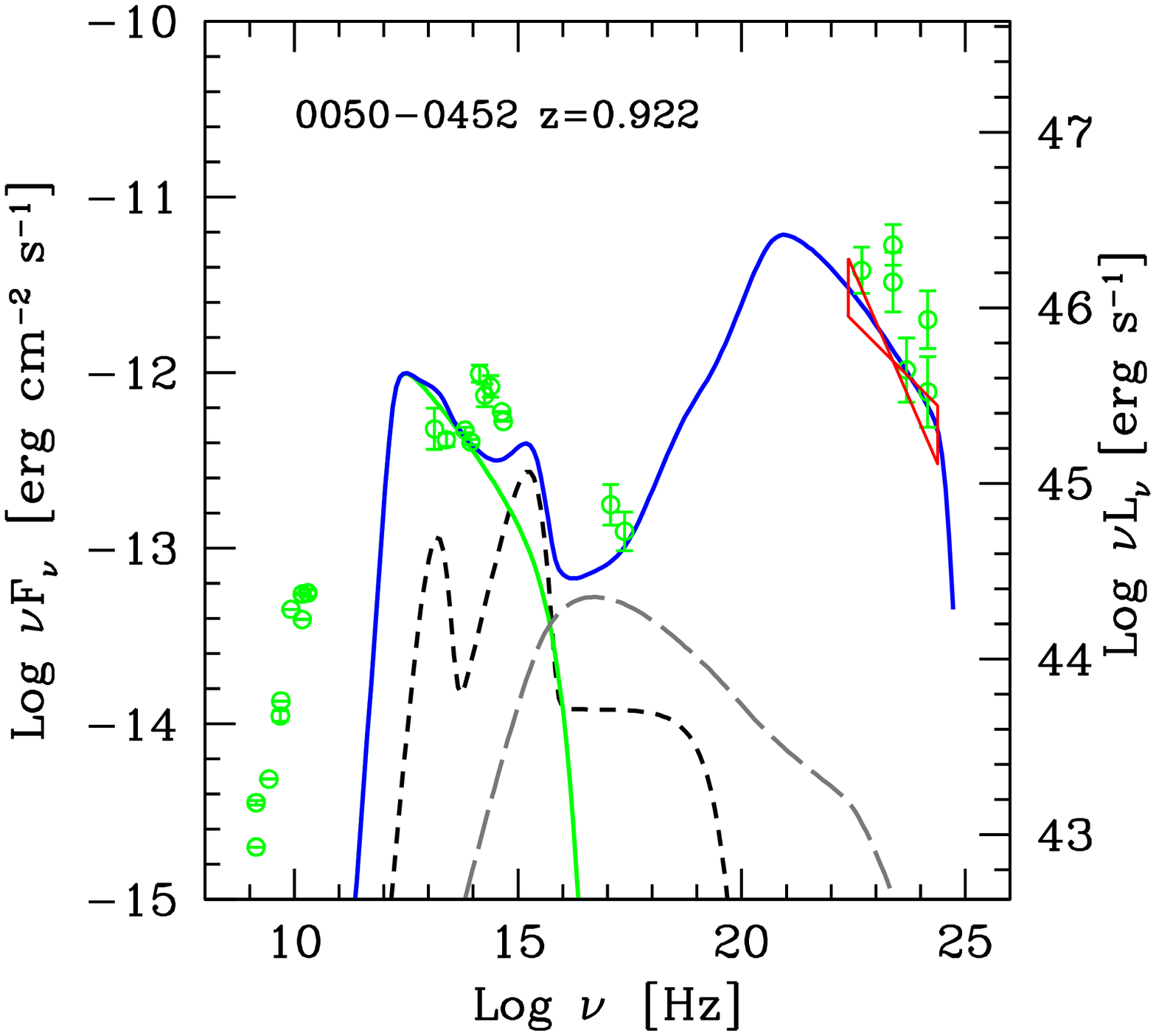,width=4.3cm,height=3.7cm} 
&\psfig{file=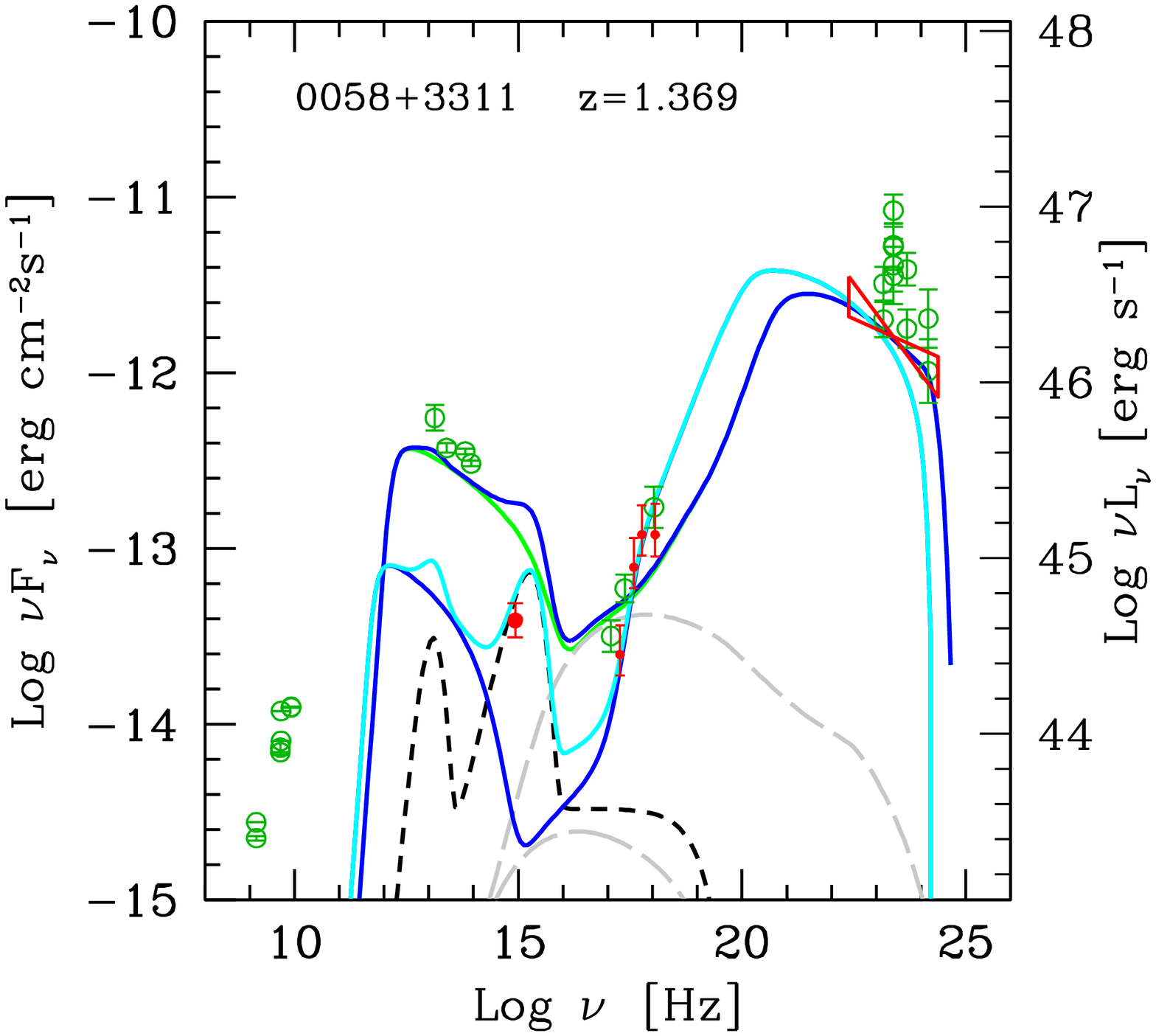,width=4.3cm,height=3.7cm} \\
\psfig{file=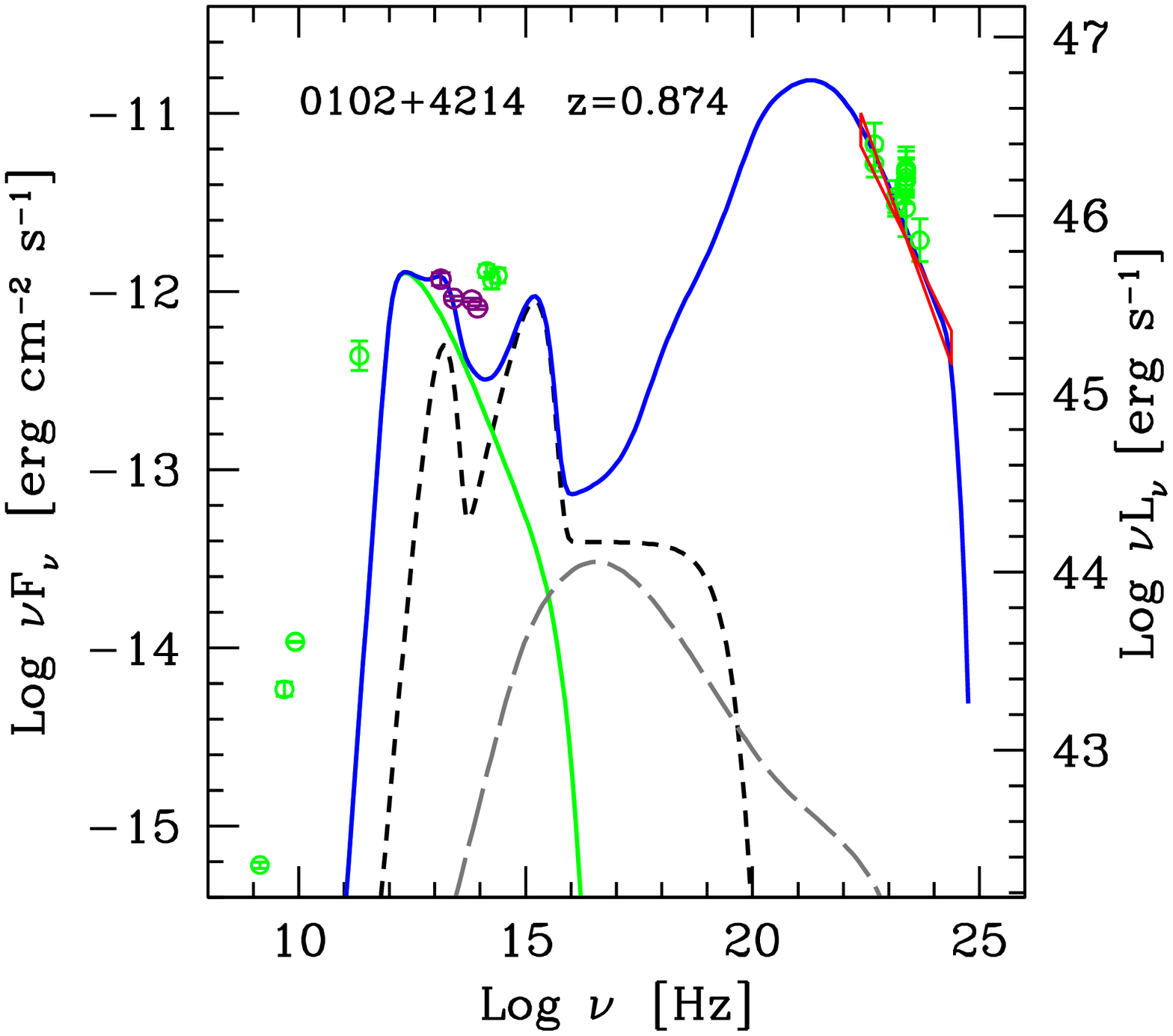,width=4.3cm,height=3.7cm} 
&\psfig{file=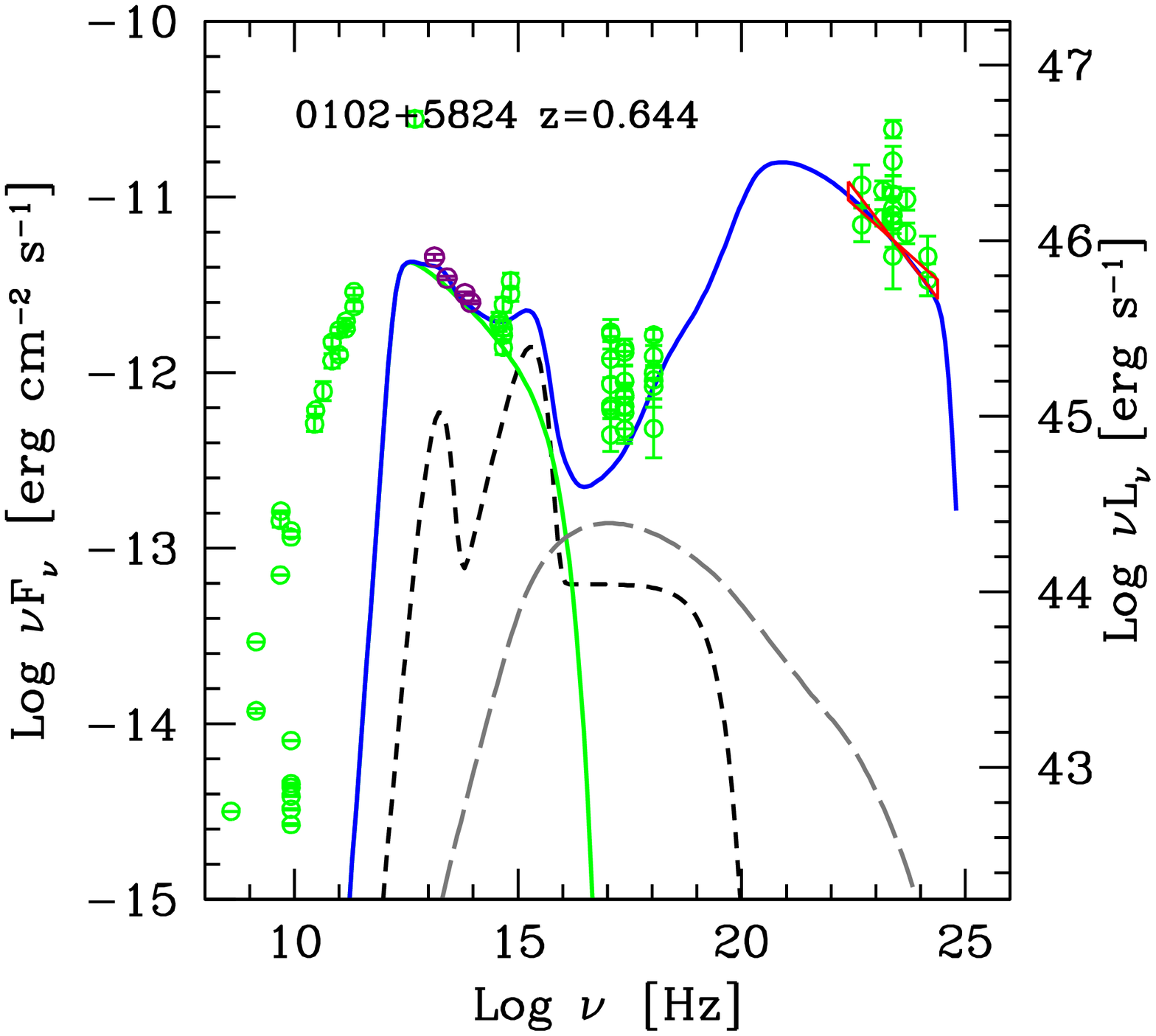,width=4.3cm,height=3.7cm} 
&\psfig{file=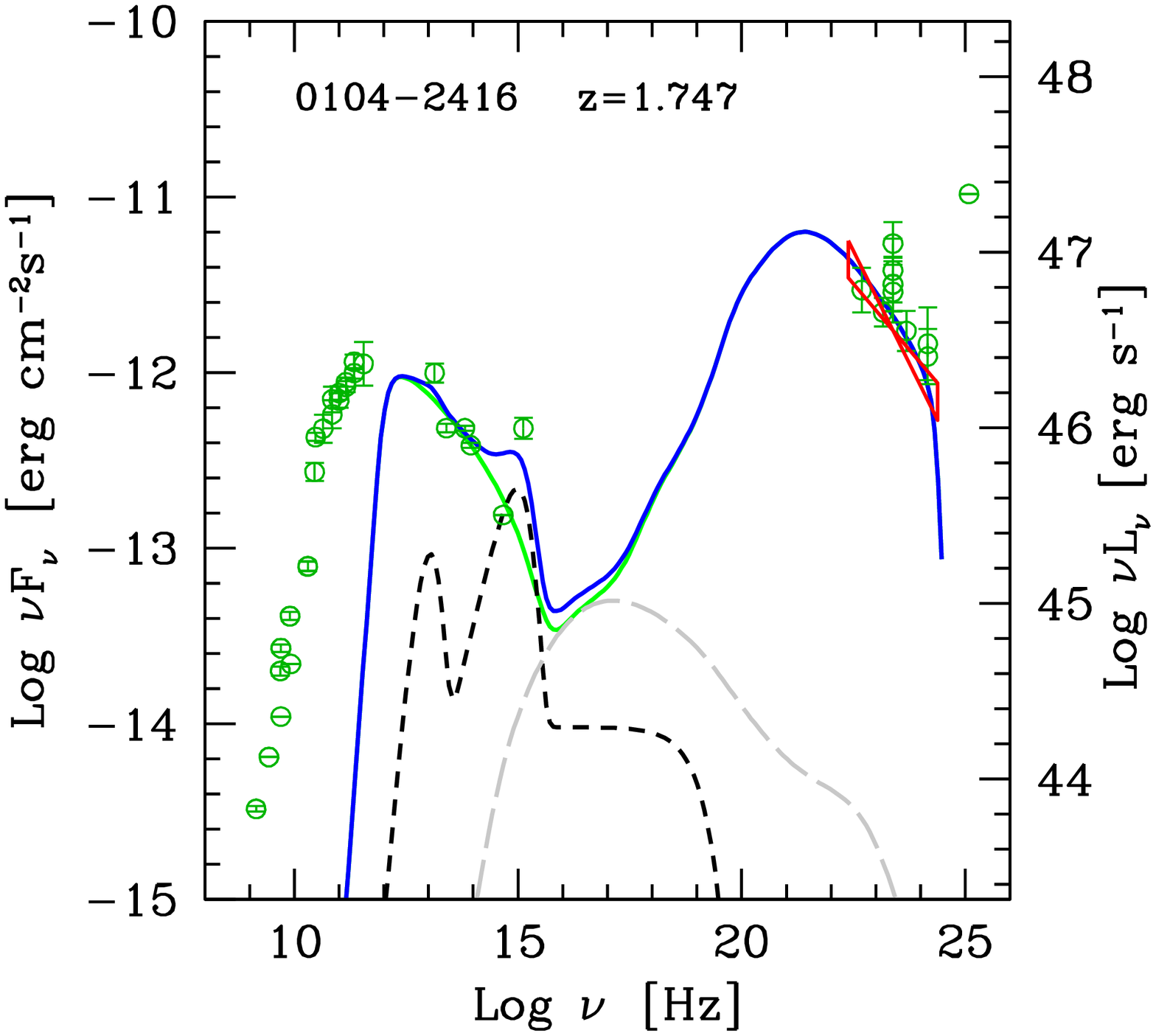,width=4.3cm,height=3.7cm} 
&\psfig{file=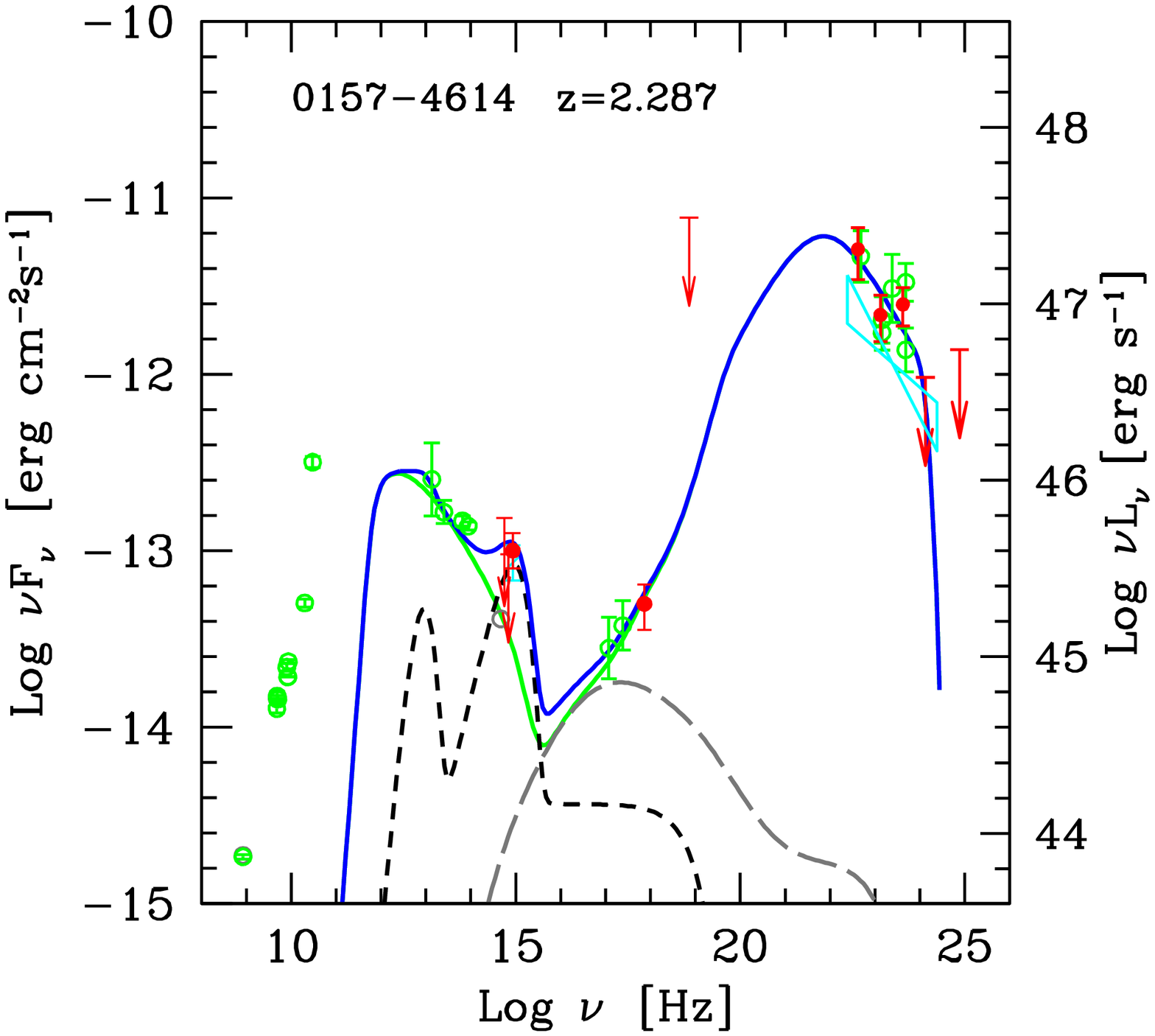,width=4.3cm,height=3.7cm} \\
\psfig{file=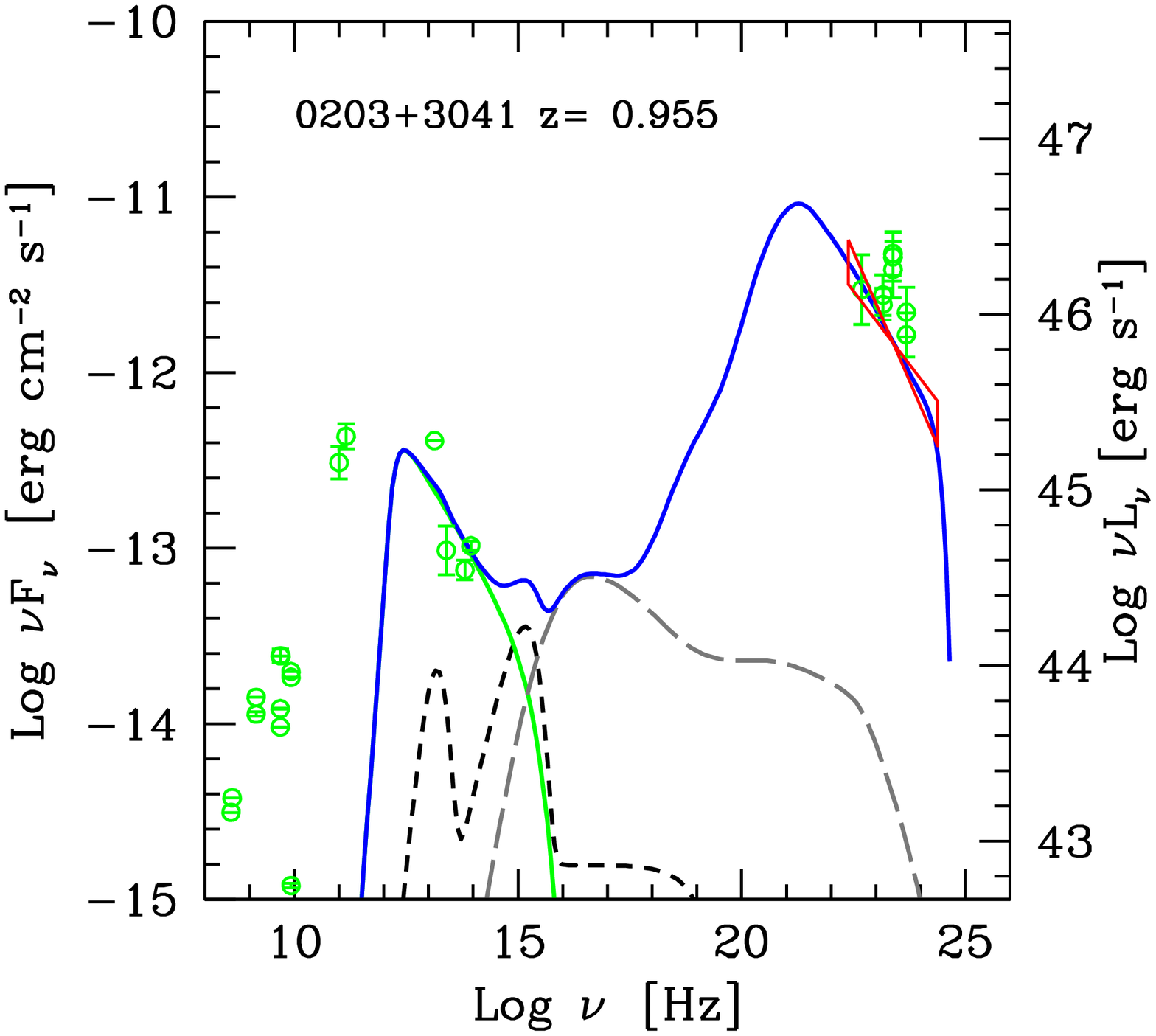,width=4.3cm,height=3.7cm} 
&\psfig{file=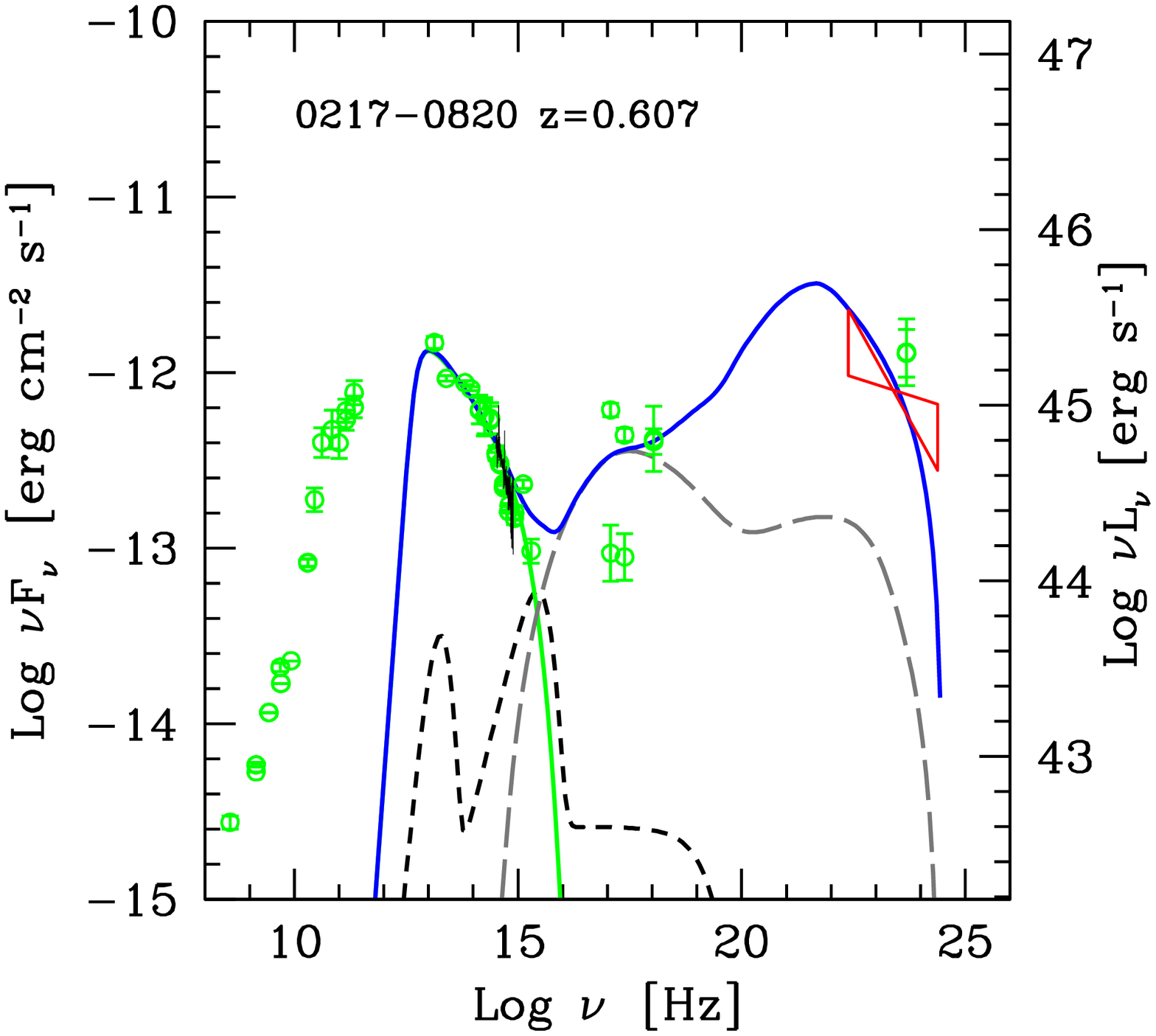,width=4.3cm,height=3.7cm}  
&\psfig{file=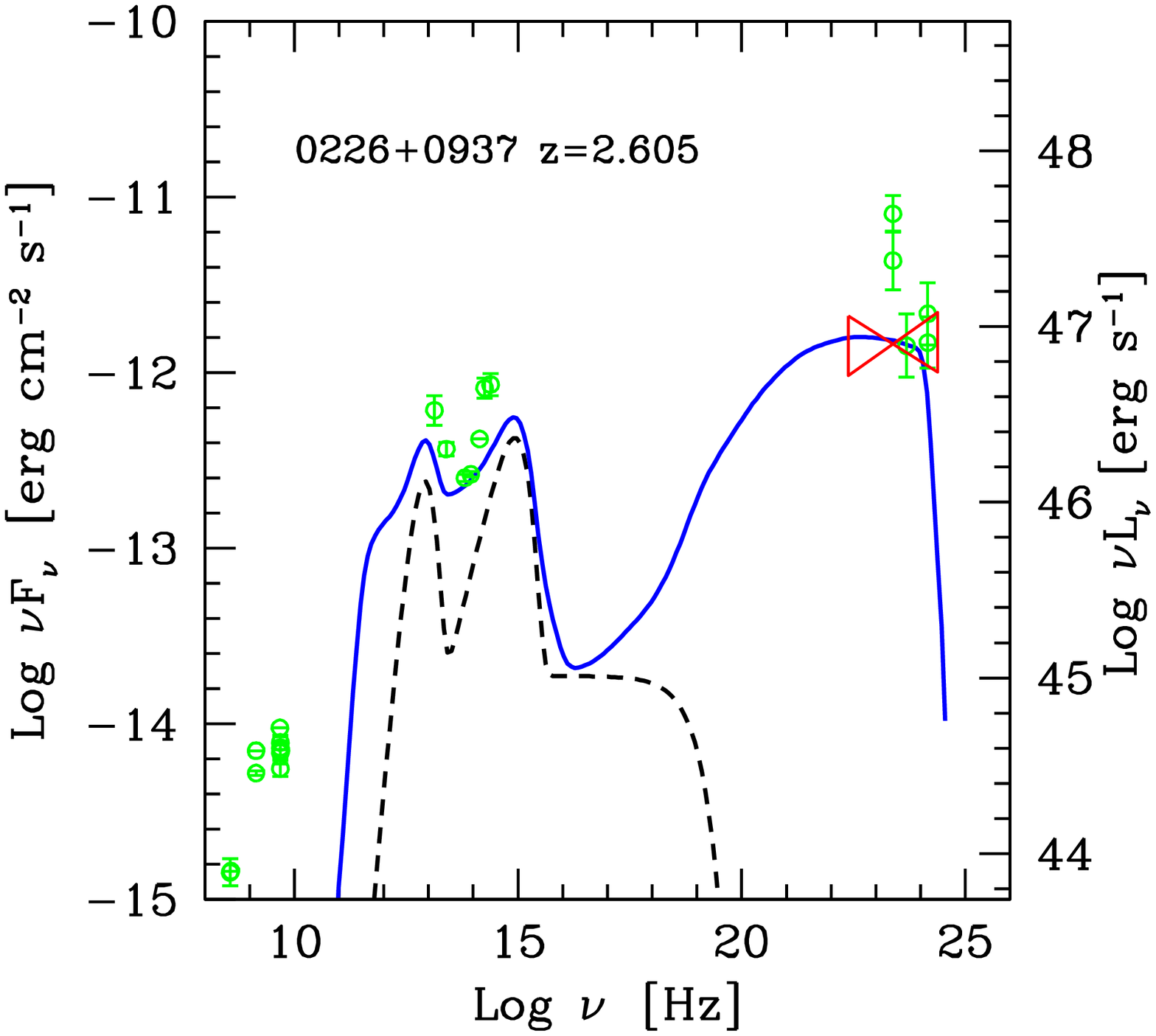,width=4.3cm,height=3.7cm} 
&\psfig{file=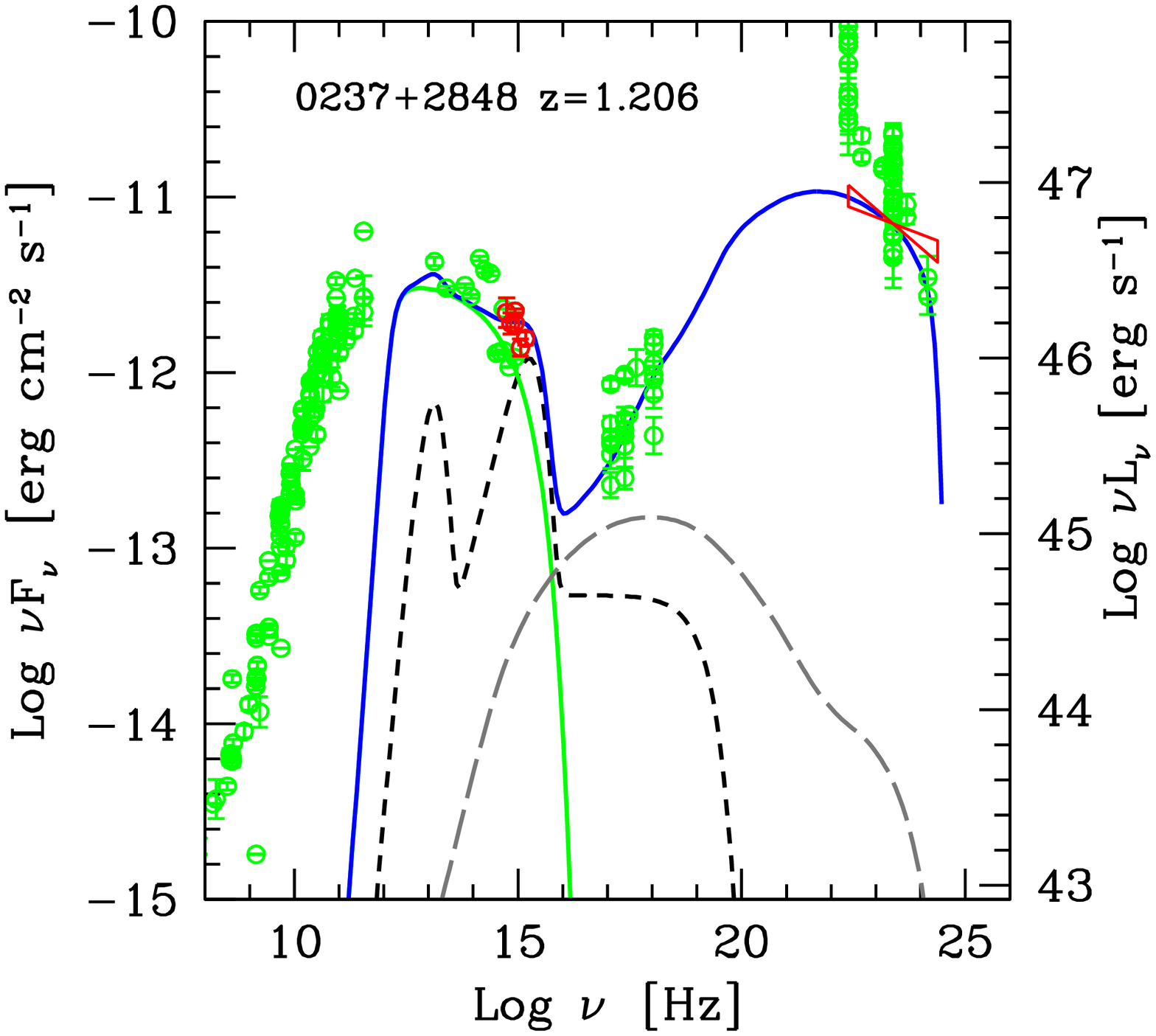,width=4.3cm,height=3.7cm} \\
\psfig{file=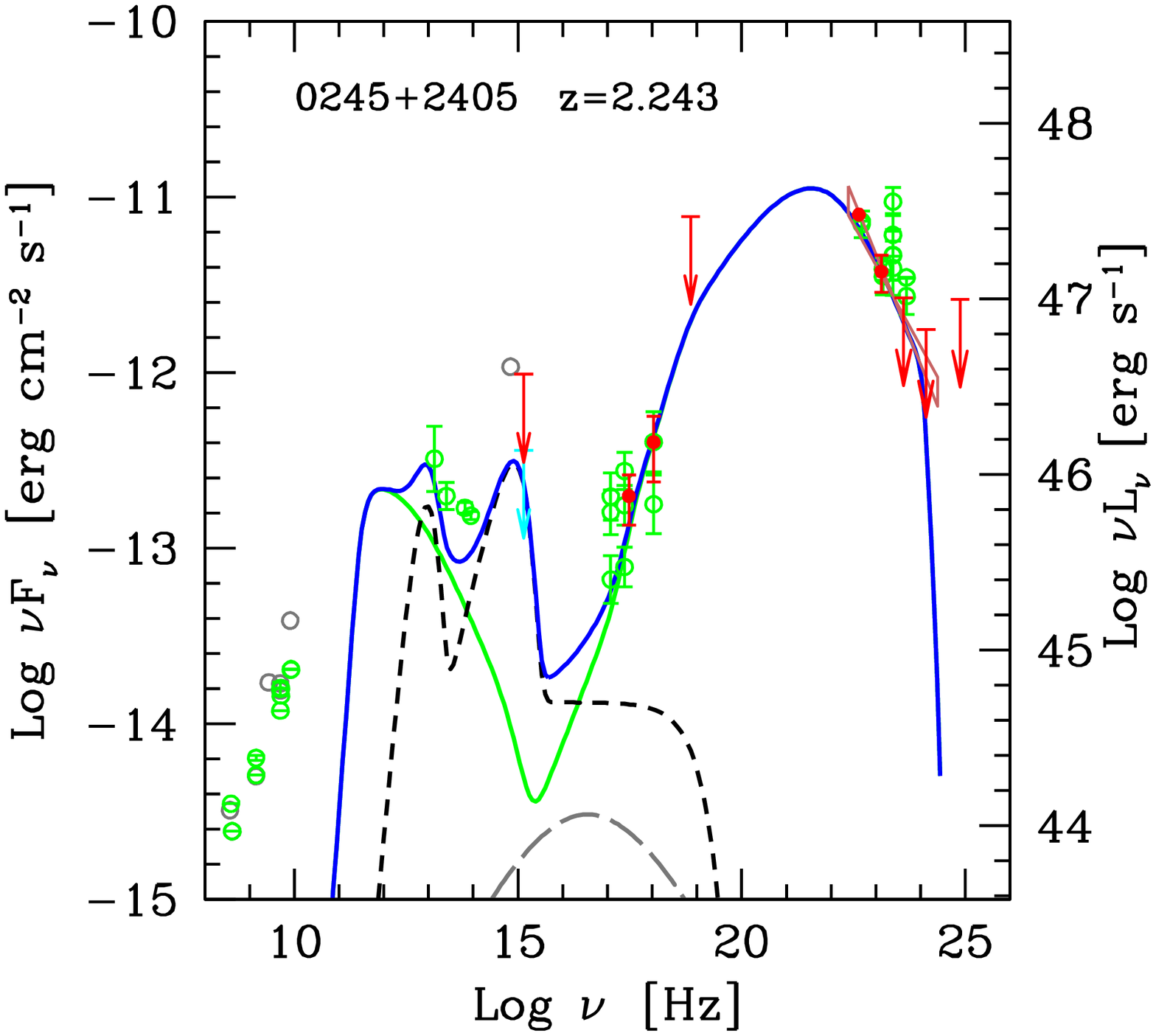,width=4.3cm,height=3.7cm} 
&\psfig{file=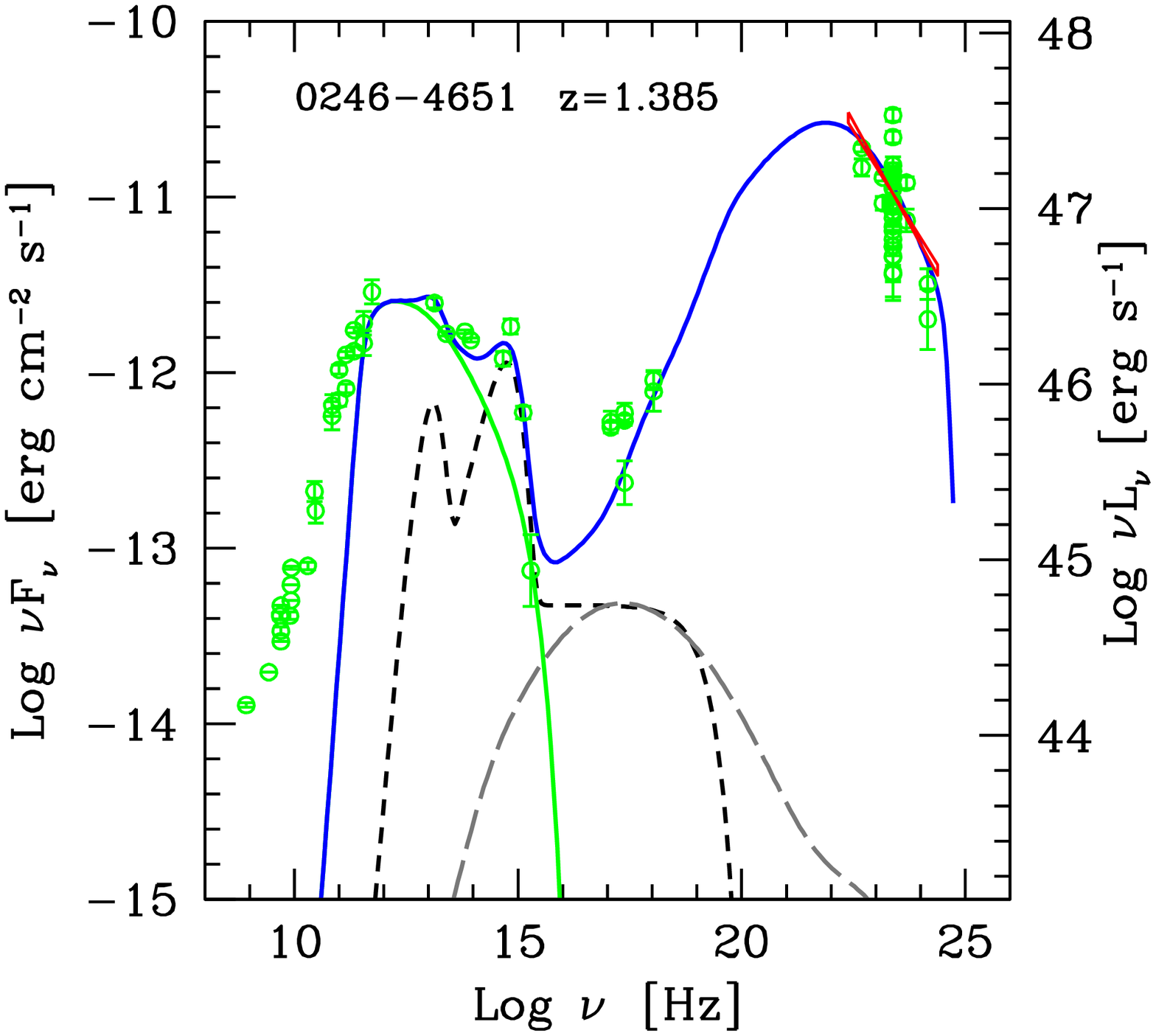,width=4.3cm,height=3.7cm}
&\psfig{file=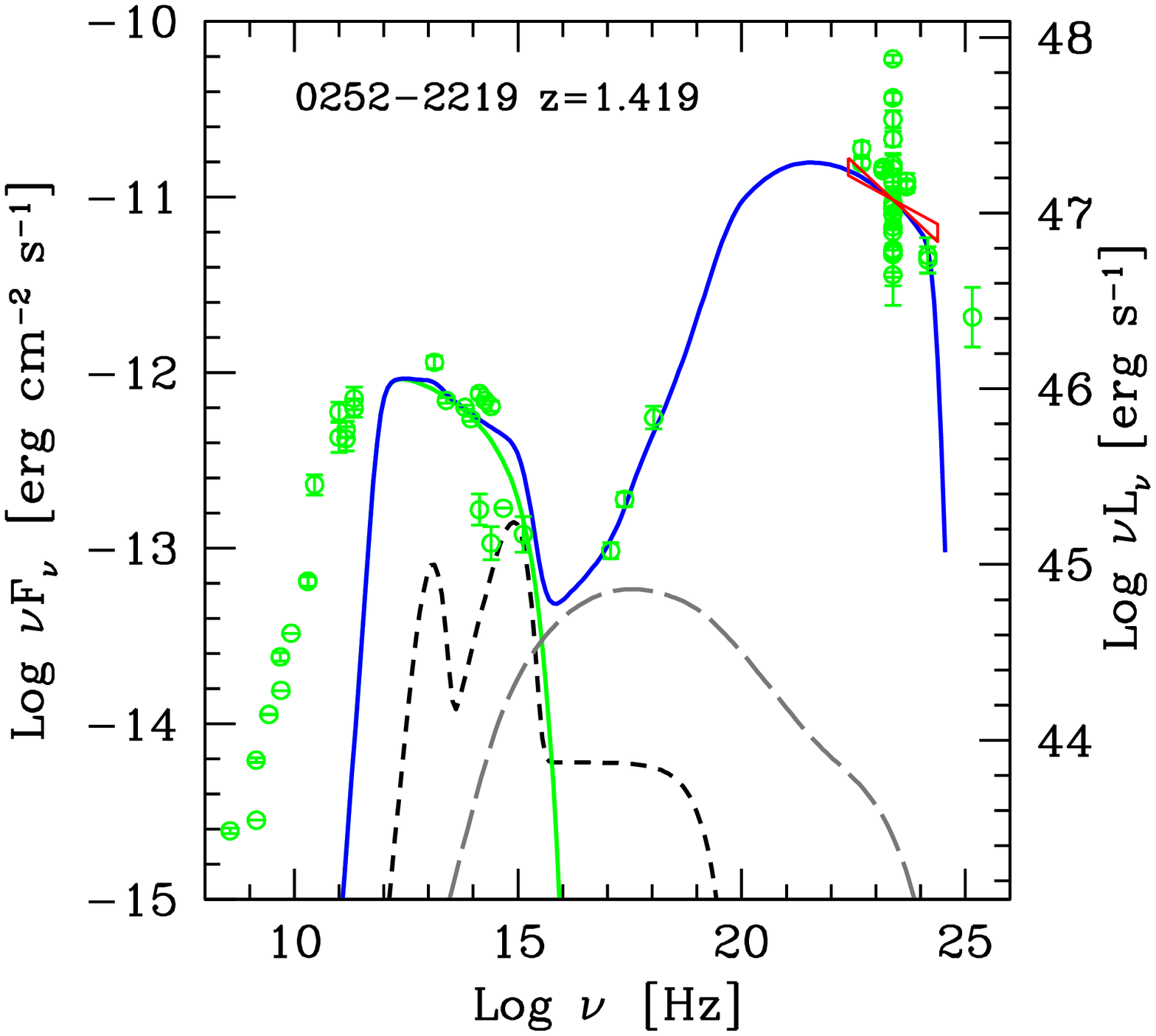,width=4.3cm,height=3.7cm} 
&\psfig{file=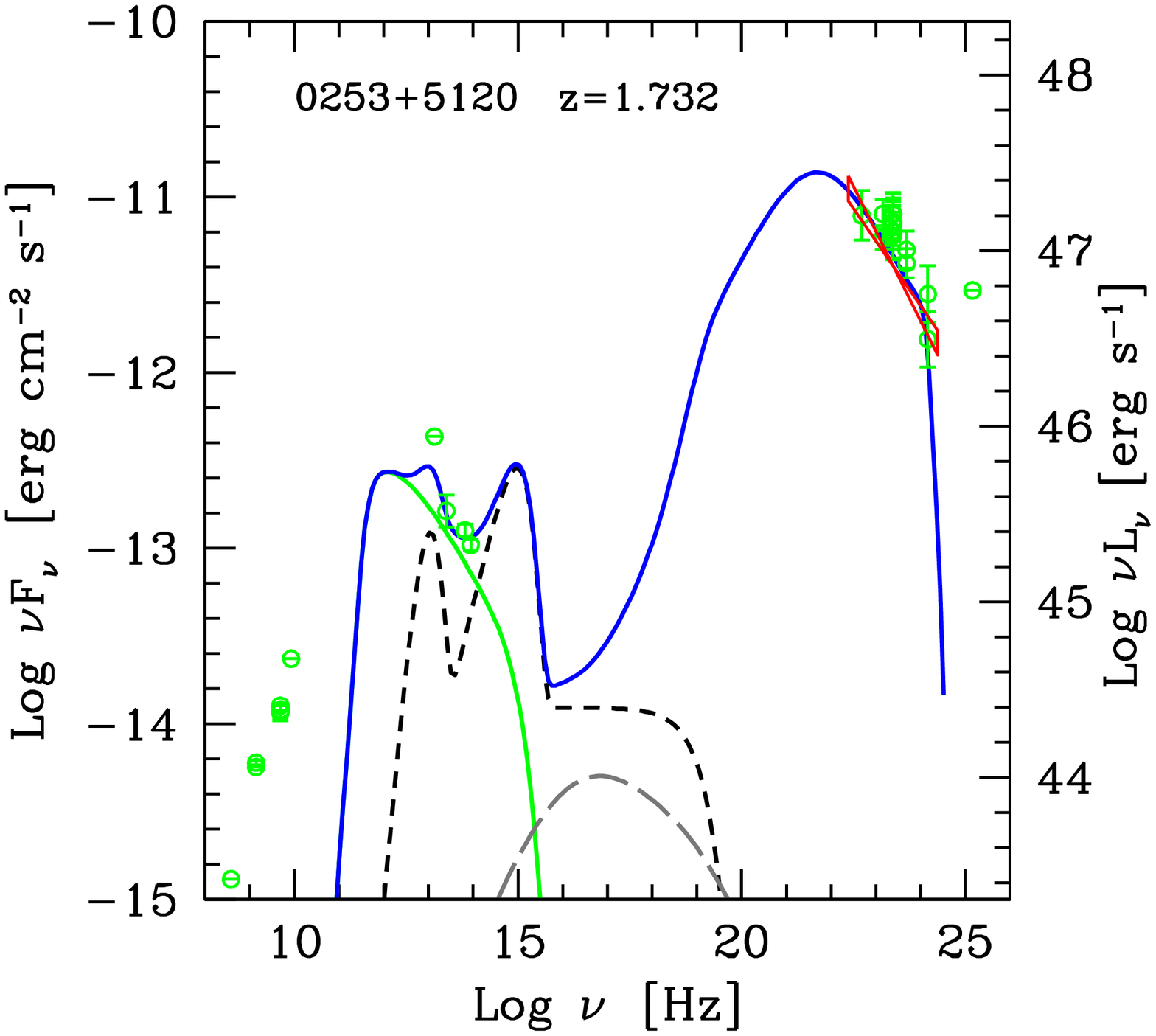,width=4.3cm,height=3.7cm} 
\end{tabular}
\caption{SED of the FSRQs studied in this paper. The blue solid line
refers to the applied model, the dotted line is the torus+accretiondisc+X--ray
corona component, the long dashed line is the SSC component.
Green symbols: archival data from ASDC, red tie: {\it Fermi}/LAT data;
black line: optical SDSS spectrum.
}
\label{sed1} 
\end{figure*}

\vskip 0.5 cm
\noindent
This appendix contains the list of the objects in our sample (FSRQs in Tab. \ref{sample}
and BL Lacs in Tab. \ref{bllacs}), the list of the parameters used for the fitting models
(FSRQs in Tab. \ref{para} and BL Lacs in Tab. \ref{parabl}) and the figures reporting
the SEDs (with models) for all FSRQ (Fig. \ref{sed1}). 
The SEDs (and models) for BL Lacs are reported in Sbarrato et al. (2014).

\begin{table*} 
\centering
\begin{tabular}{llllllllllll}
\hline
\hline
Name      &Alias &$z$  &$\log M_{\rm H\beta}$  &$\log M_{\rm MgII}$ &$\log M_{\rm CIV}$ &$\log M_{\rm fit}$ &Note\\
\hline 
0004 --4736 &CRATES J0004--4736 &0.880 &...  &7.85 &...  &(8)    &---\\
0011 +0057  &PMN J0011+0057     &1.493 &...  &8.80 &8.09 &8.5    &   \\
0015 +1700  &CRATES J0015+1700  &1.709 &...  &9.36 &9.15 &(9.18) &---\\
0017 --0512 &CRATES J0017--0512 &0.226 &7.55 &...  &...  &8      &    \\
0023 +4456  &TXS 0020+446       &2.023 &...  &...  &7.78 &8.3    &*  \\
0024 +0349  &GB6 J0024+0349     &0.545 &...  &7.76 &...  &8      &* \\
0042 +2320  &CRATES J0042+2320  &1.426 &...  &9.01 &...  &8.6    &*  \\
0043 +3426  &GB6 J0043+3426     &0.966 &...  &8.01 &...  &(8.18) &---\\
0044 --8422 &CRATES J0044--8422 &1.032 &...  &8.68 &...  &(8.5)  &---  \\
0048 +2235  &CLASS J0048+2235   &1.161 &...  &8.43 &8.25 &(8.5)  &---  \\
0050 --0452 &CRATES J0050--0452 &0.922 &...  &8.20 &...  &(8.3)  &---  \\
0058 +3311  &CRATES J0058+3311  &1.369 &...  &8.01 &7.97 &(8)    &---  \\
0102 +4214  &GB6 J0102+4214     &0.874 &7.92 &8.49 &...  &(8.6)  &---  \\
0102 +5824  &TXS 0059+581       &0.644 &...  &8.57 &...  &8.5    & \\
0104 --2416 &PKS 0102--245      &1.747 &...  &8.85 &8.97 &8.8    &* \\
0157 --4614 &PMN J0157--4614    &2.287 &...  &7.98 &8.52 &8.6    & \\
0203 +3041  &TXS 0200+304       &0.955 &...  &8.02 &...  &(8)    &---  \\
0217 --0820 &PMN J0217--0820    &0.607 &...  &6.53 &...  &(7.5)  &1  \\
0226 +0937  &TXS 0223+093       &2.605 &...  &...  &9.65 &9      & \\
0237 +2848  &CRATES J0237+2848  &1.206 &...  &9.22 &...  &(8.6)  &2 \\
0245 +2405  &B2 0242+23         &2.243 &...  &9.02 &9.18 &9      &* \\
0246 --4651 &PKS 0244--470      &1.385 &...  &8.48 &8.32 &9.6    & \\
0252 --2219 &CRATES J0252--2219 &1.419 &...  &9.40 &...  &8.85   &  \\
0253 +5102  &TXS 0250+508       &1.732 &...  &9.11 &8.37 &8.85   &* \\
0257 --1212 &CRATES J0257--1212 &1.391 &...  &9.22 &...  &9      &  \\
0303 --6211 &CRATES J0303--6211 &1.348 &...  &9.76 &...  &9.3    & \\
0309 --6058 &CRATES J0309--6058 &1.479 &...  &8.87 &...  &(8.7)  &--- \\
0315 --1031 &PKS 0313--107      &1.565 &...  &7.17 &8.33 &(8.6)  &--- \\
0325 --5629 &CRATES J0325--5629 &0.862 &...  &8.68 &...  &8.6    & \\
0325 +2224  &TXS 0322+222       &2.066 &...  &9.50 &9.16 &9      & \\
0407 +0742  &CRATES J0407+0742  &1.133 &...  &8.65 &...  &(8.6)  &--- \\
0413 --5332 &CRATES J0413--5332 &1.024 &...  &7.83 &...  &8.3    & \\
0422 --0643 &CRATES J0422--0643 &0.242 &7.47 &...  &...  &(7.85) &--- \\
0430 --2507 &CRATES J0430--2507 &0.516 &...  &6.51 &...  &(7.5)  &1 \\
0433 +3237  &CRATES J0433+3237  &2.011 &...  &9.17 &9.20 &8.9    & \\
0438 --1251 &CRATES J0438--1251 &1.285 &...  &8.66 &...  &8.5    &* \\
0442 --0017 &CRATES J0438--1251 &0.845 &7.74 &8.46 &...  &8.6    & \\
0448 +1127  &CRATES J0448+1127  &1.370 &...  &9.44 &...  &8.6    & \\
0449 +1121  &CRATES J0449+1121  &2.153 &...  &...  &7.89 &(8.8)  &1 \\
0456 --3136 &CRATES J0456--3136 &0.865 &7.78 &8.61 &...  &8.3    &* \\
0507 --6104 &CRATES J0507--6104 &1.089 &...  &8.74 &...  &(8.7)  &--- \\
0509 +1011  &CRATES J0509+1011  &0.621 &8.03 &8.52 &...  &(8)    &2 \\
0516 --6207 &PKS 0516--621      &1.30  &...  &7.93 &8.52 &(8.3)  &--- \\
0526 --4830 &PKS 0524--485      &1.30  &...  &9.15 &8.46 &(8.3)  &2  \\
0532 +0732  &CRATES J0532+0732  &1.254 &...  &8.43 &...  &(8.6)  &--- \\
0533 --8324 &CRATES J0533--8324 &0.784 &...  &7.40 &...  &7.7    & \\
0533 +4822  &TXS 0529+483       &1.160 &...  &9.25 &...  &8.85   & \\
0541 --0541 &PKS 0539--057      &0.838 &...  &8.74 &...  &8.7    &*  \\
0601 --7036 &CRATES J0601--7036 &2.409 &...  &...  &7.36 &(8.7)  &1 \\
0607 --0834 &PKS 0605--08       &0.870 &8.63 &9.02 &...  &8.8    & \\
0608 --1520 &CRATES J0608--1520 &1.094 &...  &8.09 &...  &(8.3)  &---  \\
0609 --0615 &PMN J0609--0615    &2.219 &...  &...  &8.89 &9      & \\
\hline
\hline 
\end{tabular}
\vskip 0.1 true cm
\caption{
The FSRQs in our sample. The name is the same of Shaw et al. (2012), the alias
helps to find the object in NED.
The third column is the redshift, the next three columns report the
black hole masses calculated through the virial method by Shaw et al. (2012) 
using the H$\beta$, MgII and the CIV line. 
The penultimate column reports the black hole mass estimated in this paper through the disc fitting method.
Values in parenthesis could not be evaluated through the disc--fitting method.
In most cases these values are equal to the virial masses (within a factor 2, denoted by ``---" in the
last column).
In a minority of cases they differ from the virial values for the following reasons, as flagged
in the last column:
1) a value larger than the virial one has been adopted to avoid super Eddington or nearly Eddington
disc luminosities;
2) a value smaller than the virial one has been adopted to avoid to overproduce the NIR--optical flux;
3) a value larger than the virial one has been adopted to avoid to overproduce the optical--UV flux.
When the disc--fitting method could use only one point to find a value for the black hole mass, the
last column reports an asterisk. This is the value used for the jet model.
}
\label{sample}
\end{table*}

\setcounter{table}{2}
\begin{table*} 
\centering
\begin{tabular}{lllllllllllll}
\hline
\hline
Name      &Alias &$z$  &$\log M_{\rm H\beta}$  &$\log M_{\rm MgII}$ &$\log M_{\rm CIV}$ &$\log M_{\rm fit}$ &Note\\
\hline 
0625 --5438 &CRATES J062 5--5438 &2.051 &...  &8.40 &9.07 &9      & \\
0645 +6024  &BZU J0645+6024     &0.832 &9.09 &9.56 &...  &8.85   & \\
0654 +4514  &CRATES J0654+4514  &0.928 &...  &8.17 &...  &(8.3)  &--- \\
0654 +5042  &CRATES J0654+5042  &1.253 &...  &7.86 &8.79 &(8.5)  &---  \\
0713 +1935  &CLASS J0713+1935   &0.540 &7.33 &7.91 &...  &(8)  &--- \\
0721 +0406  &PMN J0721+0406     &0.665 &8.49 &9.12 &...  &9     \\
0723 +2859  &GB6 J0723+2859     &0.966 &...  &8.40 &...  &8.52  \\
0725 +1425  &PKS 0722+145       &1.038 &...  &8.31 &...  &8.8   \\
0746 +2549  &CRATES J0746+2549  &2.979 &...  &...  &9.23 &9     \\
0805 +6144  &CRATES J0805+6144  &3.033 &...  &...  &9.07 &9     \\
0825 +5555  &TXS 0820+560       &1.418 &...  &9.10 &...  &9     \\
0830 +2410  &TXS 0827+243       &0.942 &...  &8.70 &...  &9     \\
0840 +1312  &PKS 0838+13        &0.680 &8.37 &8.62 &...  &8.6   \\
0847 --2337 &CRATES J0847--2337 &0.059 &8.30 &...  &...  &(8)   &--- \\
0856 +2111  &CRATES J0856+2111  &2.098 &...  &9.96 &9.77 &8.7   \\
0909 +0121  &CRATES J0909+0121  &1.026 &...  &9.14 &...  &9.18  \\
0910 +2248  &CRATES J0910+2248  &2.661 &...  &...  &8.70 &8.6   \\
0912 +4126  &TXS 0908+416       &2.563 &...  &...  &9.32 &9.3   \\
0920 +4441  &CRATES J0920+4441  &2.189 &...  &...  &9.29 &9.3   \\
0921 +6215  &CRATES J0921+6215  &1.453 &...  &8.93 &...  &8.9   \\
0923 +2815  &CRATES J0923+2815  &0.744 &8.61 &9.04 &...  &8.3   \\
0923 +4125  &B3 0920+416        &1.732 &...  &7.68 &8.16 &(8)   &--- \\
0926 +1451  &CLASS J0926+1451   &0.632 &...  &7.11 &...  &(7.6) &--- \\
0937 +5008  &CRATES J0937+5008  &0.276 &7.50 &...  &...  &(7.6) &--- \\
0941 +2778  &CRATES J0941+2728  &1.305 &...  &8.63 &...  &8.6    \\
0946 +1017  &CRATES J0946+1017  &1.006 &...  &8.47 &...  &8.95   \\
0948 +0022  &PMN J0948+0022     &0.585 &7.46 &7.73 &...  &8.18   \\
0949 +1752  &CRATES J0949+1752  &0.693 &...  &8.10 &...  &8.1    \\
0956 +2515  &CRATES J0956+2515  &0.708 &8.30 &8.63 &...  &8.7    \\
0957 +5522  &CRATES J0957+5522  &0.899 &...  &8.45 &...  &(9)   &---  \\
1001 +2911  &CRATES J1001+2911  &0.558 &7.31 &7.64 &...  &(8)   &---  \\
1012 +2439  &CRATES J1012+2439  &1.800 &...  &7.73 &7.86 &8.8    \\
1016 +0513  &CRATES J1016+0513  &1.714 &...  &8.34 &7.64 &(8.8) &3    \\
1018 +3542  &CRATES J1018+3542  &1.228 &...  &9.10 &...  &9.18     \\
1022 +3931  &B3 1019+397        &0.604 &...  &8.95 &...  &8.7   &* \\
1032 +6051  &CRATES J1032+6051  &1.064 &...  &8.75 &...  &8.3     \\
1033 +4116  &CRATES J1033+6051  &1.117 &...  &8.61 &...  &8.3     \\
1033 +6051  &CRATES J1033+6051  &1.401 &...  &9.09 &...  &(8.3) &2    \\
1037 --2823 &CRATES J1037--2823 &1.066 &...  &8.99 &...  &(8.8) &---   \\
1043 +2408  &CRATES J1043+2408  &0.559 &...  &8.09 &...  &(7.6) &3   \\
1058 +0133  &CRATES J1058+0133  &0.888 &...  &8.37 &...  &(8.3) &---   \\
1106 +2812  &CRATES J1106+2812  &0.843 &...  &8.85 &...  &9.54  &* \\
1112 +3446  &CRATES J1112+3446  &1.956 &...  &8.74 &8.82 &8.85  \\
1120 +0704  &CRATES J1120+0704  &1.336 &...  &8.83 &...  &(8.8) &---  \\
1124 +2336  &CRATES J1124+2336  &1.549 &...  &8.79 &...  &(8.5) &---  \\
1133 +0040  &CRATES J1133+0040  &1.633 &...  &8.80 &...  &8.7  \\
1146 +3958  &CRATES J1146+3958  &1.088 &...  &8.93 &...  &8.8  \\
1152 --0841 &CRATES J1152--0841 &2.367 &...  &...  &9.38 &(9.3) &---  \\
1154 +6022  &CRATES J1154+6022  &1.120 &...  &8.94 &...  &(8.8) &---  \\
1155 --8101 &PMN J1155--8101    &1.395 &...  &8.30 &...  &(8.6) &---  \\
1159 +2914  &CRATES J1159+2914  &0.725 &8.14 &8.61 &...  &8.5  \\
1208 +5441  &CRATES J1208+5441  &1.344 &...  &8.40 &...  &8.6  \\
1209 +1810  &CRATES J1209+1810  &0.845 &8.26 &8.77 &...  &8.6  \\
1222 +0413  &CRATES J1222+0413  &0.966 &...  &8.37 &...  &9  \\
1224 +2122  &CRATES J1224+2122  &0.434 &8.89 &8.91 &...  &8.9  \\
1224 +5001  &CLASS J1224+5001   &1.065 &...  &8.66 &...  &8.9  \\
1226 +4340  &B3 1224+439        &2.001 &...  &8.64 &9.01 &9  \\
1228 +4858  &TXS 1226+492       &1.722 &...  &8.28 &8.23 &8.5  \\
1228 +4858  &TXS 1226+492       &1.722 &...  &8.28 &8.23 &8.5  \\
\hline
\hline 
\end{tabular}
\caption{{\it continue}
}
\label{sample}
\end{table*}
\setcounter{table}{2}
\begin{table*} 
\centering
\begin{tabular}{lllllllllllll}
\hline
\hline
Name      &Alias &$z$  &$\log M_{\rm H\beta}$  &$\log M_{\rm MgII}$ &$\log M_{\rm CIV}$ &$\log M_{\rm fit}$ &Note\\
\hline 
1239 +0443  &CRATES J1239+0443  &1.761 &...  &8.46 &8.68 &8.7  \\
1239 +0443  &CRATES J1239+0443  &1.761 &...  &8.46 &8.68 &8.7  \\
1257 +3229  &CRATES J1257+3229  &0.806 &7.89 &8.62 &...  &8.6   \\
1303 --4621 &CRATES J1303--4621 &1.664 &...  &7.95 &8.21 &(8.3) &--- \\
1310 +3220  &CRATES J1310+3220  &0.997 &...  &8.57 &...  &(8.7) &---   \\
1317 +3425  &CRATES J1317+3425  &1.055 &...  &9.14 &...  &8.8   \\
1321 +2216  &CRATES J1321+2216  &0.943 &7.87 &8.76 &...  &9.6   &*  \\
1327 +2210  &CRATES J1327+2210  &1.403 &...  &9.25 &...  &8.7   \\
1332 --1256 &CRATES J1332--1256 &1.492 &...  &8.96 &8.61 &(9) &---   \\
1333 +5057  &CLASS J1333+5057   &1.362 &...  &7.95 &...  &(8.3) &---   \\
1343 +5754  &CRATES J1343+5754  &0.933 &...  &8.42 &...  &8.7  \\
1344 --1723 &CRATES J1344--1723 &2.506 &...  &...  &9.12 &(8.9) &---  \\
1345 +4452  &CRATES J1345+4452  &2.534 &...  &...  &8.98 &9.3  \\
1347 --3750 &CRATES J1347--3750 &1.300 &...  &7.95 &8.62 &(8.3) &---   \\
1350 +3034  &CRATES J1350+3034  &0.712 &8.21 &8.33 &...  &8.3   &* \\
1357 +7643  &CRATES J1357+7643  &1.585 &...  &8.34 &8.17 &(8) &---   \\
1359 +5544  &CRATES J1359+5544  &1.014 &...  &8.00 &...  &8.18 \\
1423 --7829 &CRATES J1423--7829 &0.788 &8.14 &8.32 &...  &(8.3) &---   \\
1436 +2321  &PKS B1434+235      &1.548 &...  &8.12 &8.50 &8.85   \\
1438 +3710  &CRATES J1438+3710  &2.399 &...  &...  &8.58 &(9.7) &1   \\
1439 +3712  &CRATES J1439+3712  &1.027 &...  &9.08 &...  &9   \\
1441 --1523 &CRATES J1441--1523 &2.642 &...  &...  &8.49 &8.9   \\
1443 +2501  &CRATES J1443+2501  &0.939 &7.42 &7.84 &...  &8.6   \\
1504 +1029  &TXS 1502+106       &1.839 &...  &8.98 &8.90 &8.8   \\
1504 +1029  &TXS 1502+106       &1.839 &...  &8.98 &8.90 &8.8   \\
1505 +0326  &CRATES J1505+0326  &0.409 &...  &7.41 &...  &7.95   \\
1514 +4450  &BZQ J1514+4450     &0.570 &7.72 &7.62 &...  &(8) &---   \\
1522 +3144  &CRATES J1522+3144  &1.484 &...  &8.92 &...  &9   &*   \\
1539 +2744  &CRATES J1539+2744  &2.191 &...  &8.43 &8.51 &(8.5) &---   \\
1549 +0237  &CRATES J1549+0237  &0.414 &8.62 &8.72 &...  &8.6   \\
1550 +0527  &CRATES J1550+0527  &1.417 &...  &8.98 &...  &(8.6) &2    \\
1553 +1256  &PKS 1551+130       &1.308 &...  &8.64 &...  &9.3   \\
1608 +1029  &CRATES J1608+1029  &1.232 &...  &8.77 &...  &8.8   \\
1613 +3412  &CRATES J1613+3412  &1.400 &...  &9.08 &...  &9.18 \\
1616 +4632  &CRATES J1616+4632  &0.950 &...  &8.28 &...  &8.6   \\
1617 --5848 &PMN J1617--5848    &1.422 &...  &9.81 &9.01 &(9.3) &---   \\
1624 --0649 &PMN J1624-0649     &3.037 &...  &...  &8.23 &9.08 \\
1628 --6152 &PMN J1628-6152     &2.578 &...  &...  &8.92 &(8.85) &---  \\
1635 +3808  &CRATES J1635+3808  &1.813 &...  &9.30 &8.85 &9.6  \\
1635 +3808  &CRATES J1635+3808  &1.813 &...  &9.30 &8.85 &9.6  \\
1636 +4715  &BZQ J1636+4715     &0.823 &8.11 &8.38 &...  &8.6   \\
1637 +4717  &CRATES J1637+4717  &0.735 &8.61 &8.52 &...  &8.5   \\
1639 +4705  &CRATES J1639+4705  &0.860 &...  &8.95 &...  &8.8  \\
1642 +3940  &CRATES J1642+3948  &0.593 &8.73 &9.03 &...  &8.8  \\
1703 --6212 &CRATES J1703--6212 &1.747 &...  &8.65 &8.55 &9   \\
1709 +4318  &CRATES J1709+4318  &1.027 &...  &7.92 &...  &(8) &---   \\
1734 +3857  &CRATES J1734+3857  &0.975 &...  &7.97 &...  &(8.5) &---   \\
1736 +0631  &CRATES J1736+0631  &2.387 &...  &8.82 &9.39 &8.6  \\
1802 --3940 &PMN J1802--3940    &1.319 &...  &8.60 &8.59 &(9) &---  \\
1803 +0341  &CRATES J1803+0341  &1.420 &...  &7.79 &...  &(8) &---  \\
1818 +0903  &CRATES J1818+0903  &0.354 &7.30 &7.50 &...  &8   &*\\
1830 +0619  &TXS 1827+062       &0.745 &8.69 &8.86 &...  &(9) &---  \\
1848 +3219  &CRATES J1848+3219  &0.800 &7.87 &8.21 &...  &8.5  \\
1903 --6749 &CRATES J1903--6749 &0.254 &7.51 &...  &...  &(8) &---  \\
1916 --7946 &CRATES J1916--7946 &0.204 &7.82 &...  &...  &8   &* \\
1928 --0456 &CRATES J1928--0456 &0.587 &...  &9.07 &...  &8.6  \\
1954 --1123 &CRATES J1954--1123 &0.683 &...  &6.73 &...  &(8) &1   \\
1955 +1358  &NVSS J195511+135816 &0.743 &8.17 &8.39 &... &(8.5) &---  \\
1959 --4246 &PMN J1959--4246    &2.178 &...  &8.55 &9.41 &(8.9) &2   \\
\hline
\hline 
\end{tabular}
\caption{{\it continue}
}
\label{sample}
\end{table*}
\setcounter{table}{2}
\begin{table*} 
\centering
\begin{tabular}{lllllllllllll}
\hline
\hline
Name      &Alias &$z$  &$\log M_{\rm H\beta}$  &$\log M_{\rm MgII}$ &$\log M_{\rm CIV}$ &$\log M_{\rm fit}$ &Note\\
\hline 
2017 +0603  &1FGL J2017.2+0602  &1.743 &...  &9.39 &9.67 &9.85   \\
2025 --2845 &CRATES J2025--2845 &0.884 &...  &8.34 &...  &(8.4) &--- \\
2031 +1219  &CRATES J2031+1219  &1.213 &...  &7.99 &7.19 &(7.9) &---   \\
2035 +1056  &CRATES J2035+1056  &0.601 &7.74 &8.26 &...  &8.3   &* \\
2110 +0809  &CRATES J2110+0809  &1.580 &...  &8.82 &...  &9.18 \\
2118 +0013  &CRATES J2118+0013  &0.463 &7.60 &7.89 &...  &8   \\
2121 +1901  &CRATES J2121+1901  &2.180 &...  &...  &7.75 &8.7   \\
2135 --5006 &CRATES J2135--5006 &2.181 &...  &8.31 &8.40 &8.6   \\
2139 --6732 &CRATES J2139--6732 &2.009 &...  &8.49 &8.93 &8.7   \\
2145 --3357 &CRATES J2145--3357 &1.361 &...  &8.31 &...  &(2.8.7) &--- \\
2157 +3127  &CRATES J2157+3127  &1.448 &...  &8.89 &...  &(8.6) &2   \\
2202 --8338 &CRATES J2202--8338 &1.865 &...  &9.02 &9.16 &9     &* \\
2212 +2355  &PKS 2209+236       &1.125 &...  &8.46 &...  &8.6   \\
2219 +1806  &CRATES J2219+1806  &1.071 &...  &7.65 &7.66 &(8.3) &1   \\
2229 --0832 &PKS 2227--08       &1.560 &...  &8.70 &8.54 &8.9  \\
2236 +2828  &CRATES J2236+2828  &0.790 &...  &8.35 &...  &(8.6) &---   \\
2237 --3921 &CRATES J2237--3921 &0.297 &7.77 &7.95 &...  &8  \\
2244 +4057  &CRATES J2244+4057  &1.171 &...  &8.28 &...  &(8.3) &---   \\
2315 --5018 &CRATES J2315--5018 &0.808 &...  &7.68 &...  &(7.85) &---   \\
2321 +3204  &CRATES J2321+3204  &1.489 &...  &8.66 &8.75 &(8.6) &---   \\
2327 +0940  &CRATES J2327+0940  &1.841 &...  &8.70 &9.35 &9   \\
2331 --2148 &CRATES J2331--2148 &0.563 &7.53 &7.63 &...  &8     &* \\
2334 +0736  &CRATES J2334+0736  &0.401 &8.37 &...  &...  &8.6   \\
2345 --1555 &CRATES J2345--1555 &0.621 &8.16 &8.48 &...  &8.3  \\
2357 +0448  &PMN J2357+0448     &1.248 &...  &8.41 &8.45 &8.95  \\
\hline
\hline 
\end{tabular}
\caption{{\it continue}
}
\label{sample}
\end{table*}


\begin{table*} 
\centering
\begin{tabular}{llllllllll}
\hline
\hline
Name    &Alias  &$z$ &$\log M_{\rm fit}$ \\
\hline 
0013 +1907  &CRATES J0013+1910  &0.477 &8.3\\
0203 +3042  &NVSS J020344+304238 &0.761 &8.8\\
0334 --4008 &PKS 0332--403      &1.357 &8.6 \\
0434 --2014 &CRATES J0434--2015 &0.928 &8 \\
0438 --4521 &PKS 0437--454      &2.017 &8.5 \\
0516 --6207 &PKS 0516--621      &1.300 &8.6 \\
0629 --2001 &PKS 0627--199      &1.724 &8.5 \\
0831 +0429  &CRATES J0831+0429  &0.174 &8.5 \\
1117 +2013  &CRATES J1117+2014  &0.138 &8.6 \\
1125 --3559 &CRATES J1125--3557 &0.284 &8.8 \\
1203 +6030  &CRATES J1203+6031  &0.065 &8.6 \\
1221 +2814  &W Com              &0.103 &8.6 \\
1221 +3010  &1ES 1218+304       &0.184 &8.6 \\
1420 +5422  &OQ +530            &0.153 &8.3 \\
1534 +3720  &1RXS J153446.6+371610 &0.144 &8.6 \\
1540 +1438  &PKS 1538+149       &0.606 &8.5 \\
1755 --6423 &CRATES J1754--6423 &1.255 &8.5 \\
1824 +5651  &CRATES J1824+5651  &0.664 &8.5 \\
2015 +3709  &TXS 2013+370       &0.859 &8.7 \\
2152 +1735  &CRATES J2152+1734  &0.874 &8.8 \\
2206 --0029 &CRATES J2206--0031 &1.053 &8.5 \\
2206 +6500  &TXS 2206+650       &1.121 &8.85 \\
2236 +2828  &B2 2234+28A        &0.790 &8.85 \\
2247 --0002 &PKS 2244--002      &0.949 &8.8 \\
2315 --5018 &PKS 2312--505      &0.811 &8 \\
2353 --3034 &PKS 2351--309      &0.737 &8 \\
\hline
\hline 
\end{tabular}
\caption{The BL Lac objects considered in this paper, drawn from the sample of S13.
The first column is the name on S13, the alias helps to find the source in NED, the third
column is the redshift, the fourth column is the mass used in the fitting model.
}
\label{bllacs}
\end{table*}

\begin{table*} 
\centering
\begin{tabular}{llllllllllllllllll}
\hline
\hline
Name  &$z$  &$L_{\rm d, 45}$ &$P^\prime_{\rm e, 45}$ &$B$ &$R_{\rm diss}$ &$\Gamma$  &$\gamma_{\rm max}$ &$\gamma_{\rm b}$ &$\gamma_{\rm peak}$ &$\gamma_{\rm cool}$                                                       
                                                   &$U^\prime$   &$s_1$  &$s_2$  &$\dot M_{\rm in}$ &$\dot M_{\rm out}$  &$\log L_{\rm d, S12}$  \\
~[1]     &[2]   &[3]   &[4]   &[5]   &[6]   &[7]   &[8]   &[9]   &[10]   &[11]   &[12]   &[13]   &[14]   &[15]  &[16]  &[17]\\
\hline 
0004 --4736 &0.880 &2.1  &8.e--4  &13.3 &600  &14     &4.e3 &70  &63.7  &28.1 &14.6  &0   &2.8  &0.46   &3.8e--3 &45.11   \\ 
0011 +0057  &1.493 &3.6  &5.e--3  &2.98 &900  &13     &4e3  &100 &41.5  &17.  &5.3  &1.5  &2.8  &0.79   &0.13    &45.60  \\ 
0015 +1700  &1.709 &22.5 &4.5e--3 &1.82 &800  &12     &4e3  &100 &82.9  &4.7  &4.3   &0   &2.8  &4.93   &0.021   &46.26  \\ 
0017 --0512 &0.226 &0.45 &1.e--3  &5.55 &500  &12     &6.e3 &10  &82.2  &82.2 &5.96  &1   &2.9  &0.10   &0.023   &45.30   \\ 
0023 +4456  &2.023 &2.7  &7.3e--3 &7.02 &700  &15     &3.e3 &80  &87.5  &19.4 &9.03  &0   &2.5  &0.59   &0.034   &45.28 \\ 
0024 +0349  &0.545 &0.6  &1.5e--4 &7.69 &600  &13     &5.e3 &70  &99.6  &55.6 &7.64  &0   &2.4  &0.13   &3.9e--4 &44.80   \\ 
0042 +2320  &1.426 &4.2  &8.e--3  &3.96 &600  &13     &5.e3 &90  &51.5  &17.6 &5.76  &1   &3.0  &0.92   &0.11    &45.62  \\
0043 +3426  &0.966 &1.0  &1.2e--3 &3.77 &700  &14     &3.e3 &70  &173   &36.5 &6.39  &0   &2.1  &0.22   &3.3e--3 &45.02   \\ 
0044 --8422 &1.032 &7.65 &4.e--3  &5.66 &700  &14     &3.e3 &80  &46.5  &16.0 &7.29  &1   &3.0  &1.68   &0.062   &45.88  \\ 
0048 +2235  &1.161 &1.8  &3.5e--3 &2.37 &800  &13     &4e3  &130 &90.6  &19.8 &5.13  &1   &2.7  &0.39   &0.034   &45.26    \\ 
0050 --0452 &0.922 &2.1  &2.8e--3 &6.90 &600  &11     &5.e3 &40  &37.3  &35.4 &5.77  &1   &2.7  &0.46   &0.030   &45.35    \\ 
0058 +3311  &1.369 &1.5  &2.4e--3 &5.77 &900  &13     &5e3  &70  &93.4  &41.4 &6.54 &1.   &2.4  &0.33   &0.020   &45.21    \\ 
0058 +3311  &1.369 &1.5  &3.5e--3 &2.63 &1600 &16     &2e3  &30  &35.7  &19.8 &7.71 &1.5  &2.4  &0.33   &0.124   &45.21    \\ 
0102 +4214  &0.874 &6.   &3.e--3  &4.7  &600  &12     &5.e3 &150 &74.0  &19.3 &5.27  &1   &3.3  &1.31   &0.031   &45.83   \\ 
0102 +5824  &0.644 &4.5  &3.2e--3 &8.31 &600  &11     &5.e3 &40  &43.0  &20.6 &6.60  &1   &2.5  &0.99   &0.037   &45.66  \\ 
0104 --2416 &1.747 &8.1  &8.e--3  &5.82 &500  &11     &4e3  &100 &97.7  &15.8 &5.15  &0   &2.6  &1.77   &0.026   &45.95   \\ 
0157 --4614 &2.287 &6.0  &9.e--3  &3.44 &700  &14     &4e3  &200 &166.7 &13.8 &6.27  &0   &2.8  &1.31   &0.026   &45.73   \\ 
0203 +3041  &0.955 &0.3  &4.e--3  &3.9  &600  &13     &4.e3 &30  &70.4  &70.4 &7.52  &1   &2.9  &0.066  &0.021   &44.41  \\ 
0217 --0820 &0.607 &0.16 &1.3e--3 &11.4 &600  &13     &3.e3 &20  &103   &103  &13.2  &0   &2.7  &0.035  &6.5e--3 &44.06   \\ 
0226 +0937  &2.605 &42.  &5.e--3  &4.98 &600  &12     &6.e3 &80  &745.  &7.5  &5.35  &0   &1.9  &9.20   &9.7e--3 &46.53 \\ 
0237 +2848  &1.206 &18   &7.3e--3 &8.52 &600  &13     &3.e3 &80  &144.3 &11.8 &8.63  &1   &2.2  &3.94   &0.067   &46.38  \\ 
0245 +2405  &2.243 &21.  &1.7e--2 &2.25 &800  &14     &4e3  &250 &118.3 &5.15 &5.88  &1   &3.2  &4.60   &0.16    &46.33   \\ 
0246 --4651 &1.385 &24.  &1.6e--2 &1.77 &500  &12     &6e3  &400 &206.8 &16.2 &0.74  &1   &3.0  &5.26   &0.089   &46.43   \\ 
0252 --2219 &1.419 &3.15 &1.2e--2 &3.66 &500  &12     &4.e3 &80  &115.0 &14.4 &4.80  &1   &2.3  &0.69   &0.11    &45.72  \\ 
0253 +5102  &1.732 &10.5 &1.e--2  &2.28 &800  &14     &4e3  &150 &124.1 &7.4  &5.89  &0   &2.8  &2.30   &0.038   &45.99   \\ 
0257 --1212 &1.391 &22.5 &3.e--3  &3.41 &700  &11     &5.e3 &60  &54.3  &8.66 &4.02  &0   &2.7  &4.93   &0.017   &46.13  \\ 
0303 --6211 &1.348 &30.  &1.e--2  &4.71 &500  &11     &3.e3 &60  &53.8  &5.48 &4.46  &0   &2.7  &6.60   &0.059   &46.64  \\ 
0309 --6058 &1.479 &7.5  &1.3e--2 &4.66 &700  &13     &3.e3 &180 &140.5 &11.8 &5.89  &0   &3.0  &1.64   &0.043   &45.88  \\ 
0315 --1031 &1.565 &4.2  &4.e--3  &3.95 &550  &13     &7e3  &40  &46.7  &19.2 &5.77  &1   &2.45 &0.92   &0.053   &45.67   \\ 
0325 --5629 &0.862 &4.2  &1.2e--3 &4.07 &700  &10     &5.e3 &20  &33.2  &24.0 &3.62  &1   &2.4  &0.92   &0.015   &45.60   \\ 
0325 +2224  &2.066 &60.  &3.e--2  &5.95 &600  &12     &4e3  &100 &88.3  &6.8  &5.96  &0   &2.7  &13.1   &0.13    &46.79   \\ 
0407 +0742  &1.133 &6.0  &4.e--3  &6.60 &600  &12     &3.e3 &60  &49.4  &16.6 &6.14  &1   &2.6  &1.31   &0.050   &45.51  \\ 
0413 --5332 &1.024 &1.5  &3.5e--3 &4.03 &700  &12     &5.e3 &90  &64.2  &35.0 &4.97  &1   &2.8  &0.33   &0.032   &45.14  \\ 
0422 --0643 &0.242 &0.31 &6.e--4  &10.4 &500  &12     &6.e3 &10  &75.0  &75.0 &9.31  &1   &2.8  &0.070  &0.014   &44.42 \\ 
0430 --2507 &0.516 &0.09 &1.8e--4 &14.1 &500  &11     &5.e3 &500 &482   &112. &14.5  &1   &2.4  &0.020  &4.6e--4 &43.81   \\ 
0433 +3237  &2.011 &36.  &2.5e--3 &1.80 &1000 &13     &1.e4 &100 &5880  &6.1  &1.383 &1.5 &2.2  &7.88   &0.052   &46.58  \\ 
0438 --1251 &1.285 &6.3  &3.3e--3 &8.43 &500  &11     &5.e3 &40  &47.7  &22.5 &7.22  &0   &2.5  &1.38   &0.015   &45.78  \\ 
0442 --0017 &0.845 &6.   &6.e--3  &4.33 &600  &13     &5.e3 &170 &130.7 &17.2 &5.90  &0   &3.0  &1.31   &0.019   &45.81  \\ 
0448 +1127  &1.370 &24.  &5.e--3  &2.5  &1200 &16     &4.e3 &100 &200.1 &6.61 &7.65  &1   &2.2  &5.26   &0.050   &46.71  \\ 
0449 +1121  &2.153 &7.2  &1.3e--2 &0.86 &2400 &13$^d$ &6.e3 &300 &885.  &157. &0.11  &0   &2.0  &1.58   &0.035   &45.92 \\ 
0456 --3136 &0.865 &3.   &1.3e--3 &6.12 &600  &13     &5.e3 &70  &50.9  &29.6 &6.89  &1   &2.8  &0.66   &0.015   &45.26   \\ 
0507 --6104 &1.089 &6.75 &8.e--3  &3.42 &700  &12     &4.e3 &70  &56.1  &14.8 &4.72  &1   &2.6  &1.48   &0.094   &45.86  \\ 
0509 +1011  &0.621 &1.8  &1.e--3  &11.2 &600  &12     &5.e3 &120 &122.5 &37.8 &10.7  &0   &2.6  &0.39   &2.4e--3 &45.35 \\ 
0516 --6207 &1.30  &5.4  &5.2e--3 &7.65 &700  &12     &4.e3 &200 &303.2 &23.9 &7.29  &0   &2.2  &1.18   &7.9e--3 &45.75  \\ 
0526 --4830 &1.30  &6.0  &3.5e--3 &8.02 &600  &14     &4.e3 &100 &138.9 &21.3 &9.54  &1   &2.3  &1.31   &0.031   &45.87  \\ 
0532 +0732  &1.254 &9.   &1.1e--2 &4.92 &600  &14     &4.e3 &70  &75.9  &14.2 &7.11  &0   &2.5  &1.97   &0.055   &45.86   \\ 
0533 --8324 &0.784 &1.2  &1.e--3  &7.21 &1400 &12     &5.e3 &60  &119.4 &53.6 &6.49  &1   &2.3  &0.26   &7.4e--3 &44.73   \\ 
0533 +4822  &1.160 &20.  &6.e--3  &7.74 &600  &13     &3.e3 &150 &140.5 &7.7  &7.55  &--1 &2.8  &4.37   &0.016   &46.26  \\ 
0541 --0541 &0.838 &10.5 &3.e--3  &3.93 &700  &13     &5.e3 &50  &39.0  &12.4 &5.63  &1   &2.75 &2.30   &0.050   &46.06   \\ 
0601 --7036 &2.409 &6.0  &2.1e--2 &0.77 &2500 &14     &7.e3 &500 &908.  &130. &0.15  &0   &2.1  &1.31   &0.013   &45.69 \\ 
0607 --0834 &0.870 &17.1 &3.5e--3 &5.29 &600  &12     &5.e3 &60  &47.7  &12.3 &5.54  &1   &2.6  &3.75   &0.045   &46.33 \\ 
0608 --1520 &1.094 &3.6  &3.1e--3 &4.14 &900  &14     &4.e3 &50  &80.9  &21.0 &6.47  &1   &2.3  &0.79   &0.036   &45.51   \\ 
0609 --0615 &2.219 &33.  &2.5e--2 &1.79 &1100 &16     &4.e3 &200 &144.  &2.9  &7.51  &0   &3.1  &7.23   &0.10    &46.53 \\ 
0625 --5438 &2.051 &36.  &2.8e--2 &1.75 &900  &15     &5.e3 &240 &144.  &4.0  &6.63  &0   &4.0  &7.88   &0.12    &46.21  \\ 
0645 +6024  &0.832 &10.5 &1.8e--3 &5.98 &500  &12     &5.e3 &50  &40.8  &12.0 &5.84  &1   &2.6  &2.30   &0.026   &46.09 \\ 
0654 +4514  &0.928 &1.8  &5.e--3  &5.15 &600  &12     &5.e3 &60  &72.1  &35.9 &5.68  &0   &2.5  &0.39   &0.017   &45.25   \\ 
0654 +5042  &1.253 &0.95 &3.9e--3 &8.32 &500  &12     &5.e3 &100 &367.5 &22.0 &7.35  &0   &2.0  &0.21   &7.0e--3 &44.97  \\ 
\hline
\hline 
\end{tabular}
\caption{
Parameters used to model the SED.
Col. [1]: name (from S12);
Col. [2]: redshift;
Col. [3]: accretion disk luminosity in units of $10^{45}$ erg s$^{-1}$;
Col. [4]: power injected in the blob calculated in the comoving frame, in units of $10^{45}$ erg s$^{-1}$; 
Col. [5]: magnetic field in Gauss;
Col. [6]: dissipation radius in units of $R_{\rm S}$;
Col. [7]: bulk Lorentz factor at $R_{\rm diss}$;
Col. [8], [9]: maximum and break random Lorentz factors of the injected electrons;
Col. [10]: random Lorentz factors of the electrons emitting at the peak of the SED and
The viewing angle is always $3^\circ$, except for:
Col. [11]: random Lorentz factors of the electrons cooling in one dynamical time $R/c$;
Col. [12]: total (magnetic+radiative) energy density in the comoving frame in erg cm$^{-3}$;
Col. [13], [14]: slopes of the injected electron distribution;
Col. [15] and [16]: mass accretion rate and mass outflowing rate, in solar masses per year;
Col. [17] disc luminosity according to S12.
a: $\theta_{\rm v}=2^\circ$; 
b: $\theta_{\rm v}=2.3^\circ$, 
c: $\theta_{\rm v}=2.4^\circ$, 
d: $\theta_{\rm v}=2.5^\circ$,
e: $\theta_{\rm v}=3.5^\circ$, 
f: $\theta_{\rm v}=4^\circ$,
g: $\theta_{\rm v}=6^\circ$,
h: $\theta_{\rm v}=10^\circ$.
}
\label{para}
\end{table*}

\newpage
\setcounter{table}{4}
\begin{table*} 
\centering
\begin{tabular}{llllllllllllllllll}
\hline
\hline
Name  &$z$  &$L_{\rm d, 45}$ &$P^\prime_{\rm e, 45}$ &$B$ &$R_{\rm diss}$ &$\Gamma$  &$\gamma_{\rm max}$ &$\gamma_{\rm b}$ &$\gamma_{\rm peak}$ &$\gamma_{\rm cool}$                                                       
                                                   &$U^\prime$   &$s_1$  &$s_2$  &$\dot M_{\rm in}$ &$\dot M_{\rm out}$  &$\log L_{\rm d, S12}$  \\
~[1]     &[2]   &[3]   &[4]   &[5]   &[6]   &[7]   &[8]   &[9]   &[10]   &[11]   &[12]   &[13]   &[14]   &[15]  &[16]  &[17]\\
\hline
0713 +1935  &0.540 &0.6  &5.e--4  &10.0 &500  &12$^c$ &2.e3 &100 &265.8 &52.9 &0.21  &0   &2.0  &0.13    &9.0e--4 &44.93  \\ 
0721 +0406  &0.665 &18.  &2.e--3  &5.1  &500  &13     &5.e3 &140 &106.3 &7.9  &6.07  &0   &3.0  &3.94    &8.0e--3 &46.33  \\ 
0723 +2859  &0.966 &5.94 &9.e--4  &7.95 &600  &12     &4.e3 &80  &139.  &17.6 &6.99  &0   &2.2  &1.30    &2.5e--3 &45.75   \\ 
0725 +1425  &1.038 &9.   &5.9e--3 &3.28 &600  &14     &4.e3 &50  &370.8 &10.8 &6.24  &1   &2.0  &1.97    &0.053   &45.95  \\ 
0746 +2549  &2.979 &30.  &1.5e--1 &2.35 &800  &16     &4.e3 &100 &66.1  &3.8  &7.75  &0   &3.6  &6.57    &1.39    &46.31 \\ 
0805 +6144  &3.033 &33.  &1.2e--1 &3.00 &700  &15     &4.e3 &120 &79.9  &4.7  &7.06  &0   &3.4  &7.23    &0.86    &46.56 \\ 
0825 +5555  &1.418 &22.5 &1.e--2  &3.11 &600  &14     &4.e3 &160 &72.2  &6.5  &6.20  &1   &3.4  &4.93    &0.134   &46.32  \\ 
0830 +2410  &0.942 &22.5 &1.e--2  &5.11 &600  &12     &5.e3 &70  &48.9  &7.6  &5.33  &0   &3.3  &4.93    &0.075   &45.97   \\ 
0840 +1312  &0.680 &5.4  &2.3e--3 &5.95 &450  &12     &5.e3 &10  &22.0  &22.  &6.23  &1   &2.6  &1.18    &0.054   &45.75  \\ 
0847 --2337 &0.059 &0.015 &1.e--3 &8.22 &400  &5$^h$  &1.e4 &100 &344.6 &126. &4.83  &1   &2.2  &3.0e--3 &1.8e--3 &43.32  \\ 
0856 +2111  &2.098 &26.3 &4.e--3  &3.06 &1100 &15     &4.e3 &200 &229.9 &6.4  &6.87  &0   &2.4  &5.75    &0.010   &47.11  \\ 
0909 +0121  &1.026 &33.75 &1e--2  &3.33 &600  &13     &4.e3 &150 &71.3  &5.1  &5.37  &1   &3.2  &7.39    &0.12    &46.47   \\ 
0910 +2248  &2.661 &15.  &1.6e--2 &3.32 &1000 &16     &5.e3 &100 &99.6  &7.61 &7.91  &0   &2.6  &3.29    &0.084   &46.21 \\ 
0912 +4126  &2.563 &22.5 &1.5e--2 &3.73 &500  &11     &6.e3 &100 &83.3  &5.9  &4.11  &0   &2.8  &4.93    &0.062   &46.36 \\ 
0920 +4441  &2.189 &51.  &6.e--2  &5.15 &500  &12     &5.e3 &80  &154.6 &4.5  &5.41  &1   &2.2  &11.2    &0.48    &46.70 \\ 
0921 +6215  &1.453 &10.2 &8.e--3  &5.76 &500  &12     &4.e3 &100 &79.9  &10.7 &5.69  &0   &2.95 &2.23    &0.038   &46.05   \\ 
0923 +2815  &0.744 &3.   &1.3e--3 &6.1  &600  &13     &5.e3 &100 &75.   &29.6 &6.89  &0   &3.2  &0.66    &5.6e--3  &45.63  \\ 
0923 +4125  &1.732 &0.75 &4.e--3  &5.25 &700  &13     &4.e3 &200 &240.2 &49.9 &6.92  &0   &2.4  &0.16    &6.6e--3  &44.75  \\ 
0926 +1451  &0.632 &0.06 &7e--4   &2.68 &900  &14     &4.e3 &170 &127.2 &112. &6.04  &0   &3.4  &0.013   &1.3e--3  &43.75   \\ 
0937 +5008  &0.276 &0.06 &1.e--3  &12.1 &450  &11     &4.e3 &10  &129   &129  &10.6  &0   &2.9  &0.013   &0.012   &43.99  \\ 
0941 +2778  &1.305 &5.1  &1.5e--3 &4.33 &600  &12     &5.e3 &100 &99.6  &20.1 &5.08  &0   &2.6  &1.12    &5.0e--3 &45.72  \\ 
0946 +1017  &1.006 &4.7  &2.e--3  &2.23 &500  &12     &5.e3 &110 &88.7  &12.3 &4.42  &0   &2.9  &1.04    &8.3e--3 &45.74  \\ 
0948 +0022  &0.585 &2.25 &2.7e--3 &6.91 &620  &15     &4.e3 &4   &29.2  &29.2 &8.65  &1   &2.4  &0.49    &0.10    &45.03  \\ 
0949 +1752  &0.693 &1.76 &1.5e--3 &4.45 &900  &14     &4.e3 &40  &34.4  &31.9 &6.60  &1   &2.75 &0.38    &0.021   &45.25   \\ 
0956 +2515  &0.708 &7.5  &3.e--3  &5.32 &600  &10     &4.e3 &40  &34.4  &19.5 &4.16  &1   &2.65 &1.64    &0.035   &45.93  \\ 
0957 +5522  &0.899 &4.5  &1.2e--2 &1.1  &900  &11$^f$ &1.5e4 &100 &8247 &65.  &0.26  &1.4 &1.7  &0.99    &0.056   &45.59   \\ 
1001 +2911  &0.558 &0.09 &2.5e--3 &1.20 &1700 &16     &7.e3 &80  &946.  &946. &0.15  &1   &2.6  &0.02    &0.016   &44.06  \\ 
1012 +2439  &1.800 &13.5 &8.e--3  &1.49 &1000 &11     &6.e3 &100 &548.5 &11.3 &3.57  &0   &2.0  &2.96    &0.014   &45.56  \\ 
1016 +0513  &1.714 &4.5  &1.e--2  &4.15 &550  &12     &4.e3 &70  &265.4 &14.7 &4.97  &0   &2.0  &0.99    &0.025   &45.55  \\ 
1018 +3542  &1.228 &22.5 &2.1e--2 &1.66 &700  &15     &5.e3 &170 &104.9 &3.5  &6.62  &0   &3.8  &4.93    &0.11    &46.34  \\ 
1022 +3931  &0.604 &7.5  &1.7e--3 &5.52 &500  &11     &1.e4 &100 &78.3  &19.4 &5.02  &0   &3.0  &1.64    &2.3e--3  &45.89   \\ 
1032 +6051  &1.064 &2.4  &1.2e--3 &2.77 &1100 &14     &5.e3 &50  &61.0  &18.6 &5.98  &1   &2.4  &0.53    &0.015   &45.35  \\ 
1033 +4116  &1.117 &6.0  &3.5e--3 &9.79 &600  &14     &4.e3 &150 &133.8 &18.6 &10.9  &--1 &2.9  &1.31    &8.0e--3   &45.92  \\ 
1033 +6051  &1.401 &5.1  &1.1e--2 &5.66 &800  &15     &3.e3 &100 &106.2 &18.2 &8.29  &1   &2.4  &1.12    &0.12    &45.66  \\ 
1037 --2823 &1.066 &10.8 &4.e--3  &4.21 &600  &12     &3.e3 &150 &110.2 &13.6 &5.00  &0   &3.1  &2.37    &0.014   &45.95  \\ 
1043 +2408  &0.559 &0.48 &1.1e--3 &9.79 &800  &9      &4.e3 &10  &200   &96.4 &7.93  &1   &2.1  &0.11    &0.011   &44.65   \\ 
1058 +0133  &0.888 &3.0  &7.e--3  &11.  &500  &14     &3.e3 &80  &117   &19.2 &11.7  &0   &2.3  &0.66    &0.024   &45.51   \\ 
1106 +2812  &0.843 &1.6  &7.e--3  &0.6  &400  &11     &4.e4 &30  &504   &504  &0.035 &0   &2.5  &0.34    &0.025   &46.26   \\ 
1112 +3446  &1.956 &18.9 &1.6e--2 &4.9  &700  &10     &2.e3 &200 &221.2 &12.4 &4.00  &0   &2.4  &4.14    &0.029   &46.13  \\ 
1120 +0704  &1.336 &0.9  &3.e--3  &0.91 &800  &12     &4.e3 &100 &653.2 &371. &0.14  &1   &2.0  &0.20    &0.011   &45.47  \\ 
1124 +2336  &1.549 &6.75 &4.e--3  &7.59 &550  &13     &3.e3 &100 &97.6  &18.5 &8.01  &0   &2.6  &1.48    &0.015   &46.05 \\ 
1133 +0040  &1.633 &7.5  &5.e--3  &2.69 &800  &14     &5.e3 &90  &125.7 &10.1 &6.00  &0   &2.3  &1.64    &0.018   &45.64   \\ 
1146 +3958  &1.088 &11.7 &6.e--3  &6.99 &500  &12     &4.e3 &90  &80.9  &12.4 &6.55  &0   &2.7  &2.56    &0.026   &46.06   \\ 
1152 --0841 &2.367 &19.5 &1.4e--2 &0.82 &1200 &15     &6.e3 &100 &595.6 &89   &0.088 &1   &2.0  &4.27    &0.072   &46.25 \\ 
1154 +6022  &1.120 &9.   &4.5e--3 &3.53 &600  &13     &5.e3 &130 &84.2  &12.3 &5.54  &0   &3.6  &1.97    &0.023   &45.97   \\ 
1155 --8101 &1.395 &3.   &5.5e--3 &2.84 &600  &14     &5.e3 &90  &75.95 &16.5 &6.12  &0   &2.85 &0.66    &0.028   &45.55   \\ 
1159 +2914  &0.725 &3.6  &4.5e--3 &4.86 &600  &12     &7.e3 &80  &82.   &25.2 &5.4   &0   &2.6  &0.79    &0.016   &45.65   \\ 
1208 +5441  &1.344 &4.2  &7.5e--3 &3.36 &600  &14     &5.e3 &120 &93.4  &716.1 &6.32 &0   &3.0  &0.92    &0.034   &45.53  \\ 
1209 +1810  &0.845 &4.8  &1.e--3  &4.21 &600  &12     &7.e3 &70  &58.6  &20.3 &5.03  &0   &2.9  &1.05    &0.005   &45.48  \\ 
1222 +0413  &0.966 &15.  &1.5e--2 &5.0  &500  &12     &7.e3 &40  &28.8  &9.1  &5.39  &0   &3.3  &3.29    &0.162   &45.97   \\ 
1224 +2122  &0.434 &12.  &3.2e--3 &3.06 &600  &13     &7.e3 &70  &99    &9.43 &5.36  &0   &2.3  &2.63    &0.012   &46.16  \\ 
1224 +5001  &1.065 &36.  &5.e--3  &3.69 &800  &14     &5.e3 &120 &82.1  &6.0  &6.29  &0   &3.3  &7.88    &0.029   &45.85  \\ 
1226 +4340  &2.001 &25.5 &1.6e--2 &0.88 &1400 &16     &5.e3 &200 &130.0 &2.3  &7.38  &0   &3.5  &5.59    &0.068   &46.14  \\ 
1228 +4858  &1.722 &4.5  &6.e--3  &6.17 &600  &12     &4.e3 &100 &316.9 &22.0 &6.14  &0   &2.0  &0.99    &0.011   &45.68  \\ 
1228 +4858  &1.722 &4.5  &6.e--3  &4.57 &600  &12     &4.e3 &100 &97.7  &25.4 &5.36  &0   &2.7  &0.99    &0.020   &45.68  \\ 
1239 +0443  &1.761 &6.75 &5.e--2  &2.38 &800  &15     &3.e3 &100 &121.5 &8.86 &6.80  &0   &2.4  &1.48    &0.22    &45.83  \\ 
1239 +0443  &1.761 &6.75 &2.e--2  &1.29 &1300 &17     &3.e3 &180 &126.8 &4.42 &8.37  &1   &2.6  &1.48    &0.22    &45.83  \\ 
1257 +3229  &0.806 &4.2  &2.e--3  &6.21 &600  &13     &7.e3 &50  &31.2  &15.5 &6.60  &1   &3.0  &0.92    &0.035   &45.26  \\ 
\hline
\hline 
\end{tabular}
\caption{ 
{\it continue.}
a: $\theta_{\rm v}=2^\circ$; 
b: $\theta_{\rm v}=2.3^\circ$, 
c: $\theta_{\rm v}=2.4^\circ$, 
d: $\theta_{\rm v}=2.5^\circ$,
e: $\theta_{\rm v}=3.5^\circ$, 
f: $\theta_{\rm v}=4^\circ$,
g: $\theta_{\rm v}=6^\circ$,
h: $\theta_{\rm v}=10^\circ$.
}
\label{para}
\end{table*}

\setcounter{table}{4}
\begin{table*} 
\centering
\begin{tabular}{llllllllllllllllll}
\hline
\hline
Name  &$z$  &$L_{\rm d, 45}$ &$P^\prime_{\rm e, 45}$ &$B$ &$R_{\rm diss}$ &$\Gamma$  &$\gamma_{\rm max}$ &$\gamma_{\rm b}$ &$\gamma_{\rm peak}$ &$\gamma_{\rm cool}$                                                       
                                                   &$U^\prime$   &$s_1$  &$s_2$  &$\dot M_{\rm in}$ &$\dot M_{\rm out}$  &$\log L_{\rm d, S12}$  \\
~[1]     &[2]   &[3]   &[4]   &[5]   &[6]   &[7]   &[8]   &[9]   &[10]   &[11]   &[12]   &[13]   &[14]   &[15]  &[16]  &[17]\\
\hline 
\hline
1303 --4621 &1.664 &1.5  &3.e--3  &3.88 &600  &13     &5.e3 &50  &73.6  &35.2 &6.35  &1   &2.4  &0.33   &0.034  &45.21  \\ 
1310 +3220  &0.997 &9.   &7.e--3  &6.05 &500  &11     &5.e3 &50  &119.4 &17.9 &5.43  &1   &2.2  &1.97   &0.059  &45.92   \\ 
1317 +3425  &1.055 &10.8 &2.e--3  &9.43 &500  &11     &5.e3 &50  &40.3  &11.1 &7.37  &1   &2.6  &2.37   &0.026  &46.09  \\ 
1321 +2216  &0.943 &2.4  &1.7e--3 &0.31 &900  &13     &2.5e4 &800 &813  &333  &0.020 &1   &2.5  &0.53   &2.6e--3 &45.99  \\ 
1327 +2210  &1.403 &10.5 &1.2e--2 &5.01 &600  &13     &5.e3 &150 &82.1  &12.9 &6.21  &1   &3.0  &2.30   &0.13   &45.96   \\ 
1332 --1256 &1.492 &18.  &1.e--2  &1.73 &900  &15     &5.e3 &200 &140.8 &4.15 &6.55 &0.5  &2.9  &3.94   &0.050  &46.26  \\ 
1333 +5057  &1.362 &2.4  &2.8e--3 &4.05 &700  &15     &3.e3 &200 &190.3 &23.6 &7.42  &0   &2.6  &0.53   &7.3e--3 &45.36  \\ 
1343 +5754  &0.933 &6.   &2.e--3  &3.76 &600  &12$^f$ &5.e3 &50  &119.4 &16.7 &4.85  &1   &2.2  &1.31   &0.018  &45.65   \\ 
1344 --1723 &2.506 &10.8 &1.e--2  &0.89 &2100 &17$^c$ &6.e3 &1e3 &1332. &73.  &0.2   &--1 &2.0  &2.37   &4.6e--3 &46.02 \\ 
1345 +4452  &2.534 &10.5 &3.5e--2 &1.49 &600  &14     &5.e3 &180 &144.9 &9.4  &2.16  &0   &2.9  &2.30   &0.120  &46.12  \\ 
1347 --3750 &1.300 &5.4  &5.e--3  &7.65 &700  &12     &4.e3 &80  &67.25 &24.8 &7.06 &1.3  &2.5  &1.18   &0.074  &45.67  \\ 
1350 +3034  &0.712 &3.5  &2.e--3  &6.65 &600  &12     &4.e3 &40  &33.1  &31.9 &6.41  &1   &2.8  &0.66   &0.025  &45.26  \\ 
1357 +7643  &1.585 &1.5  &4.5e--3 &5.28 &850  &15     &4.e3 &90  &79.4  &35.8 &8.06  &1   &2.6  &0.33   &0.047  &45.20  \\ 
1359 +5544  &1.014 &1.35 &3.e--3  &4.69 &700  &13     &5.e3 &130 &92.2  &38.3 &6.05  &0   &3.4  &0.30   &0.011  &44.99  \\ 
1423 --7829 &0.788 &1.8  &8.3e--4 &3.86 &800  &12     &4.e3 &10  &96.5  &31.9 &4.82  &1.5 &2.15 &0.39   &0.021  &45.30  \\ 
1436 +2321  &1.548 &14.7 &5.2e--3 &7.43 &520  &11     &3.e3 &60  &58.7  &11.3 &5.96 &--1  &2.7  &3.22   &0.018  &45.66  \\ 
1438 +3710  &2.399 &15.  &3.e--2  &0.26 &950  &17     &1.e4 &400 &449.4 &126. &0.04  &1   &2.4  &3.29   &0.11   &46.36  \\ 
1439 +3712  &1.027 &37.5 &2.5e--3 &3.01 &800  &14     &5.e3 &70  &70.2  &5.1  &6.01  &0   &2.55 &8.21   &0.015  &46.16  \\ 
1441 --1523 &2.642 &15.6 &5.e--2  &3.24 &700  &12     &5.e3 &150 &70.3  &9.1  &4.71  &1   &3.3  &3.42   &0.58   &46.20 \\ 
1443 +2501  &0.939 &1.8  &1.7e--3 &5.30 &600  &10     &4.e3 &1   &55.4  &25.1 &4.07  &2.1 &2.1  &0.39   &0.046  &45.28  \\ 
1504 +1029  &1.839 &15.3 &3.2e--2 &4.21 &800  &13$^a$ &5.5e3 &120 &183  &8.9  &5.70 &--1  &2.2  &3.35   &0.074  &46.17   \\ 
1504 +1029  &1.839 &15.3 &9.e--2  &0.93 &2600 &18$^a$ &5.5e3 &500 &3512 &9.3  &1.15 &--1  &1.6  &3.35   &0.093  &46.17   \\ 
1505 +0326  &0.409 &0.675 &6.e--4 &5.13 &700  &14     &5.e3 &100 &61.8  &55.  &7.05  &1   &3.15 &0.15   &5.4e--3 &44.72   \\ 
1514 +4450  &0.570 &0.2  &1.3e--3 &7.6  &500  &11     &4.e3 &20  &78.9  &78.9 &6.24  &1   &2.6  &0.04   &0.017  &44.33  \\ 
1522 +3144  &1.484 &4.5  &1.e--2  &1.95 &500  &12     &4.e3 &90  &170.  &11.1 &5.41  &1   &2.2  &0.99   &0.078  &45.90  \\ 
1539 +2744  &2.191 &4.5  &7.e--3  &9.14 &500  &12     &3.e3 &300 &519.8 &18.7 &8.68  &0   &2.0  &0.99   &8.1e--3 &45.63  \\ 
1549 +0237  &0.414 &6.0  &1.e--3  &6.12 &500  &11     &4.e3 &70  &48.9  &22.7 &5.40  &1   &2.8  &1.31   &0.011  &45.83  \\ 
1550 +0527  &1.417 &12.  &8.e--3  &9.40 &500  &12     &3.e3 &80  &71.1  &13.8 &8.83  &1   &2.5  &2.63   &0.086  &46.08 \\ 
1553 +1256  &1.308 &60.  &4.3e--3 &1.19 &1500 &12     &1.e4 &120 &5880  &7.81 &0.59  &0   &2.0  &13.1   &6.8e--3 &46.18 \\ 
1608 +1029  &1.232 &18.  &8.e--3  &7.61 &600  &12     &5.e3 &50  &47.1  &9.99 &6.81  &1   &2.5  &3.94   &0.108  &46.07 \\ 
1613 +3412  &1.400 &54.  &6.e--3  &8.50 &500  &10     &5.e3 &40  &35.4  &5.44 &5.99  &1   &2.5  &11.8   &0.081  &46.46 \\ 
1616 +4632  &0.950 &3.   &1.7e--3 &2.83 &600  &14     &5.e3 &200 &91.0  &16.5 &7.37  &1   &3.5  &0.66   &0.019  &45.42   \\ 
1617 --5848 &1.422 &105. &2.e--2  &3.21 &650  &15     &3.e3 &150 &88.2  &2.7  &6.72 &0.5  &3.3  &23.0   &0.14   &47.00    \\ 
1624 --0649 &3.037 &36.  &8.e--3  &2.81 &700  &14     &4.e3 &90  &292.  &4.8  &6.02  &1   &2.1  &7.88   &0.065  &46.35 \\ 
1628 --6152 &2.578 &10.5 &2.e--2  &5.98 &500  &12     &5.e3 &150 &125.7 &11.5 &5.99  &1   &2.5  &2.30   &0.16   &46.03  \\ 
1635 +3808  &1.813 &60.  &4.e--1  &2.18 &600  &14     &4.e3 &150 &227.3 &1.72 &5.84  &-1  &2.2  &13.1   &0.69   &46.66  \\ 
1635 +3808  &1.813 &60.  &4.5e--2 &4.17 &400  &11     &4.e3 &150 &172.4 &3.5  &4.27  &-1  &2.4  &13.1   &0.097  &46.66  \\ 
1636 +4715  &0.823 &3.6  &4.5e--3 &1.93 &900  &15     &4.e3 &200 &85.1  &10.1 &6.65  &1   &3.8  &0.80   &0.061  &45.49  \\ 
1637 +4717  &0.735 &1.8  &3e--3   &5.78 &500  &12     &4.e3 &220 &127.  &28.1 &5.80  &0   &3.8  &0.39   &9.1e--3 &45.58  \\ 
1639 +4705  &0.860 &9.   &1.2e--3 &2.18 &900  &14     &5.e3 &190 &119.4 &7.71 &5.85  &0   &3.8  &1.97   &0.0134 &45.97   \\ 
1642 +3940  &0.593 &9.0  &1.2e--2 &9.57 &450  &11$^e$ &5.e3 &70  &57.2  &11.7 &7.75  &0   &2.9  &1.97   &0.064  &46.01 \\ 
1703 --6212 &1.747 &52.5 &3.5e--2 &6.68 &500  &12     &3.e3 &130 &113.9 &7.2  &6.69  &0   &2.7  &11.5   &0.12   &46.31   \\ 
1709 +4318  &1.027 &1.5  &3.5e--3 &7.11 &700  &15     &5.e3 &120 &111.9 &33.7 &9.90  &0   &2.7  &0.33   &0.012  &45.03 \\ 
1734 +3857  &0.975 &2.   &3.e--3  &5.14 &550  &13     &5.e3 &190 &183.  &23.7 &6.19  &0   &2.6  &0.44   &6.7e--3 &45.01   \\ 
1736 +0631  &2.387 &12.  &2.2e--2 &2.64 &1200 &15     &3.e3 &120 &91.0  &7.39 &6.79  &0   &3.0  &2.63   &0.12   &46.21  \\ 
1802 --3940 &1.319 &15   &2.2e--2 &4.34 &500  &12     &5.e3 &90  &173.7 &9.5  &5.13 &--1  &2.1  &3.29   &0.044  &46.11  \\ 
1803 +0341  &1.420 &1.2  &8.e--3  &5.74 &700  &15     &5.e3 &200 &183.9 &35.4 &9.66  &0   &2.7  &0.26   &0.020  &45.02 \\ 
1818 +0903  &0.354 &0.75 &7.5e--4 &6.65 &600  &12     &5.e3 &70  &82.1  &64.2 &6.38  &0   &2.6  &0.16   &8.0e--3 &44.93 \\ 
1830 +0619  &0.745 &22.5 &2.e--3  &4.04 &700  &13     &5.e3 &40  &43.6  &6.3  &5.59  &1   &2.4  &4.93   &0.031  &46.45 \\ 
1848 +3219  &0.800 &4.5  &3.5e--3 &7.02 &600  &13     &5.e3 &45  &35.9  &18.8 &7.22  &1   &2.7  &0.99   &0.053  &45.58 \\ 
1903 --6749 &0.254 &0.225 &1.1e--3 &7.9 &500  &11     &4.e3 &7   &77.9  &77.9 &6.26  &0   &2.9  &0.05   &0.019  &44.42  \\ 
1916 --7946 &0.204 &0.54 &3.e--4  &3.11 &900  &13     &4.e3 &80  &56.1  &51.4 &5.31  &0   &3.7  &0.12   &1.5e--3 &44.90  \\ 
1928 --0456 &0.587 &4.2  &2.3e--3 &1.83 &800  &15     &5.e3 &120 &77.9  &11.4 &6.67  &0   &3.6  &0.92   &0.015  &45.62   \\ 
1954 --1123 &0.683 &0.3  &1.5e--3 &5.2  &600  &11     &5.e3 &120 &150.  &80.  &5.08  &0   &2.5  &0.066  &2.3e--3 &44.37   \\ 
1955 +1358  &0.743 &4.5  &5.e--3  &9.98 &500  &11     &5.e3 &70  &52.2  &19.8 &8.25  &1   &2.7  &0.99   &0.053  &45.73 \\ 
1959 --4246 &2.178 &14.4 &1.6e--2 &7.6  &450  &12     &3.e3 &200 &225.6 &9.6  &7.04 &--1  &2.4  &3.15   &0.031  &46.13  \\ 
\hline
\hline 
\end{tabular}
\caption{{\it continue.}
a: $\theta_{\rm v}=2^\circ$; 
b: $\theta_{\rm v}=2.3^\circ$, 
c: $\theta_{\rm v}=2.4^\circ$, 
d: $\theta_{\rm v}=2.5^\circ$,
e: $\theta_{\rm v}=3.5^\circ$, 
f: $\theta_{\rm v}=4^\circ$,
g: $\theta_{\rm v}=6^\circ$,
h: $\theta_{\rm v}=10^\circ$.
}
\label{para}
\end{table*}

\setcounter{table}{4}
\begin{table*} 
\centering
\begin{tabular}{llllllllllllllllll}
\hline
\hline
Name  &$z$  &$L_{\rm d, 45}$ &$P^\prime_{\rm e, 45}$ &$B$ &$R_{\rm diss}$ &$\Gamma$  &$\gamma_{\rm max}$ &$\gamma_{\rm b}$ &$\gamma_{\rm peak}$ &$\gamma_{\rm cool}$                                                       
                                                   &$U^\prime$   &$s_1$  &$s_2$  &$\dot M_{\rm in}$ &$\dot M_{\rm out}$  &$\log L_{\rm d, S12}$  \\
~[1]     &[2]   &[3]   &[4]   &[5]   &[6]   &[7]   &[8]   &[9]   &[10]   &[11]   &[12]   &[13]   &[14]   &[15]  &[16]  &[17]\\
\hline 
\hline
2017 +0603  &1.743 &157.5 &8.e--3 &0.53 &800  &15     &2.e4 &1e3 &12552 &10.9 &0.097 &1.4 &1.8  &34.5   &0.041   &47.42  \\ 
2025 --2845 &0.884 &1.12 &8.e--3  &2.57 &700  &13     &5.e3 &100 &49.6  &26.7 &5.22  &0   &3.2  &0.25   &0.032   &45.01   \\ 
2031 +1219  &1.213 &0.72 &1.e--3  &8.35 &800  &15$^b$ &2.e3 &150 &130   &39.6 &6.92  &0   &2.8  &0.16   &3.1e--3 &44.76  \\ 
2035 +1056  &0.601 &1.5  &1.9e--3 &6.59 &600  &12     &5.e3 &130 &73.1  &32.8 &6.19  &1   &3.1  &0.33   &0.017   &45.11  \\ 
2110 +0809  &1.580 &11.25 &5.e--3 &2.22 &600  &13     &5.e3 &100 &119.4 &5.30 &5.09  &0   &2.4  &2.46   &0.018   &46.08 \\ 
2121 +1901  &2.180 &6.   &1.e--2  &2.59 &800  &13     &5.e3 &250 &208.  &11.7 &5.18  &1   &2.5  &1.31   &0.064   &45.26  \\ 
2135 --5006 &2.181 &13.2 &1.8e--3 &2.83 &1100 &16     &5.e3 &120 &91.   &7.04 &7.75  &0   &3.0  &2.89   &0.11    &46.36  \\ 
2139 --6732 &2.009 &7.5  &3.e--2  &2.51 &800  &15     &4.e3 &250 &101.4 &8.94 &6.77  &1   &4.1  &1.64   &0.36    &45.77  \\ 
2145 --3357 &1.361 &2.25 &3.5e--3 &4.13 &700  &13$^a$ &3.e3 &120 &142.2 &24.4 &5.71  &0   &2.4  &0.49   &9.6e--3 &45.17 \\ 
2157 +3127  &1.448 &5.4  &7.e--3  &6.04 &600  &13$^c$ &5.e3 &120 &99.6  &15.3 &6.63  &0   &2.9  &1.18   &0.028   &45.74  \\ 
2118 +0013  &0.463 &0.6  &3.e--4  &7.79 &500  &12$^f$ &2.e4 &100 &5833  &65.3 &4.62  &1.3 &1.8  &0.13   &1.3e--3 &44.93 \\ 
2202 --8338 &1.865 &15   &2.1e--3 &3.98 &600  &14     &5.e3 &80  &109.1 &6.35 &6.39  &1.  &2.3  &3.29   &0.22    &46.18  \\ 
2212 +2355  &1.125 &6.0  &1.5e--3 &4.33 &600  &13     &6.e3 &100 &76.0  &17.3 &5.87  &2.05 &2.05 &1.31  &0.059   &45.78 \\ 
2219 +1806  &1.071 &1.5  &5.e--4  &4.01 &600  &14     &5.e3 &100 &119.4 &31.1 &6.54  &1   &2.4  &0.33   &4.0e--3 &45.07  \\ 
2229 --0832 &1.560 &30.  &2.4e--2 &5.07 &600  &13     &5.e3 &120 &99.6  &8.01 &6.26  &0   &2.8  &6.57   &0.10    &46.45  \\ 
2236 +2828  &0.790 &2.4  &4.e--3  &10.8 &500  &11     &3.e3 &60  &204.7 &14.3 &8.56  &0   &2.0  &0.53   &0.011   &45.37   \\ 
2237 --3921 &0.297 &0.6  &3.e--3  &6.68 &600  &13$^g$ &9.e3 &50  &66.7  &53.5 &7.58  &2   &2.3  &0.13   &0.045   &44.87 \\ 
2240 +4057  &1.171 &2.4  &4.e--3  &7.73 &550  &14$^d$ &4.e3 &90  &75.   &25.1 &6.89  &1   &2.6  &0.53   &0.042   &45.31 \\ 
2315 --5018 &0.808 &0.45 &1.8e--3 &2.19 &1700 &13     &4.e3 &60  &230.  &230. &0.32  &0   &2.3  &0.099  &3.9e--3 &44.63   \\ 
2321 +3204  &1.489 &5.4  &5.5e--3 &6.52 &500  &12     &5.e3 &90  &109.1 &19.1 &6.40  &0   &2.4  &1.18   &0.017   &45.71  \\ 
2327 +0940  &1.841 &30   &3.5e--2 &3.15 &800  &12     &5.e3 &120 &63.4  &6.52 &4.63  &1   &3.0  &6.57   &0.42    &46.20  \\ 
2331 --2148 &0.563 &0.75 &1.7e--3 &3.74 &900  &12     &5.e3 &50  &56.4  &56.4 &4.8   &0   &2.8  &0.16   &5.7e--3 &44.82 \\ 
2334 +0736  &0.401 &5.4  &4.e--4  &5.84 &500  &12     &4.e3 &50  &39.5  &20.8 &5.92  &1   &2.7  &1.18   &5.2e--3 &45.76   \\ 
2345 --1555 &0.621 &2.1  &2.e--3  &8.13 &500  &12     &6.e3 &70  &98.9  &31.8 &7.68  &0   &2.35 &0.46   &5.6e--3 &45.30  \\ 
2357 +0448  &1.248 &2.7  &5.5e--3 &0.96 &700  &15     &5.e3 &90  &47.1  &22.2 &1.74  &1   &3.3  &0.59   &0.086   &46.02  \\ 
\hline
\hline 
\end{tabular}
\caption{
{\it continue.}
a: $\theta_{\rm v}=2^\circ$; 
b: $\theta_{\rm v}=2.3^\circ$, 
c: $\theta_{\rm v}=2.4^\circ$, 
d: $\theta_{\rm v}=2.5^\circ$,
e: $\theta_{\rm v}=3.5^\circ$, 
f: $\theta_{\rm v}=4^\circ$,
g: $\theta_{\rm v}=6^\circ$,
h: $\theta_{\rm v}=10^\circ$.
}
\label{para}
\end{table*}

\begin{table*} 
\centering
\begin{tabular}{llllllllllllllllll}
\hline
\hline
Name  &$z$  &$L_{\rm d, 45}$ &$P^\prime_{\rm e, 45}$ &$B$ &$R_{\rm diss}$ &$\Gamma$  &$\gamma_{\rm max}$ &$\gamma_{\rm b}$ &$\gamma_{\rm peak}$ &$\gamma_{\rm cool}$                                                       
                                                   &$U^\prime$   &$s_1$  &$s_2$  &$\dot M_{\rm in}$ &$\dot M_{\rm out}$  &$\log L_{\rm d, S12}$  \\
~[1]     &[2]   &[3]   &[4]   &[5]   &[6]   &[7]   &[8]   &[9]   &[10]   &[11]   &[12]   &[13]   &[14]   &[15]  &[16]  &[17]\\
\hline 
\hline
0013 +1907  &0.477 &0.06  &3.3e--4 &2.67 &600  &11     &5.e4 &100. &478   &478  &0.42  &1   &2.3 &0.013   &1.1e--3 &43.70    \\ 
0203 +3042  &0.761 &5.4   &1.2e--3 &8.46 &600  &11     &4.e3 &100  &67.3  &10.6 &6.41  &1   &2.7 &1.18    &0.012   &45.75    \\ 
0334 --4008 &1.357 &0.72  &8.e--3  &1.51 &1100 &17$^d$ &6.e3 &200  &396   &207  &0.27  &1   &2.3 &0.16    &0.035   &44.83    \\ 
0434 --2014 &0.928 &0.15  &1.e--3  &10.2 &500  &11     &6.e3 &300  &267   &57   &8.59  &1   &2.5 &0.033   &3.8e--3 &44.15    \\ 
0438 --4521 &2.017 &1.8   &1.e--2  &7.7  &500  &11     &4.e3 &300  &240   &23.8 &6.8   &1   &2.5 &0.39    &0.048   &45.24    \\ 
0516 --6207 &1.300 &3.0   &7.5e--3 &7.45 &500  &11     &6.e3 &200  &204   &19.6 &6.2   &1   &2.4 &0.66    &0.040   &45.43    \\ 
0629 --2001 &1.724 &1.35  &1.7e--2 &10.8 &500  &11     &5.e3 &200  &229   &15.5 &10.3  &0   &2.4 &0.30    &0.029   &45.05    \\ 
0831 +0429  &0.174 &0.045 &1.3e--3 &0.8  &600  &10     &6.e3 &800  &2692  &2692 &0.051 &2   &2.1 &0.010   &4.0e--4 &43.62    \\ 
1117 +2013  &0.138 &0.009 &2.e--5  &1.5  &700  &18$^d$ &1.e5 &5.e3 &3.9e4 &778  &0.111 &1   &1.5 &2.0e--3 &6.2e--6 &42.83    \\ 
1125 --3559 &0.284 &0.27  &6.e--5  &1.   &600  &15     &5.e4 &200  &3934  &1026 &0.066 &1   &2.  &0.059   &1.1e--4 &44.34    \\ 
1203 +6030  &0.065 &0.01  &7.e--6  &1.8  &600  &15     &3.5e5 &1e3 &7022  &710  &0.144 &1.3 &1.8 &2.2e--3 &1.7e--5 &43.00    \\ 
1221 +2814  &0.103 &0.012 &2.6e--4 &0.65 &500  &10     &8.e5 &3.e3 &4708  &4708 &0.026 &1.5 &2.7 &2.6e--3 &5.3e--5 &43.11    \\ 
1221 +3010  &0.184 &0.012 &2.5e--4 &0.85 &500  &10     &5.e5 &2.e4 &1.1e5 &2519 &0.038 &1.3 &1.9 &2.6e--3 &1.5e--5 &43.05    \\ 
1420 +5422  &0.153 &0.018 &9.e--4  &1.99 &450  &10$^f$ &5.e3 &200  &878   &858  &0.319 &1   &2.1 &3.9e--3 &3.5e--4 &43.20    \\ 
1534 +3720  &0.144 &0.006 &1.1e--4 &0.75 &600  &11$^i$ &8.e4 &1.e3 &3428  &2948 &0.035 &0   &2.2 &1.3e--3 &1.1e--5 &42.72    \\ 
1540 +1438  &0.606 &0.36  &2.2e--3 &2.18 &1000 &10     &8.e3 &100  &288   &288  &0.285 &1   &2.7 &0.079   &8.3e--3 &44.57    \\ 
1755 --6423 &1.255 &1.5   &7.e--3  &10.9 &500  &10     &5.e3 &100  &65    &20   &8.21  &1   &2.8 &0.33    &0.061   &45.16    \\ 
1824 +5651  &0.664 &0.81  &5.5e--3 &7.3  &500  &10     &5.e3 &100  &68.7  &30.3 &5.36  &1   &2.8 &0.18    &0.042   &44.91    \\ 
2015 +3709  &0.859 &4.5   &2.e--2  &3.97 &600  &12     &3.e3 &200  &107   &16.3 &4.97  &1   &3.0 &0.99    &0.17    &45.18    \\ 
2152 +1735  &0.874 &1.53  &2.e--3  &2.33 &1200 &10     &1.e4 &500  &661   &320  &0.107 &0   &2.4 &0.34    &7.0e--4 &45.15    \\ 
2206 --0029 &1.053 &0.675 &3.e--3  &1.7  &1400 &11     &3.5e3 &70  &315   &286  &0.206 &1   &2.2 &0.15    &0.014   &44.80    \\ 
2206 +6500  &1.121 &42.   &8.e--3  &2.02 &1200 &15     &5.e3 &200  &82.5  &4.33 &6.64  &1   &3.8 &9.2     &0.103   &46.67    \\ 
2236 +2828  &0.790 &4.2   &5.e--3  &1.61 &1500 &10.5   &4.e3 &400  &551.2 &137  &0.171 &1   &2.2 &0.92    &0.012   &45.64    \\ 
2247 --0002 &0.949 &1.35  &2.e--3  &1.08 &1100 &11     &8.e3 &70   &838.5 &423  &0.088 &0   &2.  &0.30    &2.3e--3 &45.10    \\ 
2315 --5018 &0.811 &0.47  &1.2e--3 &14   &600  &14     &3.e3 &60   &72    &37.5 &11.82 &0   &2.5 &0.11    &3.5e--3 &44.62   \\ 
2353 --3034 &0.737 &0.48  &8.5e--4 &15.8 &450  &10     &5.e3 &80   &205   &37.4 &7.49  &0   &2.5 &0.10    &4.7e--3 &44.63    \\ 
\hline
\hline 
\end{tabular}
\caption{
Parameters for BL Lac objects in S13. 
a: $\theta_{\rm v}=2^\circ$; 
b: $\theta_{\rm v}=2.3^\circ$, 
c: $\theta_{\rm v}=2.4^\circ$, 
d: $\theta_{\rm v}=2.5^\circ$,
e: $\theta_{\rm v}=3.5^\circ$, 
f: $\theta_{\rm v}=4^\circ$,
g: $\theta_{\rm v}=6^\circ$,
h: $\theta_{\rm v}=10^\circ$.
}
\label{parabl}
\end{table*}


\setcounter{figure}{15}
\begin{figure*}
\begin{tabular}{cccc}
\psfig{file=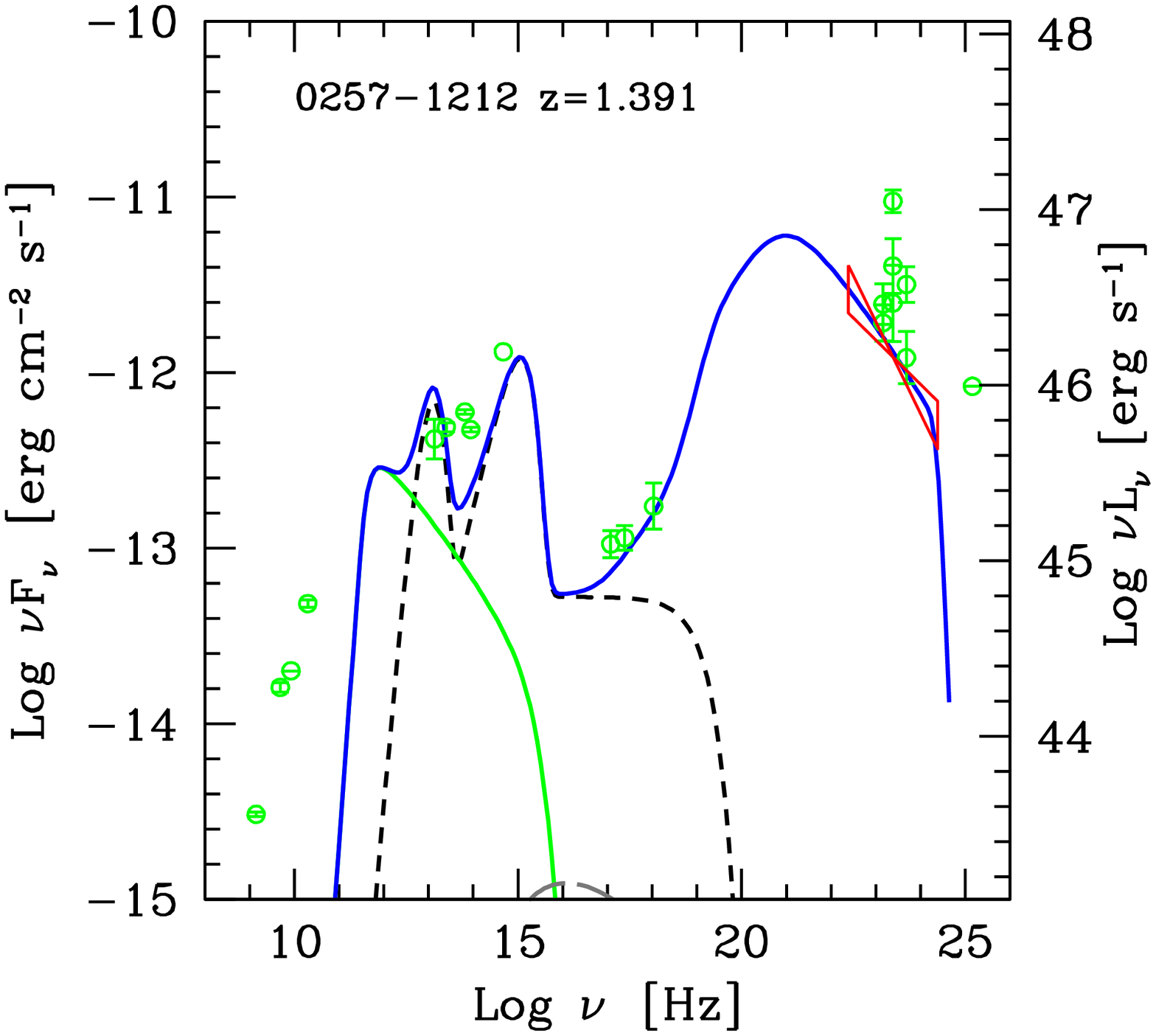,width=4.3cm,height=3.7cm}  
&\psfig{file=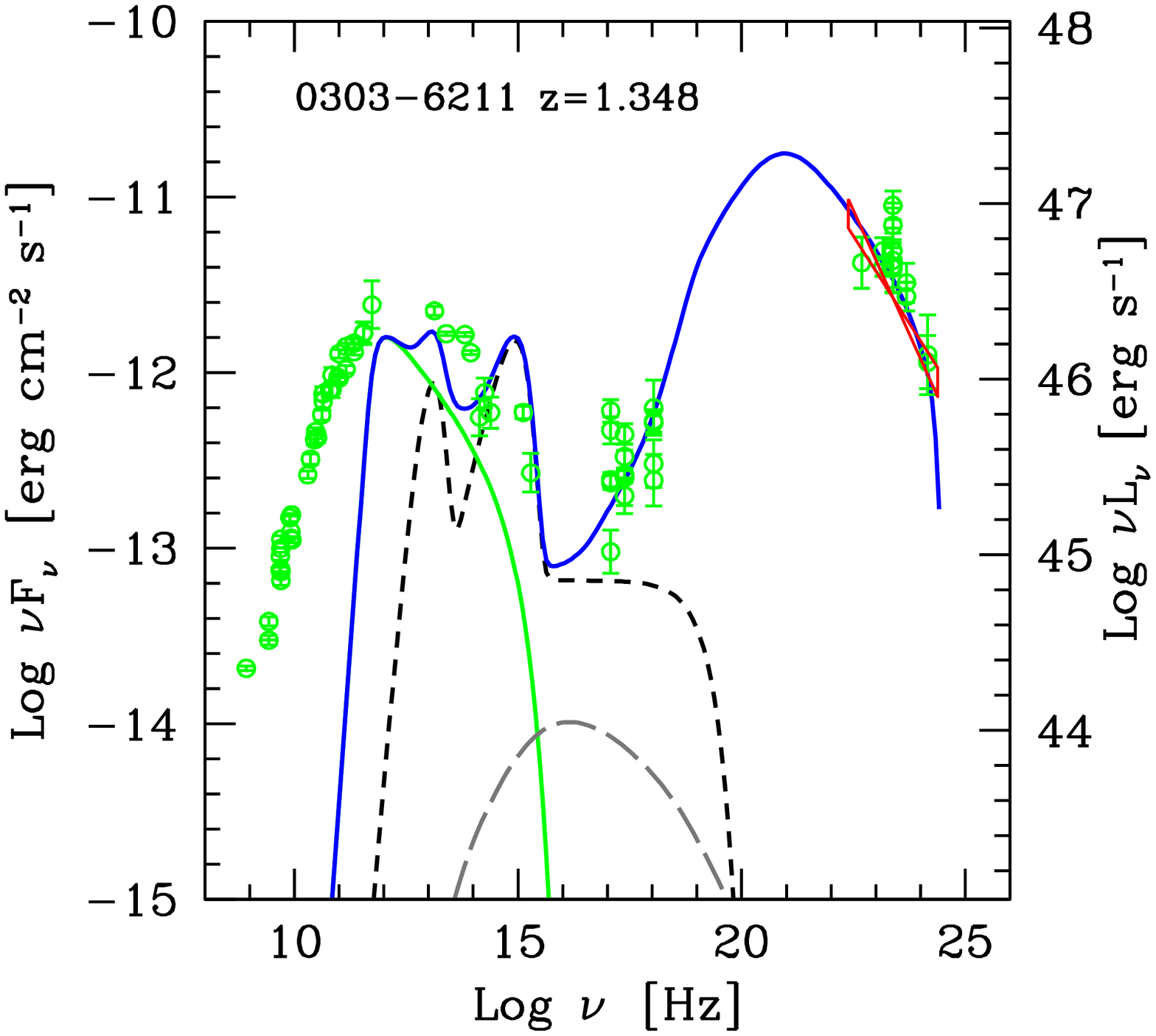,width=4.3cm,height=3.7cm} 
&\psfig{file=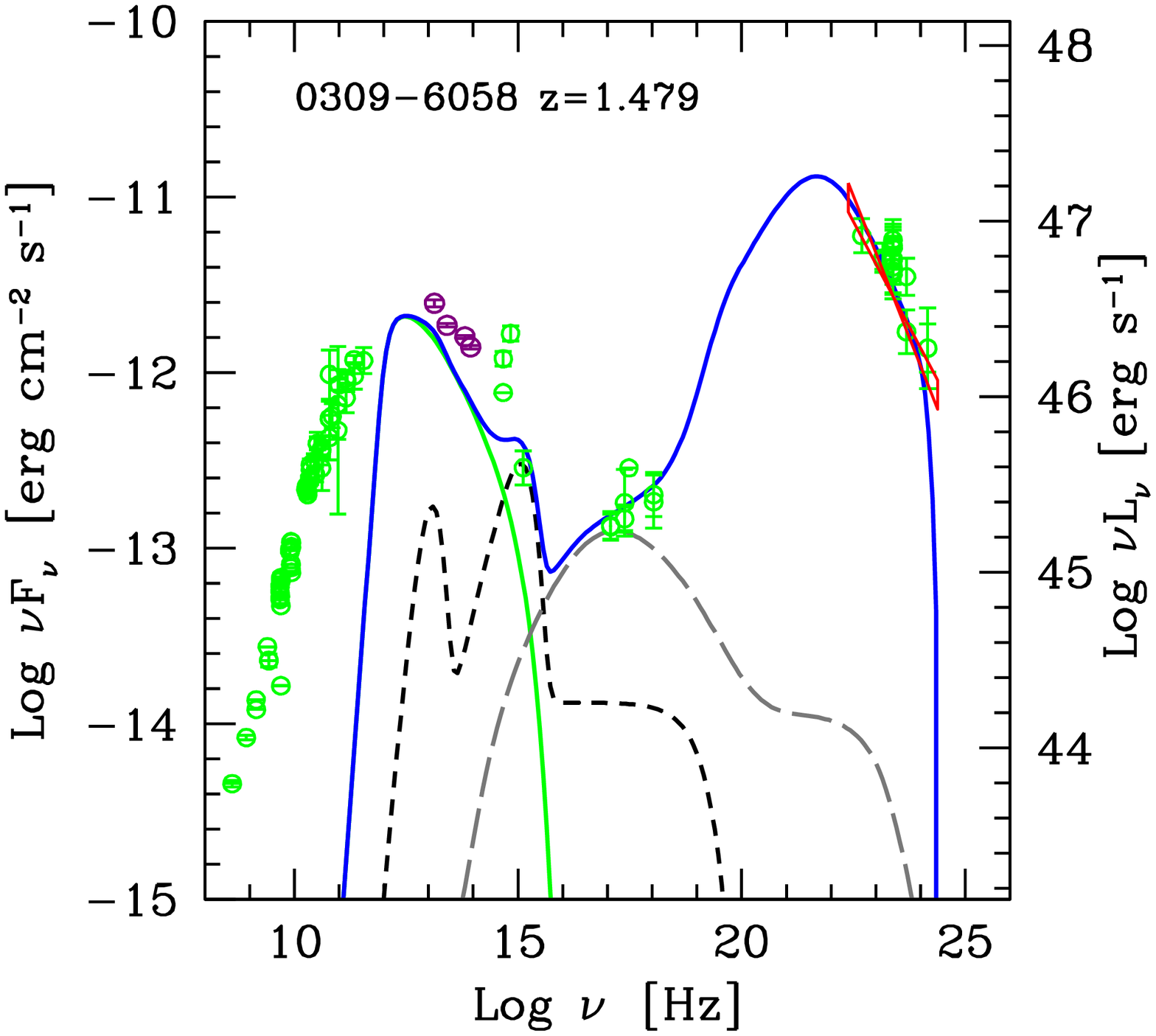,width=4.3cm,height=3.7cm}  
&\psfig{file=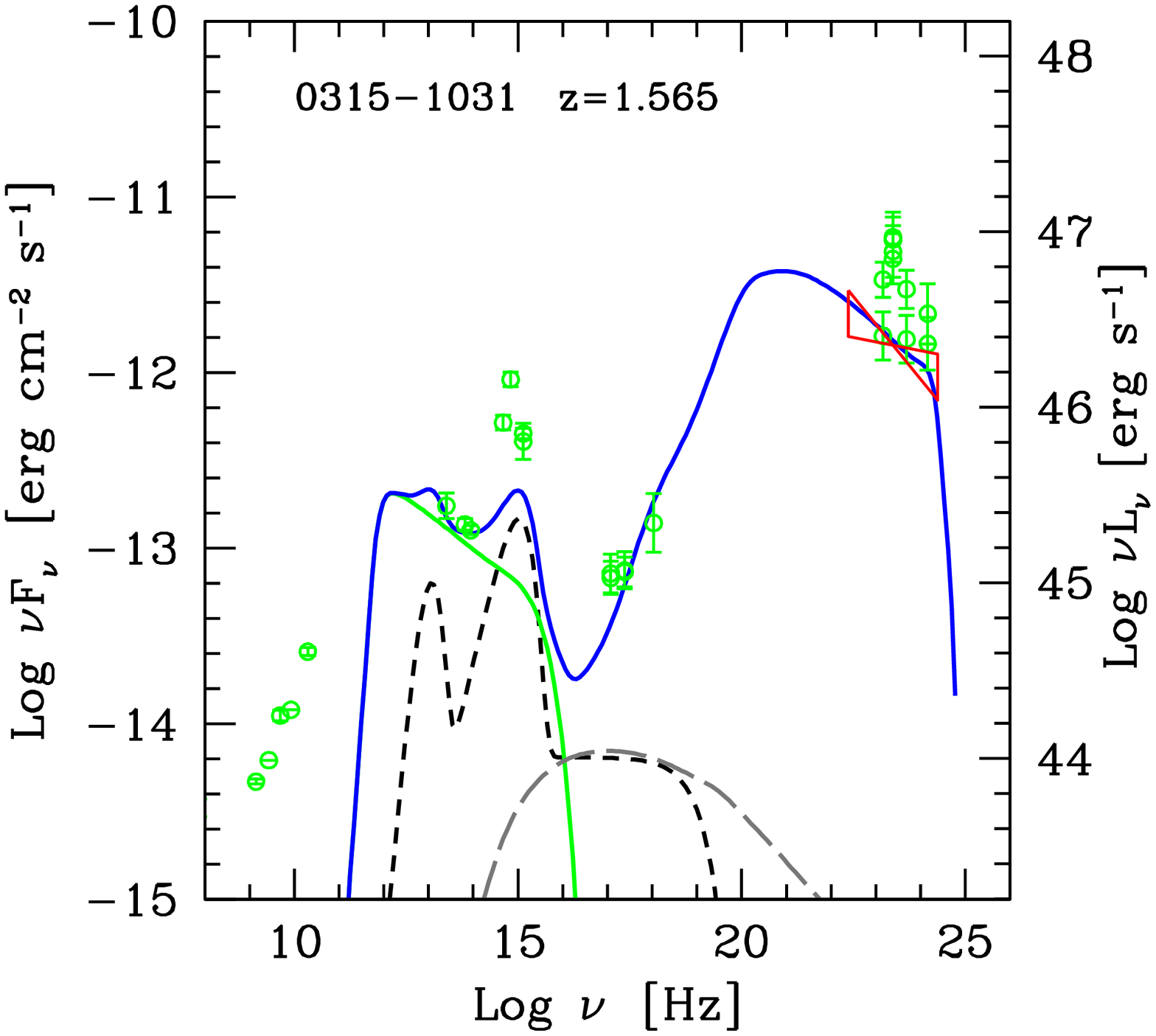,width=4.3cm,height=3.7cm} \\
\psfig{file=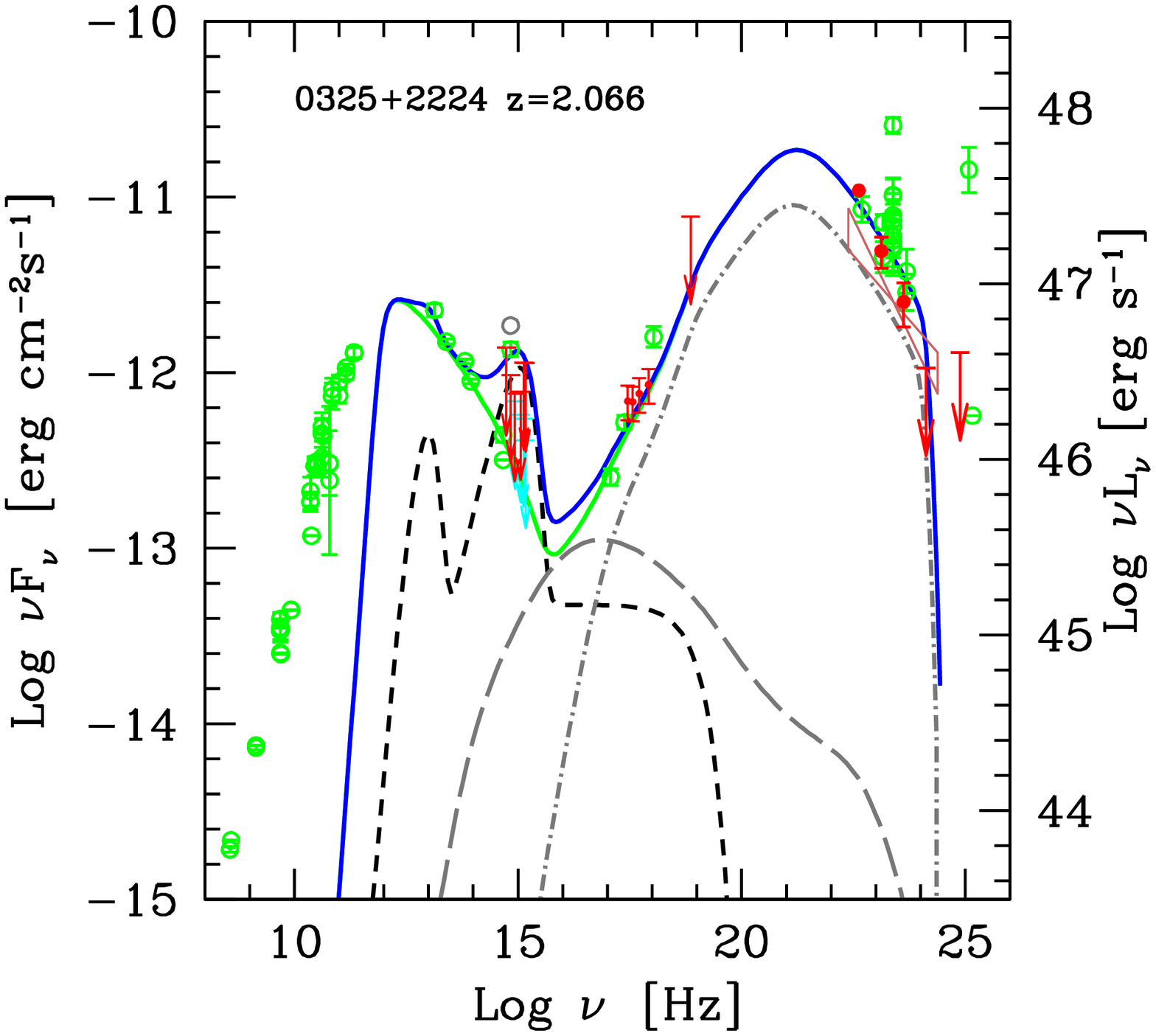,width=4.3cm,height=3.7cm} 
&\psfig{file=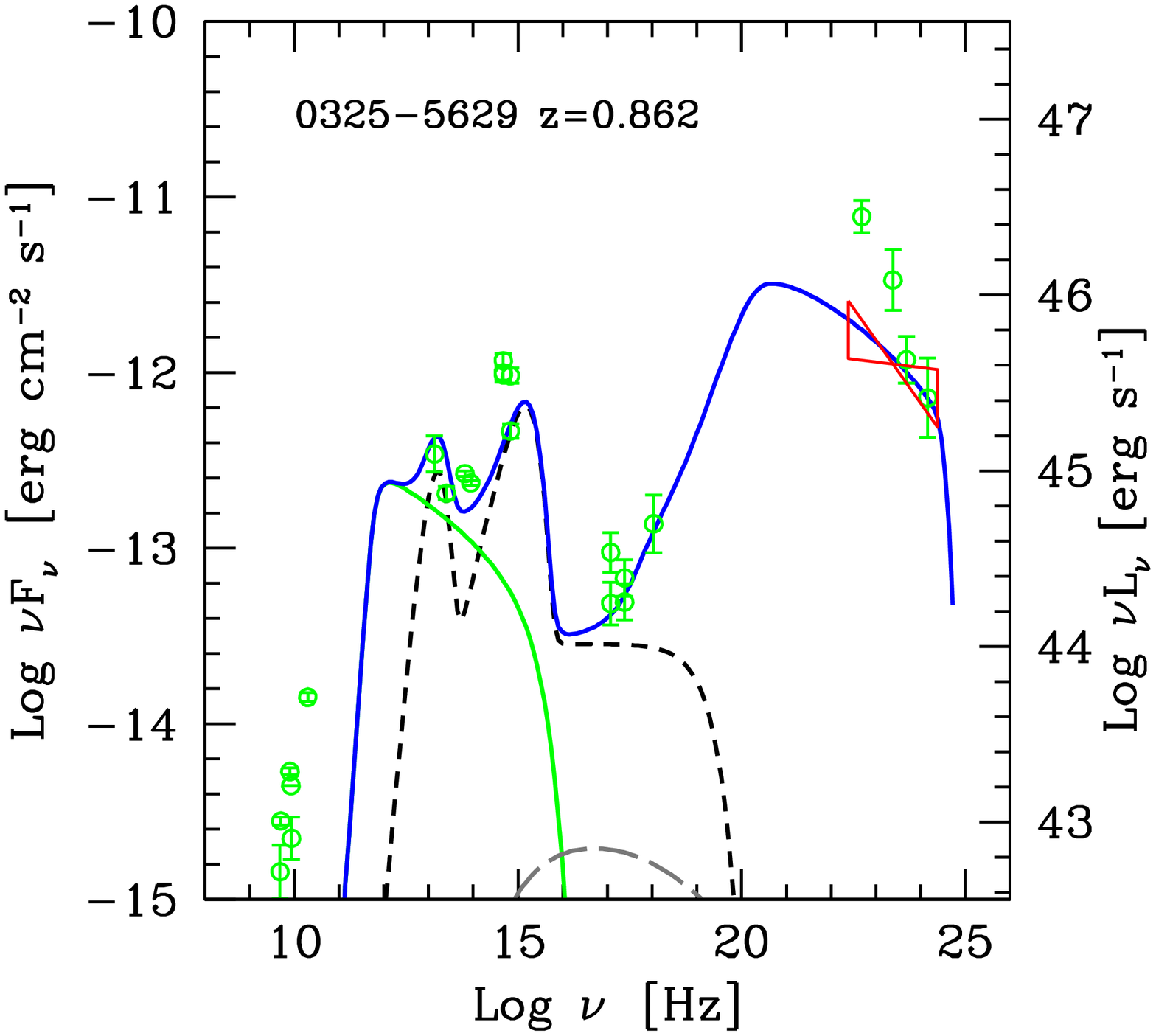,width=4.3cm,height=3.7cm} 
&\psfig{file=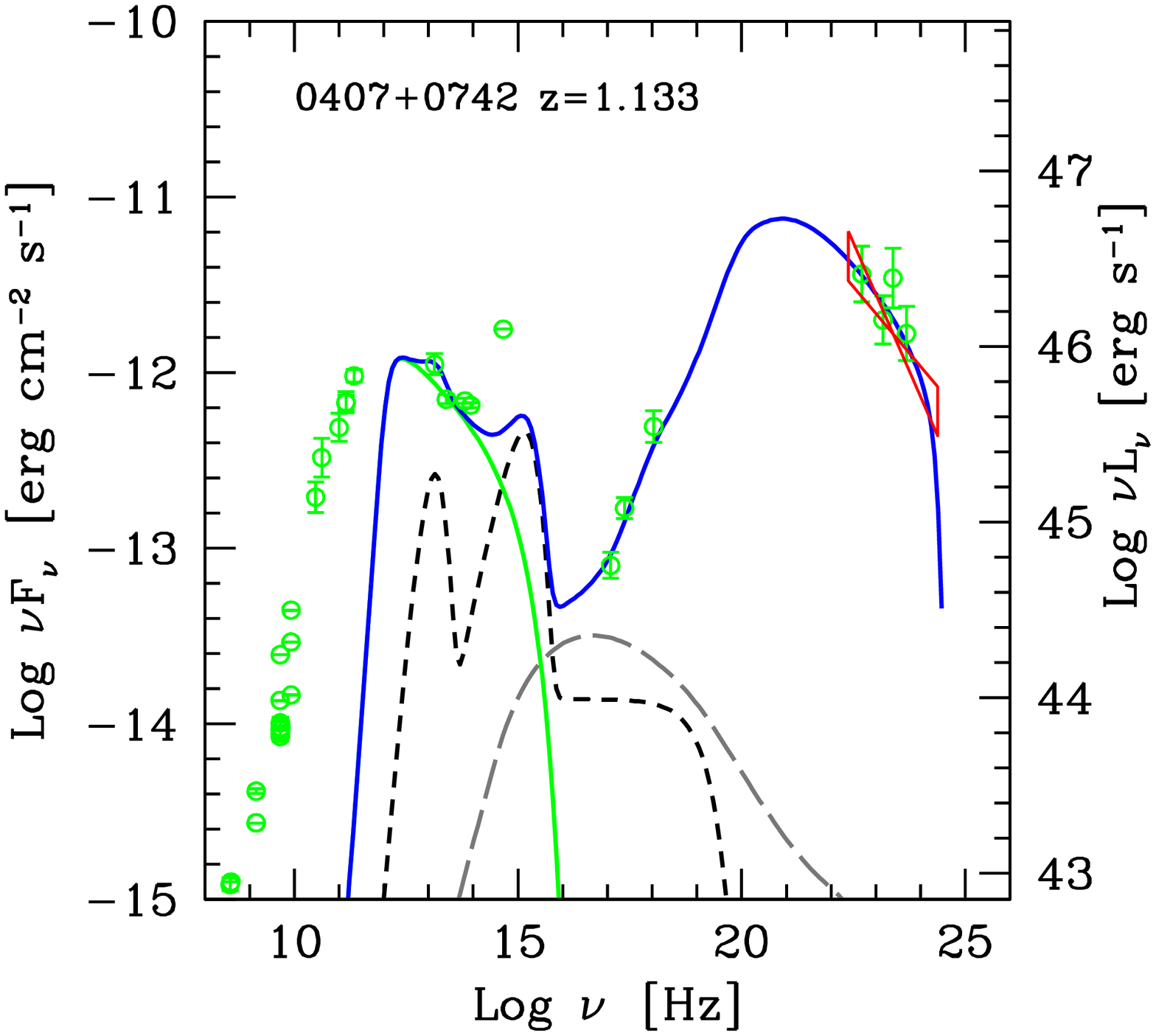,width=4.3cm,height=3.7cm} 
&\psfig{file=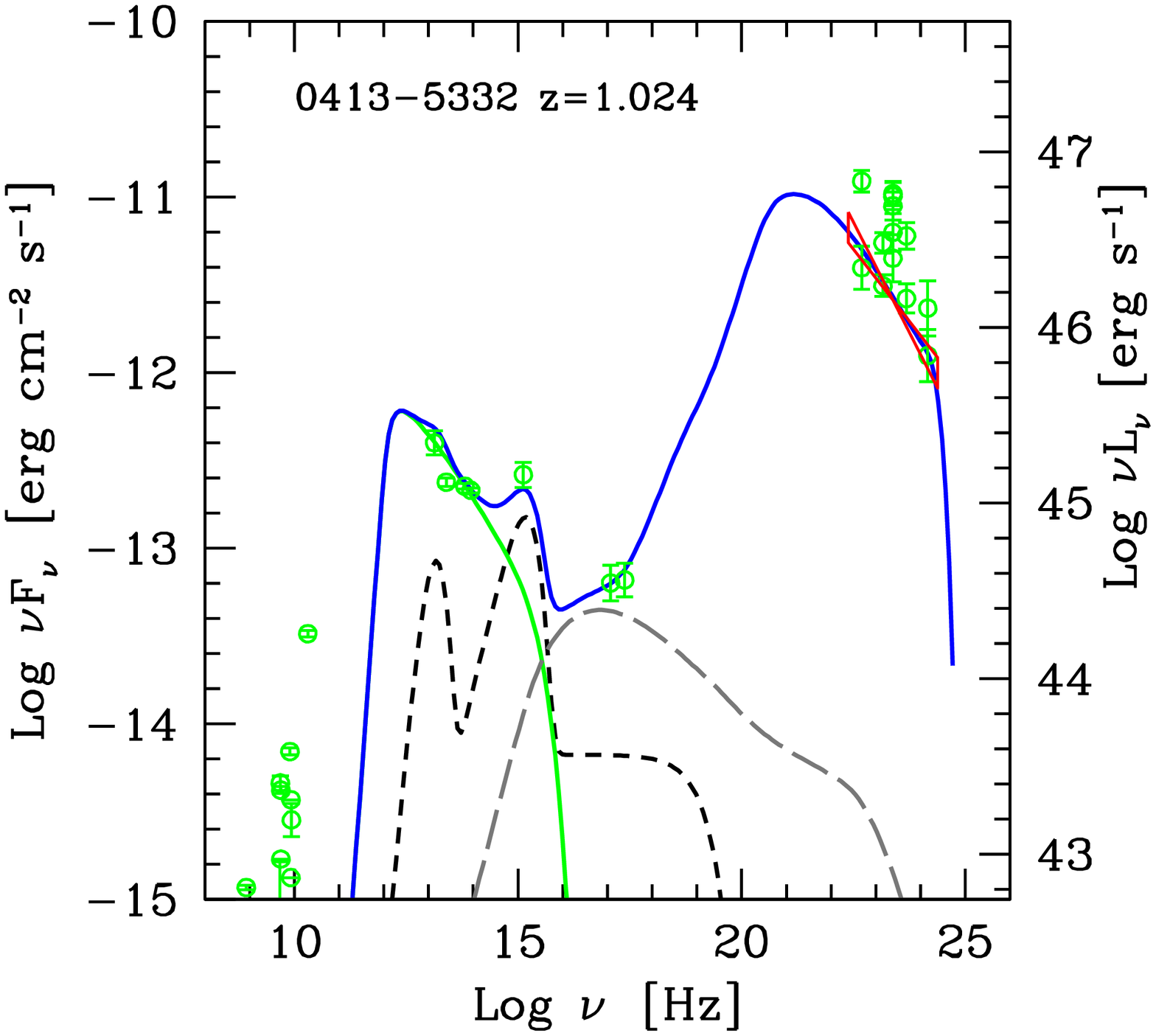,width=4.3cm,height=3.7cm} \\
\psfig{file=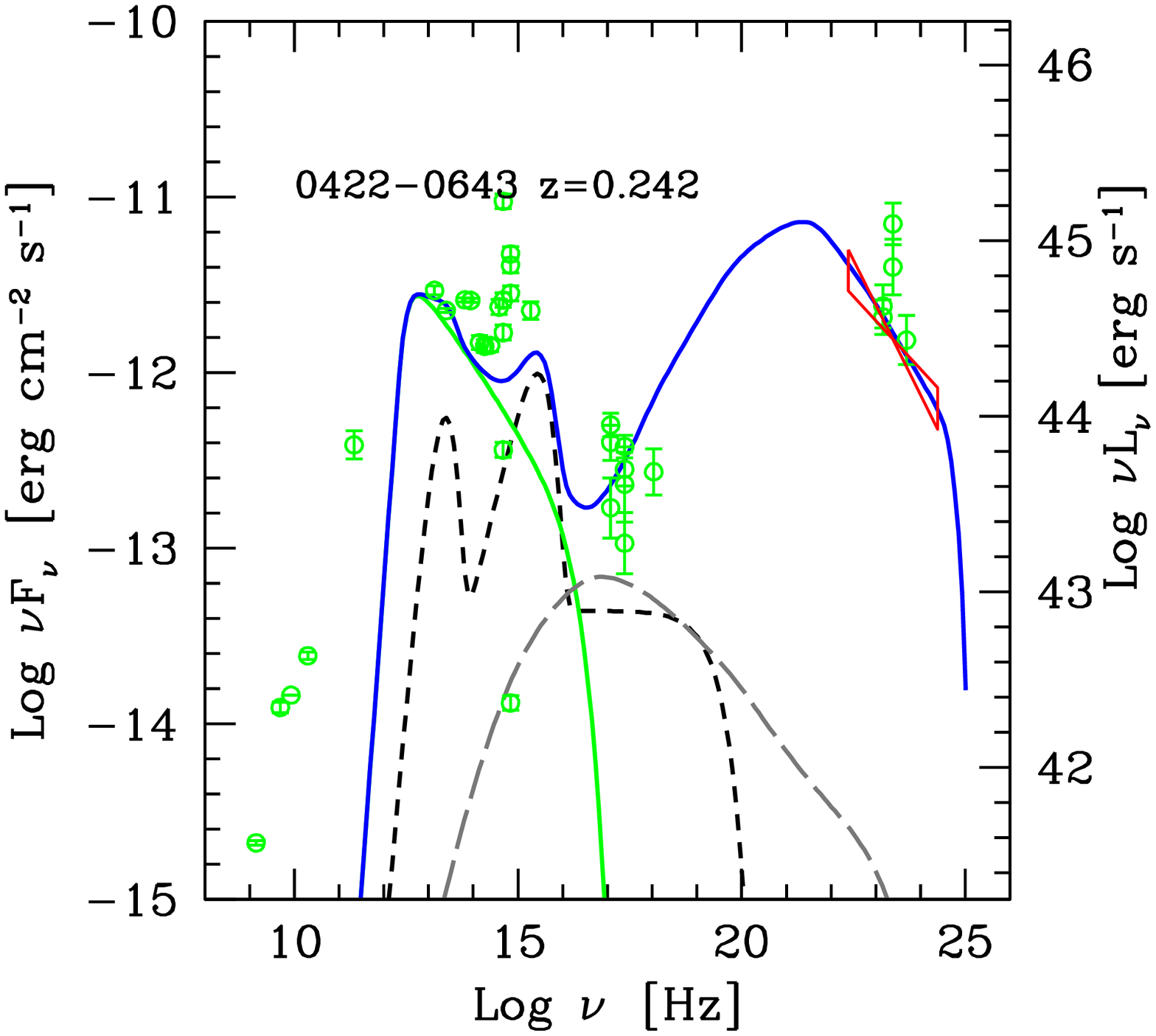,width=4.3cm,height=3.7cm}  
&\psfig{file=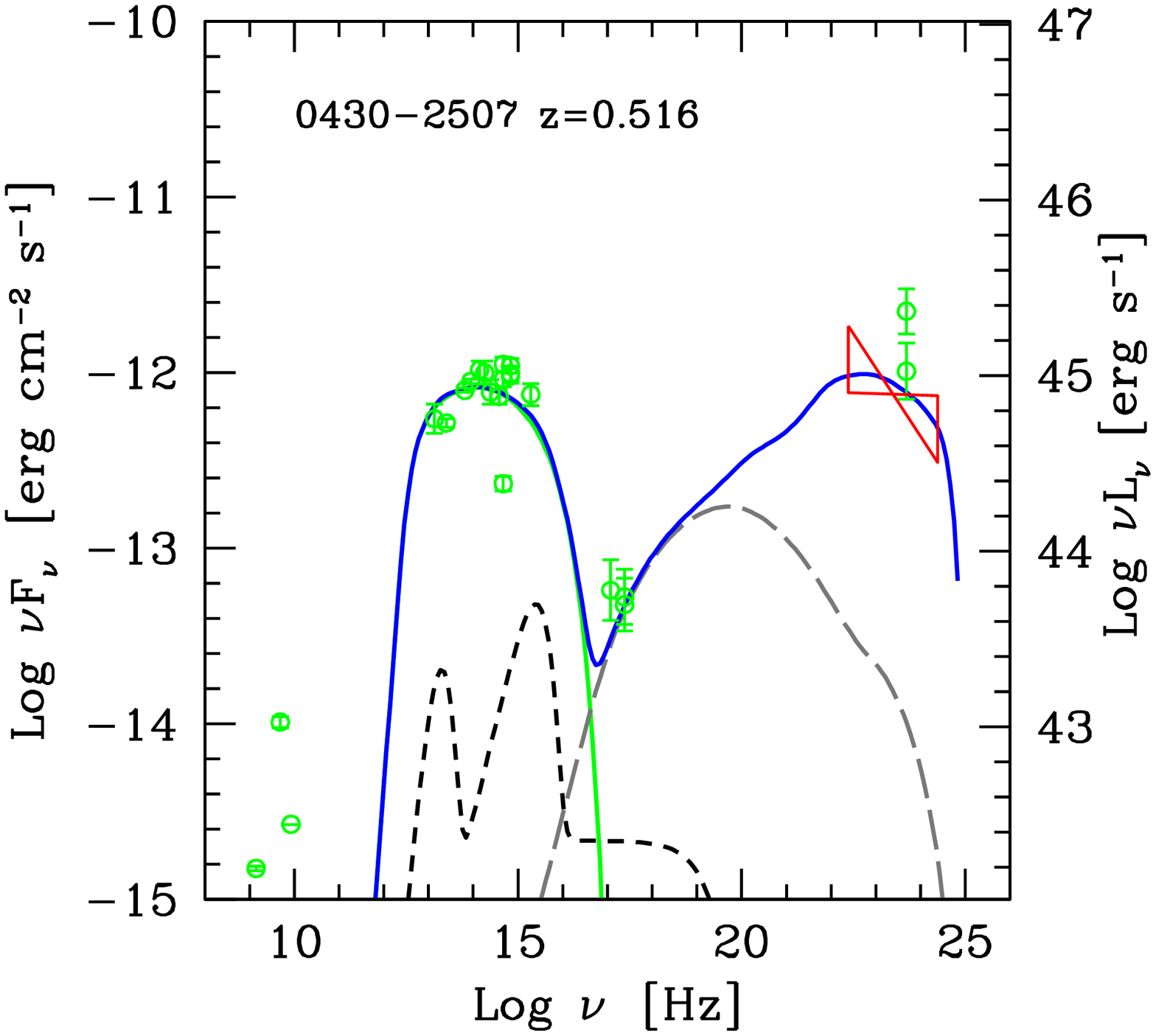,width=4.3cm,height=3.7cm} 
&\psfig{file=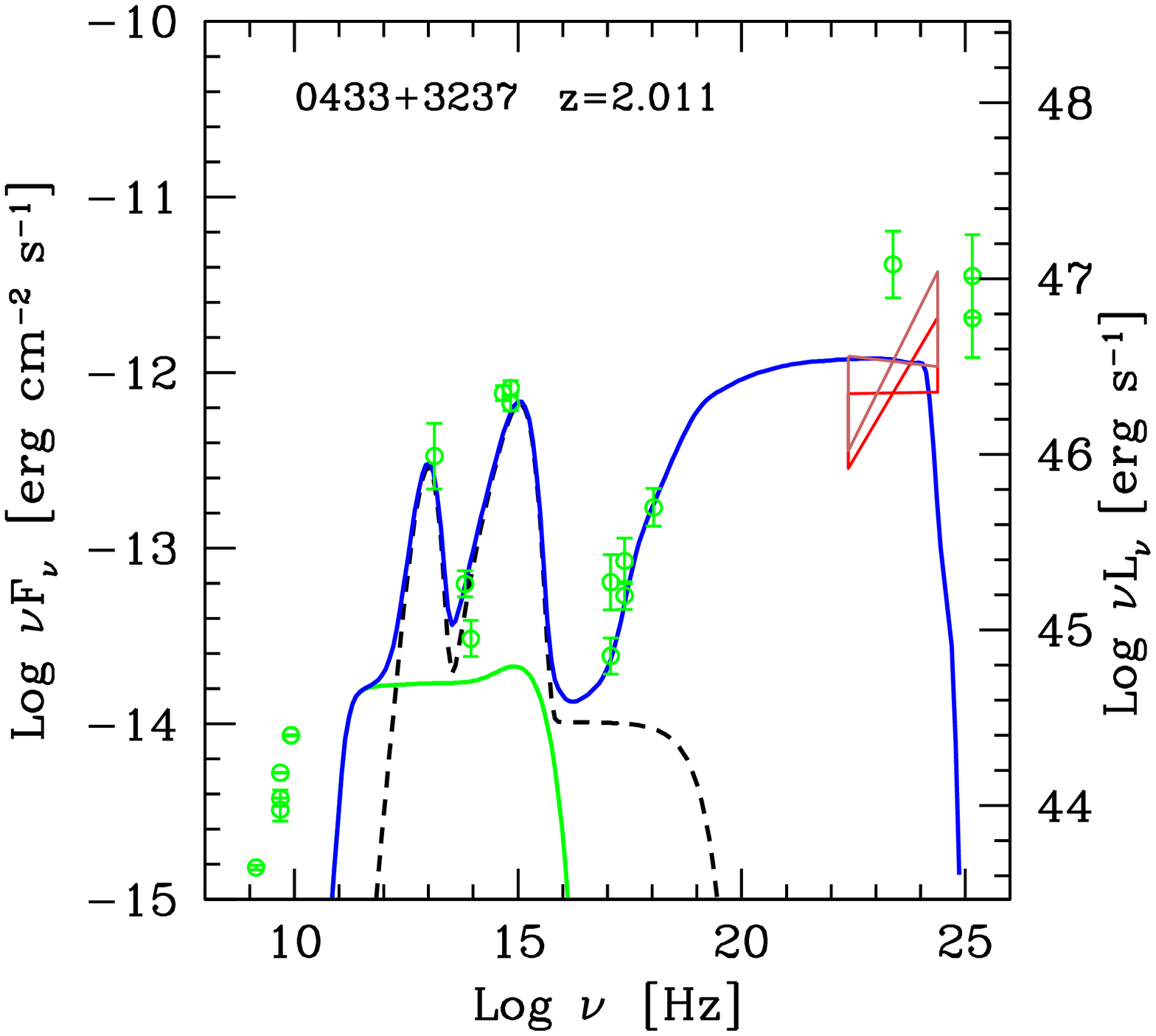,width=4.3cm,height=3.7cm}  
&\psfig{file=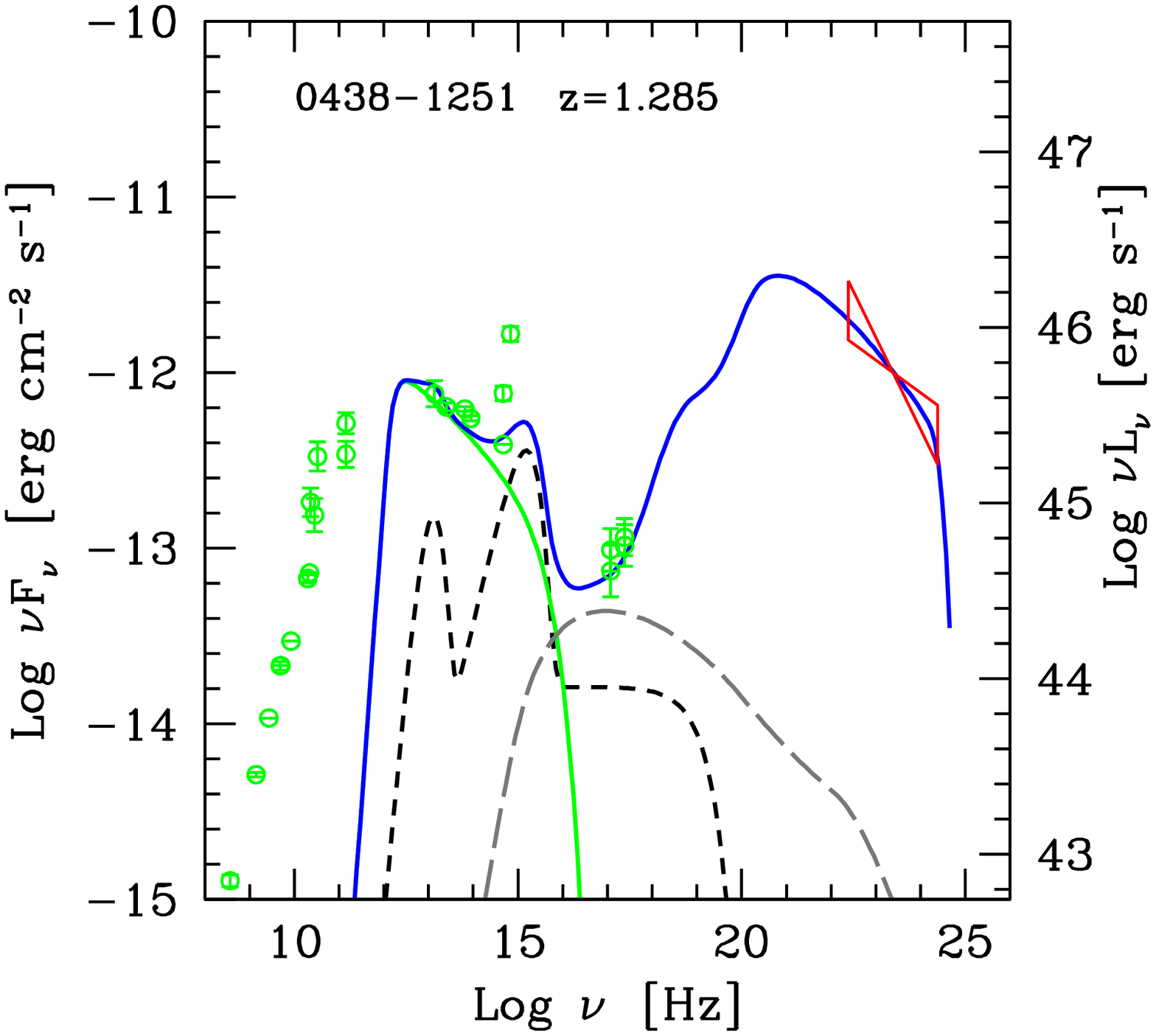,width=4.3cm,height=3.7cm} \\
\psfig{file=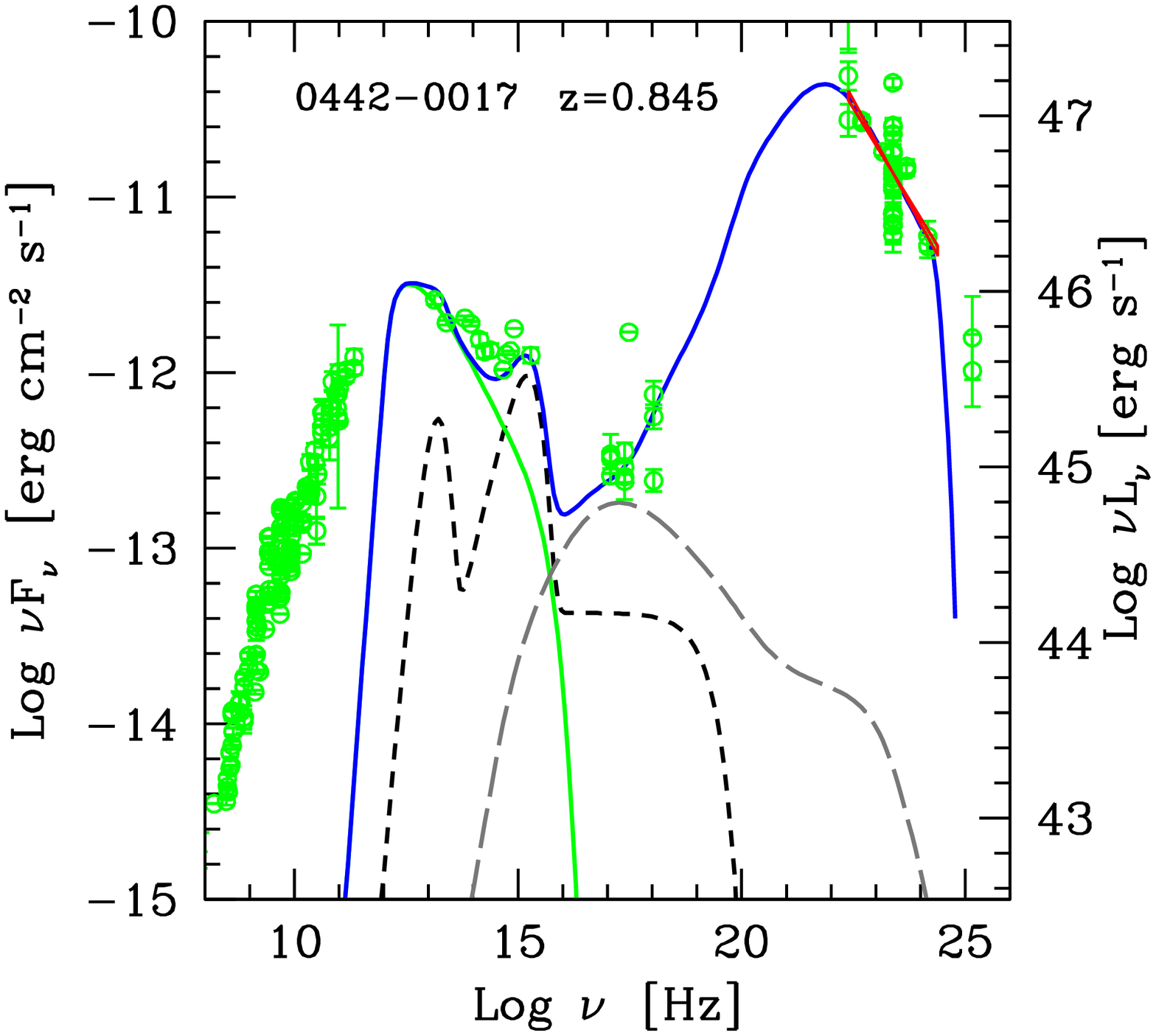,width=4.3cm,height=3.7cm}  
&\psfig{file=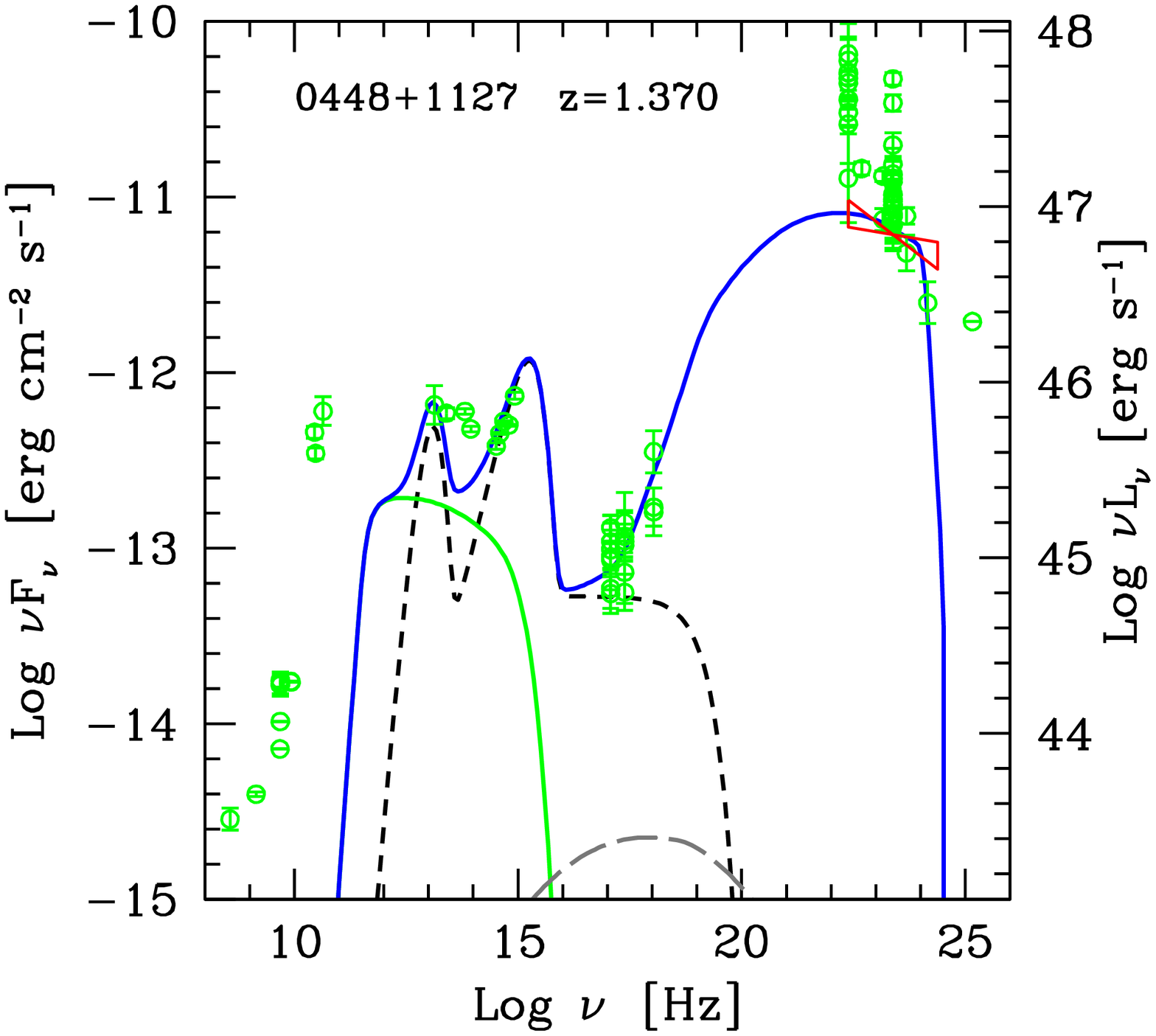,width=4.3cm,height=3.7cm} 
&\psfig{file=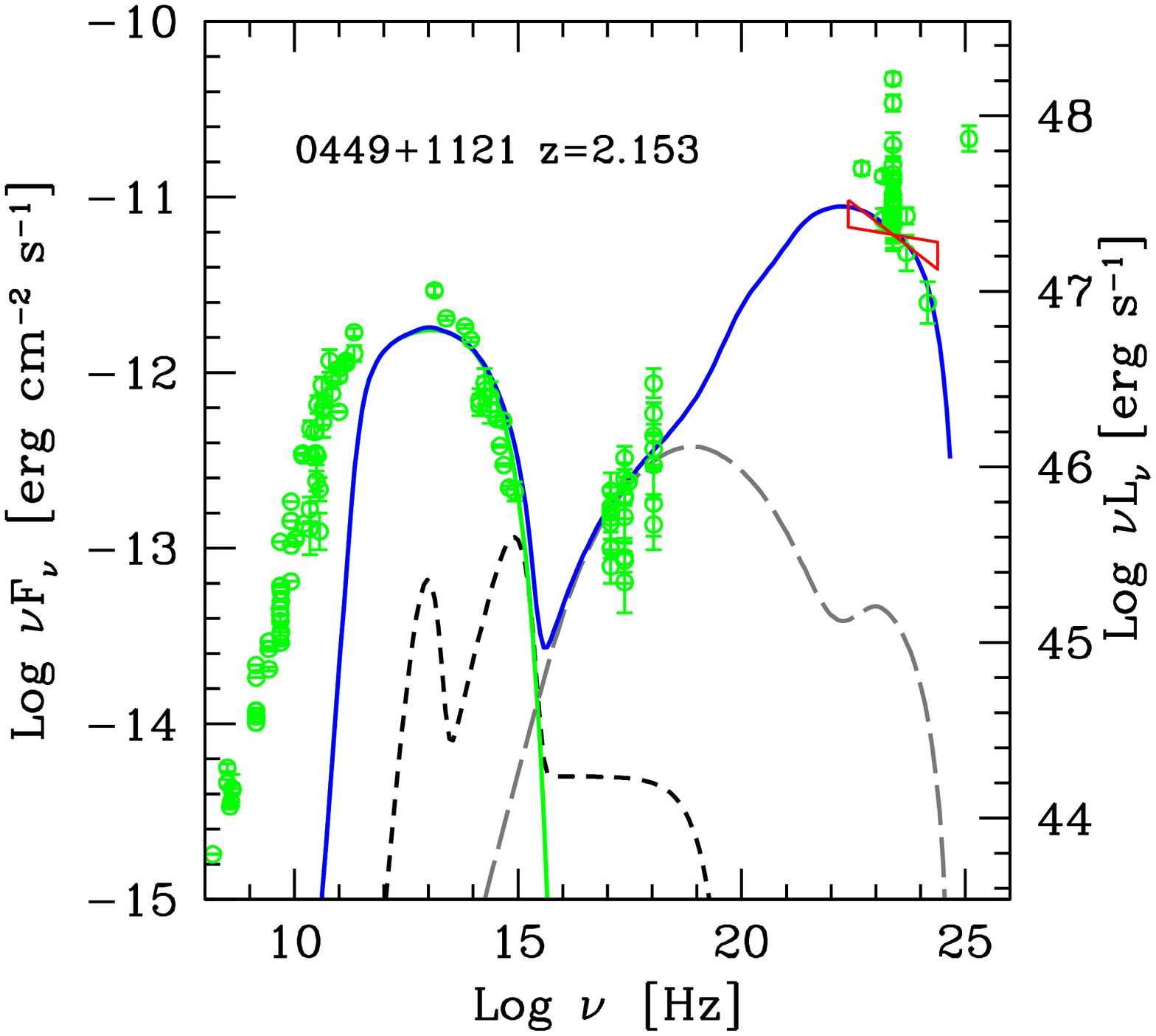,width=4.3cm,height=3.7cm } 
&\psfig{file=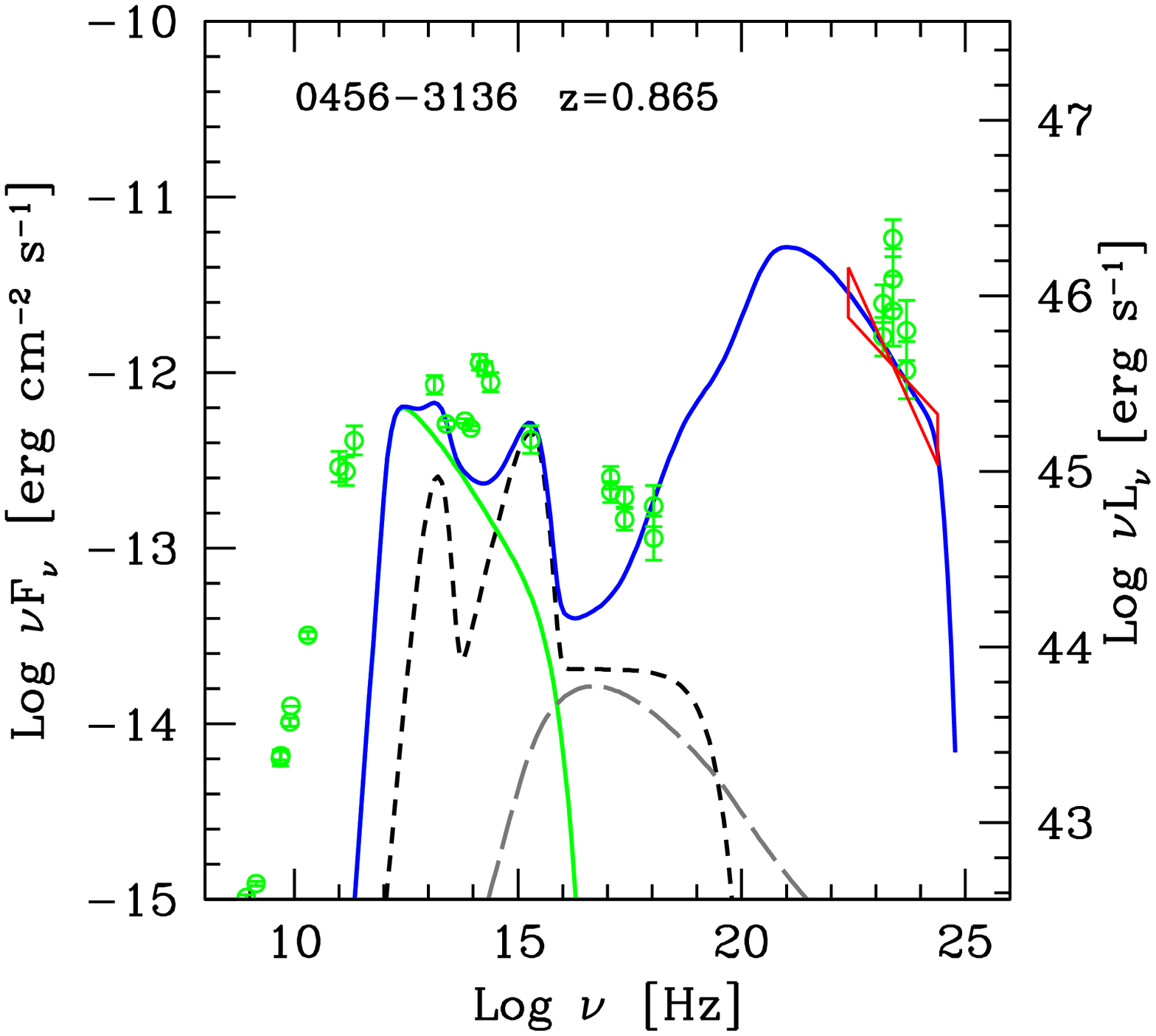,width=4.3cm,height=3.7cm } \\
\psfig{file=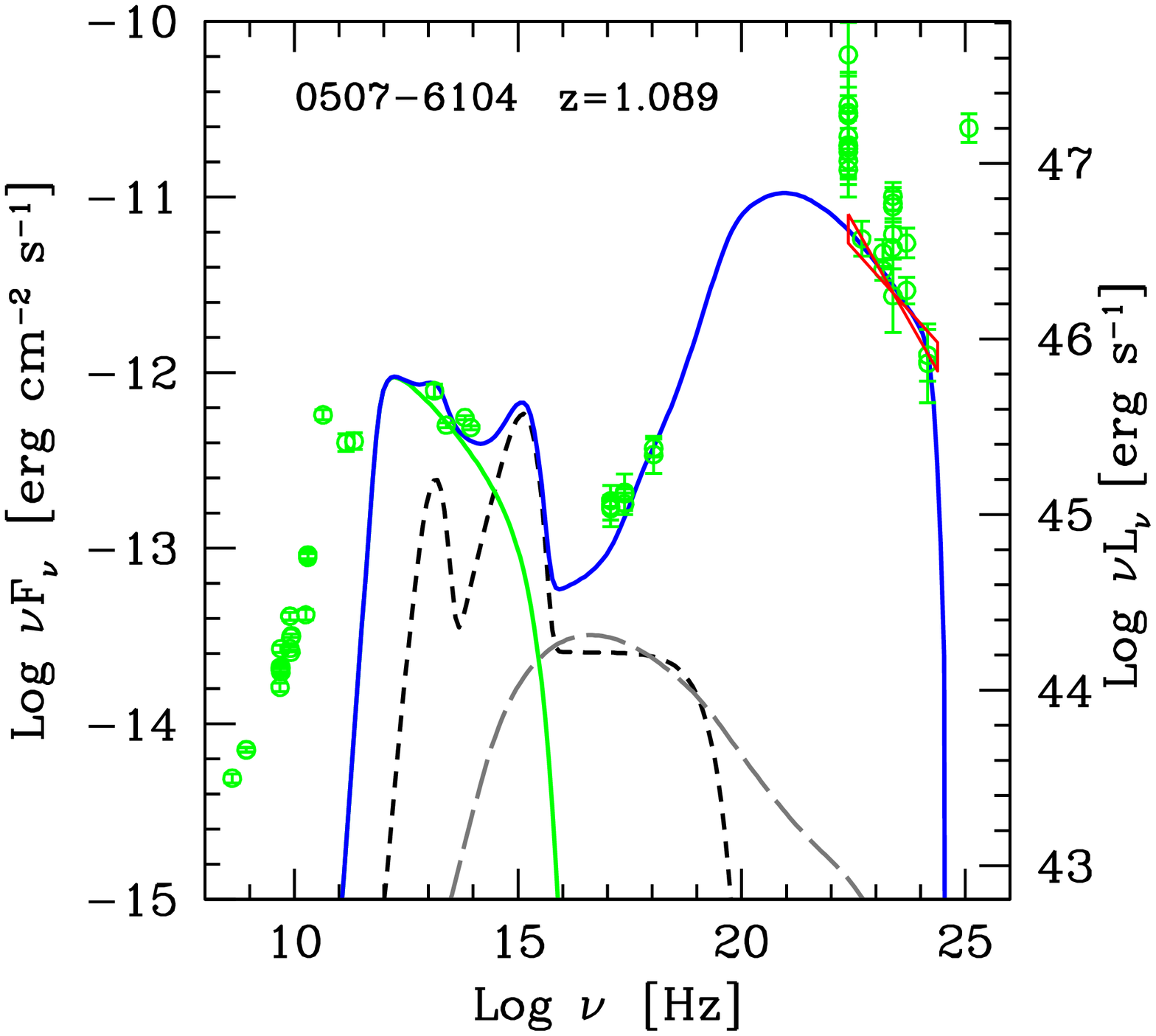,width=4.3cm,height=3.7cm }  
&\psfig{file=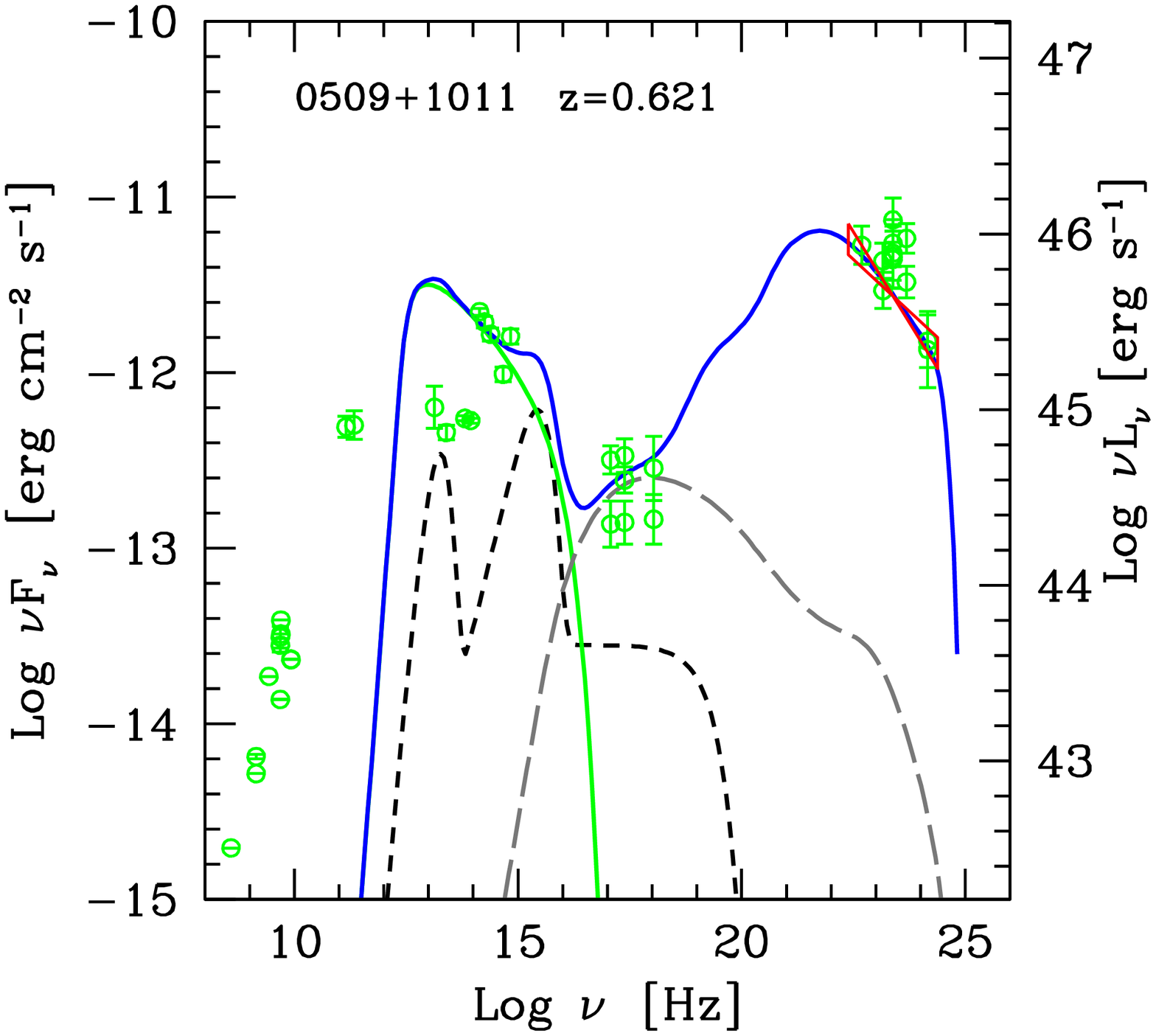,width=4.3cm,height=3.7cm } 
&\psfig{file=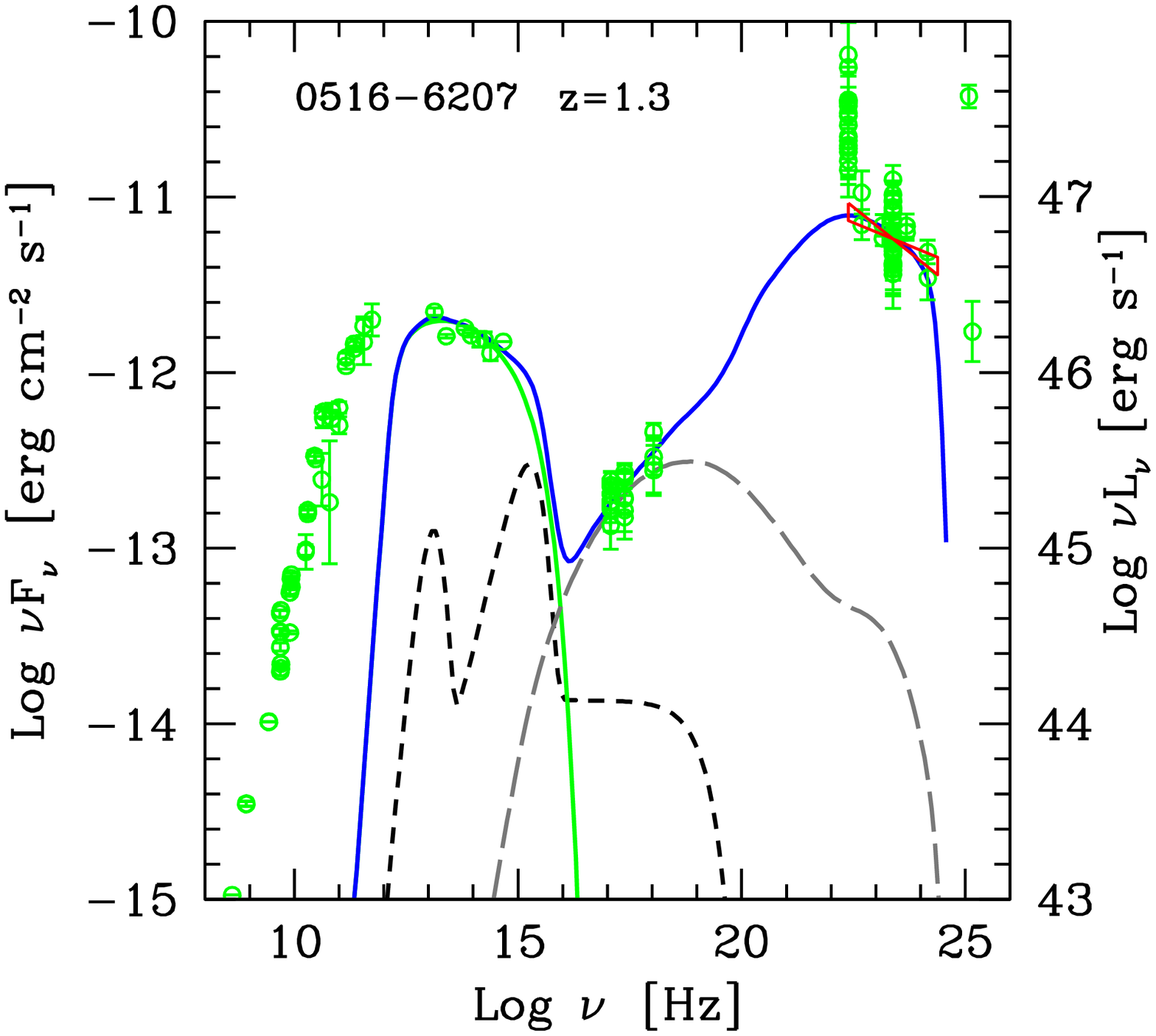,width=4.3cm,height=3.7cm }  
&\psfig{file=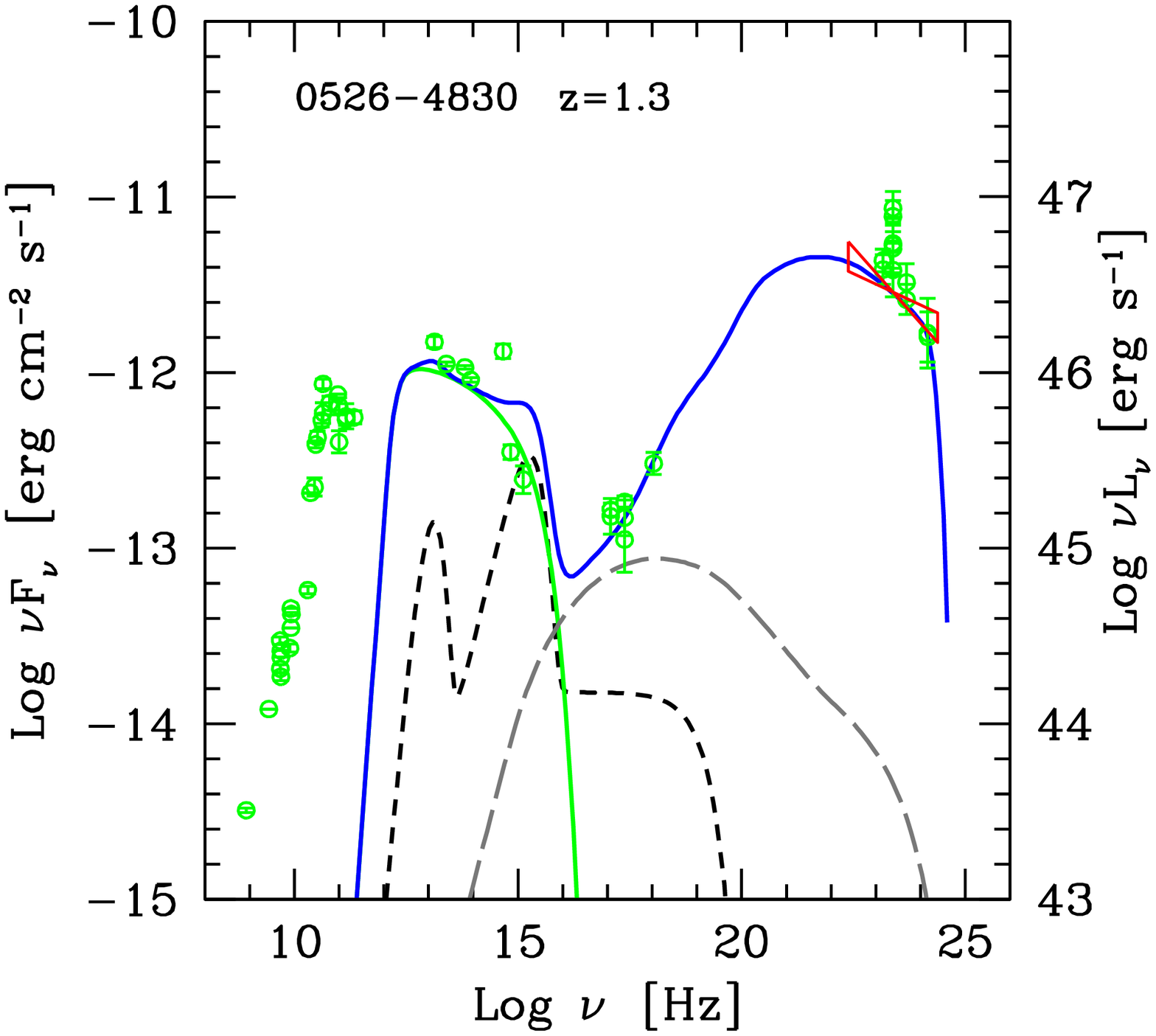,width=4.3cm,height=3.7cm } \\
\psfig{file=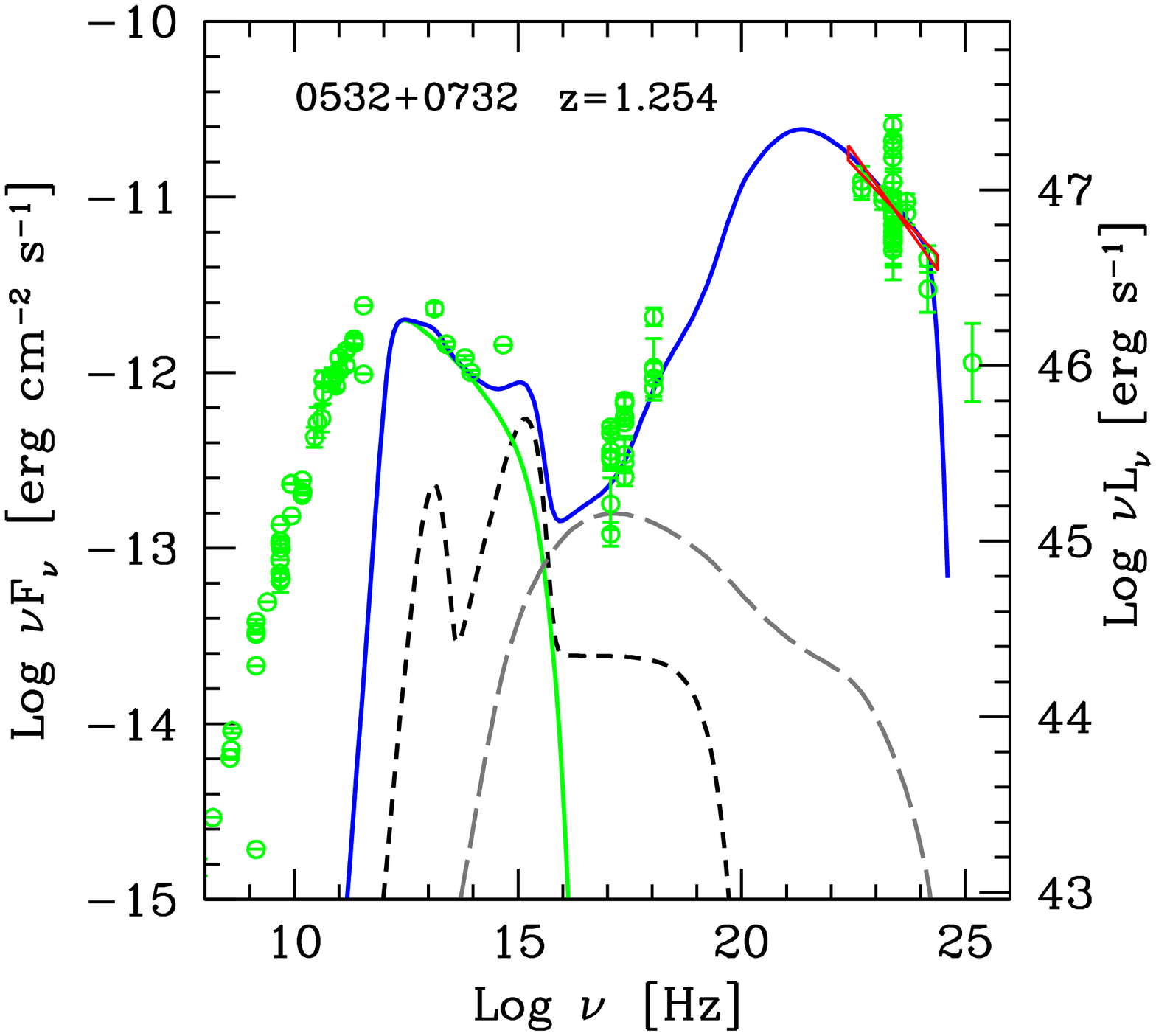,width=4.3cm,height=3.7cm }  
&\psfig{file=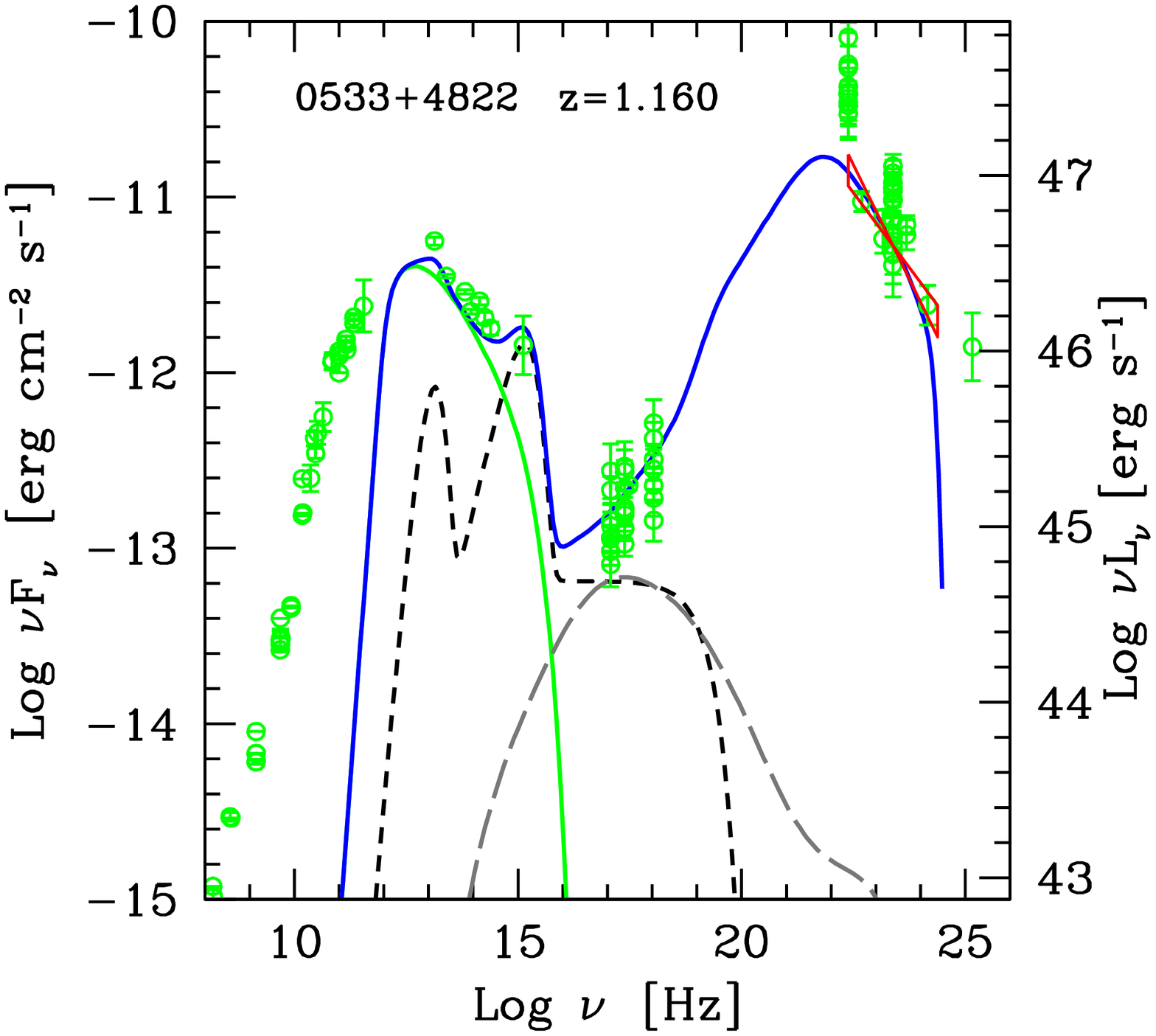,width=4.3cm,height=3.7cm } 
&\psfig{file=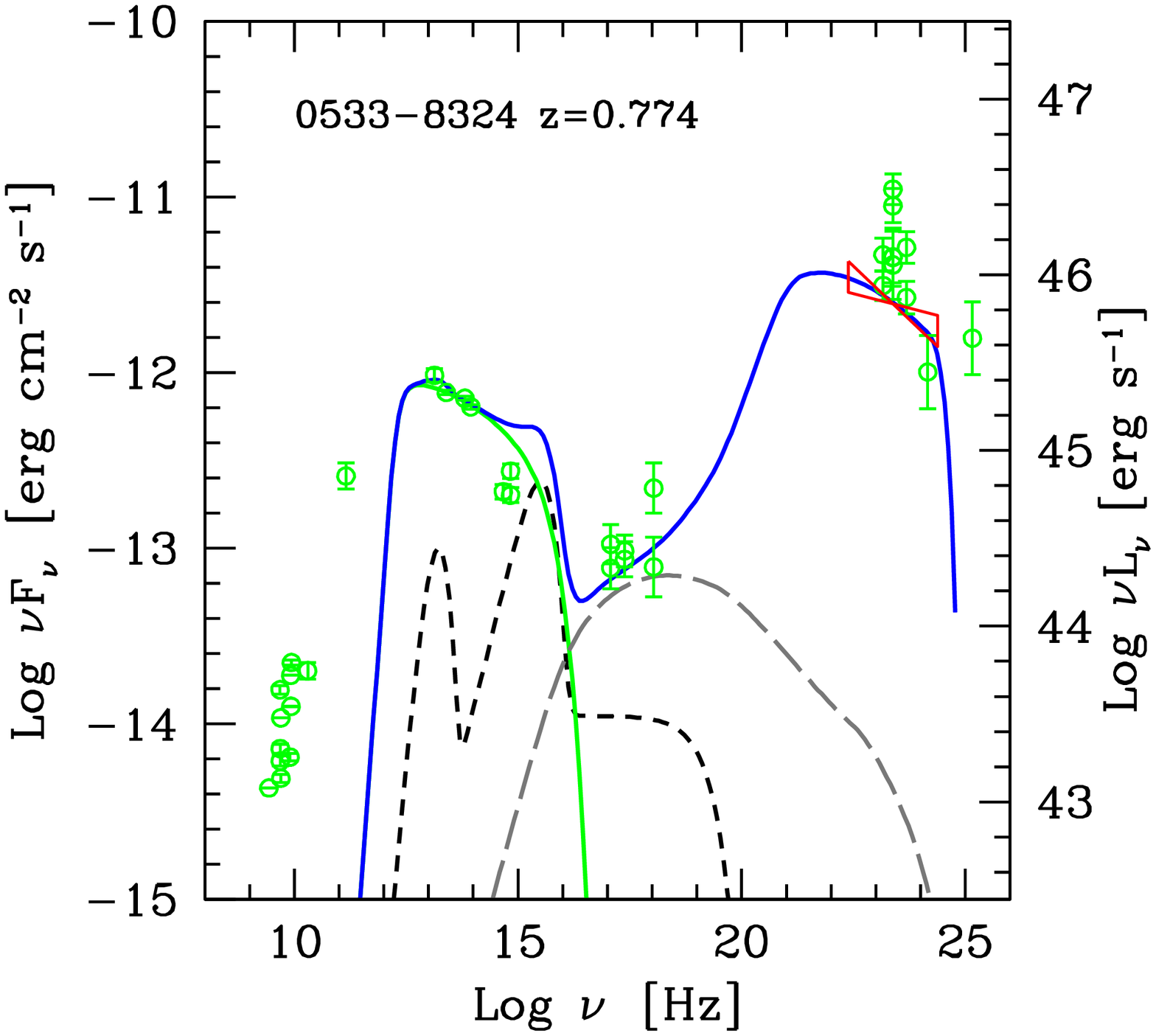,width=4.3cm,height=3.7cm }  
&\psfig{file=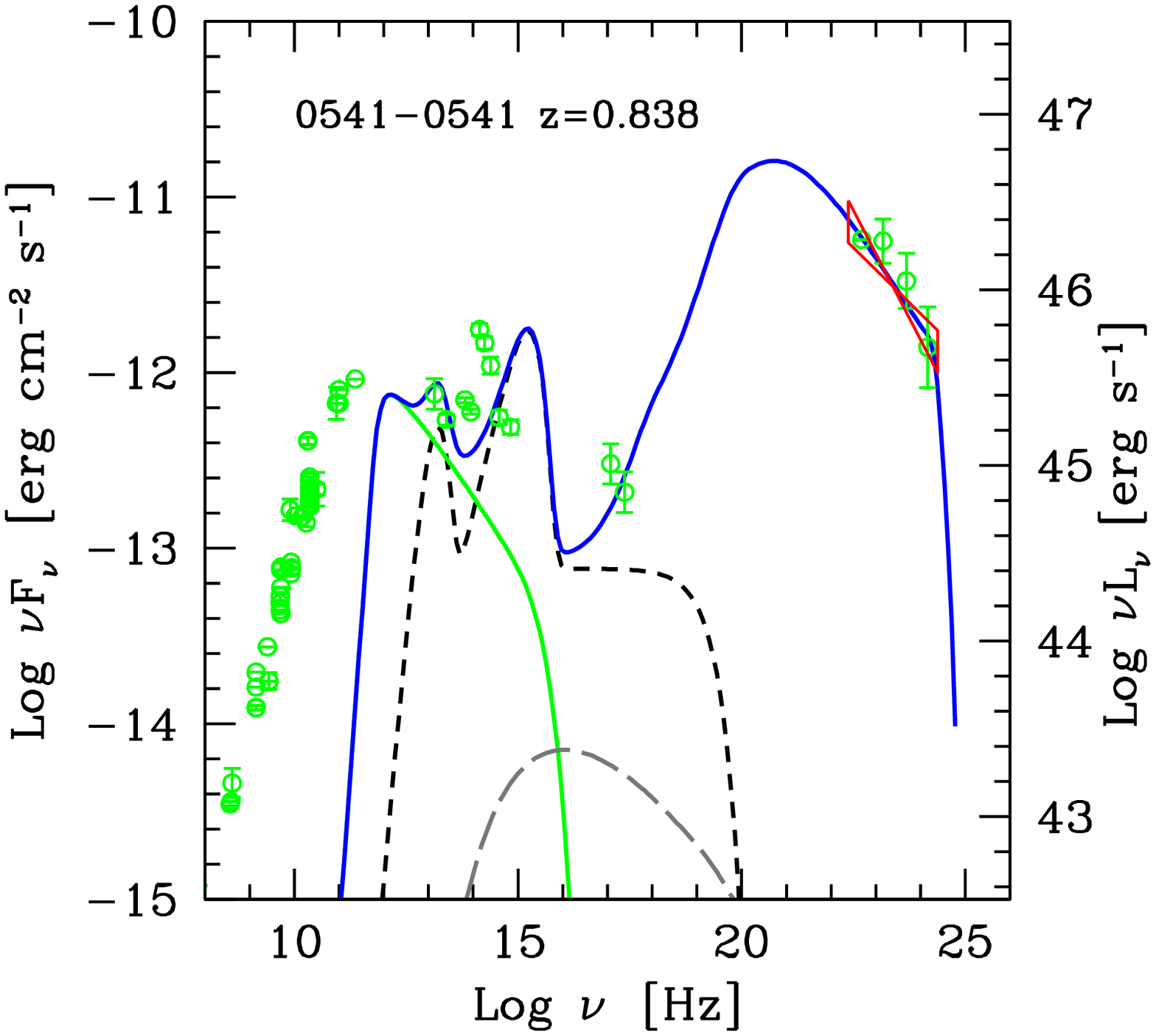,width=4.3cm,height=3.7cm } 
\end{tabular}
\caption{{\it continue.} SED of the FSRQs studied in this paper.}
\end{figure*} 

\setcounter{figure}{15}
\begin{figure*}
\begin{tabular}{cccc}
\psfig{file=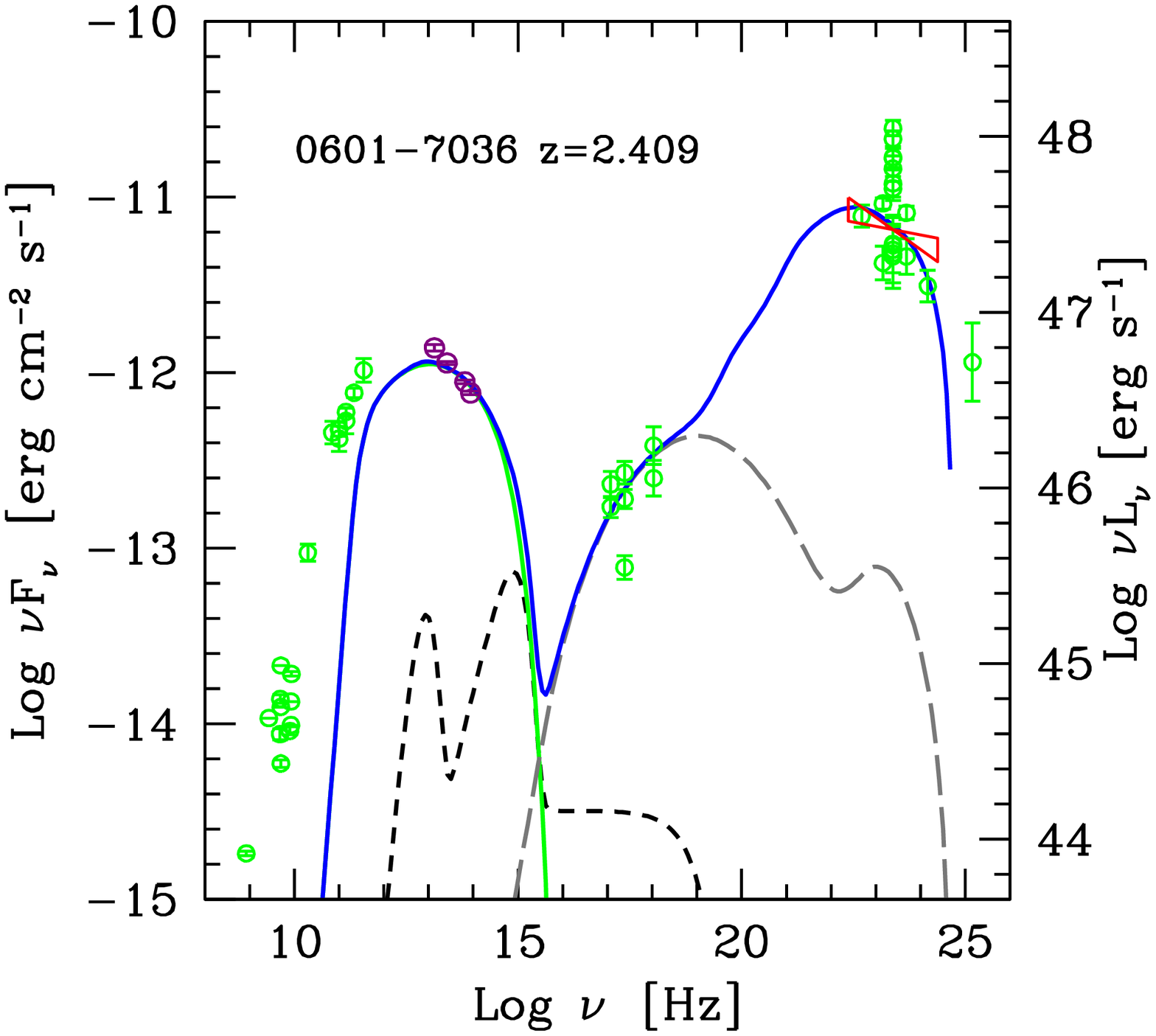,width=4.3cm,height=3.7cm } 
&\psfig{file=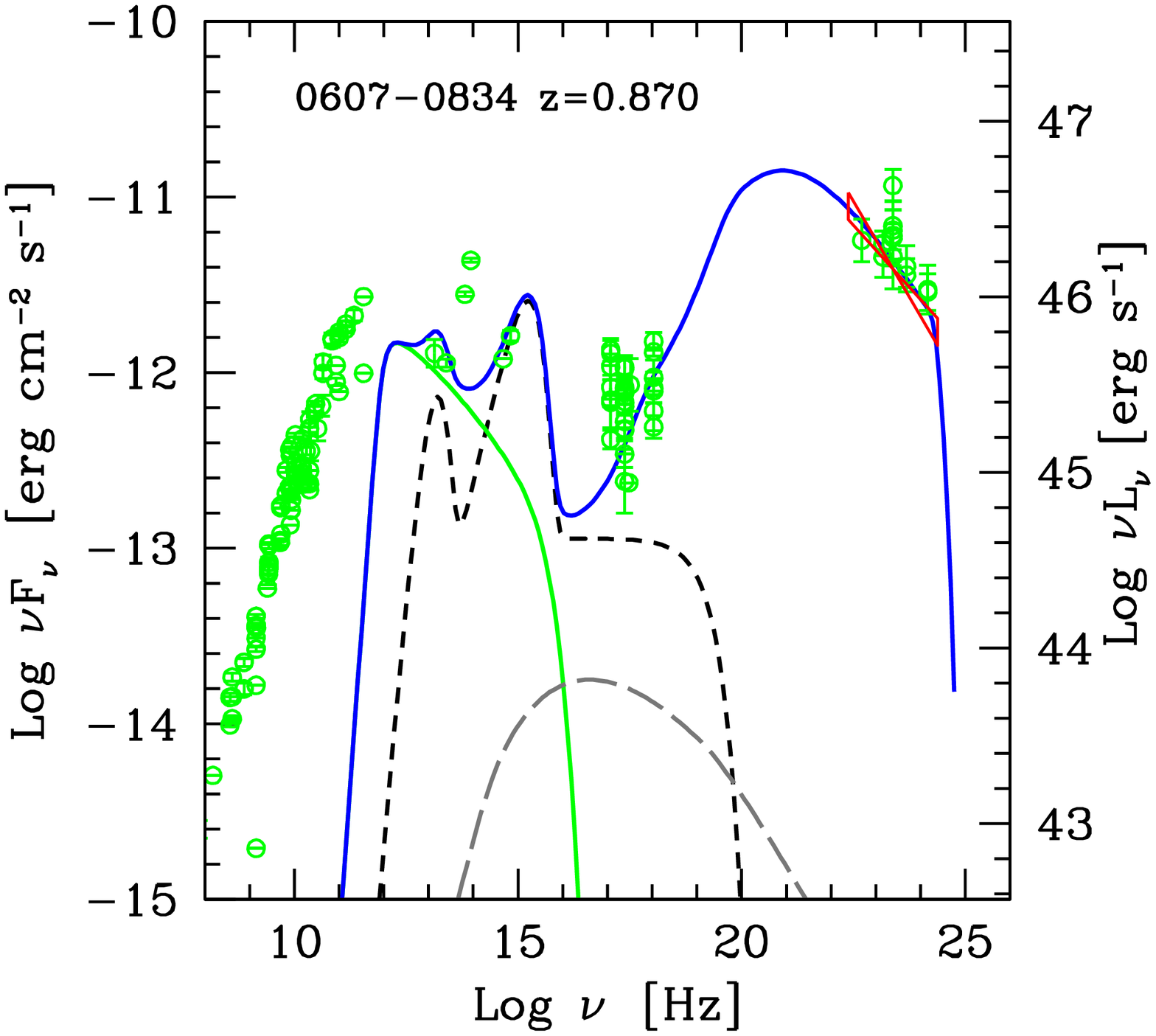,width=4.3cm,height=3.7cm }  
&\psfig{file=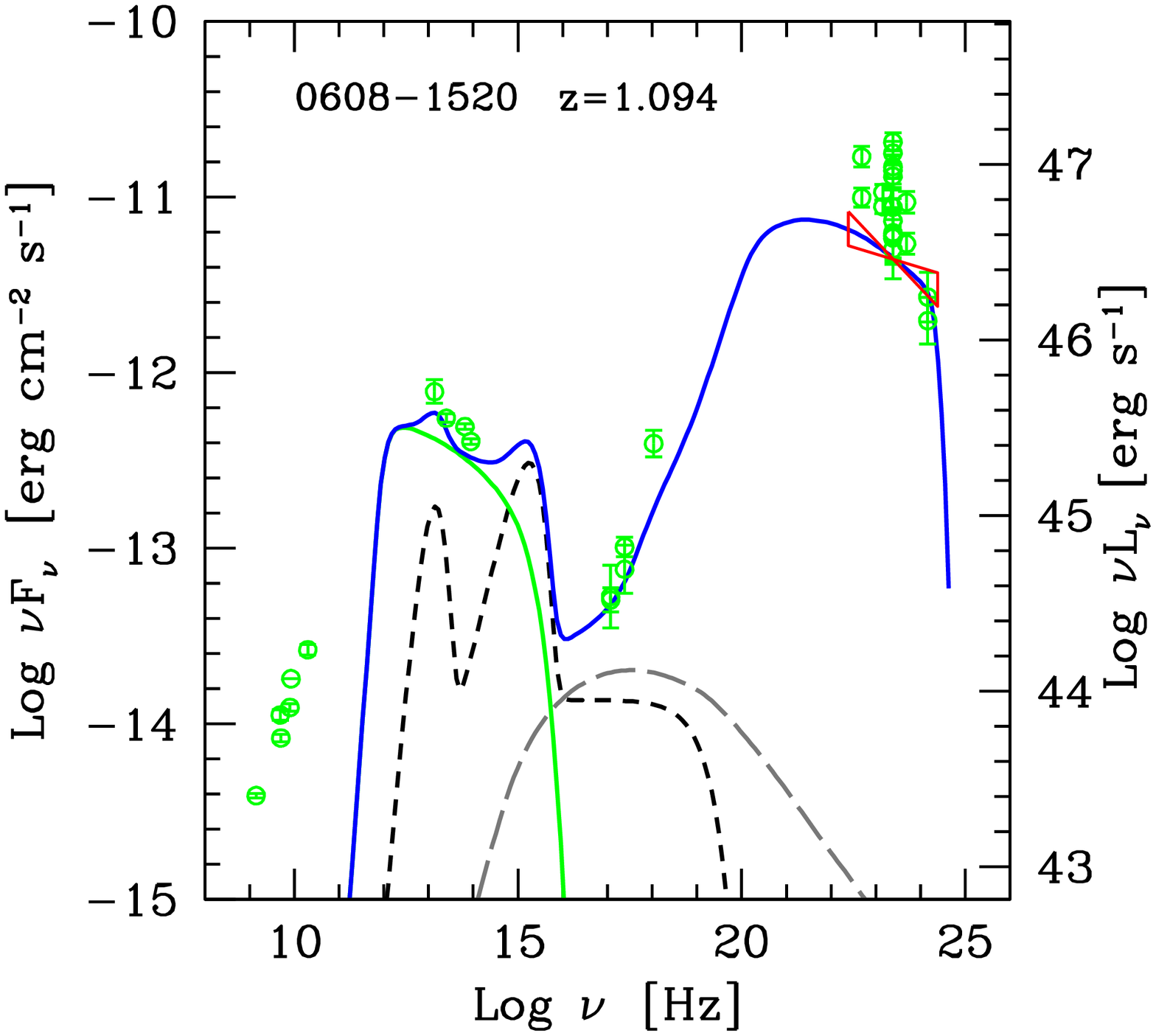,width=4.3cm,height=3.7cm }  
&\psfig{file=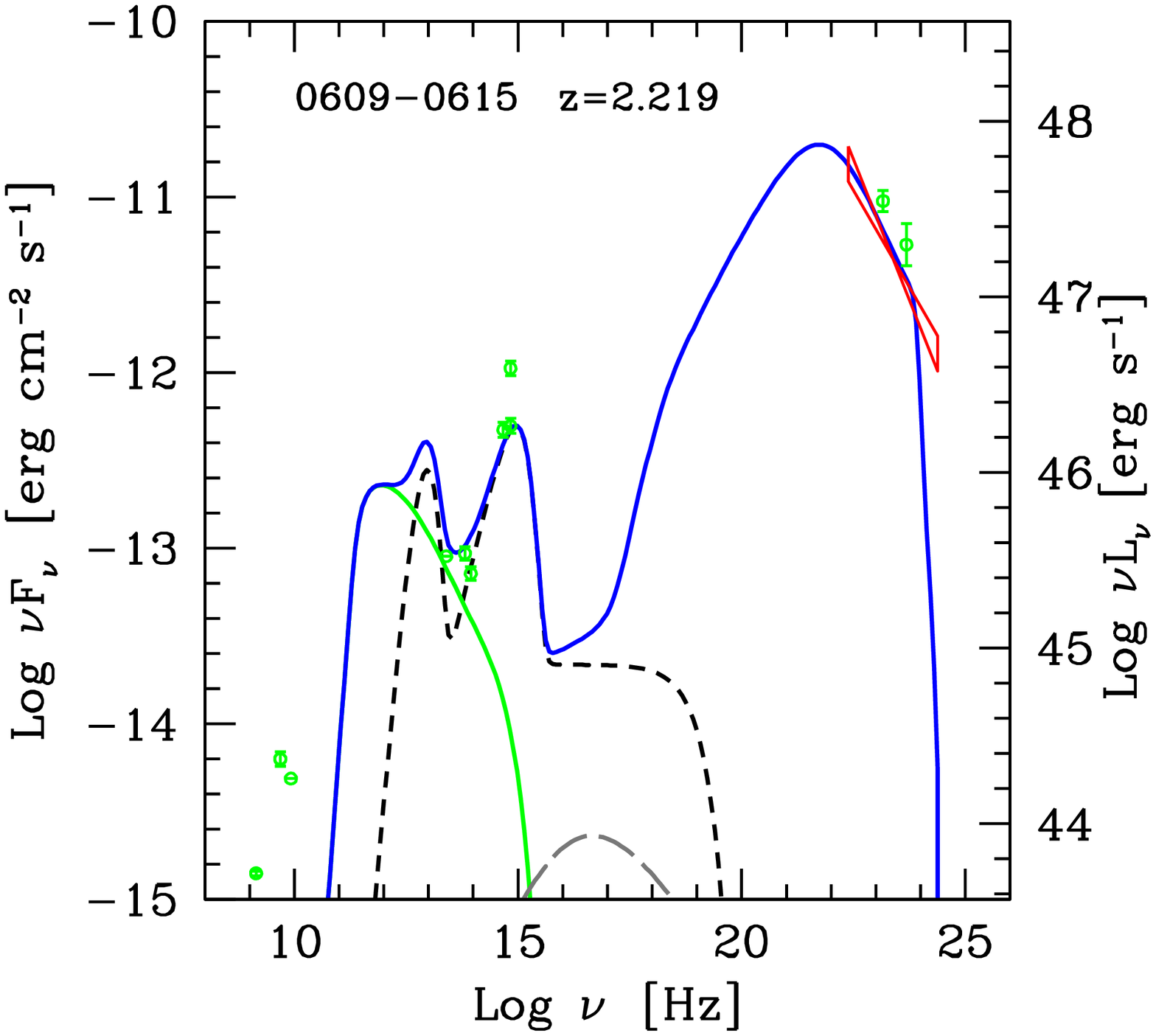,width=4.3cm,height=3.7cm }   \\
\psfig{file=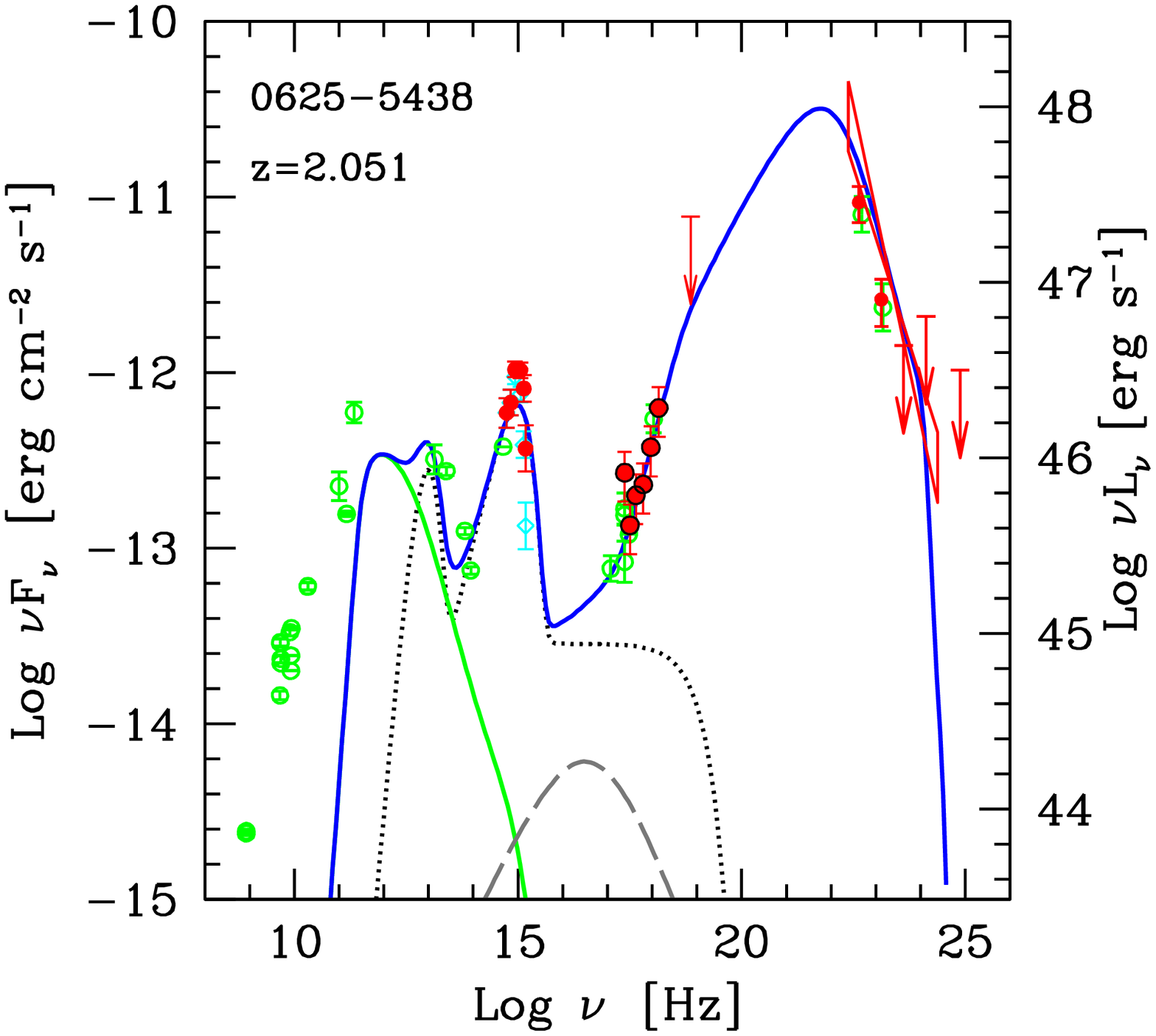,width=4.3cm,height=3.7cm }  
&\psfig{file=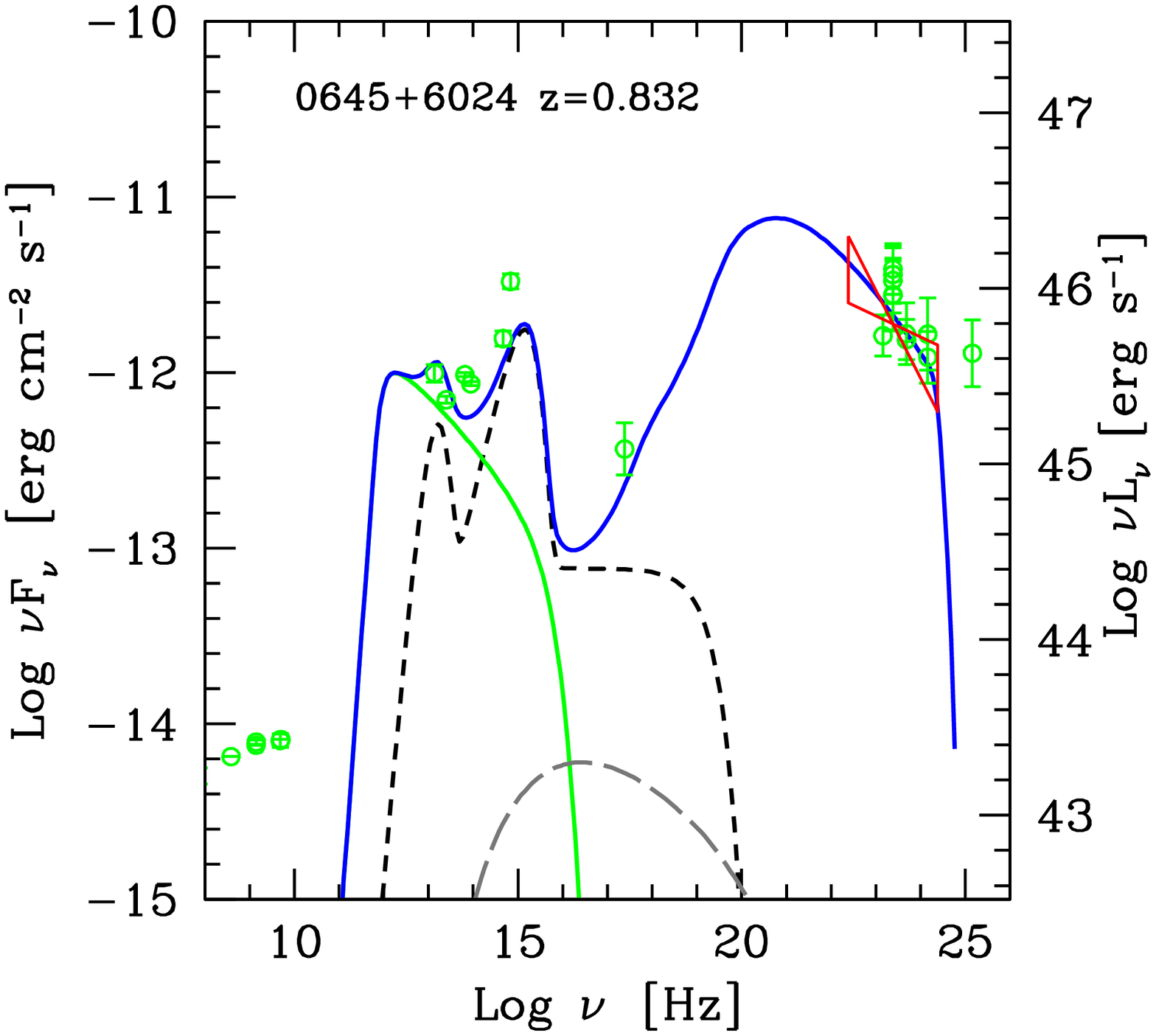,width=4.3cm,height=3.7cm } 
&\psfig{file=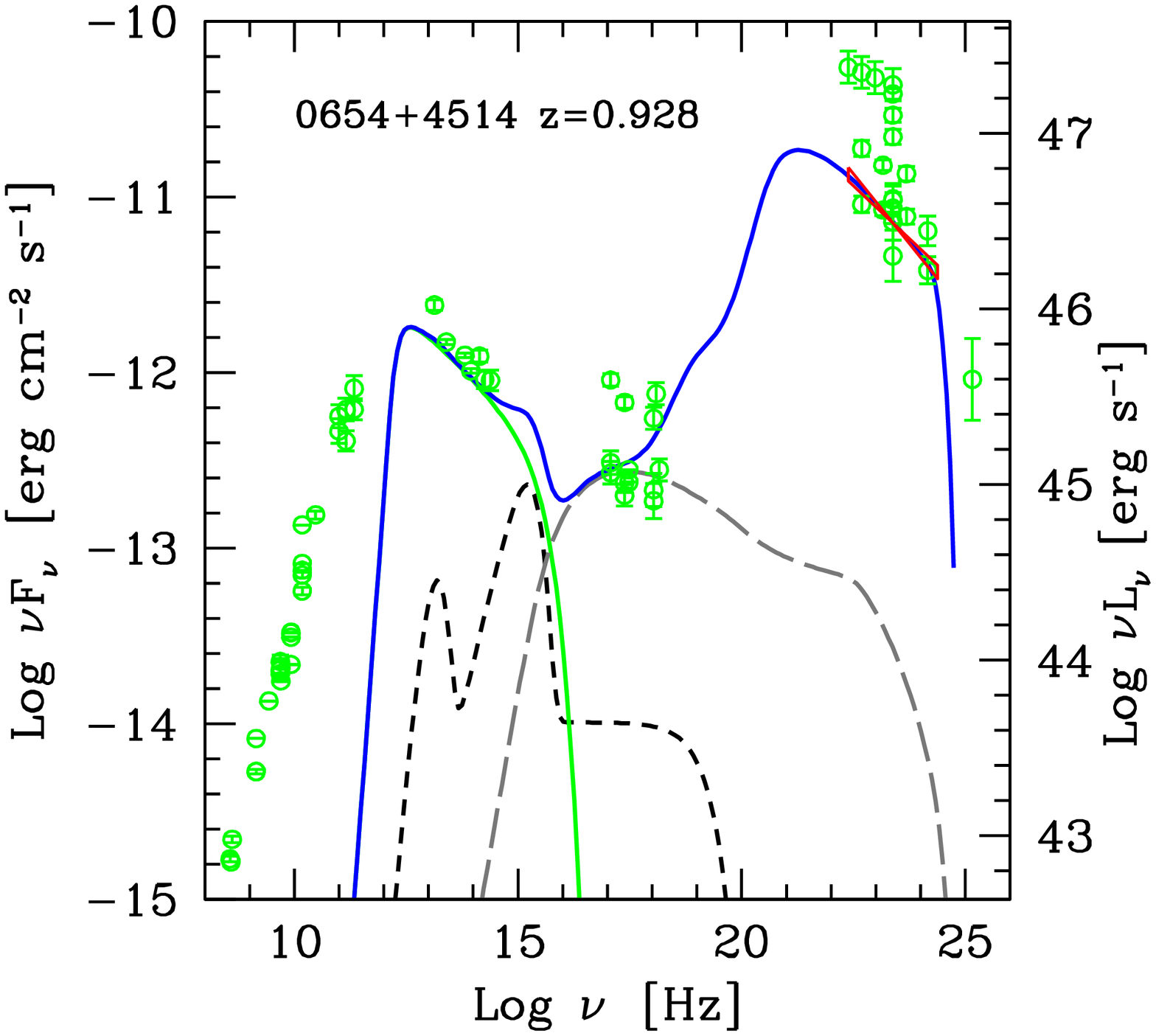,width=4.3cm,height=3.7cm } 
&\psfig{file=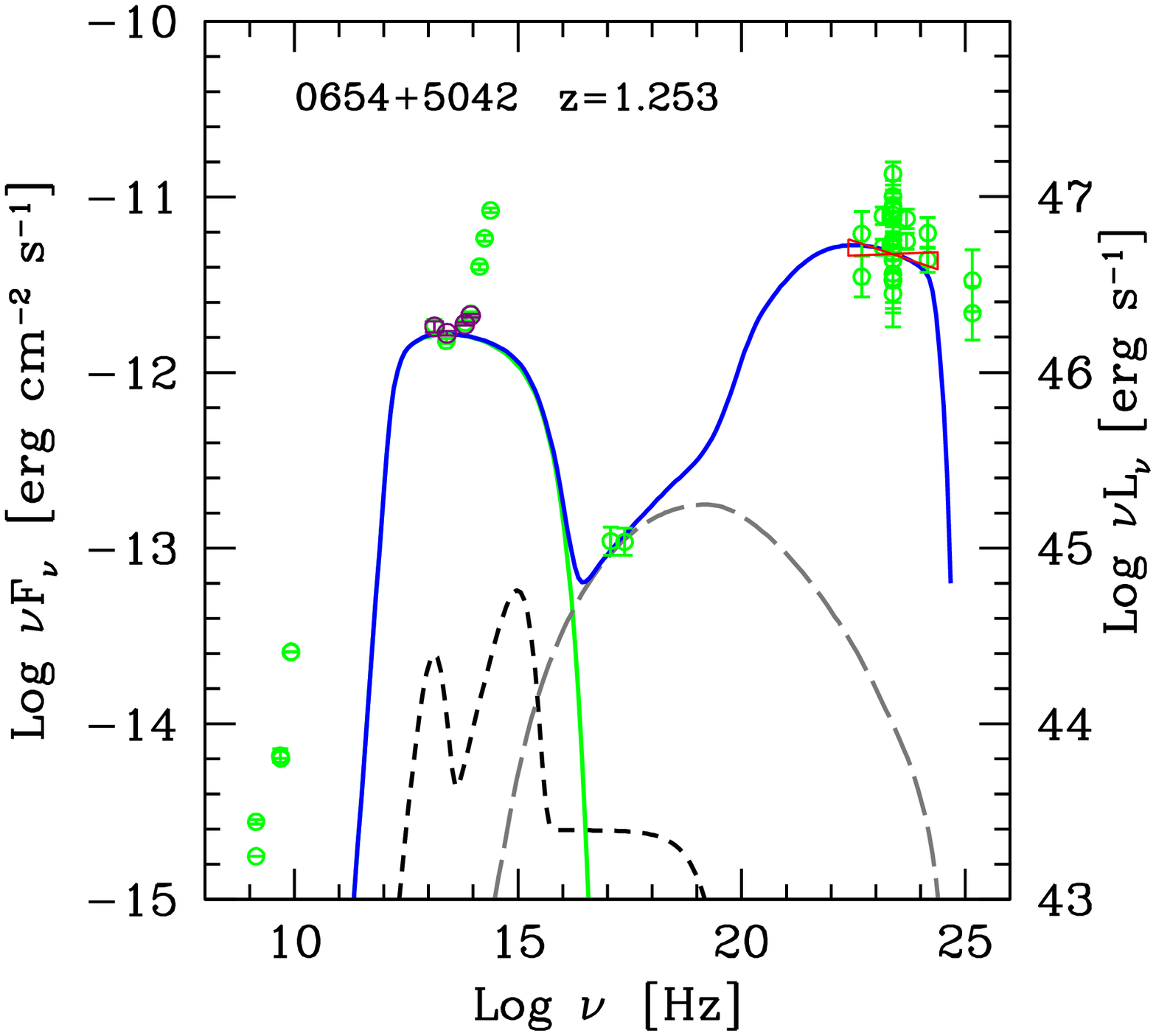,width=4.3cm,height=3.7cm }   \\
\psfig{file=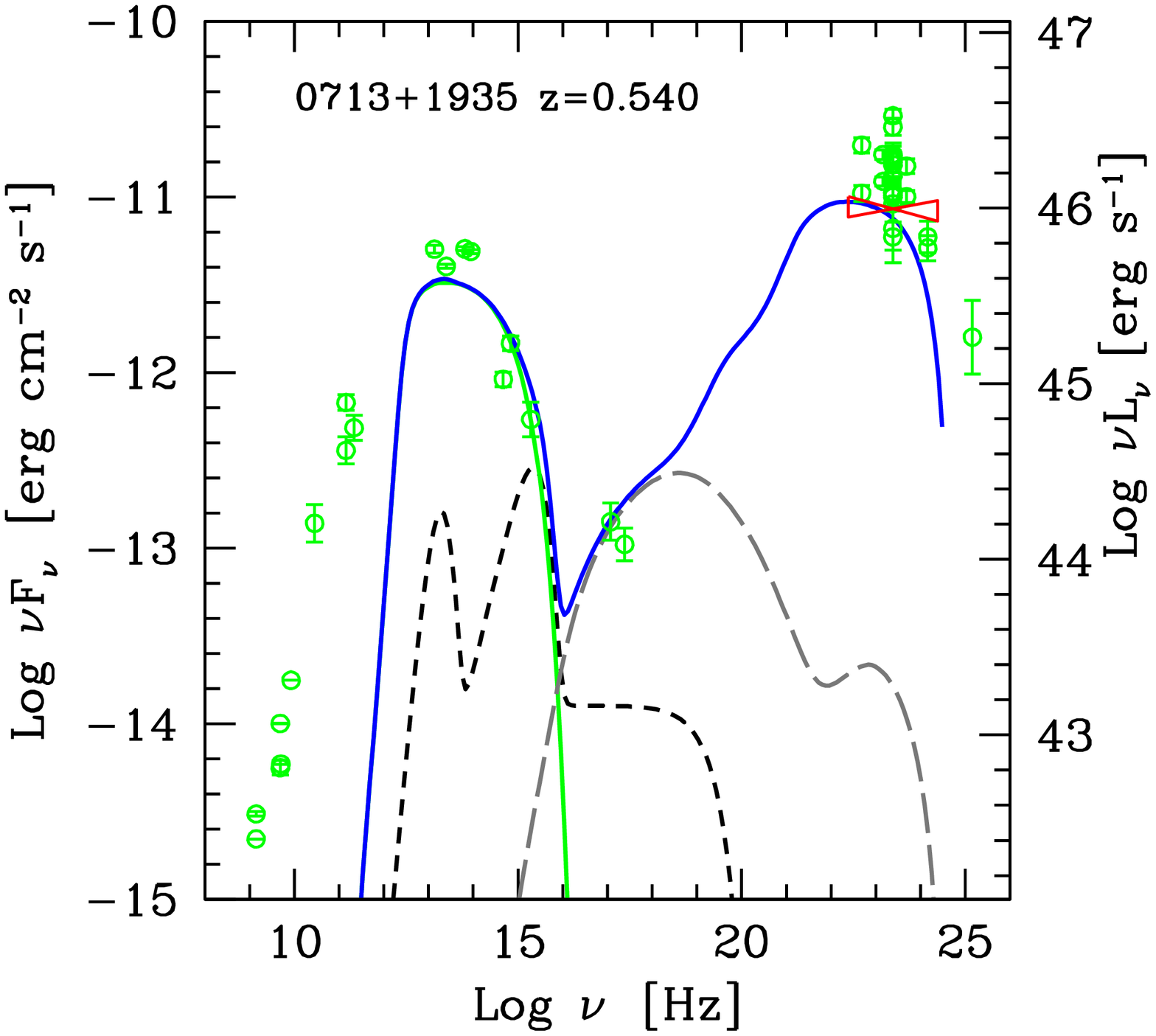,width=4.3cm,height=3.7cm }  
&\psfig{file=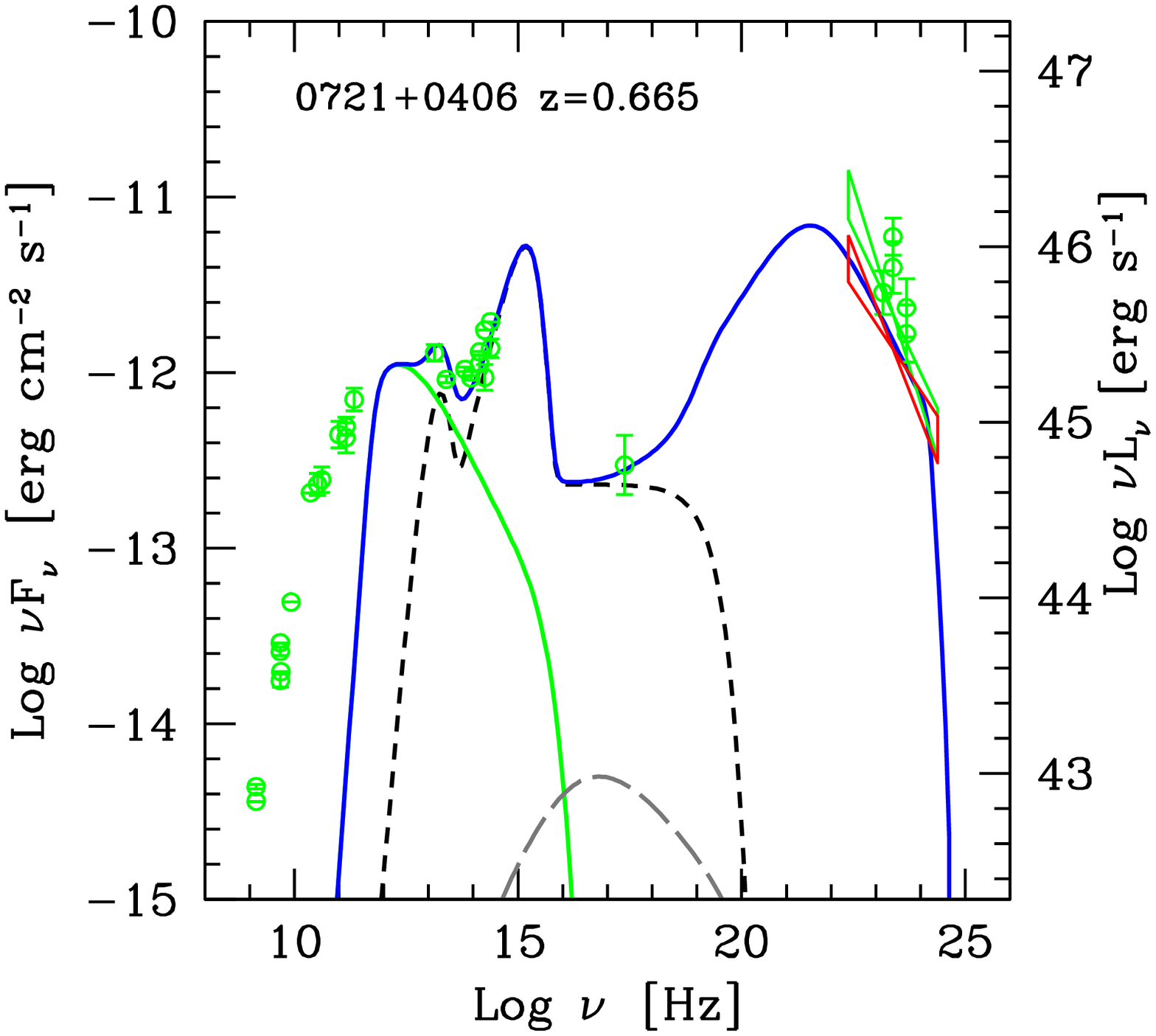,width=4.3cm,height=3.7cm }  
&\psfig{file=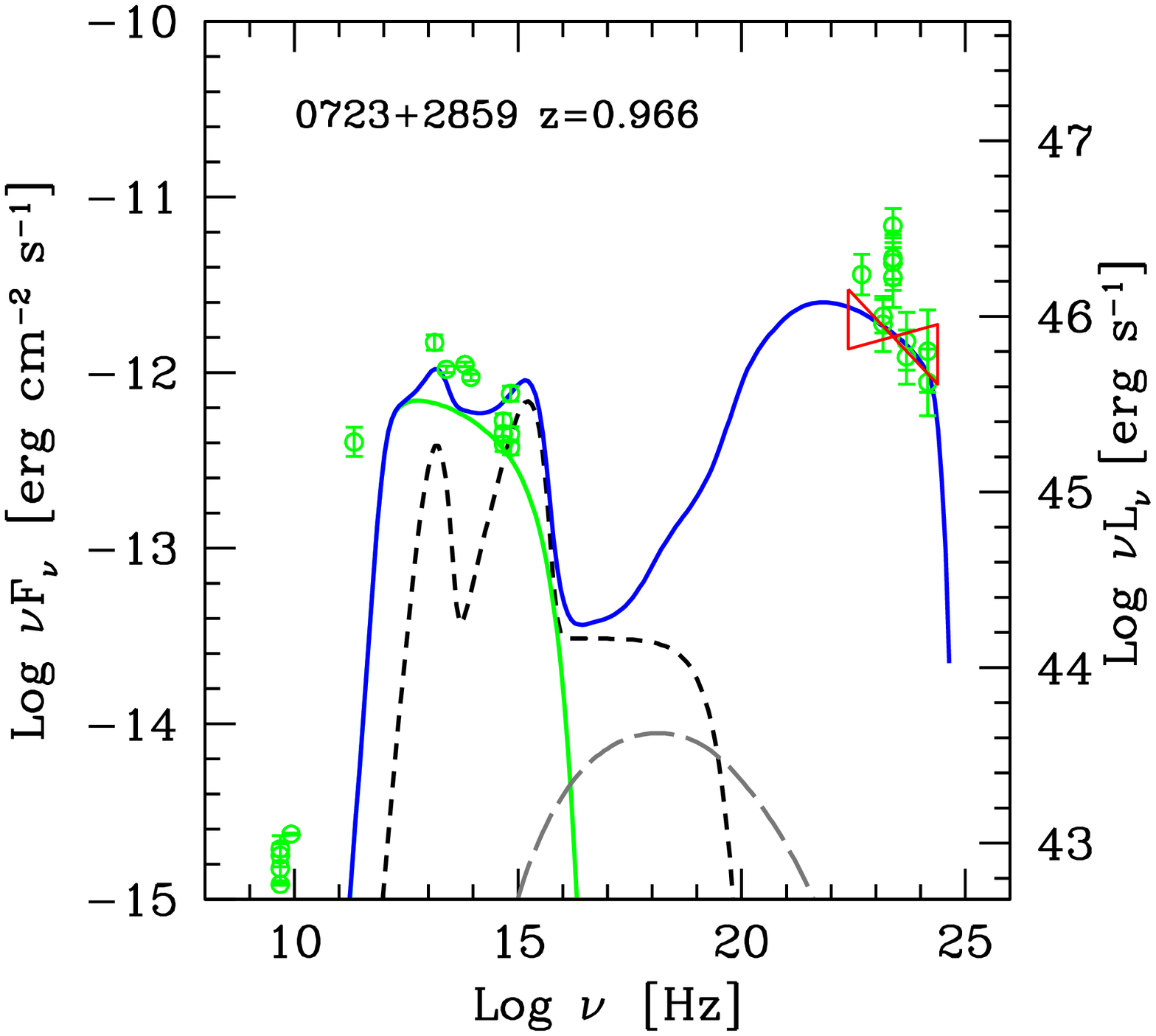,width=4.3cm,height=3.7cm } 
&\psfig{file=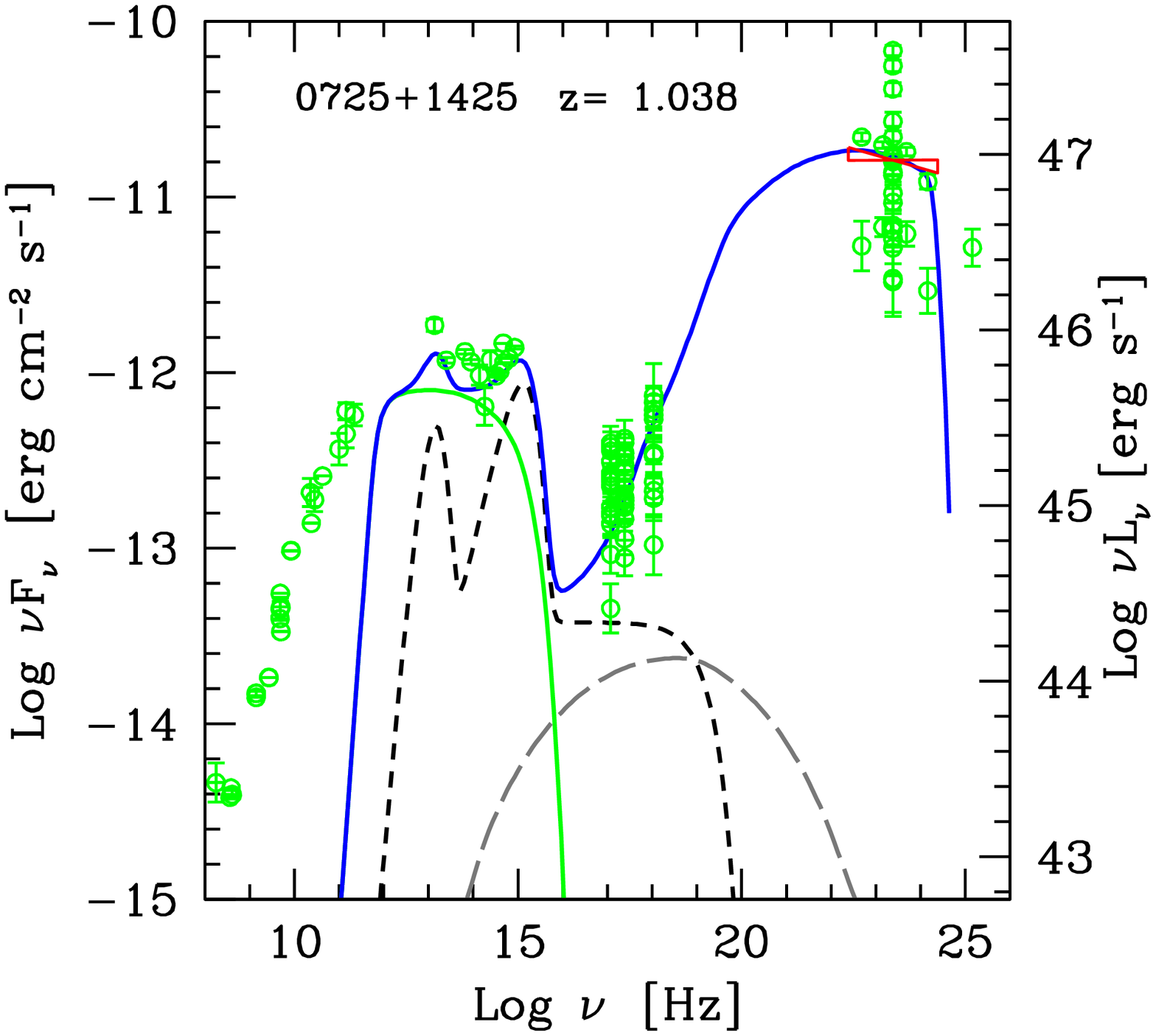,width=4.3cm,height=3.7cm }   \\
\psfig{file=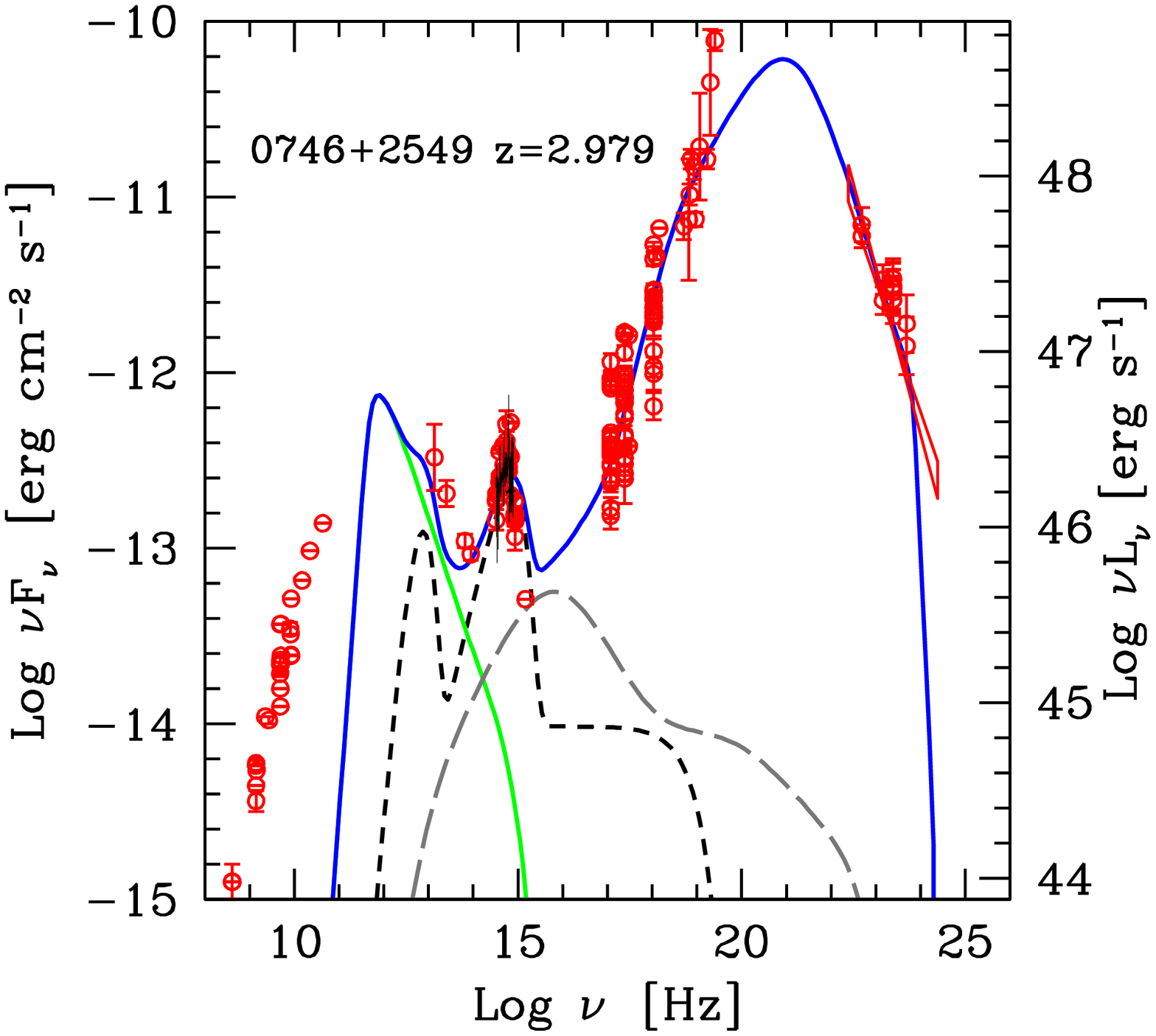,width=4.3cm,height=3.7cm }  
&\psfig{file=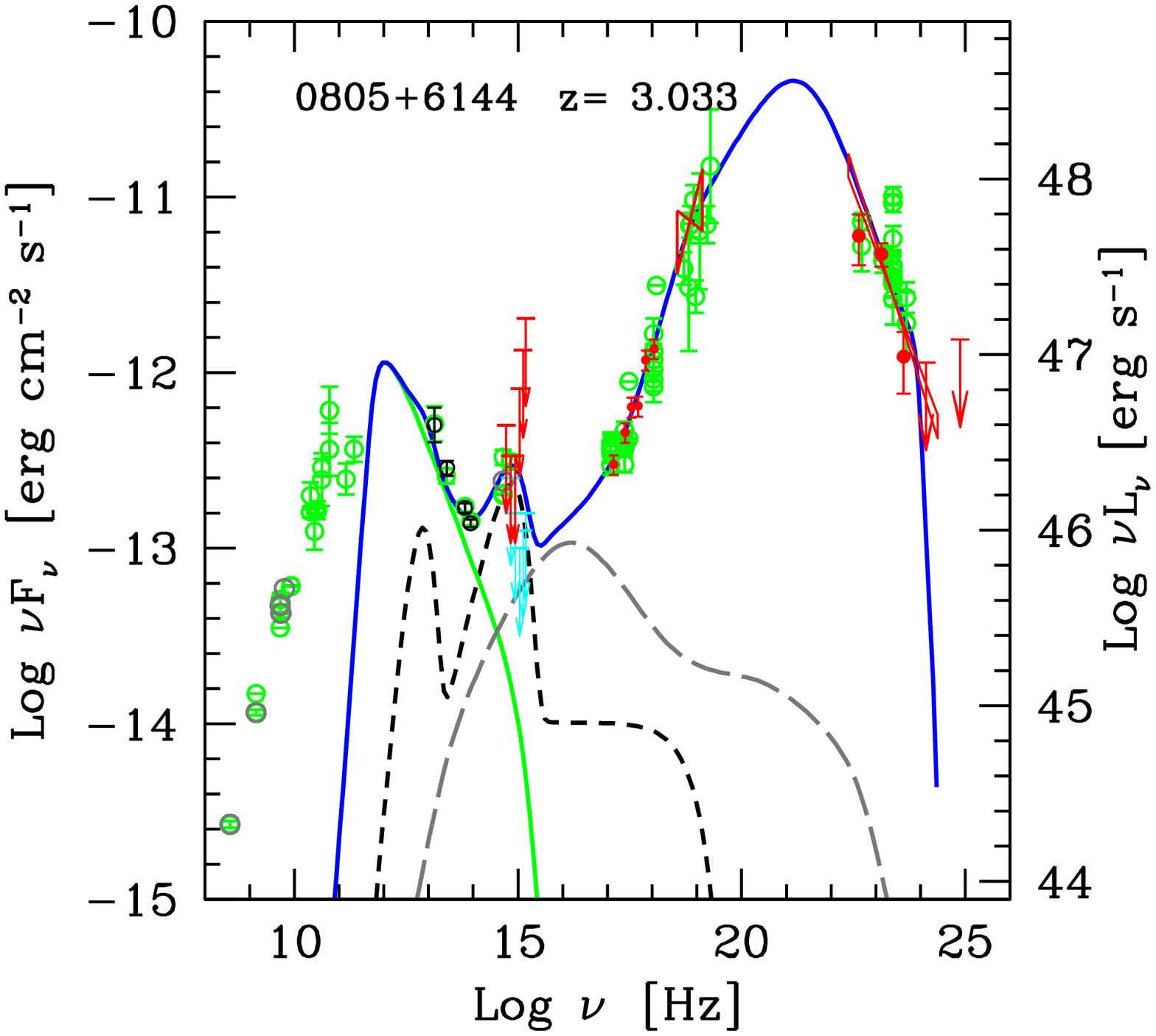,width=4.3cm,height=3.7cm } 
&\psfig{file=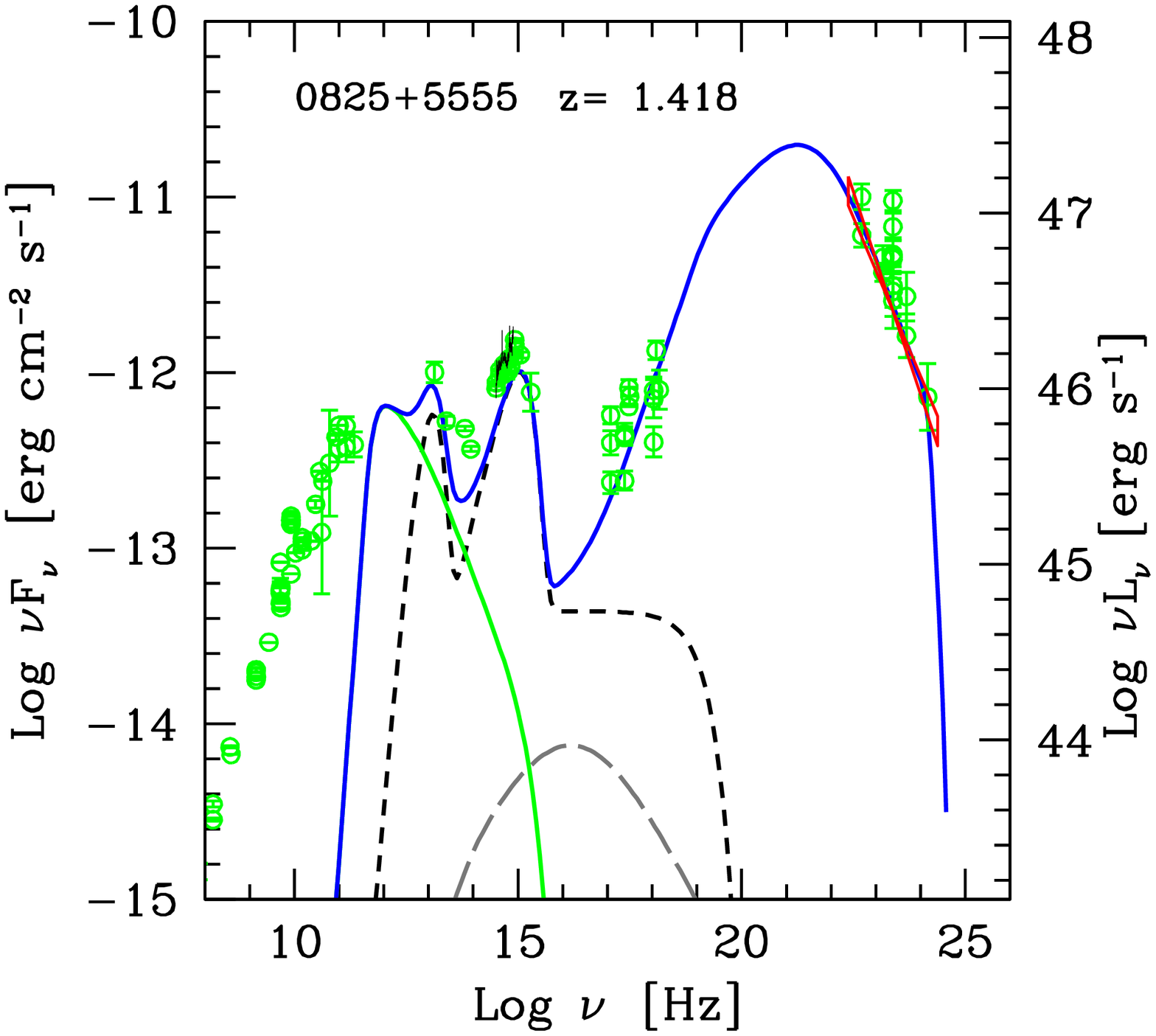,width=4.3cm,height=3.7cm }  
&\psfig{file=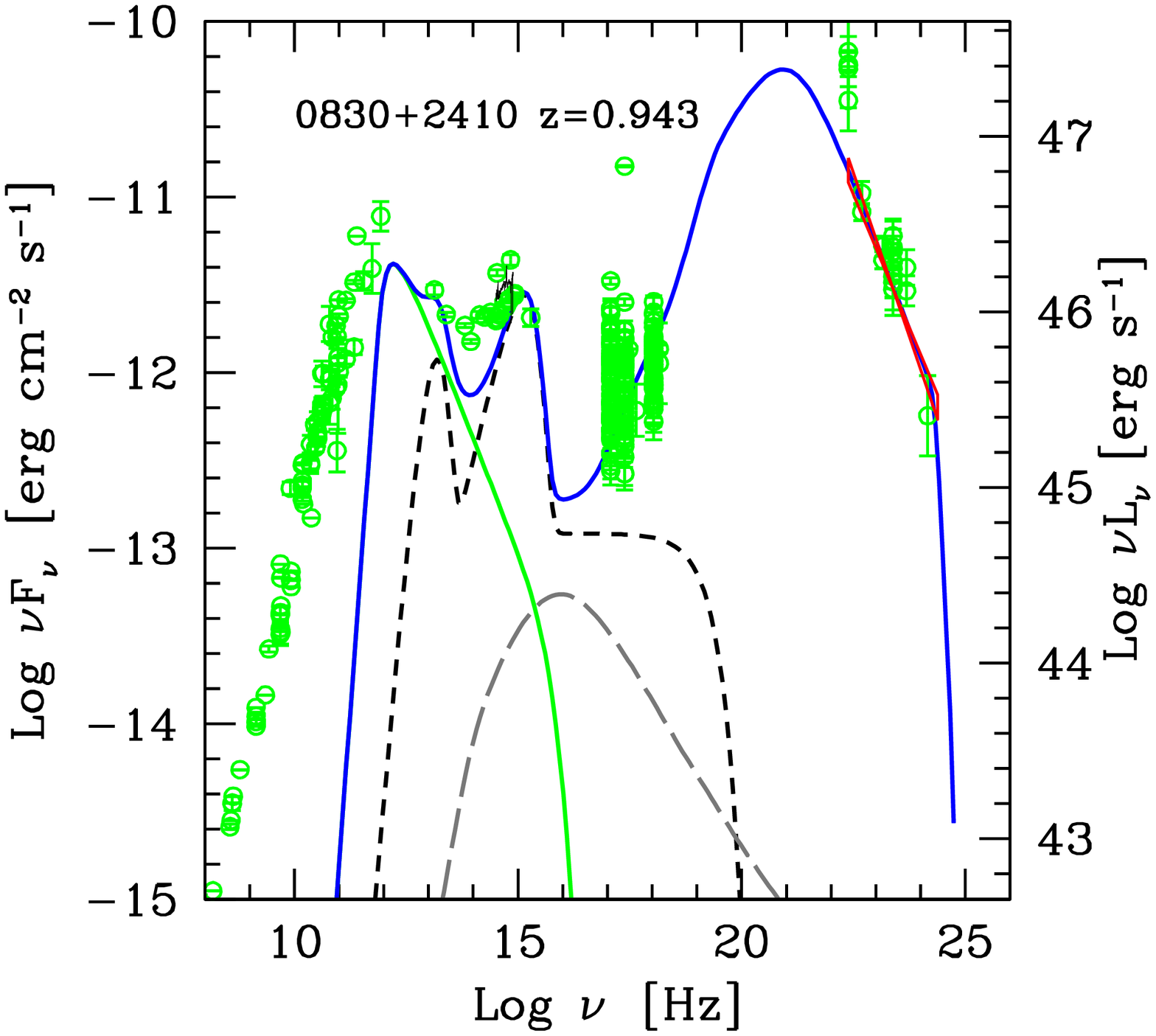,width=4.3cm,height=3.7cm }  \\
\psfig{file=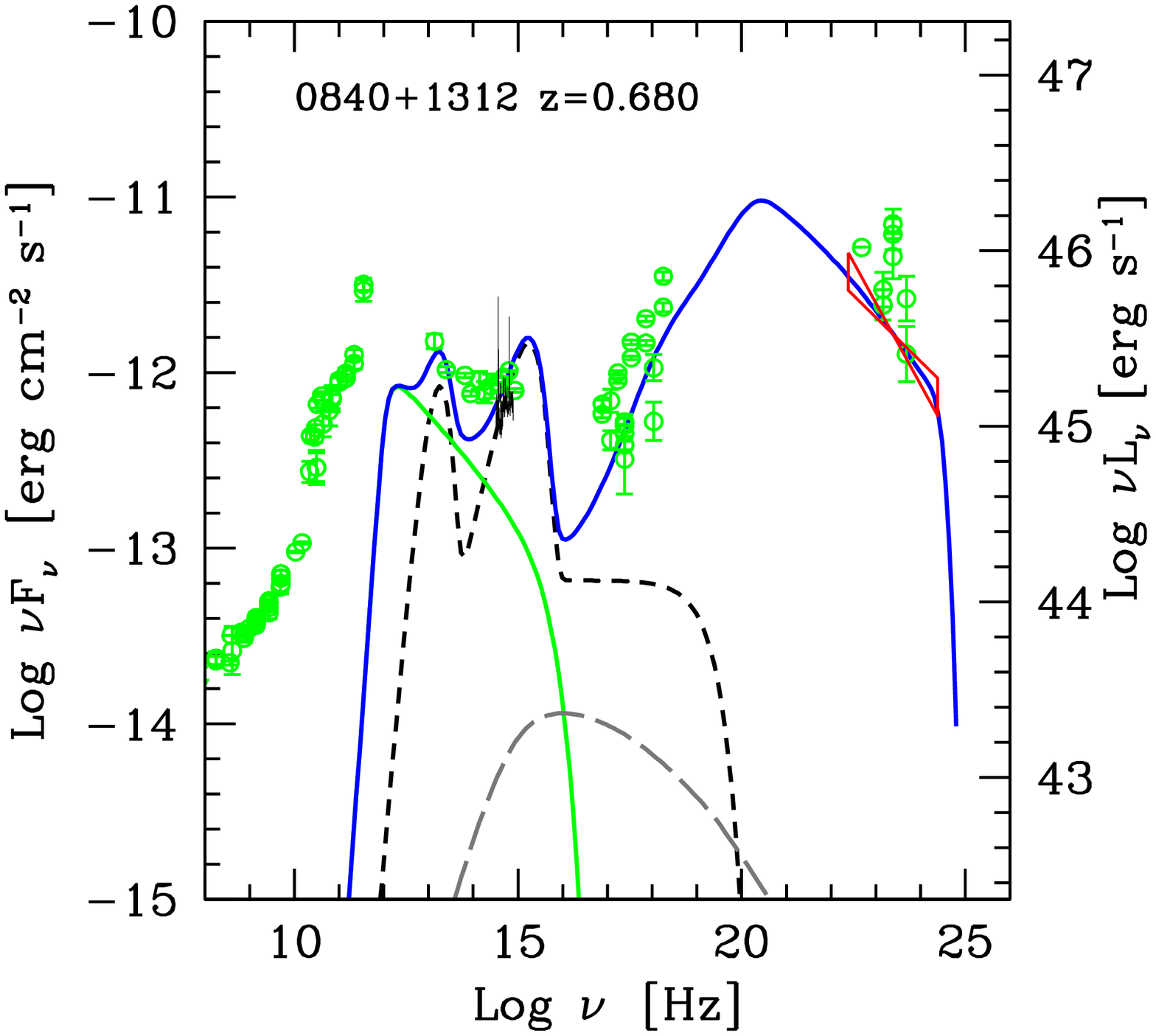,width=4.3cm,height=3.7cm }   
&\psfig{file=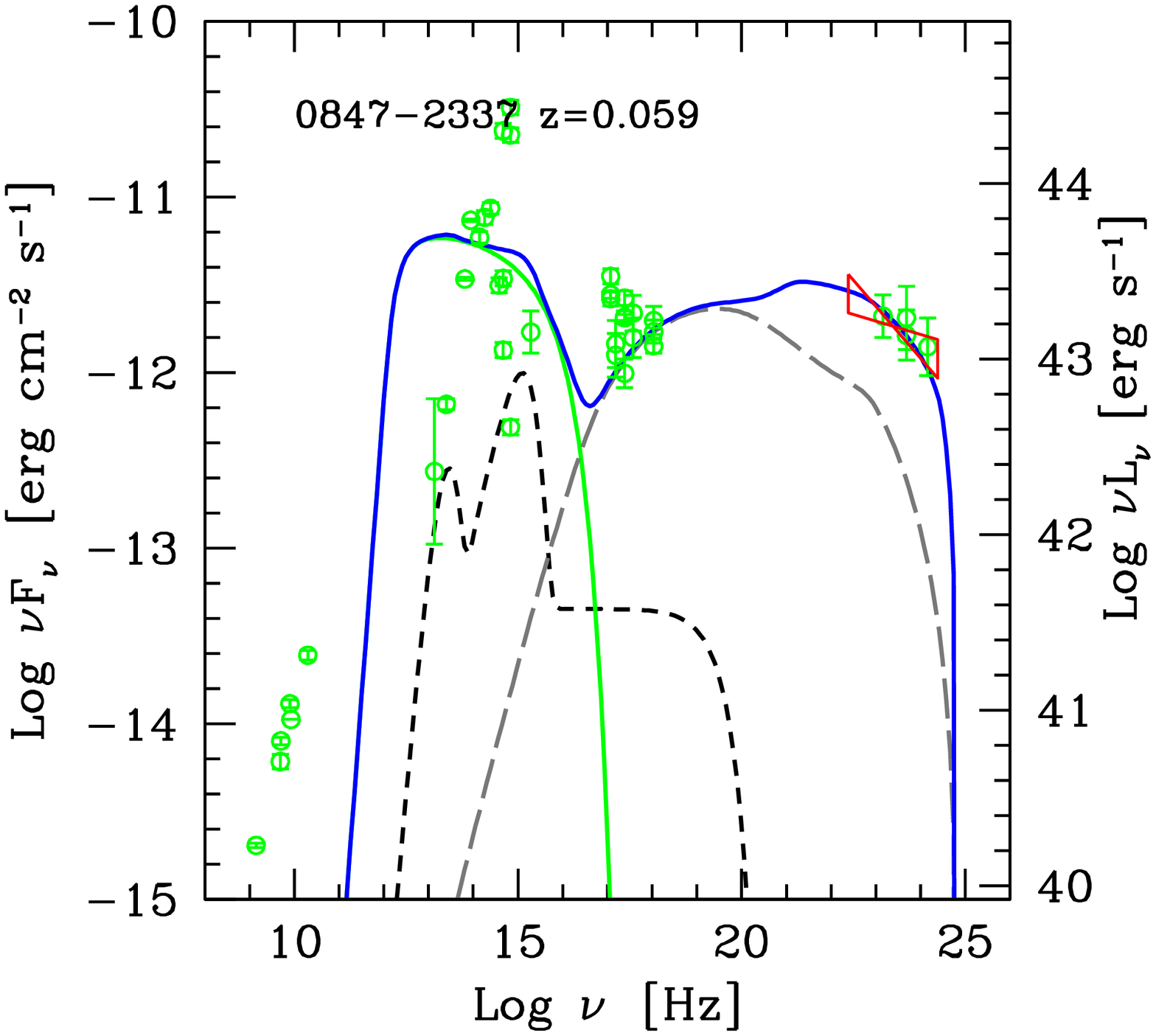,width=4.3cm,height=3.7cm } 
&\psfig{file=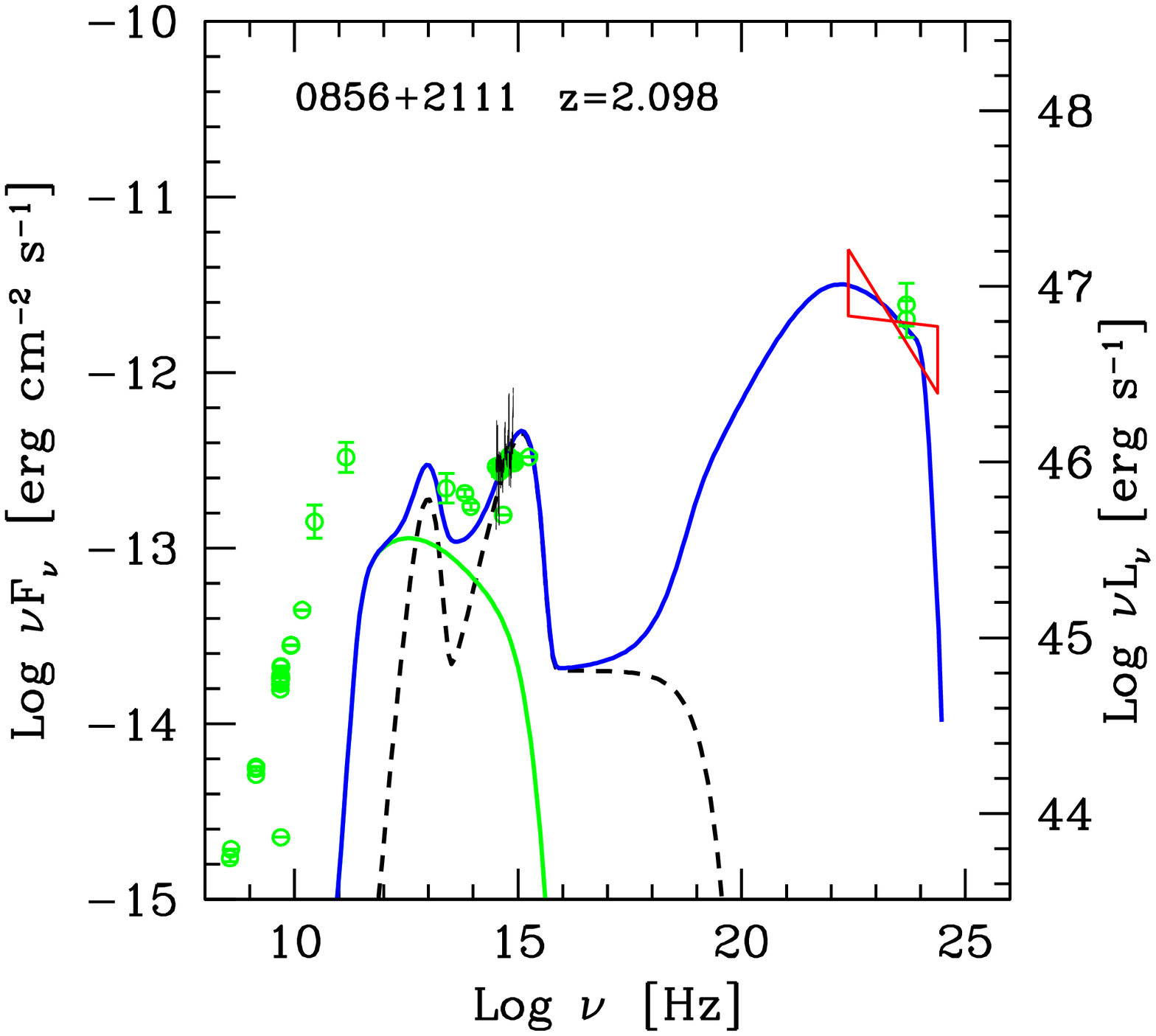,width=4.3cm,height=3.7cm } 
&\psfig{file=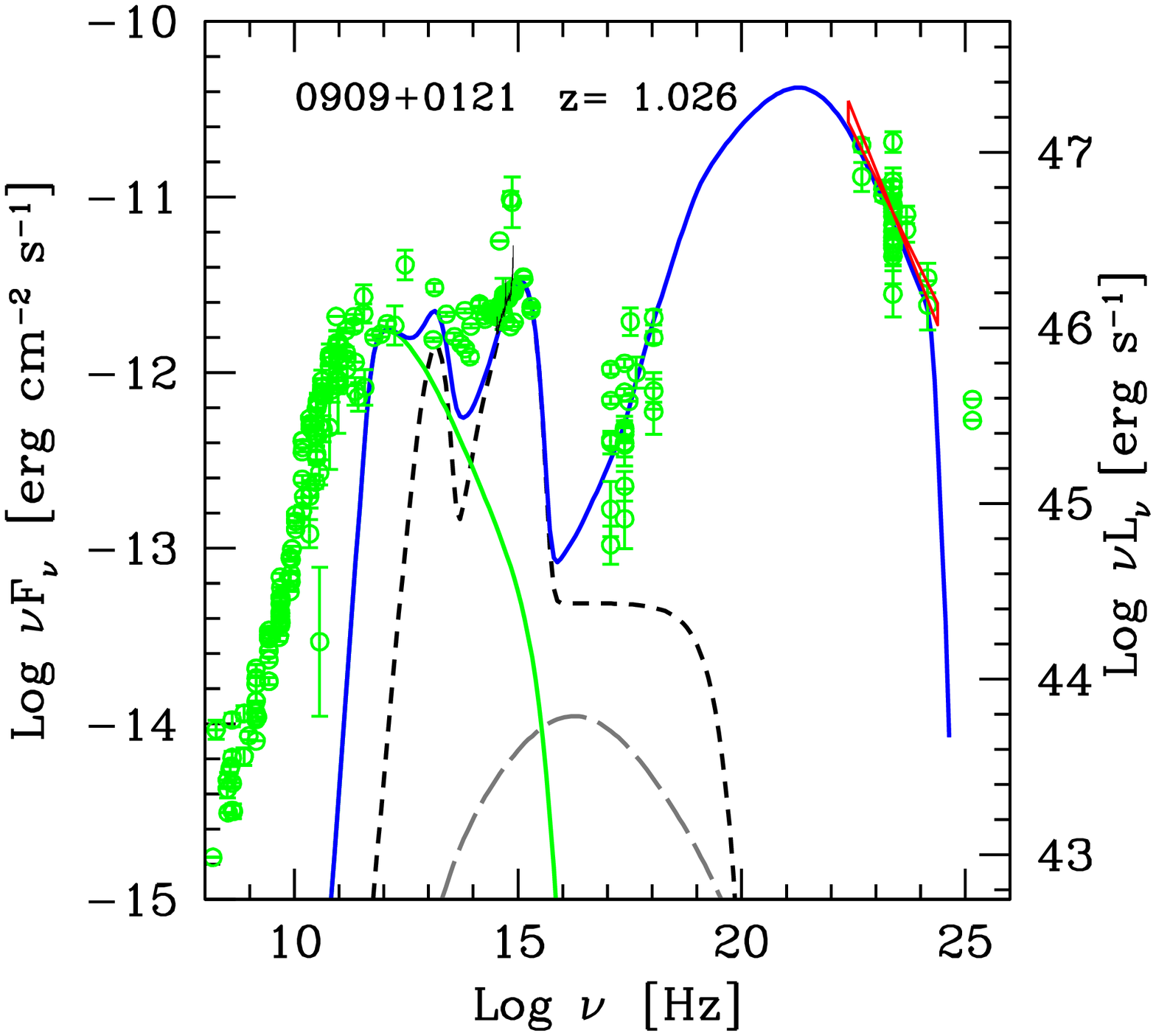,width=4.3cm,height=3.7cm }  \\
\psfig{file=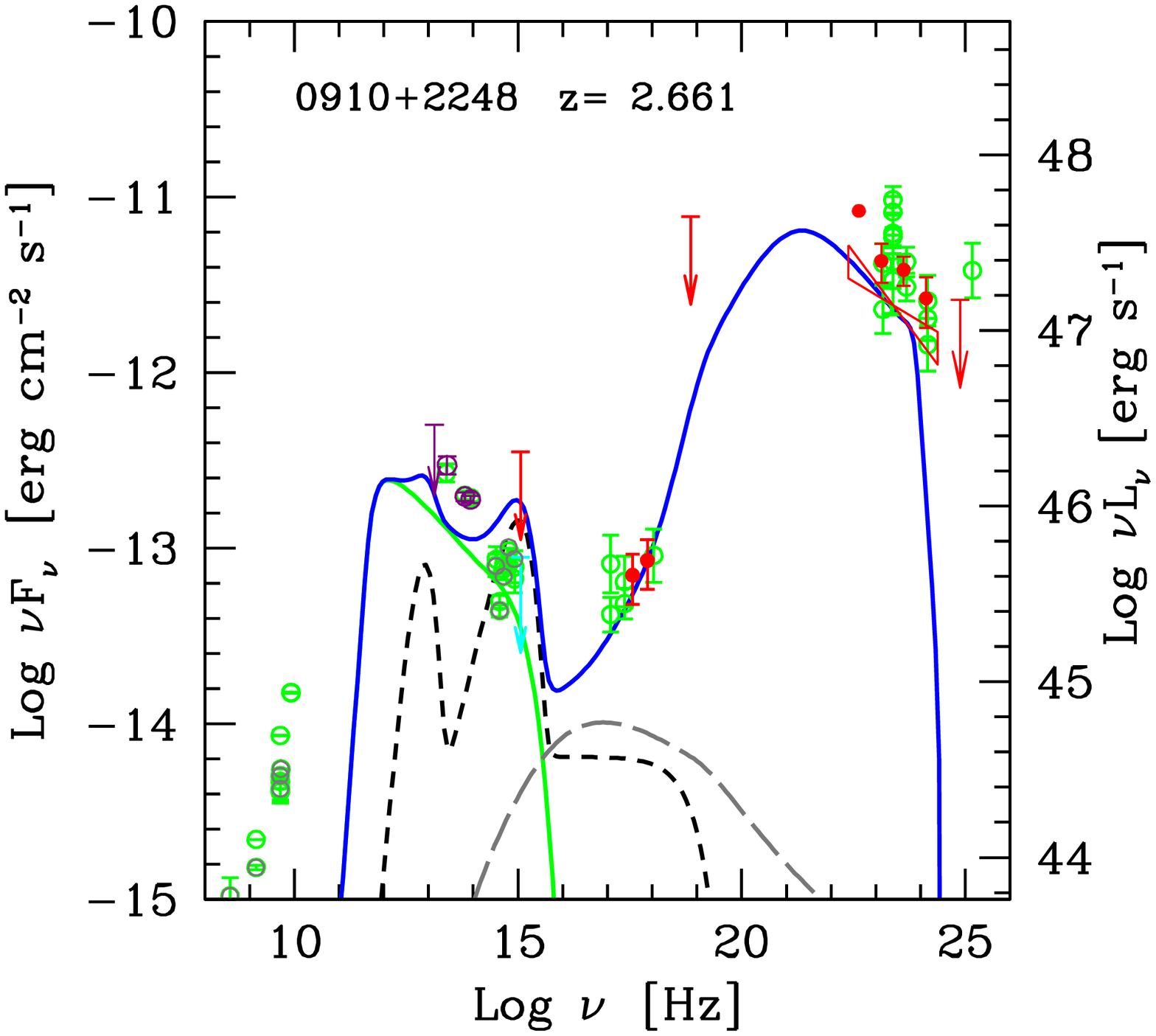,width=4.3cm,height=3.7cm } 
&\psfig{file=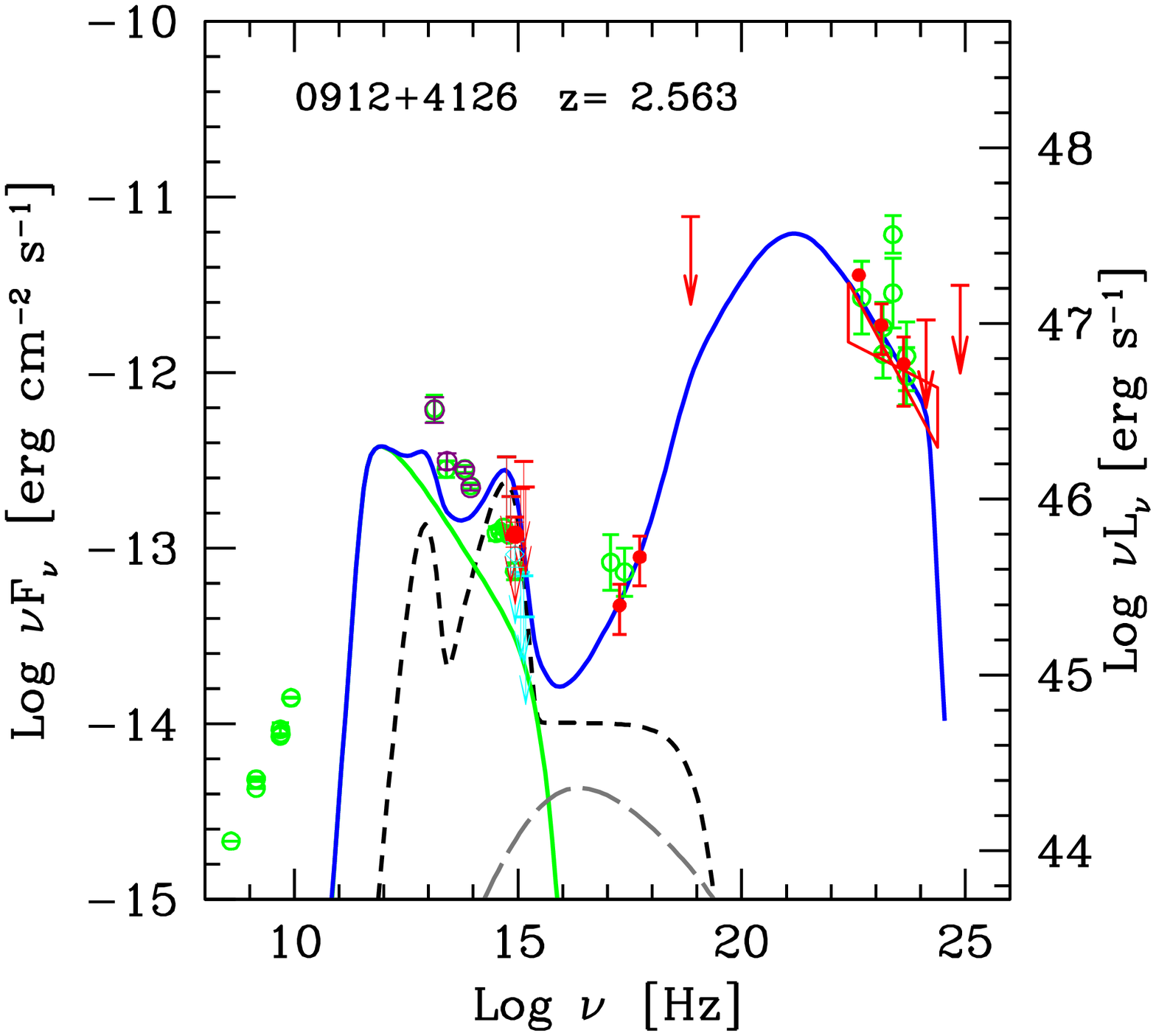,width=4.3cm,height=3.7cm } 
&\psfig{file=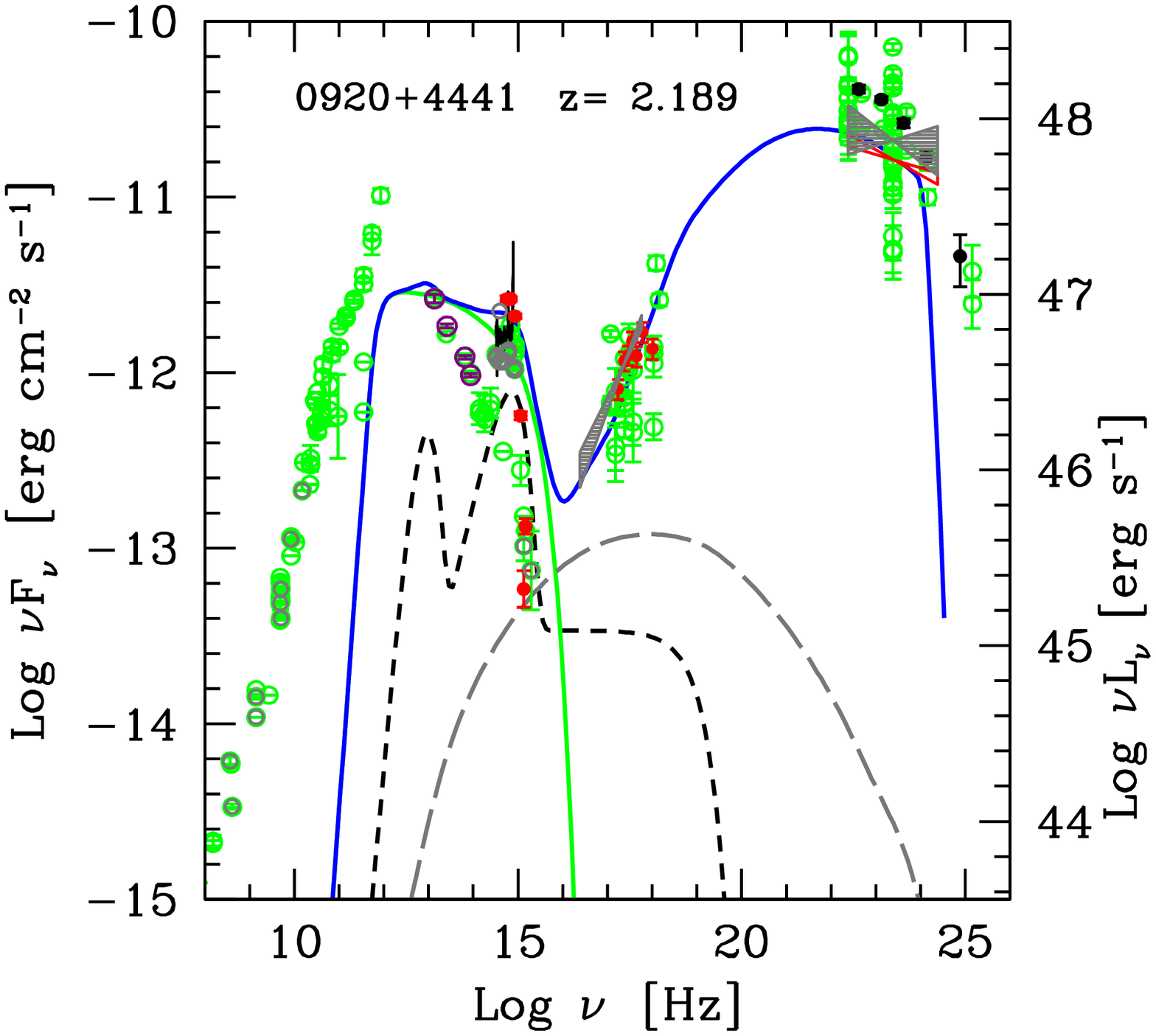,width=4.3cm,height=3.7cm }  
&\psfig{file=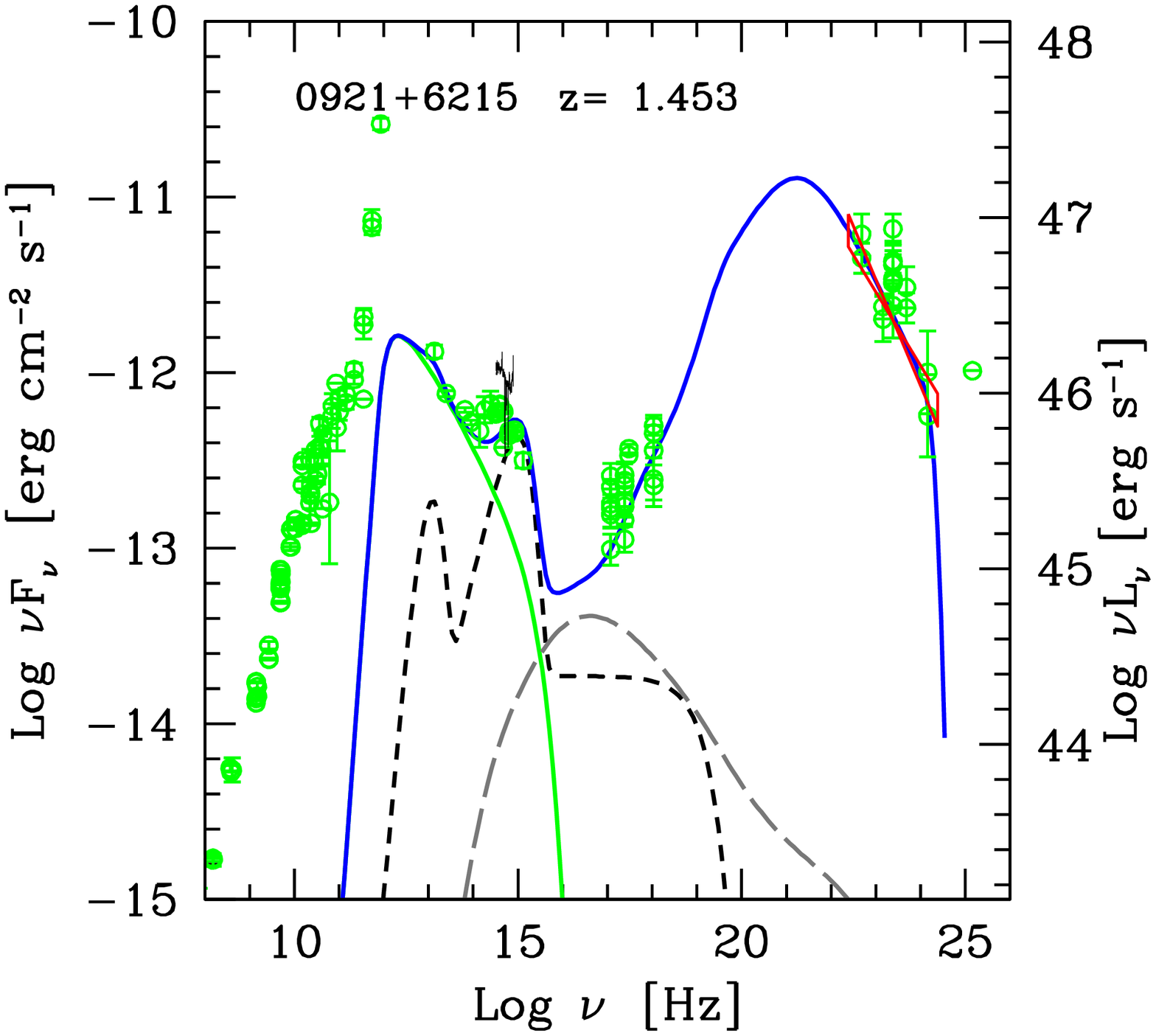,width=4.3cm,height=3.7cm } 
\end{tabular}
\caption{{\it continue.} SED of the FSRQs studied in this paper.}
\end{figure*} 

\setcounter{figure}{15}
\begin{figure*}
\begin{tabular}{cccc}
\psfig{file=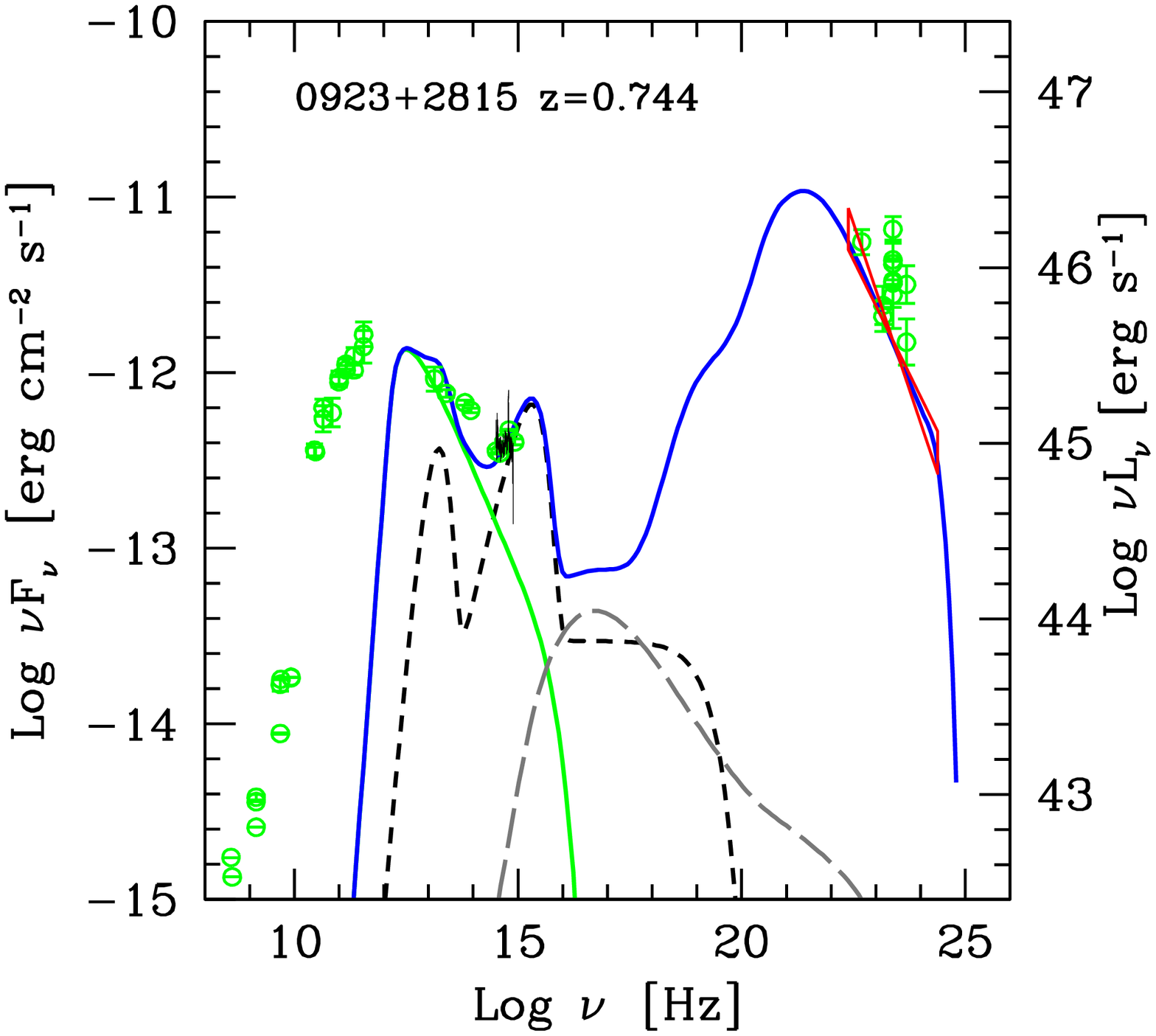,width=4.3cm,height=3.7cm }  
&\psfig{file=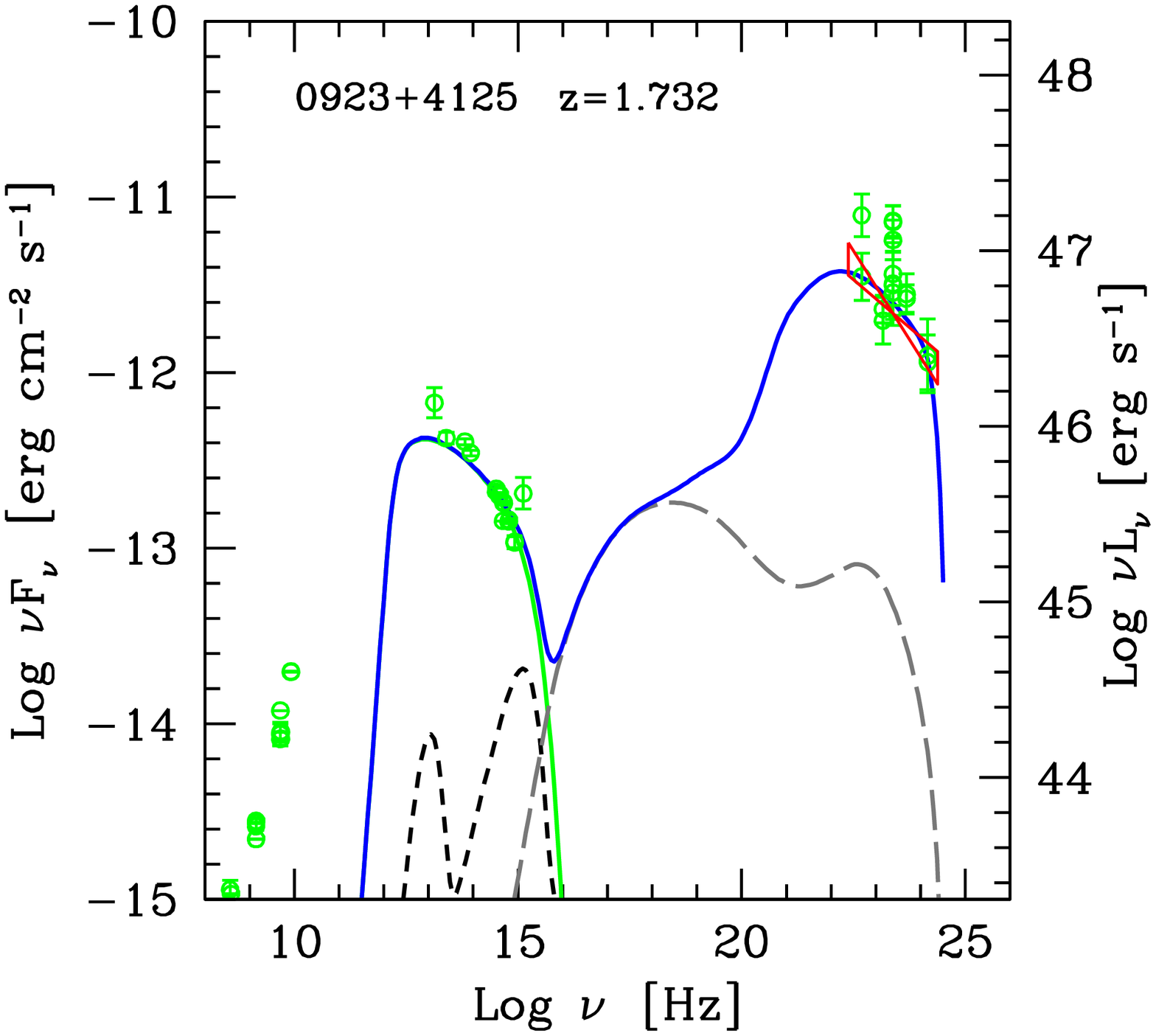,width=4.3cm,height=3.7cm } 
&\psfig{file=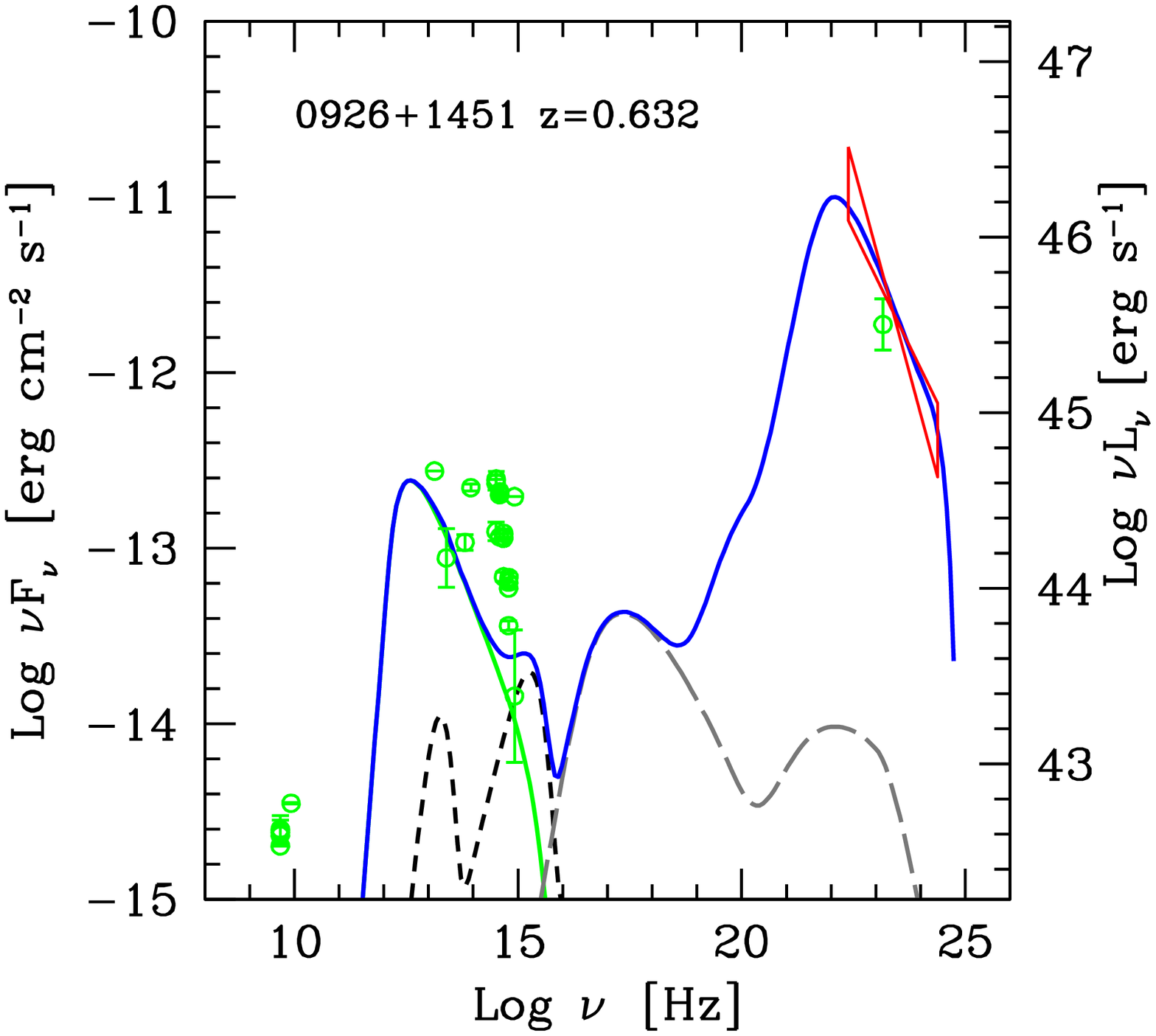,width=4.3cm,height=3.7cm } 
&\psfig{file=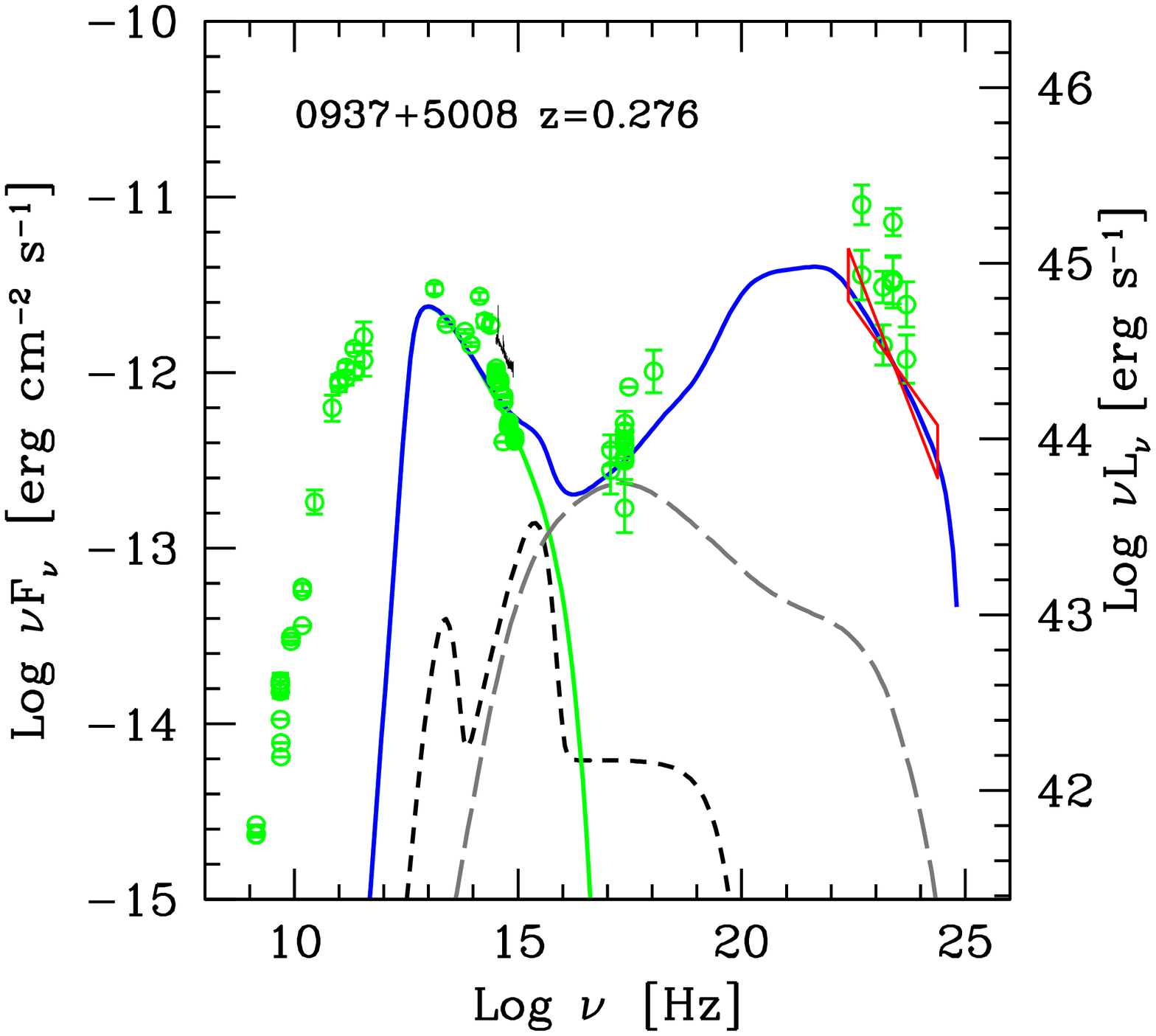,width=4.3cm,height=3.7cm } \\
\psfig{file=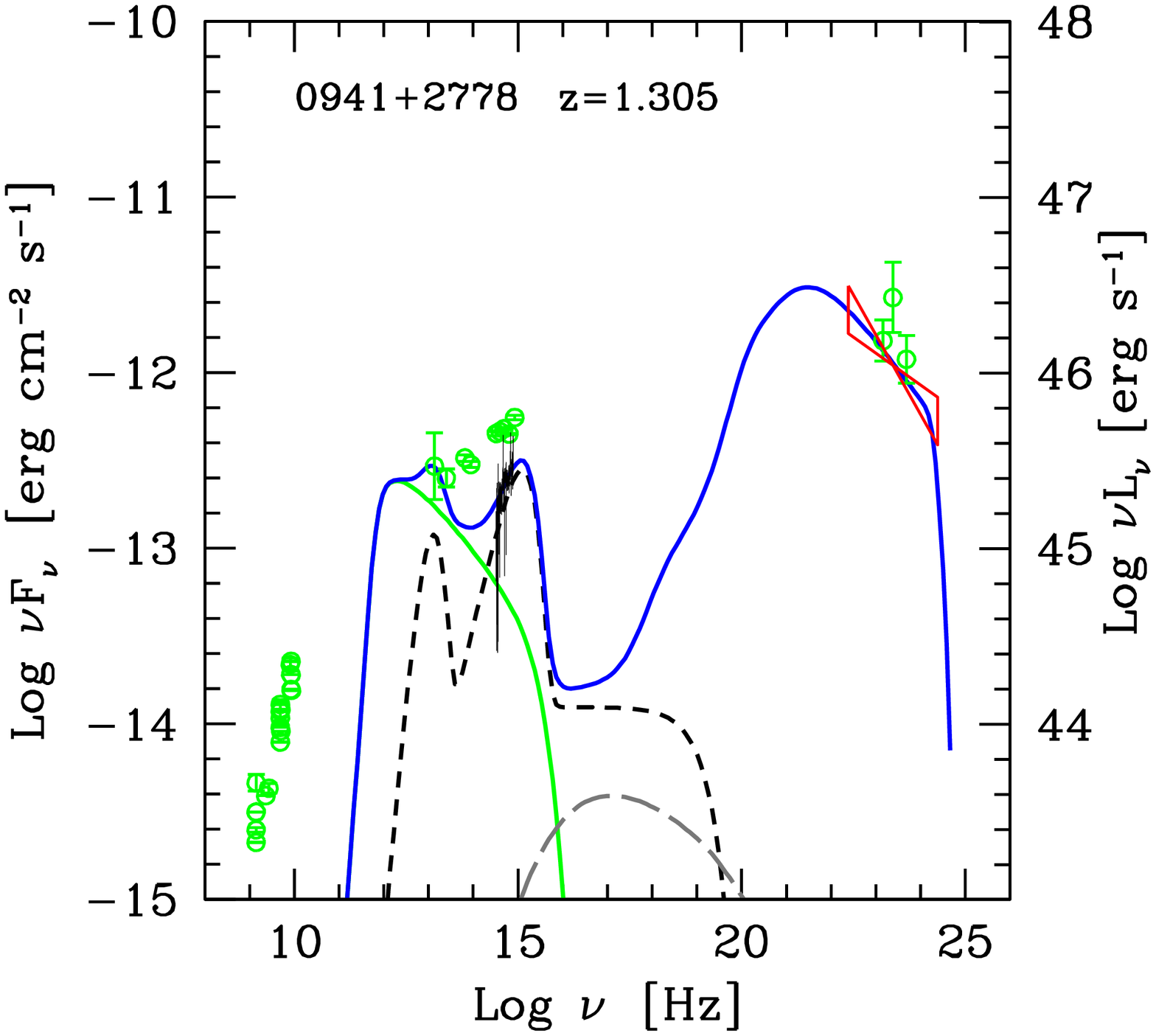,width=4.3cm,height=3.7cm } 
&\psfig{file=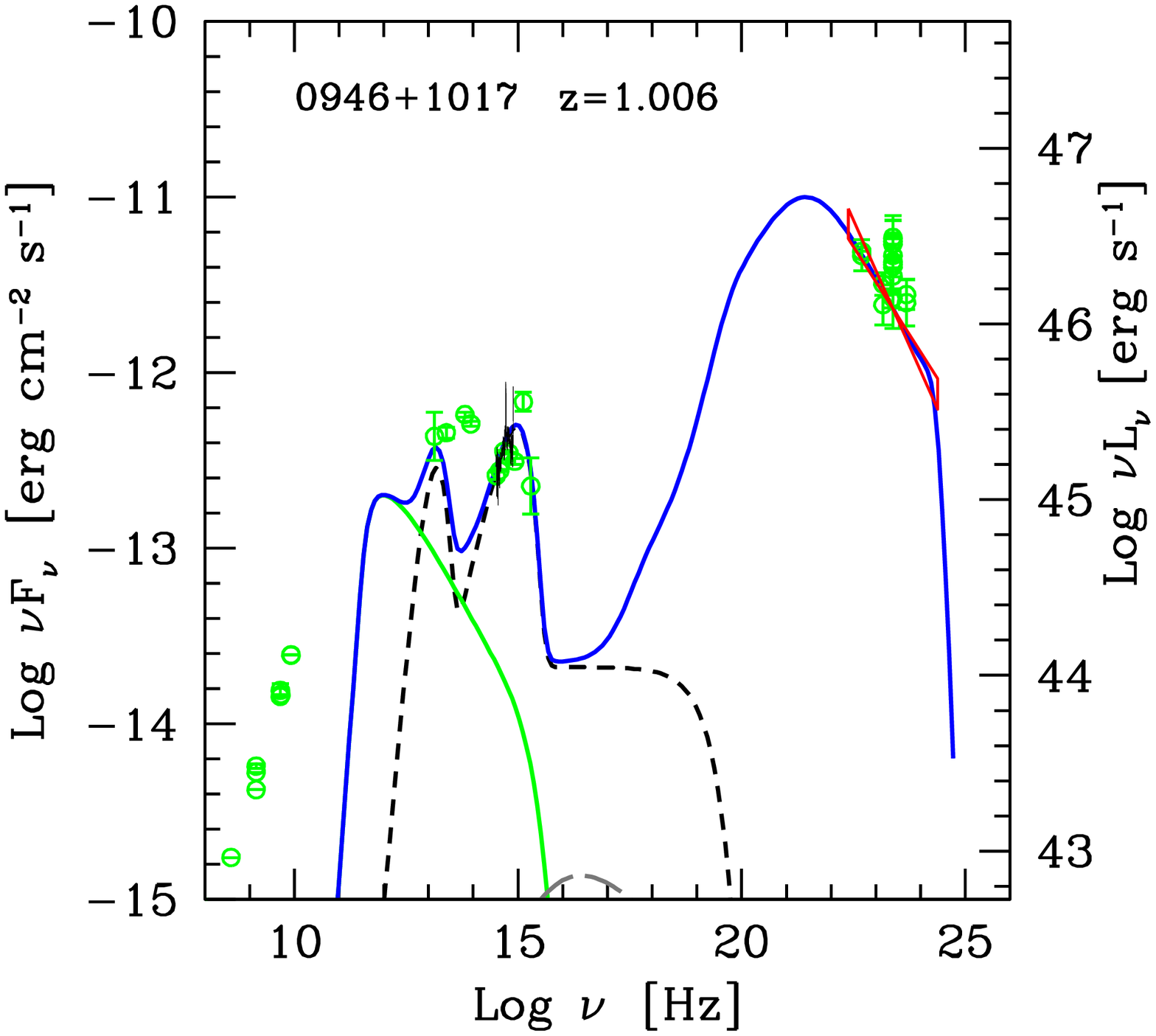,width=4.3cm,height=3.7cm } 
&\psfig{file=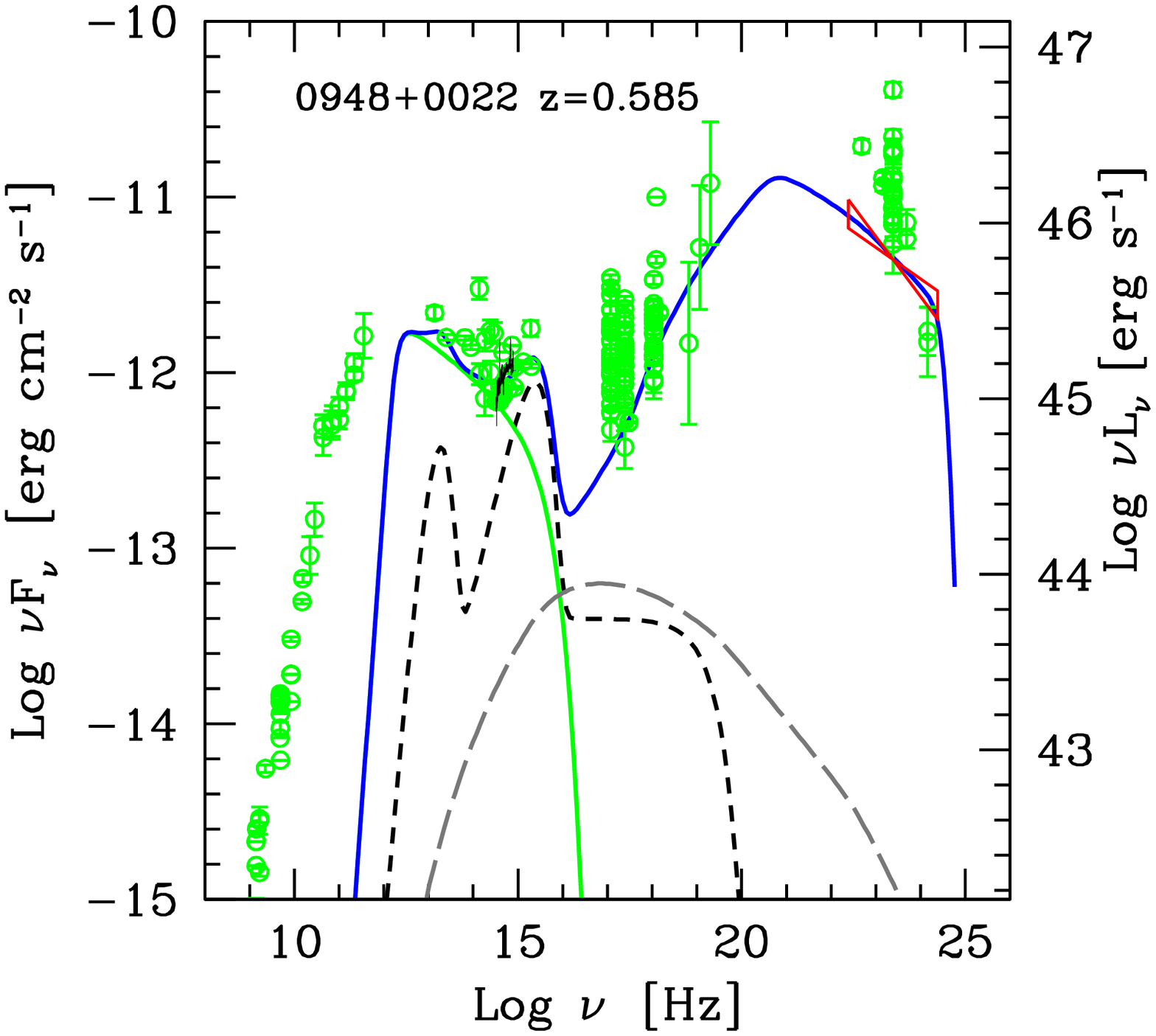,width=4.3cm,height=3.7cm }  
&\psfig{file=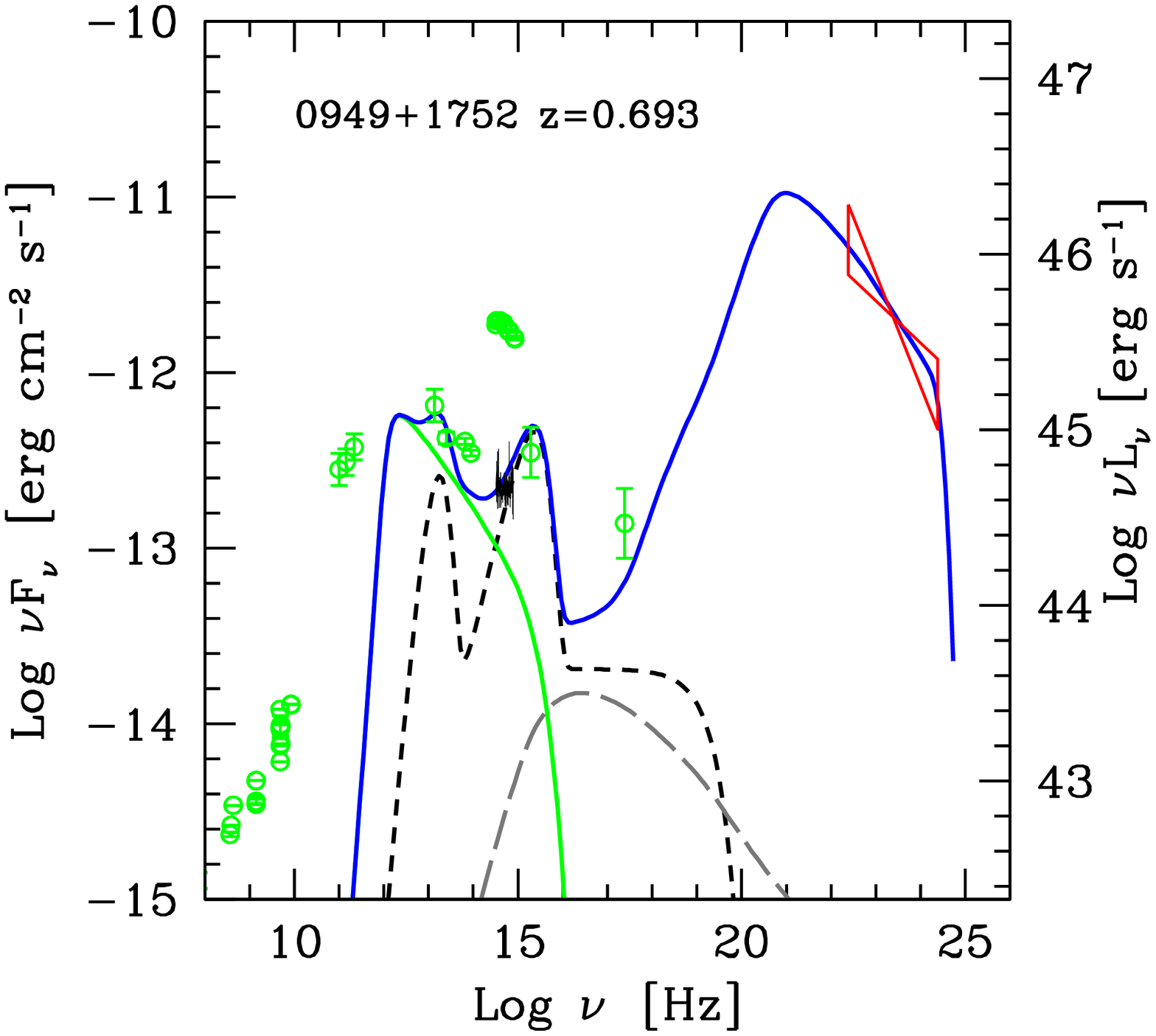,width=4.3cm,height=3.7cm } \\ 
\psfig{file=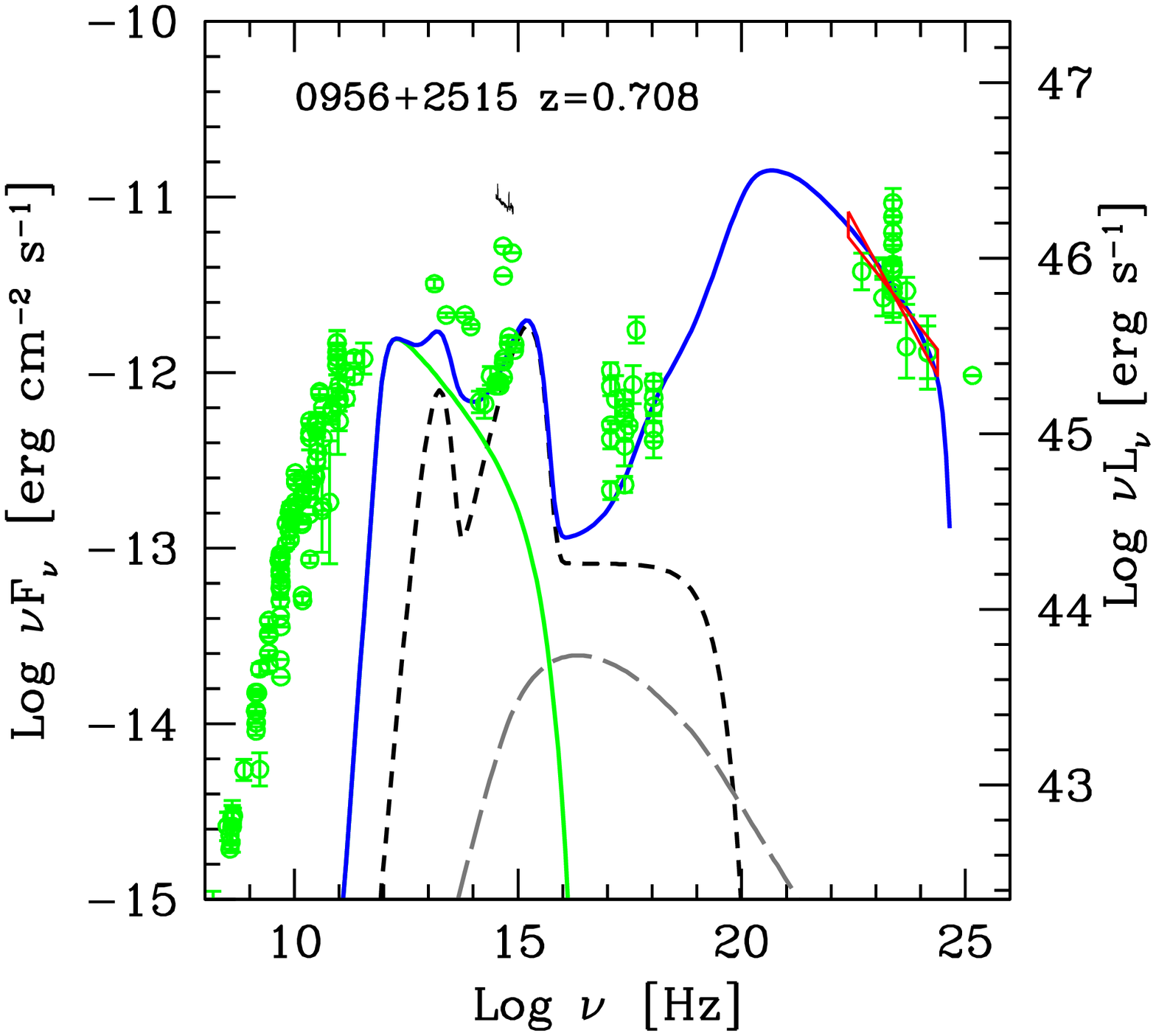,width=4.3cm,height=3.7cm }  
&\psfig{file=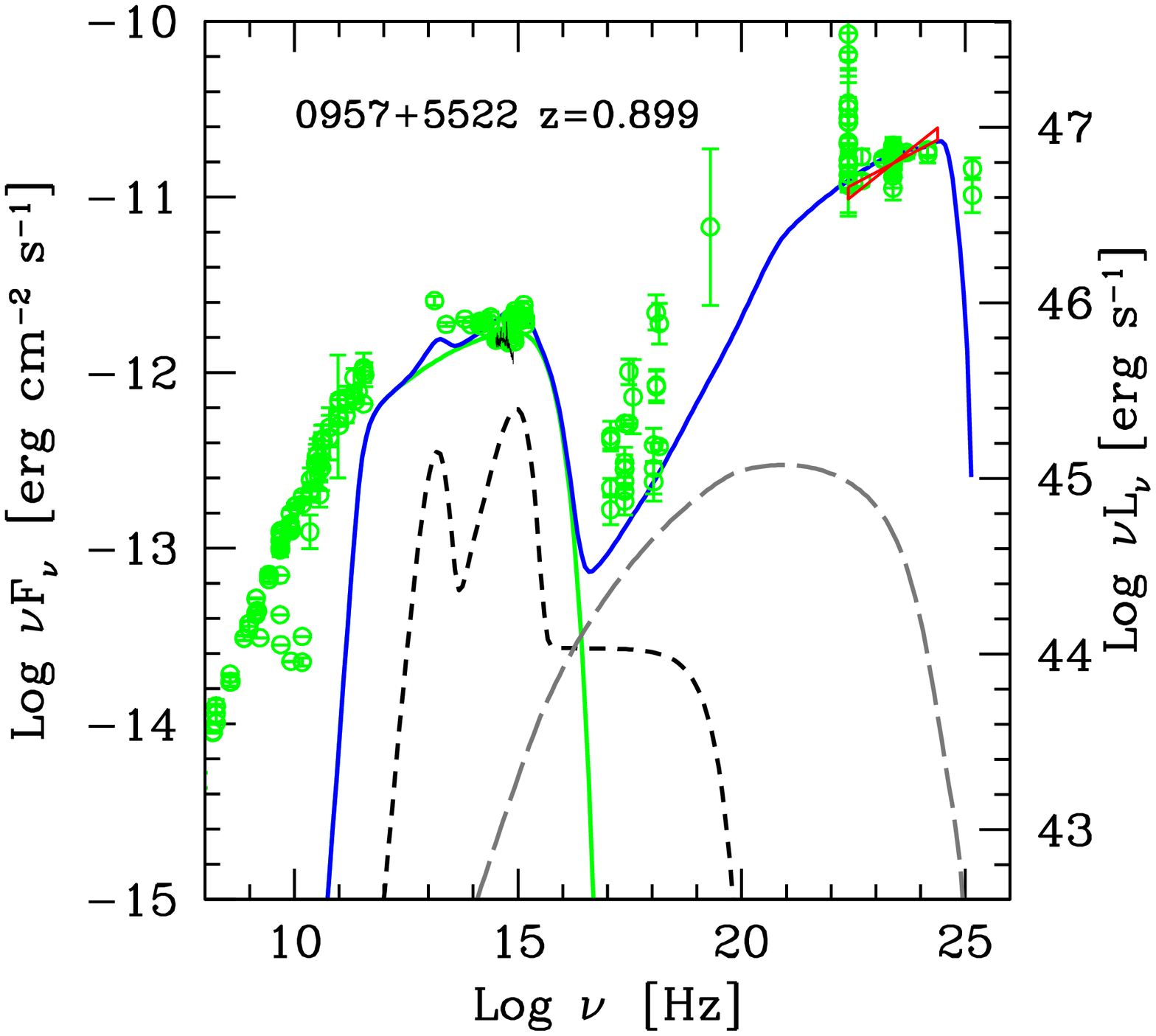,width=4.3cm,height=3.7cm } 
&\psfig{file=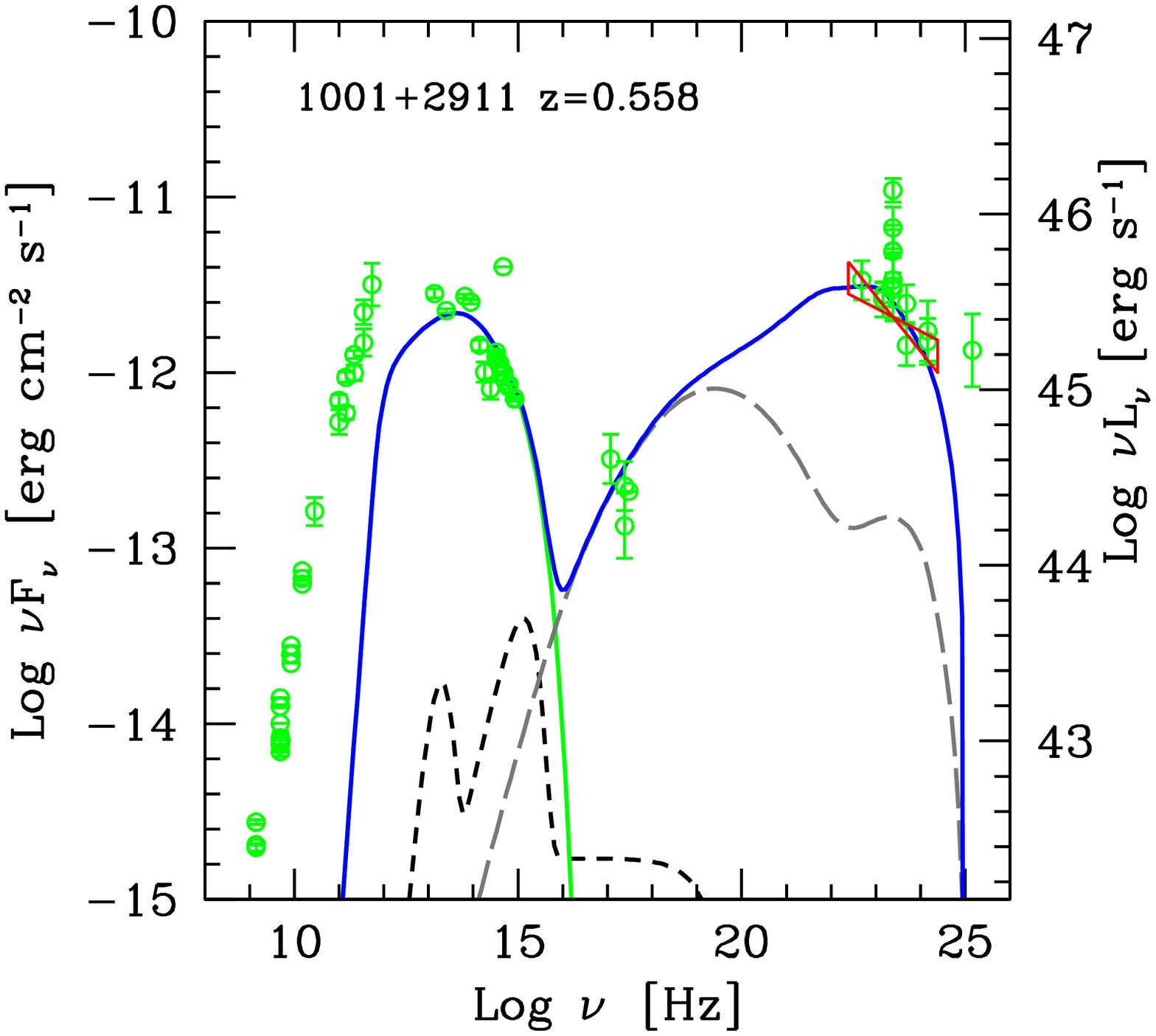,width=4.3cm,height=3.7cm } 
&\psfig{file=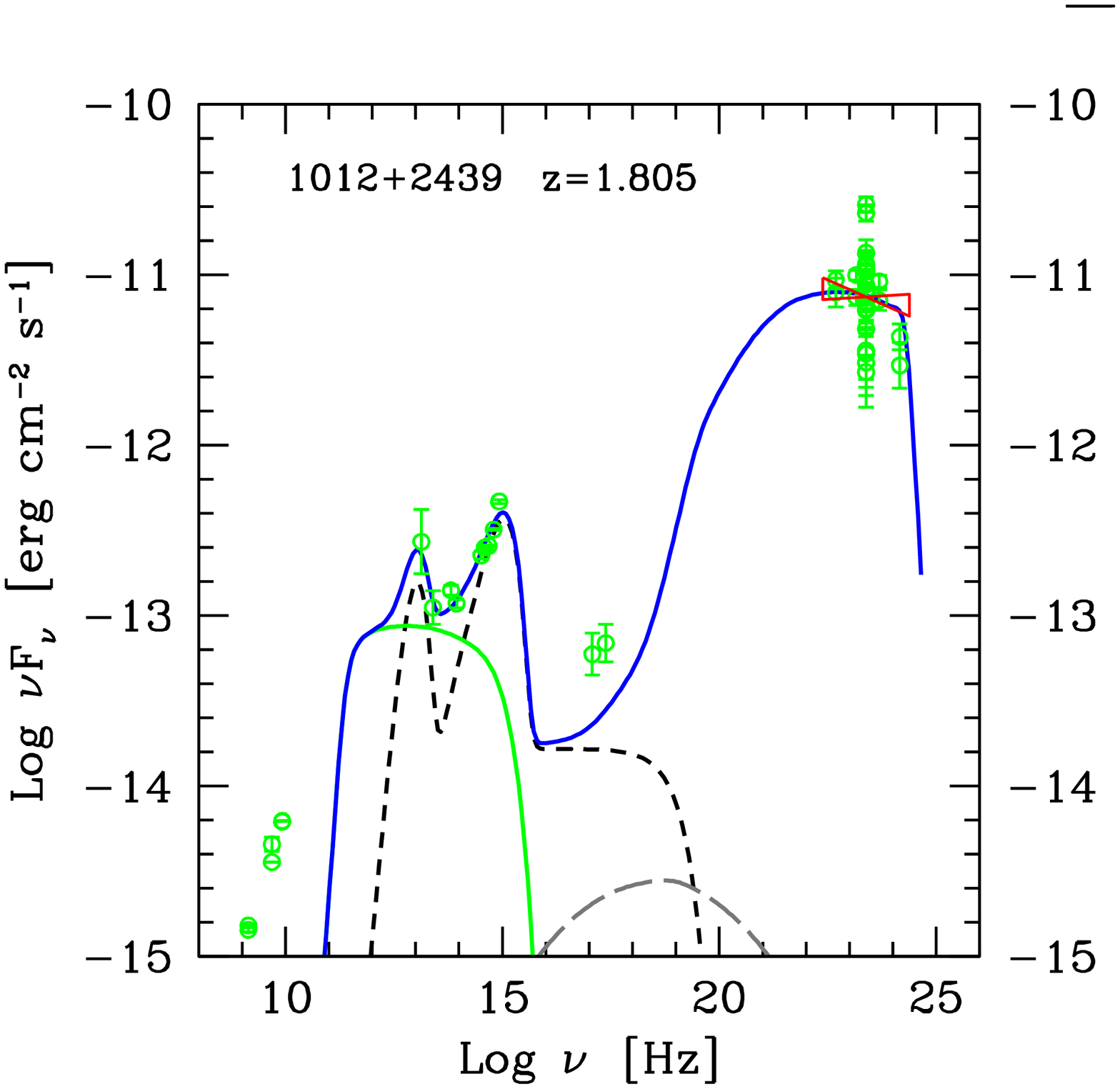,width=4.3cm,height=3.7cm }  \\
\psfig{file=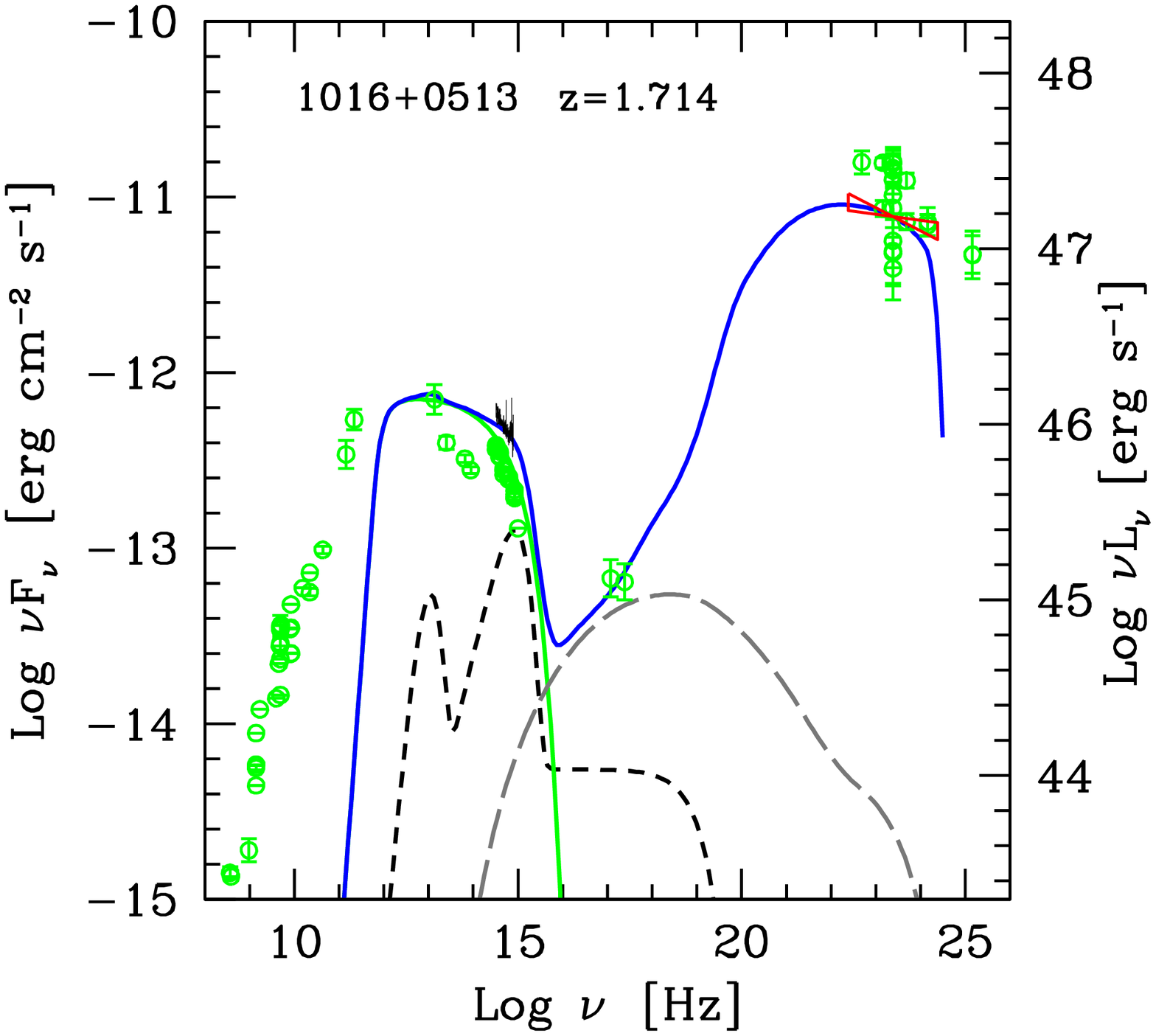,width=4.3cm,height=3.7cm } 
&\psfig{file=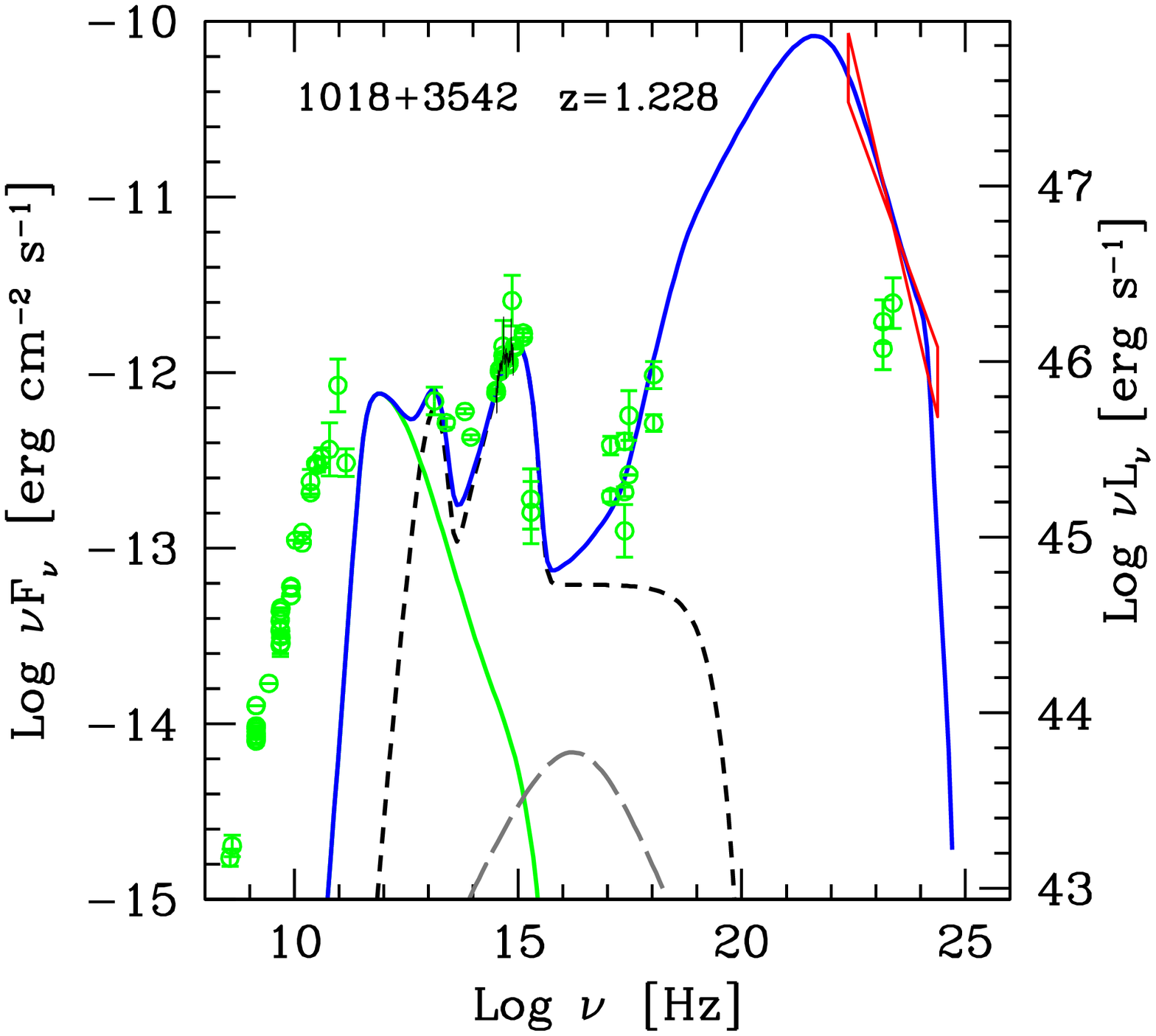,width=4.3cm,height=3.7cm } 
&\psfig{file=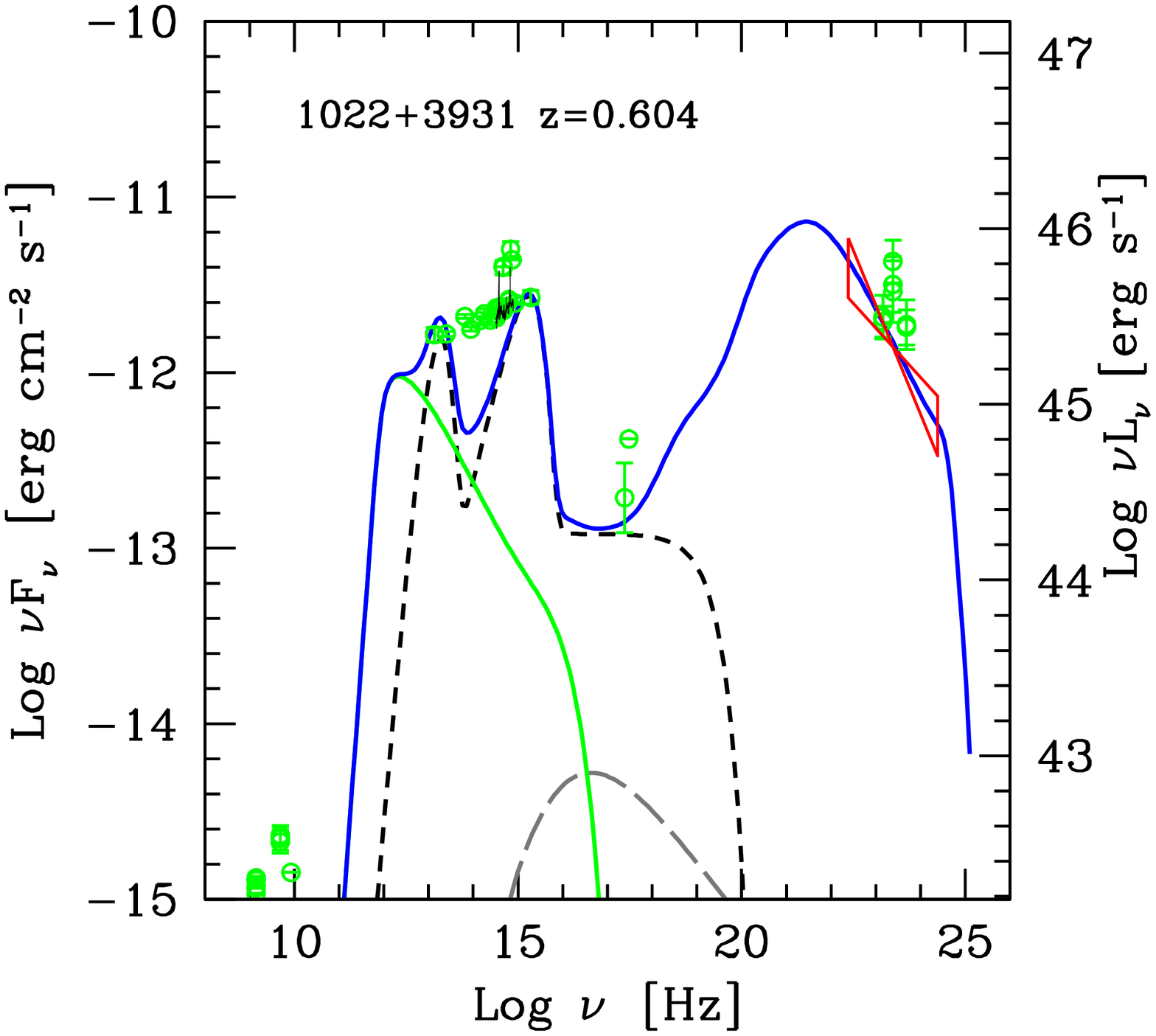,width=4.3cm,height=3.7cm } 
&\psfig{file=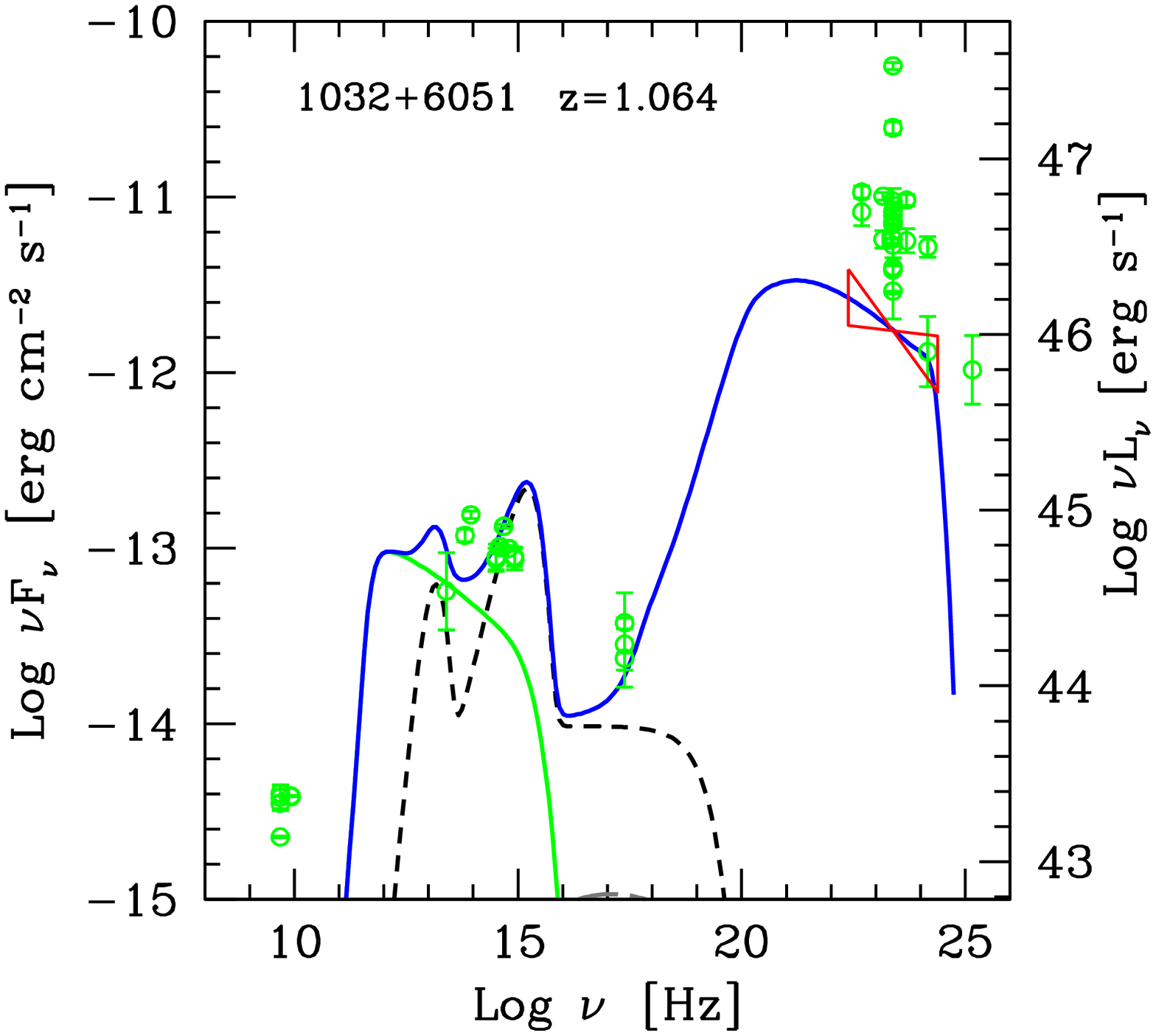,width=4.3cm,height=3.7cm } \\ 
\psfig{file=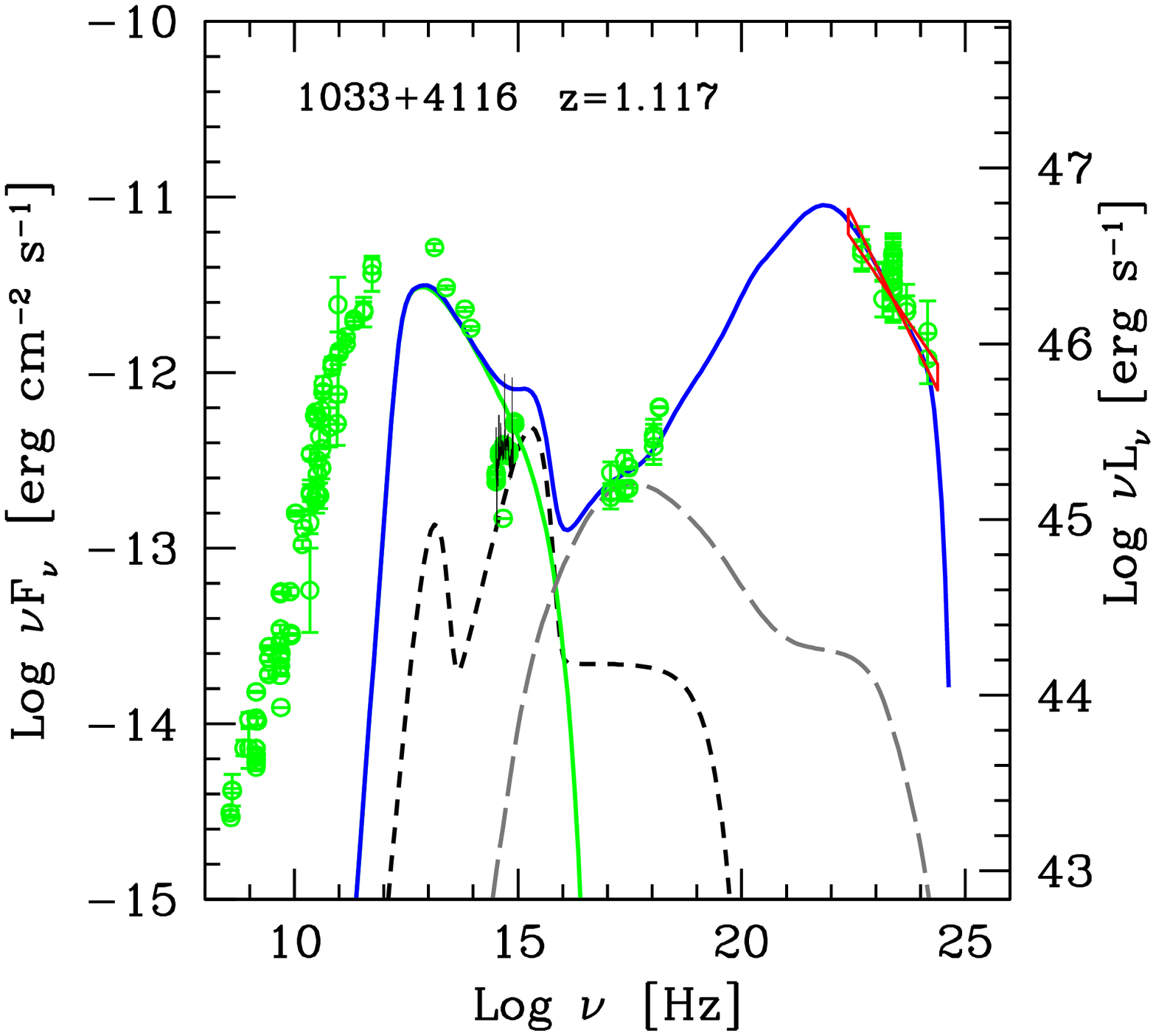,width=4.3cm,height=3.7cm }  
&\psfig{file=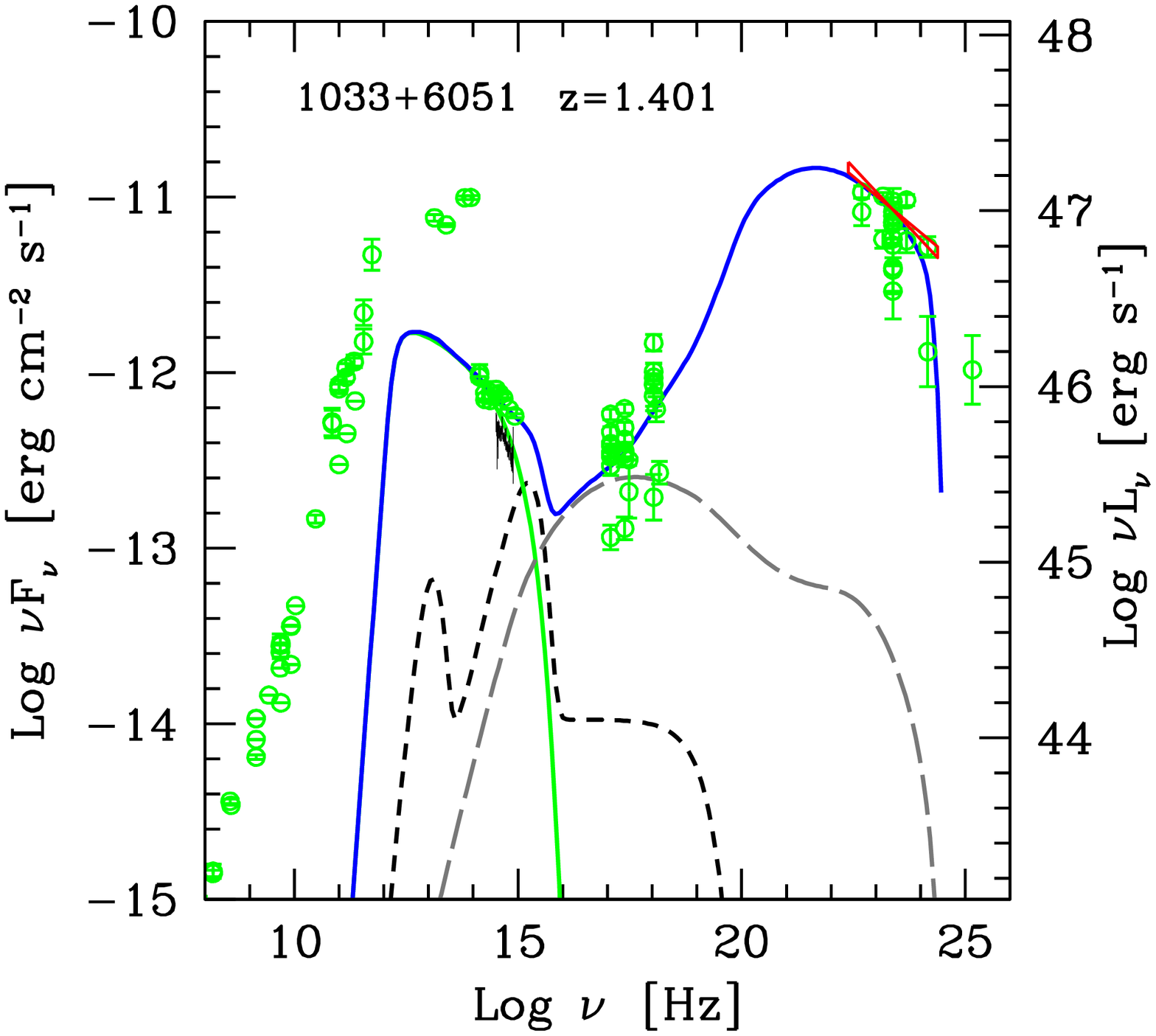,width=4.3cm,height=3.7cm } 
&\psfig{file=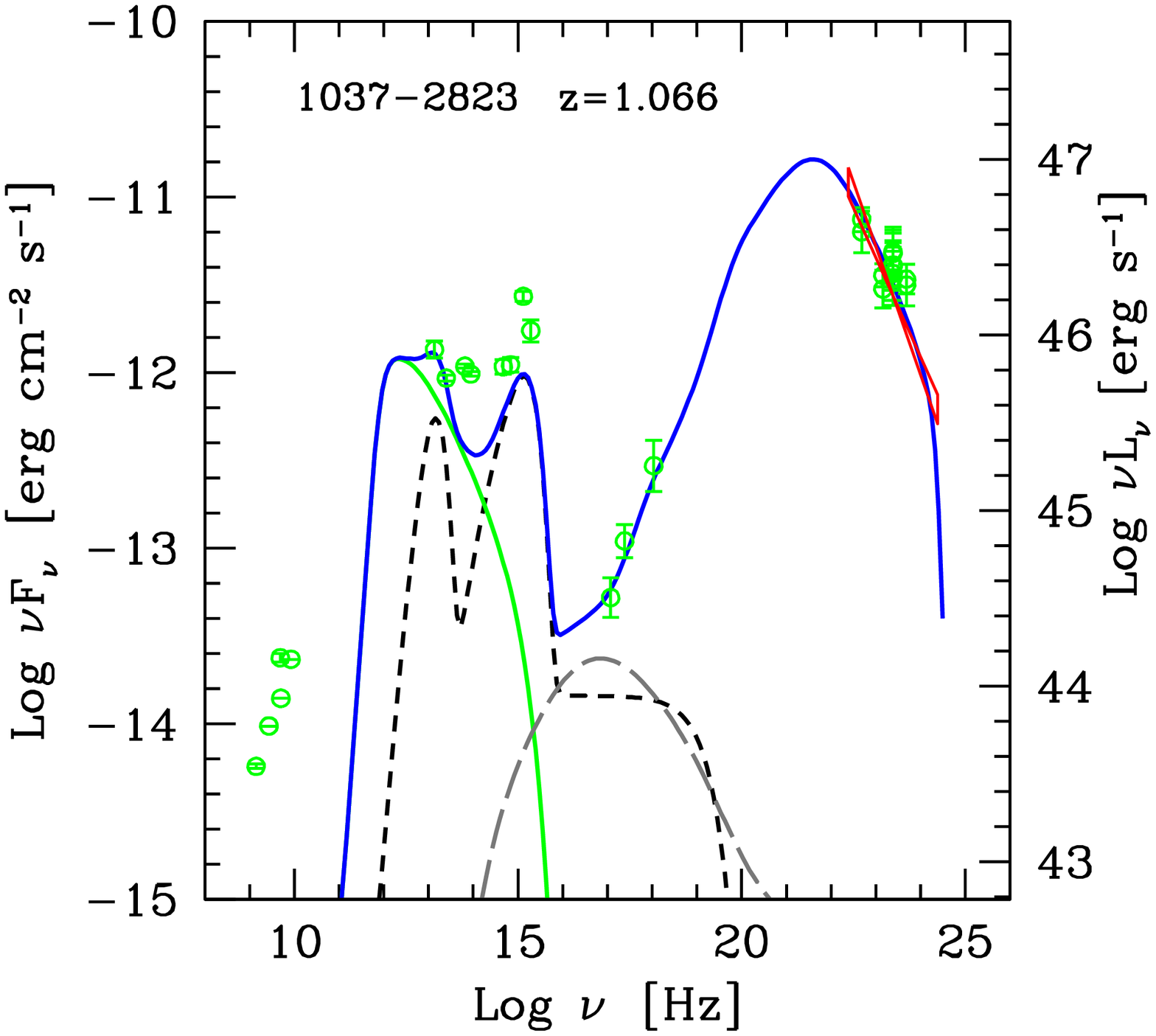,width=4.3cm,height=3.7cm }  
&\psfig{file=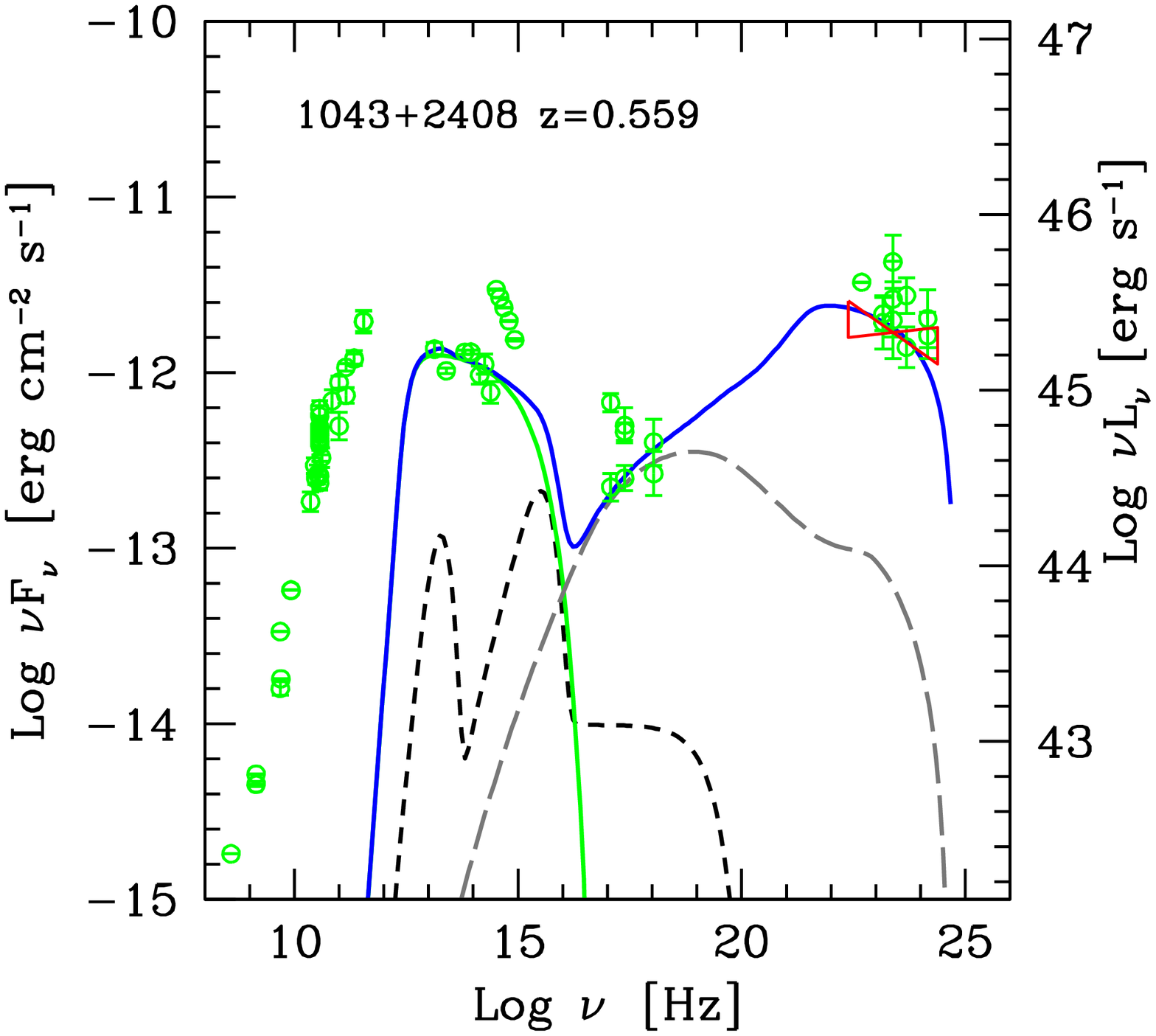,width=4.3cm,height=3.7cm }  \\
\psfig{file=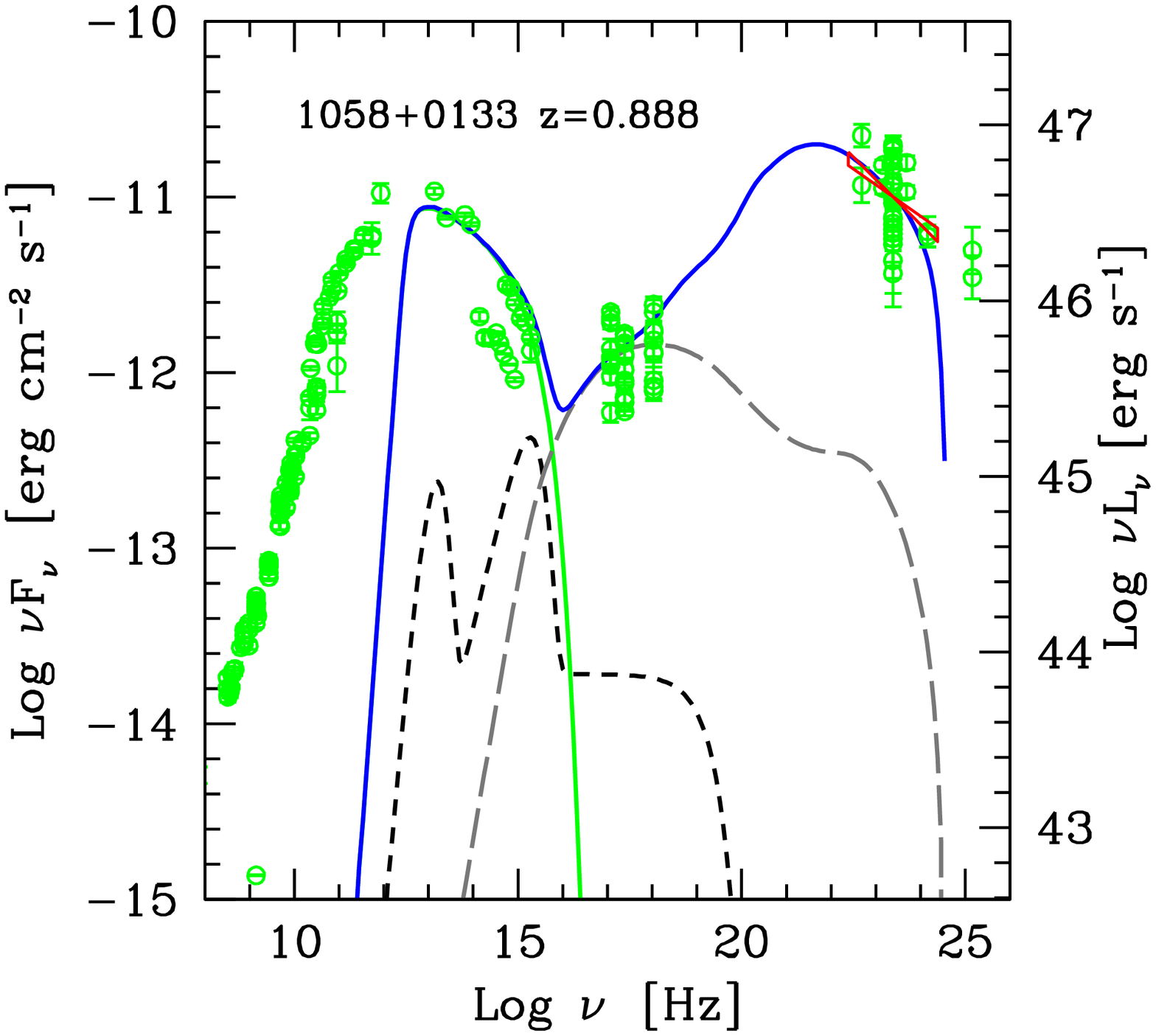,width=4.3cm,height=3.7cm }  
&\psfig{file=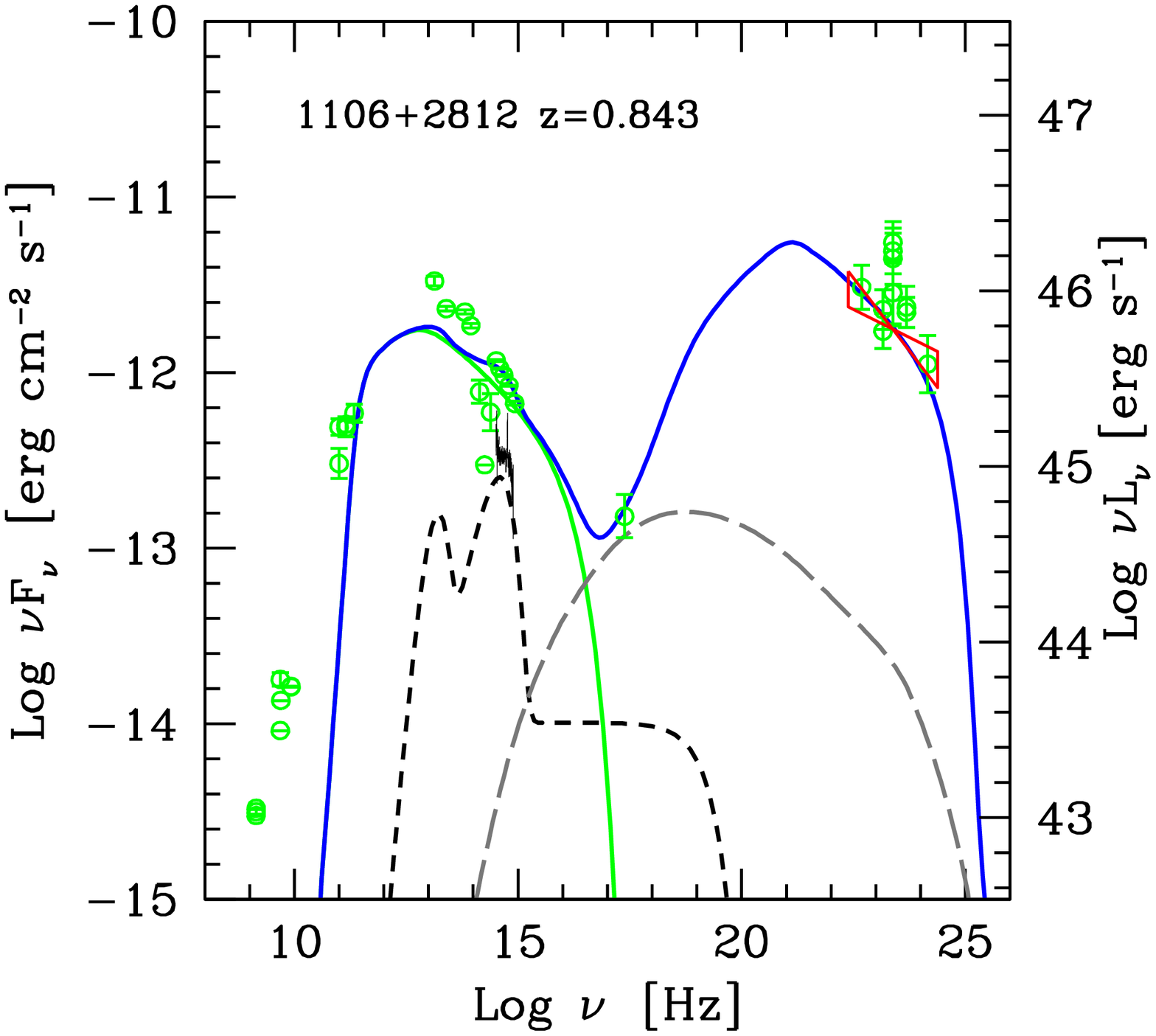,width=4.3cm,height=3.7cm } 
&\psfig{file=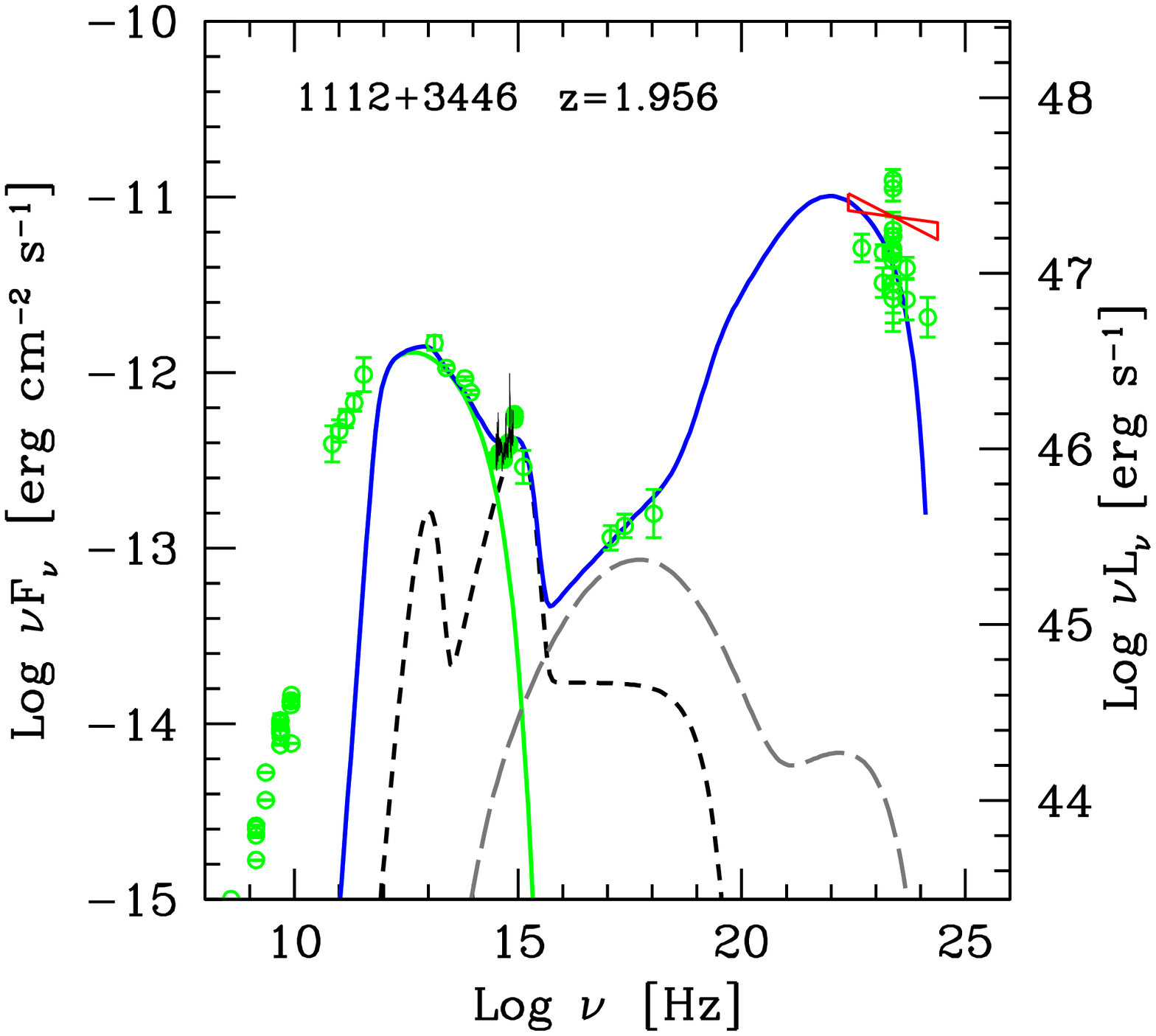,width=4.3cm,height=3.7cm }  
&\psfig{file=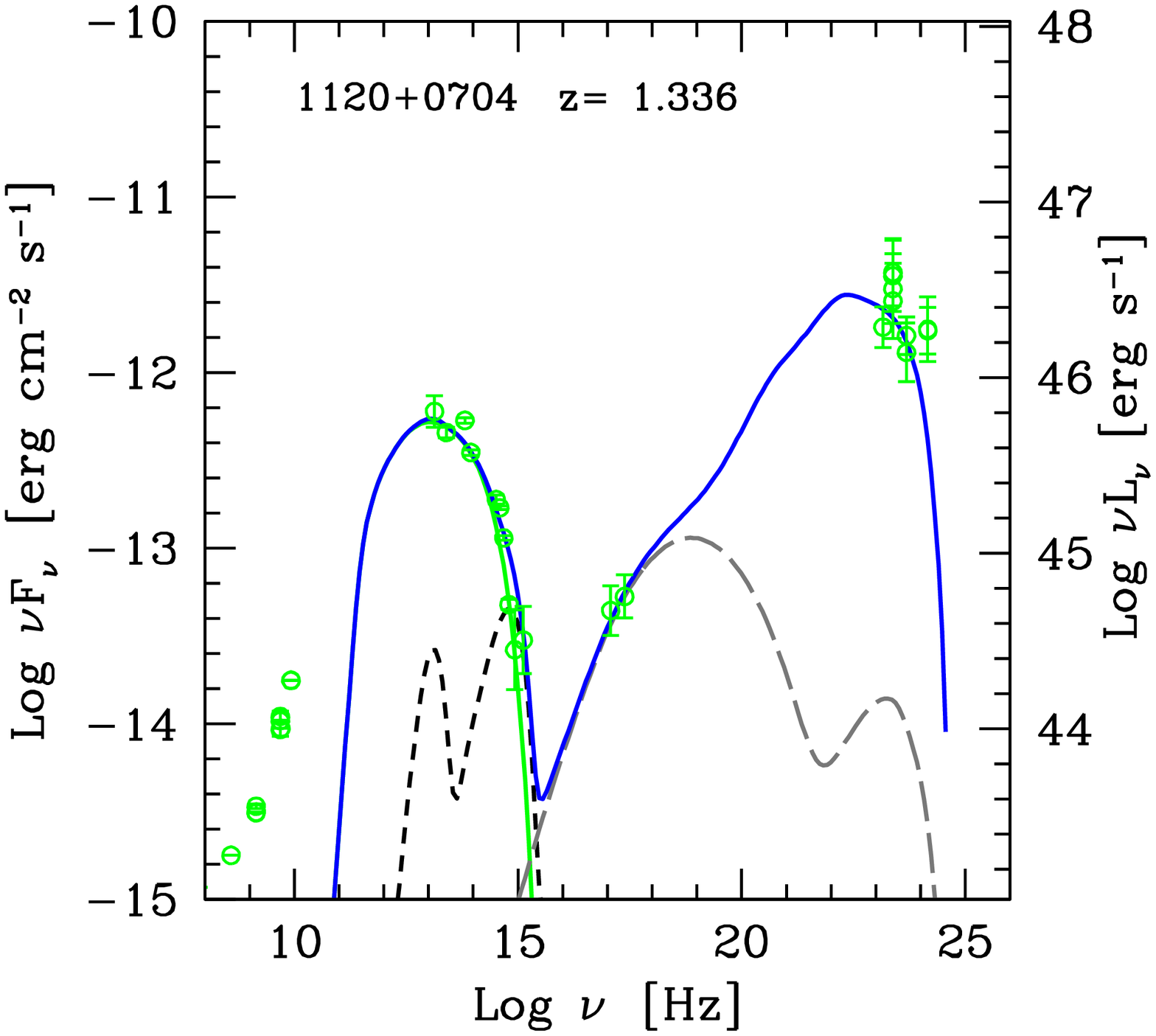,width=4.3cm,height=3.7cm } 
\end{tabular}
\caption{{\it continue.} SED of the FSRQs studied in this paper.}
\end{figure*} 

\setcounter{figure}{15}
\begin{figure*}
\begin{tabular}{cccc}
\psfig{file=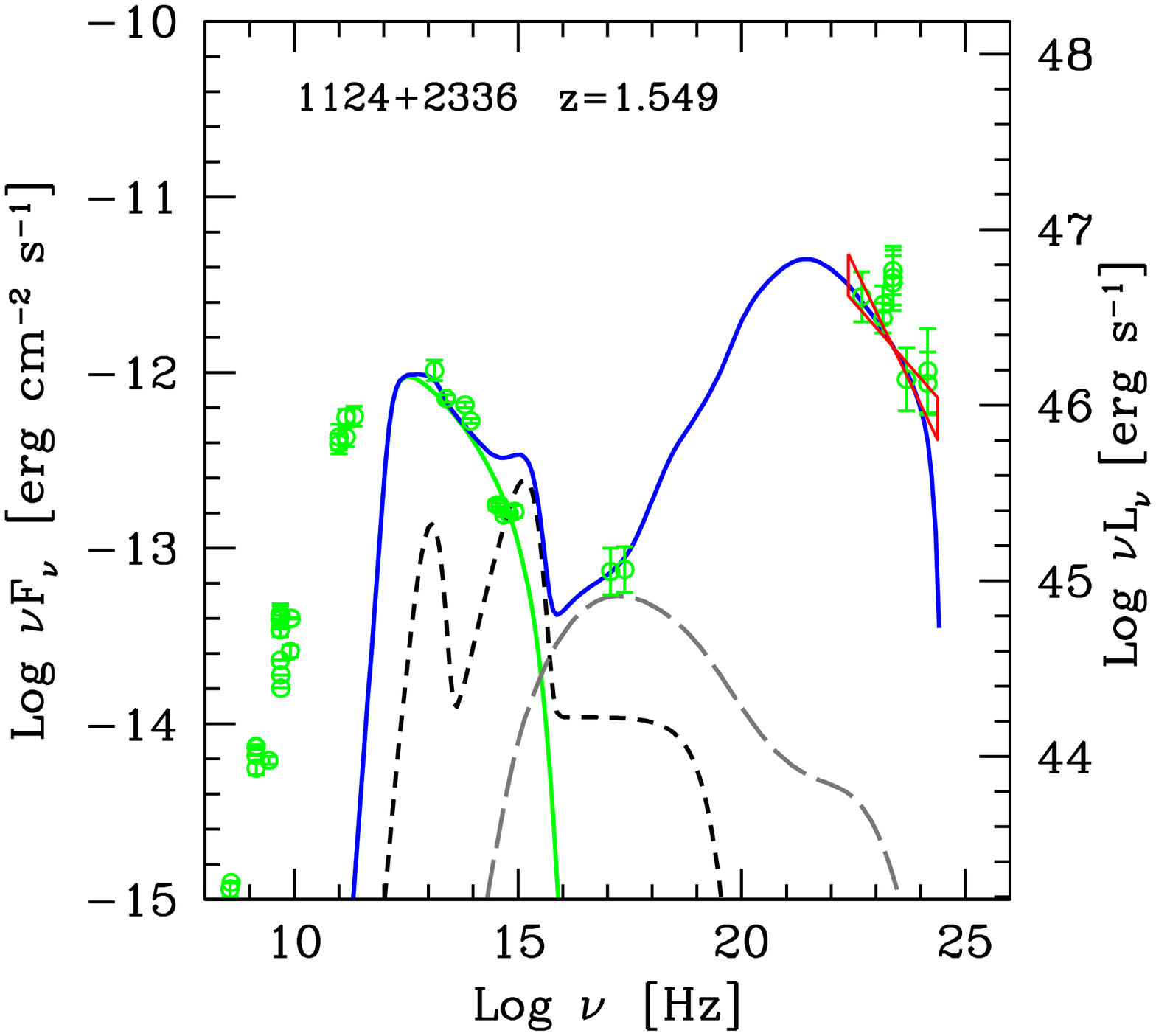,width=4.3cm,height=3.7cm }  
&\psfig{file=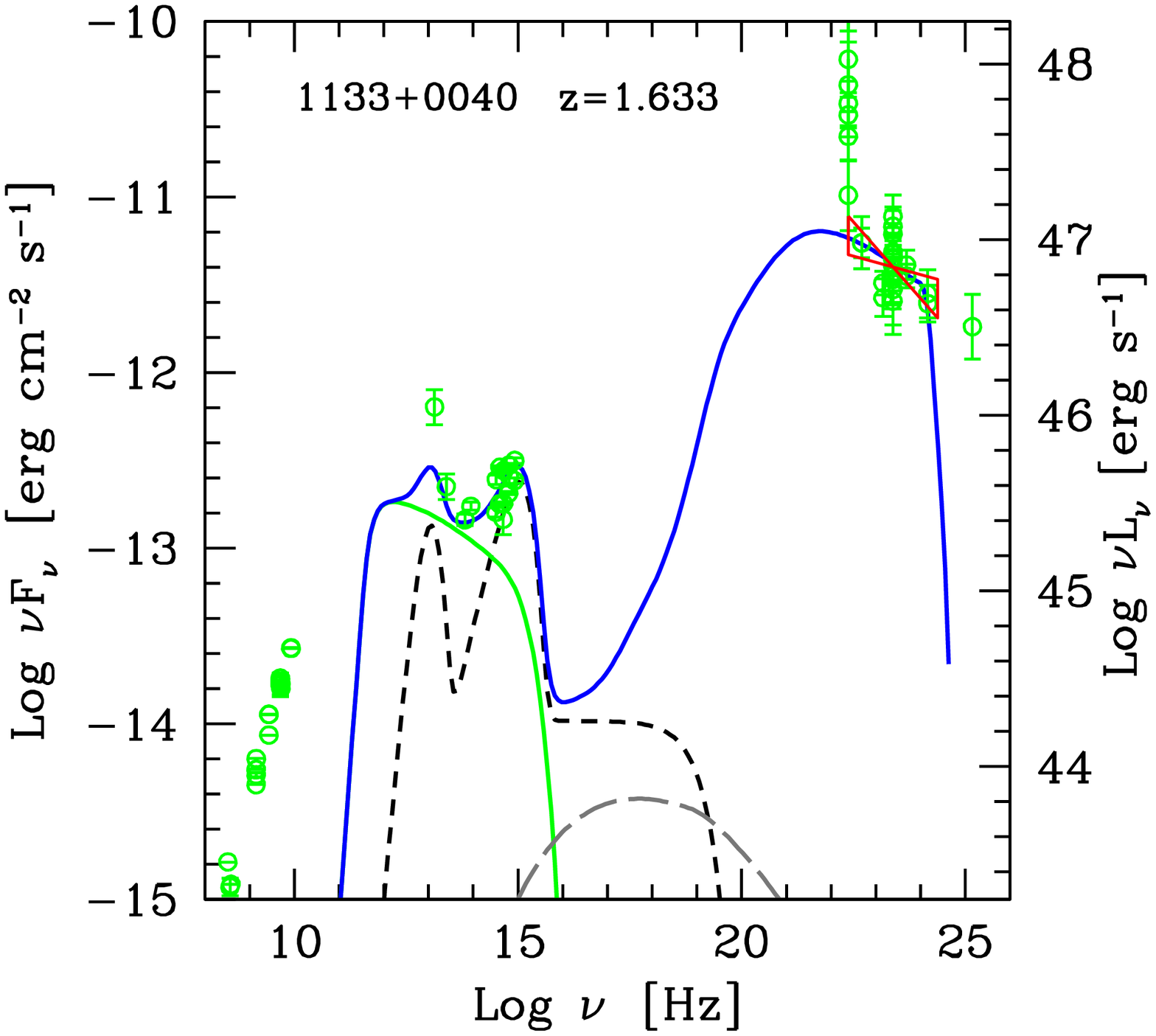,width=4.3cm,height=3.7cm } 
&\psfig{file=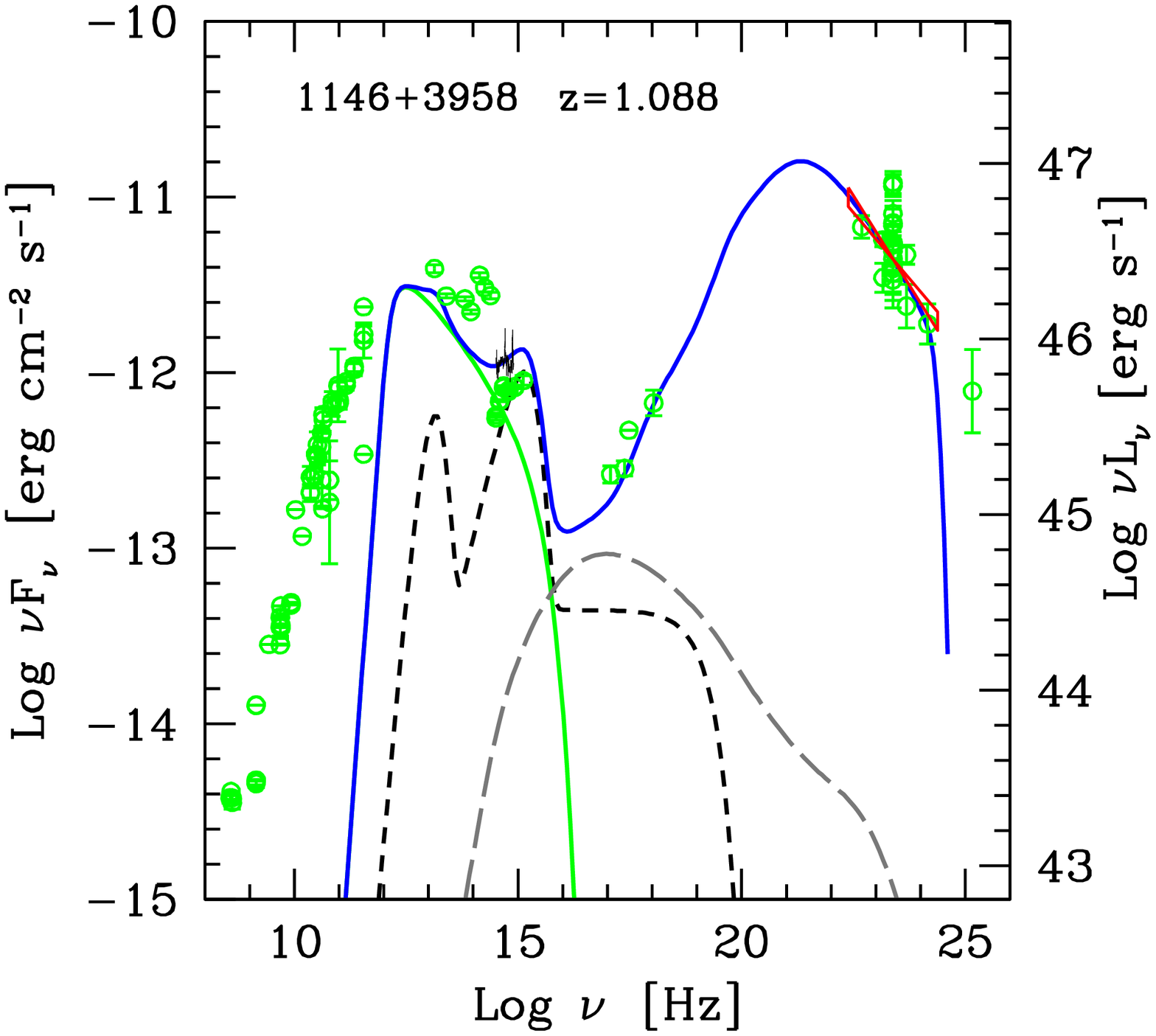,width=4.3cm,height=3.7cm }  
&\psfig{file=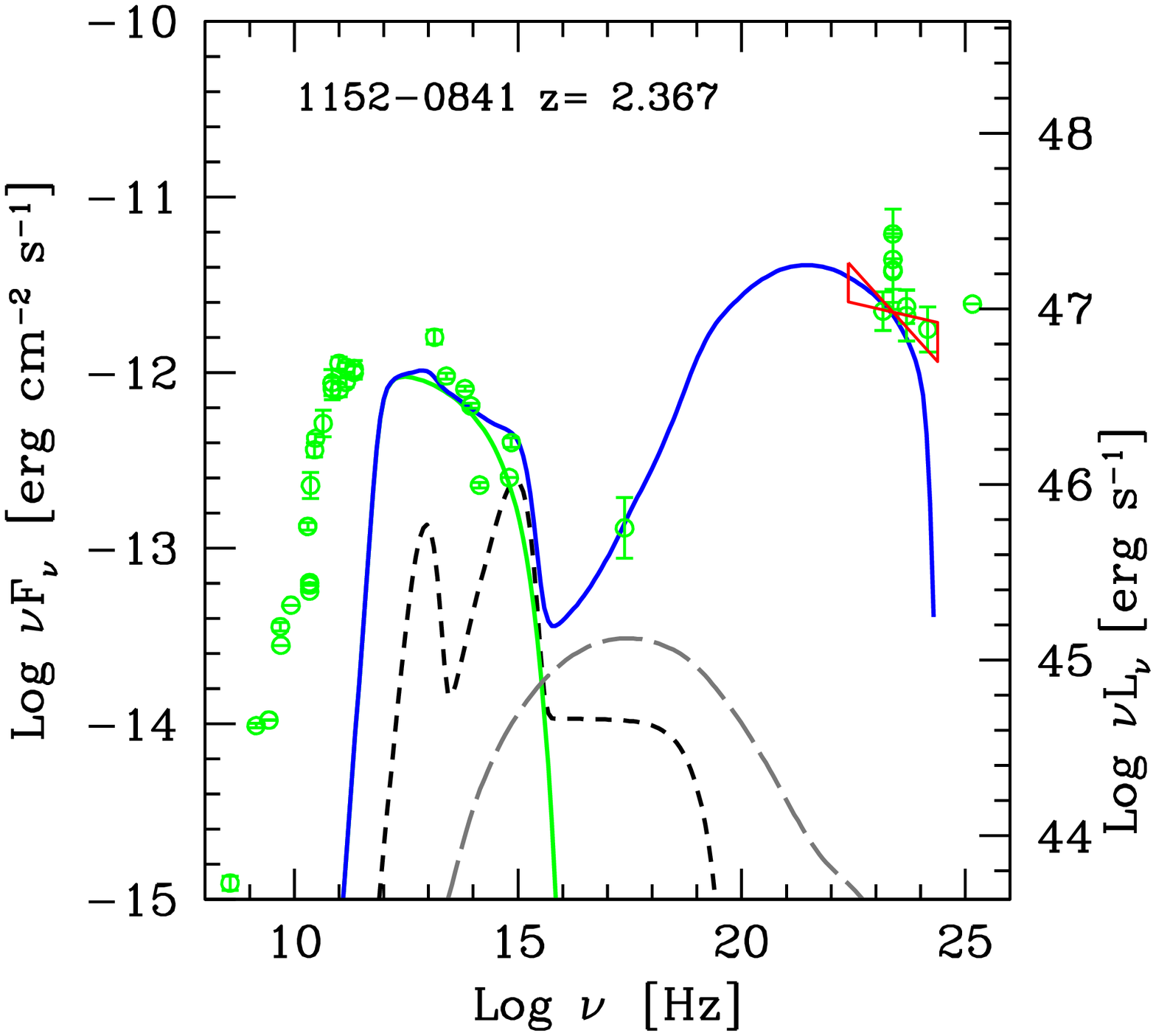,width=4.3cm,height=3.7cm } \\
\psfig{file=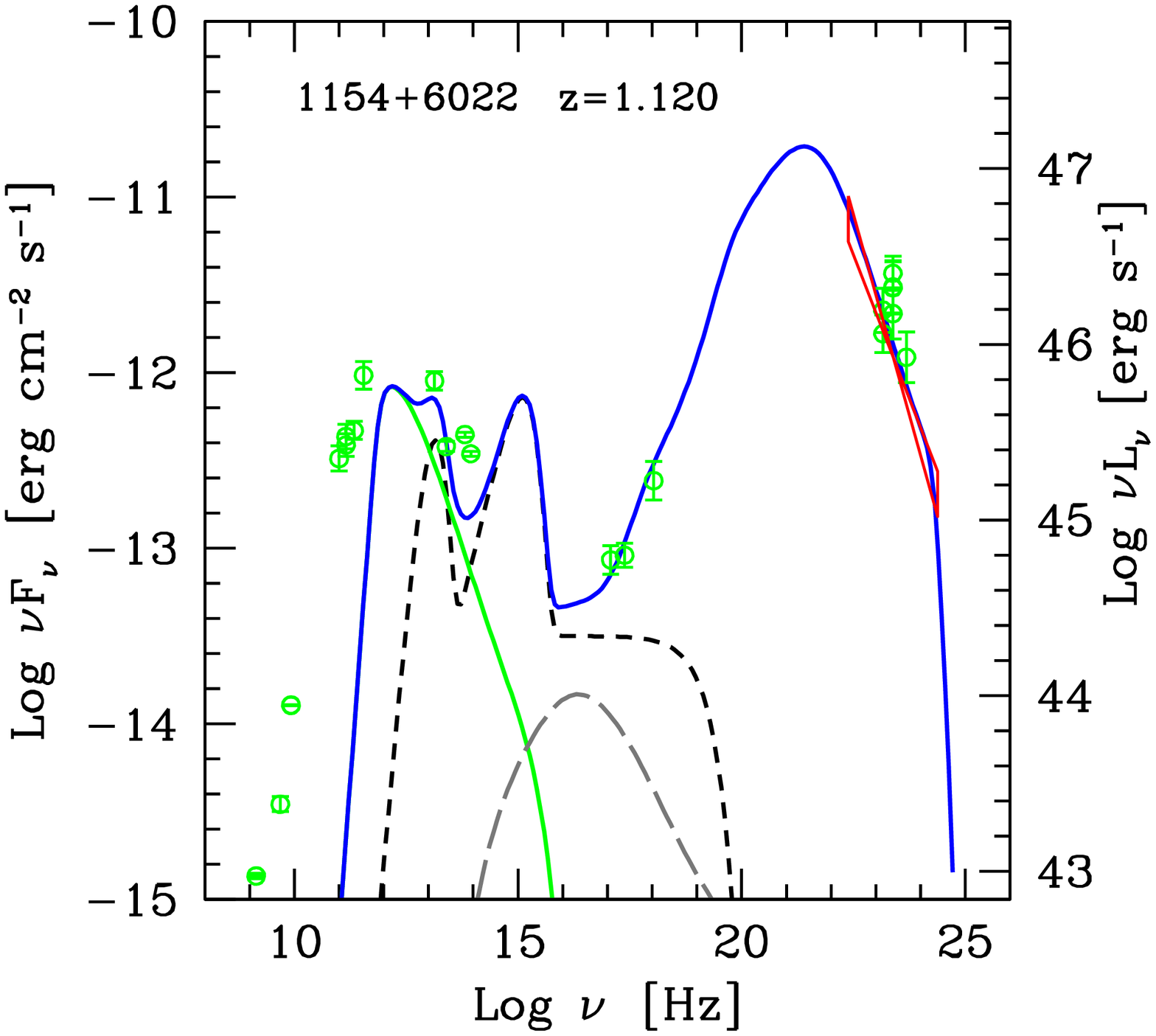,width=4.3cm,height=3.7cm } 
&\psfig{file=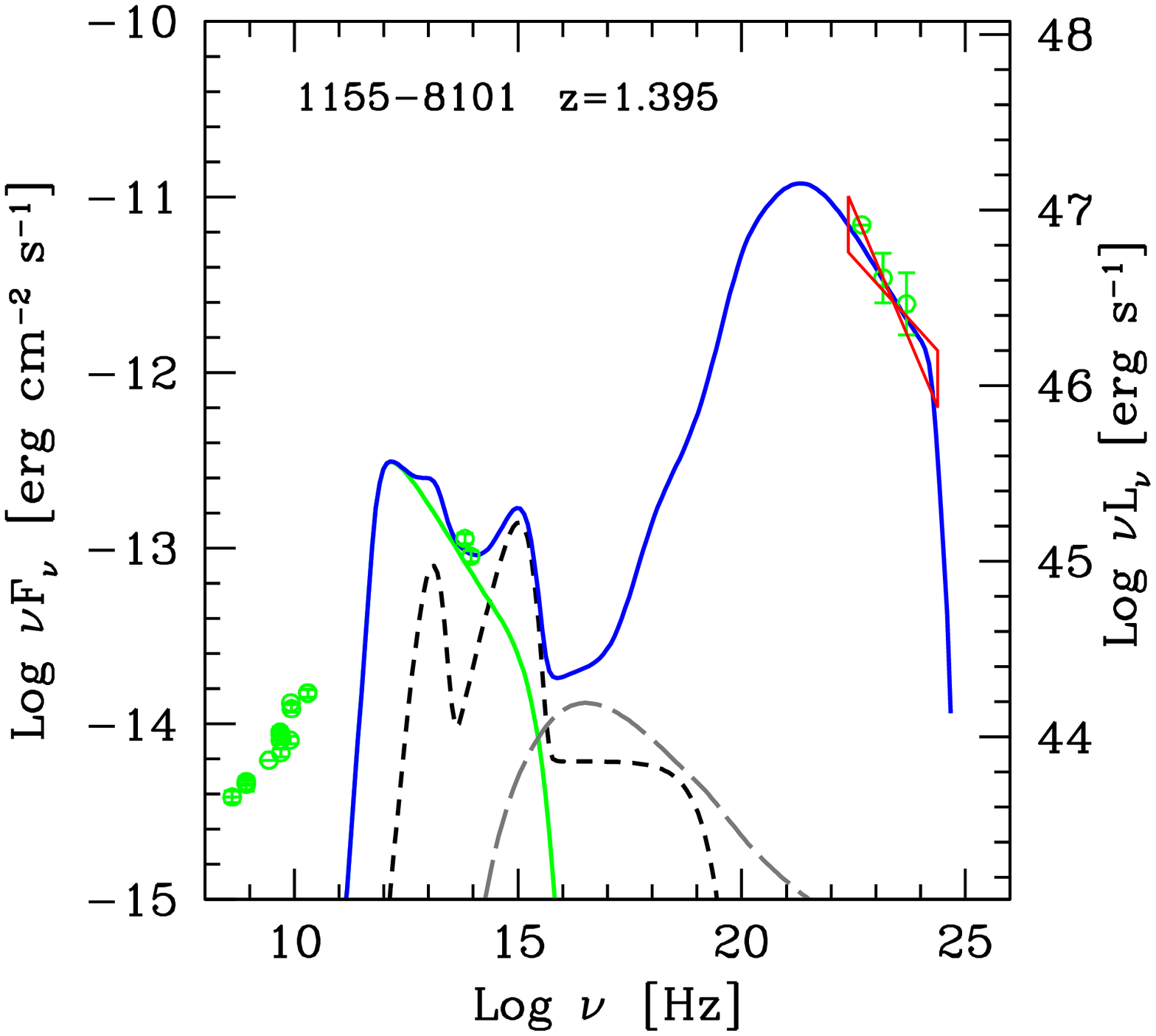,width=4.3cm,height=3.7cm } 
&\psfig{file=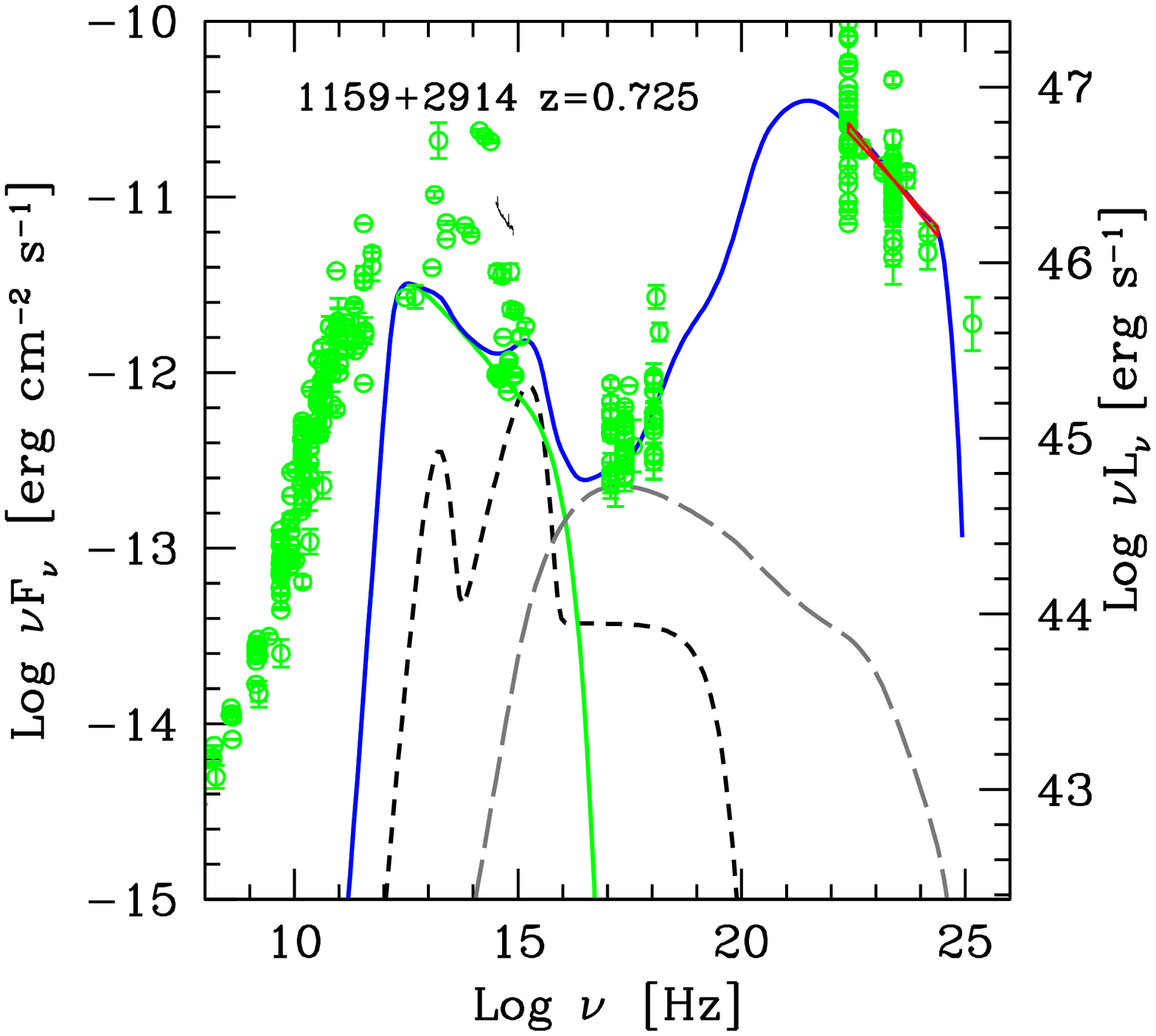,width=4.3cm,height=3.7cm } 
&\psfig{file=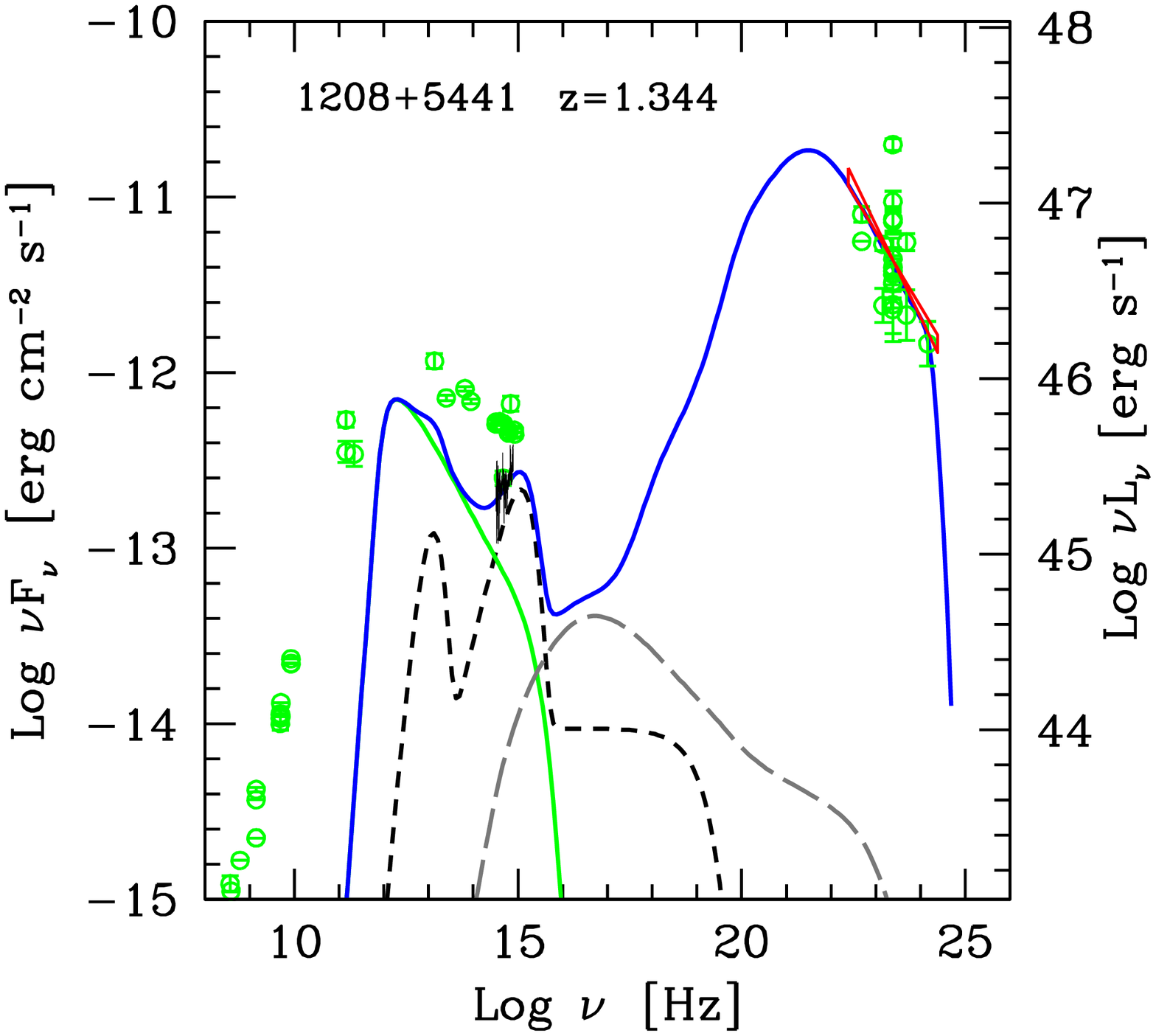,width=4.3cm,height=3.7cm } \\
\psfig{file=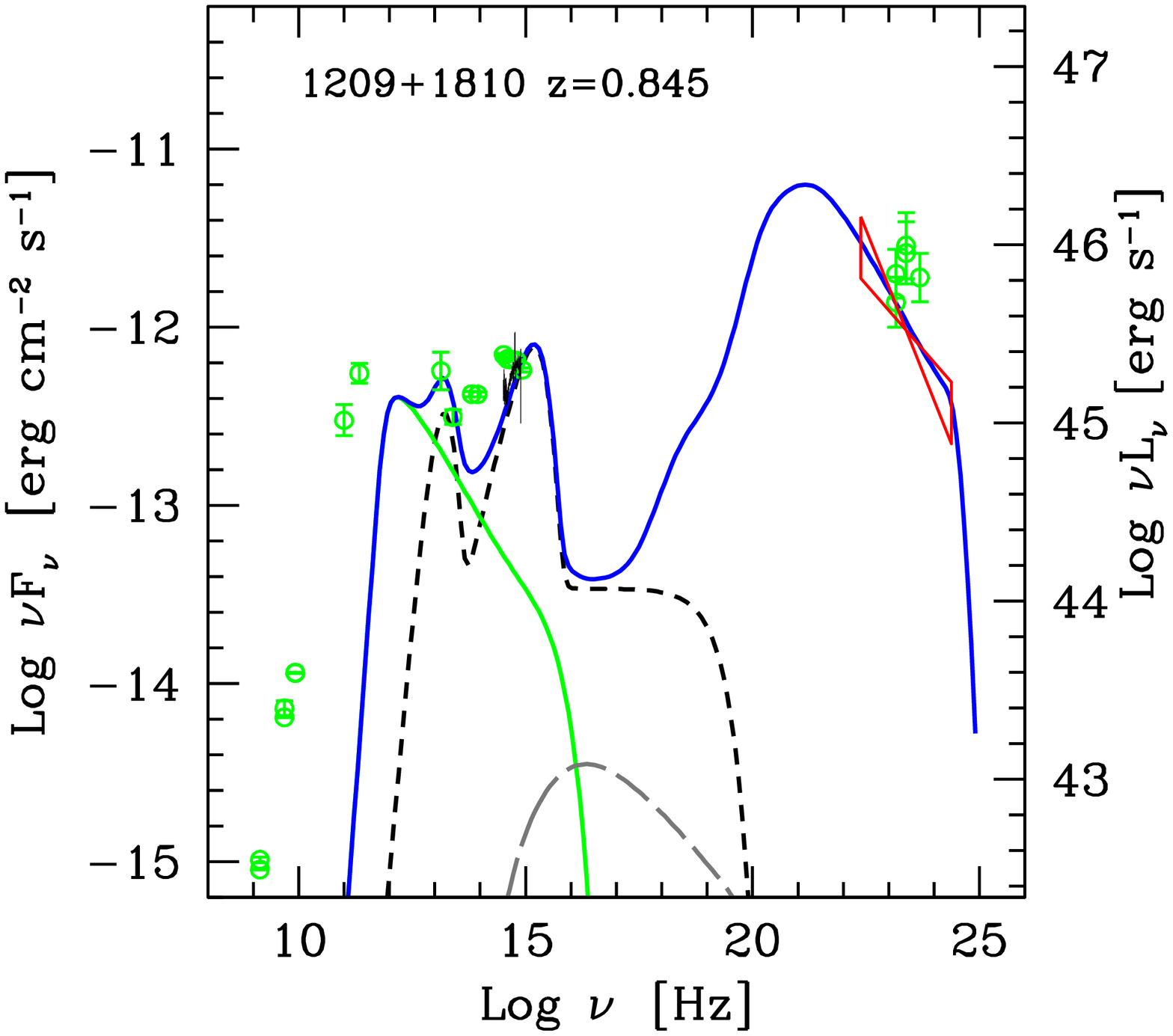,width=4.3cm,height=3.7cm } 
&\psfig{file=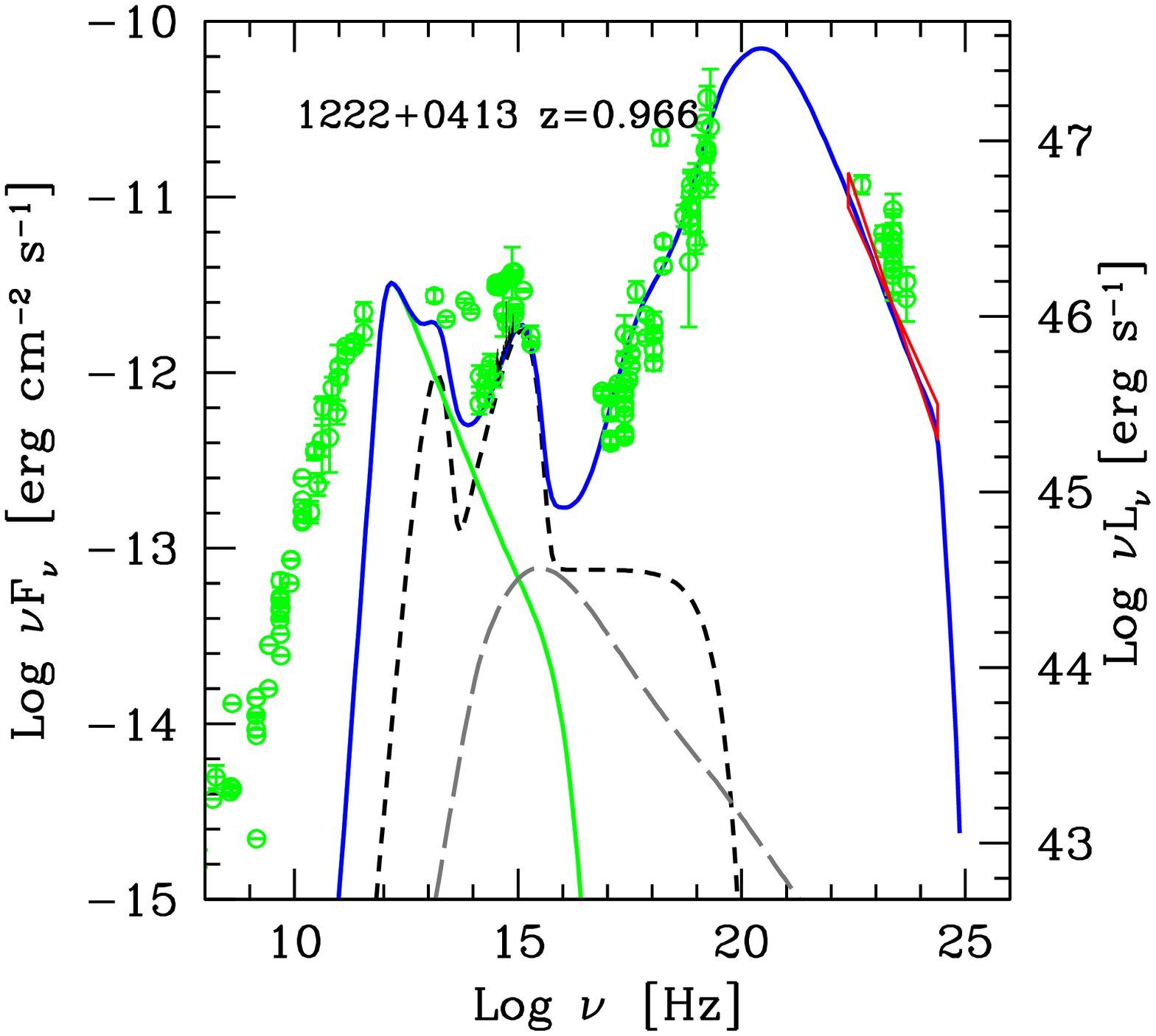,width=4.3cm,height=3.7cm } 
&\psfig{file=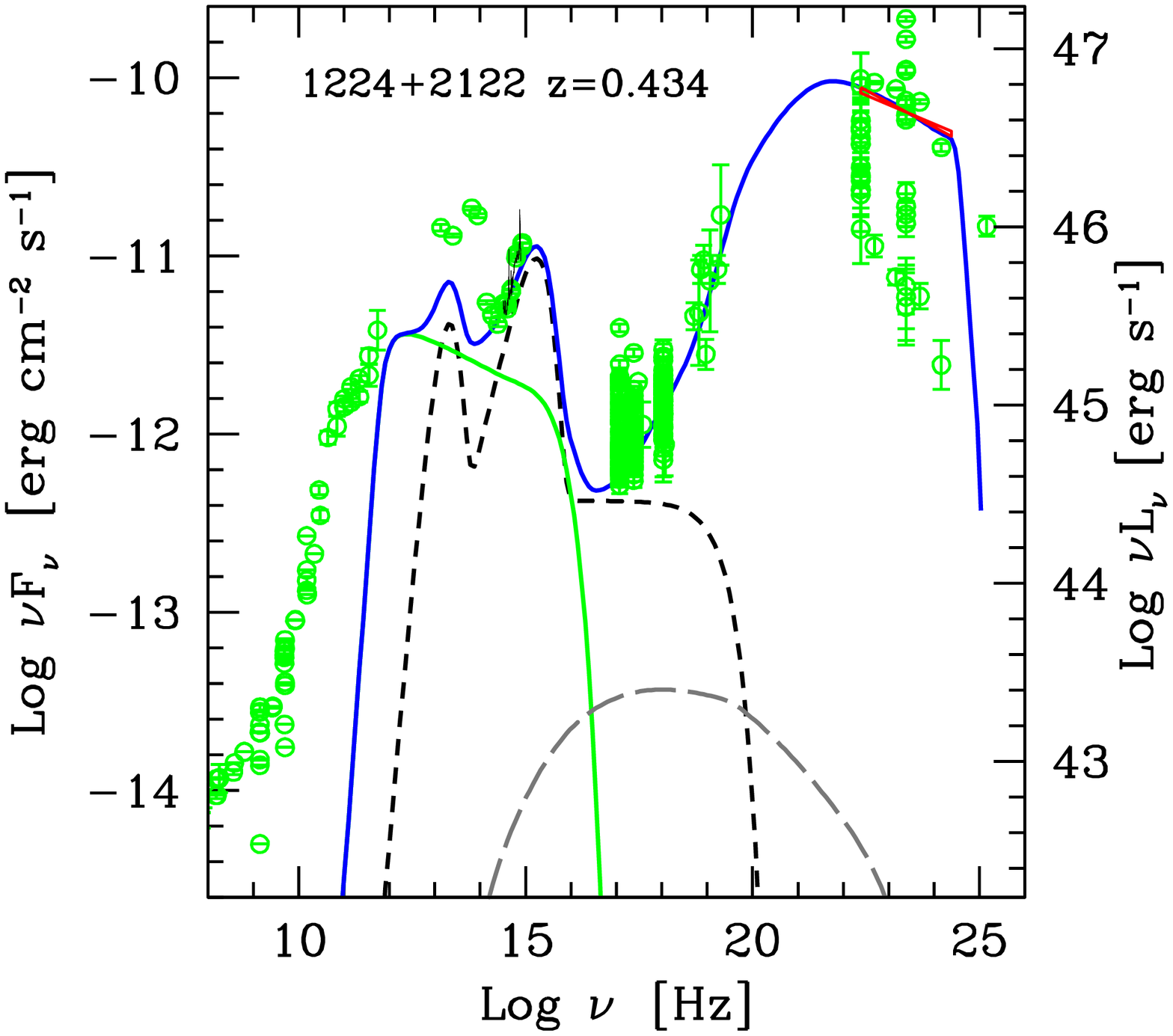,width=4.3cm,height=3.7cm }  
&\psfig{file=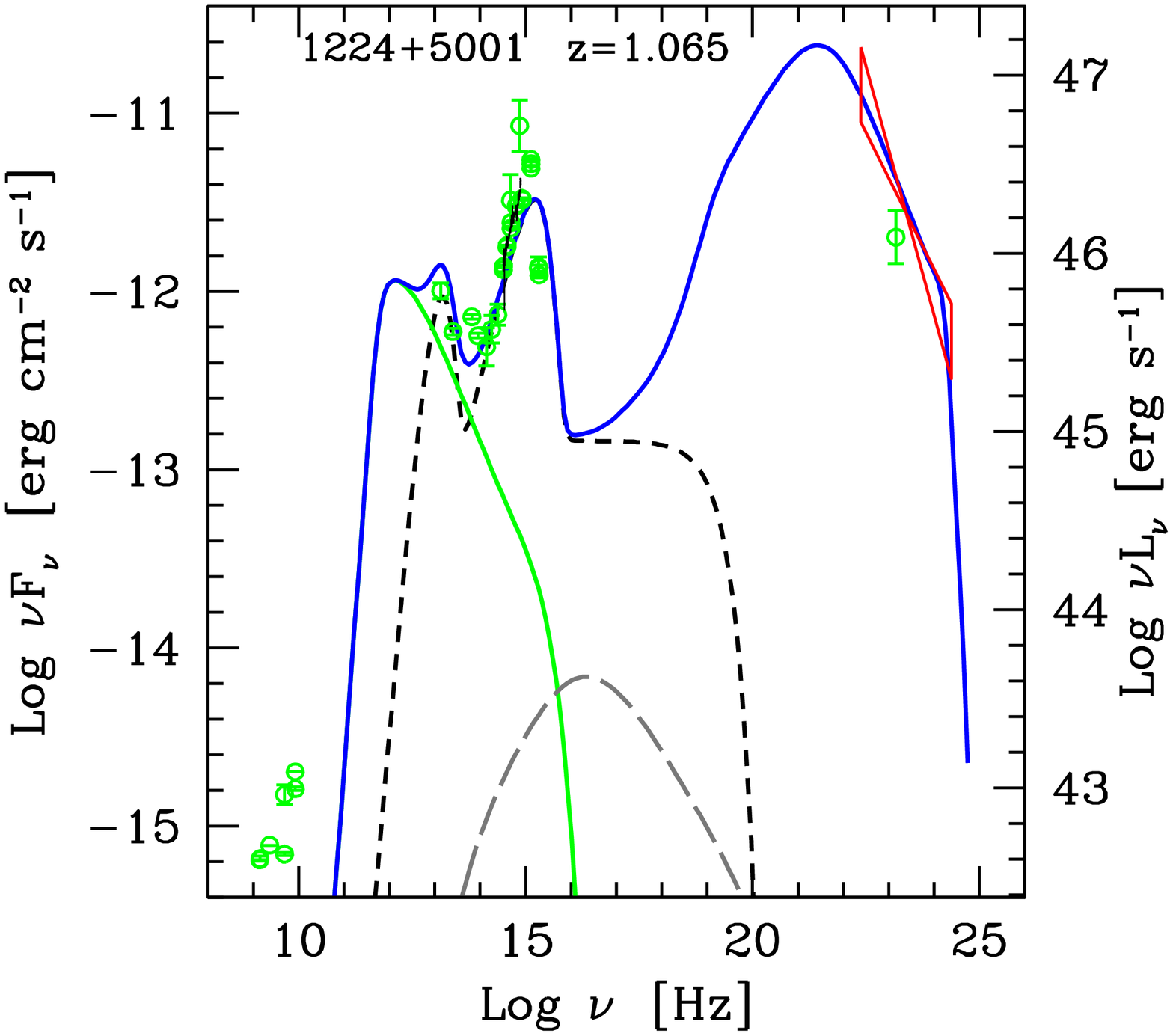,width=4.3cm,height=3.7cm } \\
\psfig{file=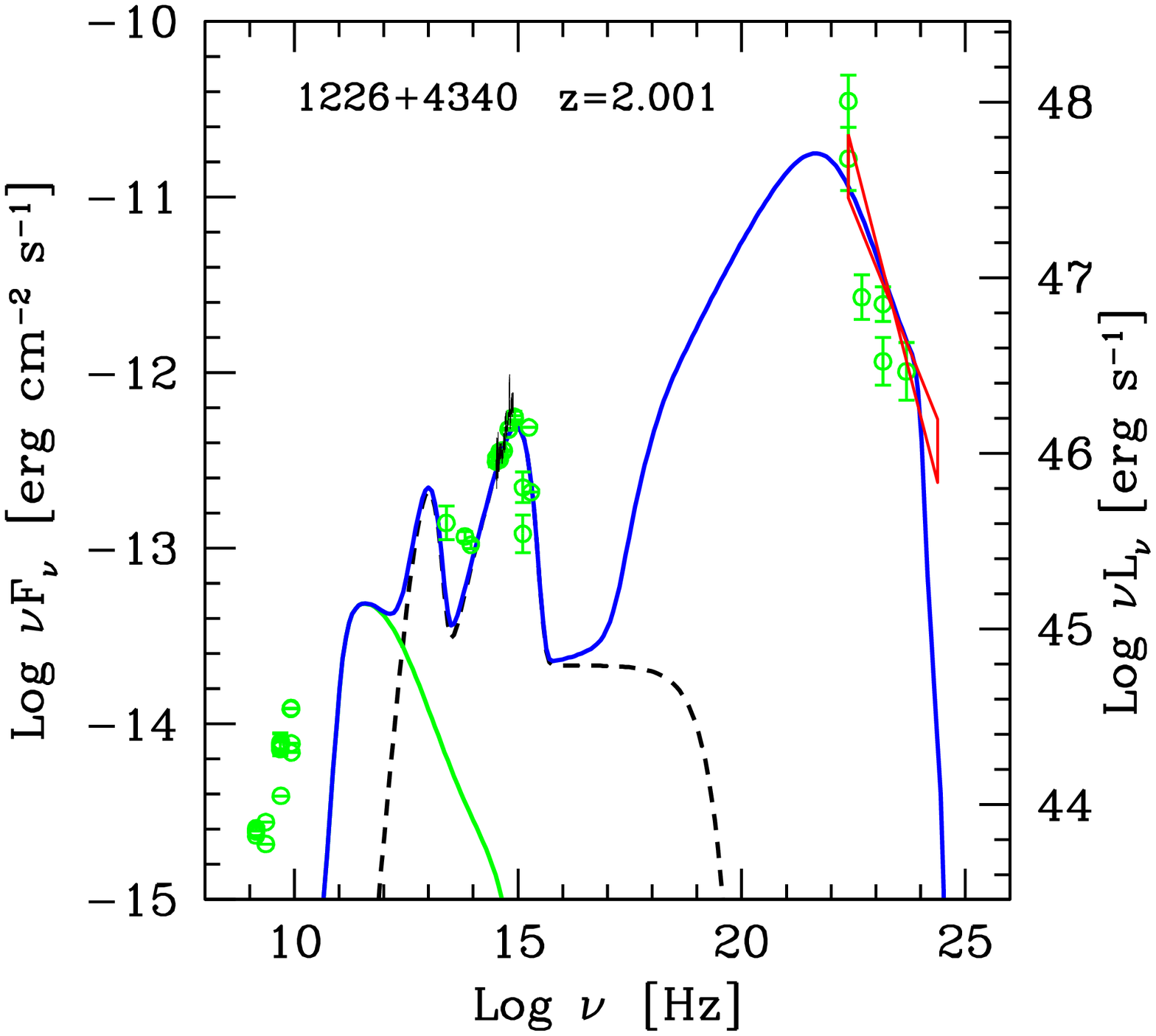,width=4.3cm,height=3.7cm } 
&\psfig{file=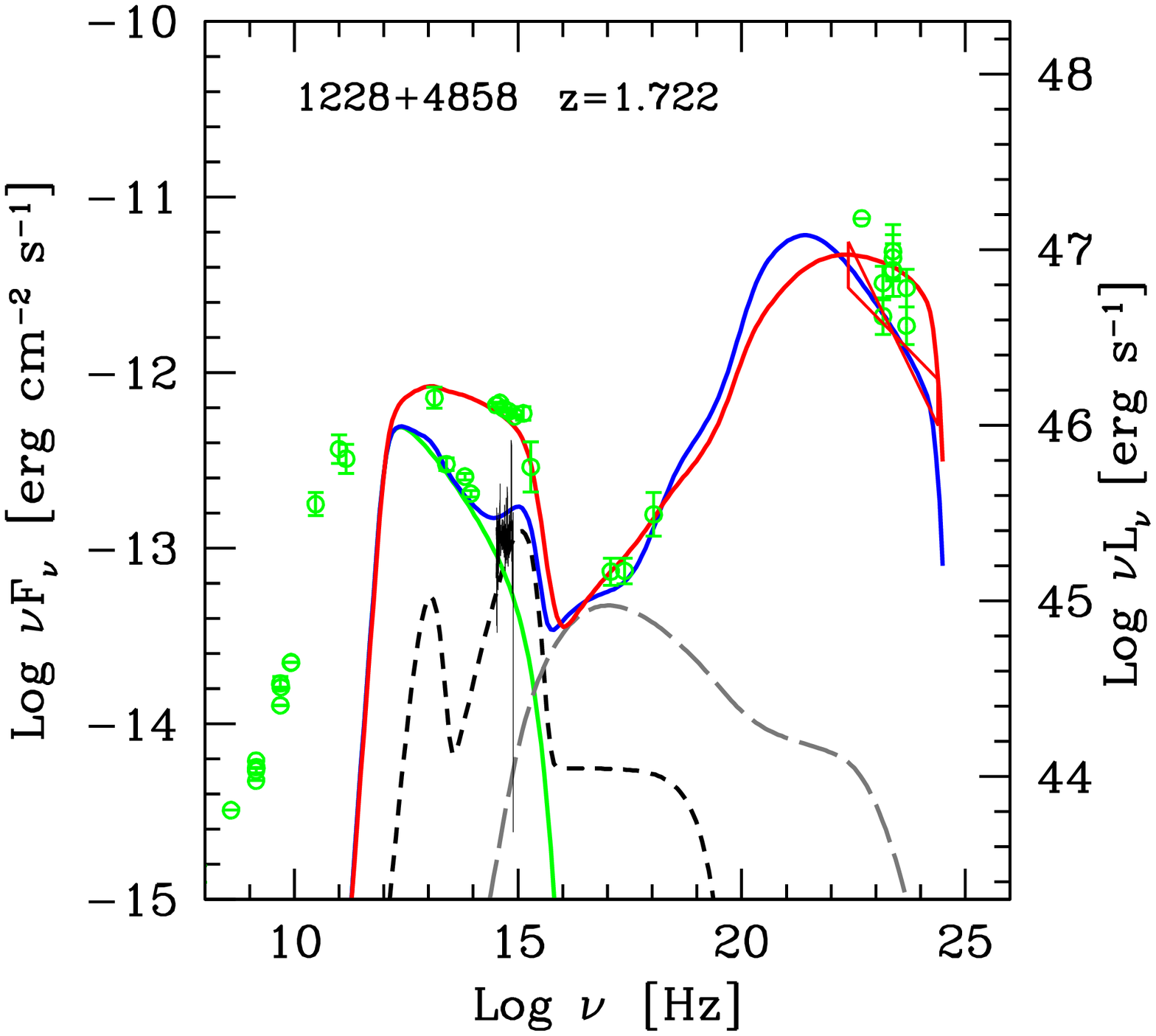,width=4.3cm,height=3.7cm } 
&\psfig{file=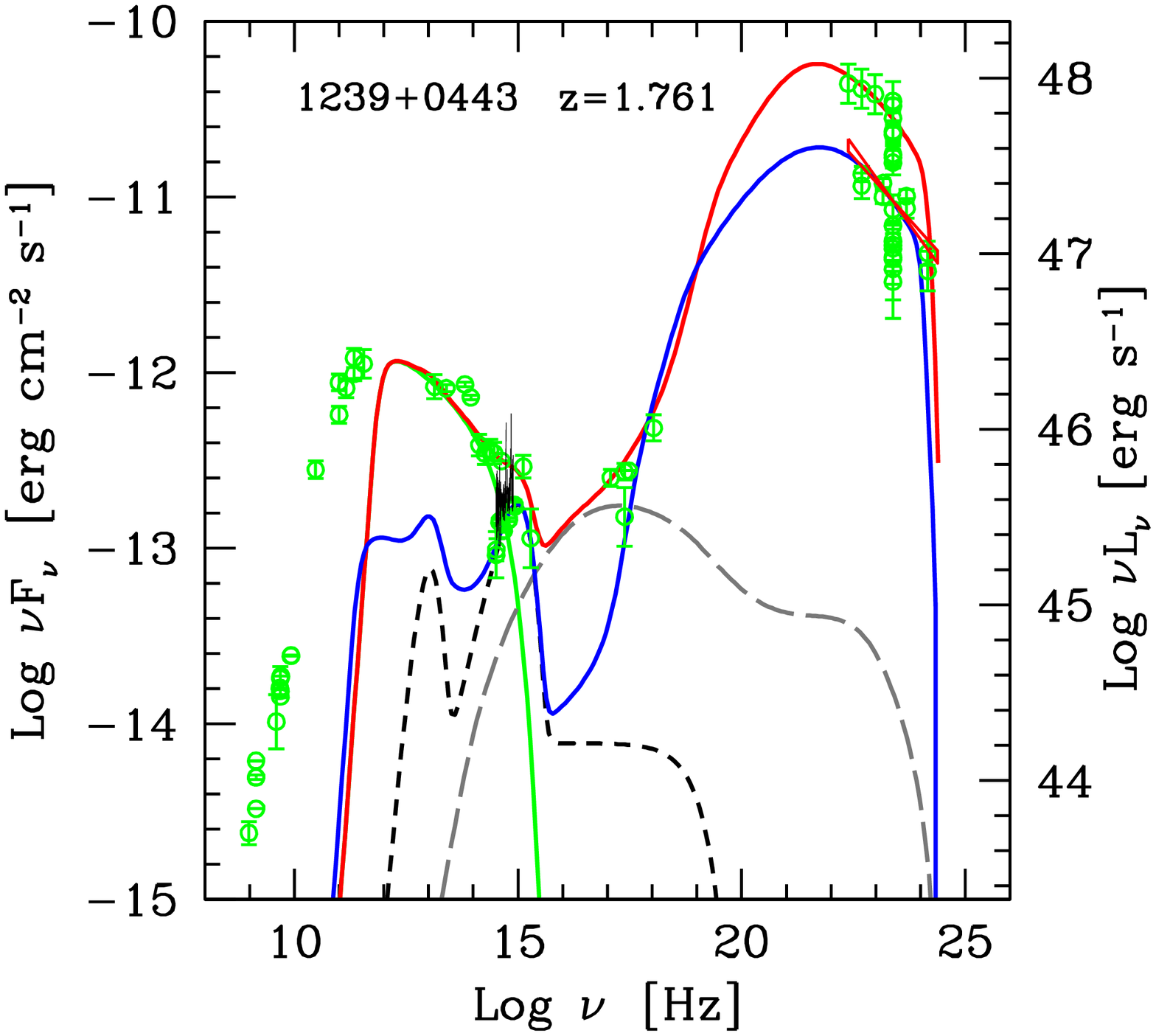,width=4.3cm,height=3.7cm } 
&\psfig{file=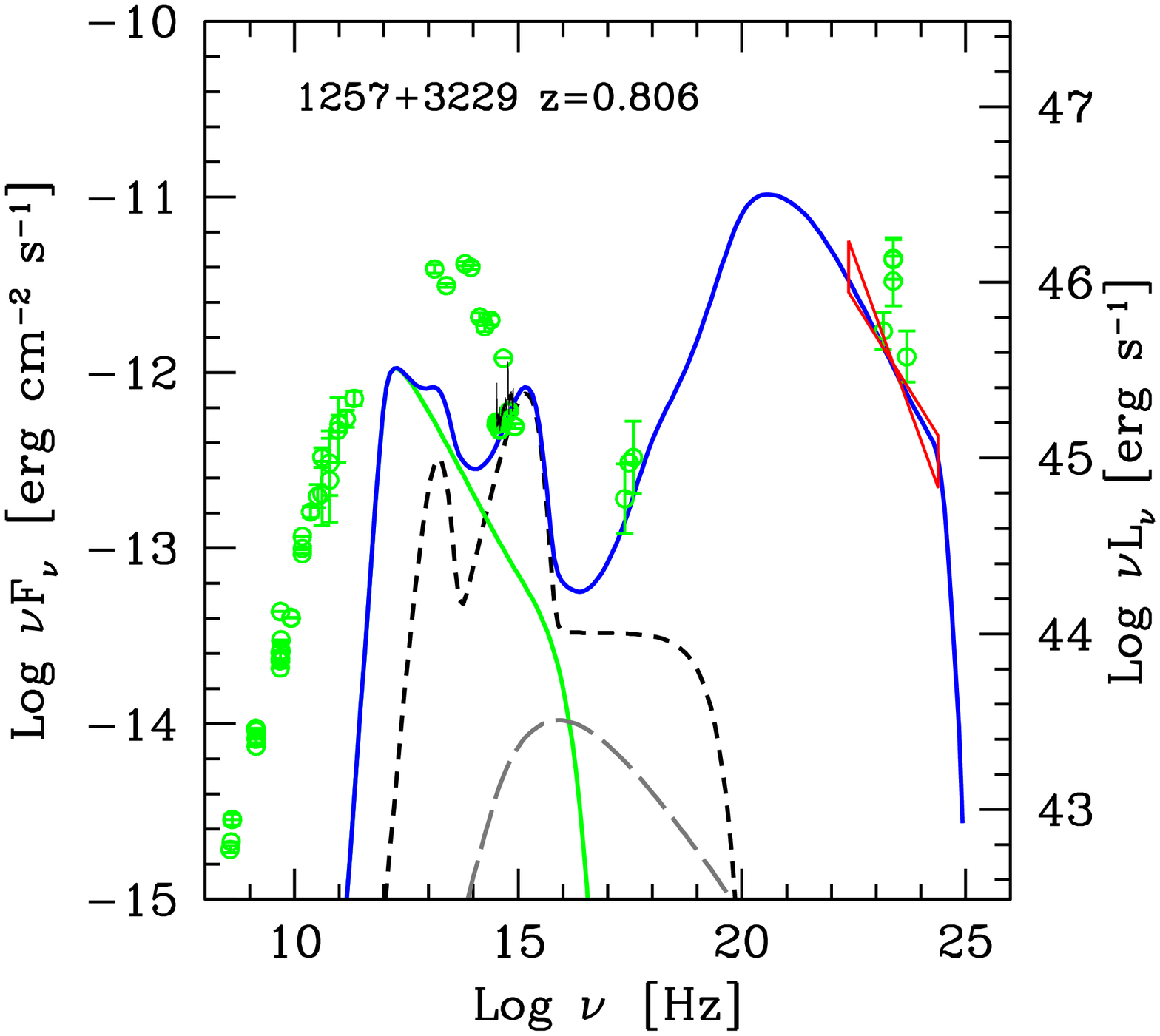,width=4.3cm,height=3.7cm } \\
\psfig{file=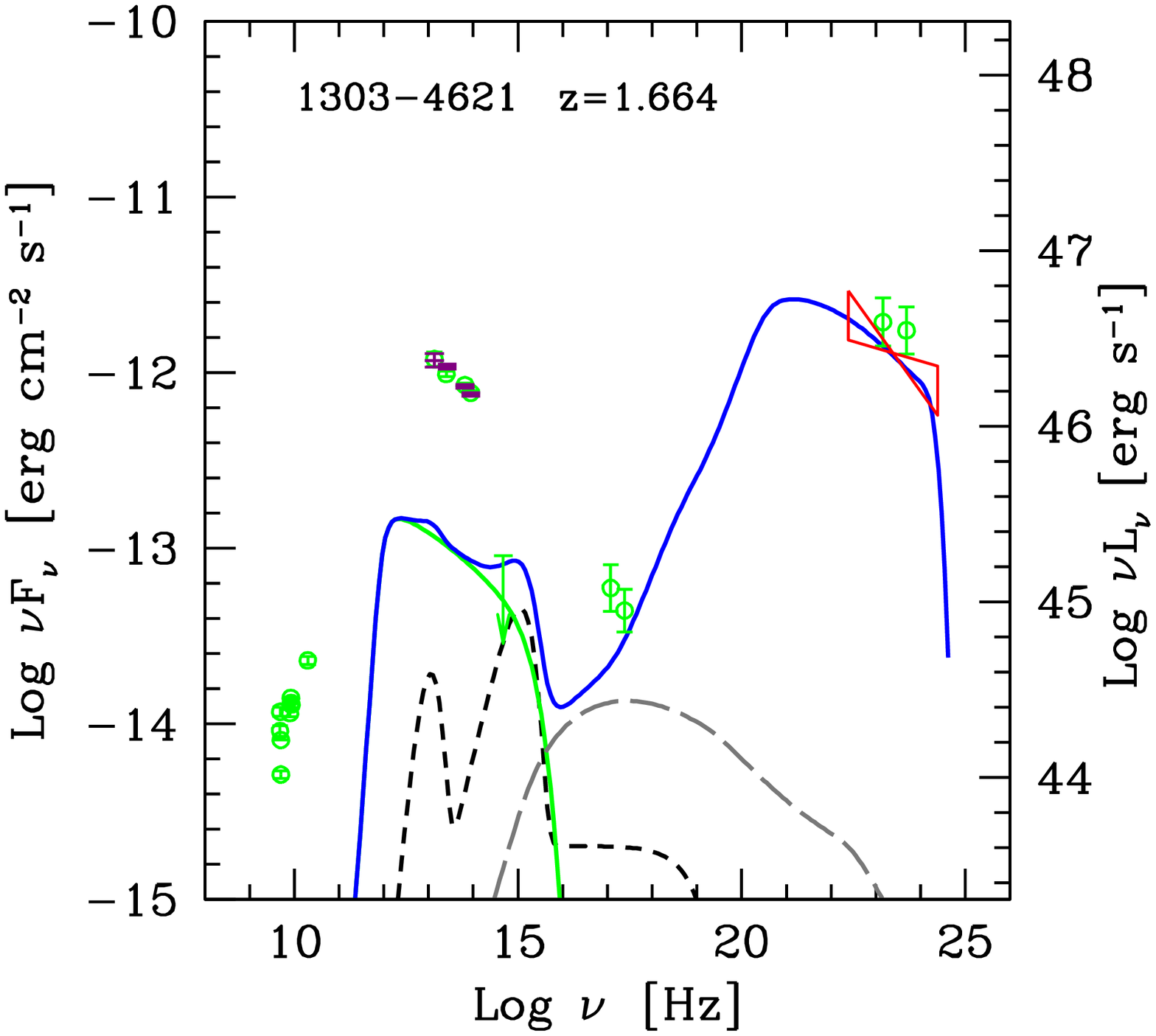,width=4.3cm,height=3.7cm } 
&\psfig{file=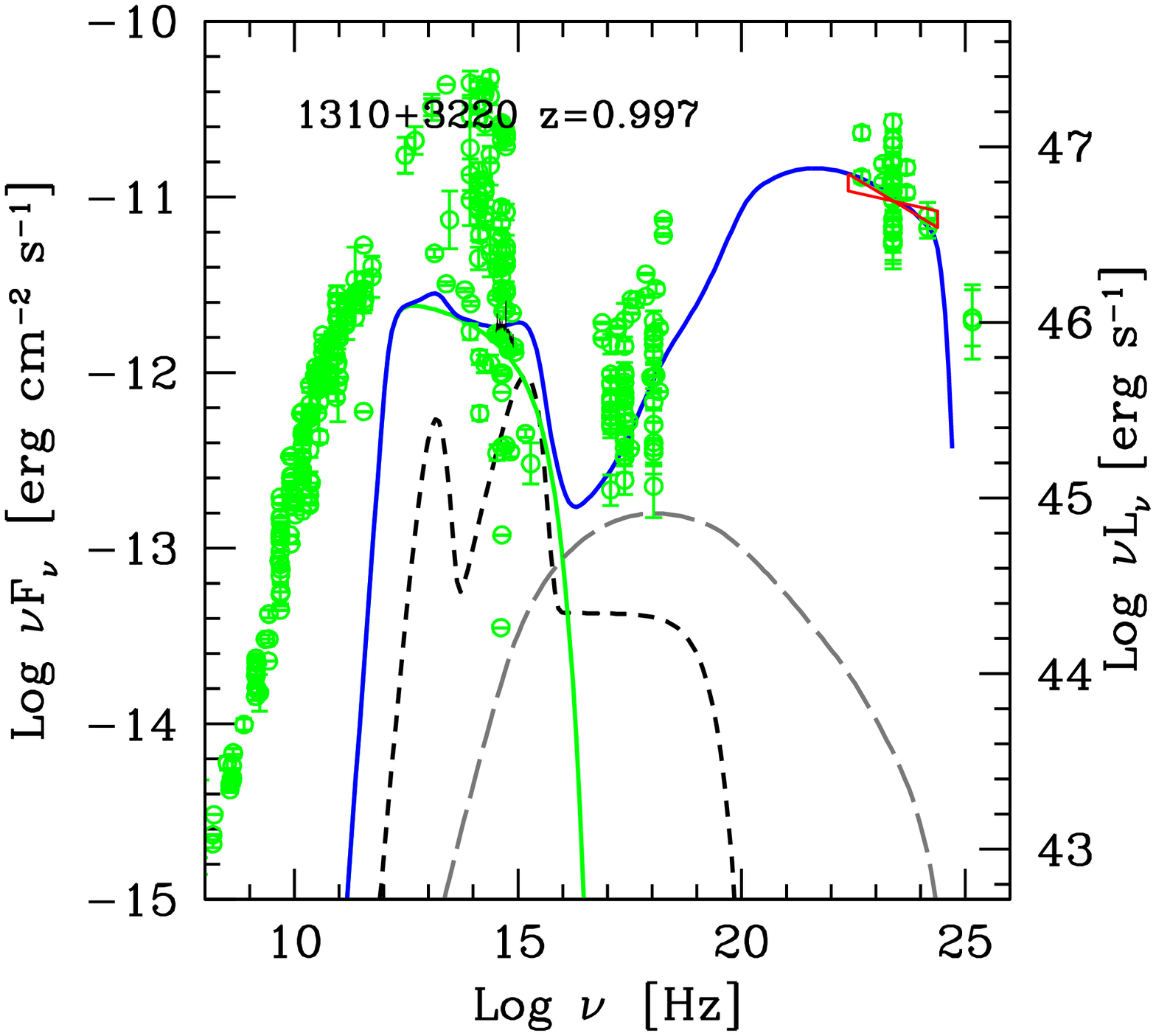,width=4.3cm,height=3.7cm } 
&\psfig{file=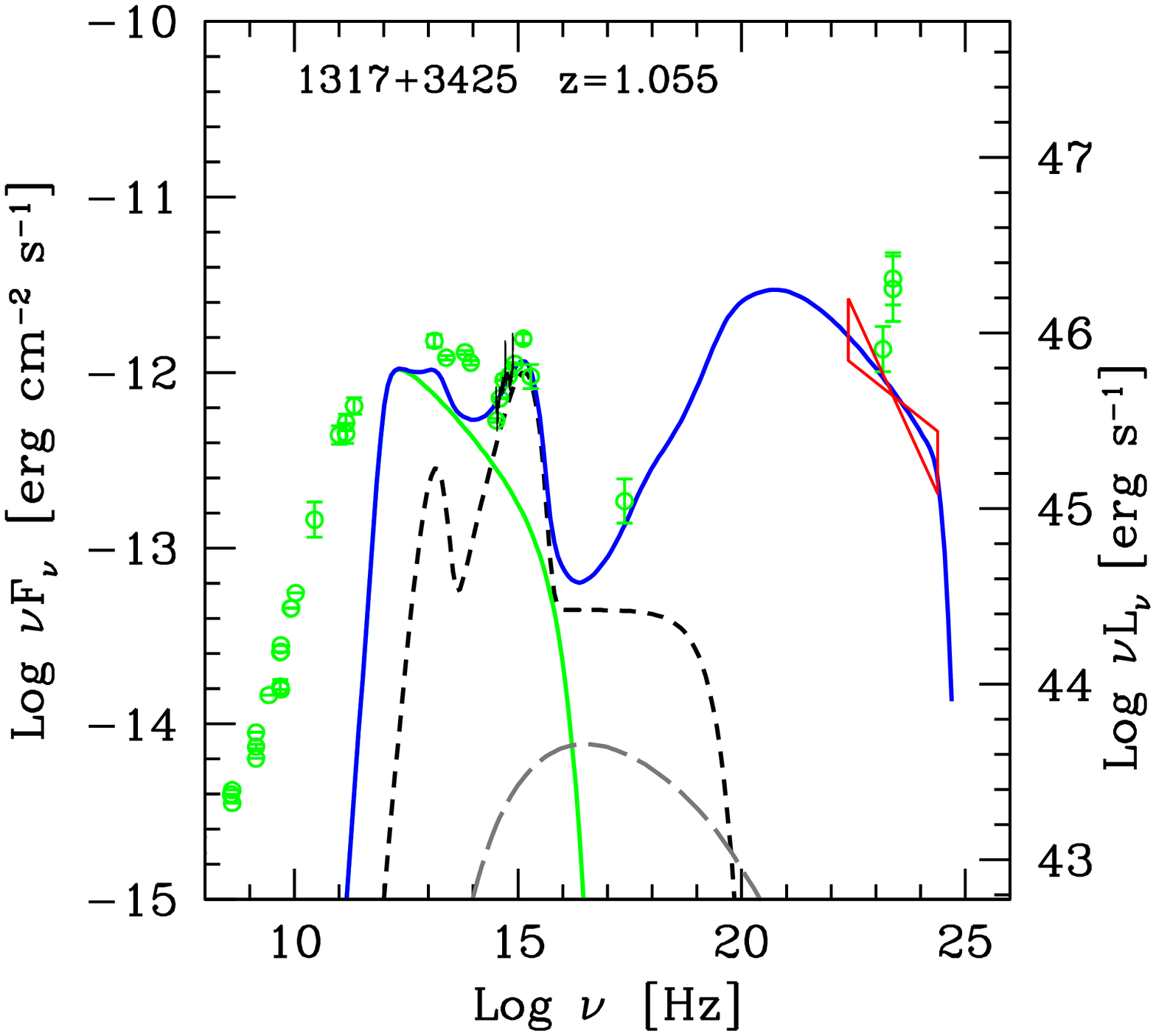,width=4.3cm,height=3.7cm }  
&\psfig{file=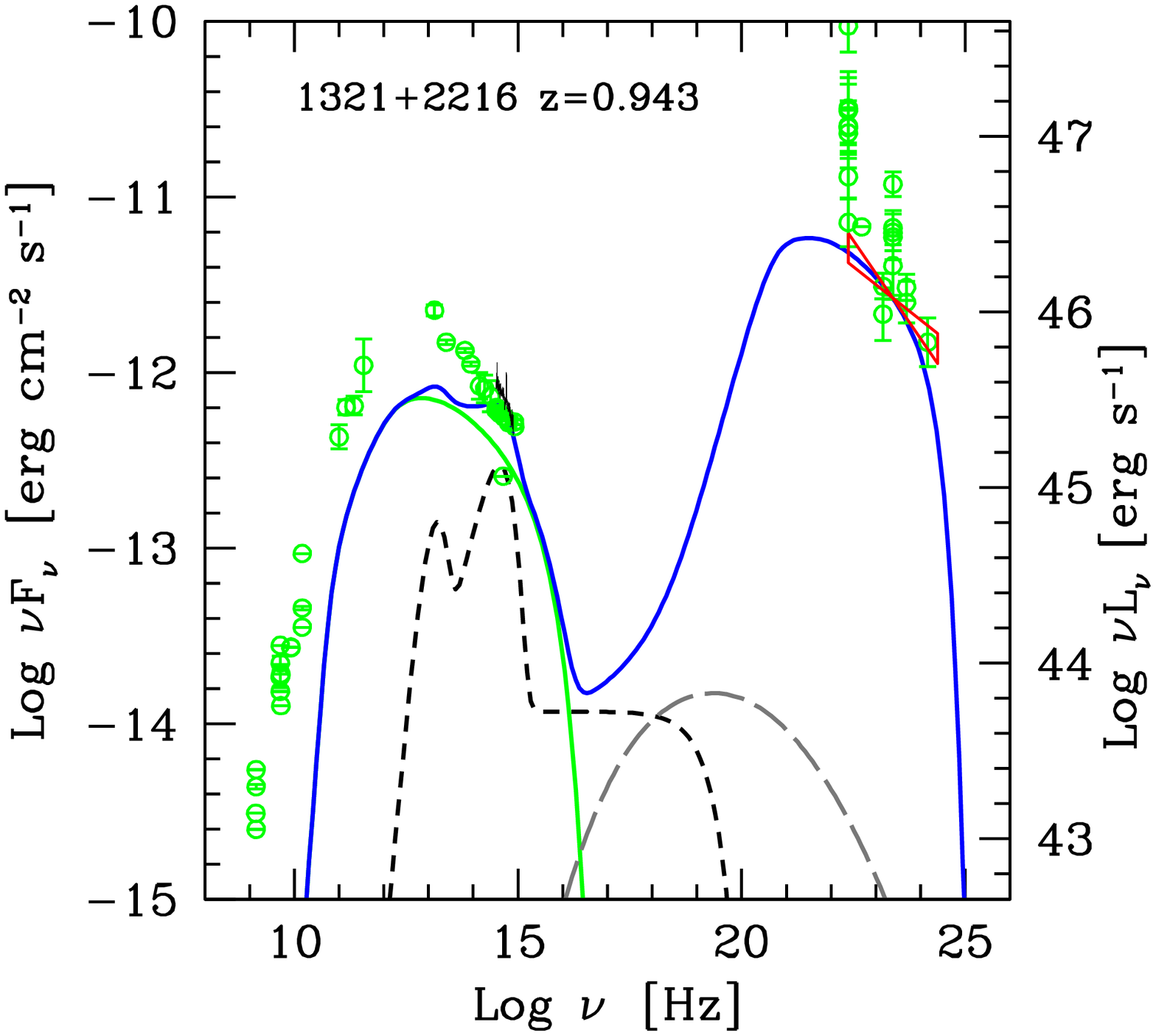,width=4.3cm,height=3.7cm } \\
\psfig{file=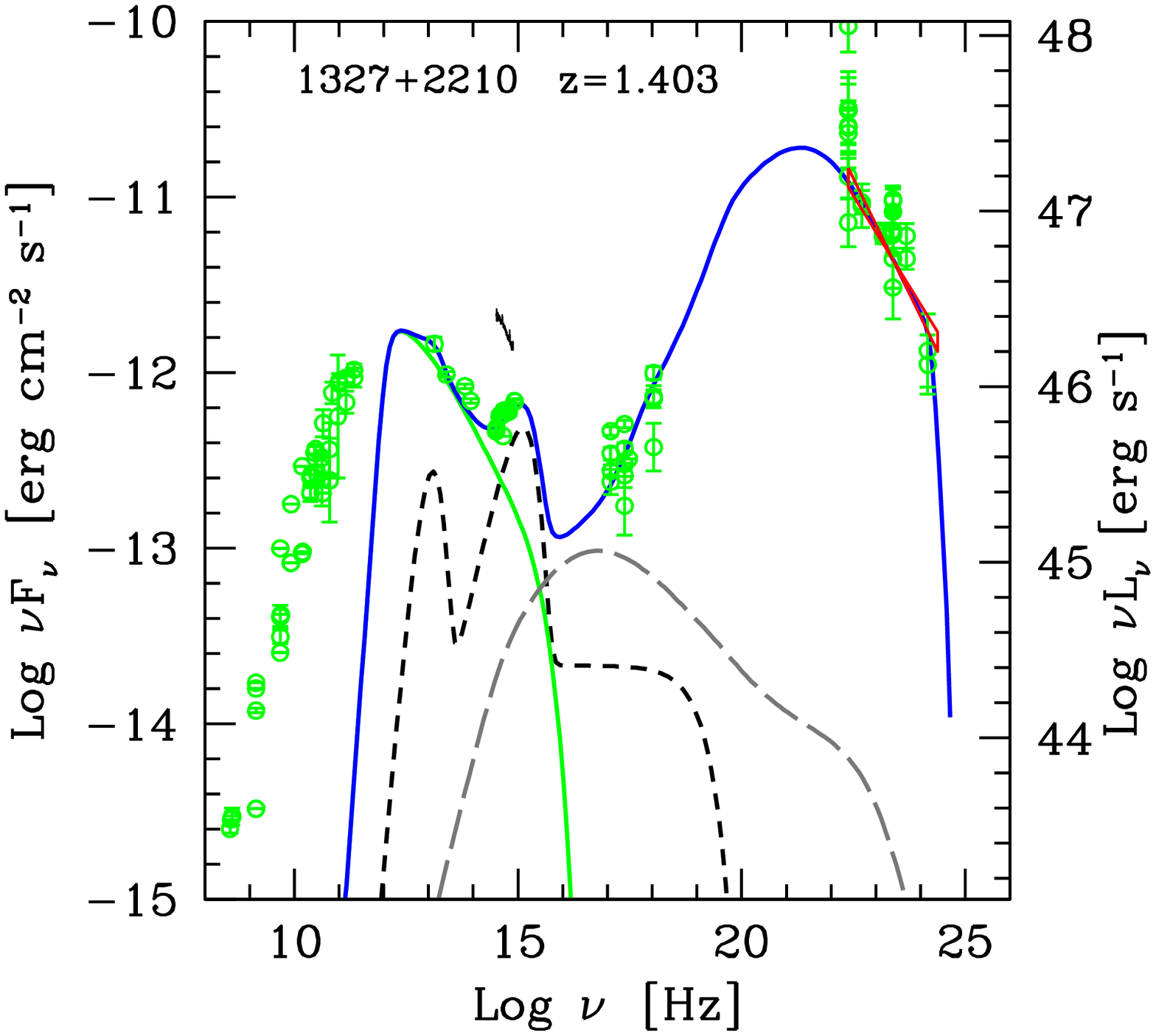,width=4.3cm,height=3.7cm } 
&\psfig{file=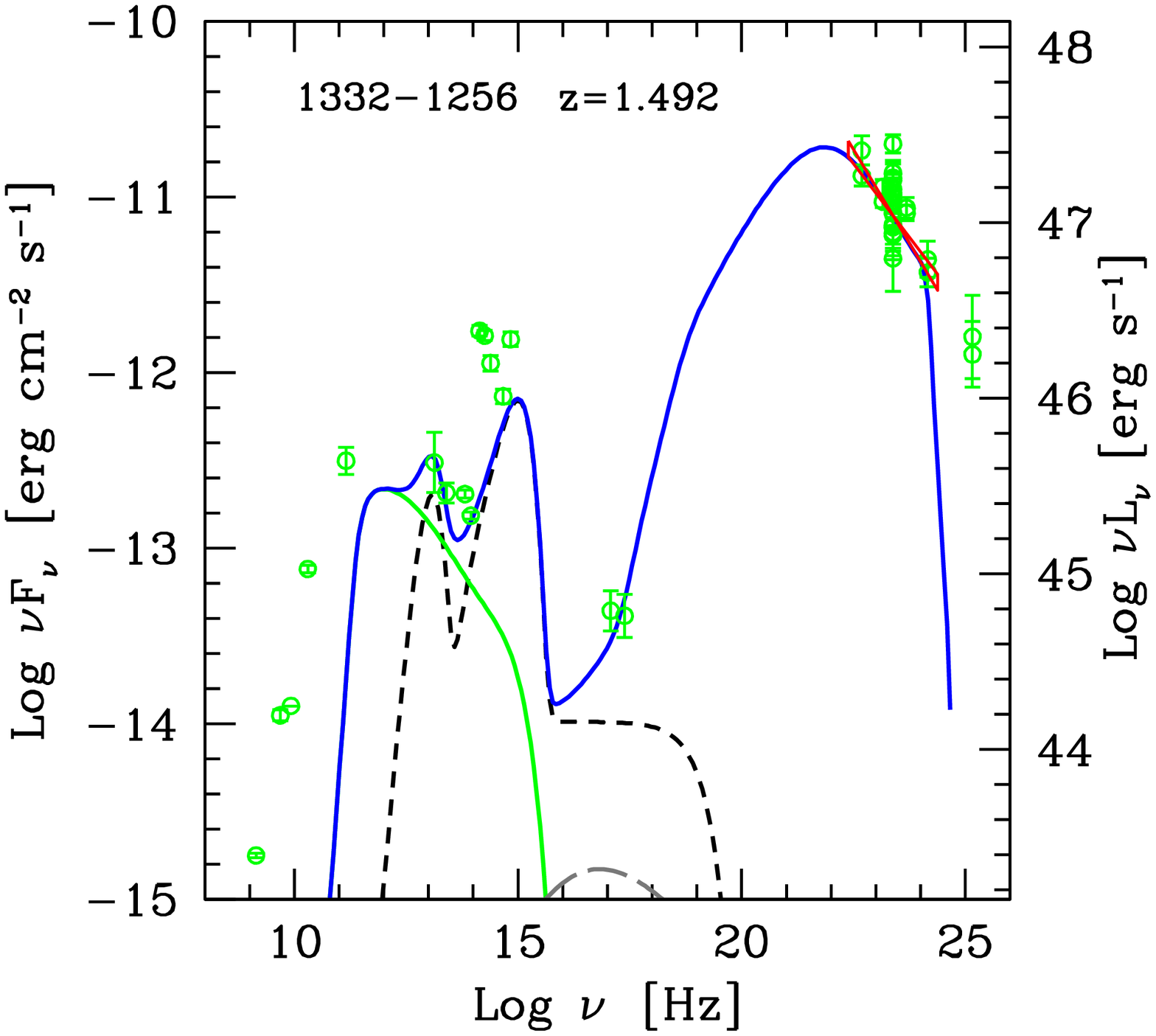,width=4.3cm,height=3.7cm } 
&\psfig{file=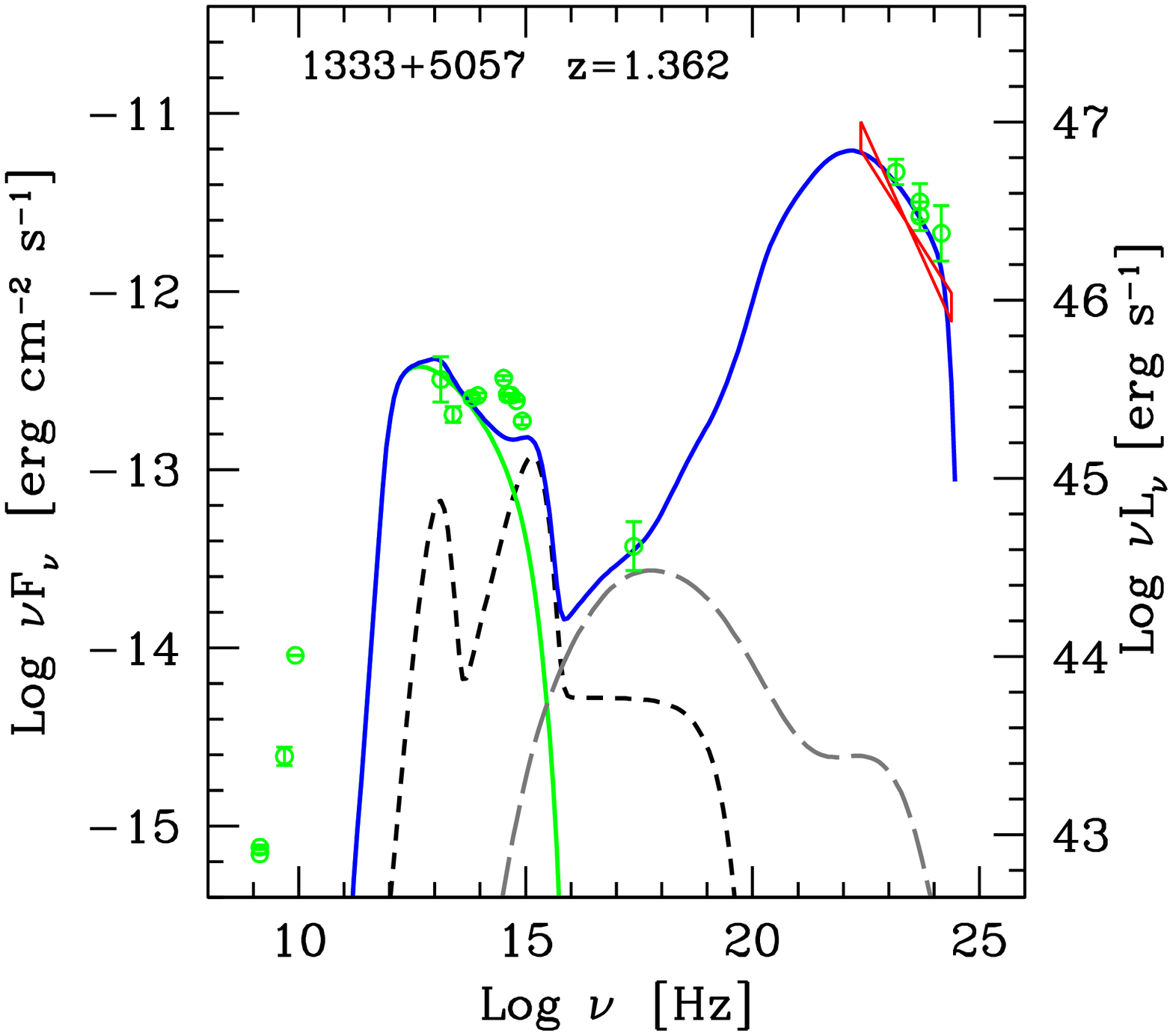,width=4.3cm,height=3.7cm }  
&\psfig{file=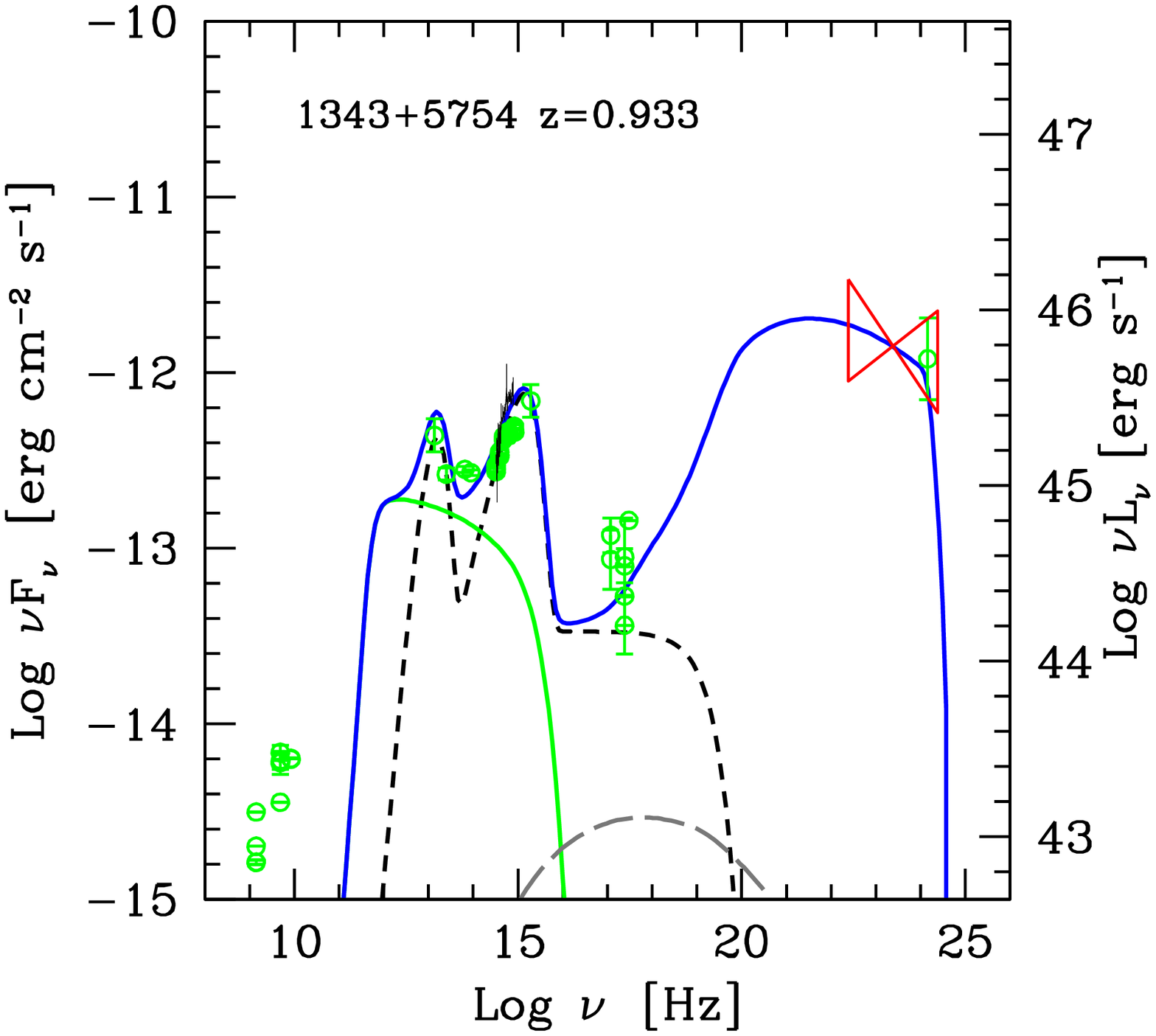,width=4.3cm,height=3.7cm } 
\end{tabular}
\caption{{\it continue.} SED of the FSRQs studied in this paper.}
\end{figure*} 

\setcounter{figure}{15}
\begin{figure*}
\begin{tabular}{cccc}
\psfig{file=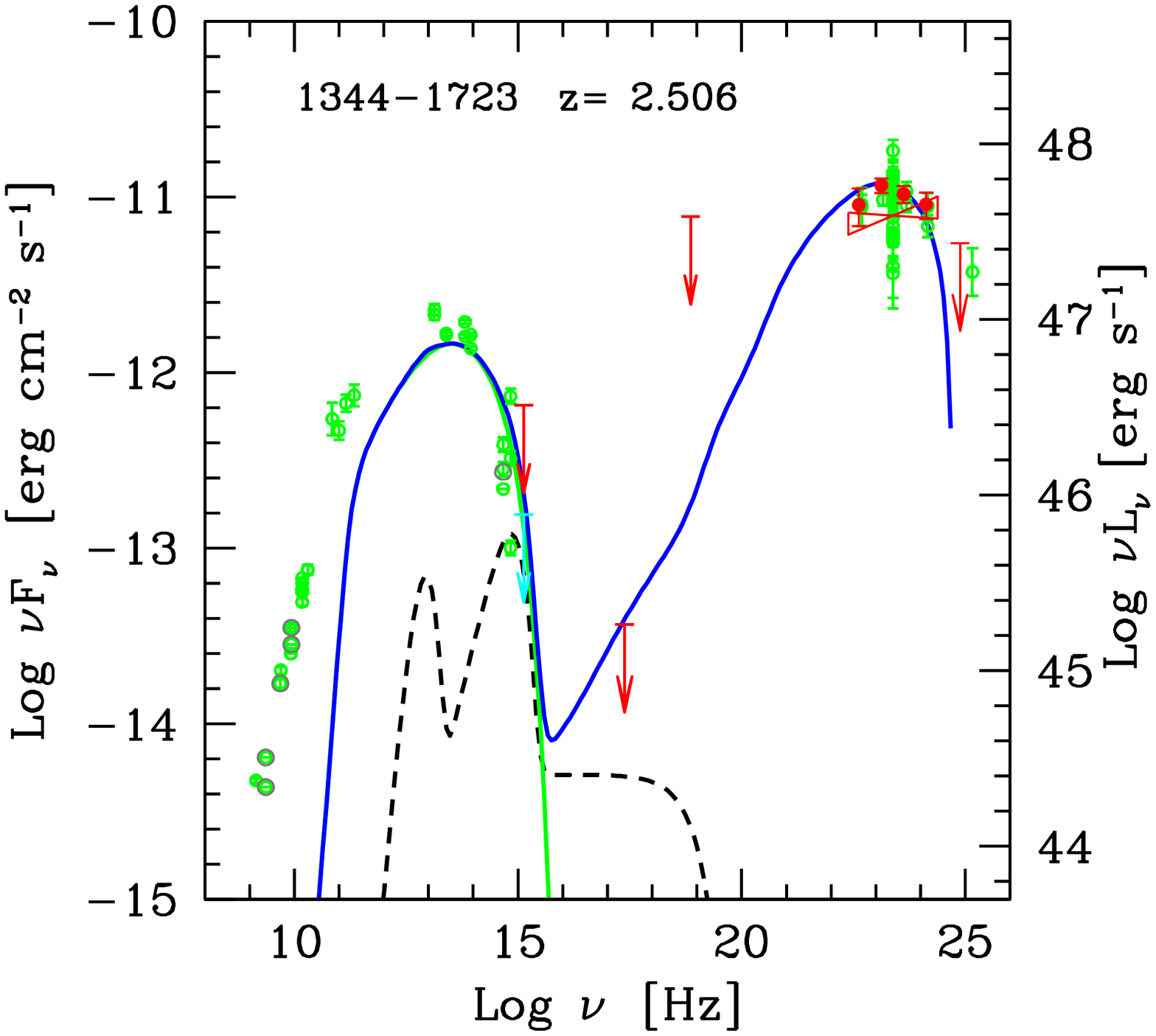,width=4.3cm,height=3.7cm }  
&\psfig{file=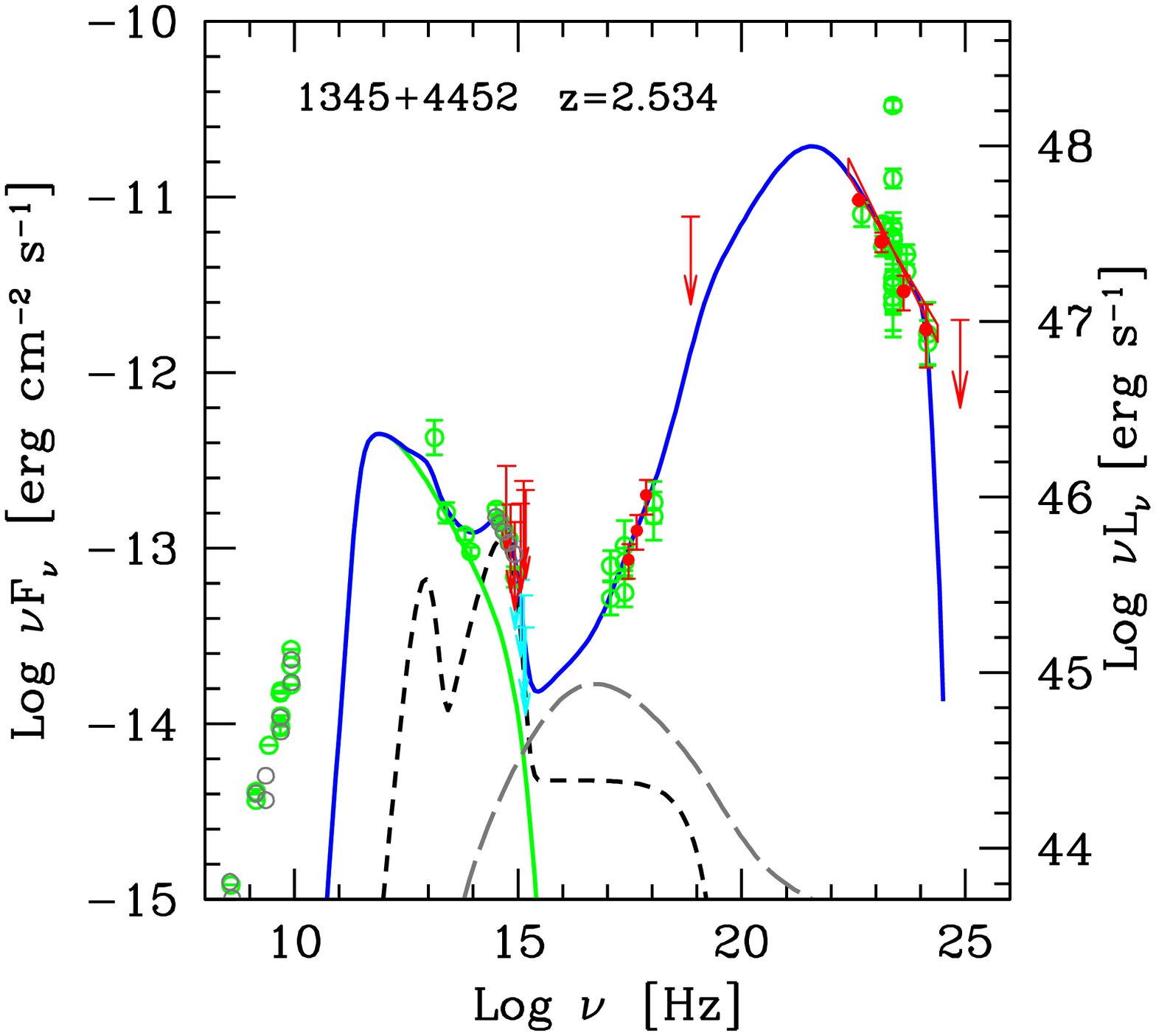,width=4.3cm,height=3.7cm } 
&\psfig{file=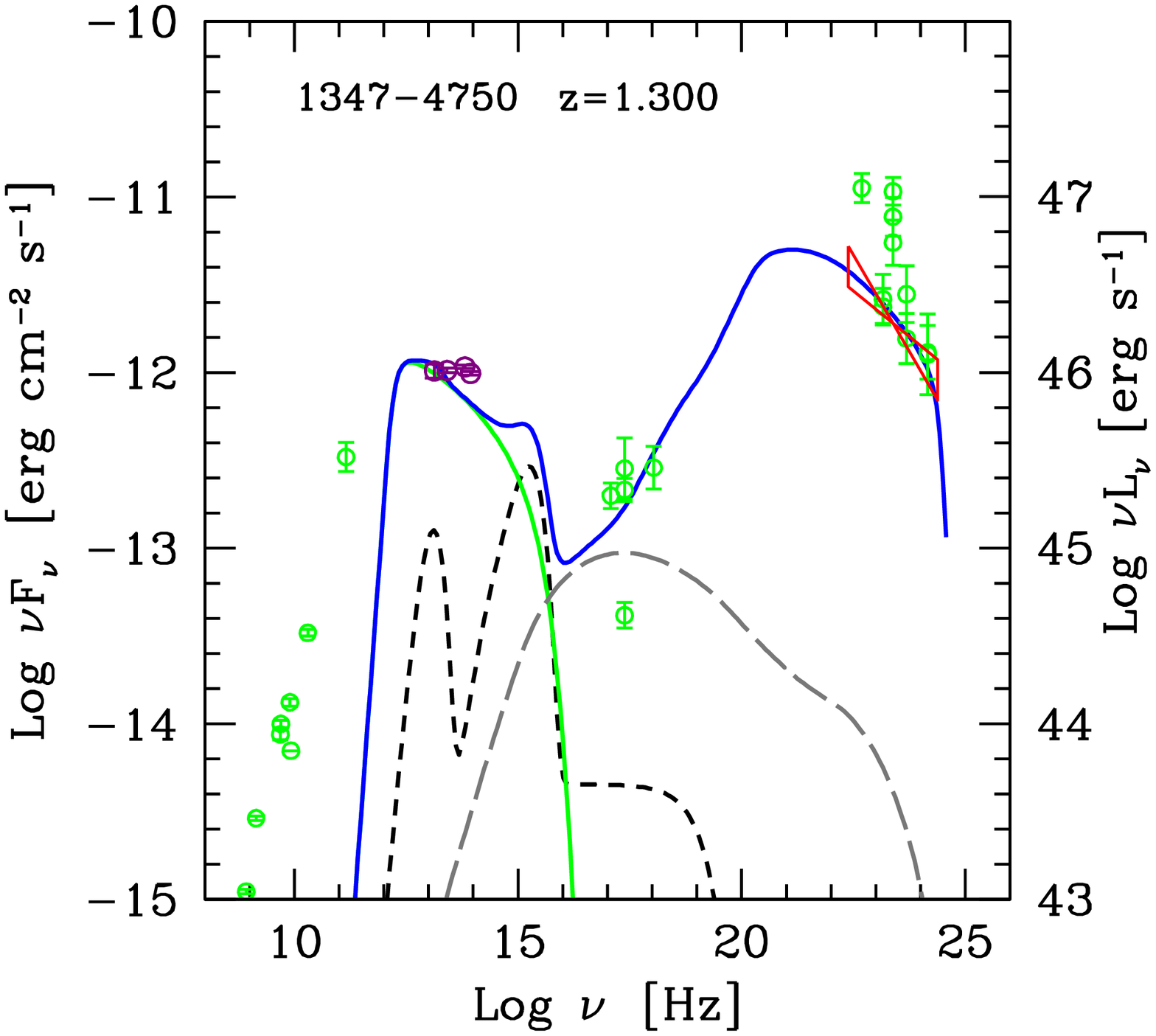,width=4.3cm,height=3.7cm }  
&\psfig{file=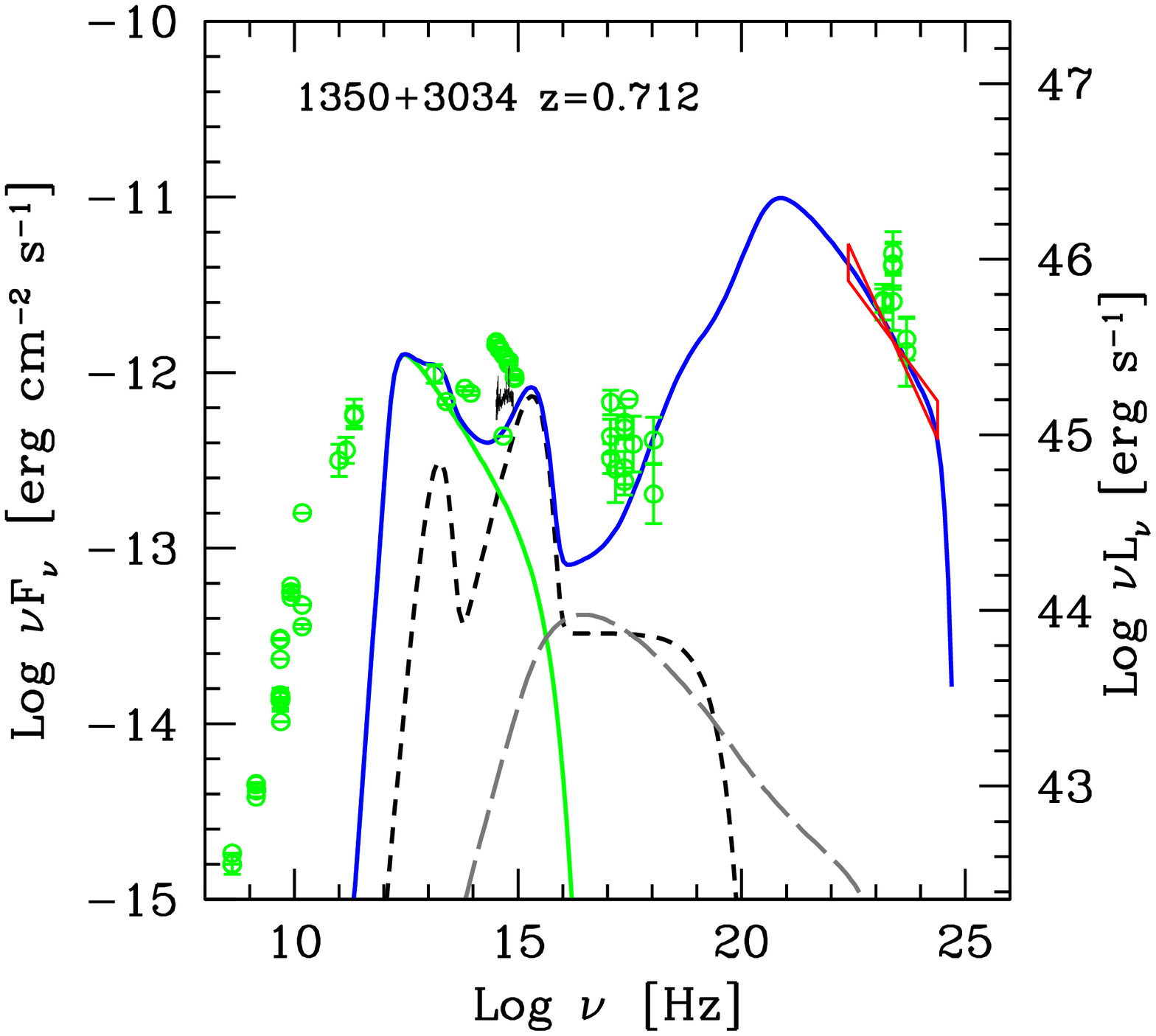,width=4.3cm,height=3.7cm } \\
\psfig{file=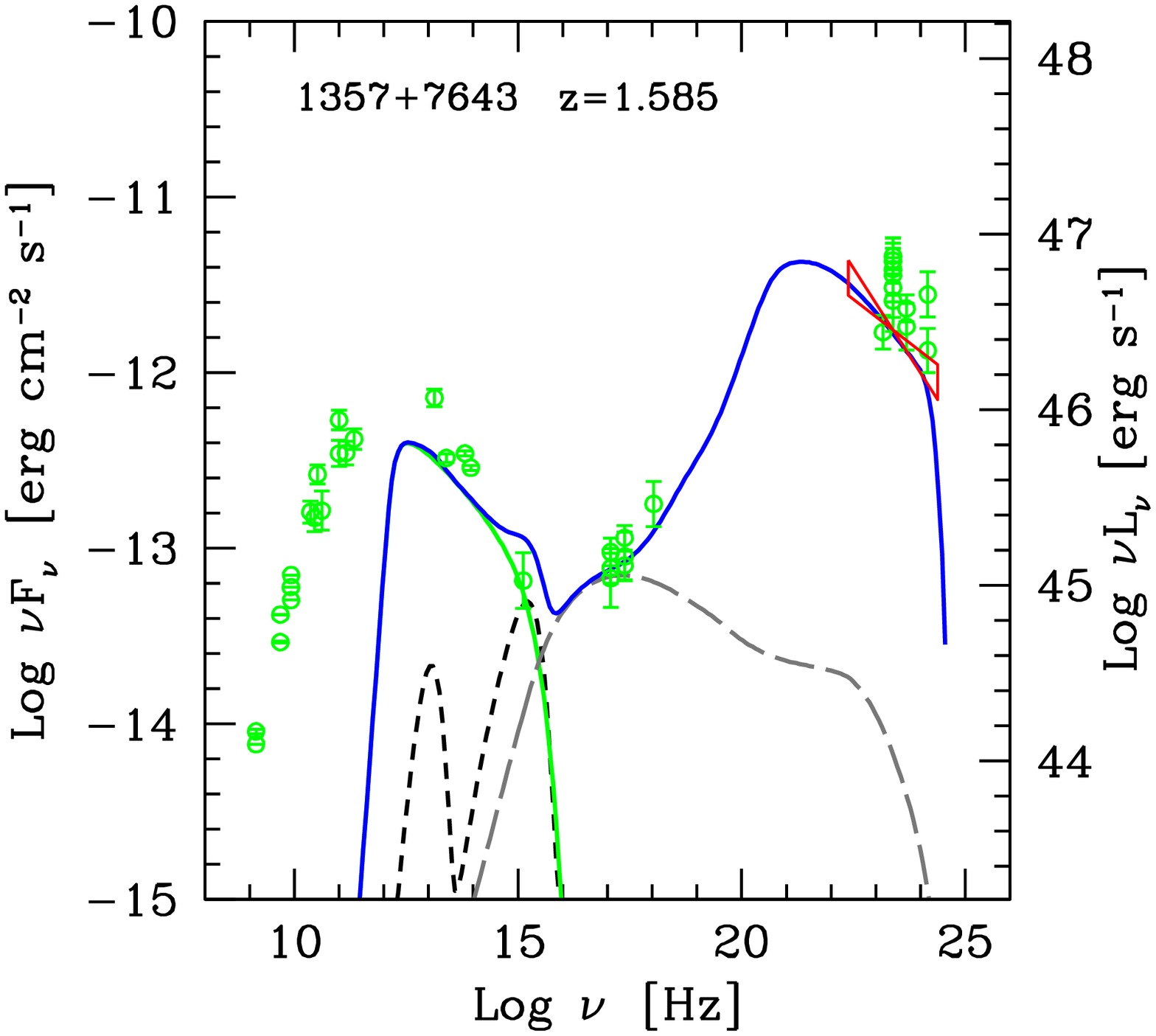,width=4.3cm,height=3.7cm } 
&\psfig{file=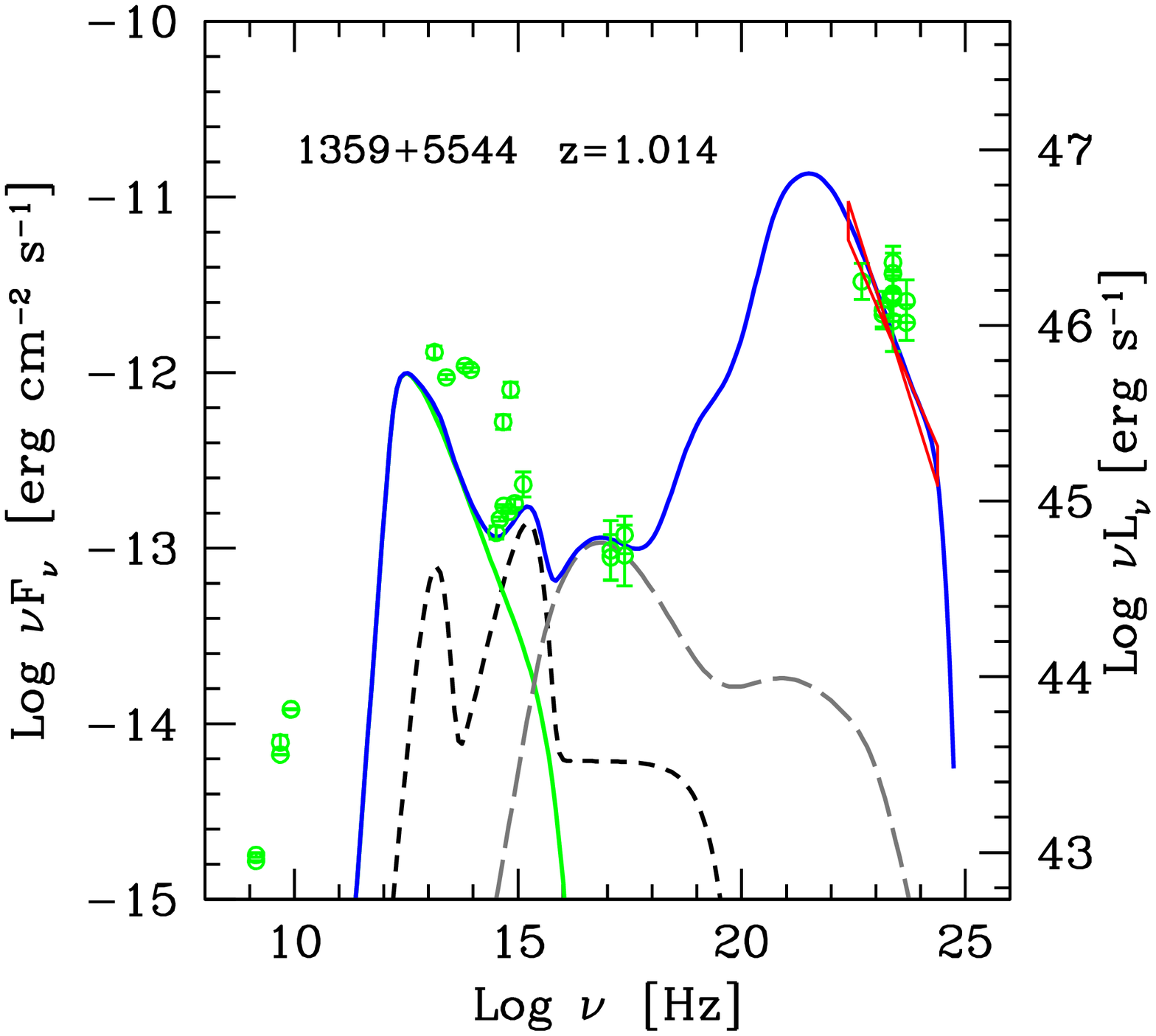,width=4.3cm,height=3.7cm } 
&\psfig{file=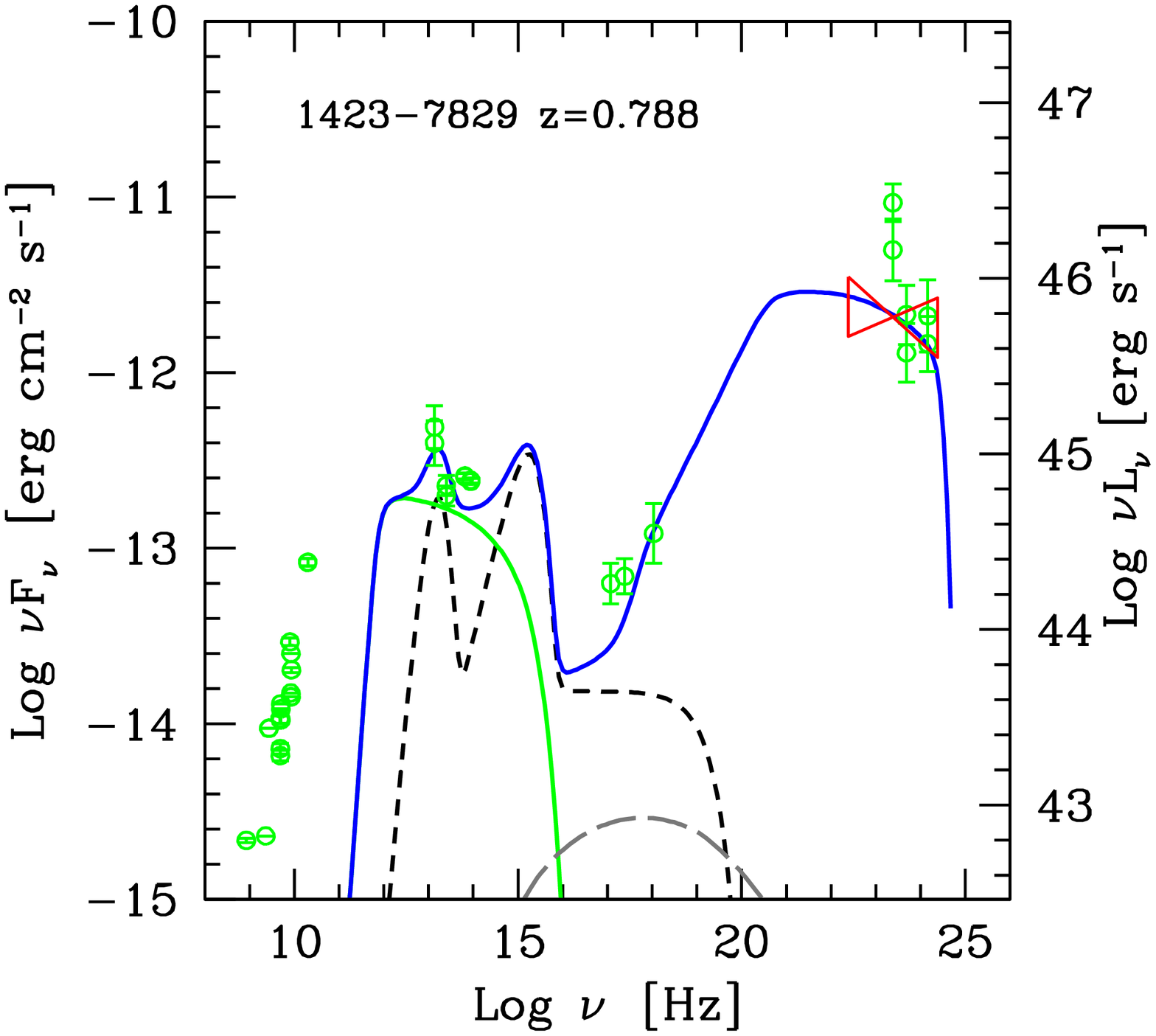,width=4.3cm,height=3.7cm } 
&\psfig{file=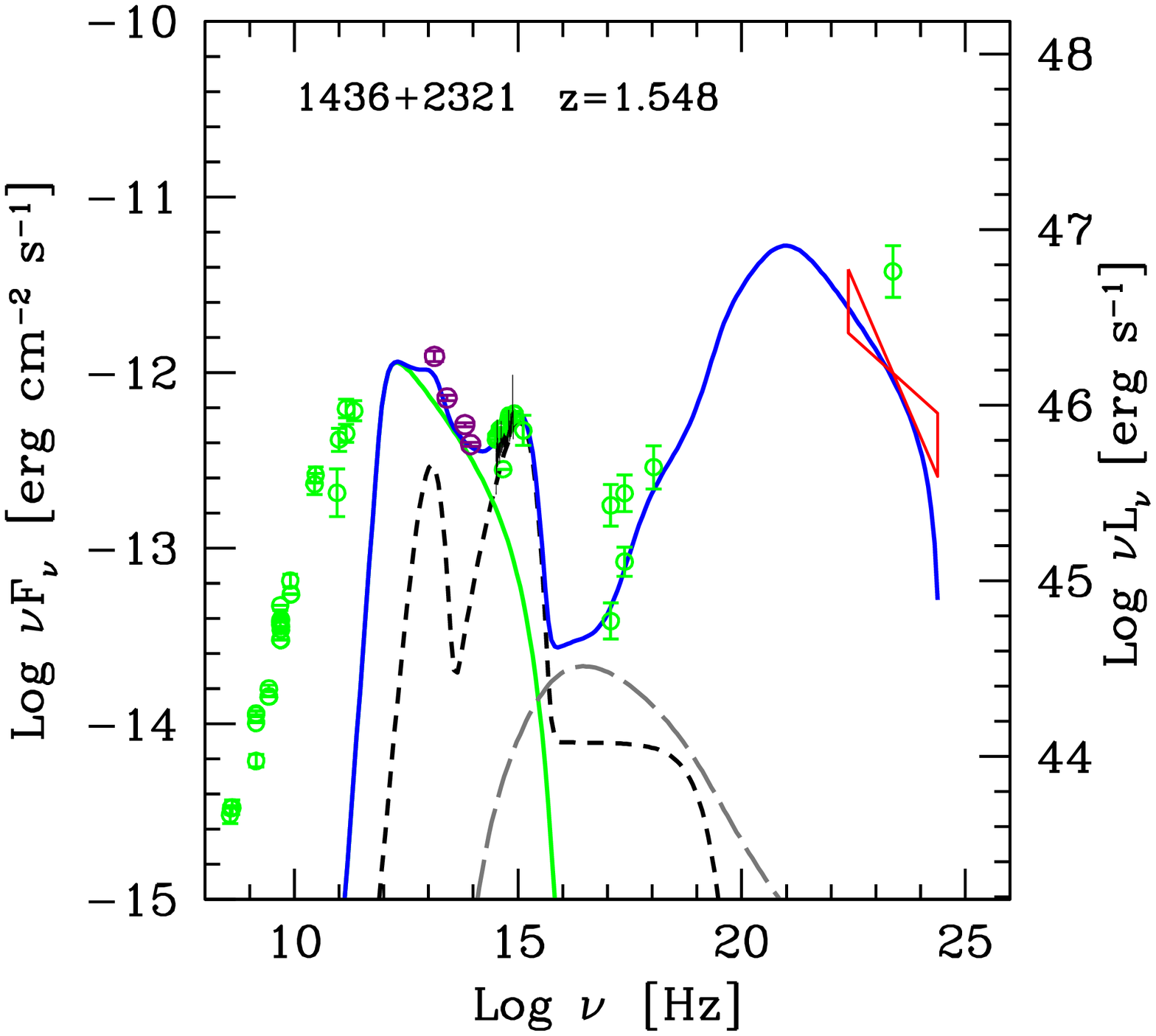,width=4.3cm,height=3.7cm }\\ 
\psfig{file=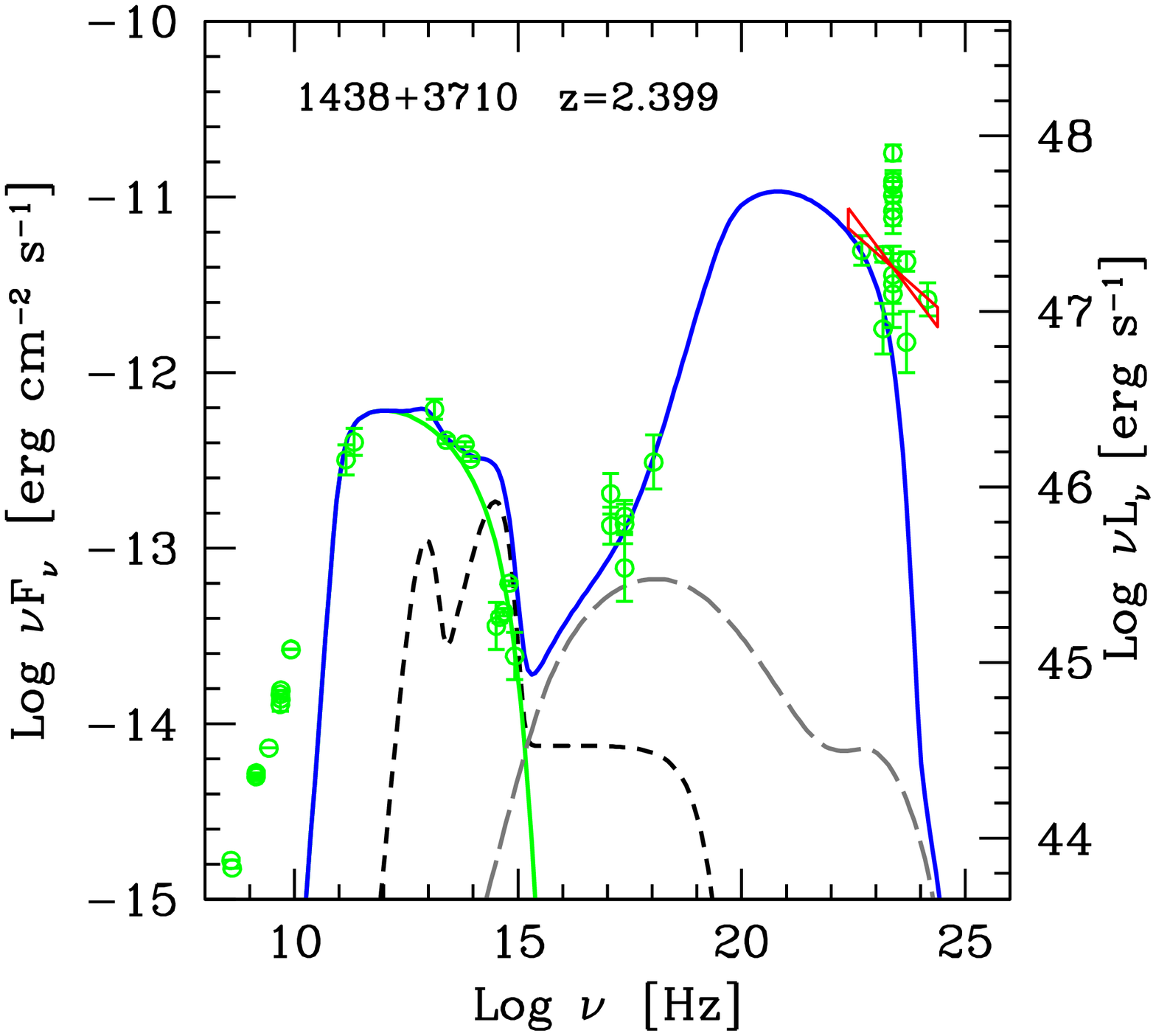,width=4.3cm,height=3.7cm } 
&\psfig{file=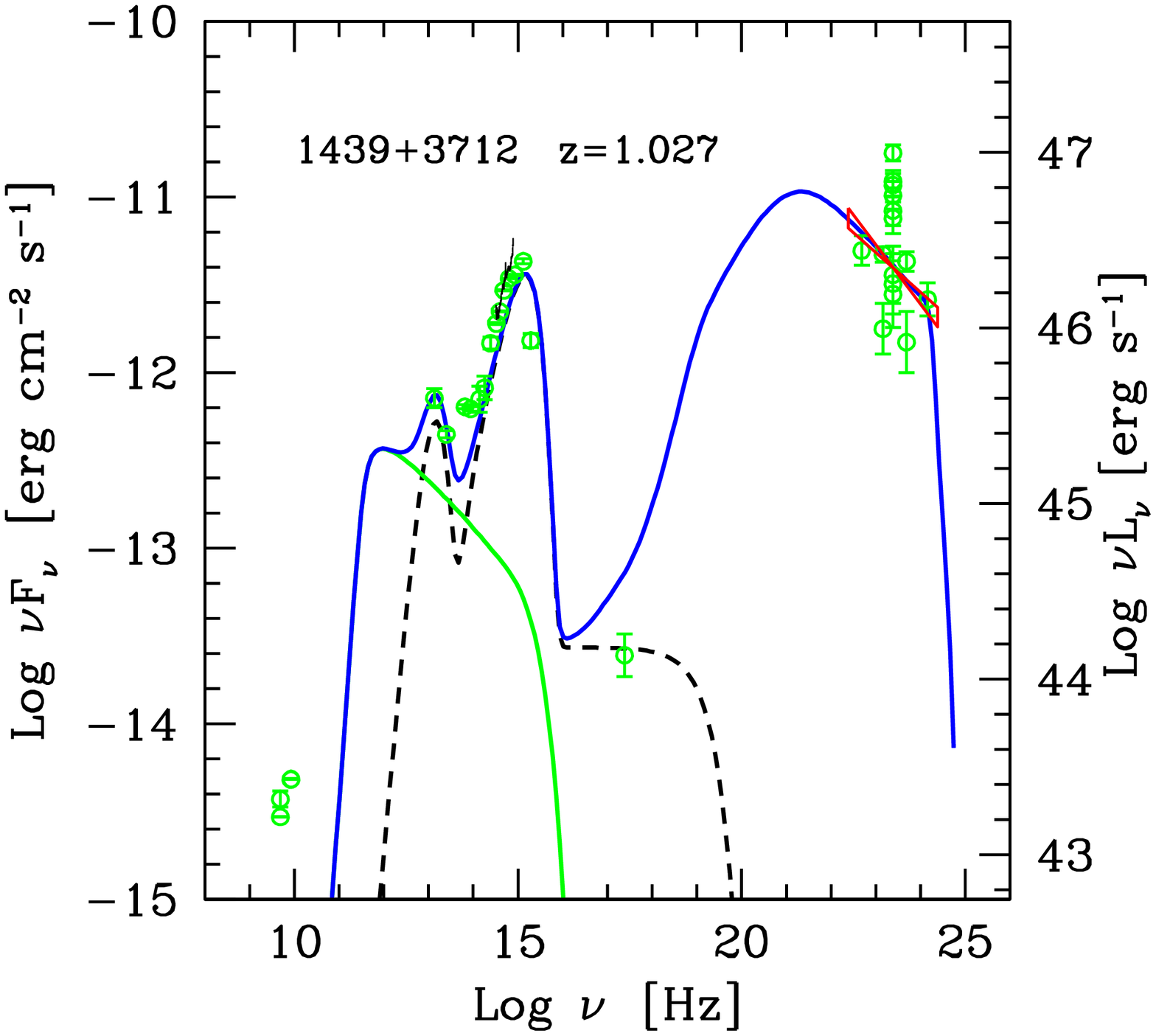,width=4.3cm,height=3.7cm } 
&\psfig{file=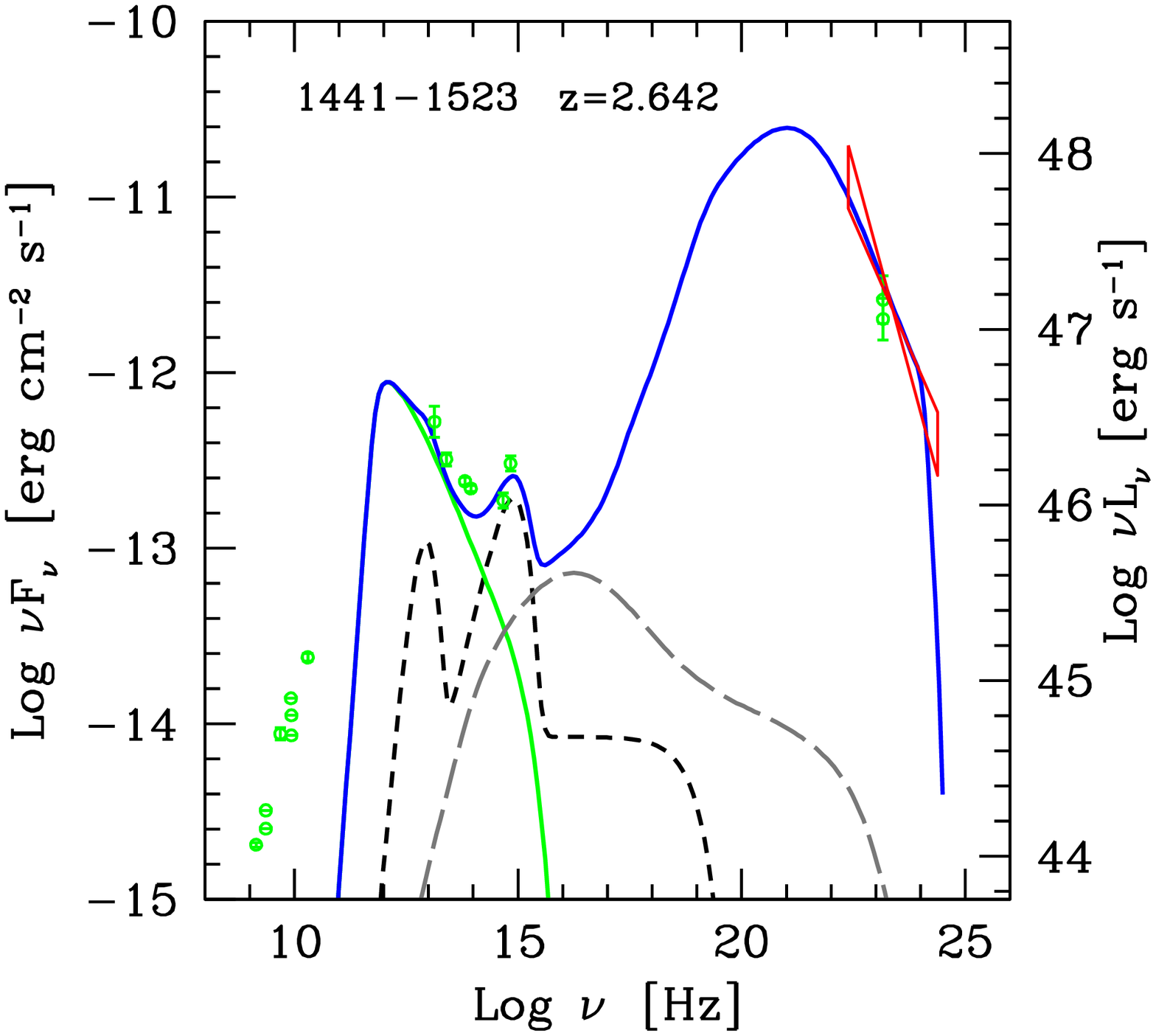,width=4.3cm,height=3.7cm }  
&\psfig{file=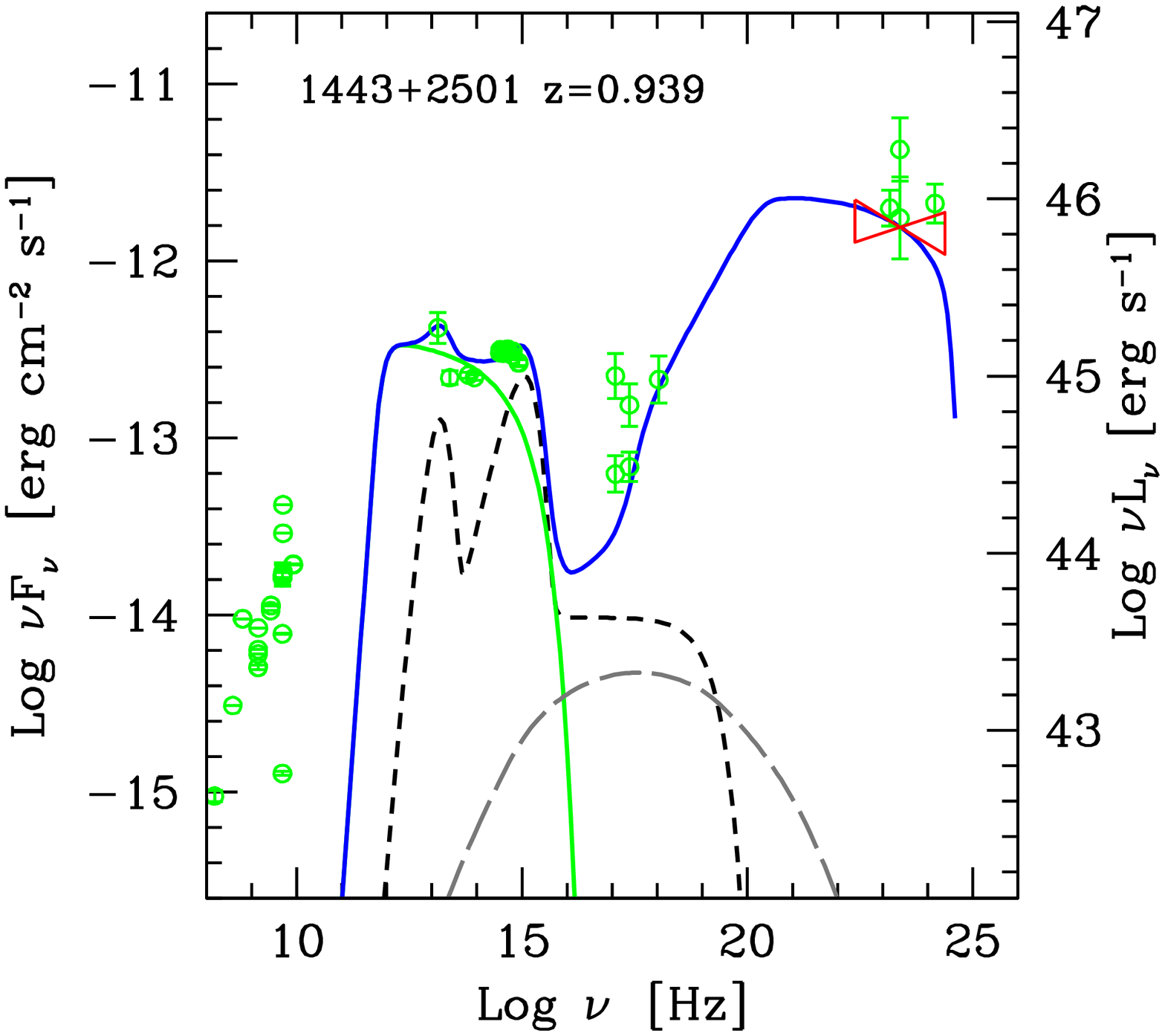,width=4.3cm,height=3.7cm }\\ 
\psfig{file=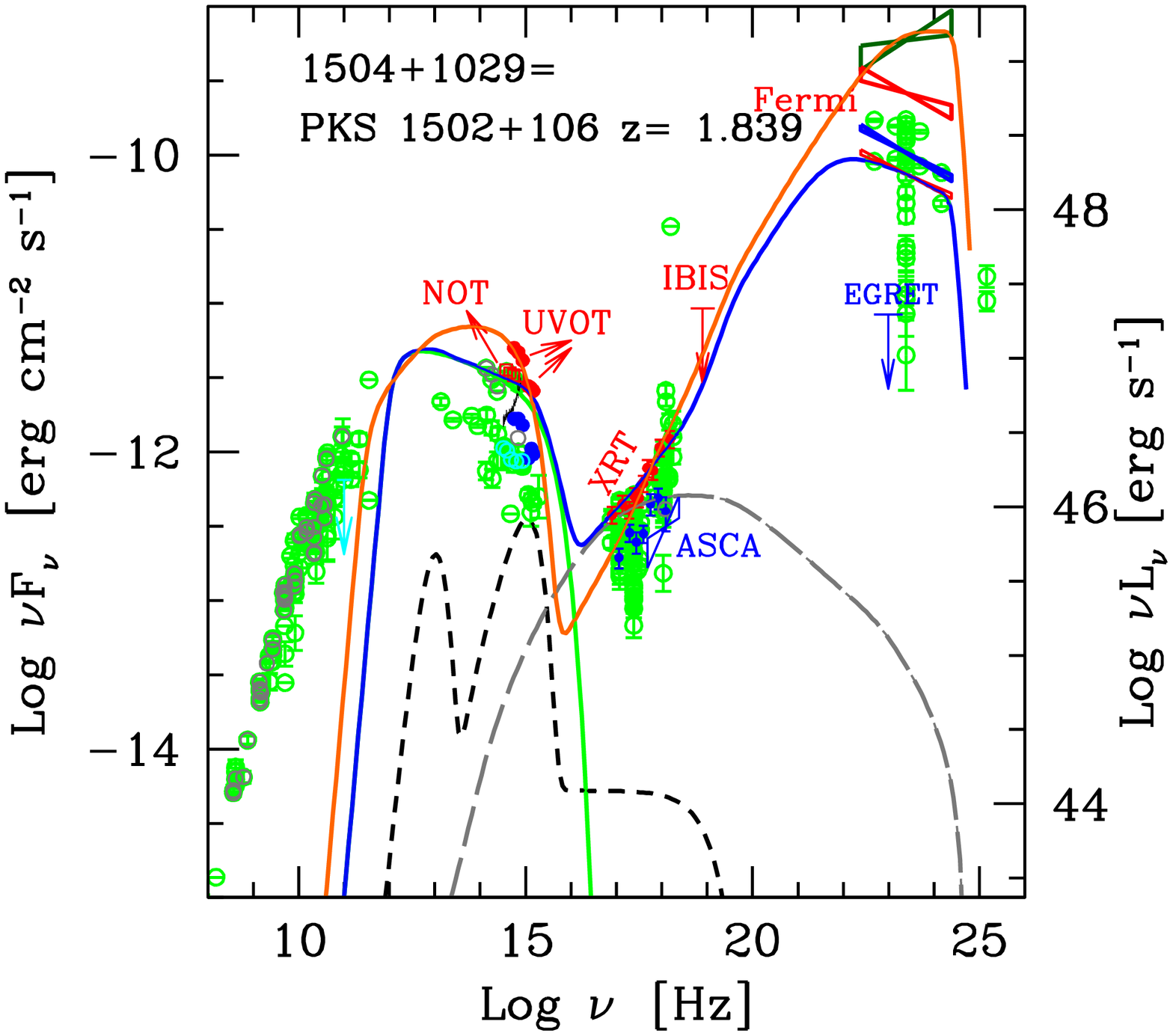,width=4.3cm,height=3.7cm } 
&\psfig{file=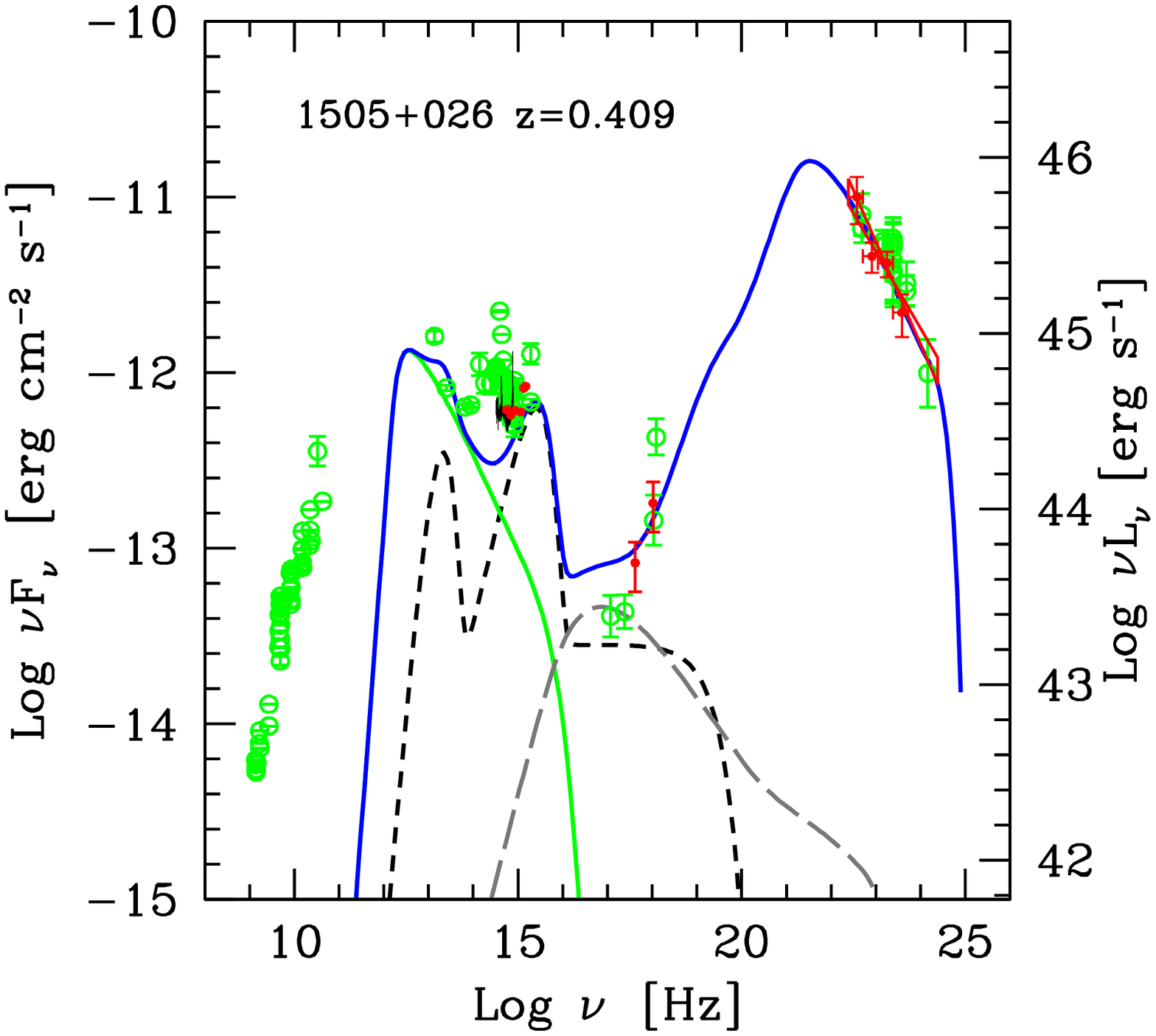,width=4.3cm,height=3.7cm } 
&\psfig{file=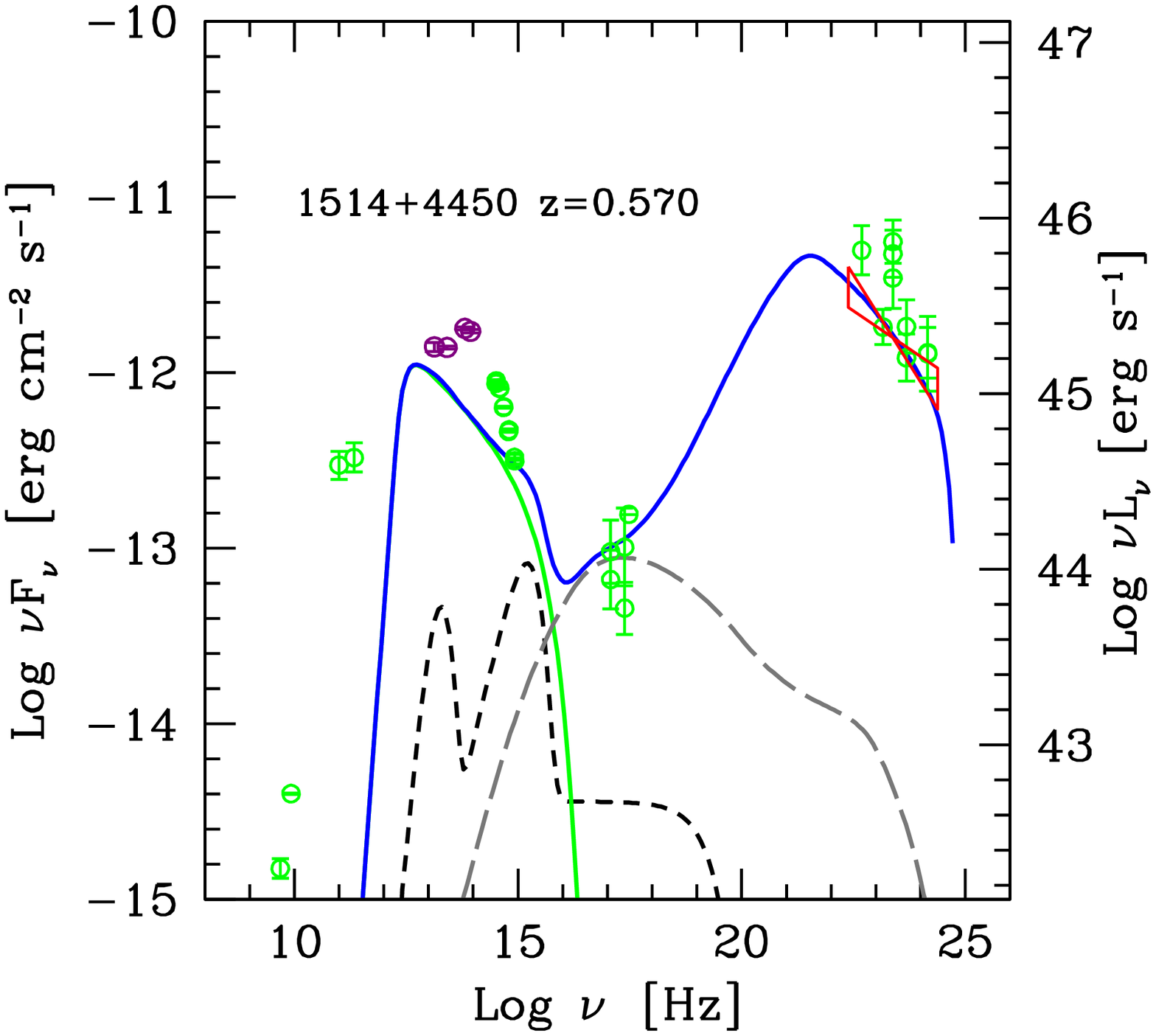,width=4.3cm,height=3.7cm } 
&\psfig{file=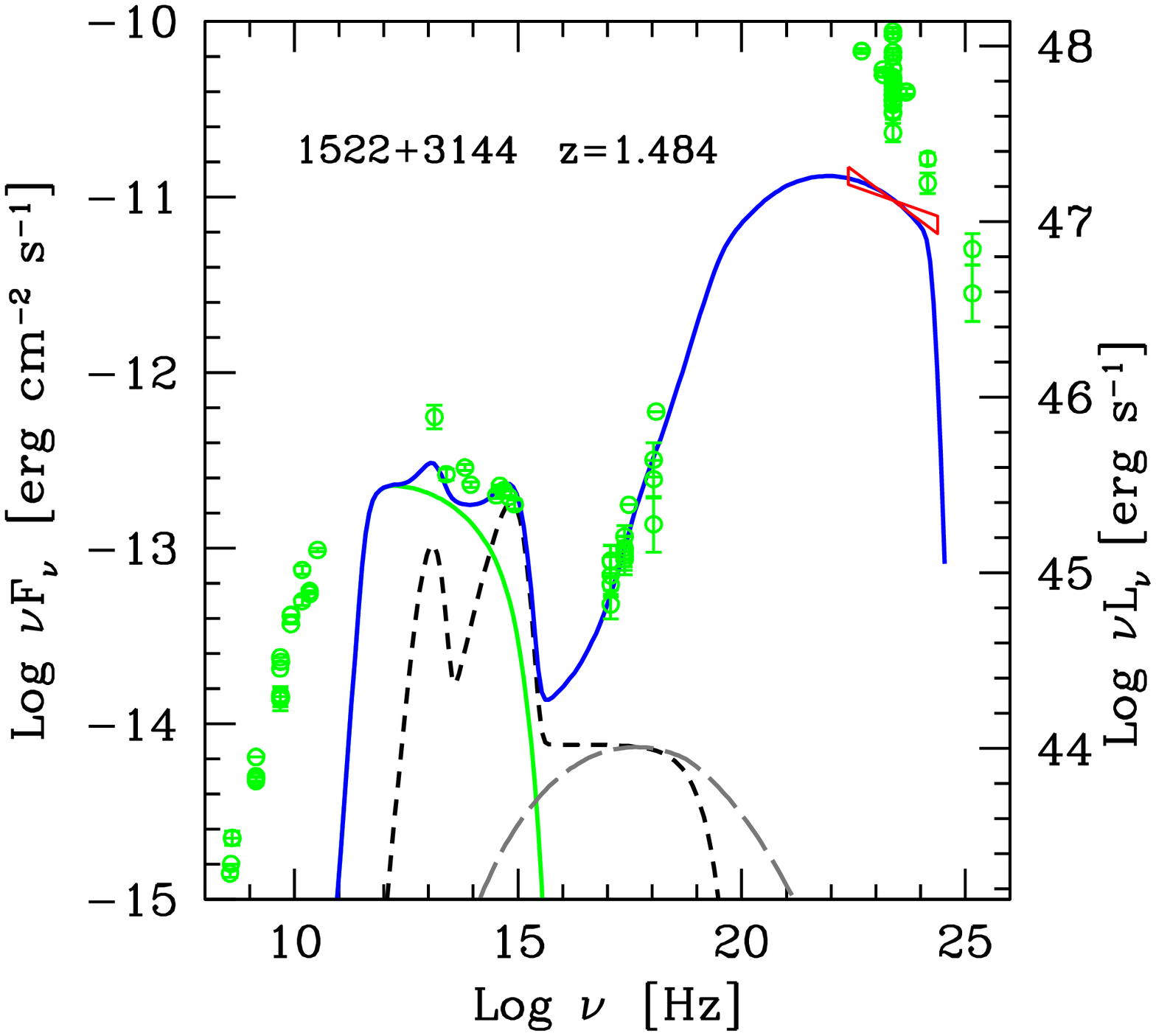,width=4.3cm,height=3.7cm }\\ 
\psfig{file=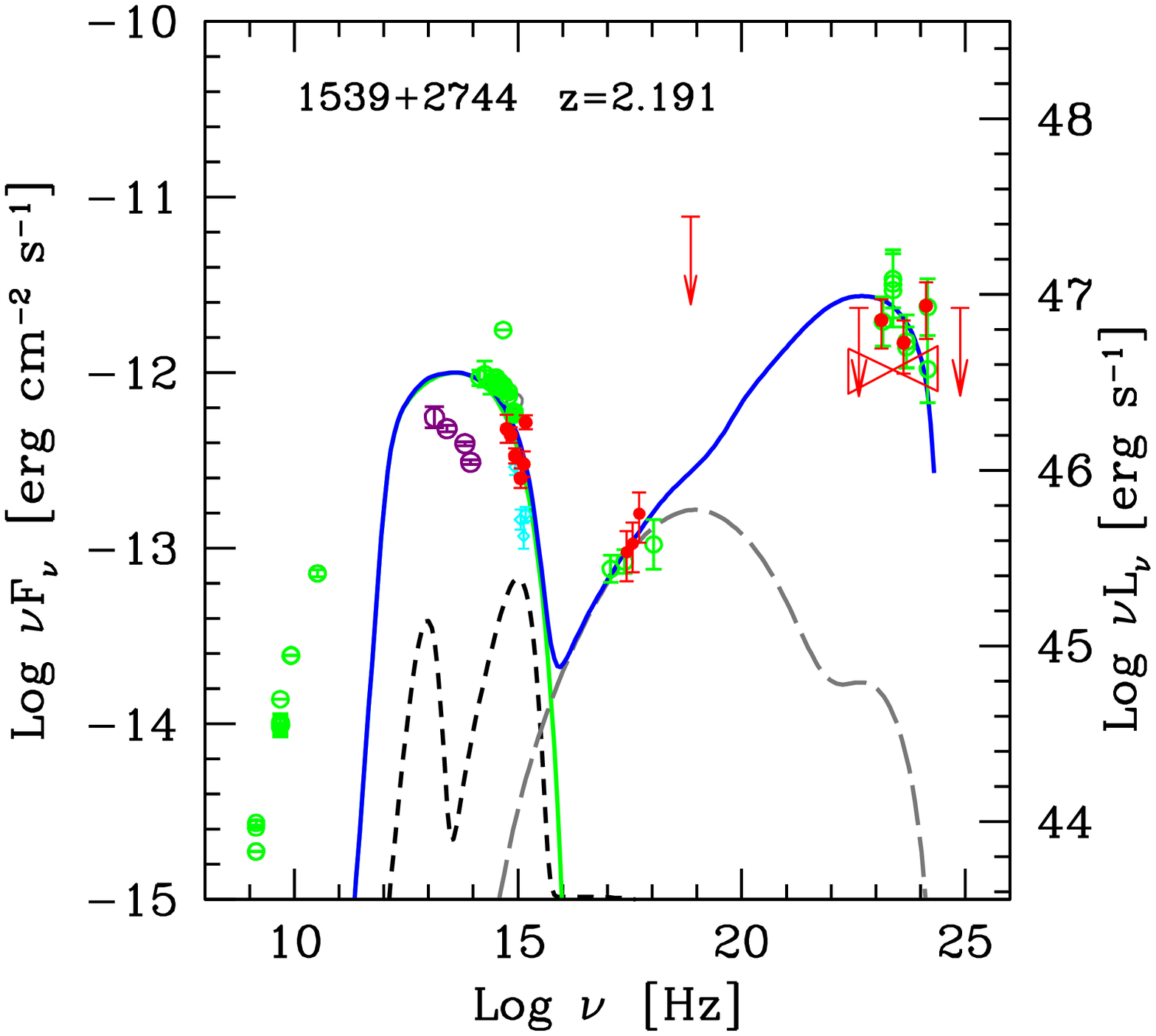,width=4.3cm,height=3.7cm } 
&\psfig{file=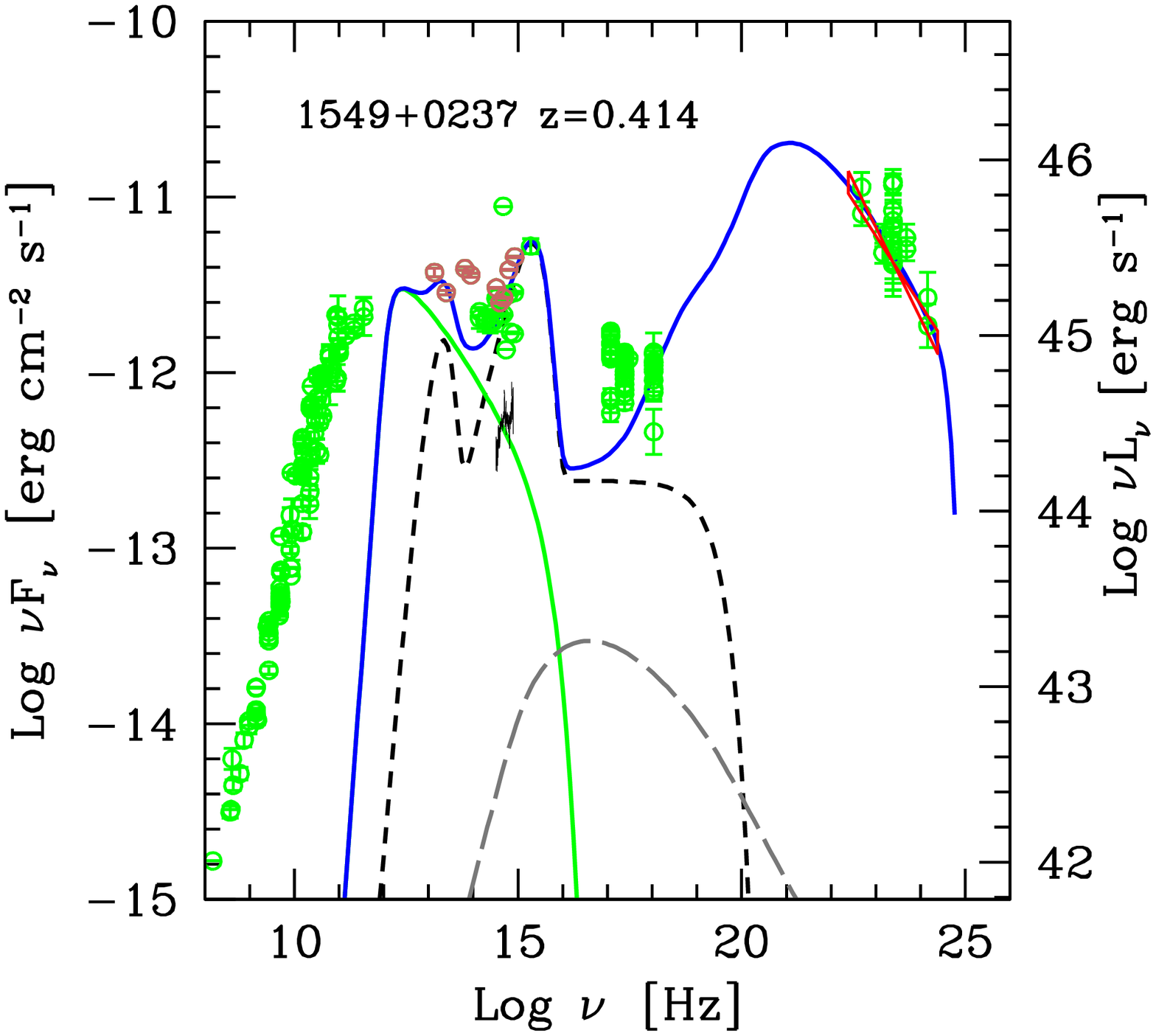,width=4.3cm,height=3.7cm } 
&\psfig{file=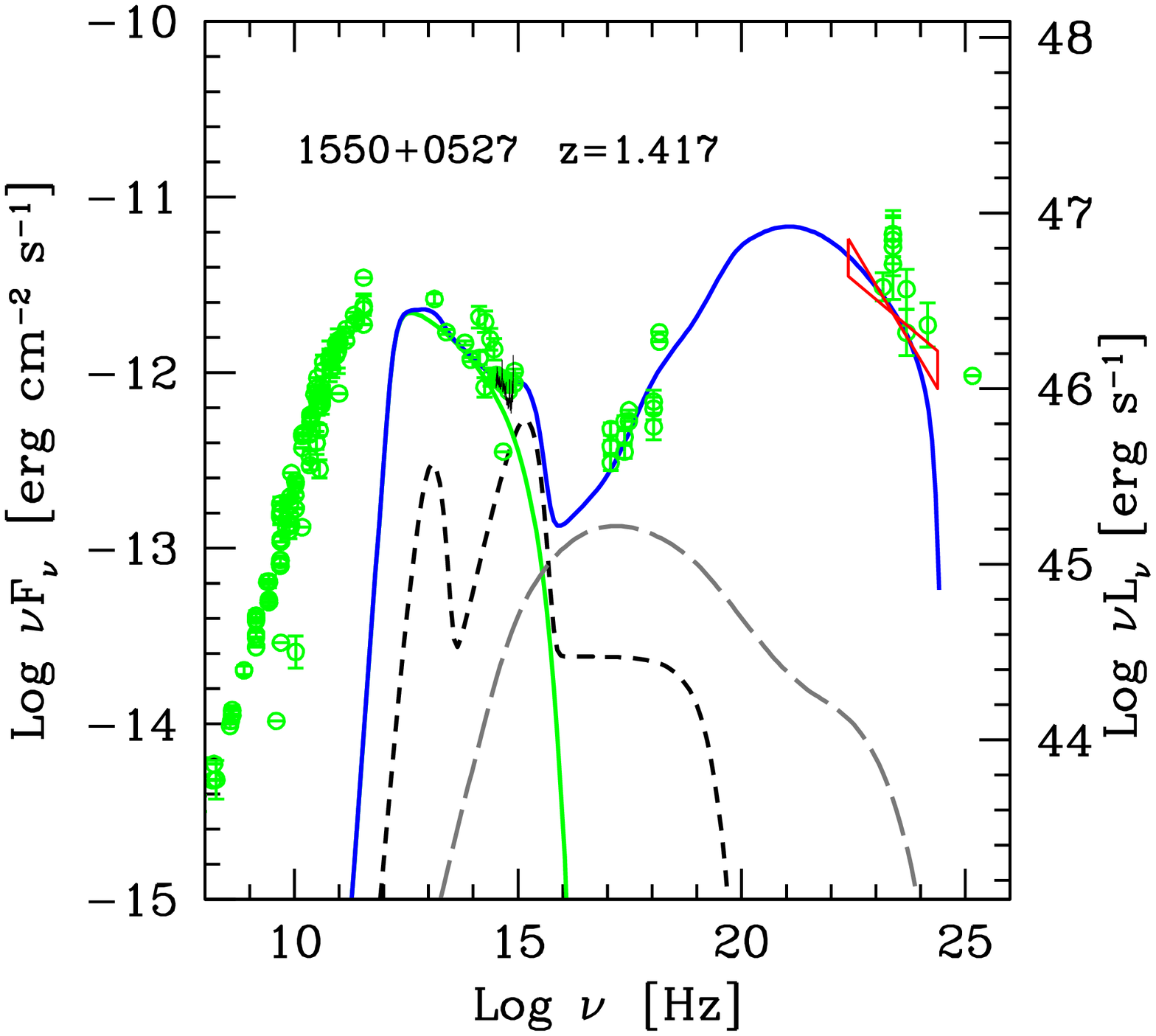,width=4.3cm,height=3.7cm }  
&\psfig{file=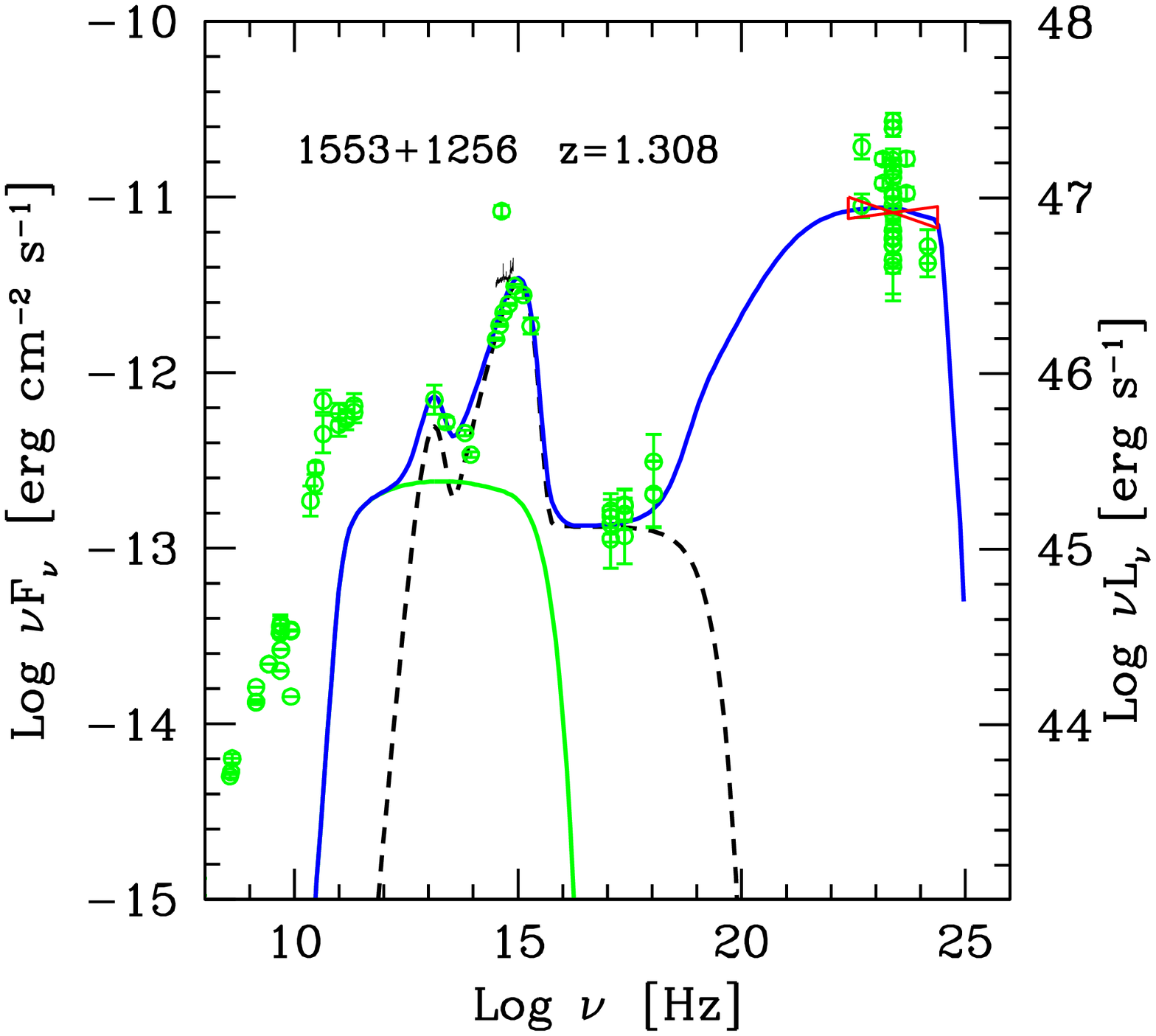,width=4.3cm,height=3.7cm }\\ 
\psfig{file=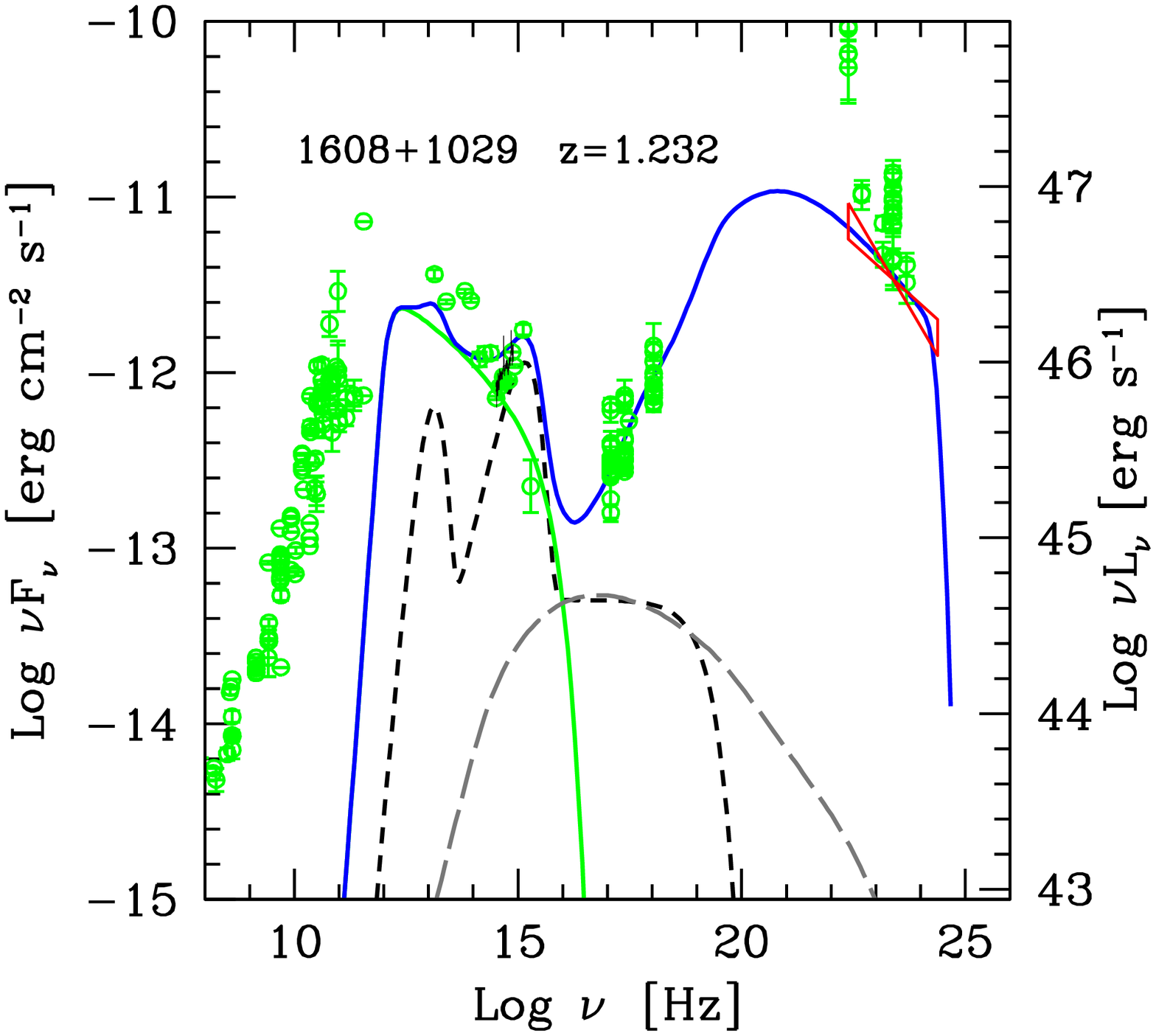,width=4.3cm,height=3.7cm } 
&\psfig{file=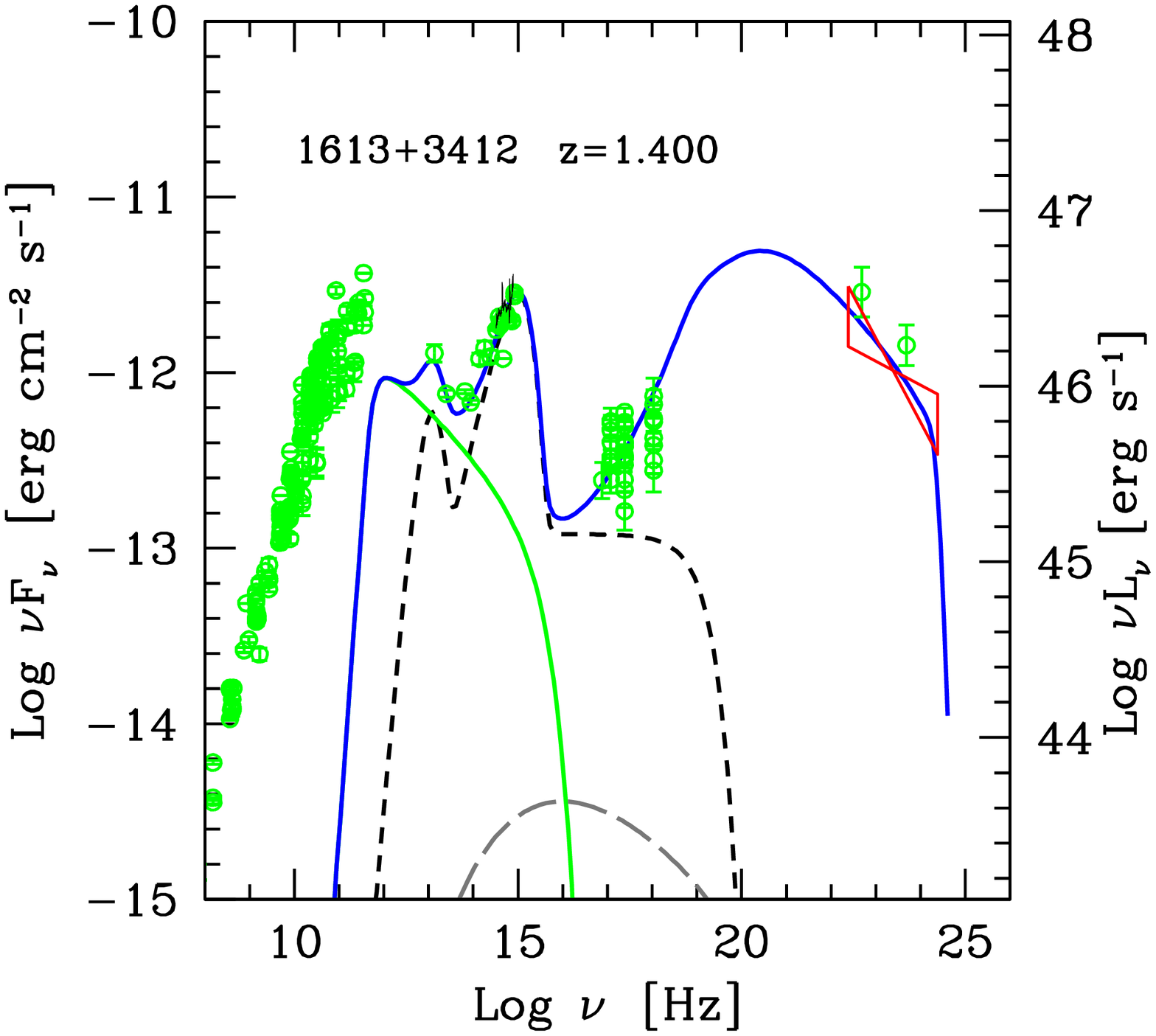,width=4.3cm,height=3.7cm } 
&\psfig{file=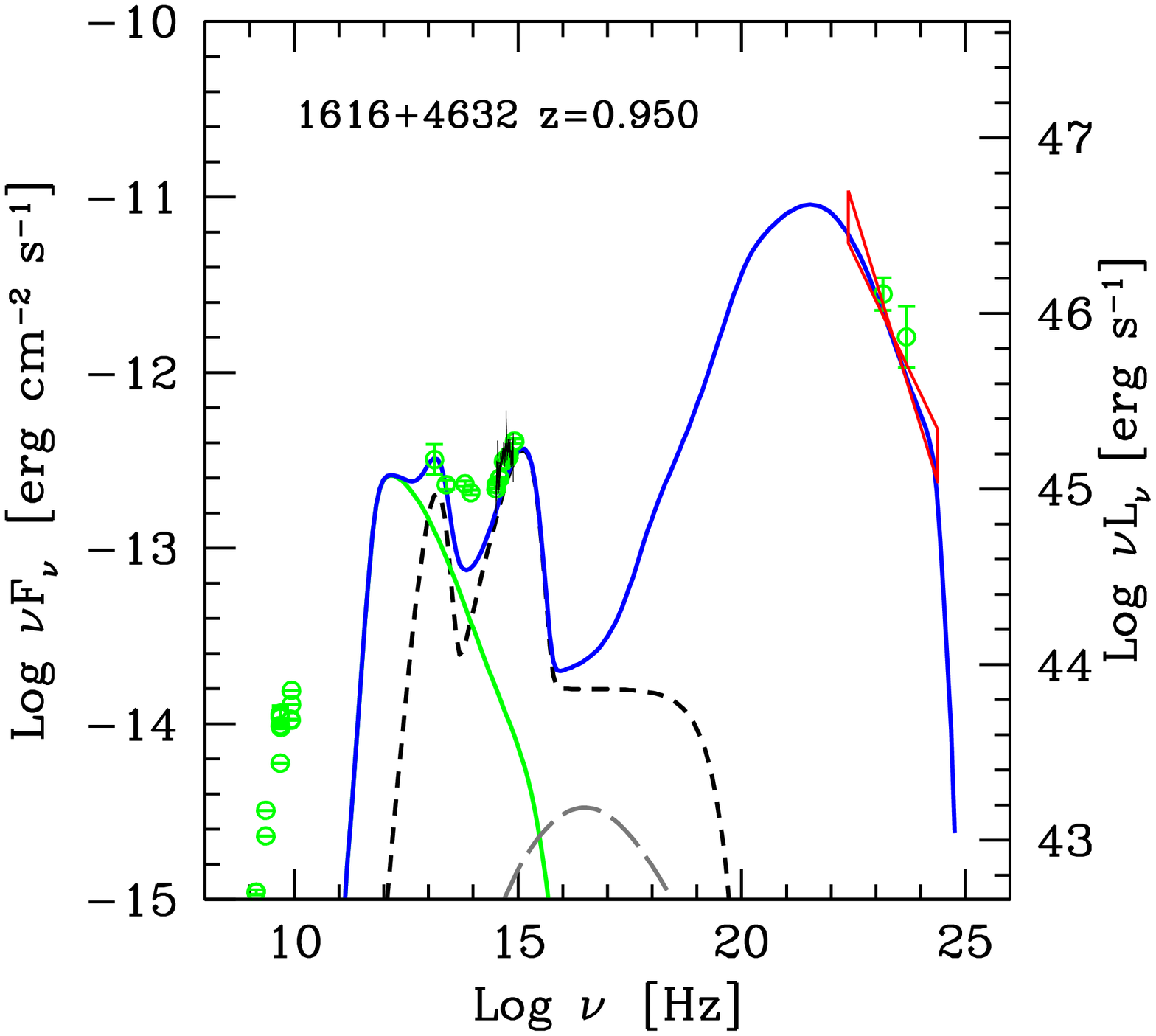,width=4.3cm,height=3.7cm }  
&\psfig{file=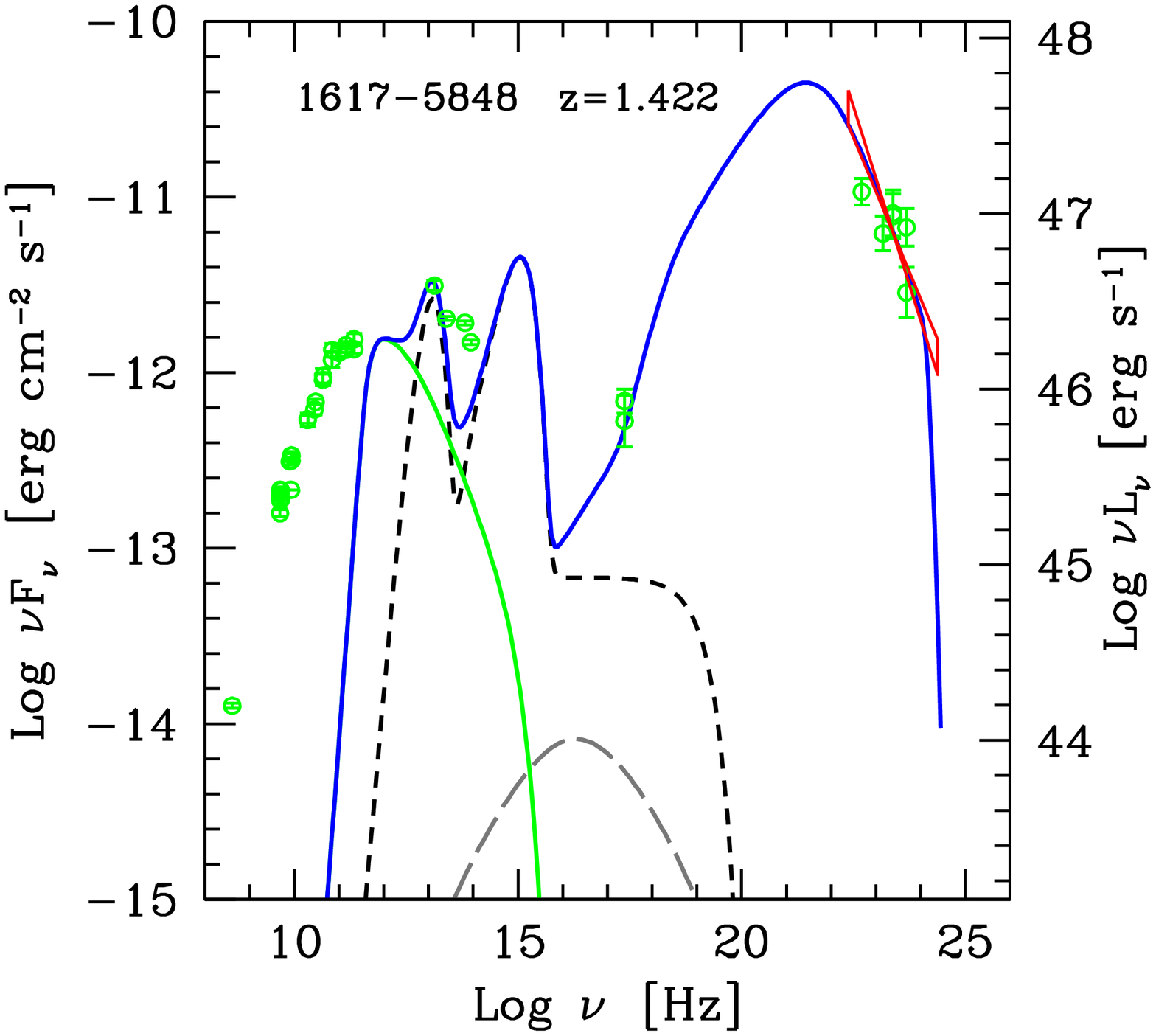,width=4.3cm,height=3.7cm } 
\end{tabular}
\caption{{\it continue.} SED of the FSRQs studied in this paper.}
\end{figure*} 

\setcounter{figure}{15}
\begin{figure*}
\begin{tabular}{cccc}
\psfig{file=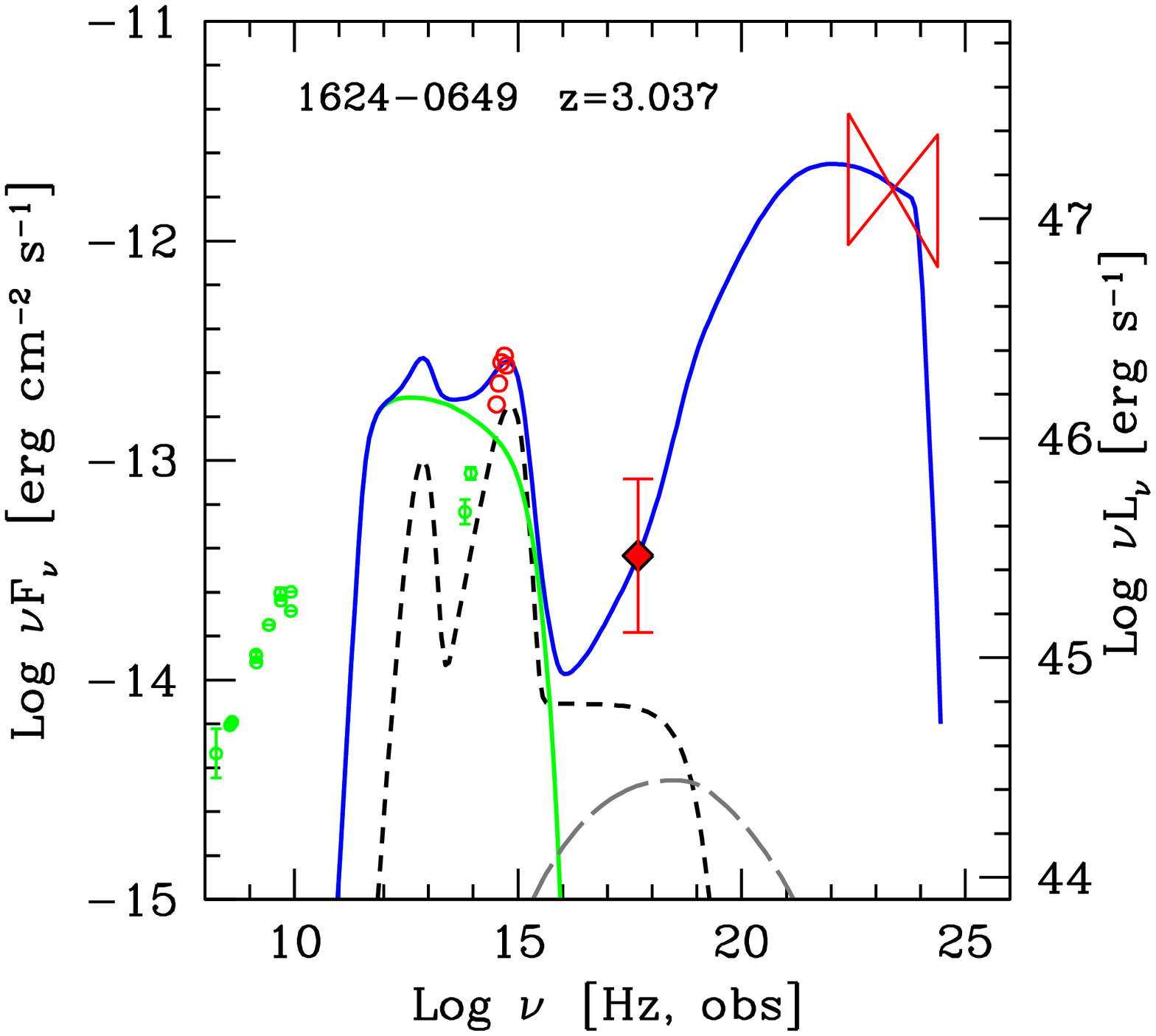,width=4.3cm,height=3.7cm } 
&\psfig{file=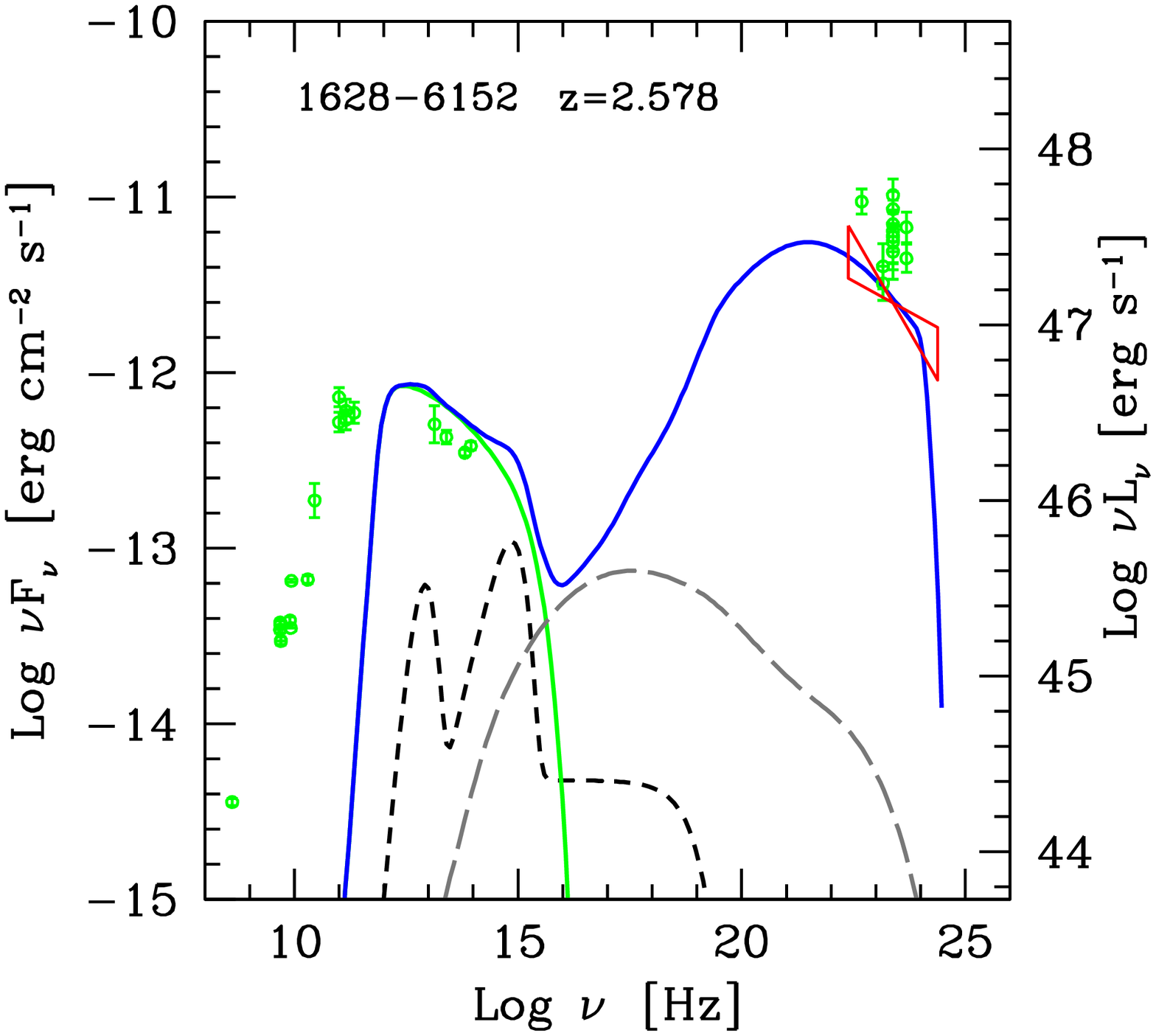,width=4.3cm,height=3.7cm } 
&\psfig{file=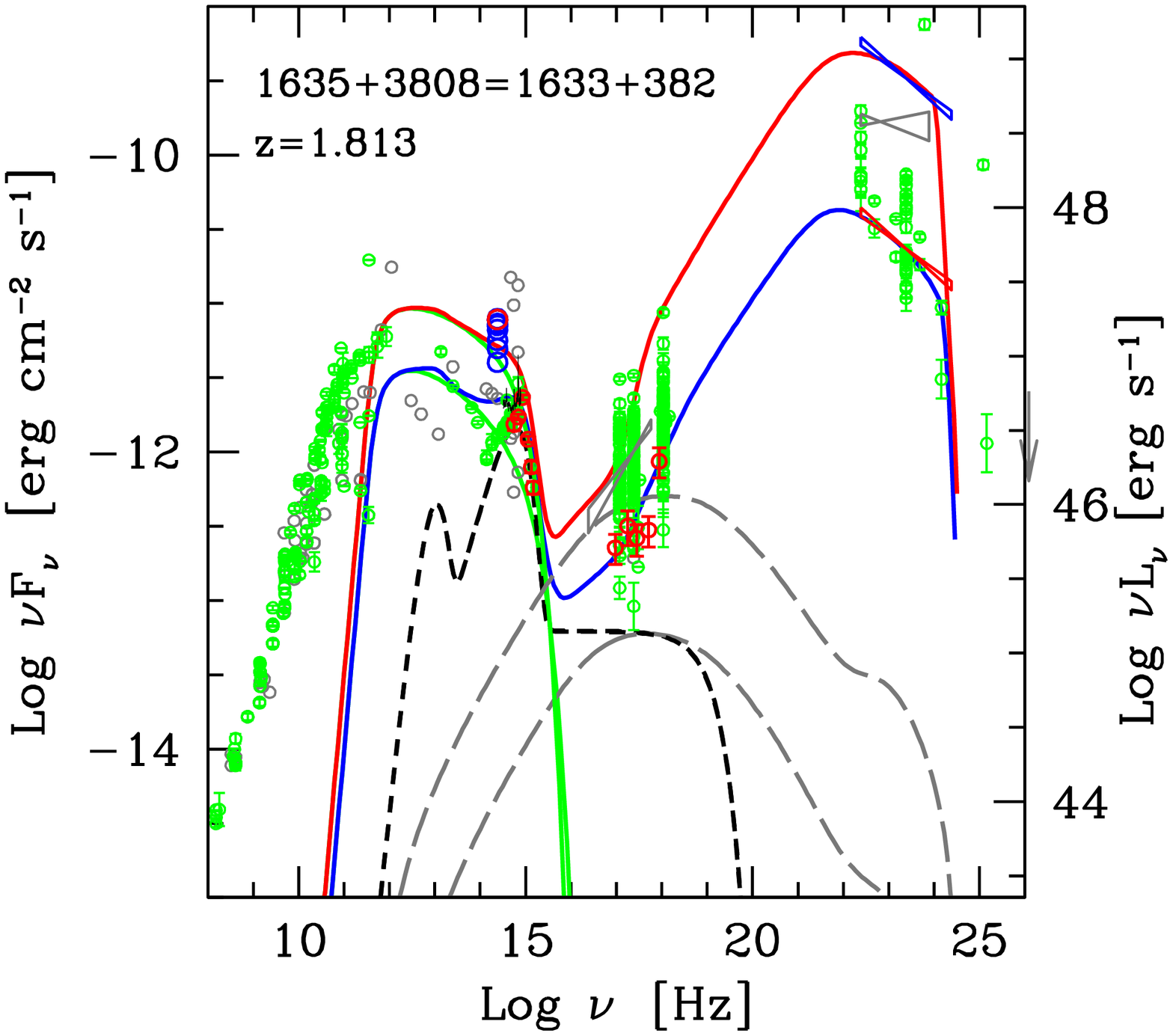,width=4.3cm,height=3.7cm }  
&\psfig{file=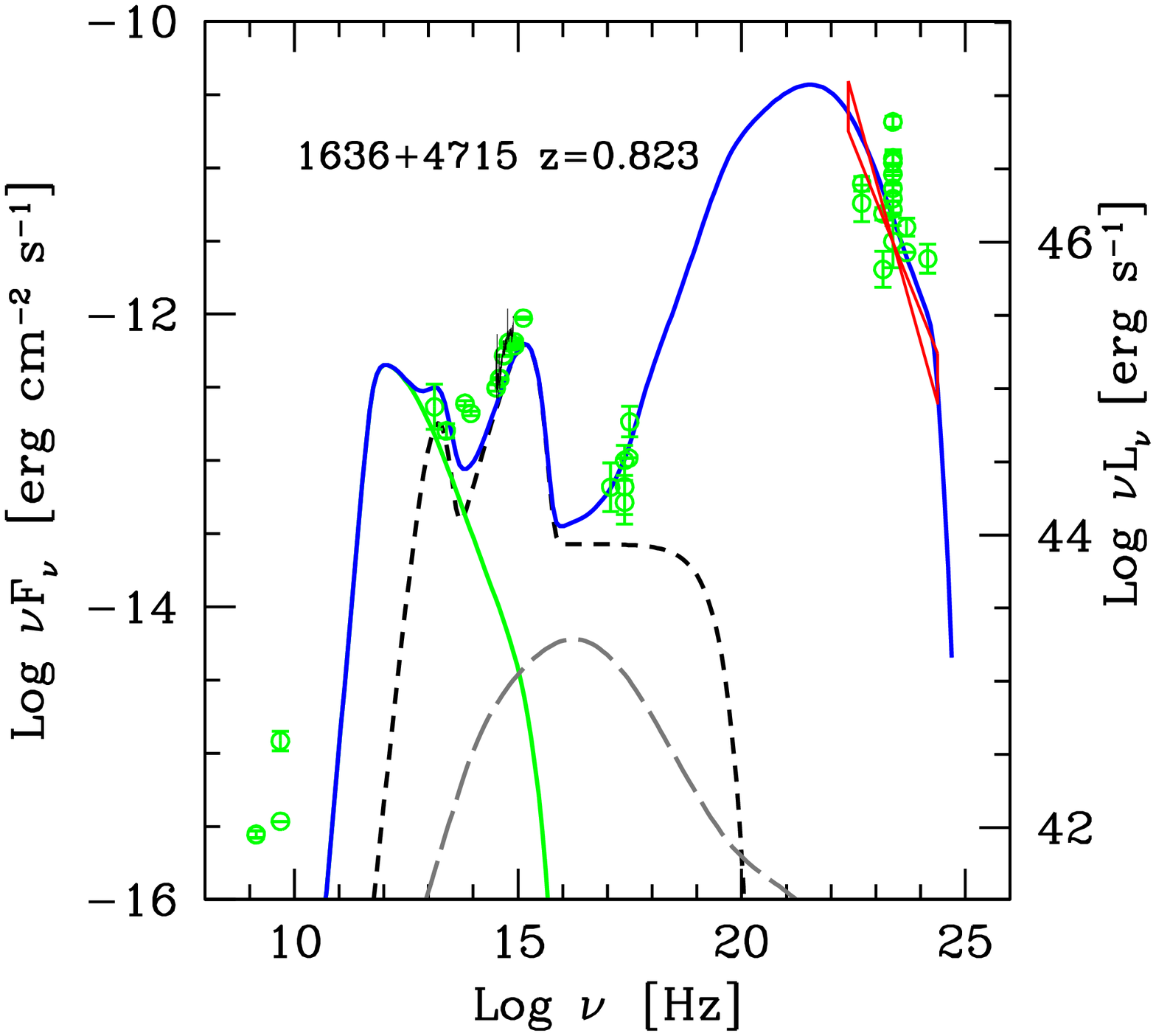,width=4.3cm,height=3.7cm }\\ 
\psfig{file=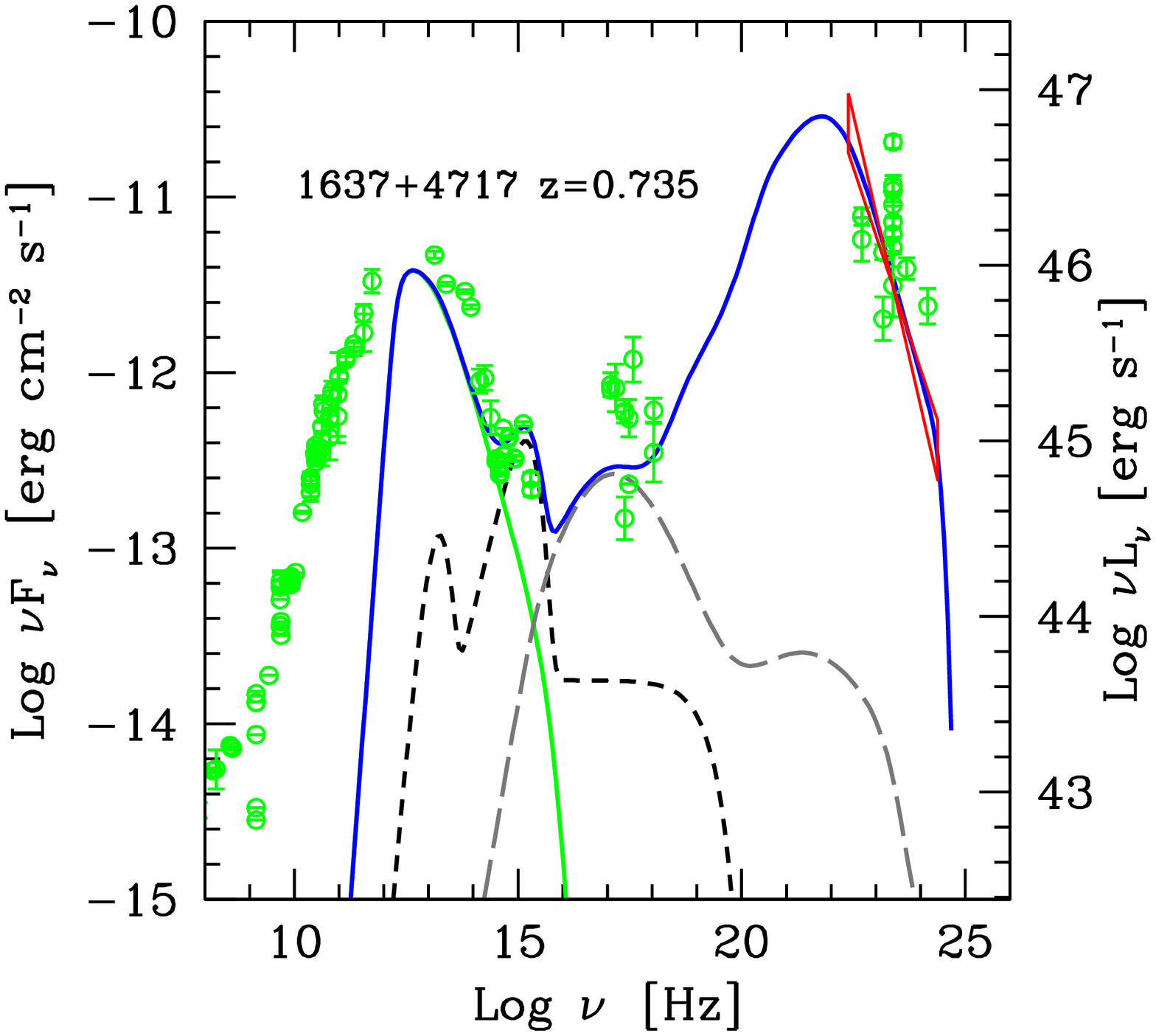,width=4.3cm,height=3.7cm } 
&\psfig{file=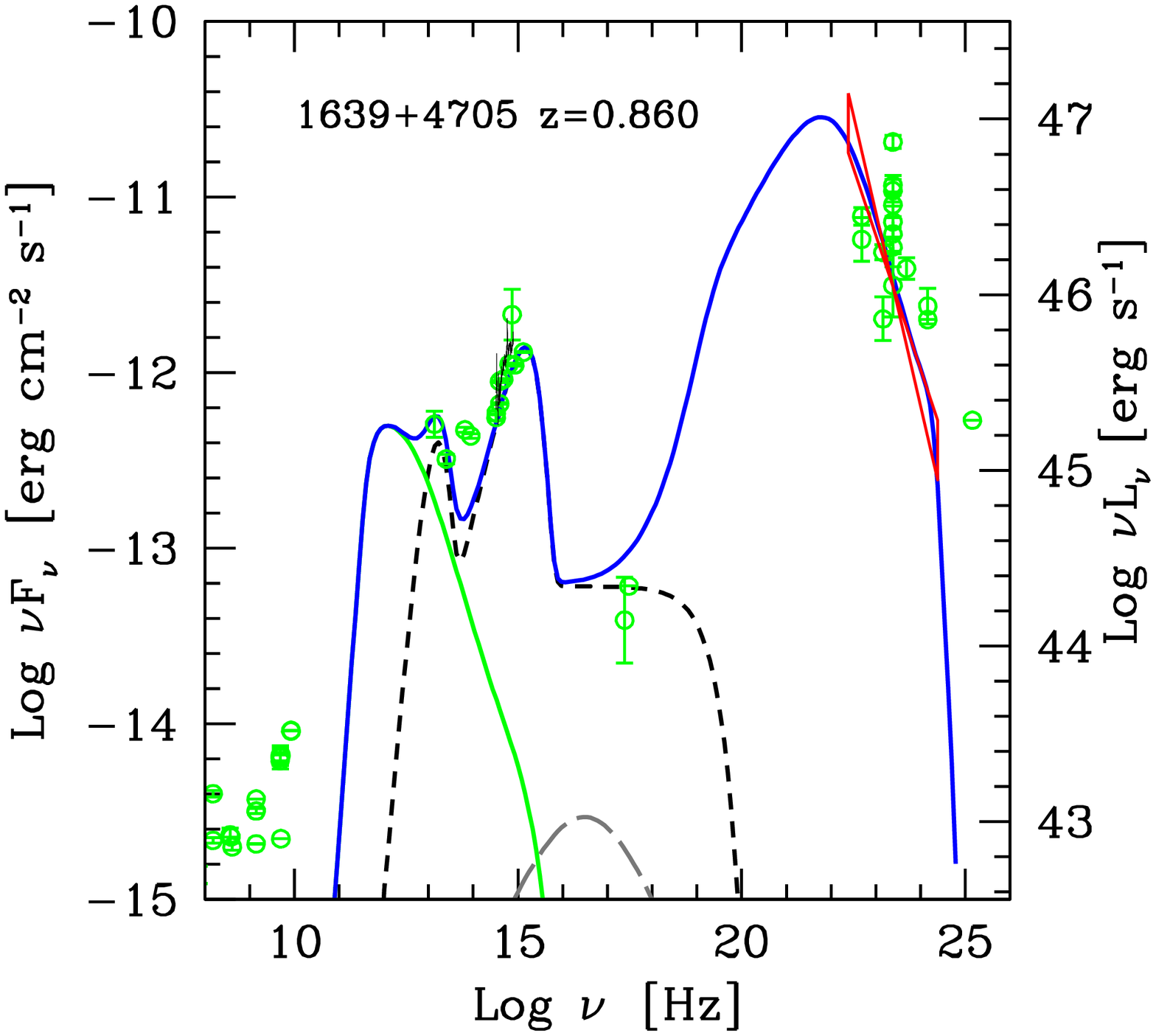,width=4.3cm,height=3.7cm } 
&\psfig{file=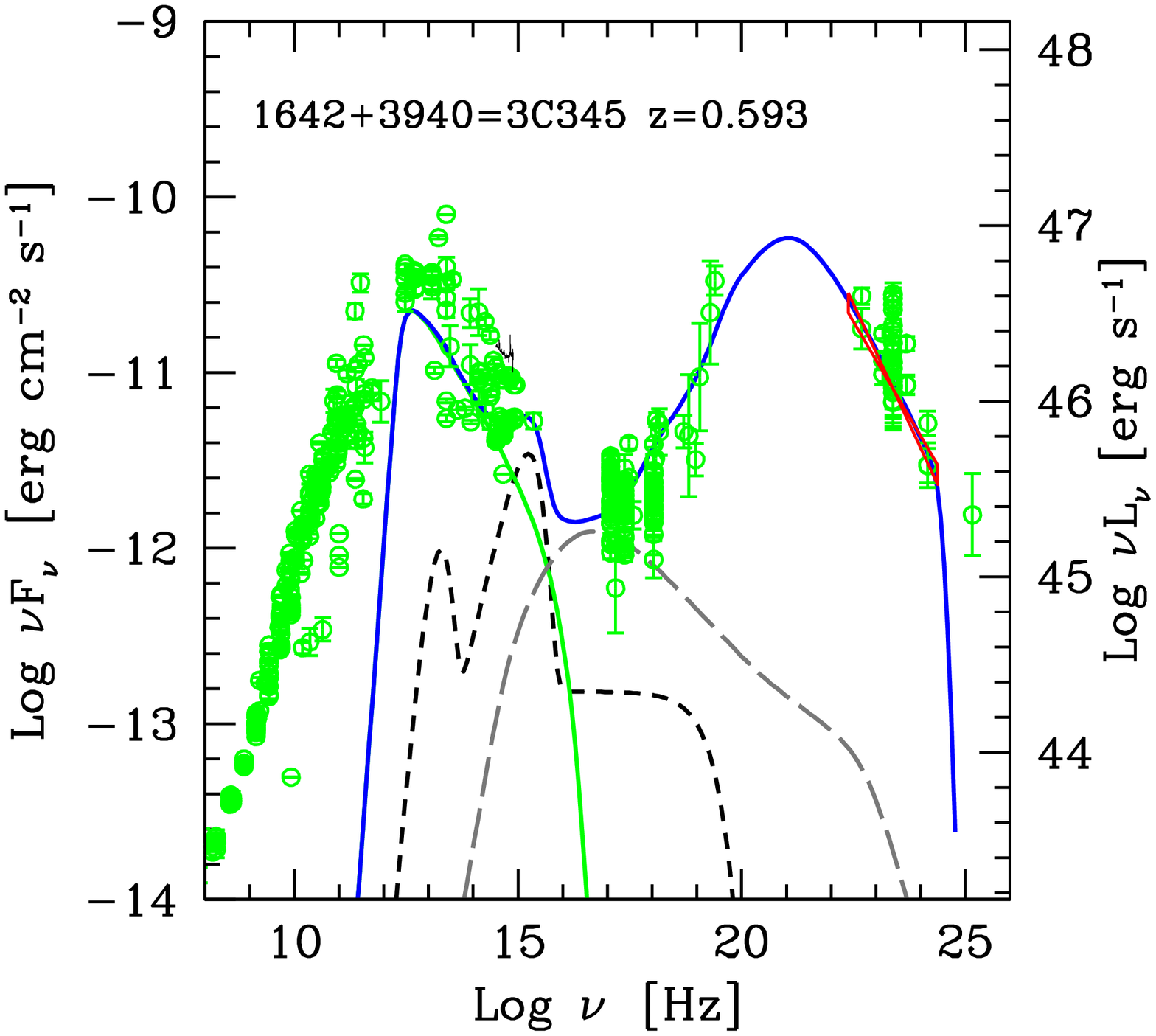,width=4.3cm,height=3.7cm } 
&\psfig{file=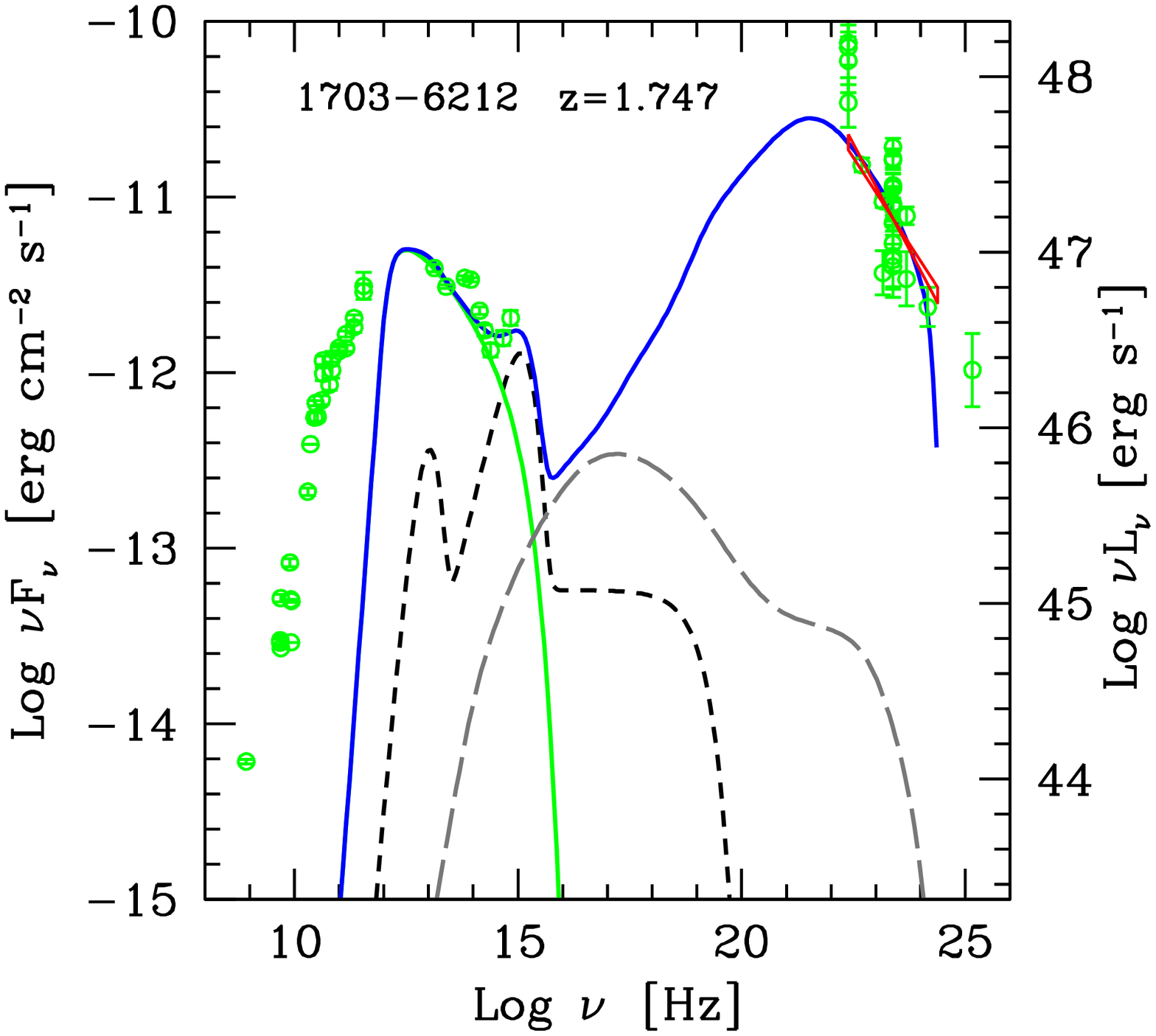,width=4.3cm,height=3.7cm } \\
\psfig{file=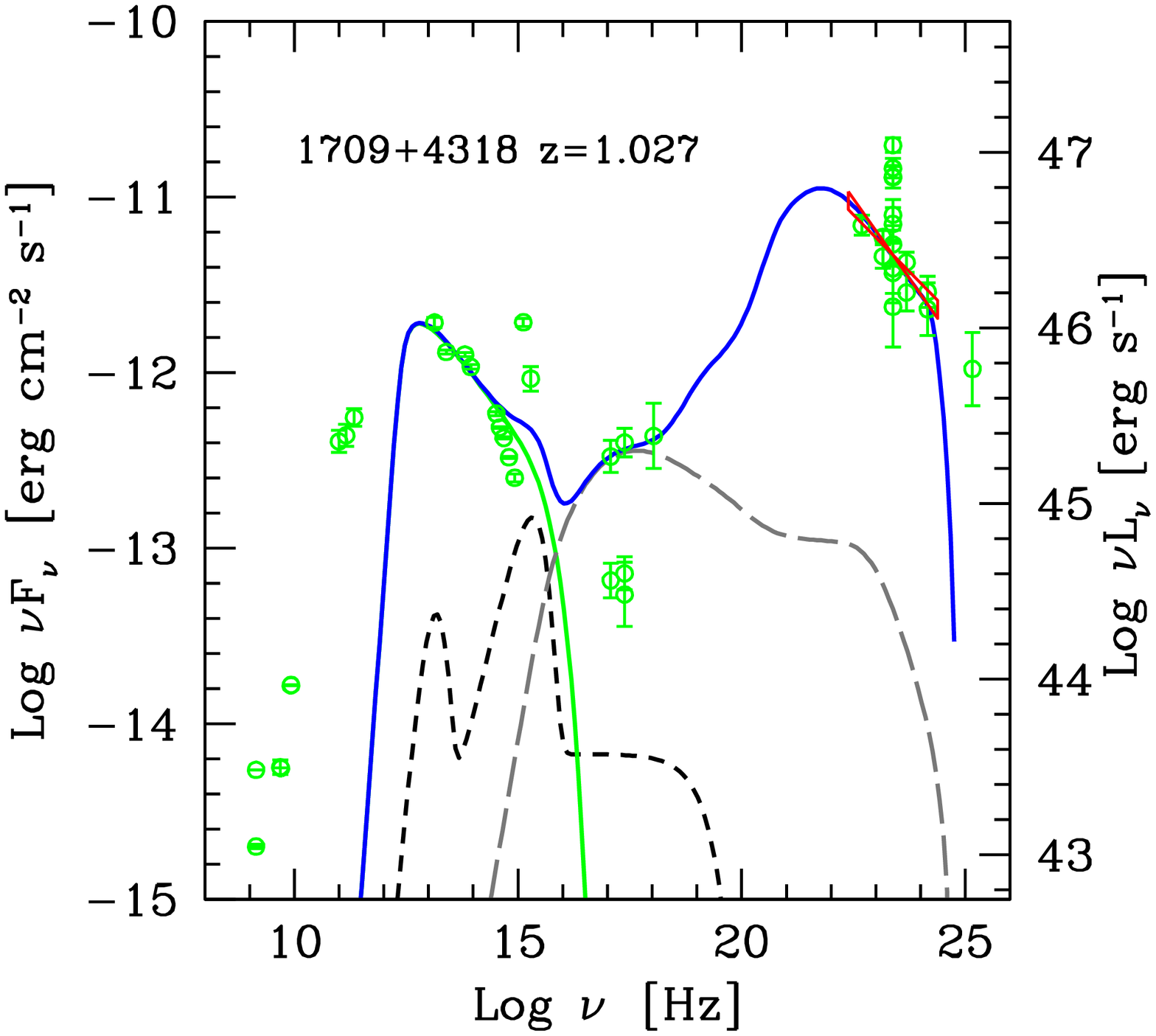,width=4.3cm,height=3.7cm } 
&\psfig{file=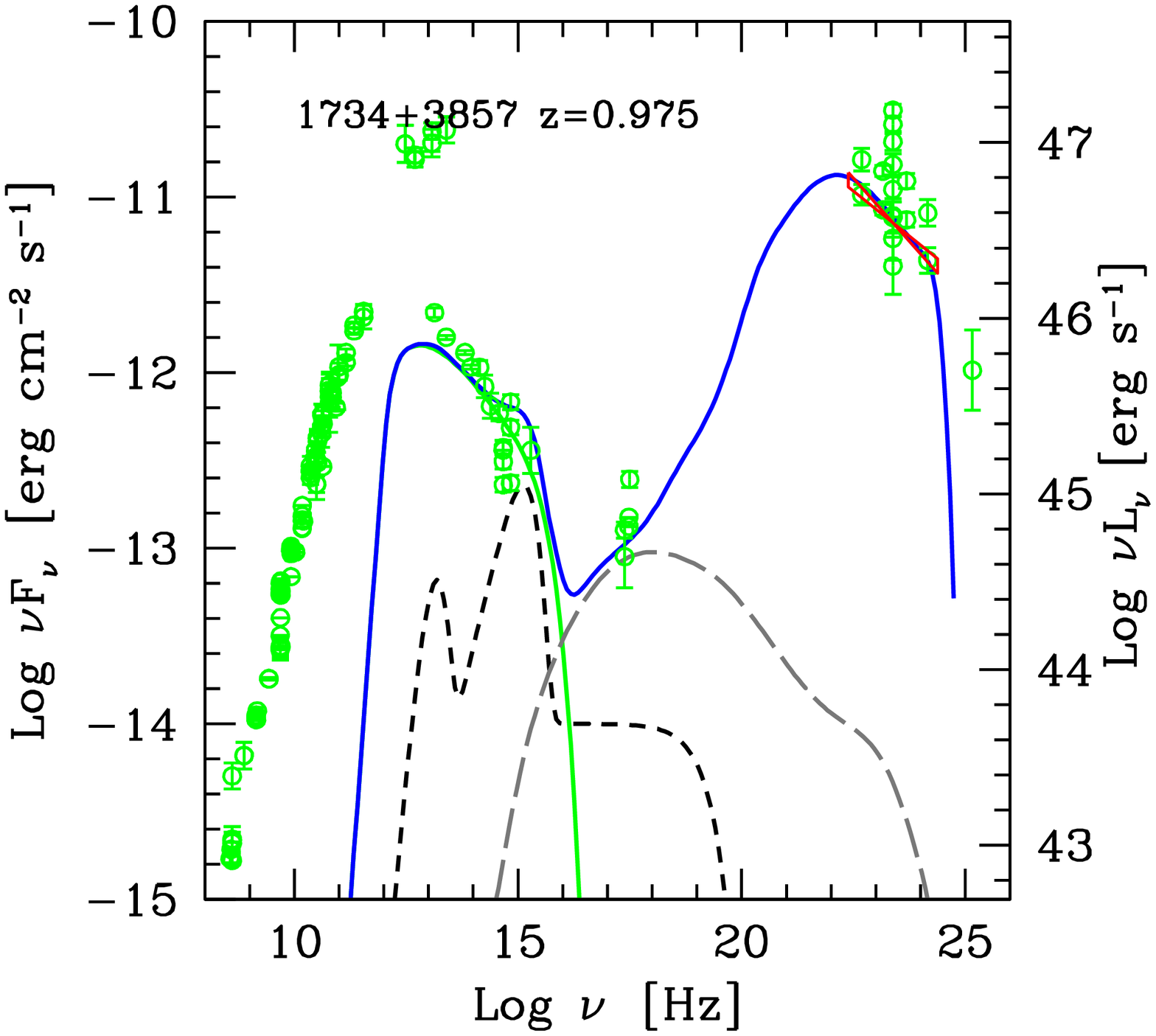,width=4.3cm,height=3.7cm } 
&\psfig{file=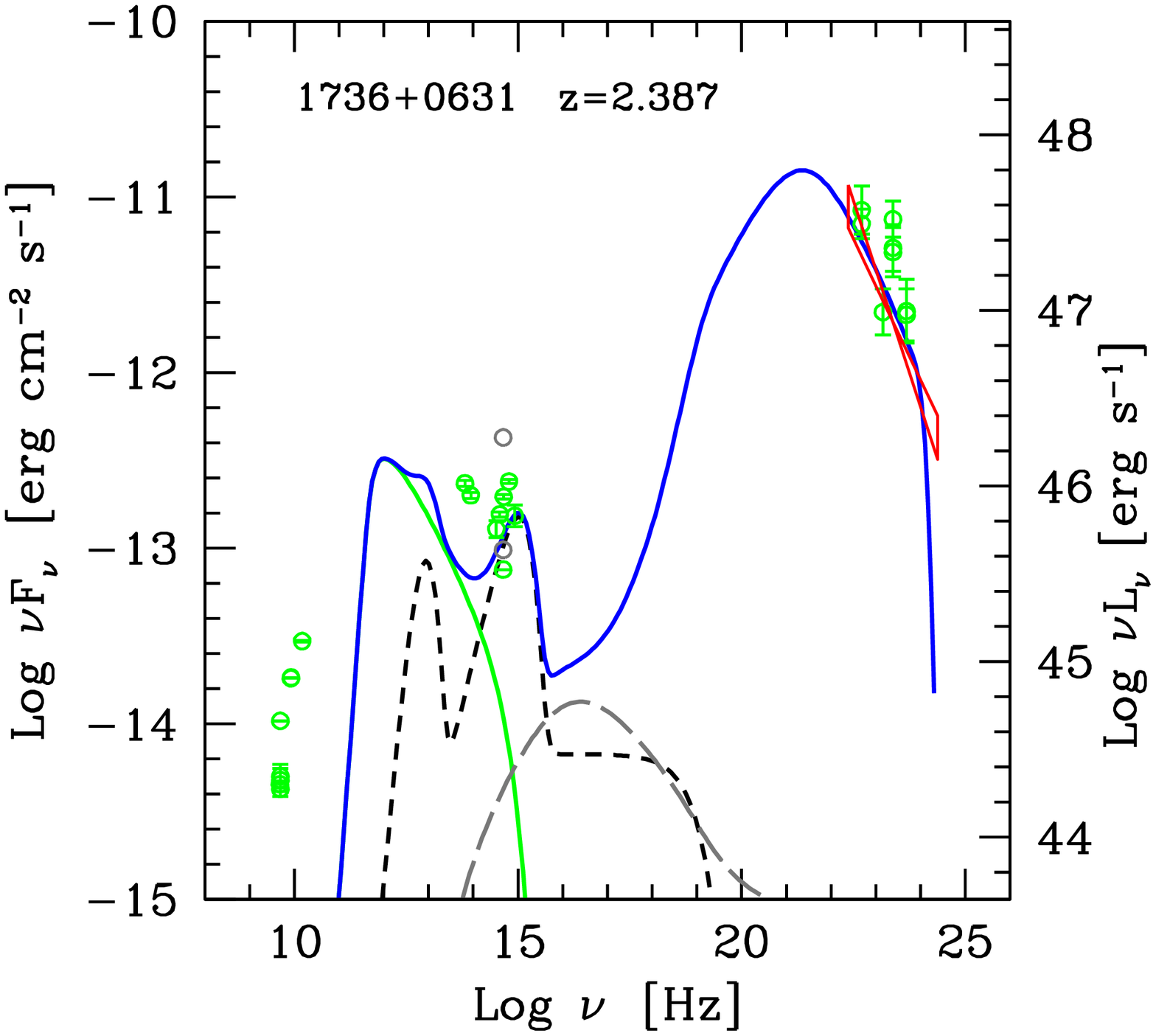,width=4.3cm,height=3.7cm }  
&\psfig{file=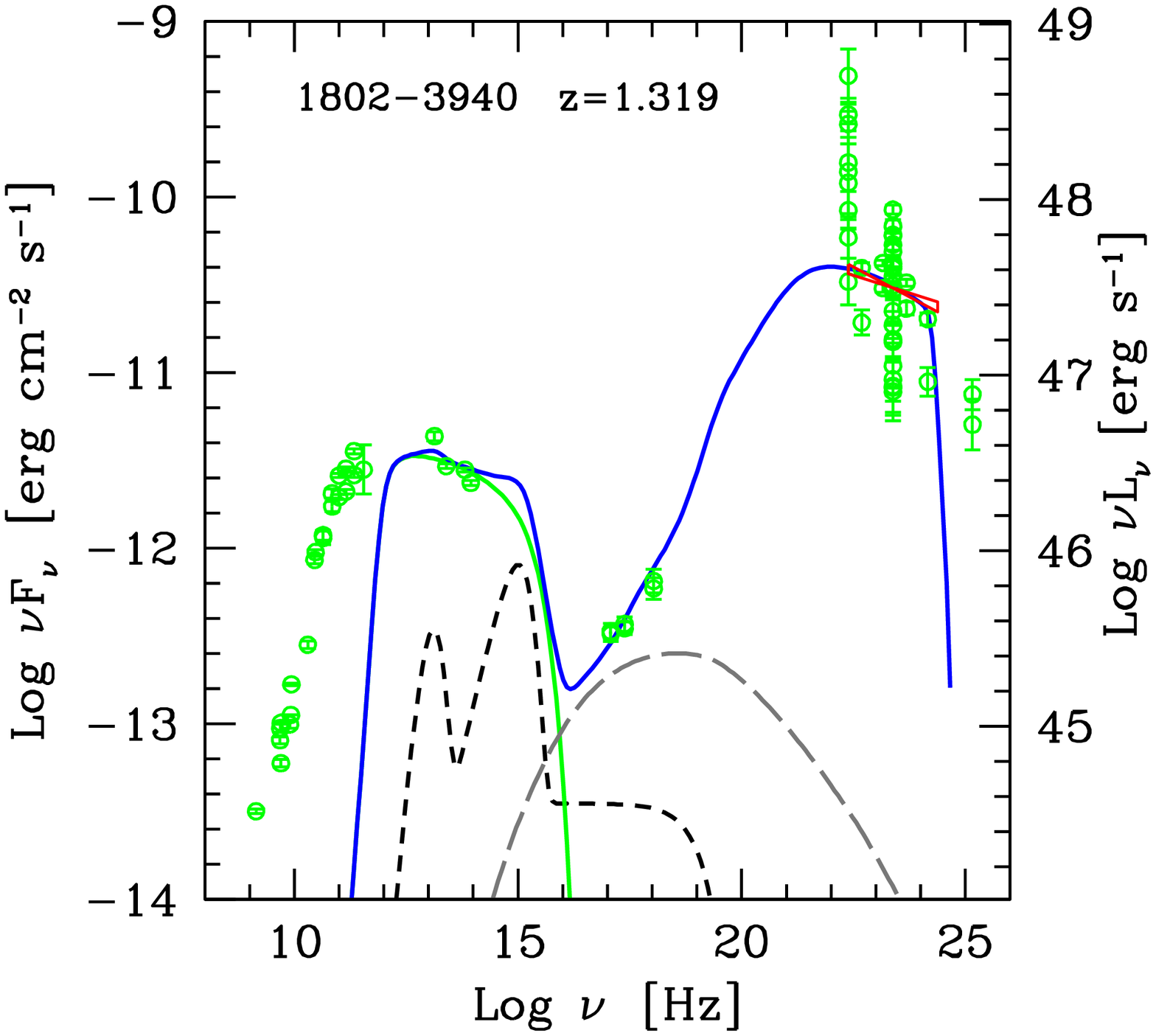,width=4.3cm,height=3.7cm }\\ 
\psfig{file=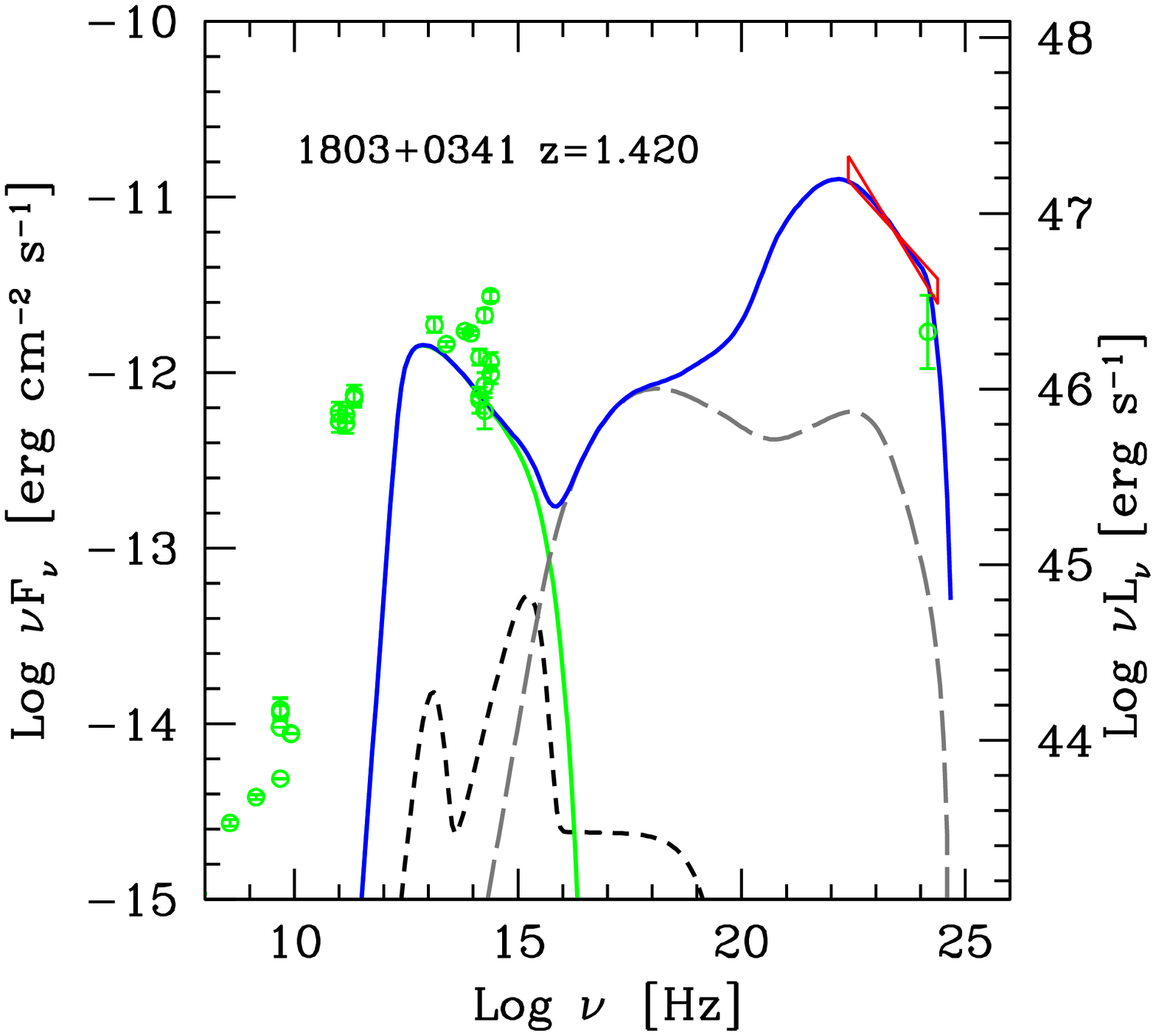,width=4.3cm,height=3.7cm } 
&\psfig{file=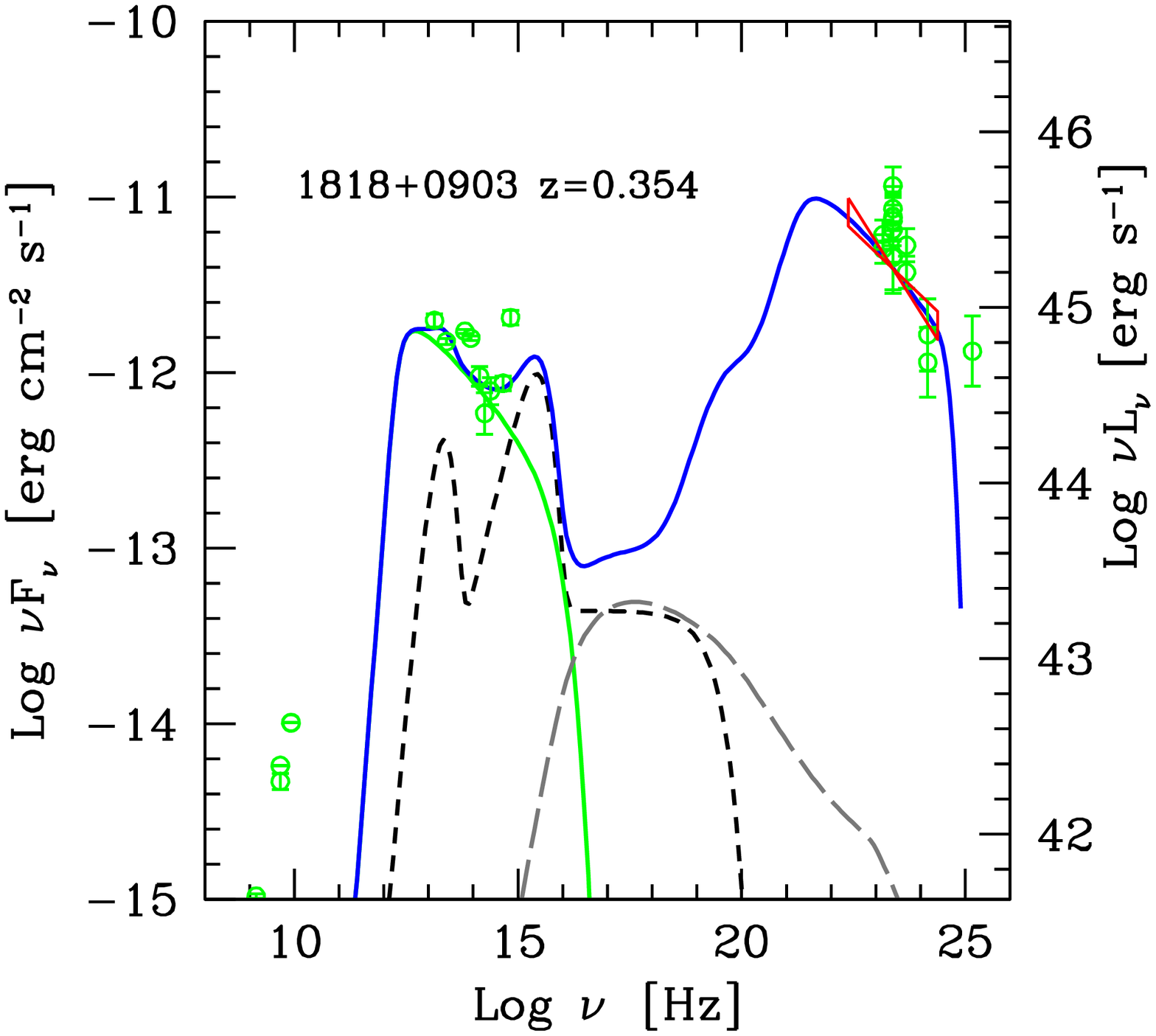,width=4.3cm,height=3.7cm } 
&\psfig{file=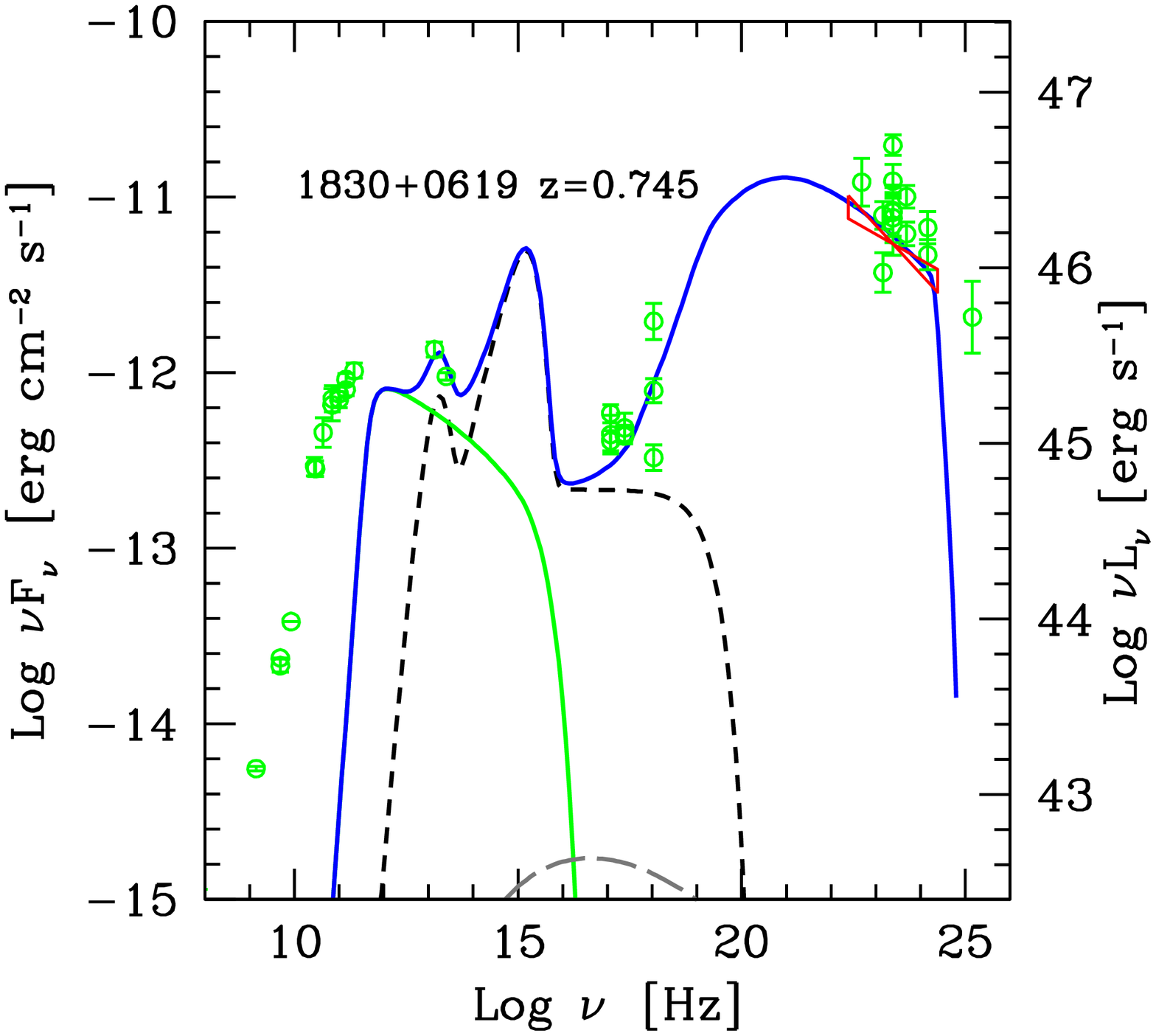,width=4.3cm,height=3.7cm } 
&\psfig{file=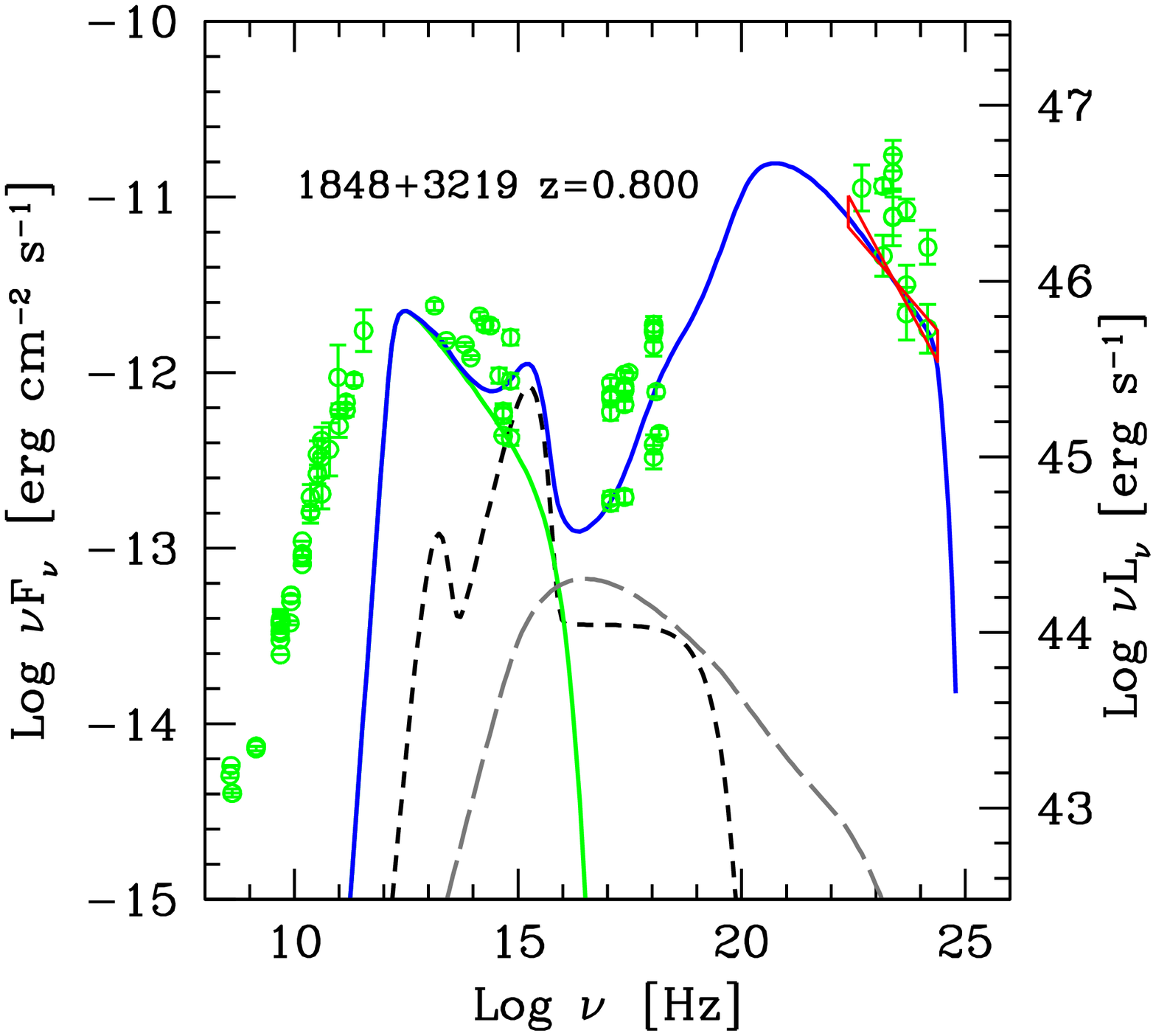,width=4.3cm,height=3.7cm } \\
\psfig{file=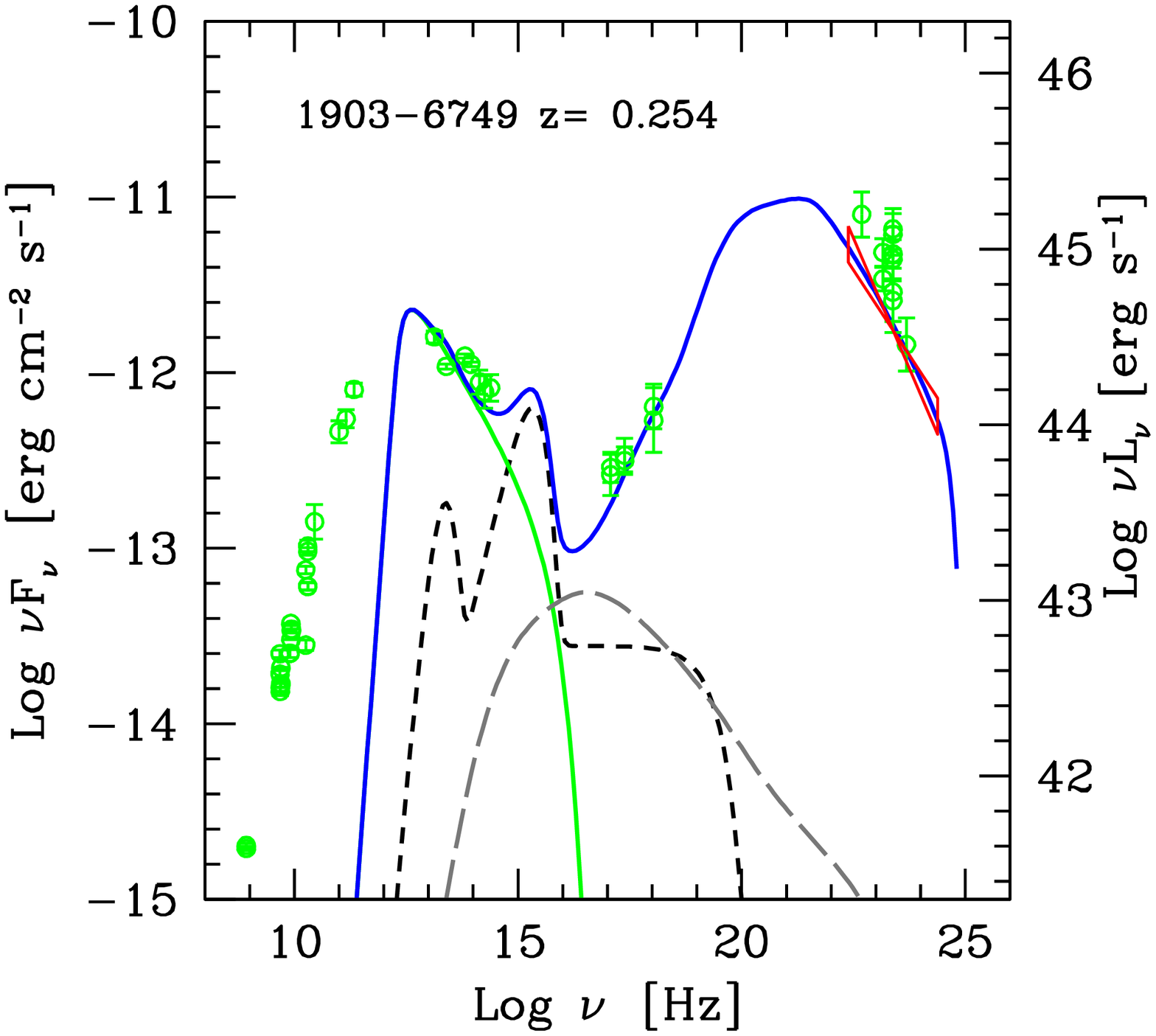,width=4.3cm,height=3.7cm } 
&\psfig{file=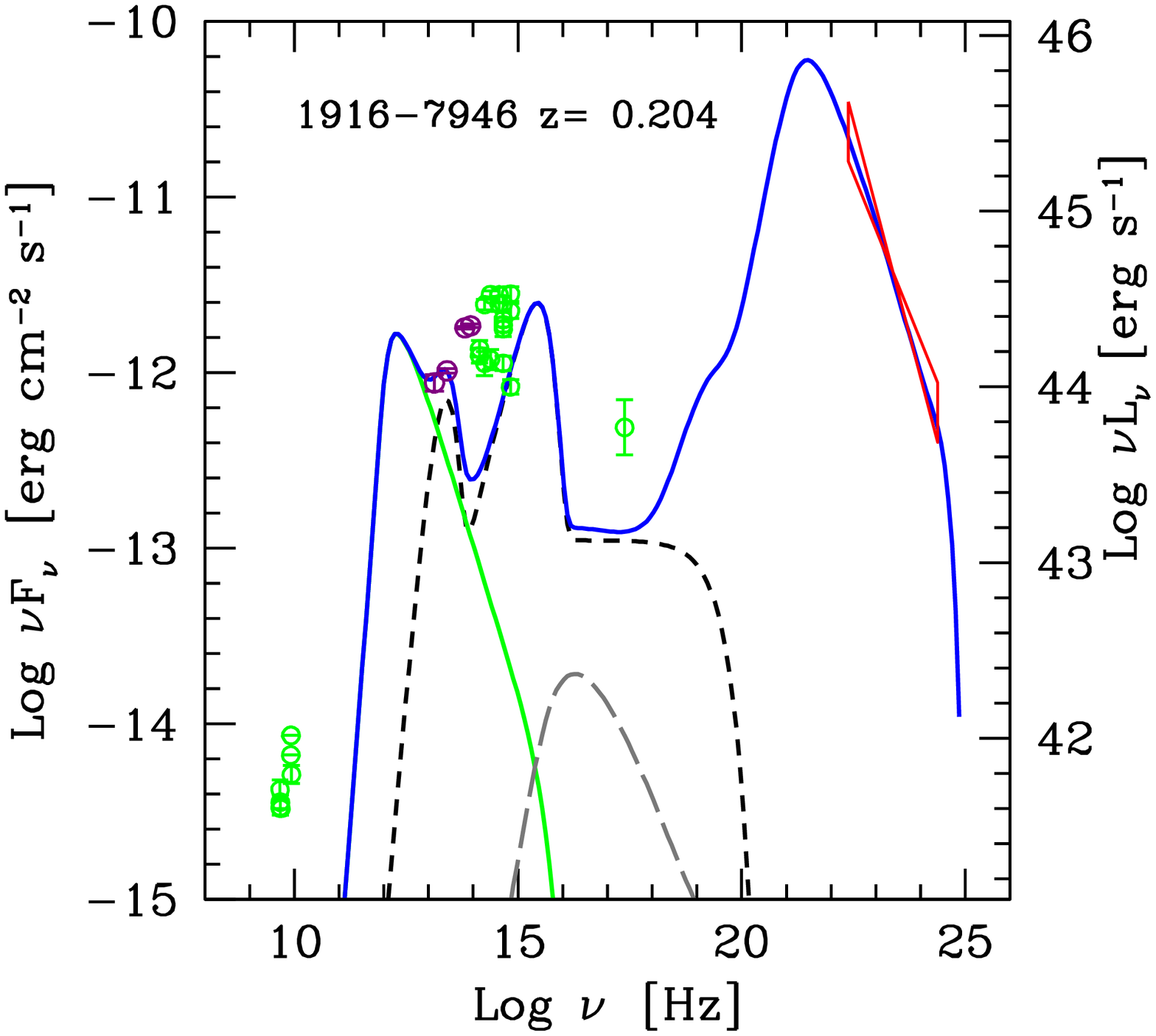,width=4.3cm,height=3.7cm } 
&\psfig{file=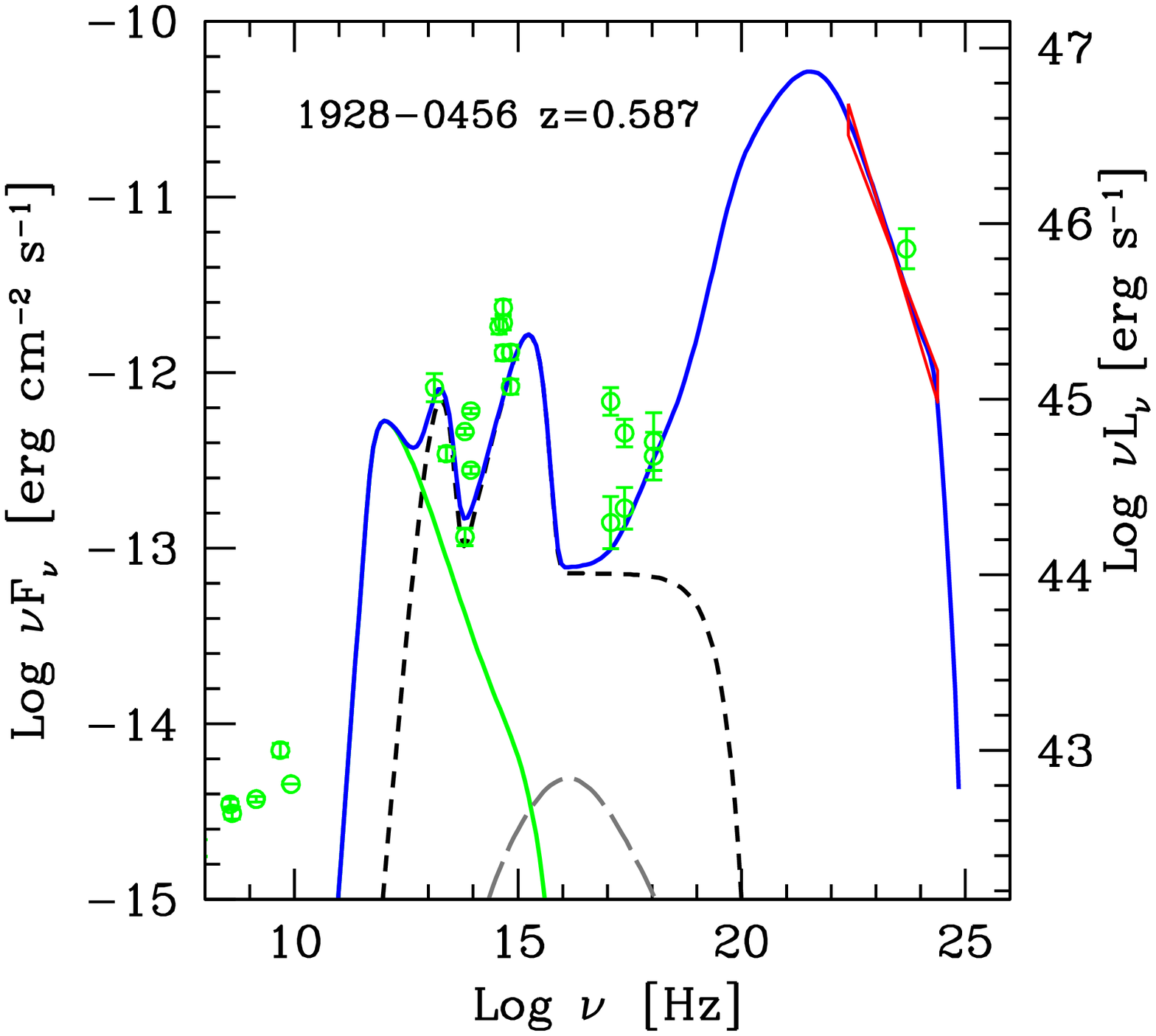,width=4.3cm,height=3.7cm }  
&\psfig{file=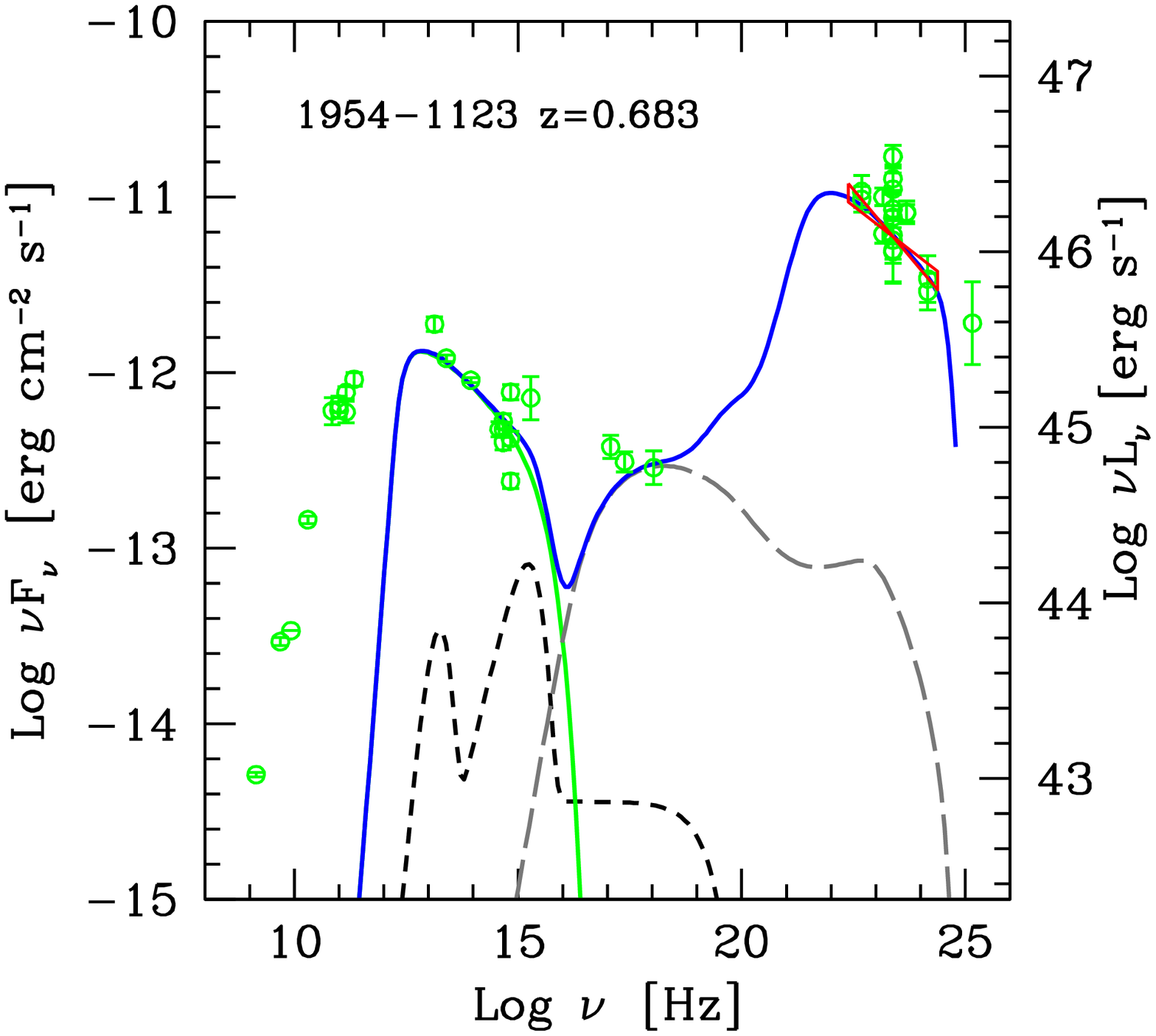,width=4.3cm,height=3.7cm } \\
\psfig{file=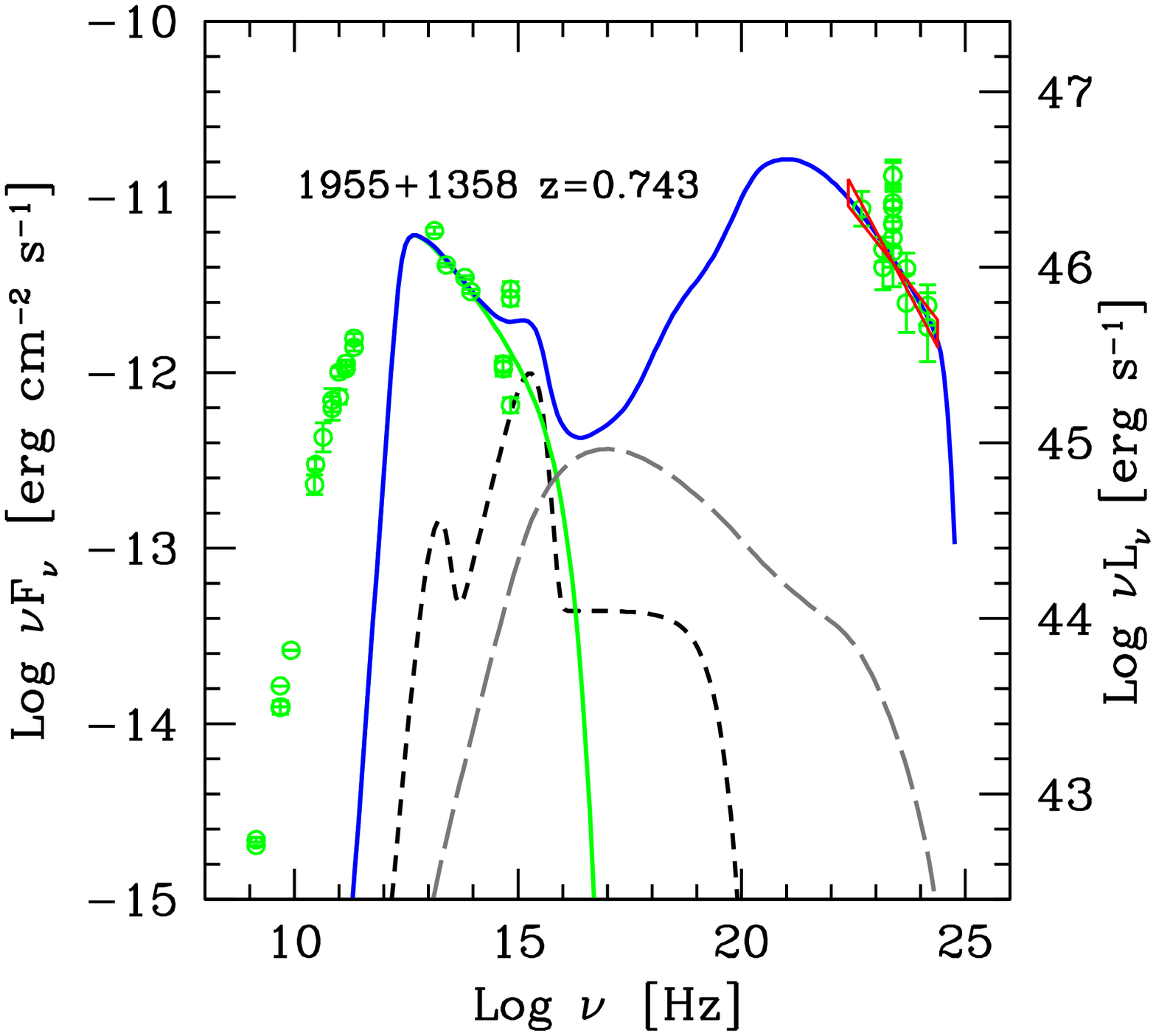,width=4.3cm,height=3.7cm }  
&\psfig{file=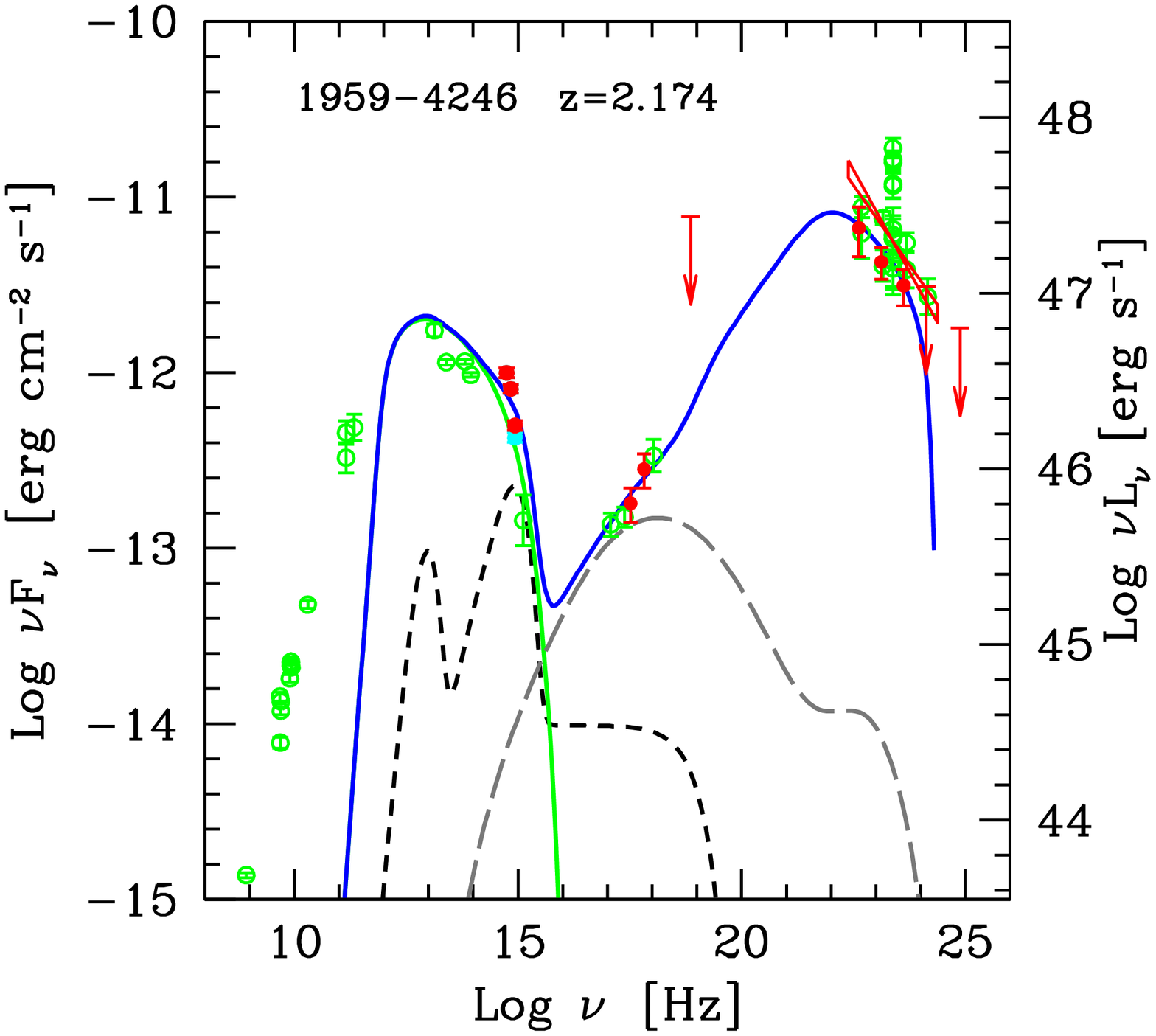,width=4.3cm,height=3.7cm } 
&\psfig{file=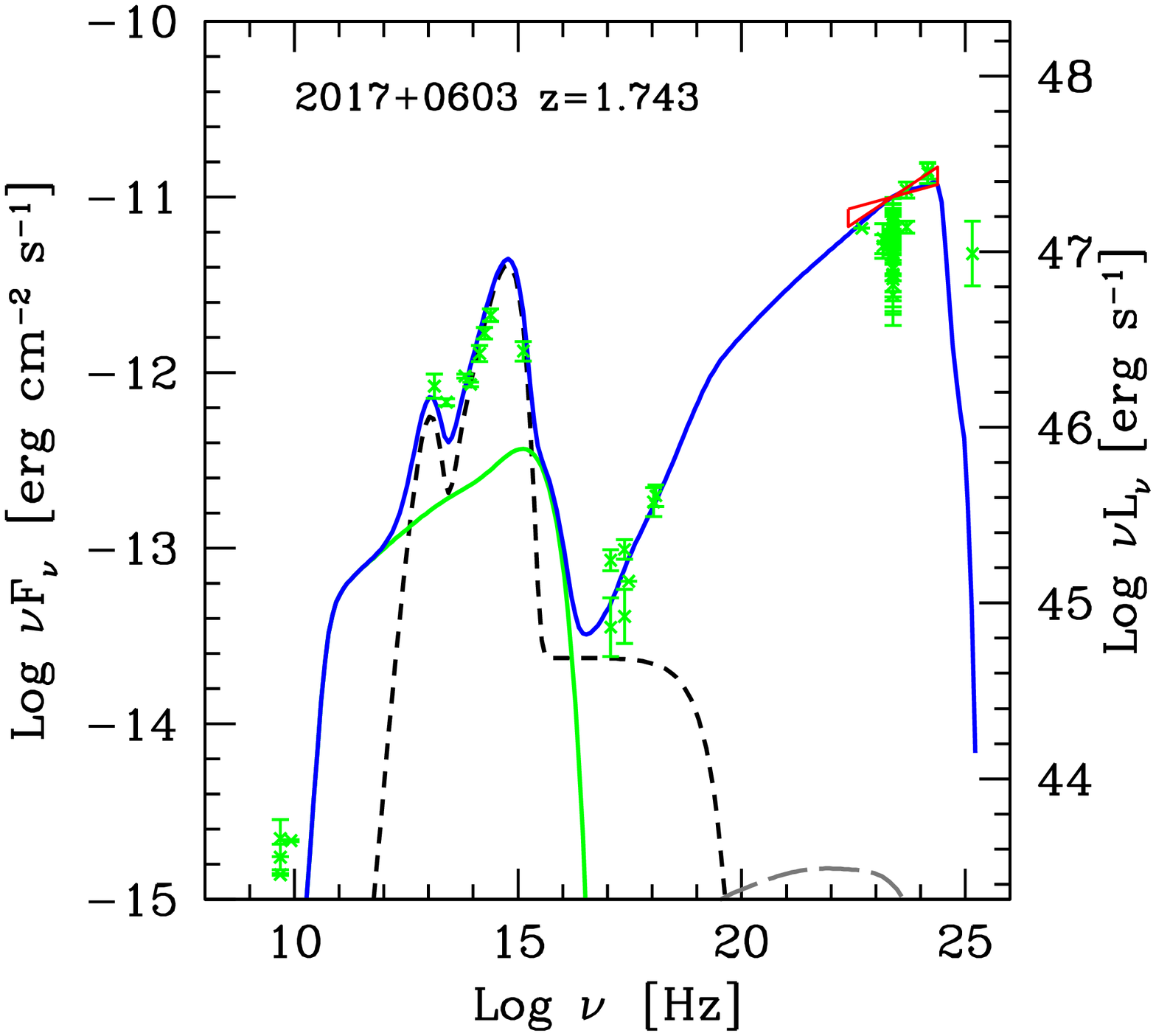,width=4.3cm,height=3.7cm }  
&\psfig{file=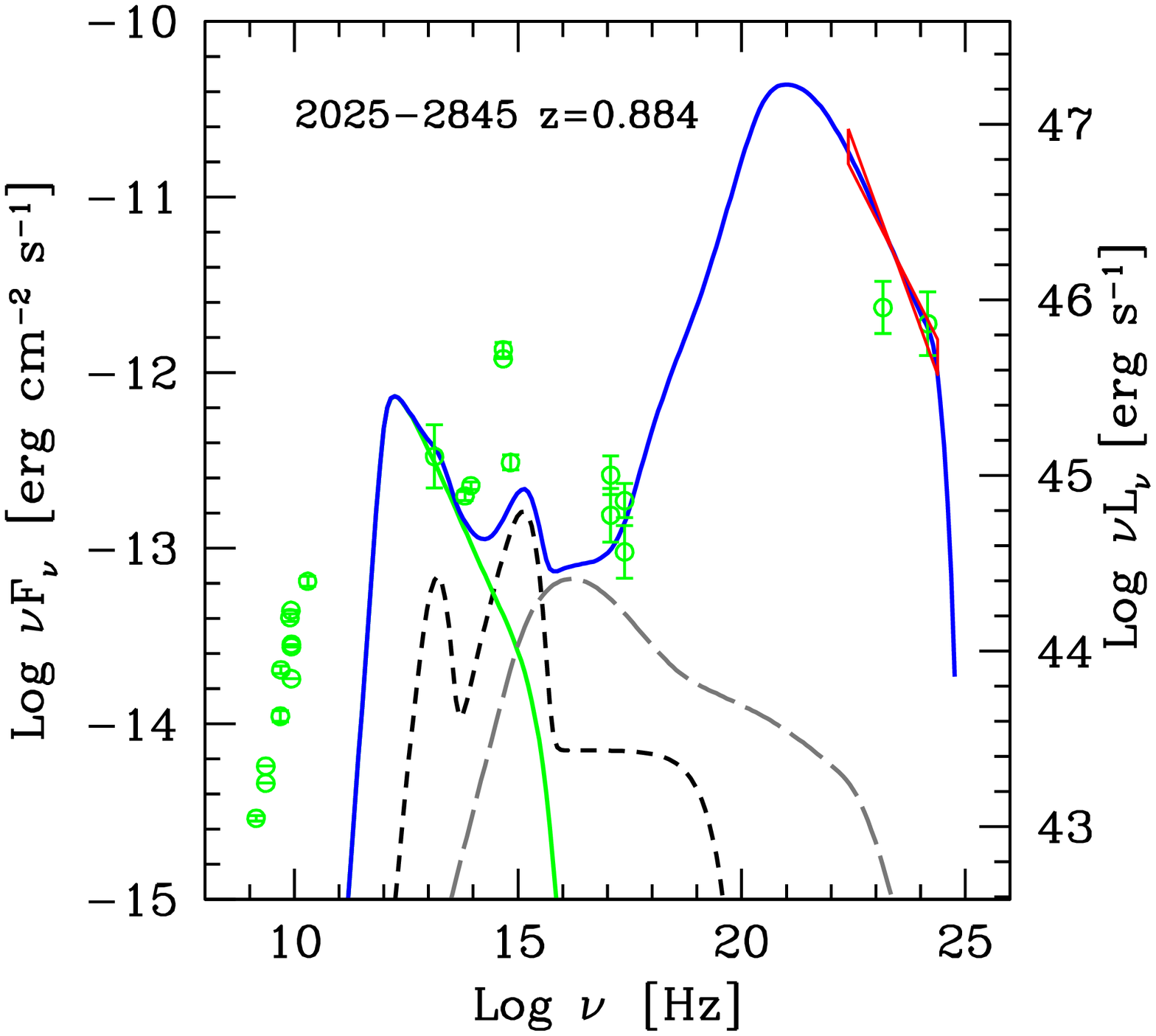,width=4.3cm,height=3.7cm } 
\end{tabular}
\caption{{\it continue.} SED of the FSRQs studied in this paper.}
\end{figure*} 

\setcounter{figure}{15}
\begin{figure*}
\begin{tabular}{cccc}
\psfig{file=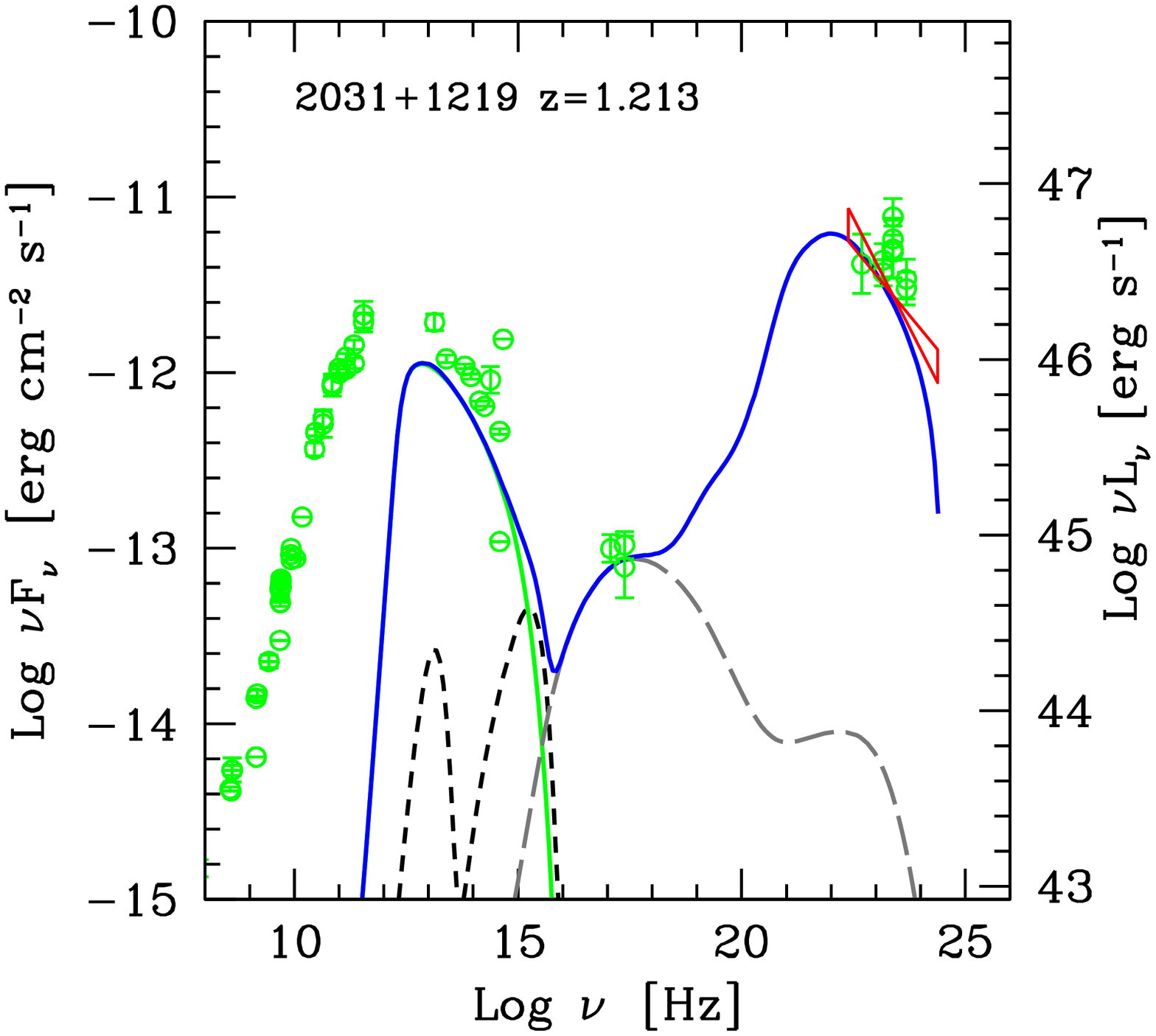,width=4.3cm,height=3.7cm } 
&\psfig{file=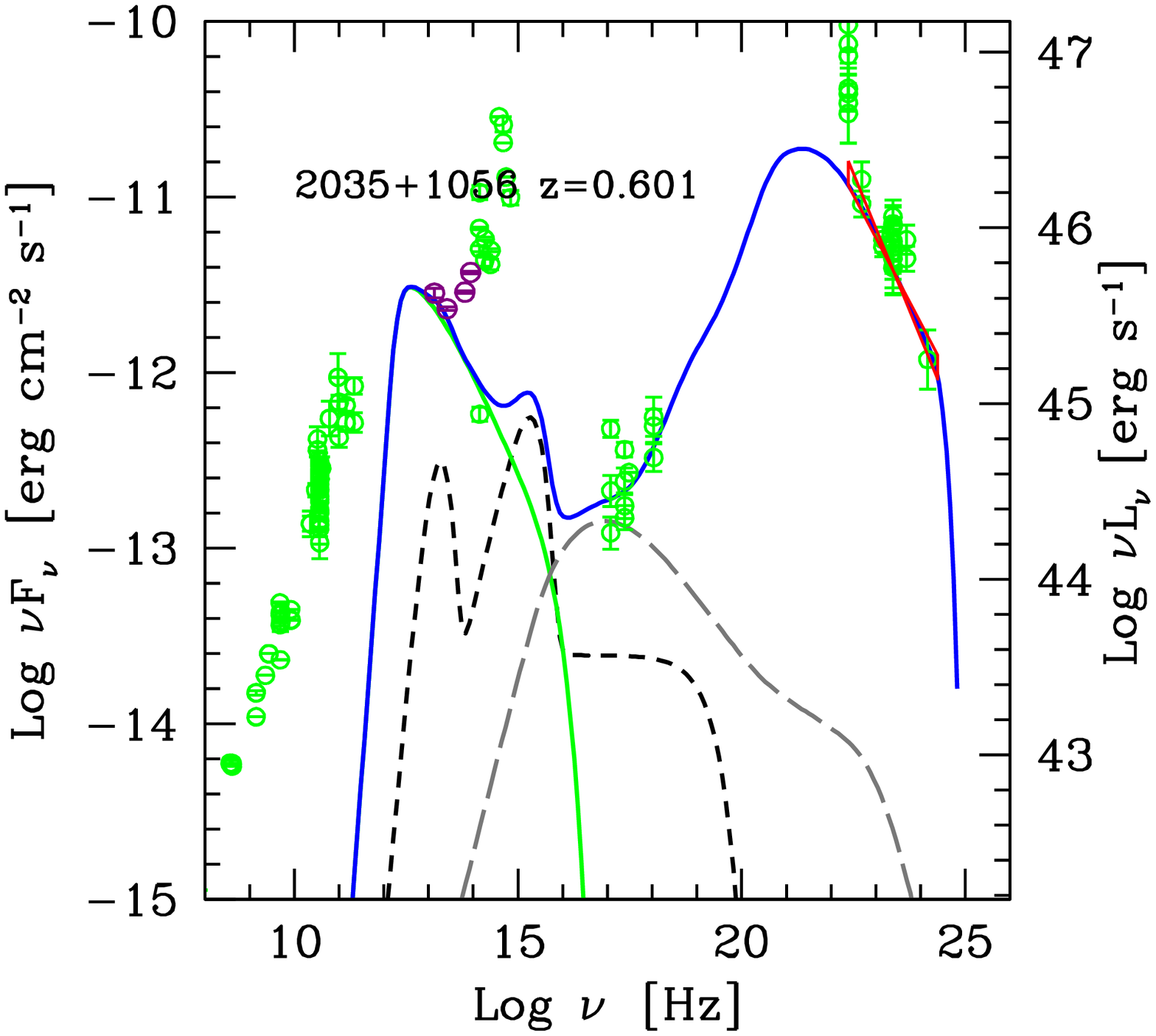,width=4.3cm,height=3.7cm } 
&\psfig{file=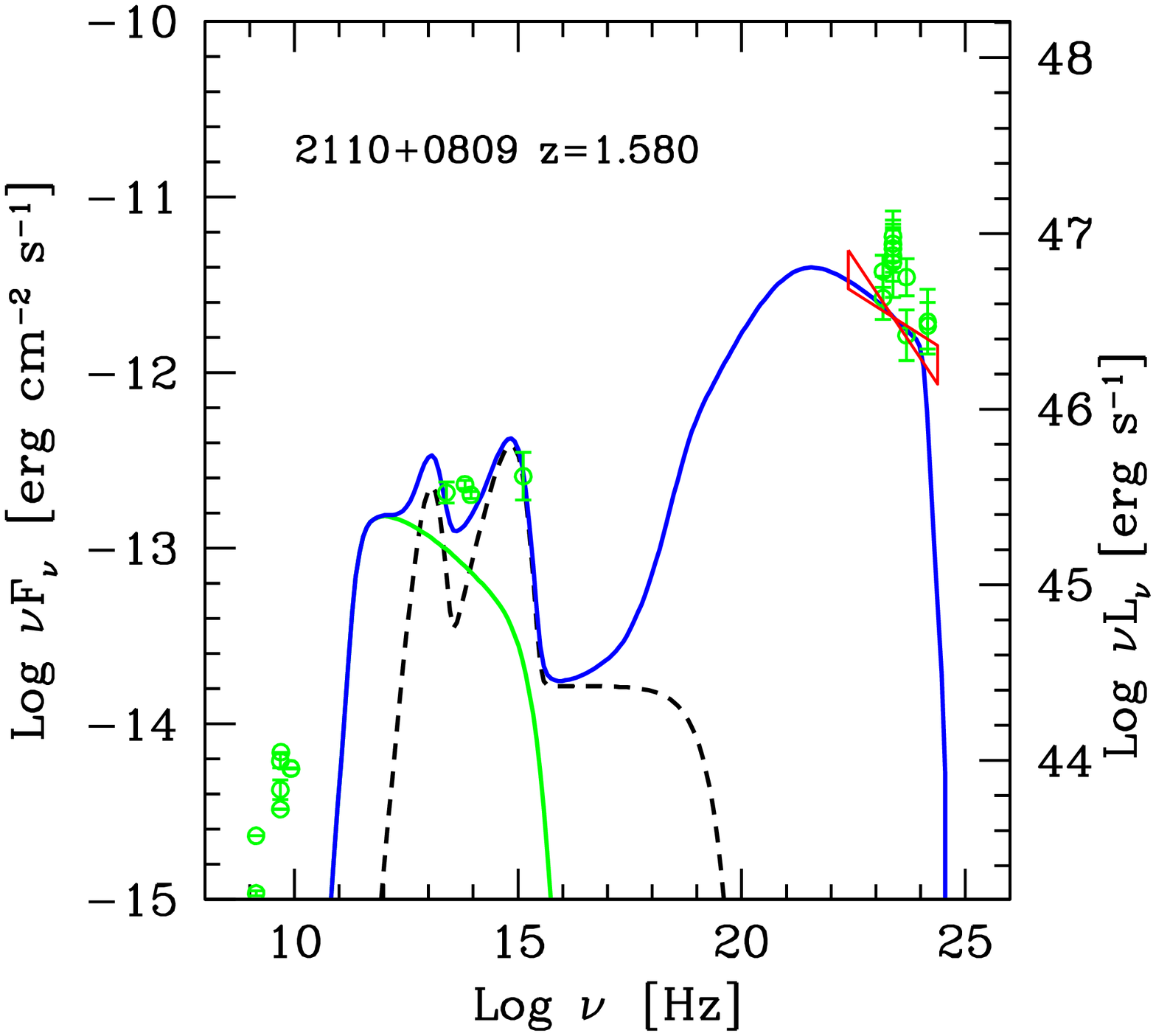,width=4.3cm,height=3.7cm }  
&\psfig{file=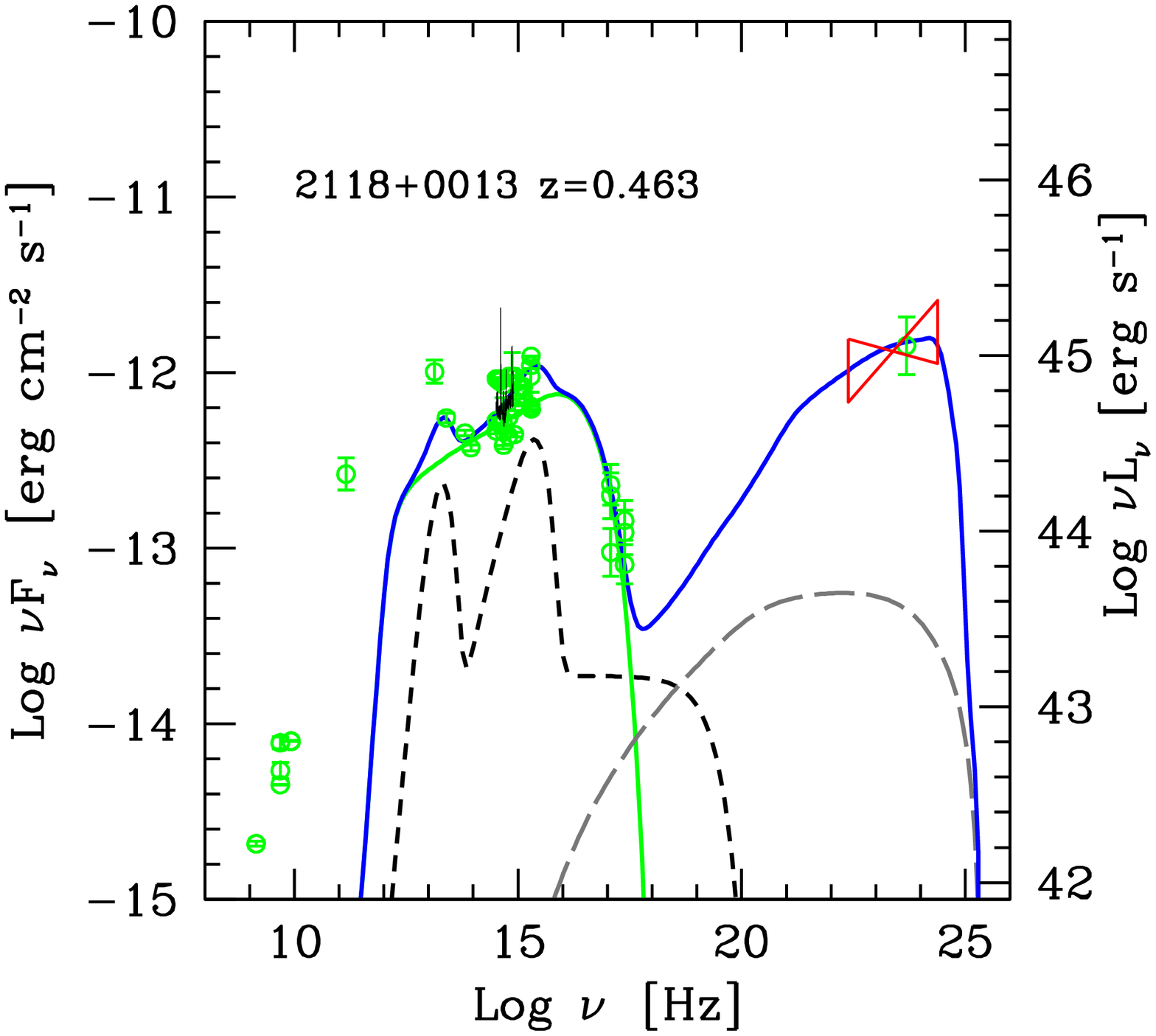,width=4.3cm,height=3.7cm } \\
\psfig{file=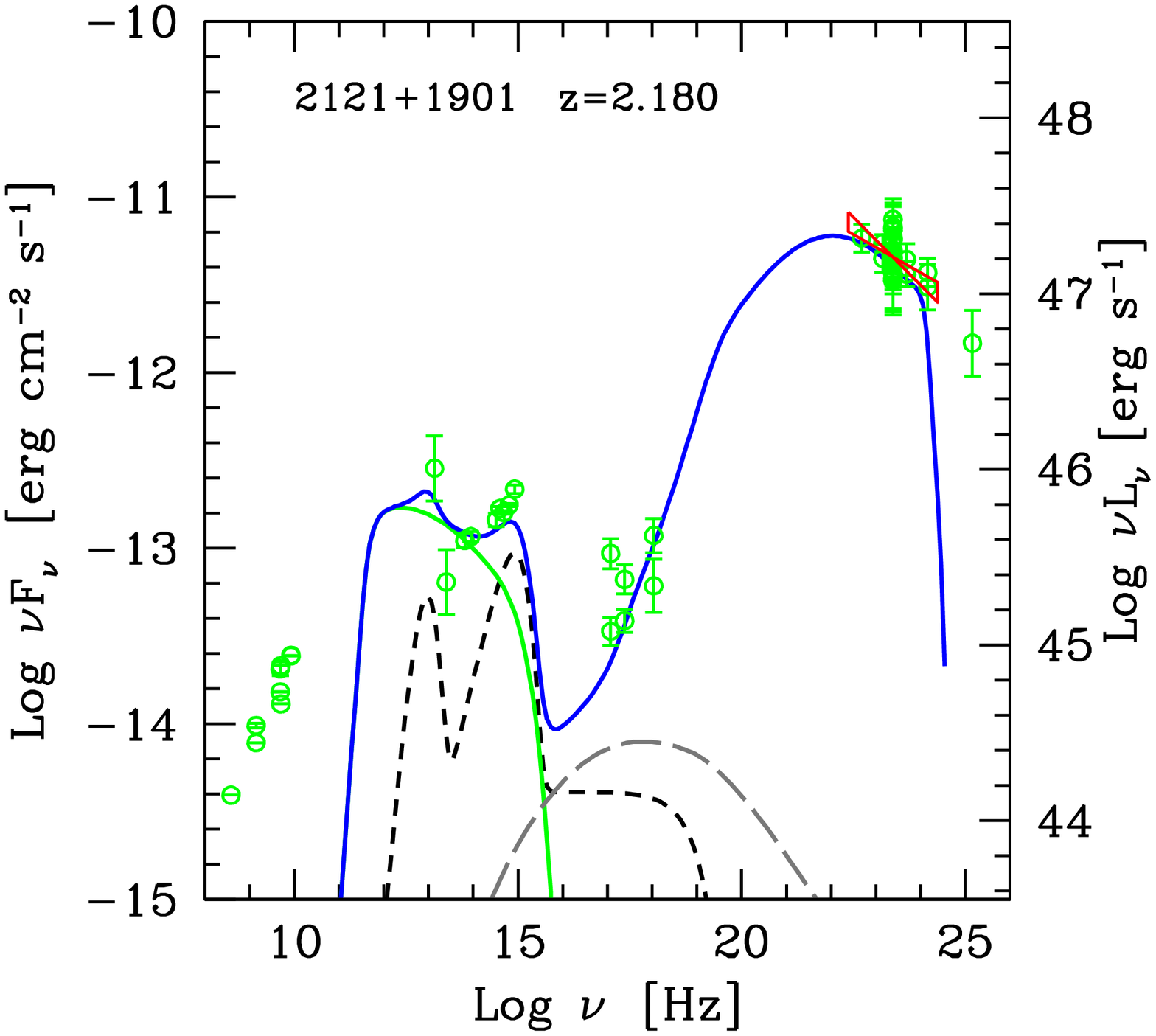,width=4.3cm,height=3.7cm } 
&\psfig{file=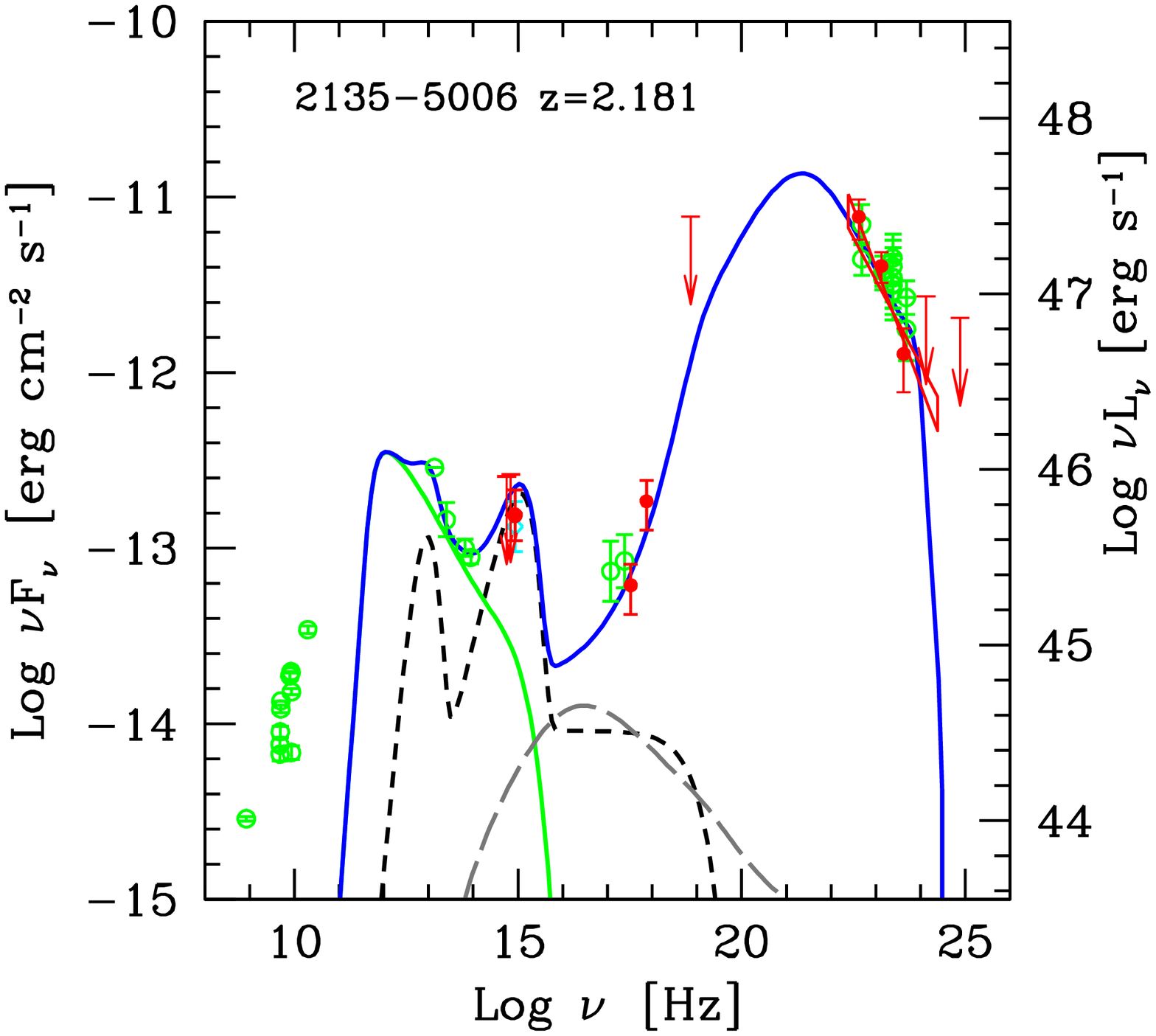,width=4.3cm,height=3.7cm } 
&\psfig{file=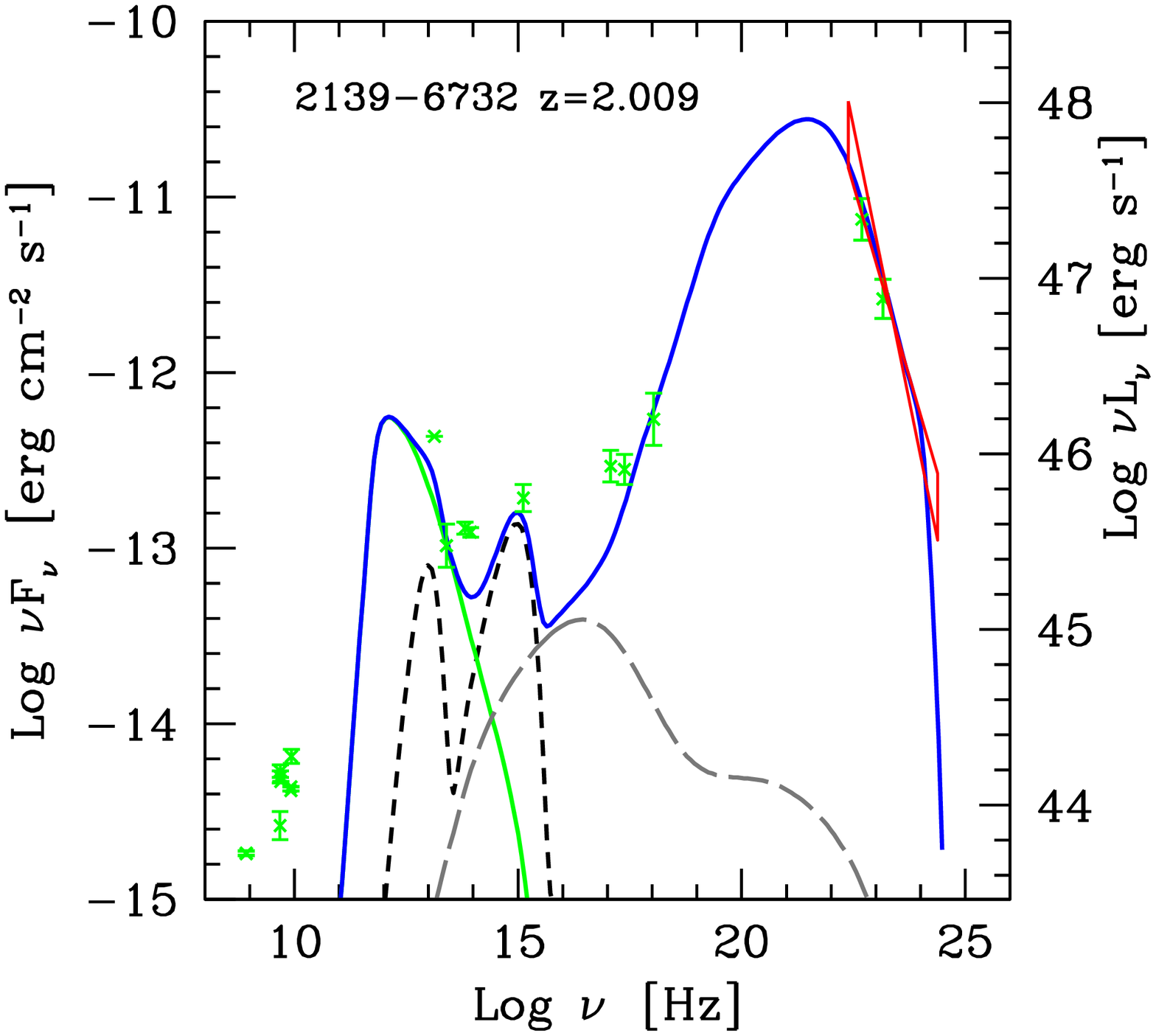,width=4.3cm,height=3.7cm } 
&\psfig{file=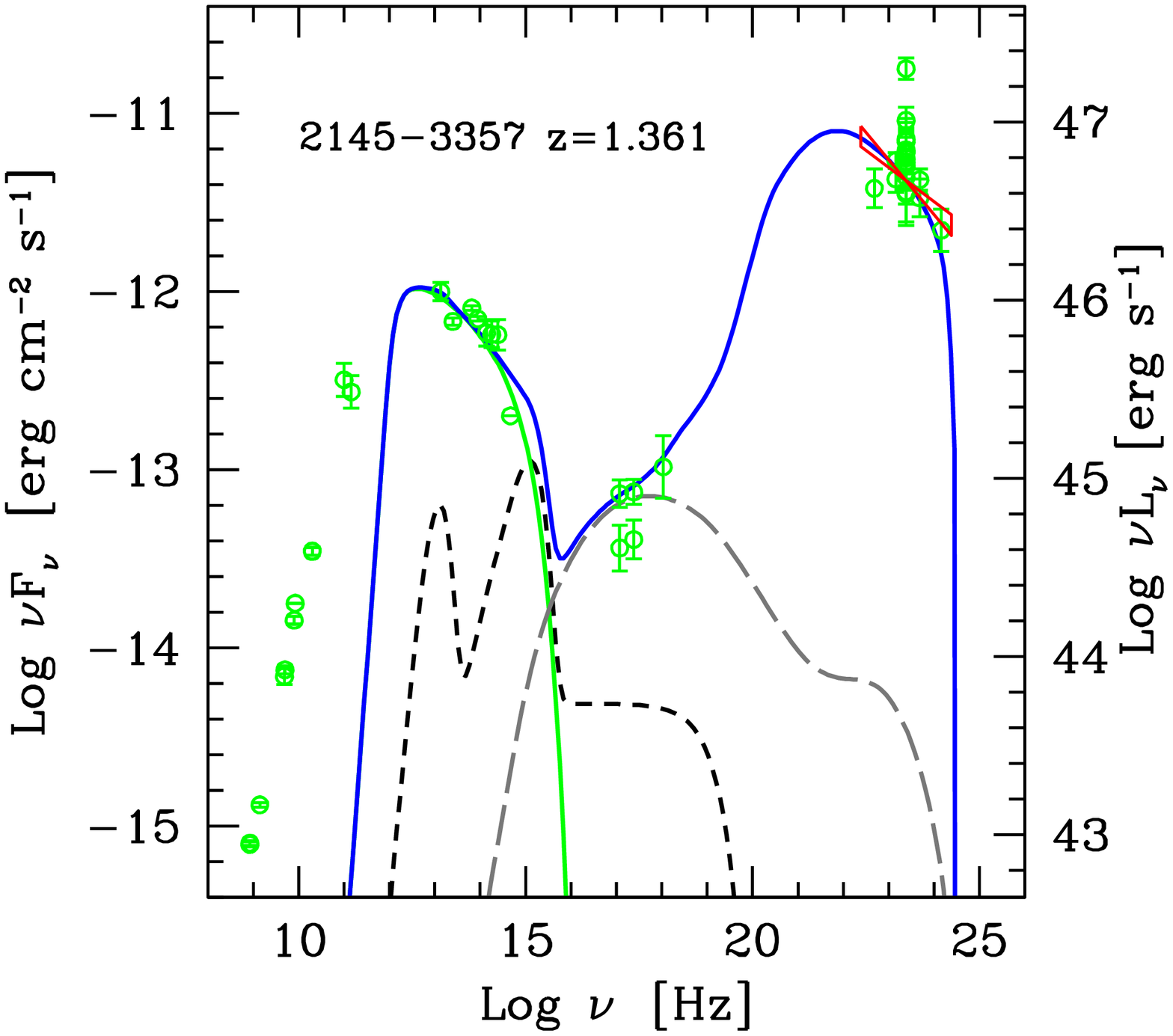,width=4.3cm,height=3.7cm } \\
\psfig{file=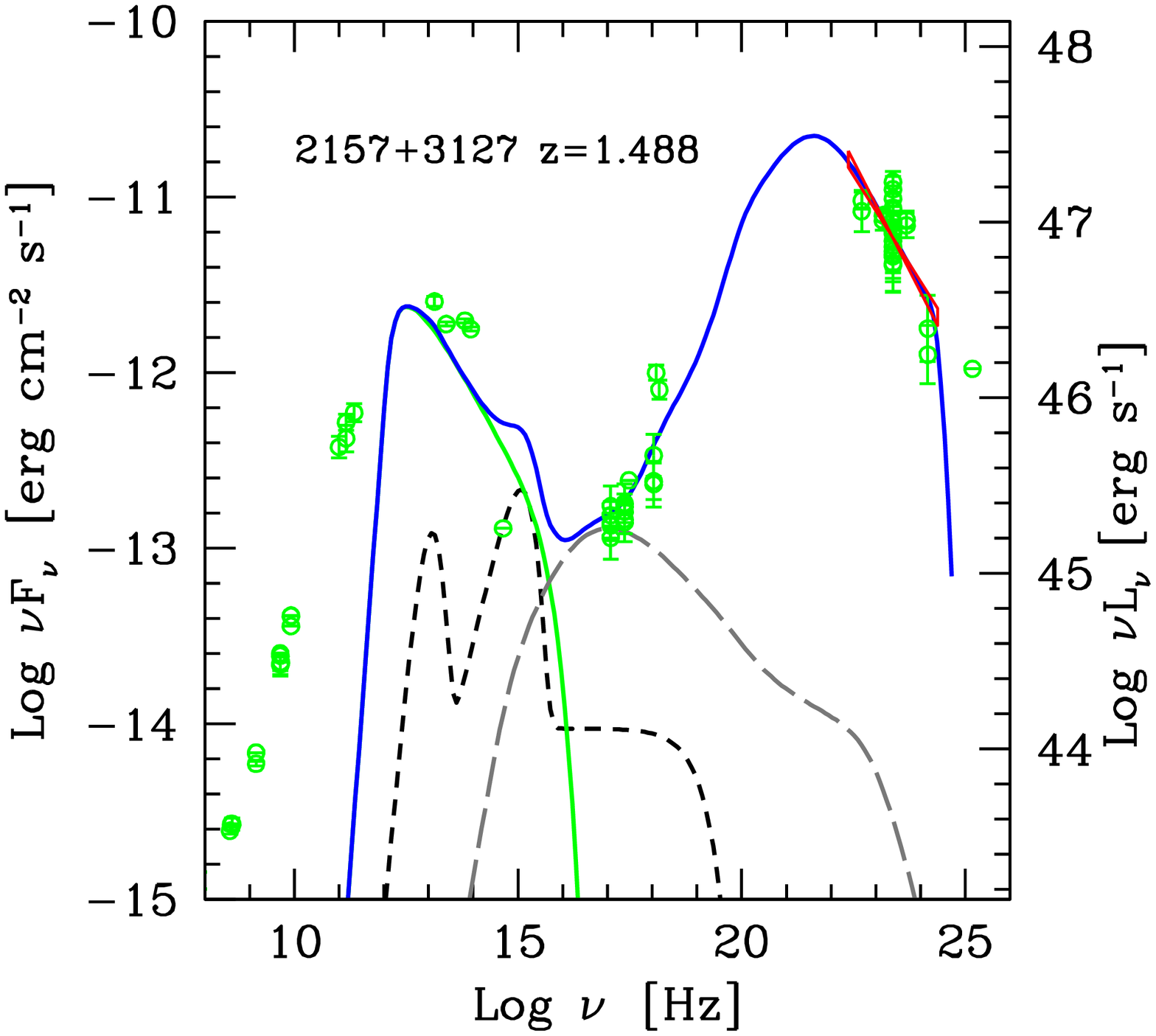,width=4.3cm,height=3.7cm } 
&\psfig{file=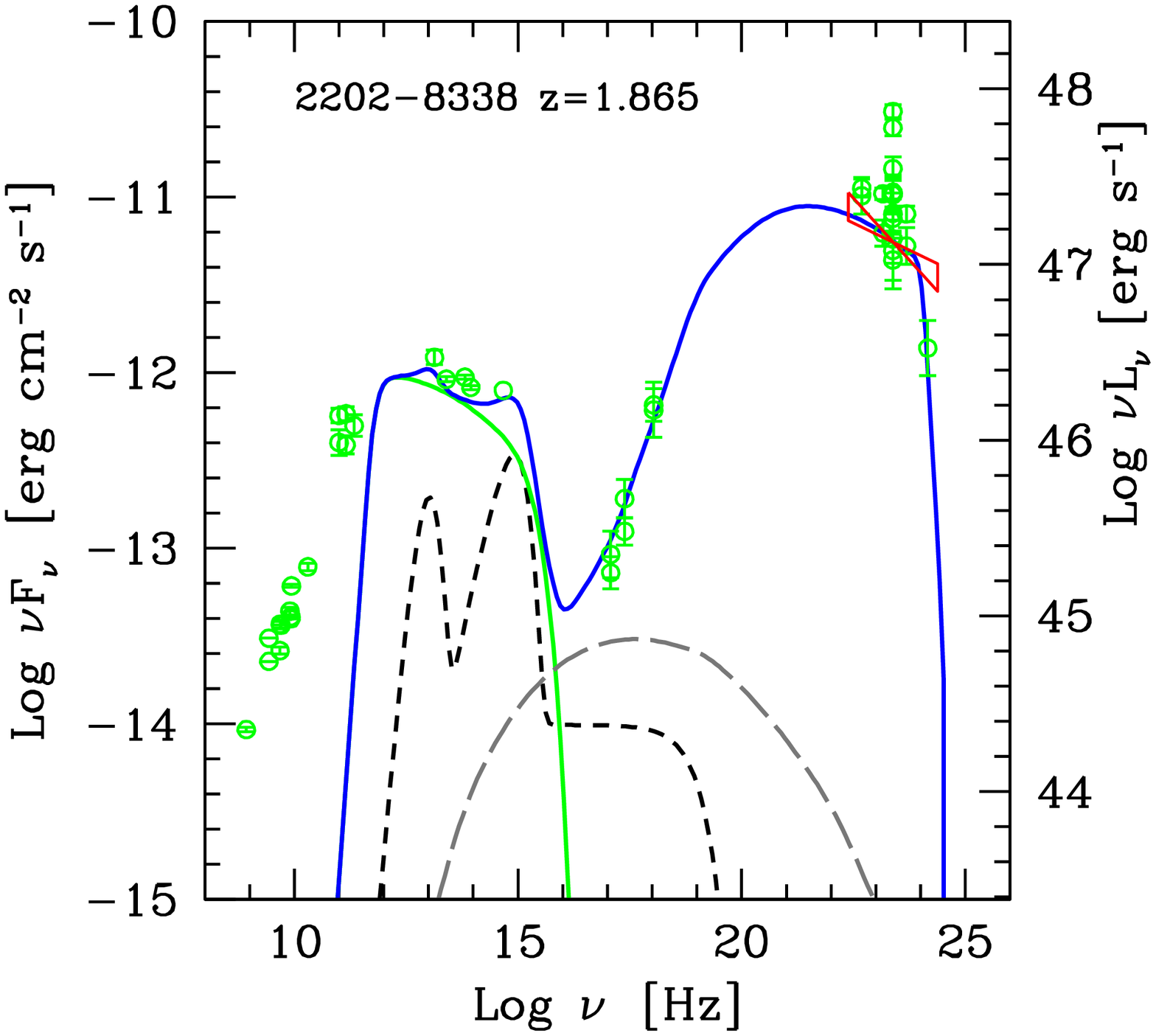,width=4.3cm,height=3.7cm } 
&\psfig{file=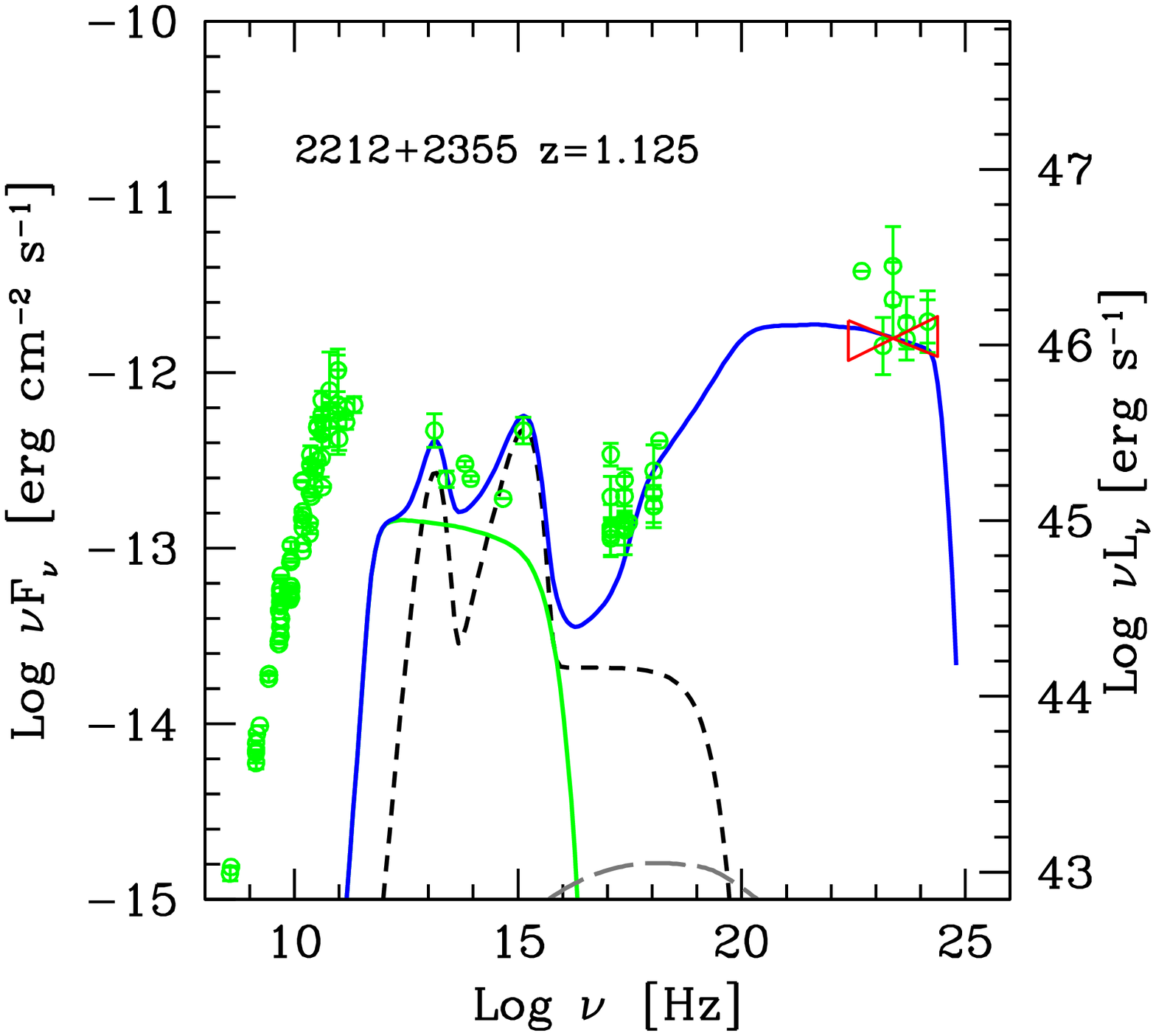,width=4.3cm,height=3.7cm }  
&\psfig{file=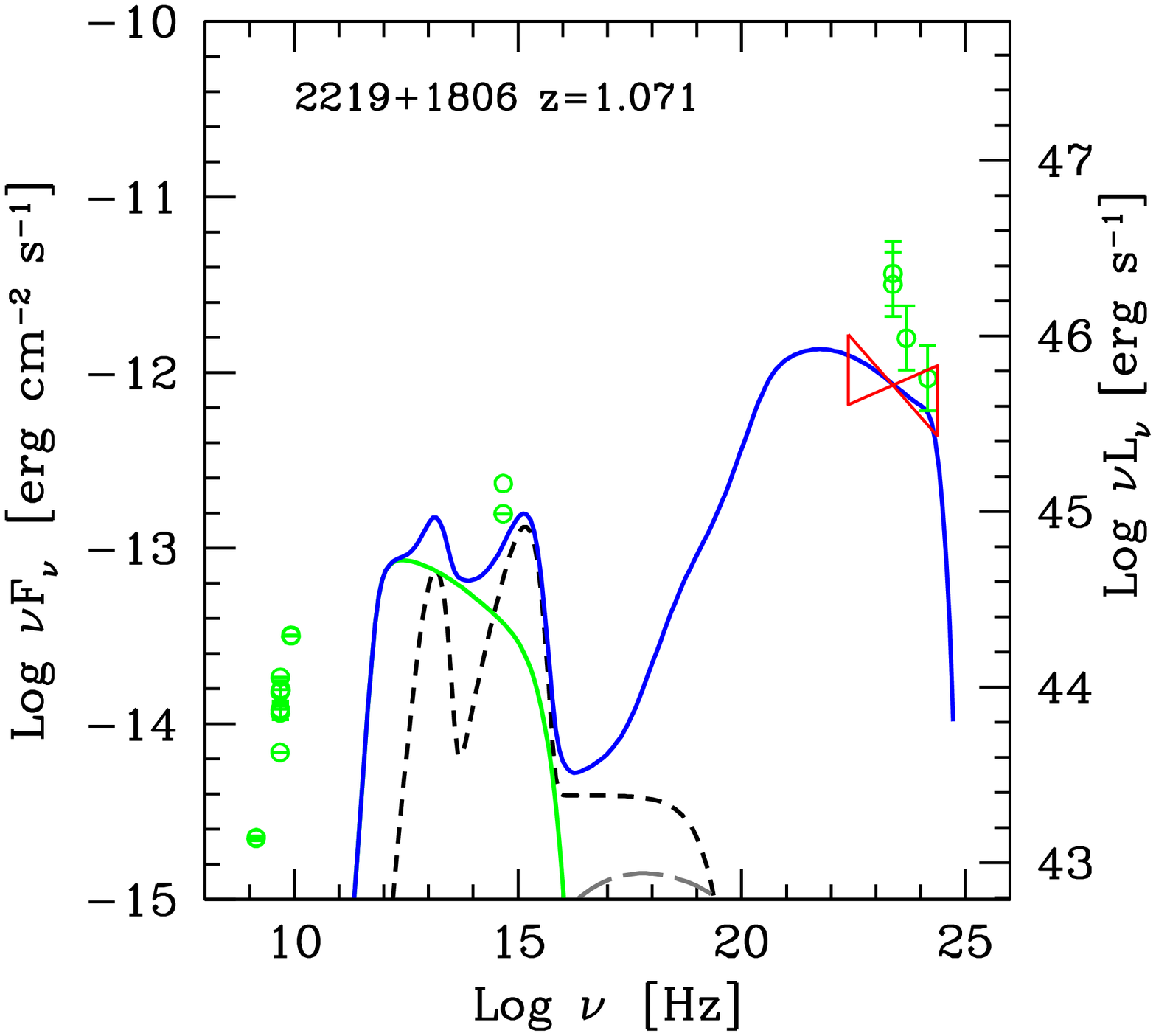,width=4.3cm,height=3.7cm } \\
\psfig{file=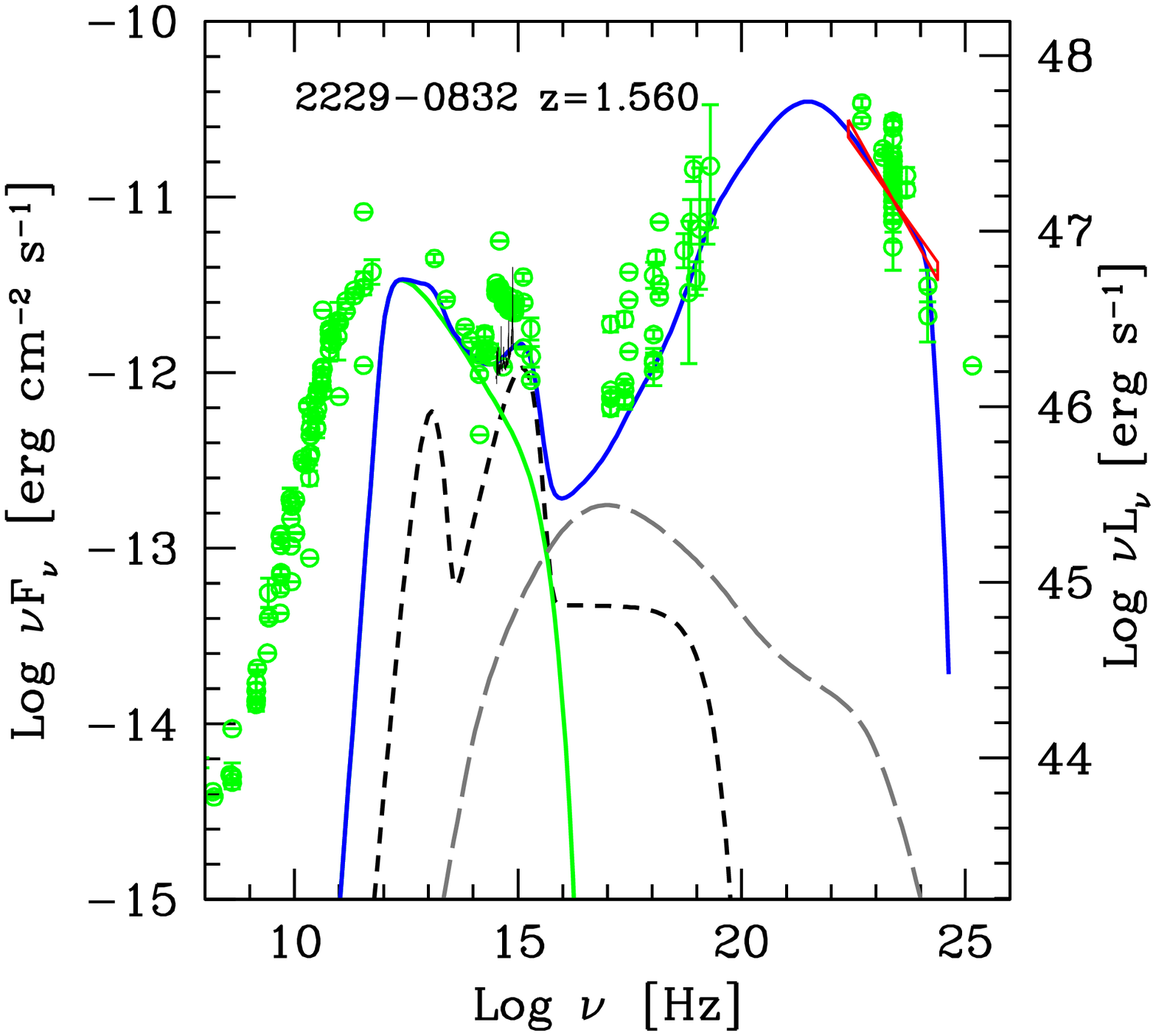,width=4.3cm,height=3.7cm } 
&\psfig{file=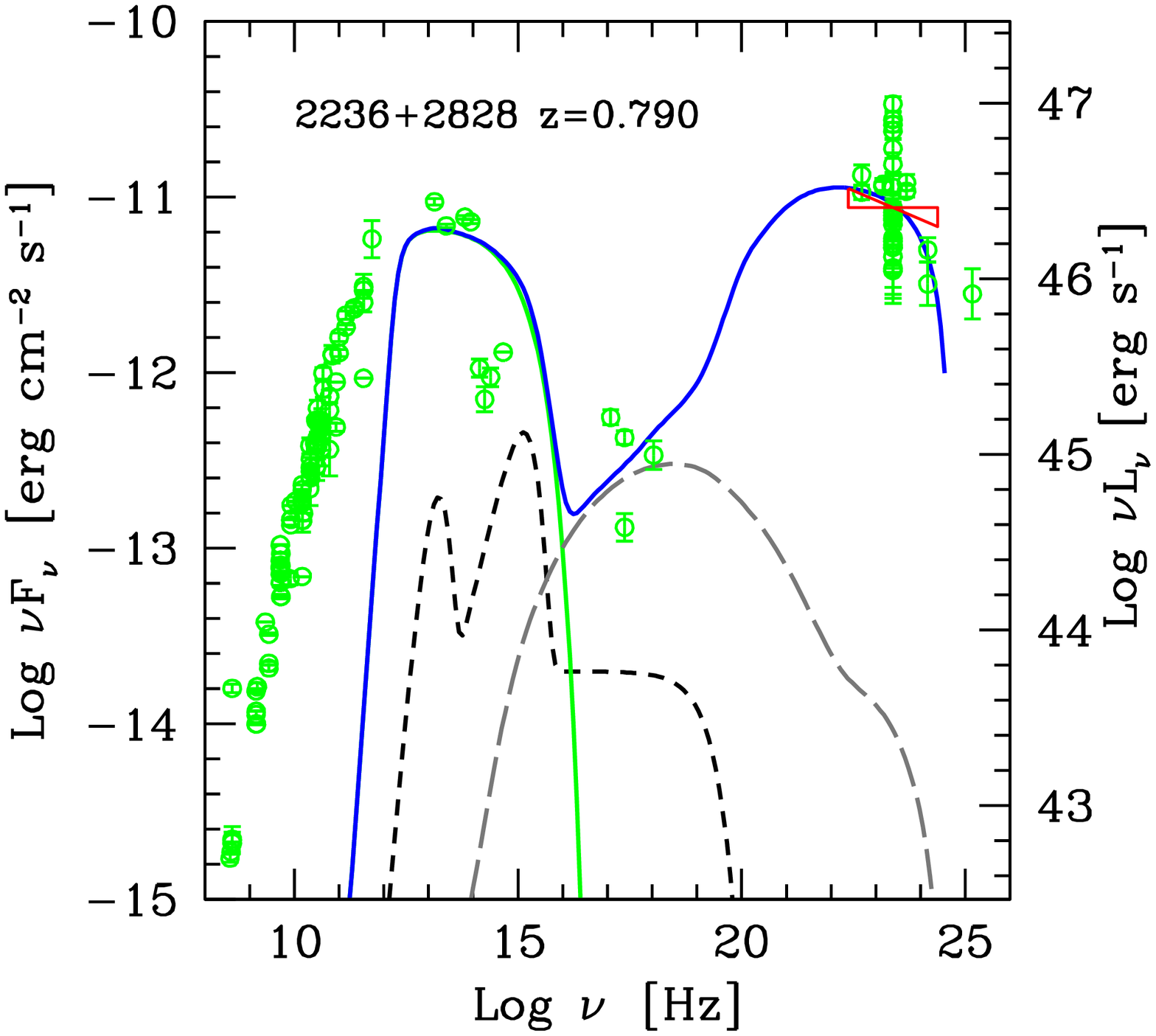,width=4.3cm,height=3.7cm } 
&\psfig{file=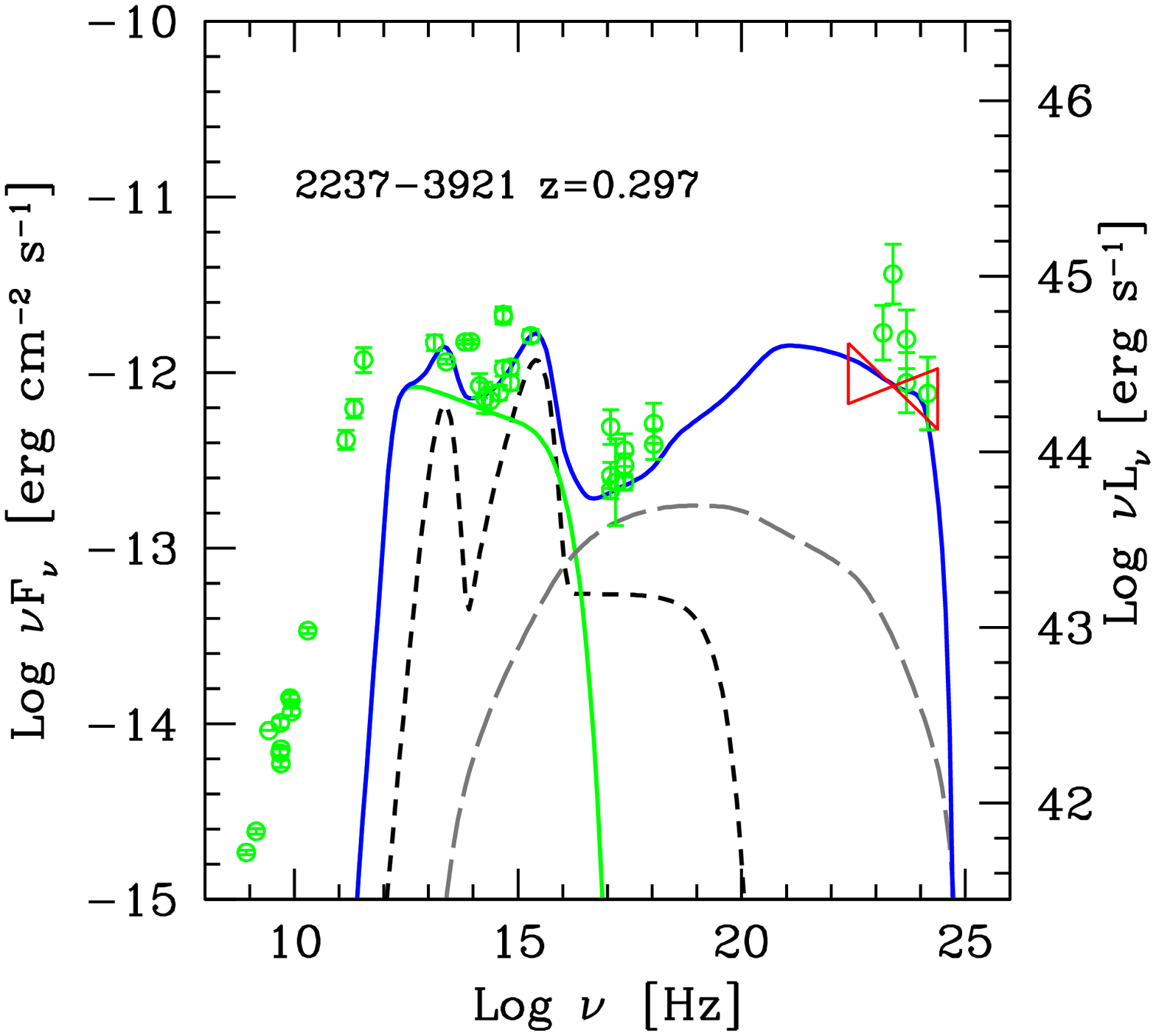,width=4.3cm,height=3.7cm } 
&\psfig{file=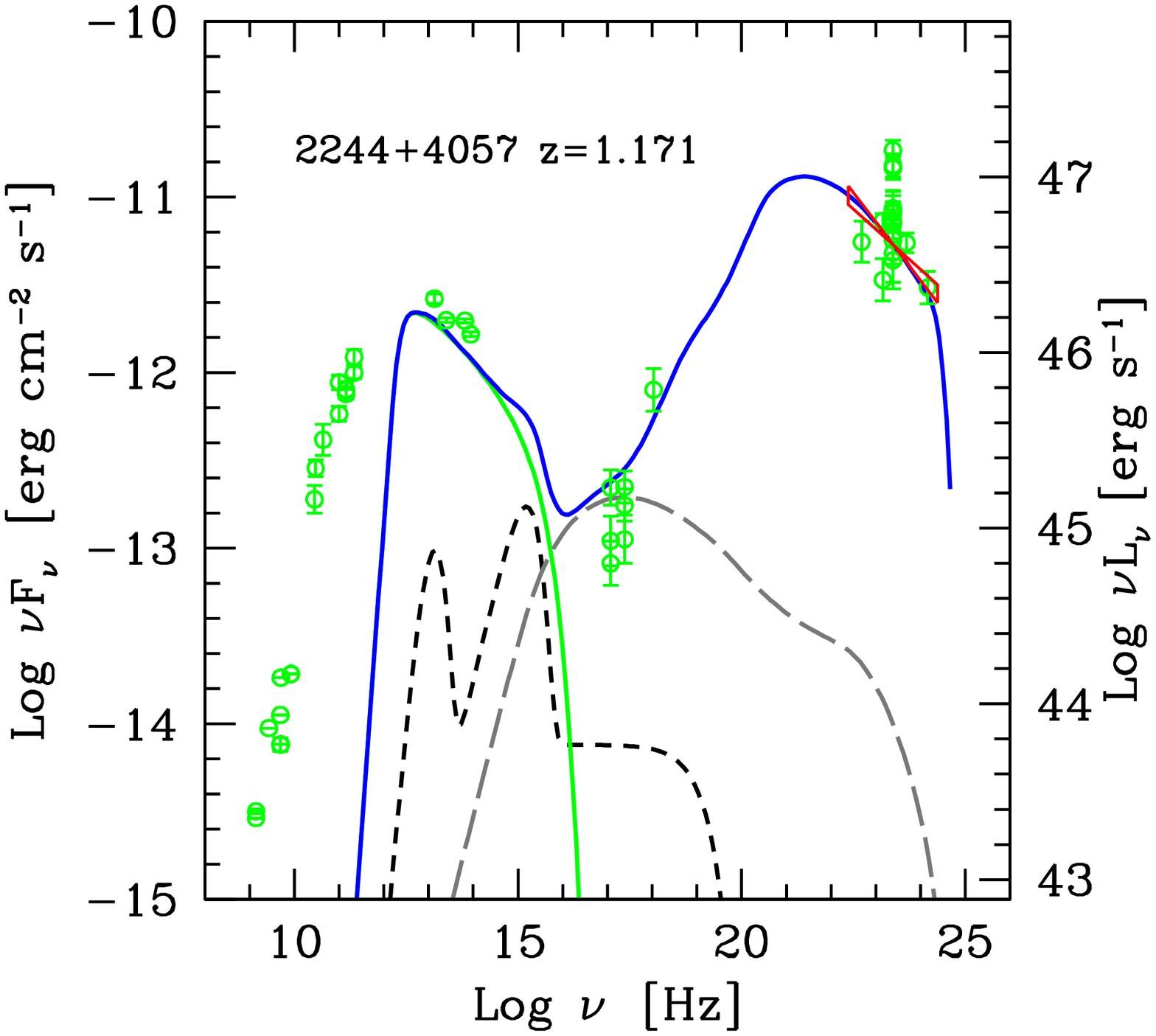,width=4.3cm,height=3.7cm } \\
\psfig{file=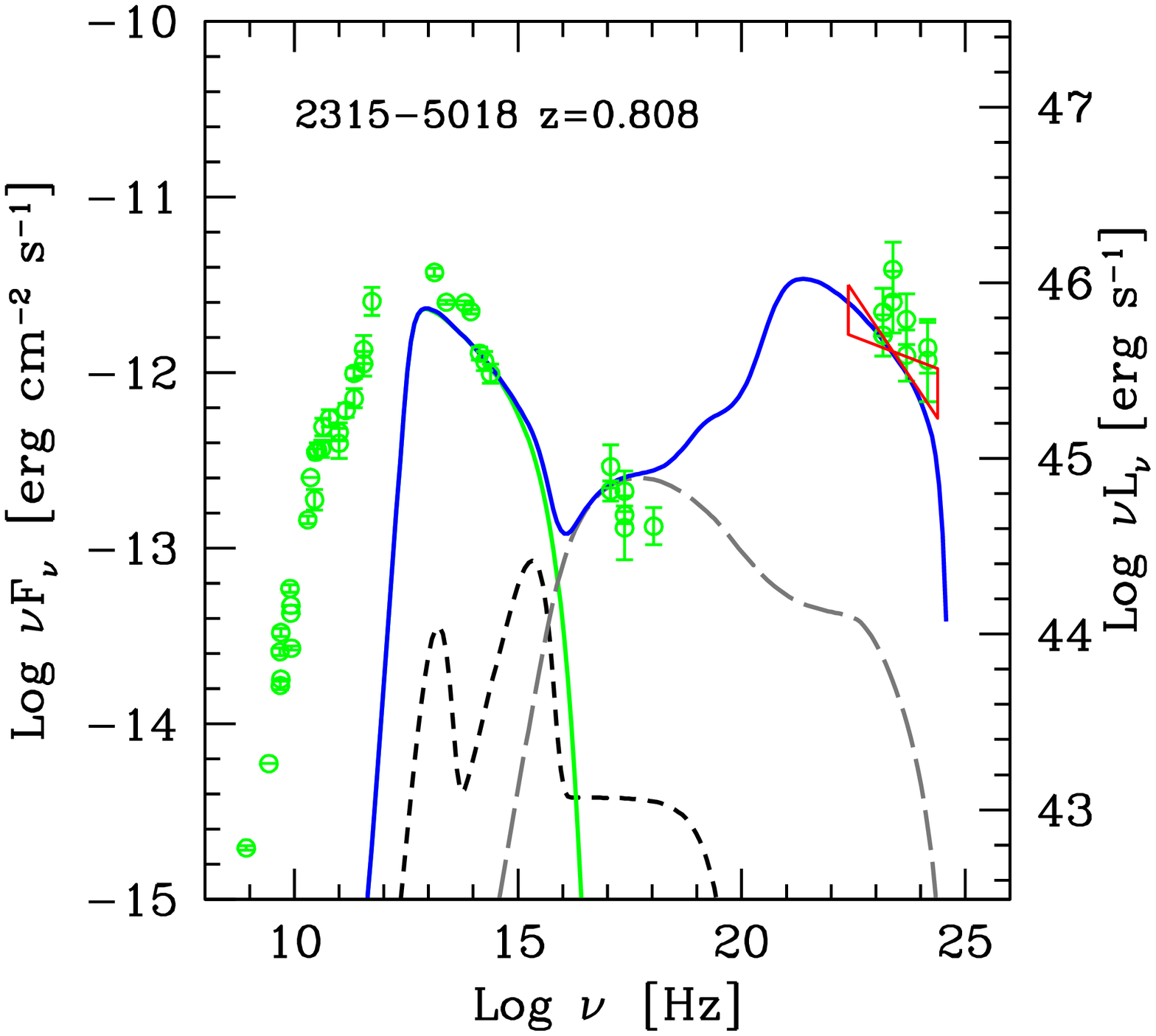,width=4.3cm,height=3.7cm } 
&\psfig{file=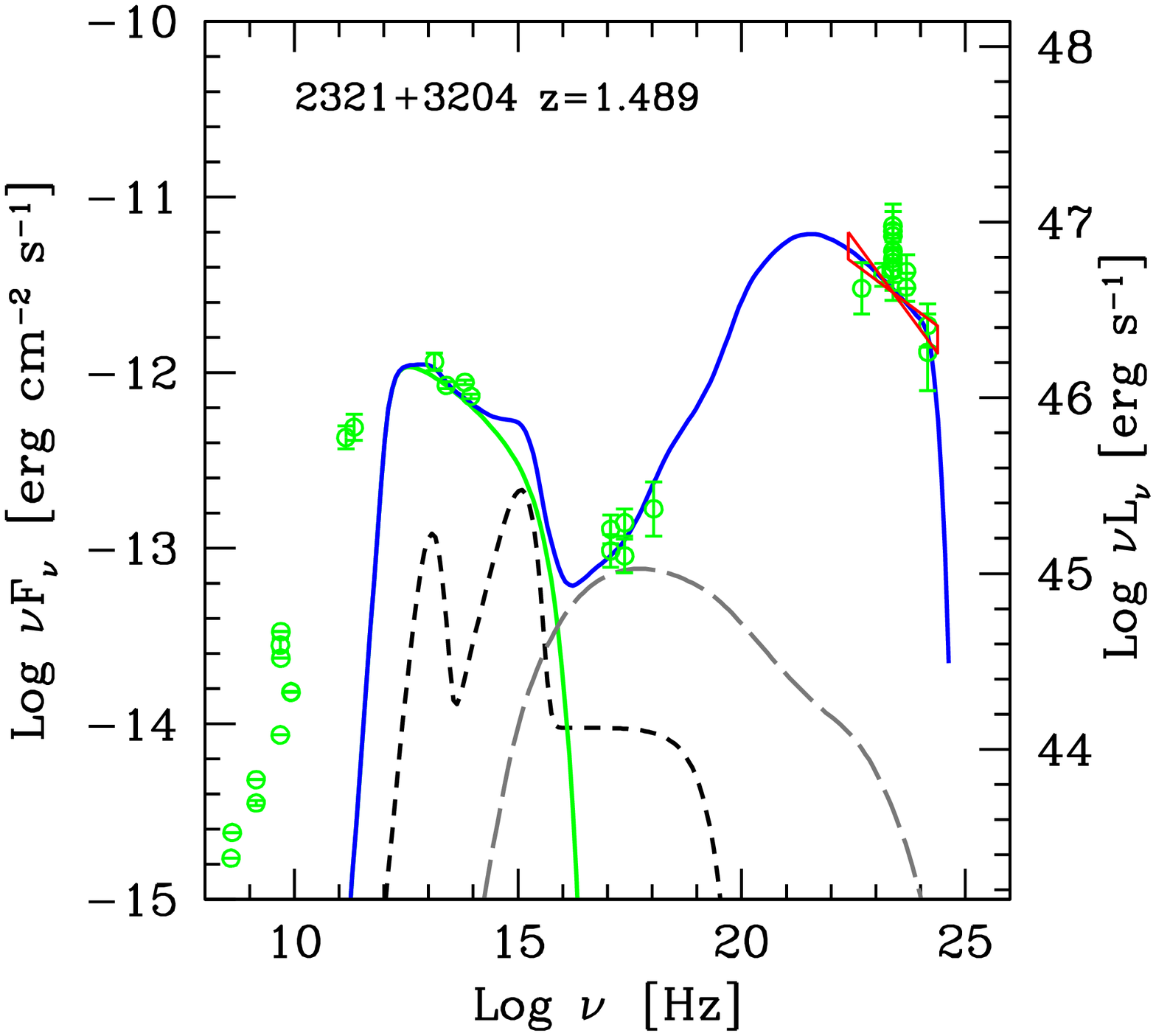,width=4.3cm,height=3.7cm } 
&\psfig{file=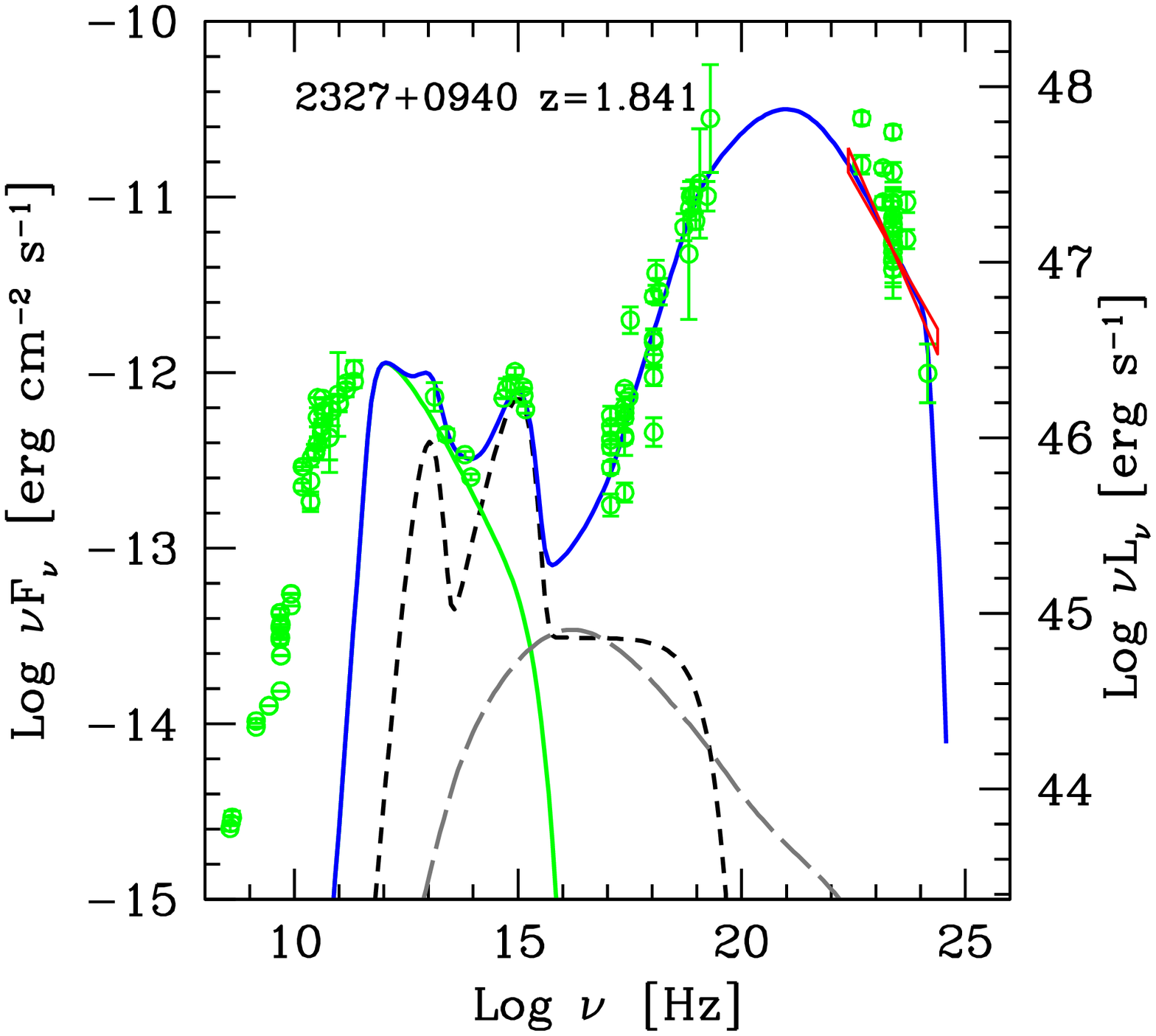,width=4.3cm,height=3.7cm }  
&\psfig{file=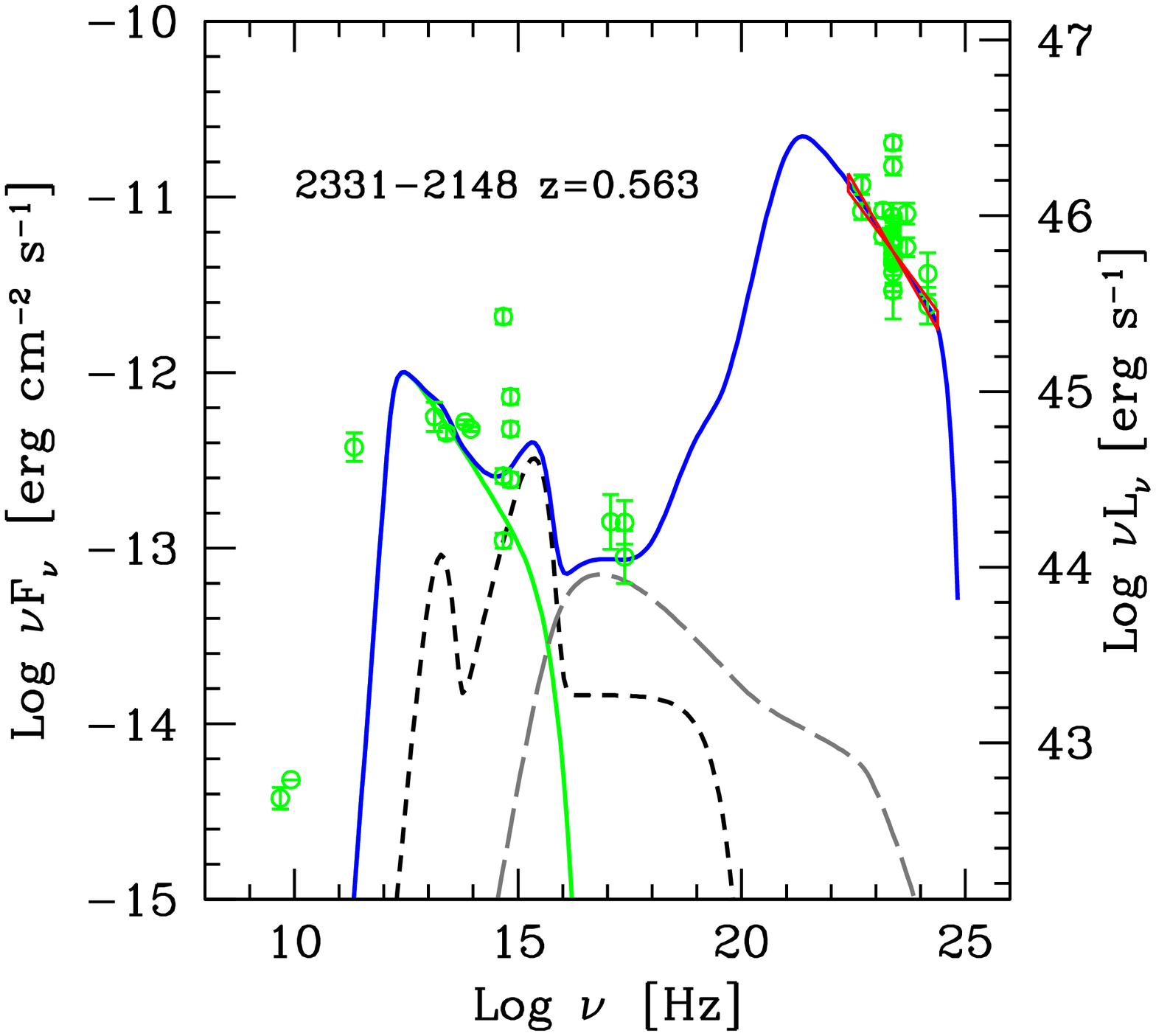,width=4.3cm,height=3.7cm } \\
\psfig{file=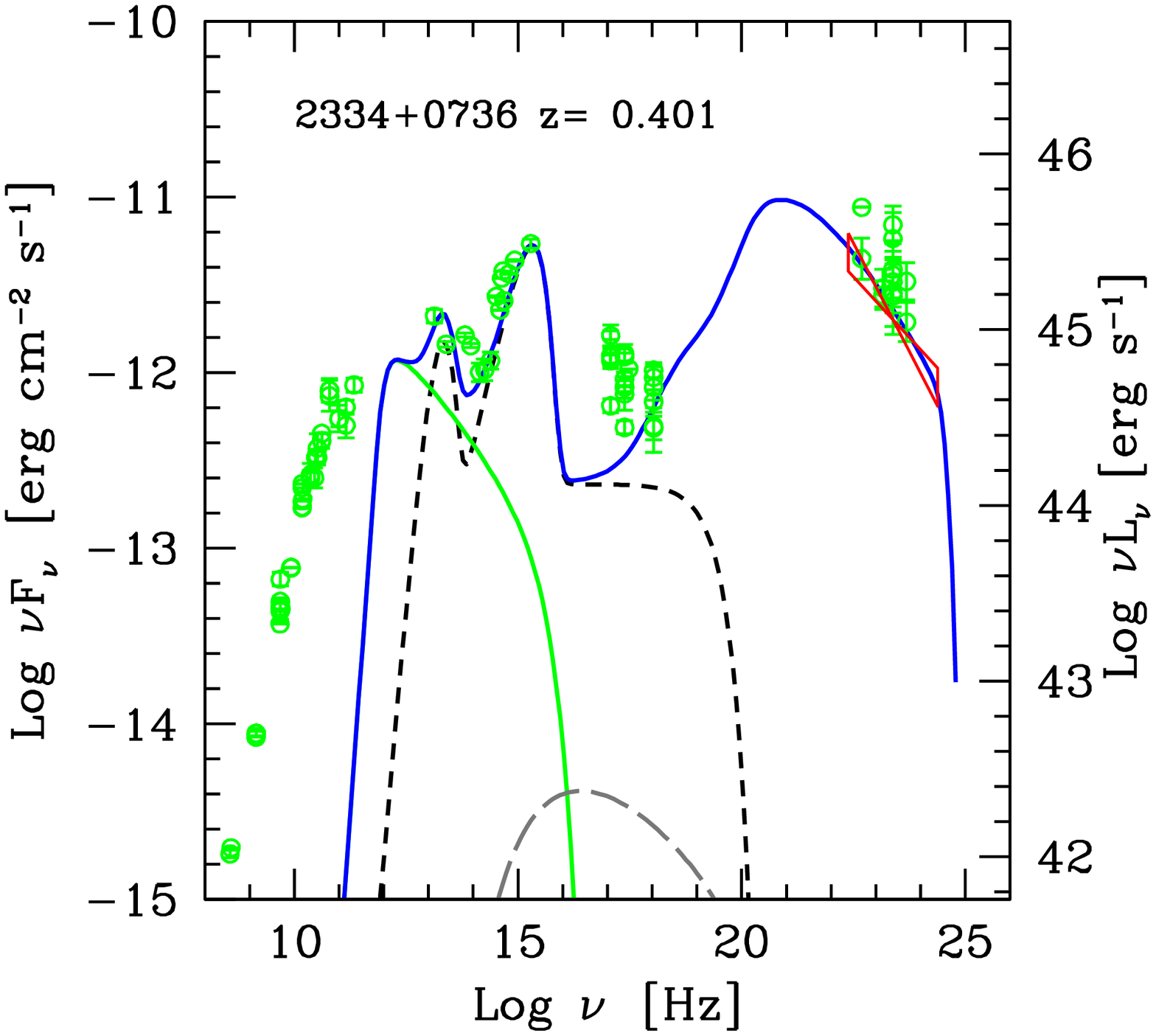,width=4.3cm,height=3.7cm }  
&\psfig{file=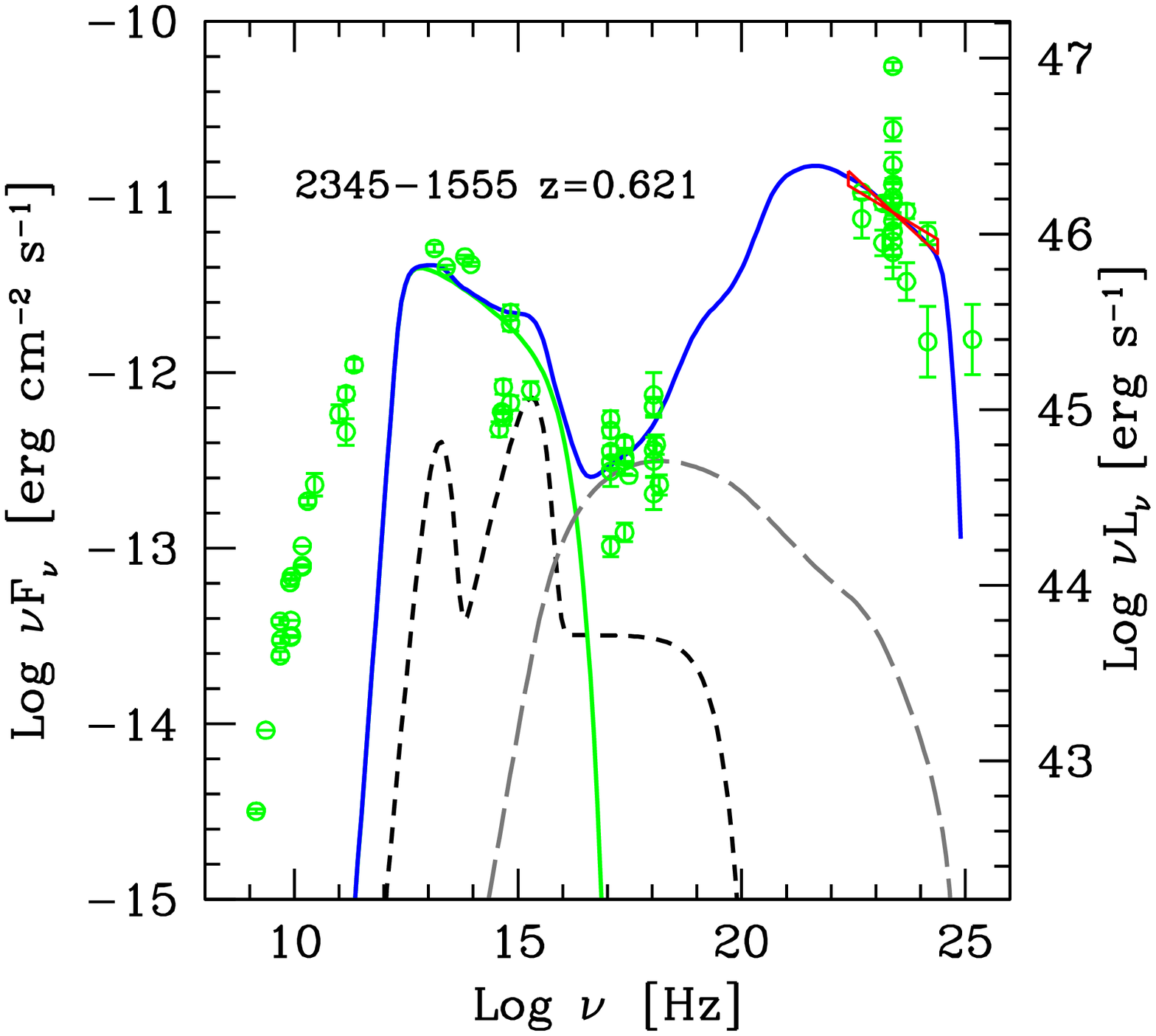,width=4.3cm,height=3.7cm } 
&\psfig{file=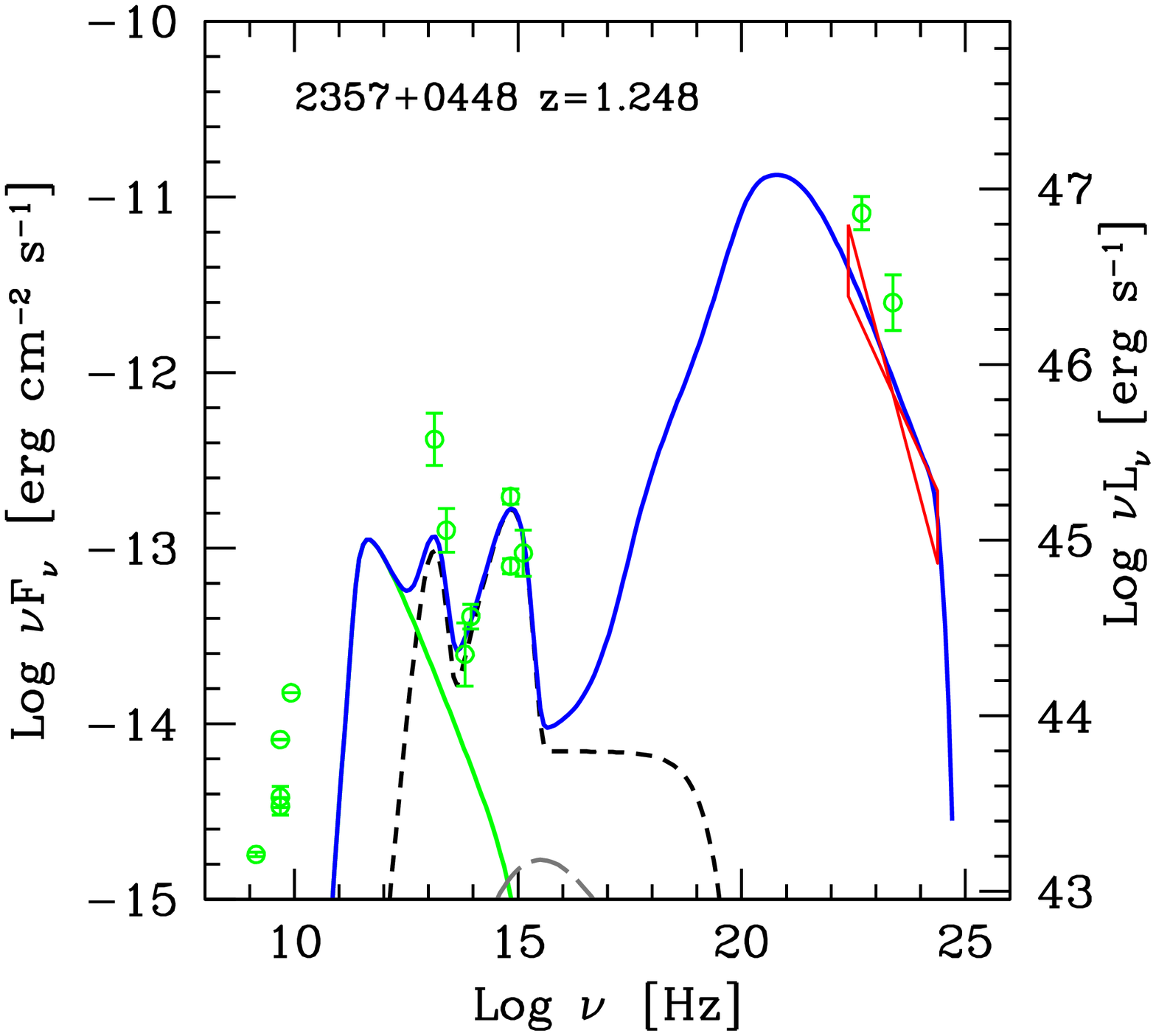,width=4.3cm,height=3.7cm }  
\end{tabular}
\caption{{\it continue.} SED of the FSRQs studied in this paper.}
\end{figure*} 

\end{document}